\documentclass[11pt,letterpaper,english]{article}

\usepackage[affil-it]{authblk} 
\usepackage{lmodern}

\usepackage{bm}
\usepackage{amsmath}
\usepackage[utf8]{inputenc}
\usepackage[T1]{fontenc} 
\usepackage{txfonts}     

\usepackage[pagestyles]{titlesec} 
\usepackage{fancyhdr} 
\usepackage{slashed}
\usepackage[top=.9in, bottom=.9in, left=.9in, right=.9in] {geometry}
\usepackage{graphicx}

\usepackage{enumitem}  

\usepackage[svgnames,x11names,table]{xcolor}
\usepackage[colorlinks=true,linkcolor=MediumBlue,citecolor=MediumBlue,urlcolor=MediumBlue]{hyperref}
\usepackage{subfigure}
\usepackage{array}
\newcolumntype{L}[1]{>{\arraybackslash}m{#1}}
\newcolumntype{C}[1]{>{\centering\arraybackslash}m{#1}}

\usepackage{pdfpages}
\usepackage{ifpdf}
\usepackage{outlines}
\usepackage[square,numbers,sort&compress]{natbib}
\usepackage{microtype}

\usepackage{pgfgantt}
\usepackage{booktabs} 
\RequirePackage{lineno}

\graphicspath{{.}{figures/}} 

\titlespacing{\section}{0pt}{20pt plus 2pt minus 2pt}{4pt plus 2pt minus 2pt}
\titlespacing{\subsection}{0pt}{20pt plus 2pt minus 2pt}{4pt plus 2pt minus 2pt}
\titlespacing{\subsubsection}{0pt}{4pt plus 0pt minus 0pt}{0pt plus 0pt minus 0pt}

\usepackage[normalem]{ulem}

\newcommand{\epb}{^{208}{\rm Pb}}
\newcommand{\eca}{^{48}{\rm Ca}}

\newcommand{\ep}{\textit{e}+\textit{p}}

\newcommand{\eA}{\mbox{\textit{e}+A}}
\newcommand{\dAu}{\textit{d}+Au}
\newcommand{\pAu}{\textit{p}+Au}
\newcommand{\pPb}{\textit{p}+Pb}
\newcommand{\pA}{\textit{p}+A}

\newcommand{\pp}{\mbox{\textit{p}+\textit{p}}}

\renewcommand{\AA}{A+A}

\newcommand{\epic}{\mbox{ePIC}}

\def\fig#1{{Fig.~\ref{#1}}}

\newcommand{\subsubsubsection}[1]{\medskip\noindent{\bf #1 }}

\newcommand{\subsubsubsubsection}[1]{\medskip\noindent{\it\underline{#1 }}}

\begin{document}

\title{The Present and Future of QCD 
\linebreak \linebreak\large{QCD Town Meeting White Paper -- An Input to the 2023 NSAC Long Range Plan}}

\date{March 1, 2023}

\renewcommand\Affilfont{\small}
\renewcommand\Authfont{\small}

\renewcommand\Affilfont{\fontsize{9}{10.5}\itshape}

\newcommand{\JLab}{1}
\newcommand{\VT}{2}
\newcommand{\GWU}{3}
\newcommand{\HISKP}{4}
\newcommand{\UMichAA}{5}
\newcommand{\YarmoukU}{6}
\newcommand{\KU}{7}
\newcommand{\UIUC}{8}
\newcommand{\ODU}{9}
\newcommand{\UniZG}{10}
\newcommand{\ANL}{11}
\newcommand{\UCR}{12}
\newcommand{\LBNL}{13}
\newcommand{\UNCW}{14}
\newcommand{\AANL}{15}
\newcommand{\BNL}{16}
\newcommand{\TempleU}{17}
\newcommand{\WilliamMary}{18}
\newcommand{\UVA}{19}
\newcommand{\HUJI}{20}
\newcommand{\UNC}{21}
\newcommand{\INFNGenova}{22}
\newcommand{\Rice}{23}
\newcommand{\BUAP}{24}
\newcommand{\YaleU}{25}
\newcommand{\IITM}{26}
\newcommand{\INFNCatania}{27}
\newcommand{\UHouston}{28}
\newcommand{\Duquesne}{29}
\newcommand{\IFUFRGS}{30}
\newcommand{\SPRACEUNESP}{31}
\newcommand{\SBU}{32}
\newcommand{\RBRC}{33}
\newcommand{\MSSTATE}{34}
\newcommand{\UCincinnati}{35}
\newcommand{\OSU}{36}
\newcommand{\CNU}       {37}
\newcommand{\UCBerkeley}{38}
\newcommand{\SLAC}      {39}
\newcommand{\MIT}       {40}
\newcommand{\Syracuse}  {41}
\newcommand{\AdiyamanU} {42}
\newcommand{\IstanbulU} {43}
\newcommand{\GSU}       {44}
\newcommand{\LamarU}    {45}
\newcommand{\FIU}       {46}
\newcommand{\UCagliari}     {47}
\newcommand{\INFNCagliari}  {48}
\newcommand{\LANL}          {49}
\newcommand{\UMD}           {50}
\newcommand{\ORNL}          {51}
\newcommand{\IAIFI}         {52}
\newcommand{\HU}            {53}
\newcommand{\FSU}           {54}
\newcommand{\UConn}         {55}
\newcommand{\ACU}           {56}
\newcommand{\BaruchC}       {57}
\newcommand{\IJCLab}        {58}
\newcommand{\SouthernUatNO} {59}
\newcommand{\NMSU}          {60}
\newcommand{\UTsukuba}      {61}
\newcommand{\UIC}           {62}
\newcommand{\UKY}           {63}
\newcommand{\UTennKnoxville}{64}
\newcommand{\ITA}           {65}
\newcommand{\TAMU}          {66}
\newcommand{\PennSUBerks}   {67}
\newcommand{\Duke}          {68}
\newcommand{\CUA}           {69}
\newcommand{\Rutgers}       {70}
\newcommand{\UnionC}        {71}
\newcommand{\USC}           {72}
\newcommand{\VanderbiltU}   {73}
\newcommand{\CUNY}          {74}
\newcommand{\UTFSM}         {75}
\newcommand{\Longwood}  {76}
\newcommand{\KoreaU}    {77}
\newcommand{\TUNL}      {78}
\newcommand{\UCLA}      {79}
\newcommand{\IowaStateU}{80}
\newcommand{\URegina}   {81}
\newcommand{\SINPMSU}   {82}
\newcommand{\NCSU}      {83}
\newcommand{\VUU}       {84}
\newcommand{\IndianaU}  {85}
\newcommand{\ChonnamU}  {86} 
\newcommand{\SJU}       {87}
\newcommand{\OhioU}     {88}
\newcommand{\CUBoulder} {89}
\newcommand{\UMassAmherst}{90}
\newcommand{\MSU}       {91}
\newcommand{\SDU}       {92}
\newcommand{\WayneStateU}{93}
\newcommand{\CCNU}      {94}
\newcommand{\UG}        {95}
\newcommand{\CMU}       {96}
\newcommand{\NISER}     {97}
\newcommand{\UNISON}    {98}
\newcommand{\SMU}       {99}
\newcommand{\VAMilitaryInst}{100}
\newcommand{\CFNS}      {101}
\newcommand{\PaviaU}    {102}
\newcommand{\INFNPavia} {103}
\newcommand{\LVC}       {104}
\newcommand{\Lehigh}    {105}
\newcommand{\INFNFrascati}{106}
\newcommand{\PSI}       {107}
\newcommand{\NSU}       {108}
\newcommand{\INFNRome}  {109}
\newcommand{\UNH}       {110}
\newcommand{\KEK}       {111}
\newcommand{\RIKEN}     {112}
\newcommand{\UWF}       {113}
\newcommand{\WashingtonU}{114}
\newcommand{\CERN}      {115}
\newcommand{\PSU}       {116}
\newcommand{\PUWL}      {117}
\newcommand{\KSU}       {118}
\newcommand{\FutureUE}  {119}
\newcommand{\UTAustin}  {120}
\newcommand{\UTuebingen}{121}
\newcommand{\LLNL}      {122}
\newcommand{\UCDavis}   {123}
\newcommand{\Canisius}  {124}
\newcommand{\CUHKSZ}    {125}
\newcommand{\THU}       {126}
\newcommand{\ColumbiaU} {127}

\author[\JLab]{P.~Achenbach}
\author[\VT]{D.~Adhikari}
\author[\GWU]{A.~Afanasev}
\author[\HISKP]{F.~Afzal}
\author[\UMichAA]{C.A.~Aidala}
\author[\YarmoukU,\KU]{A.~Al-bataineh}
\author[\UIUC]{D.K.~Almaalol}
\author[\ODU]{M.~Amaryan}
\author[\UniZG]{D.~Androić}
\author[\ANL]{W.R.~Armstrong}
% ...

\maketitle

$^\dagger = $ QCD Town Meeting Conveners 

$^\star = $ QCD Town Meeting Organizing Committee 

\newpage
\tableofcontents

\newpage
\newpage
\section{Executive Summary}\label{sec:exe_sum}

\begin{figure}[!hb]
\centering
\includegraphics[width=0.9\textwidth]{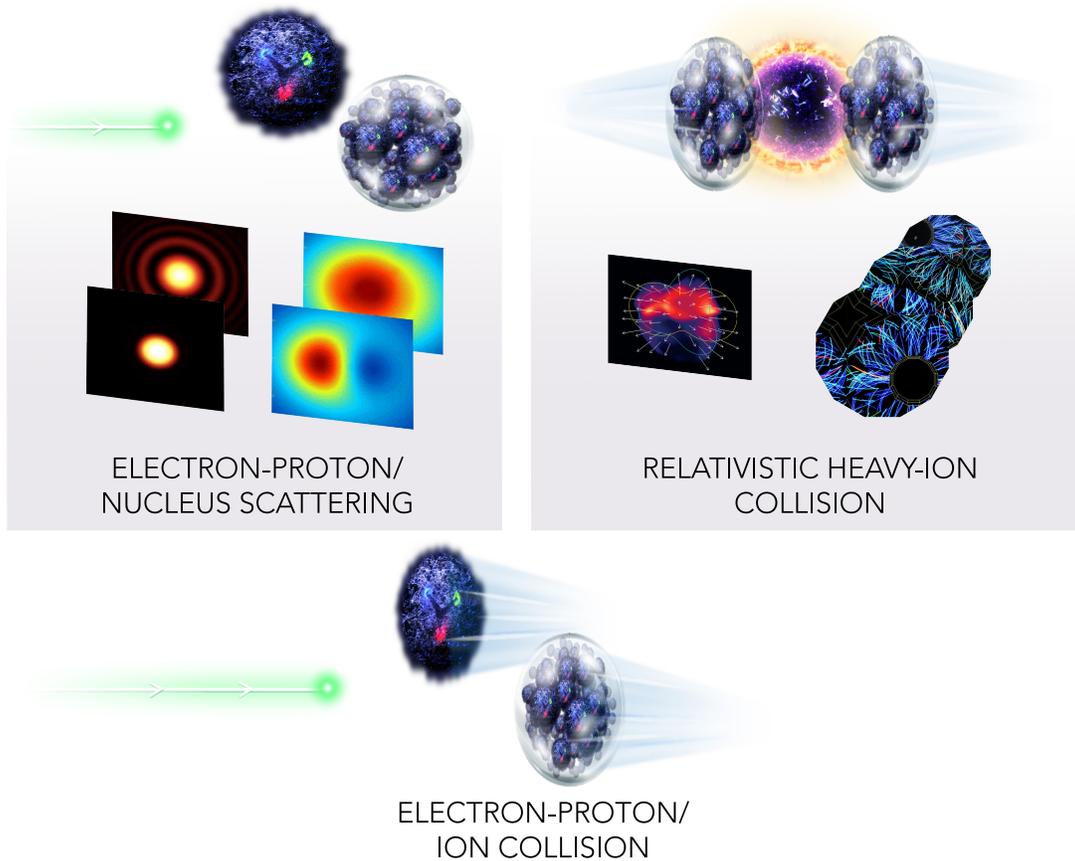}
\caption{Experimental methods to study Cold and Hot QCD using electron-nucleon (nucleus) scattering (top left) and heavy-ion collisions (top right), respectively. The Electron-Ion Collider (bottom), to be realized within the next LRP period, will bring new, exciting experimental programs to QCD research. 
}\label{fig:qcd_processes}
\end{figure}

It is currently understood that there are four fundamental forces in nature: gravitational, electromagnetic, weak and strong forces.  
The strong force governs the interactions between quarks and gluons, elementary particles whose interactions give rise to the vast majority of visible mass in the universe.
The mathematical description of the strong force is provided by the non-Abelian gauge theory Quantum Chromodynamics (QCD). While QCD is an exquisite theory, constructing the nucleons and nuclei from quarks, and furthermore explaining the behavior of quarks and gluons at all energies, remain to be complex and challenging problems. Such challenges, along with the desire to understand all visible matter at the most fundamental level, position the study of QCD as a central thrust of research in nuclear science. 
Experimental insight into the strong force can be gained using large particle accelerator facilities, which are necessary to probe the very short distance scales over which quarks and gluons interact. 
The Long Range Plans (LRPs) exercise of 1989 and 1996 led directly to the construction of two world-class facilities: the Continuous Electron Beam Accelerator Facility (CEBAF) at Jefferson Lab (JLab) that is focused on studying how the structure of hadrons emerges from QCD (cold QCD research), and the Relativistic Heavy Ion Collider (RHIC) at Brookhaven National Lab (BNL) that aims at the discovery and study of a new state of matter, the quark-gluon plasma (QGP), at extremely high temperatures (hot QCD research). 
The different collision systems used to access the incredibly rich field of hot and cold QCD in the laboratory are illustrated in Fig.~\ref{fig:qcd_processes}.

These past investments have produced major advances. 
Nucleons and nuclei are being studied with increasing precision with a unified description of the partonic structure utilizing multi-dimensional imaging. Significant progress has been made, paving the way towards a complete picture of how quarks and gluons give rise to the mass, spin, and momentum of the nucleon. 
In hot QCD, the QGP is created in the collisions of nuclei at RHIC and the Large Hadron Collider (LHC) and is observed to behave like a fluid with very low
specific shear viscosity; the current goals are to understand how the fluid behavior emerges from 
QCD and to characterize the temperature (and chemical potential) dependence of the properties of the QGP. 
As this White Paper is written, current experimental programs at CEBAF, RHIC and the LHC continue to provide exciting near term opportunities to capitalize on the investments in experimental equipment and accelerator operations. 
Most importantly, the QCD community looks forward to the construction of the Electron Ion Collider (EIC) as a major new facility to push forward QCD research in the next decades, with significant focus on exploring the properties of gluons, the mediators of the strong force.

\subsection{QCD Community Input}
This White Paper presents the community inputs and scientific conclusions from the Hot and Cold QCD Town Meeting that took place September 23-25, 2022 at MIT~\cite{MITTownMeeting}, as part of the Nuclear Science Advisory Committee (NSAC) 2023 Long Range Planning process. A total of 424 physicists registered for the meeting. The meeting agenda is included in the Appendix.
The meeting highlighted progress in QCD nuclear physics since the 2015 LRP (LRP15)~\cite{Geesaman:2015fha} and identified key questions and plausible paths to obtaining answers to those questions, defining priorities for our research over the coming decade. In defining the priority of outstanding physics opportunities for the future, both prospects for the short ($\sim$ 5 years) and longer term (5-10 years and beyond) are identified together with the facilities, personnel and other resources needed to maximize the discovery potential and maintain United States leadership in QCD physics worldwide. 

We would like to note in preparation for this white paper, numerous excellent white papers were prepared by members of our community. We drew upon these documents wherever appropriate. 
This White Paper is organized as follows: 
In the next part of this Executive Summary, we detail the Recommendations and Initiatives that were presented and discussed at the Town Meeting, and their supporting rationales. A survey was sent to all town meeting participants upon conclusion of the discussion to solicit community input. A total of 342 community members completed the survey, and the results are included here. 
Section~\ref{sec:progress} highlights major progress and accomplishments of the past seven years. It is followed, in Section~\ref{sec:future}, by an overview of the physics opportunities for the immediate future, and in relation with the next QCD frontier: the EIC. Section~\ref{sec:future_eic} provides an overview of the physics motivations and goals associated with the EIC. Section~\ref{sec:workforce} is devoted to the workforce development and support of diversity, equity and inclusion. This is followed by a dedicated section on computing in Section~\ref{sec:computing}. Section~\ref{sec:nucl_data} describes the national need for nuclear data science and the relevance to QCD research. 

\subsection{Recommendations}
We present the recommendations agreed on at the QCD Town 
Meeting along with the survey results to indicate the strength of community support for each recommendation.  The four recommendations listed here received
the consensus support of attendees at the QCD Town 
Meeting.

\bigskip
\noindent \textbf{Recommendation 1: Capitalizing on past investments} \hspace*{3cm}{\it{(Yes: 335; No: 3; No Answer: 4)}}

\medskip
\noindent
The highest priority for QCD research is to maintain U.S. world leadership in nuclear science for the next decade by capitalizing on past investments. Maintaining this leadership also requires recruitment and retention of a diverse and equitable workforce. 

\smallskip
\noindent{\bf We recommend support for a healthy base theory program, full operation of the CEBAF 12-GeV and RHIC facilities, and maintaining U.S. leadership within the LHC heavy-ion program, along with other running facilities, including the valuable university-based laboratories, and the scientists involved in all these efforts.}

\smallskip
\noindent This includes the following, unordered, programs:
\begin{itemize}
  \setlength\itemsep{-0.2em}
    \item The 12-GeV CEBAF hosts a forefront program of using electrons to unfold the quark and gluon structure of visible matter and probe the Standard Model. We recommend executing the CEBAF 12-GeV program at full capability and capitalizing on the full intensity potential of CEBAF by the construction and deployment of the Solenoidal Large Intensity Device (SoLID).
    \item The RHIC facility revolutionized our understanding of QCD, as well as the spin structure of the nucleon. To successfully conclude the RHIC science mission, it is essential to complete the sPHENIX science program as highlighted in the 2015 LRP, the concurrent STAR data taking with forward upgrade, and the full data analysis from all RHIC experiments.
    \item The LHC facility maintains leadership in the (heavy ion) energy frontier and hosts a program of using heavy-ion collisions to probe QCD at the highest temperature and/or energy scales. We recommend the support of continued U.S. leadership across the heavy ion LHC program.
    \item Theoretical nuclear physics is essential for establishing new scientific directions, and meeting the challenges and realizing the full scientific potential of current and future experiments. We recommend increased investment in the base program and expansion of topical programs in nuclear theory.
\end{itemize}

\medskip
\noindent \textbf{Recommendation 2: We recommend the expeditious completion of the EIC as the highest priority for facility construction.} \hspace*{8cm} {\it (Yes: 325; No: 10; No Answer: 7)}

\medskip
\noindent The Electron-Ion Collider (EIC) is a powerful and versatile new accelerator facility, capable of colliding high-energy beams ranging from heavy ions to polarized light ions and protons with high-energy polarized electron beams. In the 2015 Long Range Plan the EIC was put forward as the highest priority for new facility construction and the expeditious completion remains a top priority for the nuclear physics community. The EIC, accompanied by the general-purpose large-acceptance detector, \epic, will be a discovery machine that addresses fundamental questions such as the origin of mass and spin of the proton as well as probing dense gluon systems in nuclei. It will allow for the exploration of new landscapes in QCD, permitting the ``tomography", or high-resolution multidimensional mapping of the quark and gluon components inside of nucleons and nuclei. Realizing the EIC will keep the U.S. on the frontiers of nuclear physics and accelerator science technology.
\begin{itemize}
  \setlength\itemsep{-0.2em}
    \item Building on the recent EIC project CD-1 approval, the community-led Yellow-Report, and detector proposals, the QCD research community is committed to continue the development and timely realization of the EIC and its first detector, \epic. We recommend supporting the growth of a diverse and active research workforce for the \epic\  collaboration, in support of the expeditious realization of the first EIC detector.
    \item We recommend new investments to establish a national EIC theory alliance to enhance and broaden the theory community needed for advancing EIC science and the experimental program. This theory alliance will contribute to a diverse workforce through a competitive national EIC theory fellow program and tenure-track bridge positions, including appointments at minority serving institutions.
\end{itemize}

\medskip
\noindent
\textbf{Recommendation 3: Workforce and Conduct}  \hspace*{3.8cm} {\it (Yes: 296; No: 19; No Answer: 27)}

\medskip
\noindent
Increasing the U.S. QCD research workforce and participation of international collaborators is vital for the successful realization of the field’s science mission. In addition, the nuclear physics research program serves an important role in developing a diverse STEM workforce for the critical needs of the nation.  Creating and maintaining an equitable, productive working environment for all members of the community is a necessary part of this development.

\smallskip
\noindent\textbf{We recommend enhanced investment in the growth and development of a diverse, equitable workforce.}
\begin{itemize}
  \setlength\itemsep{-0.2em}
    \item Part of recruiting and maintaining a diverse workforce requires treating all community members with respect and dignity. Supporting the recent initiatives by the APS (American Physical Society) and DNP (Division of Nuclear Physics) to develop community-wide standards of conduct, we recommend that host labs and user facilities require the establishment and/or adoption of enforceable conduct standards by all of the experimental and theoretical collaborations they support. The enforcement of such standards is the combined responsibility of all laboratories, theoretical and experimental collaborations, conference organizers, and individual investigators supported by the nuclear physics research program.
    \item We recommend development and expansion of programs that enable participation in research by students from under-represented communities at National Labs and/or Research Universities, including extended support for researchers from minority-serving and non-PhD granting institutions.
    \item We recommend development and expansion of programs to recruit and retain diverse junior faculty and staff at universities and national laboratories through bridge positions, fellowships, traineeships, and other incentives. 
\end{itemize}

\medskip
\noindent\textbf{Recommendation 4: Computing}  \hspace*{6cm} {\it  (Yes: 302; No: 20; No Answer: 20)}

\medskip
\noindent
High-performance and high-throughput computing are essential to advance nuclear physics at the experimental and theory frontiers. Increased investments in computational nuclear physics will facilitate discoveries and capitalize on previous investments.
\begin{itemize}
  \setlength\itemsep{-0.2em}
    \item We recommend increased investments for software and algorithm development, including in AI/ML, by strengthening and expanding programs and partnerships, such as the DOE SciDAC and NSF CSSI and AI institutes.
    \item We recommend increased support for dedicated high-performance and high-throughput mid-scale computational hardware and high-capacity data systems, as well as expanding access to leadership computing facilities.
    \item Advanced computing is an interdisciplinary field. We recommend establishing programs to support the development and retention of a diverse multi-disciplinary workforce in high-performance computing and AI/ML.
\end{itemize}

\subsection{Initiatives}
The Initiatives listed here are the product of input from the QCD community.  They represent a broad range of projects and ideas that were proposed and discussed at the Town Meeting, but they do not necessarily have as high or as focused priority that the Recommendations have. 


\bigskip
\noindent\textbf{Initiative: We recommend targeted efforts to enable the timely realization of a second, complementary detector at the Electron-Ion Collider.} 
\hspace*{5.2cm} {\it (Yes: 262; No: 54; No Answer: 26)}

\medskip
\noindent The EIC is a transformative accelerator that will enable studies of nuclear matter with unprecedented precision. The EIC encapsulates a broad physics program with experimental signatures ranging from exclusive production of single particles in ep scattering to very high multiplicity final states in eA collisions. Two detectors will expand the scientific opportunities, draw a more complete picture of the science, and mitigate the inherent risks that come with exploring uncharted territory by providing independent confirmation of discovery measurements. High statistical precision matched with a similar or better level of systematic precision is vital for the EIC and this can only be achieved with carefully optimized instrumentation. A natural and efficient way to reduce systematic errors is to equip the EIC with two complementary detectors using different technologies. The second detector effort will rely heavily on the use of generic detector R$\&$D funds and accelerator design effort to integrate the detector into the interaction region. The design and construction of such a complementary detector and interaction region are interwoven and must be synchronized with the current EIC project and developed in the context of a broad and engaged international EIC community.

\bigskip
\noindent
{\bf 
 Initiative: 
We recommend the allocation of necessary resources to develop high duty-cycle polarized positron beams at CEBAF.
} \hspace*{7cm} {\it (Yes: 192; No: 91; No Answer: 59)}

\medskip
\noindent
Using the existing 12 GeV CEBAF and capitalizing on innovative concepts for a positron source developed at Jefferson Lab, a high duty-cycle polarized positron beam will enable a unique science program at the luminosity and precision frontier. It will comprise the mapping of two-photon exchange effects as well as essential measurements of the 3D structure of hadrons. It will also offer new opportunities to investigate electroweak physics and physics beyond the standard model.

\smallskip
The PEPPo experiment (2012) demonstrated a new technique for the production of polarized positrons (PRL 116, 2016) at the CEBAF injector. Since then, an extensive physics program has been developed. First presented in 2018 to the Jefferson Lab Program Advisory Committee (PAC), it was then expanded and summarized in 20 peer-reviewed publications (EPJ A58, 2022). Two experiments were already approved by the Jefferson Lab PAC in 2020. The PAC has encouraged a vigorous effort to explore the technical feasibility of such a unique facility. A positron injector concept has emerged with the help of FY21 LDRD funds and an upcoming FY23 LDRD project will study the efficiency of transporting a beam with emittance comparable to the one expected in a positron beam through CEBAF. Following these advances over the last decade, expeditious development of this outstanding worldwide capability now appears achievable.

\bigskip
\noindent
{\bf  Initiative: 
Capitalizing on recent science insights and US-led accelerator science and technology innovations, we recommend a targeted effort to develop a cost-effective technical approach for an energy upgrade of CEBAF. This would provide capabilities to enable a worldwide unique nuclear science program at the luminosity frontier. 
}\hspace*{8cm} {\it 
(Yes: 140; No: 147; No Answer: 55)
}

\medskip
\noindent
The last decade has provided multiple science surprises such as the discovery of exotic states in the charmonium sector at facilities worldwide, the so-called “XYZ” states. Studies of the 3D structure of hadrons and hadronization provided deeper access to quark-gluon dynamics and opened new opportunities for understanding QCD in its full complexity. In addition, mysteries of the visible matter around us remain unsolved, such as a small enhancement of partons found in nuclei at the interface of the quark- and gluon-dominated regions, the so-called “anti-shadowing” region, that to date lacks explanation and can only be further studied at the luminosity frontier.

\smallskip
Capitalizing on recent innovations enabled by accelerator science and technology, a cost-effective energy upgrade of the 12-GeV CEBAF at Jefferson Lab to a 22 GeV facility may become feasible. Such an upgrade would permit a worldwide unique nuclear science program with fixed targets at the luminosity frontier, roughly five decades above that possible with a collider. Beyond its nuclear science opportunities, this will further steward best-in-class accelerator technology within the US.

\bigskip
\noindent
\textbf{Initiative: 
U.S. Participation in LHC Detector Upgrades and Partnership with CERN Initiative
}\\
\hspace*{11.5cm} {\it 
(Yes: 255; No: 49; No Answer: 38)
}

\medskip
\noindent
The LHC will remain at the energy frontier of nuclear and particle physics in the coming two decades. Detector upgrades enabled by novel technologies will maximize the potential of the planned high luminosity upgrade and open new opportunities in QCD research. 

\smallskip\noindent
\textbf{To maintain U.S. leadership in the nuclear physics program at the LHC, we recommend exploring and supporting targeted detector R\&Ds and upgrades to the LHC experiments, led by U.S. groups, that provide unique capabilities.} These projects will open new physics opportunities, further stimulate the synergy between US-EIC and CERN-LHC in nuclear science, accelerator and detector technology, and also strengthen partnerships with the international community.

\bigskip
\noindent
\textbf{Initiative: Exploring opportunities for US participation in international facilities at the high baryon density frontier
}\hspace*{8.6cm} {\it (Yes: 157; No: 129; No Answer: 56)}

\medskip
\noindent
We wish to maintain US leadership in the exploration of the QCD phase diagram at high baryon density after the completion of the RHIC BES-II program and to build on the success of the BES program,  including the search for the QCD critical point, the extraction of the hyperon-nucleon interaction, and the determination of constraints on the nuclear matter equation of state at high baryon density. 

\smallskip
\noindent{\bf We recommend exploring opportunities for targeted US participation in international facilities that will probe the physics of dense baryon-rich matter and constrain the nuclear equation of state in a regime relevant to binary neutron star mergers and supernovae.} The upcoming results from RHIC BES-II will help assess which international experiments present the highest potential for new discoveries at high baryon density.

\bigskip
\noindent\textbf{Initiative: Nuclear Data
} \hspace*{7cm} {\it (Yes: 274; No: 22; No Answer: 46)}

\medskip
\noindent
Nuclear data play an essential if sometimes unrecognized role in all facets of nuclear physics. Access to accurate, reliable nuclear data is crucial to the success of important missions such as nonproliferation and defense, nuclear forensics, homeland security, space exploration, and clean energy generation, in addition to the basic scientific research underpinning the enterprise. These data are also key to innovations leading to new medicines, automated industrial controls, energy exploration, energy security, nuclear reactor design, and isotope production.  It is thus crucial to maintain effective US stewardship of nuclear data.
\begin{itemize}
  \setlength\itemsep{-0.2em}
    \item We recommend identifying and prioritizing opportunities to enhance and advance stewardship of nuclear data and maximize the impact of these opportunities.
    \item We recommend building and sustaining the  nuclear data community by recruiting, training, and retaining a diverse, equitable and inclusive workforce. 
    \item We recommend identifying crosscutting opportunities for nuclear data with other programs, both domestically and internationally, in particular with regard to facilities and instrumentation.
\end{itemize}

\section{Progress Since the Last LRP}\label{sec:progress}
\newcommand{\pt}{\mbox{$p_{T}$}}
\newcommand{\snn}{\mbox{$\sqrt{s_{\textrm{NN}}}$}}
\newcommand{\TAA}{\mbox{$T_{AA}$}}
\newcommand{\RAA}{\mbox{$R_{AA}$}}
\newcommand{\vntwo}{\mbox{$v_{n}\{ 2 \}$}}
\newcommand{\vnfour}{\mbox{$v_{n} \{ 4 \}$}}
\newcommand{\vfour}{\mbox{$v_{4}$}}
\newcommand{\vtwo}{\mbox{$v_{2}$}}
\newcommand{\vthree}{\mbox{$v_{3}$}}
\newcommand{\vn}{\mbox{$v_{n}$}}
\newcommand{\auau}{\mbox{Au$+$Au}}
\newcommand{\arar}{\mbox{Ar$+$Ar}}
\newcommand{\cucu}{\mbox{Cu$+$Cu}}
\newcommand{\oo}{\mbox{O$+$O}}
\newcommand{\po}{\mbox{$p+$O}}
\newcommand{\pbpb}{\mbox{Pb$+$Pb}}
\newcommand{\xexe}{\mbox{Xe$+$Xe}}
\newcommand{\x}{\mbox{$x$}}
\newcommand{\gammaA}{\mbox{$\gamma$+A}}
\newcommand{\HeAu}{\mbox{$^{3}$He+Au}}
\newcommand{\nb}{\mbox{nb$^{-1}$}}
\newcommand{\bjets}{\mbox{$b$-jets}}
\newcommand{\gammajets}{\mbox{$\gamma$-jets}}
\newcommand{\Zjets}{\mbox{$Z$-jets}}
\newcommand{\qhat}{\mbox{$\hat{q}$}}
\newcommand{\inb}{\mbox{$\textrm{nb}^{-1}$}}
\newcommand{\inp}{\mbox{$\textrm{pb}^{-1}$}}

\newcommand{\Ds}{${\cal D}_s$}
\def\lsim{\mathrel{\raise.3ex\hbox{$<$\kern-.75em\lower1ex\hbox{$\sim$}}}}

\subsection{Progress in Hot QCD}\label{sec:hot}

Over the last several years major advances have been made
through the experimental programs at RHIC and the LHC.  
At RHIC, this includes the successful data taking within the Beam Energy Scan II (BES-II) program, gathering unprecedented statistics on Au+Au collisions probing the QCD phase diagram from moderate to high net baryon number densities, as well as the collection and analysis of data from the isobar program, which used Ru+Ru and Zr+Zr collisions to search for the chiral magnetic effect (CME). At the LHC, the Run 2 heavy-ion program provided more than an order of magnitude increase in luminosity for Pb+Pb collisions (compared with Run 1) and explored the first experiments at LHC energies utilizing nuclei other than lead with \xexe\ collisions. Additionally, there has been a lot of progress in the relevant theory and computation, including
Bayesian analyses providing improved extraction of the QGP transport properties.  
This section highlights some of the major advances in 
this area as well as connections between studies of
the QGP and other areas of physics.

\subsubsection{Macroscopic QGP Properties}

\begin{figure}[ht]
\centering
\includegraphics[width=0.8\linewidth]{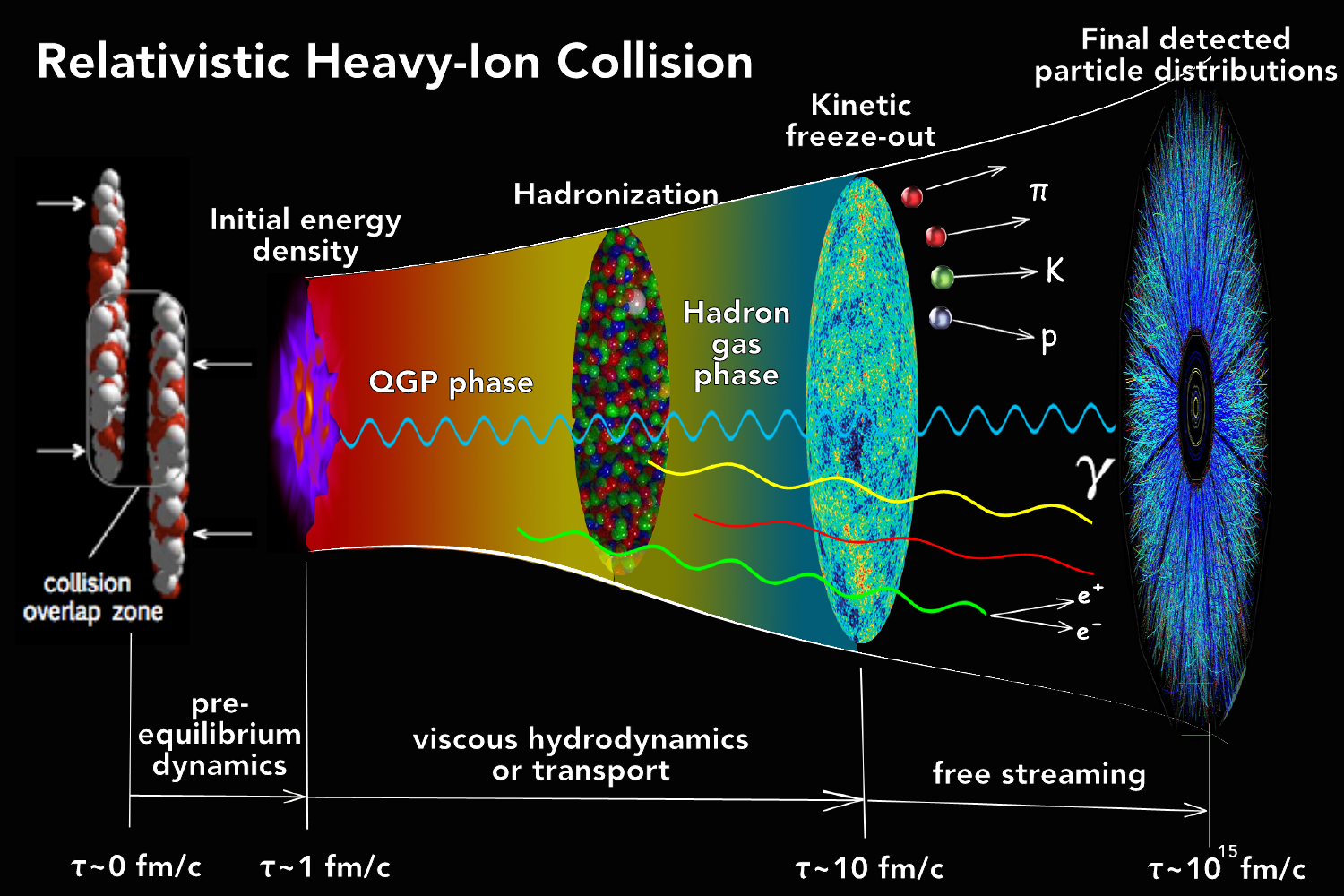}
\caption{The stages of a relativistic heavy ion collision. Figure adapted from \cite{Sorensen:2009cz,Shen:2015msa}.}
\label{fig:little_bang}
\end{figure}

A major goal of the study of heavy-ion collisions is to determine the properties of the hot and dense matter created. This includes the QCD equation of state (see Sec.\,\ref{sec:phasediagram}) and the transport properties of the QGP, including its shear and bulk viscosities and the partonic transport coefficient $\hat{q}$.
While first-principles calculations using e.g.~lattice QCD or effective models can provide results for these properties,  these methods are either extremely difficult (in case of lattice QCD, from which so far only the equation of state at zero baryon chemical potential, and the heavy quark diffusion coefficient have been reliably determined, see Sec.\,\ref{sec:theory0}) or only approximate QCD (see Sec.\,\ref{sec:eff}), such that comparison of phenomenological models to experimental measurements provides the most fruitful approach to determine QGP properties.  
A representation of the evolution of a heavy ion collision is shown in Fig.\,\ref{fig:little_bang}, indicating the different stages and time scales, and showing the various final state particles that carry all accessible information. In the following we will discuss the most important observables that carry information on the macroscopic QGP properties and then move to discussing advances in the phenomenological modeling and data-theory comparisons. 

\begin{figure}[ht]
\centering
\includegraphics[width=0.49\linewidth]{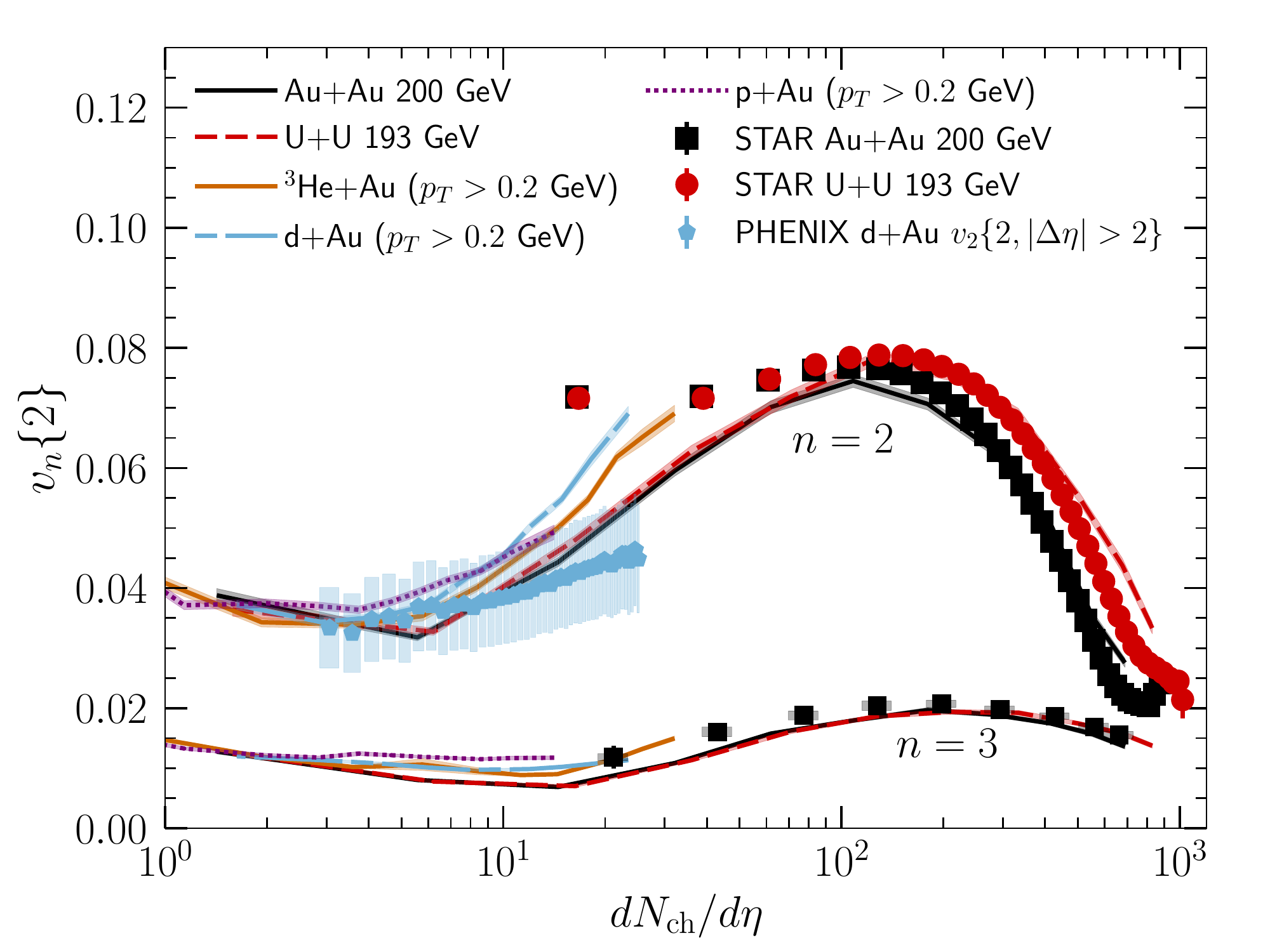}
\includegraphics[width=0.49\linewidth]{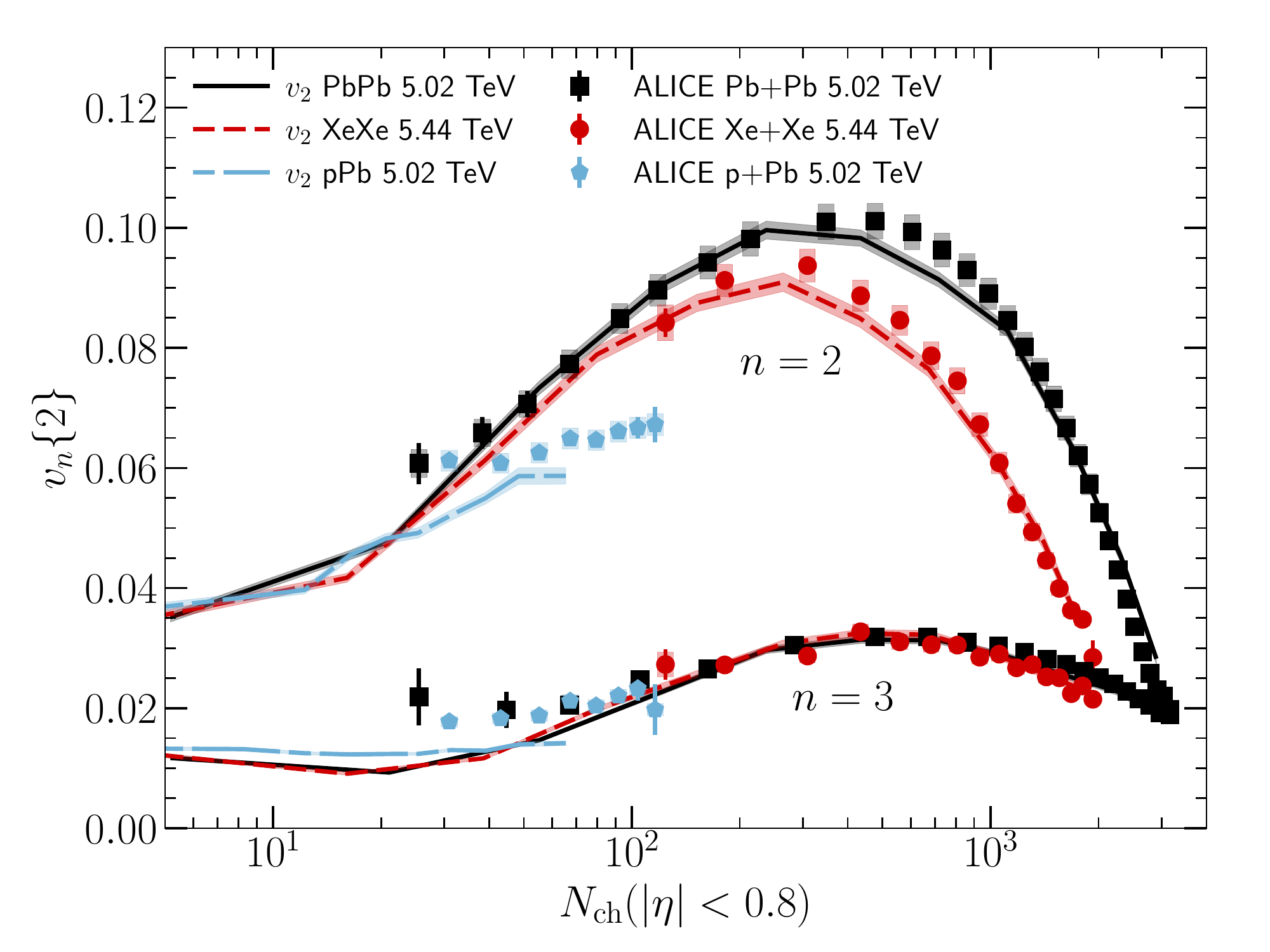}
\caption{Experimental data on $v_2\{2\}$ and $v_3\{2\}$ from the STAR \cite{Adamczyk:2015obl,Adam:2019woz}, PHENIX \cite{Aidala:2017ajz}, and ALICE \cite{ALICE:2019zfl} Collaborations, with theory results from the IPGlasma+MUSIC+UrQMD model. Figure adapted from \cite{Schenke:2020mbo}.}
\label{fig:v2v3}
\end{figure}

\subsubsubsection{Experimental observables} 
\label{sec:expflowprogress}
Over past years, experimental measurements to quantify bulk and collective properties of the QGP have achieved a new level of precision for a wide variety of differential observables. Azimuthal anisotropies of the transverse momentum distribution of produced particles, elliptic ($v_2$) and higher-order ($v_n$) flow coefficients that are particularly sensitive to the QGP's shear viscosity and equation of state, have been measured to unprecedented precision over a wide range of a wide range of initial system size, collision geometry, and phase space  ~\cite{ALICE:2016ccg,CMS:2017xgk,ALICE:2018lao,ATLAS:2018ezv, PHENIX:2018lia, CMS:2019cyz,ATLAS:2019dct,STAR:2019zaf,STAR:2022gki,ALICE:2022wpn}. 
Figure\,\ref{fig:v2v3} shows $v_2$ and $v_3$ results over a wide multiplicity (or, equivalently, centrality) range in Au+Au, U+U, and d+Au collisions at RHIC (left panel) as well as in Pb+Pb, Xe+Xe, and p+Pb collisions at the LHC (right panel), indicating collective flow of the medium, originally not expected in case of small systems (see Sec.~\ref{sec:smallprogress}). For central collisions of large systems, the $v_2$ and $v_3$ data
follow precisely the trend expected by the initial geometry and its fluctuations. Extending to very peripheral regions where the system size is diminishing, the finite system size effect and viscous corrections become significant. 
The observed trends are captured by state-of-the-art hybrid hydrodynamic calculations from the most central events down to dN/d$\eta$ $\sim$ 10--20. 

Additionally, new experimental techniques are expected to increase the precision of the QGP transport property extraction in the near future.  
The \vn\ of identified particles, especially those which contain strange quarks, can test the expected mass dependence of hydrodynamic flow and can be used to constrain the impact of the hadronic rescattering
phase on the measured anisotropies~\cite{ALICE:2018yph,STAR:2022ncy,CMS:2022bmk} 
(the extension of these measurements to hadrons containing heavy quarks
is discussed in Sec.~\ref{sec:heavy}).
Event-by-event fluctuations of the $v_2$ flow coefficient have been measured using multiparticle cumulants;
these are used to extract moments of the \vn\ distributions~\cite{ATLAS:2014qxy,ALICE:2014dwt,CMS:2013jlh,STAR:2014ofx,STAR:2015mki,ATLAS:2017rtr,CMS:2017glf,ALICE:2019zfl,ATLAS:2019peb}.
Also, symmetric cumulants measure the correlated fluctuations of different orders of flow coefficients~\cite{ALICE:2016kpq,CMS:2017kcs,ATLAS:2018ngv}. 
Mixed-higher-order flow harmonics
measured up to $7{\rm th}$ order can uniquely probe linear and nonlinear hydrodynamic responses~\cite{CMS:2019nct,ALICE:2021adw}.
The $v_n$-$\left<p_{T}\right>$ correlation, with unique sensitivity to the correlation between the system size and shape, has already been used to constrain the shape of the
xenon nucleus using LHC \xexe\ data \cite{ALICE:2021gxt,ATLAS:2022dov}; it also has the potential to disentangle different origins of momentum anisotropy \cite{Giacalone:2020byk,CMS-PAS-HIN-21-012}.
Further, femtoscopic observables are sensitive to the system size and provide additional information to disentangle medium and initial state properties \cite{PHENIX:2004yan,STAR:2004qya,ALICE:2015hvw,CMS:2017mdg,ATLAS:2017shk,CMS:2023jjt}.
More observables, such as electromagnetic probes, hard probes including jets and heavy flavors, as well as measurements at varying collision energy and for different system sizes can aide in the determination of QGP properties. We will discuss each of them below. 

\subsubsubsection{Phenomenological modeling of heavy ion collisions} \label{sec:pheno}
There has been significant progress over the last several years in the modeling of heavy ion collisions over a wide range of energies and collision systems. In particular, there has been progress in modeling the initial state (see Sec.\,\ref{sec:init}), with three-dimensional dynamic initial states, that 
progressively deposit energy into the hydrodynamic medium \cite{Shen:2022oyg}, with initial conditions for fluctuating conserved charges \cite{Carzon:2019qja}, as well as extensions to three dimensions in the color glass condensate (CGC) based models \cite{Schenke:2016ksl, Schenke:2022mjv}. 
Smoothly connecting the initial state to hydrodynamics has been significantly improved by means of QCD effective kinetic theory \cite{Kurkela:2018wud}. 

Extending the applicability of fluid dynamics can also be achieved with anisotropic fluid dynamics, which allows for larger differences between the longitudinal and transverse pressure in the system and therefore applies at earlier times than second-order viscous hydrodynamics \cite{Alqahtani:2017mhy,McNelis:2018jho}. Further extensions of fluid dynamics include spin \cite{Bhadury:2020cop} and chiral currents \cite{Shi:2019wzi, Ammon:2020rvg}, triggered by interest in chiral magnetic effect and polarization observables (see Sec.\,\ref{sec:chiralityvorticity}).

Equations of state were constructed with input from lattice QCD, in the space of temperature and chemical potentials of the conserved charges \cite{Parotto:2018pwx, Monnai:2019hkn,Noronha-Hostler:2019ayj, Monnai:2021kgu}, with some of them including a critical point (see Sec.\,\ref{sec:phasediagram} and \cite{An:2021wof} for a review). These equations of state require extrapolations, for example into the region of high baryon chemical potential where lattice QCD cannot directly provide results. Constraints in that region can be obtained from thermal perturbation theory, effective models of QCD (see Sec.\,\ref{sec:eff}), or calculations of strongly-coupled gauge theories that have known holographic duals, and are similar to QCD \cite{Casalderrey-Solana:2011dxg}.

Progress has also been made in describing the evolution of hydrodynamic and critical fluctuations by solving stochastic differential equations \cite{Nahrgang:2018afz,Nahrgang:2020yxm} or employing the hydro-kinetic formalism \cite{Akamatsu:2016llw,Stephanov:2017ghc,An:2019osr,Rajagopal:2019xwg,An:2019csj, An:2020vri, Du:2020bxp,De:2022tkb}, as well as the conversion from fluids to particles that respects local conservation laws \cite{Oliinychenko:2019zfk, Pradeep:2022mkf}. Both developments are particularly relevant for including effects related to the existence of the critical point. 
Core-corona models, in which regions of high energy density are described using hydrodynamics, while low energy density matter is described using particle degrees of freedom throughout the evolution, have also been advanced significantly in the past years. Such models allow for a unified description across systems sizes and produced particle transverse momenta \cite{Kanakubo:2021qcw}.

Purely hadronic transport simulations are essential for constraining the dense nuclear matter equation of state (EOS) and interpreting experimental results from collisions at very low to intermediate beam energies, $\sqrt{s_{ \rm{NN} } }  \approx 1.9$ to $\sqrt{s_{ \rm{NN} } } \approx 8.0\ \rm{GeV}$, where equilibrium is not typically expected to be reached, and, therefore, a hydrodynamic description is not possible. Comparisons of hadronic transport simulations with experimental data can reveal not only the EOS of symmetric nuclear matter \cite{Danielewicz:2002pu,LeFevre:2015paj,Oliinychenko:2022uvy,Steinheimer:2022gqb}, but can also help constrain the isospin-dependence of the EOS (e.g., by using meson yields \cite{Colonna:2020euy,Fuchs:2000kp,Li:2002qx,Xiao:2008vm,Yong:2022pyb,SRIT:2021gcy}, proton and neutron flow \cite{Li:2000bj,Li:2014oda,Colonna:2020euy, Xu:2019hqg,Russotto:2011hq, Cozma:2011nr, Giordano:2010pv}, or pion flow \cite{Liu:2019ags}) and help understand strange interactions (e.g., by using strange particle flow \cite{Li:1996ju,Wang:1998ew,Ko:2000cd}). The influence of the possible QCD critical point on the hadronic evolution, either within purely hadronic transport simulations or in afterburner calculations, can also be explored using hadronic potentials,  enabling description of non-trivial features at high baryon densities \cite{Sorensen:2020ygf}. Significant theoretical, conceptual, and modeling work remains, however, to ensure valid conclusions can be discerned from the heavy-ion data.

\subsubsubsection{Extracting QGP properties using Bayesian inference} \label{sec:bayesian}  
An important tool that has helped precision extraction of information is Bayesian inference, which has been increasingly used over the last few years to constrain the temperature dependence of $\eta/s$ and $\zeta/s$, as well as other quantities such as $\hat{q}$. Bayesian inference determines the probability that certain values of shear and bulk viscosity, and any other model parameter, are consistent with a set of measurements and their uncertainties. The resulting posterior probability distribution, which has the dimension of the number of parameters, can be projected to lower dimensions by marginalizing over all but one or two parameters, or by calculating credible intervals, providing interpretable constraints on the model parameters.

\begin{figure}[ht]
\centering
\includegraphics[width=0.9\linewidth]{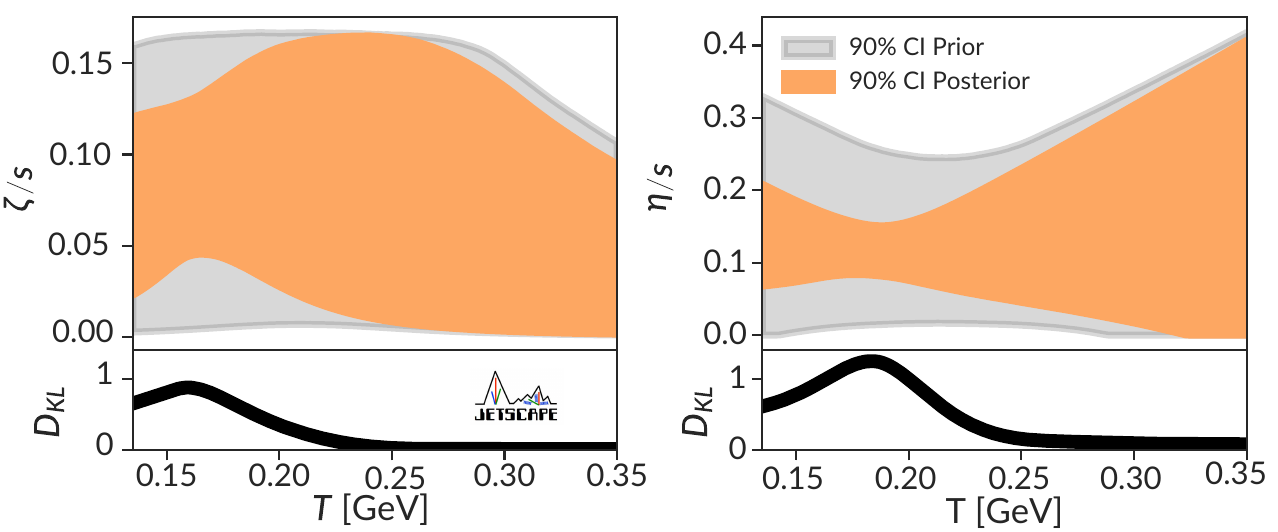}
\caption{Constraints on $\eta/s$ and $\zeta/s$ from Ref.~\cite{JETSCAPE:2020shq}, as represented by the $90$\% credible intervals for the posterior and the prior, along with their corresponding information gain (Kullback-Leibler divergence $D_{KL}$).}
\label{fig:jetscape_viscosities}
\end{figure}

A number of different constraints on the QGP shear and bulk viscosities have been obtained over the past decade~\cite{Petersen:2010zt, Novak:2013bqa, Sangaline:2015isa, Bernhard:2015hxa,Bernhard:2016tnd,Moreland:2018gsh,Bernhard:2019bmu,JETSCAPE:2020shq,JETSCAPE:2020mzn,Nijs:2020ors,Nijs:2020roc,Nijs:2021clz,Parkkila:2021tqq}. The analyses differ because of the uncertainties in (i) modelling the pre-equilibrium stage, (ii) the equation of state, (iii) the assumed functional form of shear and bulk viscosity, (iv) the higher-order transport coefficients,  and (v) the conversion from fluids to particles. 
An additional source of uncertainty is emulation, often used to circumvent the otherwise prohibitive computational requirements of running the model. Critically, the set of measurements used to calibrate the model vary considerably between the different analyses.

Reference~\cite{Bernhard:2019bmu} marked a step forward by studying simultaneously the temperature dependence of the shear and bulk viscosities with flexible parametrizations and using a state-of-the-art model. The employed software is publicly available and formed the basis of almost all Bayesian inference studies that followed. Reference~\cite{Moreland:2018gsh} and later Refs.~\cite{Nijs:2020ors,Nijs:2020roc} added data from small system (\pPb) collisions, with the later papers also including a larger set of measurements in \pbpb\ collisions, a different pre-hydrodynamic phase, and a study of second order transport coefficients. 
As an example, we show results for the posterior distributions of temperature dependent bulk and shear viscosities (compared to the assumed priors) from Refs.~\cite{JETSCAPE:2020shq,JETSCAPE:2020mzn} in Fig.~\ref{fig:jetscape_viscosities}. These studies combined RHIC and LHC measurements, and included particlization uncertainties for the first time. 
The difference between the prior and the posterior, quantified by the Kullback-Leibler divergence $D_{KL}$ in the bottom panel of Fig.~\ref{fig:jetscape_viscosities}, highlights that most information is gained at temperatures below $T=200\,{\rm MeV}$. How to improve constraints at higher temperature is one of the major questions going forward. 
Additional advances in the past years include (i) the use of closure tests~\cite{JETSCAPE:2020avt,JETSCAPE:2020mzn} as validation of Bayesian analysis and as a method to estimate the impact of future measurements~\cite{Nijs:2021clz}, (ii) non-parametric methods~\cite{Xie:2022ght,Xie:2022fak} to reduce bias when constraining model parameter that are functions rather than scalar values, and (iii) increasing attention to correlations between measurements and their uncertainties~\cite{Soltz:2020xvw,JETSCAPE:2021ehl,JETSCAPE:2020mzn,Bernhard:2019bmu,Bernhard:2018hnz}.

\subsubsection{Accessing QGP Evolution and Chiral Symmetry Breaking Using Electromagnetic Probes}
By virtue of their negligible interaction with the quark-gluon plasma, electroweak probes provide invaluable information on the physics of heavy-ion collisions. Low-energy photons and low-mass dileptons are radiated directly by the hot and dense plasma produced in the collisions
but then do not interact further with the QGP, providing a window into the thermal properties of the plasma. Additionally, high-energy photons, dileptons and weak bosons are mainly produced when the nuclei initially collide, 
and can provide important information regarding the initial properties of the collisions~\cite{Geurts:2022xmk}.

Removing the contribution from photons emerging from hadronic decays leads to a ``direct'' photon signal.  Measurements of the low-energy direct photon spectra and 
azimuthal anisotropies, $v_n^\gamma$, have been released by ALICE, PHENIX and STAR. 
At RHIC, there is tension between the STAR~\cite{STAR:2016use} and PHENIX results~\cite{PHENIX:2018for,PHENIX:2022qfp,PHENIX:2022rsx} for the photon spectra.  Measurements are also available from ALICE at 2.76~TeV~\cite{ALICE:2015xmh}, which show an enhancement over expectations based on perturbative QCD in the transverse momentum region of 2--5~GeV which is consistent with the thermal radiation from the QGP.
Values of $v_n^\gamma$ have been measured by both ALICE~\cite{ALICE:2018dti}
and PHENIX~\cite{PHENIX:2015igl}.  
The results are found to be compatible with each 
other but the measured values are systematically larger than the model results \cite{Gale:2021emg}.  The source of
this large $v_n^\gamma$ in the data is not understood.

Inclusive dileptons that include hadronic decays have been measured in ALICE~\cite{ALICE:2018ael} and found to be consistent in the low-mass limit with inclusive real photon measurements. Measurements of dileptons for collision energies between 19.6--200 GeV are  also available~\cite{STAR:2015zal, STAR:2015tnn,STAR:2018xaj,STAR:2018ldd}.
Models that include an in-medium broadening of the $\rho$-meson spectral function consistently describe the observed excess over the hadronic decay contributions~\cite{Rapp:2014hha}. 

Significant advances have been made in the theoretical description of photon and dilepton production in heavy-ion collisions. Calculations of the thermal production of photons~\cite{Paquet:2015lta,Kim:2016ylr,Dasgupta:2018pjm,Garcia-Montero:2019kjk,Monnai:2022hfs,Gale:2021emg,Chatterjee:2017akg, Shen:2013vja, Shen:2013cca} and dileptons~\cite{Vujanovic:2017psb,Vujanovic:2019yih,Vujanovic:2016anq} in viscous hydrodynamic backgrounds have been improved by including the effects of shear and bulk viscosity on the emission rates \cite{Paquet:2015lta,Liu:2017fib,Hauksson:2017udm,Gale:2021emg,Vujanovic:2019yih}, and electromagnetic emission channels have been included in the hadronic transport stage \cite{Schafer:2021slz,Schafer:2019edr}. This brings the sophistication of thermal photon and dilepton calculations on par with those of soft hadrons.
Various new calculations of photon emission rates have emerged~\cite{Holt:2015cda,Holt:2020mwf,Hidaka:2015ima,Kim:2016ylr,Schafer:2019edr,Liu:2017fib,Zakharov:2017cul,Bandyopadhyay:2015wua,Iatrakis:2016ugz}, including from lattice QCD \cite{Ding:2016hua,Ghiglieri:2016tvj,Jackson:2019yao,Ce:2020tmx,Ce:2022fot}. Results from Refs.~\cite{Bhattacharya:2015ada,Kasmaei:2018oag,Kasmaei:2019ofu} using anisotropic hydrodynamics and electromagnetic emission rates from a momentum-anisotropic quark-gluon plasma further contribute to better understanding non-equilibrium effects. 
Works on other topics include predictions for the direct photon Hanbury Brown Twiss (HBT) interferometry \cite{Garcia-Montero:2019kjk}, studying photons that originate from the hadronization of intermediate energy hadrons~\cite{Fujii:2022hxa}, additional photon production mechanisms~\cite{Tuchin:2019jxd,Ayala:2017vex}, and relativistic transport studies of electromagnetic probes~\cite{Linnyk:2015tha}.

Invariant mass spectra of dileptons also provide a unique opportunity to study the effects of chiral symmetry restoration on hadrons, such as the $\rho$ meson and its chiral partner, the $a_1$. 
Vector meson spectral functions in the medium can be computed in a variety of frameworks, including lattice QCD, massive Yang Mills, and hadronic many-body theory, or the analytically-continued functional renormalization group (FRG) method~\cite{Geurts:2022xmk}. 
Theoretical calculations predict melting of the $\rho$ meson in the medium, indicating a transition from hadronic degrees of freedom to a quark-antiquark continuum that is consistent with chiral symmetry restoration. This picture is consistent~\cite{Rapp:2014hha} with dilepton data from NA60~\cite{NA60:2008ctj} and STAR~\cite{Geurts:2012rv}.
Furthermore, chiral partners become degenerate at the ground state mass in a way that the chiral mass splitting disappears but the ground-state mass remains \cite{Hohler:2013eba,Geurts:2022xmk}.

\subsubsection{QGP Tomography with Hard Probes}
The goal of using hard probes to study the QGP is to understand the emergent phenomena which give rise
to the nearly perfect liquid QGP that has been described in previous sections.
Hard probes, such as jets, open heavy flavor and quarkonia, probe the QGP on varying short distance scales, as with a microscope.
Because the QGP is short-lived, the probes are generated
in the same nuclear collisions which create the QGP itself.
The three Upsilon states and jets are examples of important probes.
The $\Upsilon(1S)$, 
$\Upsilon(2S)$, and $\Upsilon(3S)$ states each characterize the QGP on a separate length scale that depends on its binding energy. Jets probe the QGP on a variety of length scales depending on their
energy and the characteristics of the jet structure.
LRP15 \cite{Geesaman:2015fha} discussed the importance of measurements of these observables at both RHIC and LHC in order to understand the temperature dependence of QGP properties.  Over the last several years there have been new measurements from LHC experiments and the existing RHIC detectors. Crucially, sPHENIX is about to begin its physics program, which focuses on jets and Upsilons.

\subsubsubsection{Jets} 
\label{sec:jetprogress}
QCD jets arise from the hard scattering of quarks and gluons (collectively, \textit{partons}) in hadronic and nuclear collisions.  This is a process that can be well described by perturbative QCD (pQCD)  calculations~\cite{CMS:2016jip,ATLAS:2017ble}.
Jets are measured as a collimated spray of particles carrying approximately the energy of the scattered parton. These particles and/or their energies is clustered together to form measured jets.
Some of the earliest measurements at RHIC and LHC in heavy-ion collisions were about the reduction in the rate of these jets in heavy-ion collisions compared to expectations from \pp\ 
collisions~\cite{PHENIX:2001hpc,STAR:2002svs,ATLAS:2010isq,CMS:2011iwn,CMS:2016uxf}.  This phenomenon is
called \textit{jet quenching}. Our understanding of how jets are quenched in heavy-ion collisions has evolved dramatically in the last several years driven by increasingly precise and differential measurements from the LHC and RHIC and improvements in theory.  Additionally, the techniques used to measure jet substructure have advanced and the number of jet substructure measurements available has increased dramatically. 
The current focus is on understanding how jet quenching depends on the structure of the parton shower and the length of the QGP the jet travels though. In addition to the modification and quenching of the jet itself it is of great interest to study how the QGP responds to the passage of the jet through it. Some of the highlights
are listed here.

\begin{figure}
\centering
\includegraphics[width=0.42\textwidth]{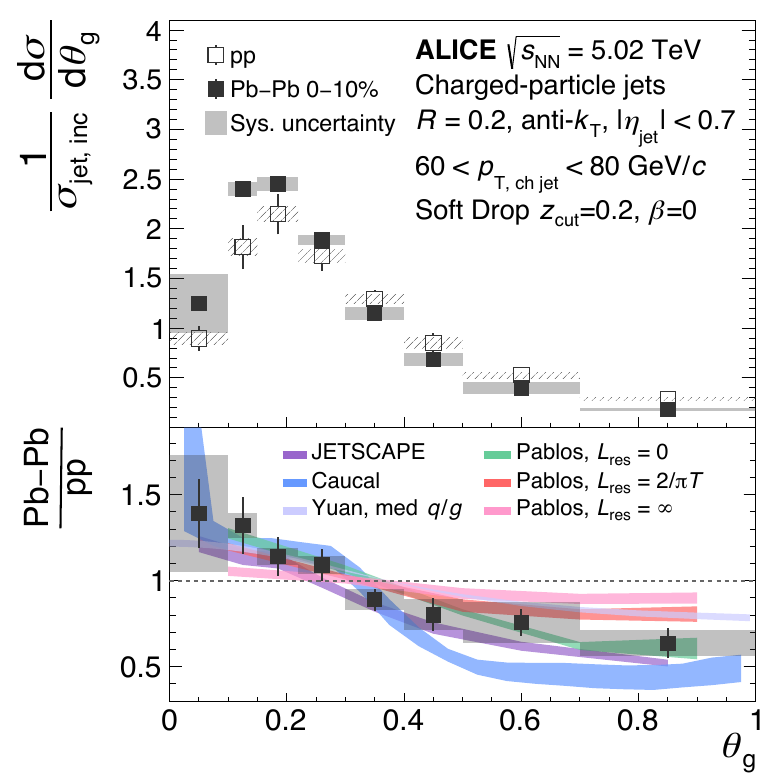}
\includegraphics[width=0.45\textwidth]{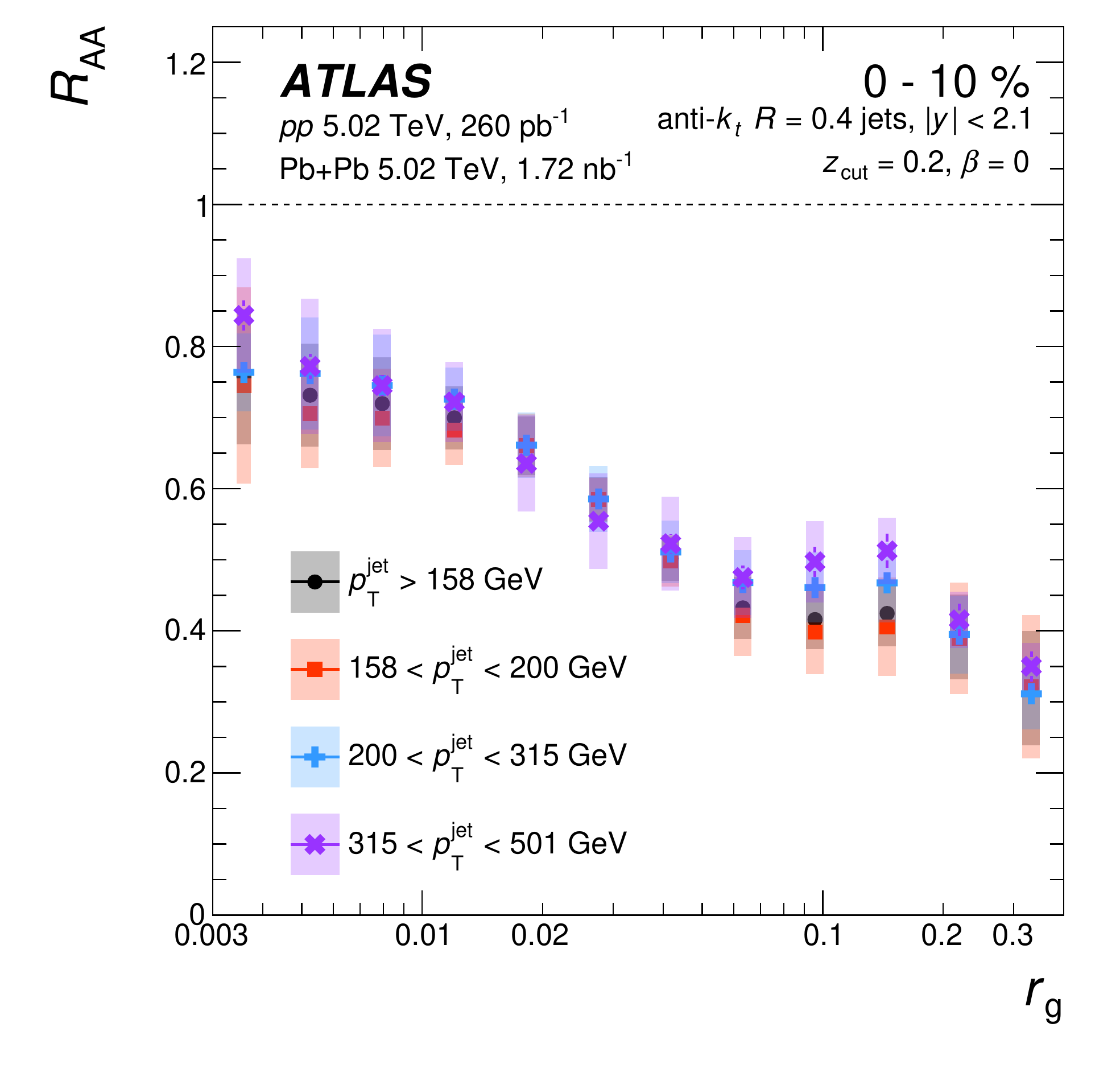}
\caption{Left: The distribution of jets as a function
of $\theta_g$ in 0--10\% central \pbpb\
collisions and \pp\ colliions (top) and the ratio
of the distributions in \pbpb\ collisions compared to \pp\ collisions
(bottom) compared to various theoretical calculations.
Figure from Ref.~\cite{ALargeIonColliderExperiment:2021mqf}.  Theory results shown include JETSCAPE~\cite{Putschke:2019yrg}, JEWEL~\cite{Zapp:2012ak, Zapp:2013vla}, Caucal et al.~\cite{Caucal:2019uvr, Caucal:2018dla}, Chien et al.~\cite{Chien:2016led}, Qin et al.~\cite{Chang:2019nrx}, and Pablos et al.~\cite{Casalderrey-Solana:2014bpa, Hulcher:2017cpt, Casalderrey-Solana:2019ubu}. Right: The nuclear
suppression factor \RAA\ as a function of $r_{g}$ for
0--10\% central \pbpb\ collisions for four selections on 
$p_T^{\rm jet}$. Figure from Ref.~\cite{ATLAS:2022vii}.  }
\label{fig:ATLASrg}
\end{figure}

The population of jets in heavy-ion collisions has been measured to have different internal 
structure~\cite{CMS:2014jjt,ATLAS:2014dtd,ATLAS:2018bvp,CMS:2018fof}
and substructure~\cite{CMS:2017qlm,ALargeIonColliderExperiment:2021mqf} than jets in \pp\ collisions.  The distribution of jets as a function of the angle between the two hardest subjets in the event, $\theta_g$ (or $r_g \equiv R \theta_g$, where $R$ is the jet cone size) has been measured and is shown in Fig.~\ref{fig:ATLASrg}.
These studies have shown that wider fragmenting jets are suppressed more by the QGP than narrower fragmenting jets, providing possible new connections to the color coherence length scale of the QGP~\cite{ Hulcher:2017cpt,Caucal:2019uvr,Casalderrey-Solana:2019ubu,Caucal:2020uic}:
Only structures in the parton 
shower that are larger than the coherence scale are 
seen by the QGP as separate color charges and thus
quenched separately.  

Measurements have been made of the interjet angular  correlations in \pbpb\
collisions~\cite{ATLAS:2010isq,CMS:2017ehl,ATLAS:2018gwx,STAR:2017hhs}.
Recent measurements have been inspired by
calculations of potential quasiparticles in the QGP~\cite{DEramo:2018eoy,DEramo:2018ydo, Barata:2020rdn,Ce:2020wgg, Hulcher:2022kmn}.
No evidence of this has been found to date.

In addition to measuring the jet substructure directly,
it is possible to change the quark and gluon fractions in a 
particular jet sample with respect to the inclusive sample by looking at jets balanced by a photon or Z-boson rather
than another jet. At leading order, the dominant process
for photon-jet production is $q + g \to q + \gamma$, which
selects on quark jets.  Additionally, it is possible to
select on $b$-jets \cite{CMS:2013qak,ATLAS:2022fgb}, which also provides an enhanced quark
sample of jets.
Due to the larger color charge of the gluon compared to the quark, gluon jets are broader and expected to lose more energy on average than quark jets. Several measurements of $b$-jets found the suppression to be consistent with inclusive jets \cite{CMS:2013qak,CMS:2018dqf}, while recent results for samples of jets with increased quark fractions are found to have reduced jet 
quenching compared to inclusive 
jets~\cite{CMS:2018dqf,ATLAS:2022fgb}.  We note that recent theoretical developments point to possible future methods for data-driven extraction of quark and gluon jet modification in heavy ion collisions \cite{Brewer:2020och, Ying:2022jvy}.
Additional insights into the mechanisms of energy loss were also obtained from studying the semi-inclusive distribution of jets recoiling from a high-$p_T$ trigger hadron \cite{ALICE:2015mdb,STAR:2017hhs}.

\begin{figure}
\centering
\includegraphics[width=0.43\textwidth]{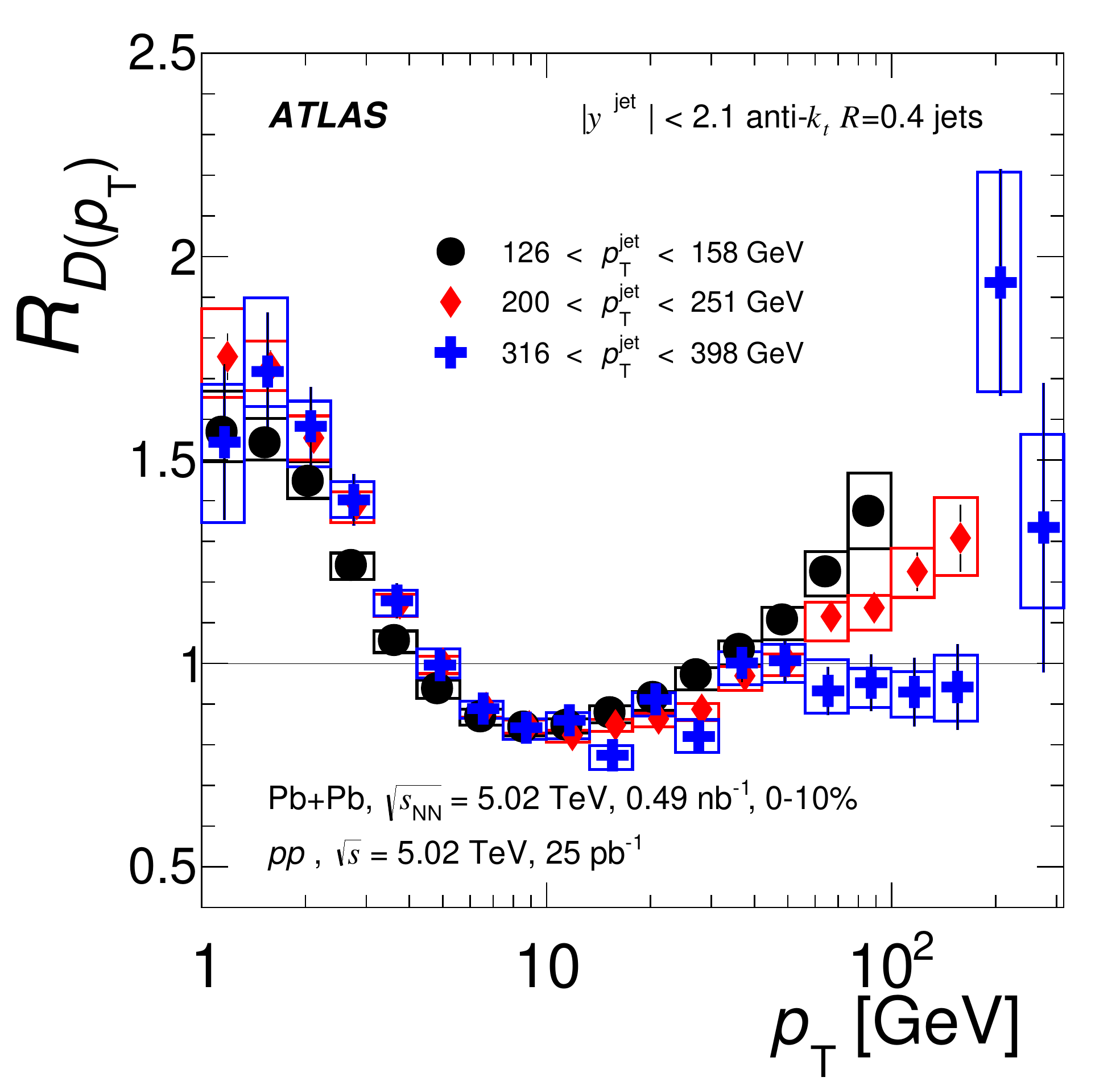}
\includegraphics[width=0.43\textwidth]{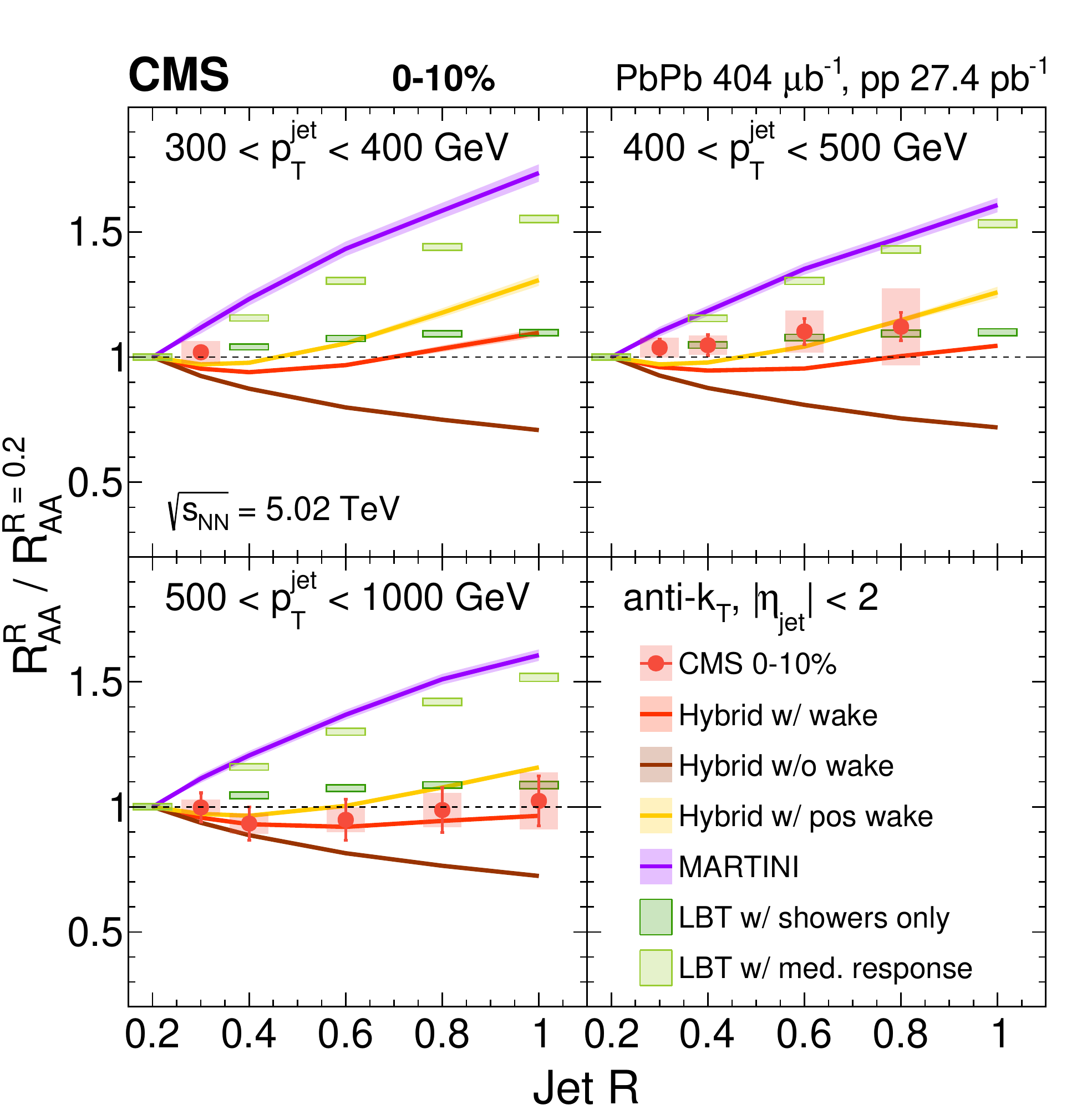}
\caption{Left: Ratio of the jet fragmentation functions in 0--10\% 
central
\pbpb\ collisions to those in \pp\ collisions 
for three jet $p_T^{\rm jet}$ as a function
of the $p_T$ of the particles in the jet. From Ref.~\cite{ATLAS:2018bvp}. Right: Ratio of the 
jet \RAA\ in 0--10\% central \pbpb\ collisions
for jets of radius $\textrm{R}$  to 
the \RAA\ of ${\rm R} = 0.2$ jets as a function of $\textrm{R}$ for
three $p_T^{\rm jet}$.  Comparisons
to a variety of theoretical calculations are shown.
Figure from 
Ref.~\cite{CMS:2021vui}. Theory results from three models~\cite{Schenke:2009gb,Hulcher:2017cpt,He:2018xjv}.}
\label{fig:jetcombo}
\end{figure}

The response of the QGP to the passage of a jet
is characterized by an increased amount of low momentum 
particles within and around the 
jet~\cite{CMS:2011iwn,CMS:2016qnj,CMS:2016cvr,CMS:2018zze,ATLAS:2018bvp,ATLAS:2019pid}. 
Reference~\cite{ATLAS:2018bvp} found an excess of particles in
\pbpb\ collisions relative to \pp\ collisions
below 4~GeV inside the jet cone (of size $R=0.4$).  The size of this excess
was largely independent of the jet transverse momentum,
suggesting that its properties were characteristic of
the QGP and not of the jet itself (see Fig.~\ref{fig:jetcombo}).
The $p_T>4\,{\rm GeV}$ component of the fragmentation function was found to be qualitatively similar to that for jets in \pp\ collisions \cite{CMS:2012nro}. 
Measurements of the angular distribution of the low 
momentum particles near the jet have shown that they have
a wider distribution than those from the jet 
itself~\cite{CMS:2018zze,ATLAS:2019pid}.
This suggests that measuring jets with increasingly large
radii might allow recovery of the energy lost by the jet
and incorporated into the QGP.
Measurements of the jet cone size dependence of
the jet yields in heavy-ion collisions
have been made at RHIC~\cite{STAR:2020xiv} and the 
LHC~\cite{ATLAS:2012tjt,CMS:2016uxf,ALICE:2019qyj,CMS:2021vui}.
Jets with a large cone, $R=$~1, were measured for the first time
in \pbpb\ collisions~\cite{CMS:2021vui} (see Fig.~\ref{fig:jetcombo}).
This measurement
is sensitive to the interplay between the angular 
dependence of the energy lost by the jet 
and the energy incorporated into the medium as \textit{medium
response}. The current measurement does not
show a strong cone size dependence to the jet quenching,
in contrast to many, but not all, theoretical models.  
Tagged (e.g.~with Z-bosons)jets can provide further information on the parton medium interactions \cite{CMS:2021otx}.

\begin{figure}[htb]
\begin{minipage}{0.57\textwidth}
\includegraphics[width=\columnwidth]{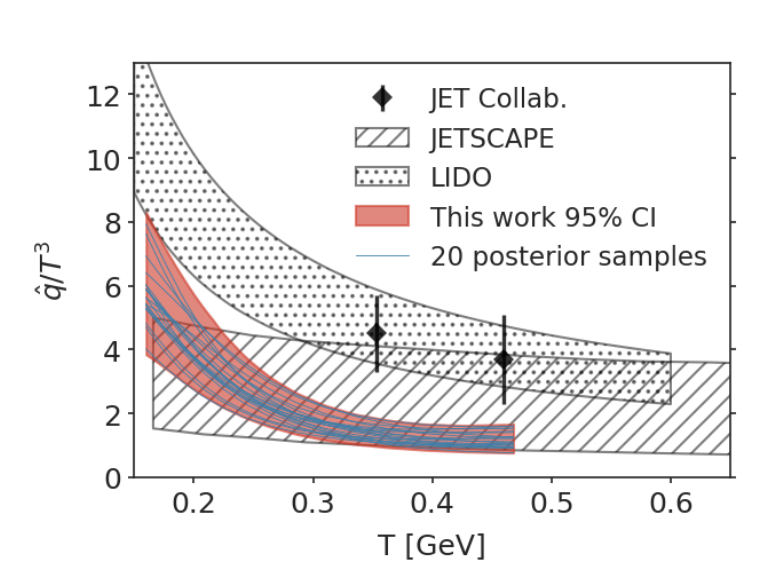}
\end{minipage}
\begin{minipage}{0.35\textwidth}
\linespread{1.0}\selectfont{} 
\caption{
Four extractions of the strength of the jet 
quenching parameter $\qhat / T^3$ as a function of the temperature
$T$ of the QGP. The figure is from Ref.~\cite{Xie:2022ght} and 
the calculations are from
Refs.~\cite{JETSCAPE:2021ehl,Ke:2020clc,Xie:2022ght,JET:2013cls}. }
\label{fig:qhat}
\end{minipage}\end{figure}

The azimuthal anisotropy, \vn\ of 
jets~\cite{ALICE:2015efi,ATLAS:2021ktw}
and hadrons from jets~\cite{CMS:2017xgk} has been measured to
be non-zero in \pbpb\ collisions. 
The values of \vtwo\ are measured
to be significantly larger than zero over a wide range of 
centrality and to have a centrality dependence that is similar
to that seen from hydrodynamic flow for lower momentum particles.
This is expected
if the amount of jet quenching depends on the path
length of the jet through the QGP.  
In \cite{CMS:2022nsv} \vthree\ was found to be consistent with zero, while a non-zero \vthree\ of jets was observed in Ref.~\cite{ATLAS:2021ktw}. The \vthree\ component can be explained by sensitivity to the geometrical fluctuations in the initial state of the collision.
The values of \vfour\ are 
consistent with zero over all measured centralities \cite{ATLAS:2021ktw,CMS:2022nsv}.

Over the last several years, there 
have been significant advances in the extraction of the
parameter controlling the strength of jet
quenching in the QGP, \qhat.  Extractions from 
a number of groups use hydrodynamical models of the 
QGP combined with state-of-the-art jet quenching calculations
to extract \qhat\ from experimental data on the \RAA\
of jets and hadrons at the LHC and 
RHIC~\cite{JETSCAPE:2021ehl,Ke:2020clc,Xie:2022ght,Xie:2022fak}
using Bayesian techniques.  Improvements in the models
and the experimental data have contributed to stronger
constraints on the temperature dependence of~\qhat.
Figure~\ref{fig:qhat} shows the $\qhat / T^3$ extractions
from four 
models~\cite{JETSCAPE:2021ehl,Ke:2020clc,Xie:2022ght,JET:2013cls}.
Reference~\cite{JET:2013cls} was published in 2013 and has large uncertainties
and a small temperature range.  The more recent 
calculations~\cite{JETSCAPE:2021ehl,Ke:2020clc,Xie:2022ght} provide much more information, however there is still tension between 
the three model results. 

\subsubsubsection{Open heavy flavor and quarkonia}\label{sec:heavy}
Heavy-flavor particles, charm and bottom quarks and the hadrons they constitute, are versatile probes of the QCD medium formed in nuclear collisions~\cite{Dong:2019byy,He:2022ywp}.  The heavy quark mass, $m_Q$, provides a large scale relative to typical temperatures in nuclear collisions, providing unique opportunities to investigate the short distance scale behavior of the QGP~\cite{Akiba:2015jwa,Geesaman:2015fha}. 
Suppression of quarkonia, heavy quark-antiquark bound states, is sensitive to the temperature of the medium: different states dissociate at different temperatures, depending on the size of the bound state \cite{Matsui:1986dk}. Quarkonia may also be (re-)generated by uncorrelated $Q$ and $\overline Q$ coalescence when multiple $Q \overline Q$ pairs are produced in a heavy-ion collision. 

Calculational advances in the description of quarkonium and open heavy flavor production have been made in a number of directions, including transport calculations, effective field theory approaches, and lattice QCD calculations. 
The understanding of quarkonium dynamics inside the QGP was greatly advanced since LRP15 by the application of the open quantum system framework (recent reviews can be found in Refs.~\cite{Rothkopf:2019ipj,Akamatsu:2020ypb,Sharma:2021vvu,Yao:2021lus}). 
Very recently, the first $1/m_Q$-correction to the heavy quark diffusion coefficient has been worked out~\cite{Bouttefeux:2020ycy}. Significant noise reduction was obtained in quenched QCD using gradient flow~\cite{Altenkort:2020fgs,Mayer-Steudte:2021hei}. Heavy quark diffusion has been implemented in different transport approaches and used to constrain the QGP transport coefficients~\cite{Rapp:2018qla,Cao:2018ews}. 

\begin{figure}
\centering
\includegraphics[width=0.32\textwidth]{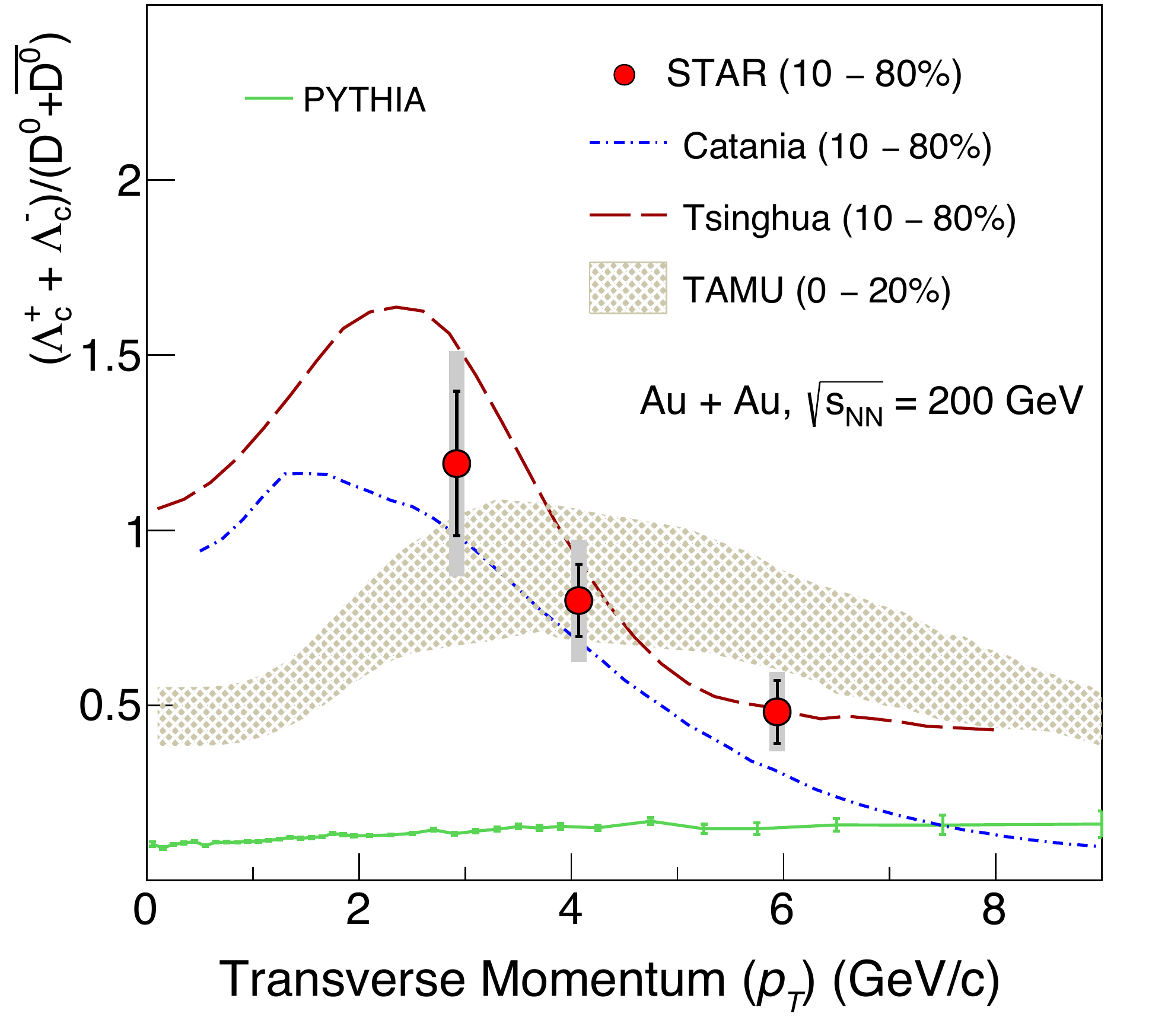} 
\includegraphics[width=0.67\textwidth]{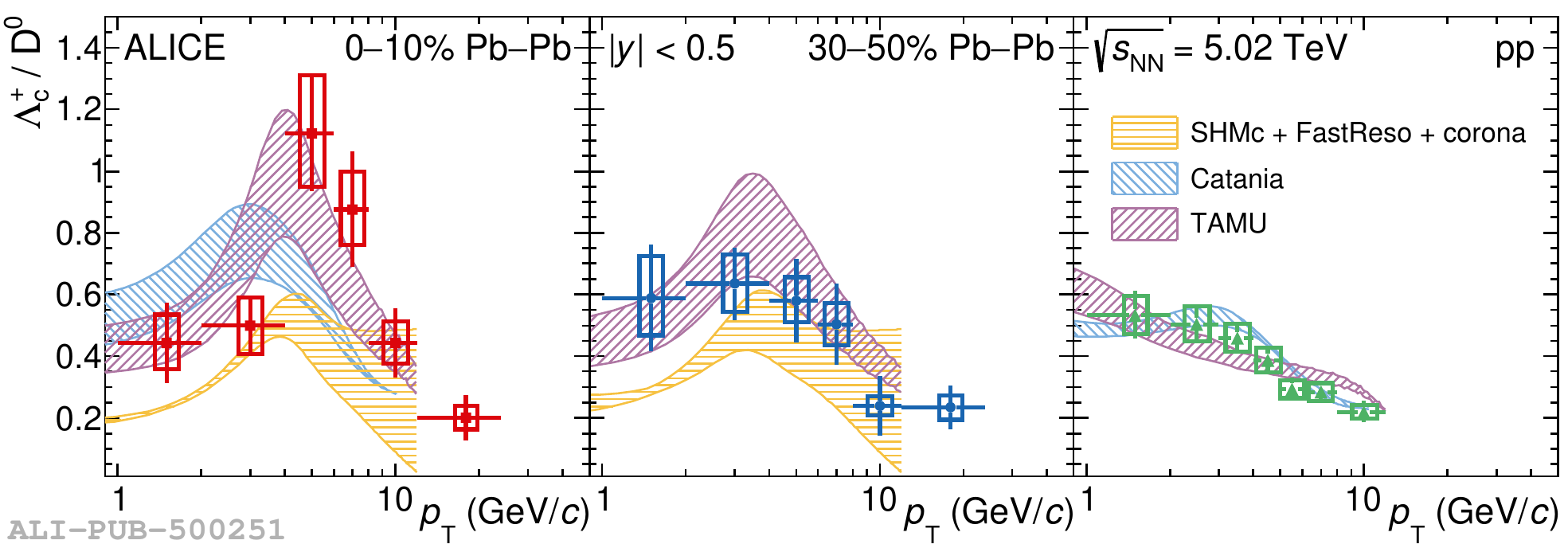}
\caption{$\Lambda_c/D^0$ ratios in \auau\  \cite{STAR:2019ank} collisions (left) and \pbpb\ \cite{ALICE:2021bib} collisions (right) as a function
of transverse momentum compared to several models as
discussed in the text.}
\label{fig:lambdacDratio}
\end{figure}

Measurements of the $\Lambda_c/D^0$ ratio in heavy-ion collisions at RHIC~\cite{STAR:2019ank} and  LHC~\cite{CMS:2019uws,ALICE:2021bib,LHCb:2022ddg} have provided
new constraints on models of hadron formation.
Two coalescence models \cite{He:2019vgs,Plumari:2017ntm},
have been compared to data from both RHIC and LHC  \cite{STAR:2019ank,ALICE:2021bib,CMS:2019uws}, see Fig.~\ref{fig:lambdacDratio}. 
Other models have been compared to one of the data sets~\cite{Andronic:2021erx,Zhao:2018jlw,Cho:2019lxb,Cao:2019iqs}. More precise data, necessary to fully constrain heavy-flavor hadron formation in the QGP, will be available in the future, see Sec.~\ref{sec:hf_future}.
In addition, recent data on the ratios $D_s^+/D^0$ at both RHIC and LHC~\cite{STAR:2021tte,ALICE:2021kfc} show significant enhancement at intermediate $p_T$ relative to more elementary collisions, suggesting that hadronization proceeds via coalescence in heavy-ion collisions.  Measurements of $B_s^0/B^+$ in
\pbpb\ collisions are also available which hint at an enhancement of this
ratio in \pbpb\ collisions compared to \pp\ collisions~\cite{CMS:2021mzx}.

Effective theories for charm and bottom quark jets in QCD matter have been developed~\cite{Kang:2016ofv,Li:2018xuv} and used to improve the description of heavy flavor parton showers and advanced the understanding of heavy flavor jet propagation in medium.   
At high $p_T$, QCD predicts an energy loss hierarchy: $\Delta E_b<\Delta E_c<\Delta E_q<\Delta E_g$~\cite{Buzzatti:2011vt}.
Heavy quark jet substructure  can provide clean information on 
the ``dead-cone'' effect~\cite{Dokshitzer:2001zm}. 
Following upon lower energy results by CMS~\cite{CMS:2013qak}, ATLAS made the first observation of a larger \RAA\ for $b$-quark initiated jets than light quark jets \cite{ATLAS:2022fgb}. 
The \RAA\ of $D$ mesons at RHIC and the LHC shows a suppression pattern similar to that of light hadrons while $R_{\rm AA}$ data from $B$ decays show less suppression than charm, revealing the anticipated mass hierarchy of parton energy loss~\cite{STAR:2021uzu,PHENIX:2022wim,CMS:2018bwt,CMS:2017uuv,CMS:2017qjw,ALICE:2019nuy,CMS:2022sxl}. 
Recent calculations have predicted that, for $p_T < 30$~GeV, the QGP-induced modification is largest for bottom quark jets~\cite{Li:2017wwc}. This inversion of the mass hierarchy of jet quenching relative to QCD expectations~\cite{Dokshitzer:2001zm} can be explored by sPHENIX. 

Lattice QCD-based studies determine the in-medium properties of hadrons and their dissolution through correlation functions, the Laplace transform of the spectral function.   
The main challenge of reconstructing the spectral functions is the limited Euclidean time direction extent. 
Lattice calculations of heavy flavor probes have matured significantly since LRP15 %
For example, lattice calculations with $N_{\tau}=12$ \cite{Bala:2021fkm} determined that the real part of the potential is not screened and is, instead, about the same as that in vacuum \cite{Bala:2021fkm}. The imaginary part of the potential, on the other hand, is quite sizable and increases with both temperature and the quark-antiquark separation \cite{Bala:2021fkm}.
Current lattice data on charm fluctuations and charm baryon number correlations hint at the existence of charm mesons and baryons above the crossover temperature \cite{Mukherjee:2015mxc}, and bottomonium spatial correlation functions provide constraints on the melting temperature of different bottomonium states \cite{Petreczky:2021zmz}. 

\begin{figure}[th]
\centering
\begin{minipage}[t]{0.32\linewidth}
\includegraphics[width=1\textwidth,height=5.8cm]{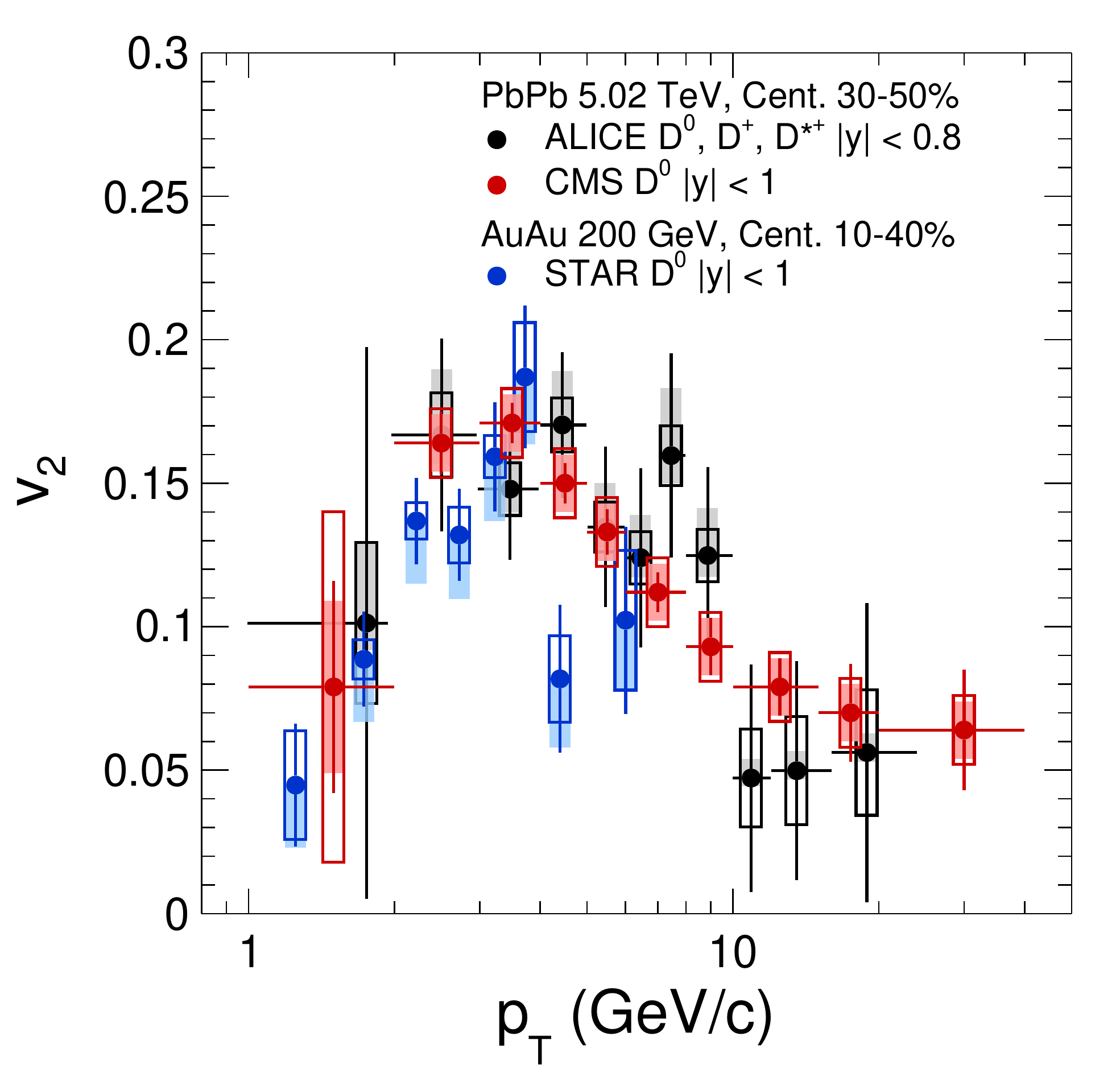}
\end{minipage}
\hspace{-0.1in}
\begin{minipage}[t]{0.33\linewidth}
\includegraphics[width=1\textwidth,height=5.8cm]{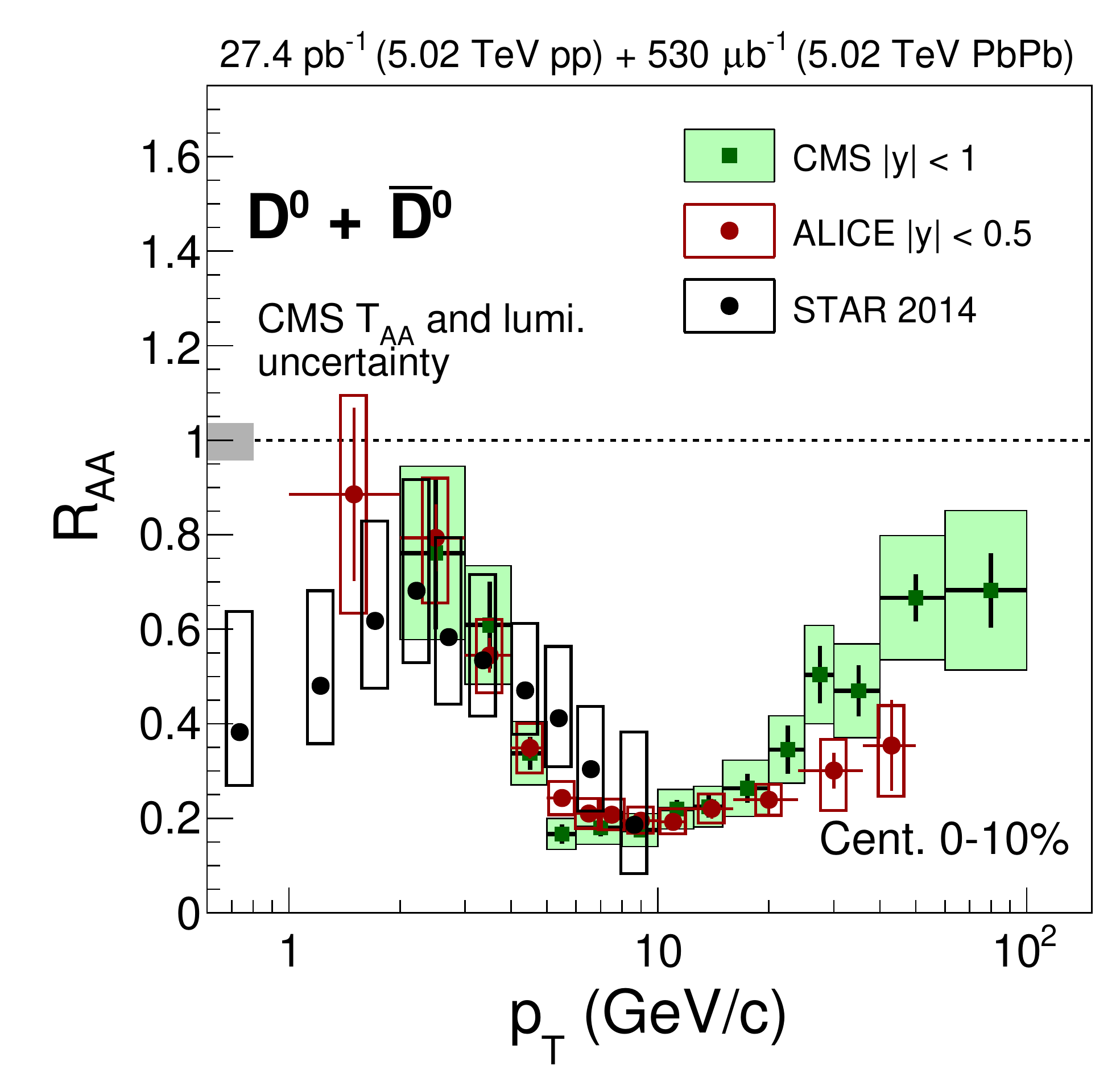}
\end{minipage}
\begin{minipage}[t]{0.33\linewidth}
\includegraphics[width=1\textwidth,height=5.6cm]{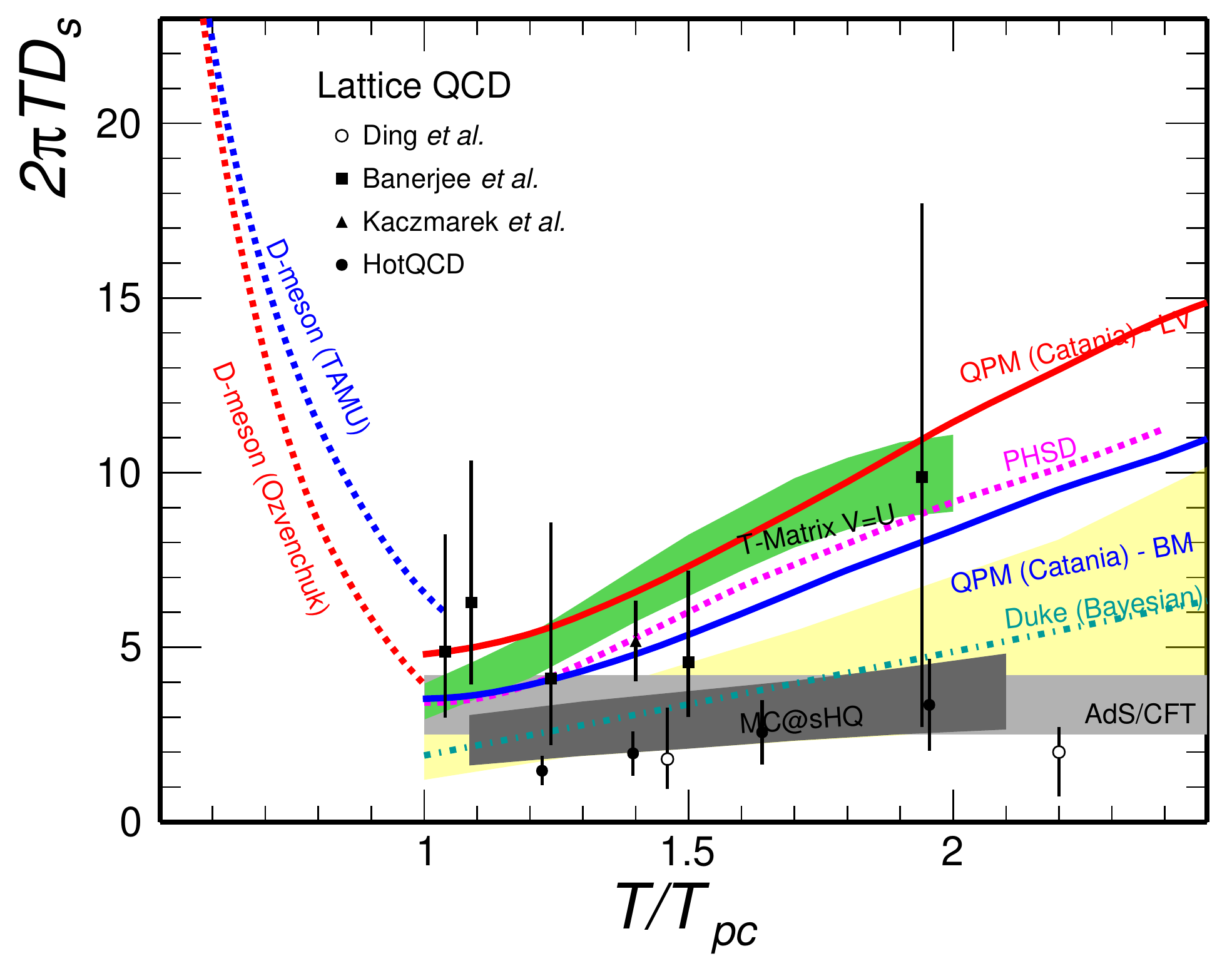}
\end{minipage}
\caption{
Left: $D$ meson \vtwo\  for \pbpb\ \cite{ALICE:2018lyv,CMS:2017vhp} and \auau\ ~\cite{Adamczyk:2017xur} collisions. 
Center: The nuclear modification factor of $D$-mesons for \pbpb\ \cite{ALICE:2018lyv,CMS:2017qjw} \auau\ ~\cite{Adam:2018inb}
collisions. 
Right: The temperature dependence of the charm quark spatial diffusion coefficient, $2\pi T$\Ds~\cite{Dong:2019byy}.
} 
\label{fig:v2-RAA}
\end{figure}

To address medium thermalization and extract the heavy quark diffusion coefficient, high precision data on the collective behavior of open heavy flavor hadrons, especially at low $p_T$, are needed. At the time of LRP15, 
little was known about charm quark diffusion due to the lack of experimental data so that the value of the charm quark scaled diffusion coefficient, $2\pi T$\Ds, extracted from various models, varied widely~\cite{Akiba:2015jwa}.
Recent data provide precision measurements of $D$-meson $R_{\rm AA}$ and $v_2$ over a wide $p_T$ region at both RHIC and LHC (left and center panels of Fig.~\ref{fig:v2-RAA})~\cite{Adamczyk:2017xur,Adam:2018inb,ALICE:2018lyv,CMS:2017vhp,CMS:2017qjw,CMS:2020bnz}. 
The right-hand side of Fig.~\ref{fig:v2-RAA} shows the temperature dependence of $2\pi T$\Ds~\cite{Dong:2019byy}, constrained to be $\sim$2--5 near $T_c$.
The spatial diffusion coefficient is proportional to the charm relaxation time, which can give a hint as to why charm participates in the flow of the system, as it was extracted to be smaller than the system lifetime from $D^0$ measurements \cite{ALICE:2021rxa,ALICE:2022wpn}.

Quarkonium measurements have been carried out at RHIC and the LHC in a variety of small and large systems.  
At $\sqrt{s_{NN}}=200$~GeV, the $J/\psi$ is suppressed in the most central collisions by a factor of 4-5 at both mid and forward rapidity \cite{PHENIX:2011img}. Precise $J/\psi$ data from d+Au~\cite{PHENIX:2012czk} collisions at $\sqrt{s_{NN}}=200$~GeV showed a $\sim 60$\% suppression at forward rapidity, making it clear that cold nuclear matter (CNM) effects are important~\cite{PHENIX:2019brm}). Incorporating these CNM effects, the RHIC quarkonium data are well described by transport calculations including dissociation in the medium as well as production by coalescence. 
Transport models also provided successful predictions for the energy dependence of $J/\psi$ production at RHIC \cite{PHENIX:2012xtg,STAR:2016utm} and LHC \cite{ALICE:2019lga}, with coalescence playing an important role at the highest energies.  
Coalescence may also be responsible for $J/\psi$ $v_2$, small at RHIC ~\cite{STAR:2012jzy} but significant at the LHC~\cite{ALICE:2020pvw}. The RHIC $J/\psi$ $v_2$ measurements in Au+Au collisions will still improve through analysis of the final PHENIX data and data from future STAR runs.

Comparative studies of quarkonia production in small systems at RHIC~\cite{PHENIX:2013pmn,PHENIX:2016vmz,PHENIX:2022nrm} and at the LHC~\cite{CMS:2016wgo,LHCb:2016vqr,ATLAS:2017prf,CMS:2018gbb,ALICE:2020tsj} found a factor of two greater suppression of the $\psi$(2S) compared to the $J/\psi$ at backward rapidity (where the final-state multiplicity is highest) while the modifications are similar at forward rapidity. 
The strong $\psi$(2S) suppression at backward rapidity may be due to the formation of small QGP droplets in $p+A$ collisions.  Similarly, CMS has
found a sequential suppression pattern of upsilon states in \pPb\
collisions~\cite{CMS:2022wfi}.
Measurements in large systems \cite{ALICE:2022jeh,CMS:2017uuv} and comparisons with calculations of transport and statistical models provide insight into the existence and properties of charmonium states in the QGP at the LHC.

Detailed studies of the modifications of the three 
$\Upsilon$ states at the energy densities produced at RHIC and LHC can provide strong constraints on models. At the LHC, CMS can fully resolve the three $\Upsilon$ states and has measured their modification in \pbpb\ collisions~\cite{CMS:2018zza}.  The existing $\Upsilon$ data from RHIC~\cite{PHENIX:2014tbe,STAR:2016pof,STAR:2022rpk} are more limited due to a combination of smaller production rates, acceptance, and mass resolution. 

\subsubsection{Initial State and Small-x} \label{sec:init}
The description of the initial state in heavy ion collisions has improved due to the use of high precision measurements of new observables and theoretical advances, for example the development of sophisticated three dimensional and dynamical initial state models, and extensive Bayesian analyses.

Nucleon substructure has been found to play a crucial role, in particular in small collision systems, where its inclusion is required to produce sufficient fluctuations to reproduce the anisotropy coefficients \cite{Moreland:2018gsh, Schenke:2021mxx}. Constraints on the subnucleon size scales have been obtained from both diffractive vector meson production at the Hadron-Electron Ring Accelerator (HERA)~\cite{Mantysaari:2016ykx} as well as studies of 
$p+A$ collisions \cite{Moreland:2018gsh}. At small $x$, within the CGC framework, direct constraints on the gluon distributions can be obtained from a variety of processes, including diffractive dijet and vector meson production, deeply virtual Compton scattering, or inclusive dijet production, and certain angular dependencies in all cases \cite{Mantysaari:2020lhf,Mantysaari:2019csc,Mantysaari:2019hkq,Mantysaari:2022sux}. Once constrained by measurements of these processes in ultraperipheral heavy ion collisions or future measurements at the EIC, the gluon distributions (Wilson lines) can be directly used in the same framework to describe the initial state in heavy ion collisions \cite{Krasnitz:1998ns,Krasnitz:1999wc,Krasnitz:2000gz,Schenke:2012wb,Schenke:2012hg}. This includes the transverse spatial distribution as well as longitudinal dependence obtained from small-$x$ evolution \cite{Schenke:2016ksl,Schenke:2022mjv}.

In addition to the initialization of the energy momentum tensor, sophisticated calculations require an initial condition for the ideally three dimensional distribution of conserved charges, including net-baryon, isospin, and strangeness densities (see e.g.~\cite{Carzon:2019qja}). Fluctuating initial net baryon distributions are particularly important when exploring net-proton fluctuations in the search for the QCD critical point \cite{An:2021wof}. We will discuss other theoretical aspects, including pre-equilibrium evolution and the transition to hydrodynamics, in Sec.\,\ref{sec:hydrotheory}.

It is also desirable to identify experimental information that isolates the impact of the initial conditions in order to determine initial conditions and transport properties of the evolving matter individually. 
For example, correlations of flow harmonics with the mean transverse momentum fluctuations have proven to be sensitive to initial state properties like the nucleon size and nuclear deformation, while being mostly insensitive to the transport properties of the medium \cite{Giacalone:2021clp,Bally:2022vgo}. 
The nucleon size, or more precisely the hot spot size in the initial energy density, has also been better constrained by using experimental information on the nuclear cross section in Bayesian analyses \cite{ALICE:2022xir,Nijs:2022rme}.

\begin{figure}[!ht]
\includegraphics[width=1\linewidth]{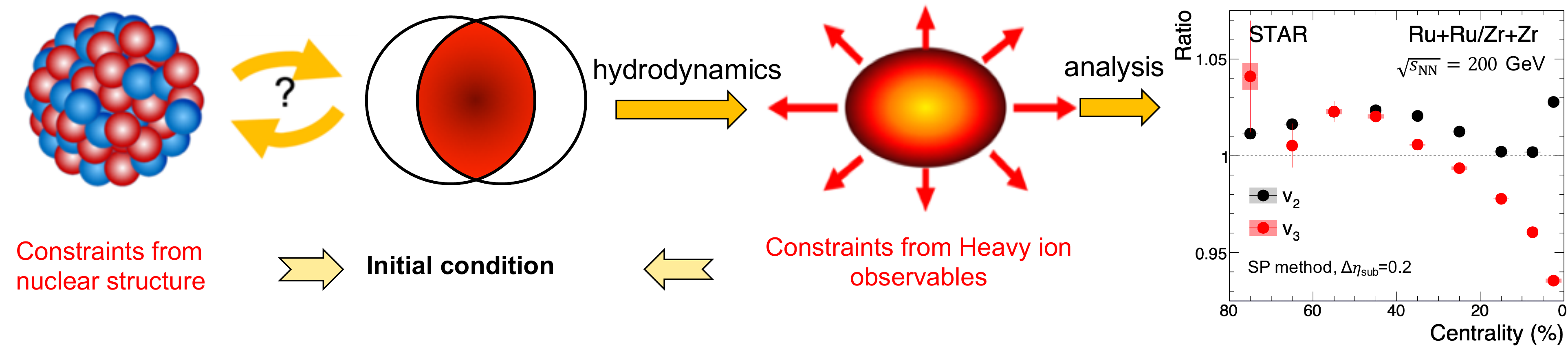}
\caption{\label{fig:ns1} 
Impact of isobar-like collisions on the initial condition of QGP. Better control on the initial condition can be achieved by exploiting the constraints from both the ratios of final-state observables ($v_2$ and $v_3$ on the right side \cite{STAR:2021mii}) and the nuclear structure knowledge (left side). 
}
\end{figure}

A powerful method to extract initial state properties is to consider collisions of systems with similar mass but different structural properties and compute the ratio of a given observable $\mathcal{O}$ in collisions of isobars $X$ and $Y$. 
Such a study was performed already using $^{96}$Ru+$^{96}$Ru and $^{96}$Zr+$^{96}$Zr collisions at RHIC~\cite{STAR:2021mii}. Ratios of more than ten observables have been measured, all displaying distinct and centrality-dependent deviations of up to 8\% from unity, two of which are reported in the right panel of Fig.~\ref{fig:ns1}~\cite{STAR:2021mii}. The ratios in central collisions are mostly impacted by deformation, while in mid-central collisions they are impacted by the nuclear radius and the surface diffuseness \cite{Jia:2021oyt,Jia:2022qgl,Xu:2021vpn,Xu:2021uar}. The behavior of $v_2$ and $v_3$ suggests a large  octupole deformation in Zr, $\beta_{3,\rm Zr}$, not predicted by mean field structure models~\cite{Cao:2020rgr}. Such rich and versatile information provides a new type of constraint on the initial conditions.

Another promising experimental tool to reveal the initial state of heavy nuclei is through photon-induced interactions, commonly known as \textit{ultra-peripheral collision} (UPCs), for which the impact parameter $b$ between the two colliding nuclei is greater than the sum of their radii - $2R_{A}$. Here, one or multiple photons emitted from one nucleus, interact with the other nucleus. Due to the large mass of the heavy nucleus, the emitted photons have very small virtualities or transverse momenta. There are generally three types of UPC physics processes studied: i) inclusive production; ii) semi-inclusive and/or jet production; iii) exclusive production. In the past decade, most of the UPC measurements focused on exclusive production, dominated by diffractive vector meson production. However, since LRP15, there has been an increasing number of studies on jet and inclusive particle photoproduction. For reviews of UPCs, see Refs.~\cite{Bertulani:2005ru,Baltz:2007kq,Contreras:2015dqa,Klein:2017nqo,Klein:2019qfb,Klein:2020fmr}. 

Exclusive vector meson (VM) photoproduction at high energy can be described as a quasi-real photon fluctuating into a quark-antiquark pair, which scatters off the target nucleus via a color neutral two-gluon exchange and then forms a VM. The sensitivity of this process to the spatial shape of the target makes it particularly valuable for constraining the initial conditions of nuclear collisions. At leading order, the cross section of this interaction scales as the square of the gluon density, which makes it a sensitive probe of the nuclear parton distribution functions (nPDFs). However, in a recent next-to-leading order (NLO) study \cite{Eskola:2022vaf}, the dependence on the gluon density is found to be different. 
Complementary studies of e.g.~photoproduction of dijets or open heavy flavor proceed via only a single gluon exchange, making them less sensitive to such theoretical uncertainties. Here, the $Q^2$ is set by the pair or dijet invariant mass, making it possible to probe parton distributions over a wide range of $Q^2$ with a single process.

Exclusive $\rho^{0}$ and $J/\psi$ production in UPCs have been systematically measured at RHIC and LHC \cite{STAR:2002caw,STAR:2007elq,STAR:2011wtm,STAR:2017enh,CMS:2016itn,ALICE:2013wjo,ALICE:2019tqa,ALICE:2015nmy,ALICE:2021gpt,CMS:2019awk,ALICE:2021tyx,LHCb:2021bfl,LHCb:2022ahs}. 
Although the systematic uncertainty related to the incoherent background is large, Ref.~\cite{STAR:2017enh} provided a first measurement of parton distributions inside a heavy nucleus by Fourier transforming the $\rho^{0}$ $|t|$ distribution to impact parameter space. A recent STAR measurement \cite{STAR:2022wfe} of azimuthal correlations of $\rho^{0}$ decays, has captured the nuclear geometry of an Au nucleus via quantum interference, linking UPC physics to quantum information science, and providing the first measurement of neutron skin from UPCs.

Measurements at the LHC~\cite{CMS:2016itn,ALICE:2013wjo,ALICE:2019tqa,ALICE:2015nmy,ALICE:2021gpt,ALICE:2021tyx,LHCb:2021bfl,LHCb:2022ahs} have shown a significant suppression of exclusive $J/\psi$ photoproduction in heavy nuclei over a wide range of rapidity, with respect to a free nucleon. This observation is qualitatively consistent with both the nuclear shadowing model in the leading twist approximation and gluon saturation models. Besides UPC VM in heavy nuclei, new experimental measurements of exclusive VM photoproduction in non-UPC heavy-ion collisions~\cite{ALICE:2015mzu,Zha:2017jch,STAR:2019yox}
can provide insight into the dynamics of photoproduction and nuclear reactions. Measurements in asymmetric collision systems~\cite{STAR:2021wwq} can probe the structure of the smaller nucleus at small $x$. 

Exclusive dijets in \pbpb\ UPCs have recently been measured by CMS \cite{CMS:2022lbi}. 
The dijet system can be used to reconstruct the initial scattering kinematics, and study the nPDFs, providing early access to some of the important physics goals of the EIC.
Dijet events have also been observed by ATLAS in events with no activity in either Zero Degree Calorimeter (ZDC) ($0n0n$), and the distributions have been found to resemble expectations from diffractive dijet production~\cite{Guzey:2016tek}. Diffractive dijet production is sensitive to the gluon distributions in nuclei, as well as their polarization, which is expected to lead to distinctive angular correlations \cite{Dumitru:2018kuw,Mantysaari:2019csc}. CMS measured angular correlations between two jets in events with rapidity gaps in both directions \cite{CMS:2022lbi}. Model comparisons \cite{Jung:1993gf,Hatta:2020bgy,Hatta:2021jcd} indicate that more work is needed to fully capture the interesting underlying physics. 

\subsubsection{Small Size Limit of the QGP}
\label{sec:smallprogress}
\begin{figure}
\centering
\includegraphics[width=0.7\textwidth]{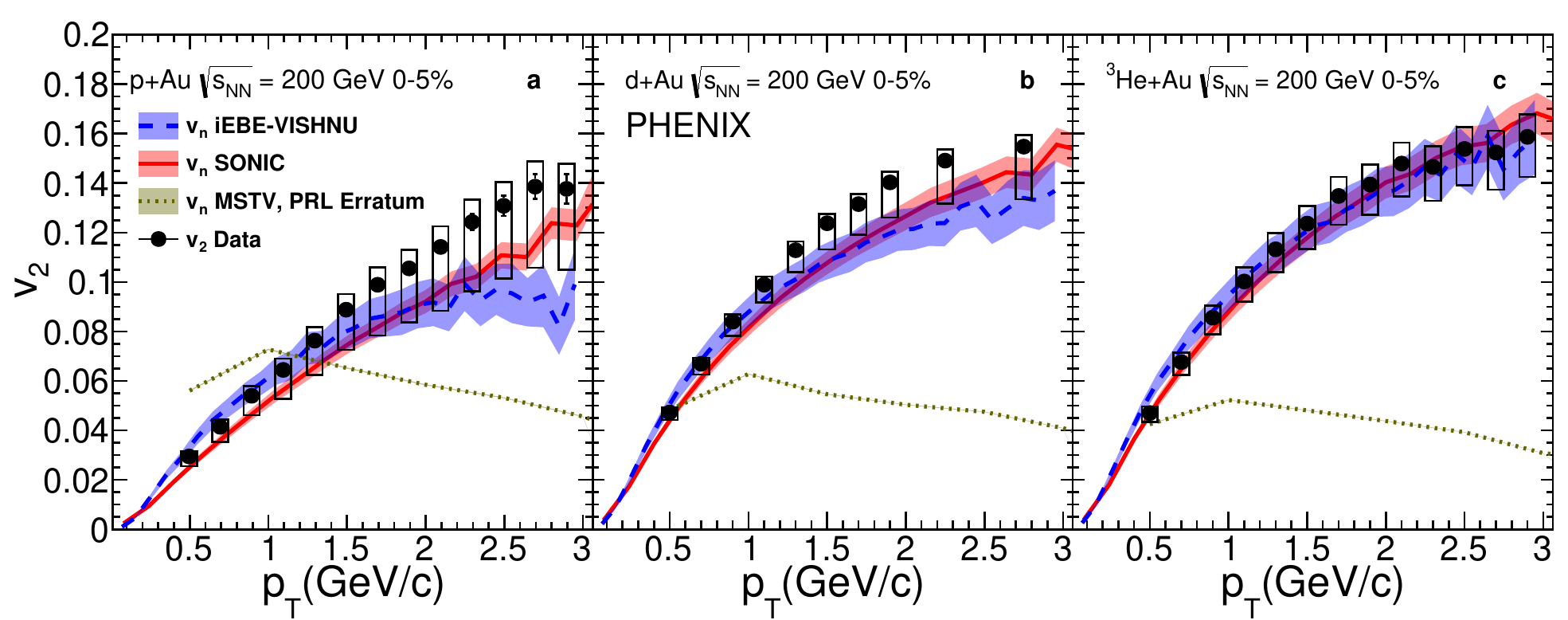}
\includegraphics[width=0.7\textwidth]{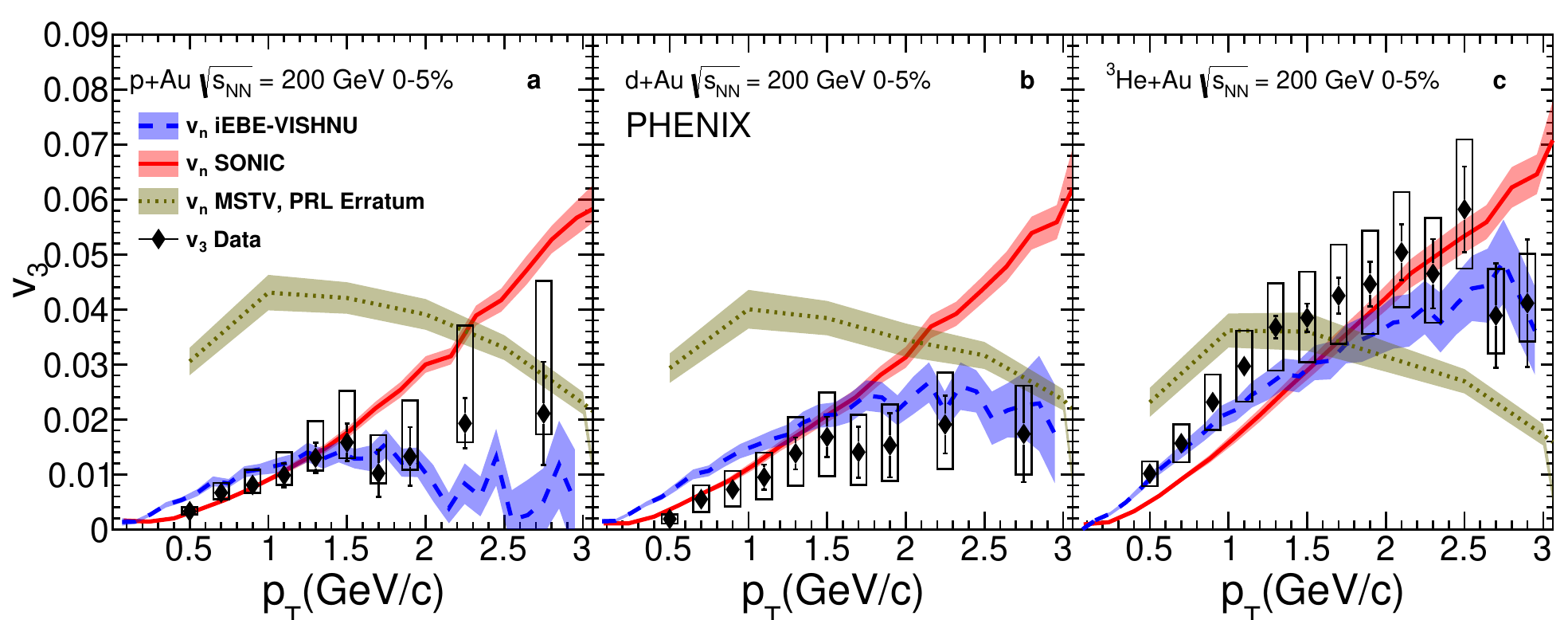}
\caption{Measurement of \vtwo\ (top) and \vthree\ (bottom) for
charged particles
in \pAu, \dAu, and \HeAu\ collisions as a function
of \pt.  
Calculations from two hydrodynamic models \cite{Shen:2016zpp,Habich:2014jna}
and a CGC based model \cite{Mace:2018vwq} are shown.
Figure from Ref.~\cite{PHENIX:2018lia}.}
\label{fig:phenix_smallvn}
\end{figure}

The discovery of flow-like signatures in \pp\ and 
\pPb\ collisions at the LHC~\cite{CMS:2010ifv,CMS:2012qk,ALICE:2012eyl,ATLAS:2012cix}
opened up a new field of study of the small size limit 
of QGP formation (for recent reviews see 
Refs.~\cite{Dusling:2015gta,Nagle:2018nvi}). Studies revealed a striking collective behavior of the measured $v_n$ for particles emitted 
in \pp\ and \pPb collisions~\cite{CMS:2014und,ATLAS:2015hzw,CMS:2015yux,CMS:2016fnw,ATLAS:2017rtr,CMS:2019lin,CMS:2019wiy,CMS:2017xnj,CMS:2017kcs,CMS:2022bmk}.
A stringent control experiment was performed at RHIC, using
three small collision systems: \pAu, \dAu, and \HeAu. The observed \vtwo\ and \vthree\ results
were found to agree with calculations in which the \vn\ values have their origin in the hydrodynamic evolution of the initial collision geometry in the three 
systems~\cite{PHENIX:2018lia}, confirming expectations that the geometry of the initial collision drives the observed \vn\ values in small systems.
The measurements in the three collision systems, shown in Fig.~\ref{fig:phenix_smallvn},
agree well with two hydrodynamic 
calculations~\cite{Habich:2014jna,Shen:2016zpp} and
disagree with a calculation based on a picture where
the observed anisotropies have non-hydrodynamic origin~\cite{Mace:2018vwq}.
However, the $v_n$ signal is found to be sensitive to the choices of $\eta$ range used in the two-particle correlation method \cite{STAR:2022pfn}, which suggests either a significant longitudinal decorrelation effect or a possible role of subnucleonic fluctuations (also see \cite{STAR:2022pfn}). 
Measurements of the production of strange hadrons in high 
multiplicity \pp\ collisions 
smoothly connect to what is seen in \pPb\
collisions~\cite{ALICE:2016fzo}, with this trend continuing towards the largest systems.
Additionally,
\vtwo\ and \vthree\ have been shown to be finite in high multiplicity photo-nuclear
\pbpb\ collisions~\cite{ATLAS:2021jhn} (but consistent with zero in $e^++e^-$ \cite{Badea:2019vey}, $e+p$ \cite{ZEUS:2019jya,ZEUS:2021qzg}, and $\gamma+p$ \cite{CMS:2022doq} collisions).  In high multiplicity photo-nuclear \pbpb\ collisions, the dominant processes are those in which the 
photon fluctuates into a vector meson such as the $\rho$
or $\omega$~\cite{Bertulani:2005ru}, which then interacts with the lead nucleus
in much the same way a proton would. First hydrodynamic calculations applied to this system show behavior consistent with experimental results \cite{Zhao:2022ayk}.
The signatures of collectivity in small systems have expanded the range of systems in which the QGP is studied and hydrodynamic models are challenged to be reliable in these smaller, more intense,
shorter lived systems. The theoretical challenges are discussed in more detail in Sec.\,\ref{sec:hydrotheory}.

\begin{figure}
\centering
\includegraphics[width=0.8\textwidth]{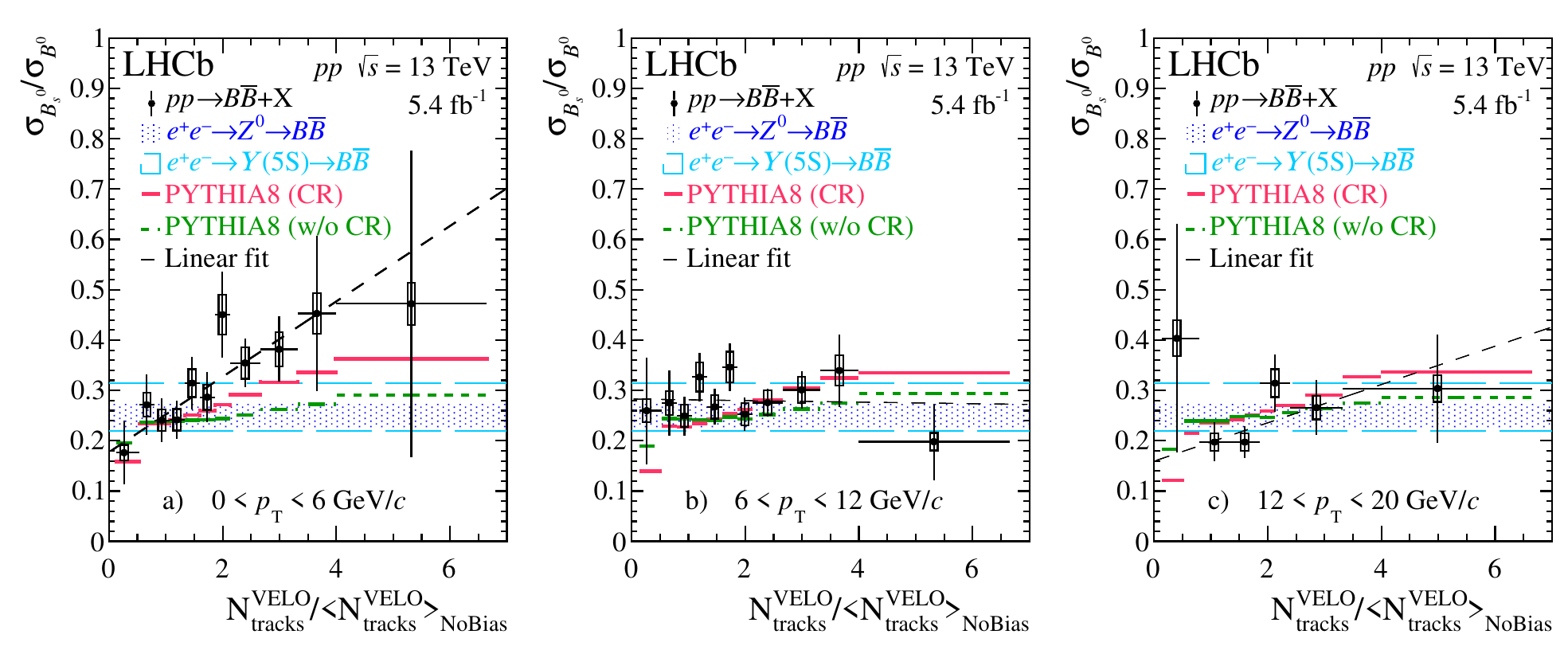}
\caption{Ratio of $B^0_s/B^0$ in \pp\ collisions as a function
of the charged particle multiplicity scaled by the average multiplicity in \pp\ 
collisions for three meson \pt\ selections:
0--6~GeV (left), 6--12 GeV (middle), 12--20~GeV (right). 
Expectations for production using the PYTHIA generator (with and without color reconnections) are also shown, along with the ranges measured previously in $e^+ e^-$ collisions.  Figure from Ref.~\cite{LHCb:2022syj}.}
\label{fig:LHCbHFmesons}
\end{figure}

In addition to hydrodynamic signatures, hints at a modification of the hadron formation process for mesons containing heavy quarks has been observed in small collision systems
relative to expectations from $e^+e^-$ collisions.  These measurements
are related to similar measurements in heavy ion collisions discussed in Sec.~\ref{sec:heavy}.  LHCb has measured the 
charged particle multiplicity dependence of the $B^0_s/B^0$ ratio as a function
of the multiplicity of charged particles in \pp\ collisions~\cite{LHCb:2022syj}.  
They have
found that the ratio shows little multiplicity dependence for high
transverse momentum mesons, but has a clear increase with the number of charged
particles for mesons with $\pt <$~6~GeV, see Fig.~\ref{fig:LHCbHFmesons}.

QGP formation is typically accompanied by evidence of jet quenching. However, no direct evidence of jet
quenching has been found in small collision systems~\cite{CMS:2016xef,CMS:2016svx}. 
 The nuclear modification factors for 
jets~\cite{PHENIX:2015fgy,ALICE:2015umm,ATLAS:2014cpa,CMS:2016wma} and hadrons~\cite{STAR:2003pjh,PHENIX:2003qdw,CMS:2016xef,CMS:2019isl} show no significant suppression at mid-rapidity, and measurements of the semi-inclusive distribution of charged jets recoiling from a high $p_T$ hadron trigger indicate little to no energy loss in small systems \cite{ALICE:2017svf}. One indirect suggestion of jet quenching is the observation 
of a non-zero \vtwo\ for charged particles in \pPb\ collisions at high \pt~\cite{ATLAS:2019vcm}.  In \pbpb\ collisions, the non-zero \vtwo\ is typically attributed to the path length dependence of jet quenching within the QGP.  However, in \pPb\ collisions, these charged particles are not significantly suppressed and no existing model has been able to explain both the overall rate and the \vtwo\ of these high \pt\ charged particles in \pPb\ collisions.
The origin of this effect in \pPb\ collisions is not known.
Observation of heavy flavor hadron \vtwo\ in \pp\ and \pPb\ also provides indirect evidence for interactions of hard probes with a QGP medium in small systems~\cite{ALICE:2017smo,CMS:2018loe,CMS:2018duw,ATLAS:2019xqc,CMS:2020qul}.

The most direct way to understand this potential tension is to measure small, symmetric collisions at RHIC and the LHC. 
\oo\ collisions have been run at RHIC and proposed
for the LHC in Run 3.
This system avoids the large theoretical and experimental uncertainties associated with jet quenching measurements in peripheral collisions and provides a system size, in terms of the number of participating nucleons, very similar to peripheral \pbpb\ collisions.
This will provide a benchmark for how much jet quenching 
(if any) is present in such small collision systems, and provide the crucial link between understanding the QGP
in large, symmetric collision systems and small asymmetric collision systems.

\subsubsection{Mapping the QCD Phase Diagram} \label{sec:phasediagram}
\begin{figure}[ht]
  \begin{center}
  \vspace{-1cm} 
    \includegraphics[width=0.54\textwidth]{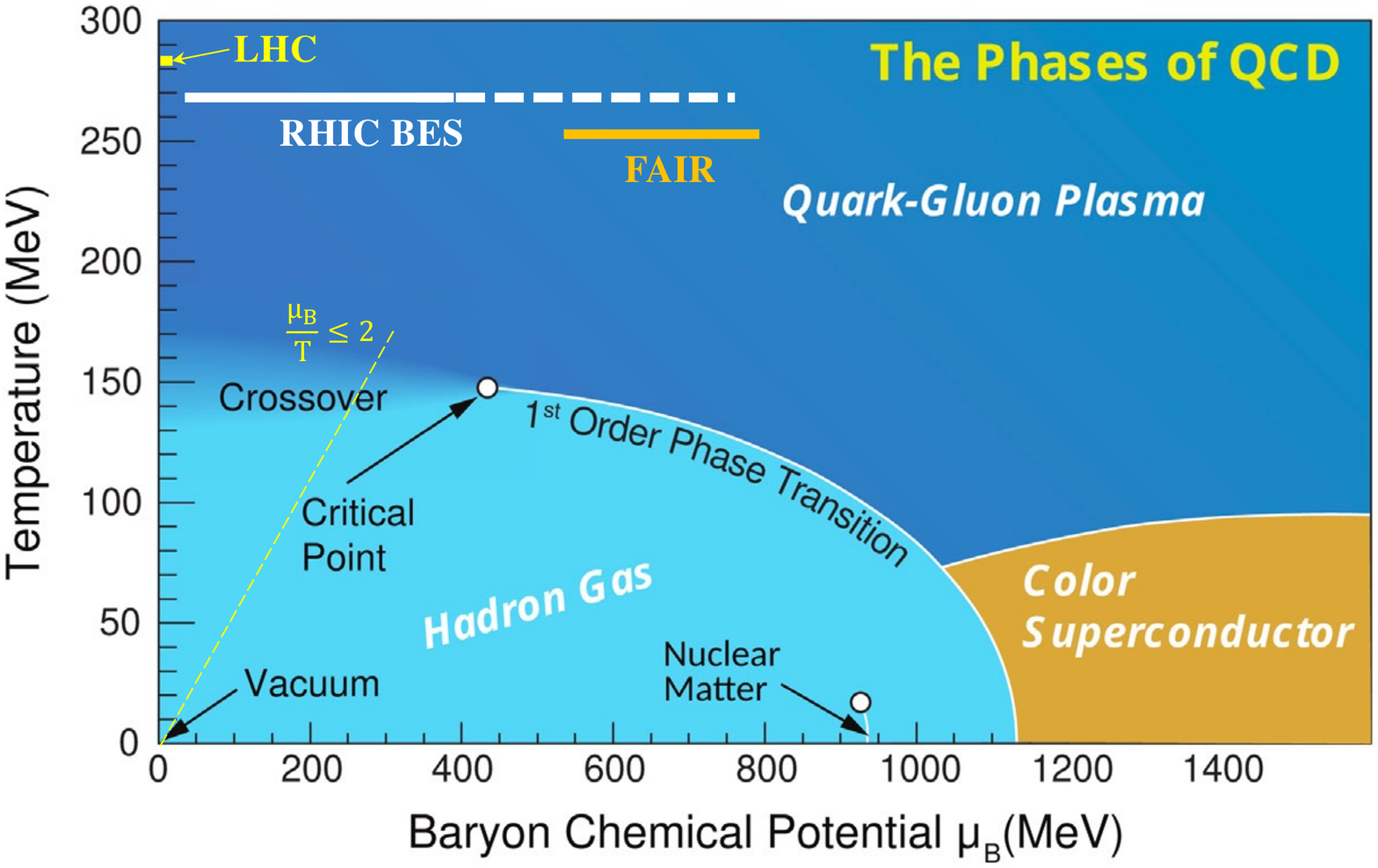}
    \includegraphics[width=0.45\textwidth]{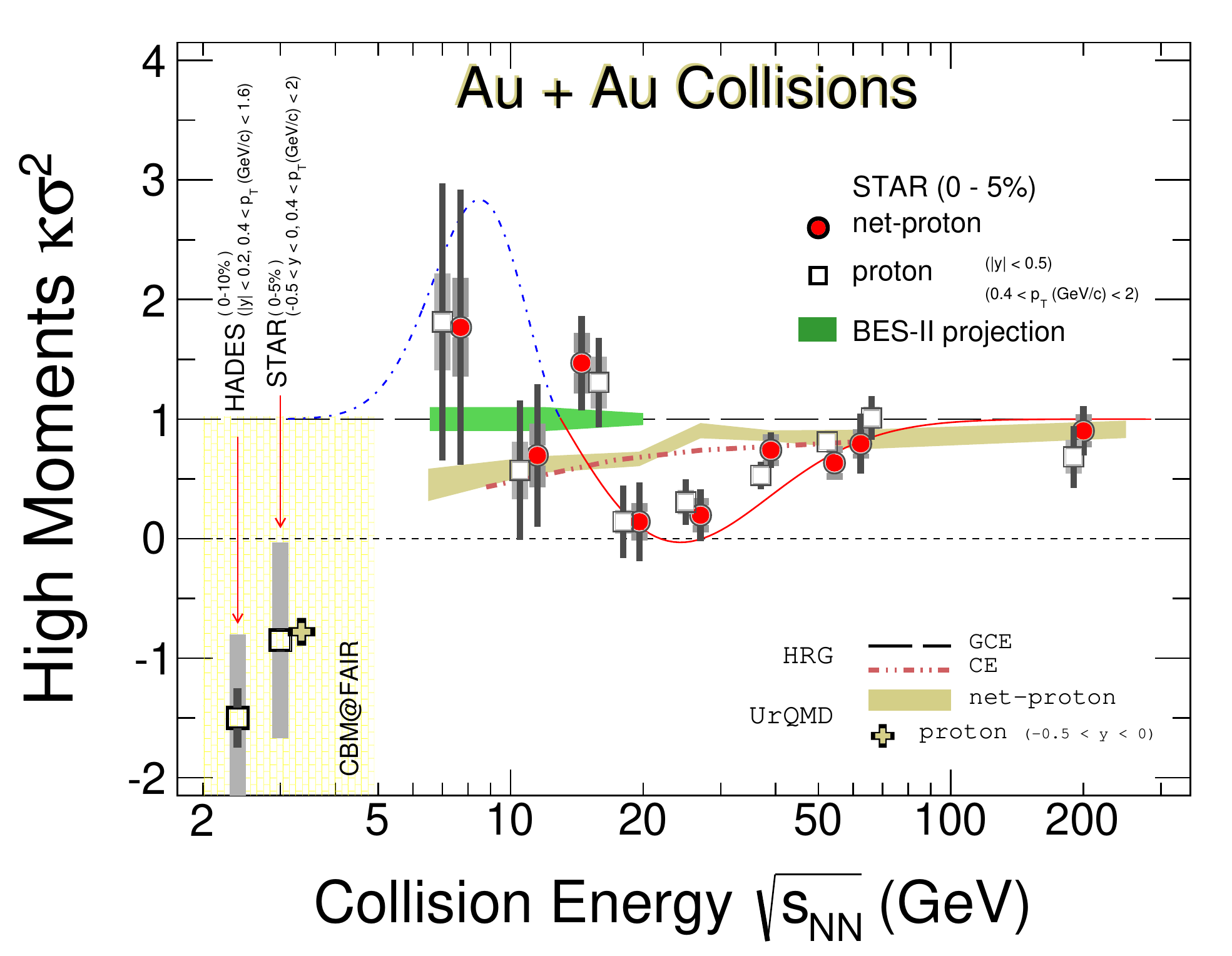}
  \end{center}
\vspace{-0.6cm}  
\linespread{1.0}\selectfont{} 
    \caption{Left: Sketch of the QCD phase diagram, incorporating a conjectured critical end point and first order transition. 
    The yellow line indicates the region of the phase diagram where lattice QCD can reliably predict the smooth crossover region of the hadron-QGP transition, up to $\mu_B/T \leq 2$. Figure adapted from \cite{Geesaman:2015fha}. 
    Right: Energy dependence of the net-proton (filled circles) and proton (open squares) high moments from Au+Au collisions ~\cite{STAR:2021iop,Adamczewski-Musch:2020slf,STAR:2021fge}. Model results from Hadron-Resonance Gas (HRG) model~\cite{Braun-Munzinger:2020jbk}, and UrQMD~\cite{Bleicher:1999xi,Bass:1998ca} are shown. The thin red and blue dot-dashed lines are qualitative predictions~\cite{Stephanov:2011pb} in the presence of a critical point. Adapted from Ref.~\cite{STAR:2021fge}.}
\label{fig:phasestructure} 
\end{figure}

Based on lattice QCD calculations, the QGP-hadron gas transition at vanishing net-baryon density is understood to be a smooth crossover with the transition temperature $T_c = 156\pm1.5$ MeV~\cite{Aoki:2006we}. 
Model studies indicate a first-order phase boundary at large net-baryon density (baryon chemical potential  $\mu_B$) \cite{Fukushima:2013rx}.
If there is a crossover and a first order transition line, they will be joined at the QCD critical point \cite{Stephanov:1998dy,Stephanov:1999zu,Bzdak:2019pkr}. 
State-of-the-art lattice calculations further predicted that the chiral crossover region extends into the finite chemical potential region $\mu_B/T \le 2$~\cite{Bazavov:2020bjn}, see Fig.~\ref{fig:phasestructure}. Precise calculations in the higher $\mu_B$ region become more difficult and experimental measurements are essential to determine if a QCD critical point exists.  

The BES program at RHIC, colliding heavy nuclei in the center of mass energy range $\sqrt{s_{NN}}$\,=\,3\,--\,200\,GeV, was initiated in 2008 in order to search for the QCD critical point and study the nuclear matter EoS in the high baryon density region~\cite{Akiba:2015jwa,Geesaman:2015fha}. 
The BES phase-I (BES-I) program was conducted during 2010--2014, covering the collision energies between 7.7 and 200 GeV (solid white line in Fig.~\ref{fig:phasestructure} (left) indicating the $\mu_B$ range). BES-II took place during 2019--2021, focusing on the center of mass energy range $\sqrt{s_{NN}}$\,=\,3\,--\,19.6\,GeV (dashed white line in Fig.~\ref{fig:phasestructure}). While data from the energy range 7.7 -- 19.6 GeV were collected in collider mode, data from fixed-target mode was also collected in the range 3 -- 13.1 GeV  (see e.g.\,\cite{STAR:2020dav}). In the overlapping energy range, the event statistics from BES-II were improved by a factor of 20 to 40 compared to that of BES-I. 
In order to reach the desired luminosity, RHIC  underwent an electron cooling upgrade, Low Energy RHIC electron Cooling (LEReC), which began operation during the BES-II RHIC Runs in 2019--2021. 
To maximize the physics output, the STAR detector has implemented a series of key subsystem upgrades: the inner Time Projection Chamber (iTPC), the Event Plane Detector (EPD) and the endcap Time-of-Flight (eTOF) Detector to enhance particle identification capabilities and extend kinematic coverages. 

All of the BES-I data have been analysed and most of the results are published. 
Evidence for the dominance of the QGP phase or the hadronic phase at different collision energies have been demonstrated in three key observations.
(i) High-$p_{T}$ Parton Energy Loss: the strong suppression in the leading hadron $R_{AA}$ at $p_T \ge$ 5 GeV/c, a signature of the formation of QGP, in central Au+Au collisions at $\sqrt{s_{NN}}$ = 200 GeV was found to gradually disappear and $R_{AA}$ became even larger than unity in central Au+Au collisions for energies lower than 19.6 GeV~\cite{STAR:2017ieb}.  (ii) Partonic Collectivity: Quark number scaling, found in the $v_2$ for all hadrons, an indication of QGP formation, has been found to persist down to 7.7 GeV Au+Au collisions~\cite{STAR:2013cow}. This implies that the partonic degrees of freedom remain dominant in these collisions. (iii) Critical Fluctuation: Moments (and their ratios) of net-baryon fluctuations are expected to be sensitive to the existence of critical point and phase boundary. High moments of net-protons (a proxy for net-baryons) from central 200 GeV in Au+Au collisions, $C_4/C_2$, $C_5/C_1$, and $C_6/C_2$, are all found to be consistent with lattice QCD predictions of a smooth crossover chiral transition~\cite{STAR:2020tga,STAR:2021rls,Aoki:2006we,STAR:2022vlo}. Hydrodynamic calculations of non-critical contributions to proton number cumulants indicate that the Au+Au data are consistent with non-critical physics at center of mass energies above 20 GeV \cite{Vovchenko:2021kxx}.
In Au+Au collisions at 3 GeV, on the other hand, hadronic interactions are evident from the measurements of moments of proton distributions, collective flow and strangeness production~\cite{STAR:2021fge,STAR:2021yiu,STAR:2021hyx}. These results imply that the QCD critical point, if it exists, should be accessible in collisions with center of mass energies between 3 and 20 GeV.

Figure~\ref{fig:phasestructure} (right) shows recent results on the fourth-order net-proton and proton high moments in central Au+Au collisions measured in BES-I~\cite{STAR:2020tga,STAR:2021fge,HADES:2020wpc} compared to models. The thin red and blue dot-dashed lines are expected from a qualitative prediction \cite{Stephanov:2011pb} 
due to critical phenomena. The hadronic transport model Ultrarelativistic Quantum Molecular Dynamics (UrQMD)~\cite{Bass:1998ca,Bleicher:1999xi}
and a thermal model with a canonical ensemble~\cite{Braun-Munzinger:2020jbk} represent non-critical baselines. Current error bars do not allow for a clear conclusion, but RHIC BES-II results will provide significantly improved statistical precision (and likely reduced systematic uncertainties), as indicated by the green band in the figure. The extended acceptance and particle identification in a larger rapidity region (from $|y|<$0.5 to $|y|<$0.8) will allow more systematic investigation into the nature of these fluctuations. 
Much progress has also been made by the BEST Collaboration \cite{An:2021wof} and others towards establishing a framework for calculations of observables sensitive to the critical point and a first order phase transition. This includes lattice QCD and effective field theory calculations, further discussed in Sec.\,\ref{sec:eff}, as well as improvements to the initial state and hydrodynamic description, especially the inclusion of propagation of stochastic and critical fluctuations, discussed in Sec.\,\ref{sec:pheno}. 

\subsubsection{Chirality and Vorticity in QCD}
\label{sec:chiralityvorticity}
\subsubsubsection{Searches for the chiral magnetic effect}
\label{sec:cme}
The creation of electric current in the direction of a magnetic field due to the imbalance of chirality is called the chiral magnetic effect (CME). A decisive experimental test of this phenomenon in a QCD medium has been among the major scientific goals of the RHIC and LHC heavy ion programs. The existence of the CME in the QCD medium formed in relativistic collisions would establish the existence of chiral fermions over sufficient timescales and therefore the restoration of chiral symmetry of QCD in these collisions. It would also indicate that such collisions form regions of space where the left-right symmetry (${\rm U}_A(1)$) is broken by local P and CP symmetry breaking in the strong interaction. Finally, it would also prove that ultra-strong electromagnetic fields are created in such collisions~\cite{Kharzeev:1999cz,Kharzeev:2004ey}. Other observables potentially sensitive to the creation of a strong magnetic field were also discussed in the literature \cite{Gursoy:2014aka,Das:2016cwd,Gursoy:2018yai,Sun:2020wkg,Oliva:2020doe}.

\begin{figure}[ht]
    \centering
    \includegraphics[width=0.8\textwidth]{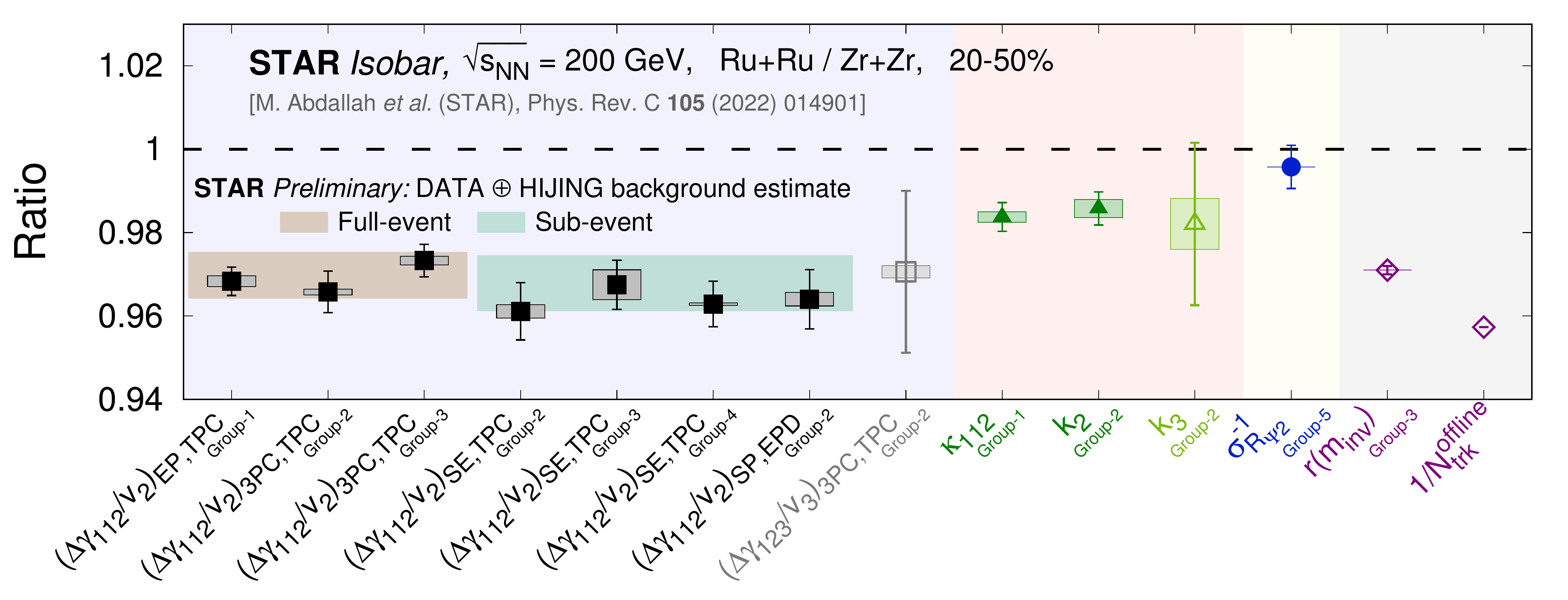}
    \caption{Ratios of observables in Ru+Ru over Zr+Zr collisions from the STAR isobar blind analysis~\cite{STAR:2021mii}. The ratios of the CME-sensitive observables (solid markers) are found to be below unity and close to the ratio of inverse multiplicities ($N_{\rm trk}^{\rm offline}$). The tan and aqua bands show background estimates calculated using data and the HIJING model~\cite{Feng:2022yus}. No significant CME signal difference between the two isobars is observed.}
    \label{fig_isobar_results}
\end{figure}
In heavy-ion experiments, a signal of the CME is the separation of charge across the reaction plane, oriented perpendicularly to the magnetic field direction in non-central collisions~\cite{Voloshin:2004vk}.  Evidence of such charge separation was first reported by the STAR collaboration in \auau\ and Cu+Cu collisions~\cite{Abelev:2009ac}. However, backgrounds driven by flow, coupled with local charge conservation  \cite{Wang:2009kd, Bzdak:2009fc, Bzdak:2010fd, Schlichting:2010qia, Pratt:2010zn} and non-flow effects, 
dominate the measurements \cite{Abelev:2009ac,Abelev:2009ad,Abelev:2012pa,Adamczyk:2013hsi, Adamczyk:2013kcb, Adamczyk:2014mzf, Adam:2015vje, Khachatryan:2016got, Acharya:2017fau,Sirunyan:2017quh,STAR:2019xzd, STAR:2020gky, STAR:2021pwb, CMS:2020bnz, ALICE:2020siw, Collaboration:2022flo}. 
A similar charge separation observed in small colliding systems, where there is no correlation between the magnetic field direction and reaction plane, at both the LHC \cite{Khachatryan:2016got} and RHIC \cite{STAR:2019xzd} confirmed the dominance of the background. 
Subsequent measurements used novel approaches to reduce or eliminate background contributions to the CME.
Using event-shape engineering techniques, the CMS and ALICE collaborations found an upper limits of 7\% and 26\%, respectively, for the CME contribution to the measured signal at 95\% confidence level in Pb+Pb collisions at the LHC~\cite{Sirunyan:2017quh,Acharya:2017fau}. Studying charge separation as a function of pair invariant mass, the STAR collaboration found an upper limit of 15\% CME contribution to the measured signal at the 95\% confidence level in Au+Au collisions at RHIC \cite{STAR:2020gky}. Using the spectator and participant planes STAR measurements indicate a possible 10\% CME signal at a significance on the order of 2 standard deviations in Au+Au collisions at 200 GeV \cite{STAR:2021pwb}.

By far the most controlled and precise CME search was performed by the STAR collaboration using the collisions of isobars Ru+Ru and Zr+Zr at RHIC~\cite{STAR:2021mii}. Ru+Ru collisions are expected to produce an about 5--9\% larger $B$ field than Zr+Zr, hence a 10--18\% larger CME signal because of its $B^2$ dependence. 
The RHIC running conditions for Zr+Zr and Ru+Ru collisions provided stringent controls on the systematic uncertainties. The STAR collaboration performed a blinded analysis. The
results are shown in Fig.~\ref{fig_isobar_results}. The ratio of CME-sensitive observables in Ru+Ru over Zr+Zr is found to be below unity with a precision down to 0.4\% indicating no pre-defined signature of CME is observed. Estimates of background using data and HIJING, shown by bands on Fig.~\ref{fig_isobar_results}, indicate no significant CME signal difference between the two isobars; a quantiative determination of an upper limit is underway~\cite{Feng:2022yus}.

\subsubsubsection{Vorticity} \label{sec:vorticity}
Despite early predictions~\cite{Liang:2004ph,Becattini:2007sr} of hadronic polarization resulting from a rotating QGP,
the first observation~\cite{STAR:2017ckg} of the phenomenon was a nonvanishing $\Lambda/\overline{\Lambda}$ polarization at midrapidity along the direction of the global angular momentum $\hat{J}$, in semi-peripheral collisions at RHIC BES energies.
Many viscous hydrodynamic calculations were 
able to reproduce the observations without special ``tuning,'' using a generalized Cooper-Frye formula~\cite{Becattini:2016gvu} to connect the fluid to 
particle degrees of freedom; in this freezeout scenario, fluid vorticity is essentially a spin chemical potential.
This achievement alone is a nontrivial confirmation of the validity of the hydrodynamic, 
local-equilibrium paradigm underlying our understanding of the bulk system created in heavy ion collisions.
This connection was strengthened by subsequent observation~\cite{STAR:2020xbm} of $\Xi$ and $\Omega$ global polarization at RHIC.

\begin{figure}[htb]
\hspace*{0.02\textwidth}
\begin{minipage}{0.5\textwidth}
\includegraphics[width=\columnwidth]{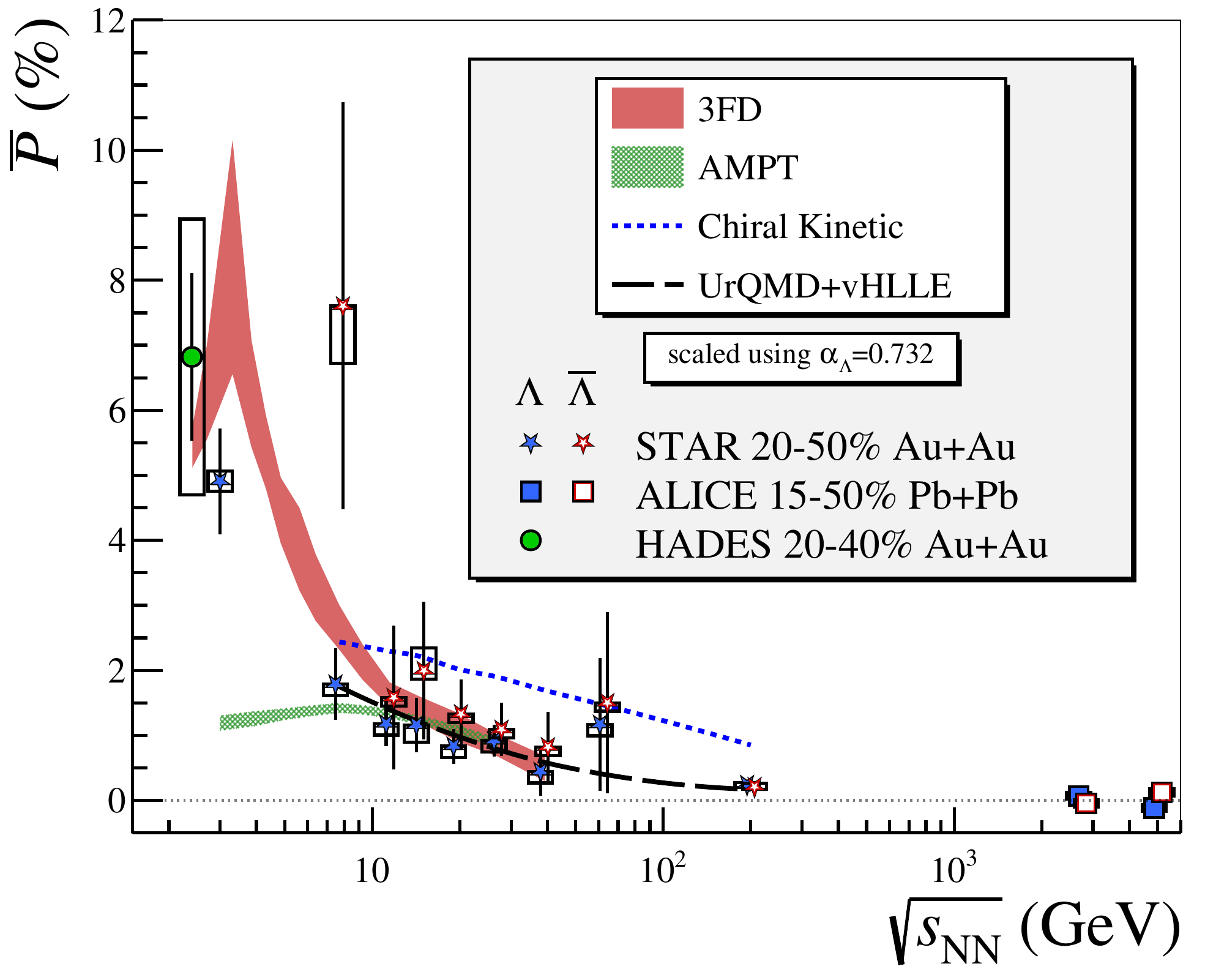} 
\end{minipage}
\hspace*{0.03\textwidth}
\begin{minipage}{0.4\textwidth}
\linespread{1.0}\selectfont{} 
\caption{
Adapted from~\cite{STAR:2021beb}.
World dataset~\cite{STAR:2007ccu,STAR:2017ckg,STAR:2018gyt,ALICE:2019onw,STAR:2021beb,HADES:2022enx} of global polarization of $\Lambda$ and $\overline{\Lambda}$ from midcentral heavy ion collisions vs.~$\sqrt{s_{ NN}}$.
Statistical uncertainties are represented with lines while systematic uncertainties are represented with boxes. 
All results are scaled~\cite{Becattini:2020ngo}
    using the decay parameter $\alpha_\Lambda=0.732$ \cite{ParticleDataGroup:2020ssz}.
    Curves are calculations with
    a hybrid hydrodynamic model \cite{Karpenko:2016jyx},
     chiral-kinetic transport \cite{Sun:2017xhx}, 
    a Monte Carlo transport model AMPT \cite{Guo:2021udq},
     and a three-fluid hydrodynamic calculation~\cite{Ivanov:2020udj}.}
\label{fig:polarization}
\end{minipage}\end{figure}
Despite the fact that higher $\sqrt{s_{NN}}$ implies 
larger system angular momentum overall, most models reproduce 
the observed trend of increasing polarization with decreasing collision energy. Recent measurements (see Fig.\,\ref{fig:polarization}) by the STAR~\cite{STAR:2021beb} and HADES~\cite{HADES:2022enx} collaborations show maximum polarization near the threshold energy for $\Lambda$ production.
It is surprising that a hydrodynamic description~\cite{Ivanov:2020udj} seems to hold at such low energies.

A ``splitting'' between the global polarization of $\Lambda$ and $\overline{\Lambda}$ may be used to estimate~\cite{Becattini:2016gvu} the magnetic field at freezeout, input highly relevant for
studies of the CME (see Sec.\,\ref{sec:cme}).
The slight but statistically insignificant tendency for $P_{\overline{\Lambda}}>P_{\Lambda}$ in the early data prompted measurements at $\sqrt{s_{ NN}}\approx20$~GeV by STAR with much higher statistics and an upgraded event-plane detector~\cite{Adams:2019fpo}.
The resulting null splitting yields a much tighter conservative 
upper bound (95\% confidence level) of $B\leq3\times10^{13}$~T, ruling out several theoretical estimates of the $B$ field.
At $\sqrt{s_{NN}}=200$~GeV, the bound is even tighter, $B\leq3\times10^{12}$~T~\cite{Muller:2018ibh}.

In non-central heavy ion collisions, anisotropic transverse flow necessarily generates vorticity in a fluid picture, leading to predictions~\cite{Becattini:2017gcx,Voloshin:2017kqp} 
of polarization along the beam direction $P_{\hat{z}}$, 
oscillating as a function of azimuthal angle.
This expectation has been borne out by the observation of longitudinal polarization relative to the second- and third-order event planes at RHIC~\cite{STAR:2019erd} and LHC~\cite{ALICE:2021pzu}.
Surprisingly, however, the observed oscillation of $P_{\hat{z}}$ was $180^\circ$ out of phase with predictions.
This has led to a realization of overlooked shear terms in the hydrodynamic equations that contribute to polarization; different treatments~\cite{Becattini:2021iol,Liu:2021uhn,Fu:2021pok,Alzhrani:2022dpi} of these terms have been proposed and there is not yet consensus on the correct formulation.

\subsection{Progress in Cold QCD}
\label{sec:coldQCD_progress} 
Hadrons, with protons and neutrons (the nucleons) the most ubiquitous, 
make up the majority of the visible matter in the universe. Thus, understanding their structure is of fundamental importance. 
The nucleon forms a frontier of subatomic physics and has been under intensive study for the last several decades. Tremendous progress has been made in mapping out the one-dimension momentum distribution of the nucleon constituents, in the form of the Feynman parton distribution functions (PDFs). These investigations not only unveil the partonic structure of the nucleon, but also provide an important opportunity to study the strong interaction.  
Still, essential questions remain to be answered. How do the spin and orbital degrees of freedom of quarks and gluons within the nucleon combine to make up its total spin?  What is the origin of the mass of the nucleon and other hadrons? Do gravitational form factors inform us about the origin of mass and can they be extracted from measurements? Where are the quarks and gluons located within the nucleon? How does the quark-gluon structure of the nucleon change when it is bound in the nucleus? What is the spectrum and structure of conventional and exotic hadrons? All these questions have stimulated further theoretical and experimental investigations in hadronic physics and major facilities have been and will be built to explore them.

Since LRP15, there has been significant progress in cold QCD research in the US and abroad. First, the CEBAF 12 GeV upgrade has been completed and the 
experimental program is in full swing. 
Second, fruitful new and exciting results have been obtained from various hadron physics facilities, including CEBAF at JLab, RHIC at BNL, and the LHC at CERN. These advances covered  static properties and partonic structure of hadrons, nuclear modifications of the structure functions and nucleon many body physics in nuclear structure, and dense medium effects in cold nuclei. These new results test the fundamental properties of QCD such as its chiral structure and predictions for new hadron states, preview the tomography imaging of the nucleon that will help to unveil the origin of the mass and spin, and deepen our understanding of nucleon-nucleon interactions to form atomic nuclei and the partonic structure of a dense cold medium. More importantly, these advances pave the way for answering the aforementioned fundamental questions. 
Meanwhile, all this progress has strong overlap with hot QCD, nuclear structure and fundamental symmetry physics. Along with the experiment achievements, theoretical developments have played a significant role, not only in the interpretation of experimental data, 
but also in stimulating the programs at these facilities. In the following, we will highlight these advances. 

\subsubsection{Properties of Hadrons} 
\label{sec:cold_highlight_longrange}

We start with long-range nucleon structure. This includes the proton (electric) charge radius, 
generalized polarizabilities and electromagnetic form factors of the nucleon. An additional relevant topic is the neutral pion lifetime measurement that tests the chiral anomaly of QCD. 

\vskip 0.3cm
\begin{figure}[hbt!]
\centerline{
\includegraphics[width=0.8\textwidth]{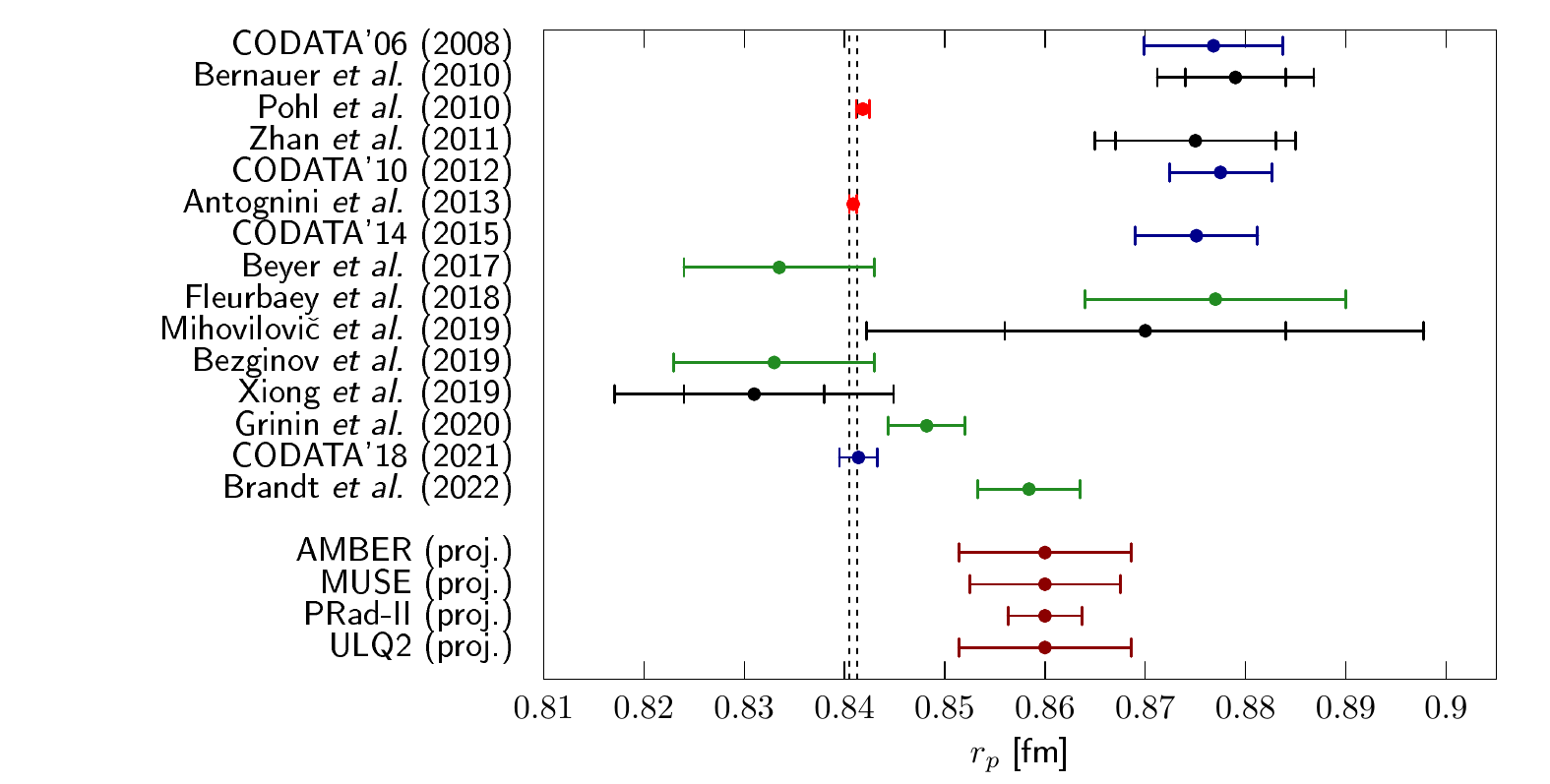}}
\caption{The PRad $r_{p}$ result shown along with the projected result for PRad-II and other measurements.}
\label{fig:fig1}
\end{figure}

\noindent
{\bf Proton charge radius} For nearly half a century the root-mean-square charge radius ($r_{p}$) of the proton had been obtained from measurements of transitions between atomic hydrogen energy levels and by scattering electrons from hydrogen atoms. Until recently, the proton charge radius obtained from these two methods agreed with one another within experimental uncertainties. In 2010, the proton charge radius was obtained for the first time by precisely measuring the Lamb shift of muonic hydrogen~\cite{Pohl:2010zza}. The charge radius of the proton obtained from muonic hydrogen was found to be significantly smaller than that obtained from ordinary hydrogen. This was called the proton charge radius puzzle and led to a rush of experimental and theoretical efforts to understand the difference in the proton size between ordinary hydrogen and muonic hydrogen. The Proton Charge Radius (PRad) experiment at JLab was one such new effort which utilized several innovations and studied electron scattering from ordinary hydrogen atoms to high precision. It was the only lepton scattering experiment to use an electromagnetic calorimeter and a windowless hydrogen gas flow target, allowing robust extraction of the proton charge radius from form factors measured in a very low $Q^2$ range of 2$\times$10$^{-4}$~to~6$\times$10$^{-2}$GeV$^2$. 
The PRad result, shown in Fig.~\ref{fig:fig1}, was found to be in agreement with the small radius measured in muonic hydrogen spectroscopy experiments as well as some of the recent ordinary hydrogen spectroscopy measurements~\cite{Xiong:2019umf}. 
The PRad result provided critical input to the recent revision of the Committee on Data of the International Science Council (CODATA) recommendation for the proton charge radius and the Rydberg constant as noted in the most recent review~\cite{Gao:2021sml}. A followup experiment, PRad-II, is being planned to reach an even smaller uncertainty, see Section~\ref{sec:cold_future_jlab}. In addition, other measurements, such as the US-led Muon Scattering Experiment (MUSE), will address the puzzle by measuring elastic muon scattering on the proton, see Section~\ref{sec:otherfacility}.

\begin{figure}[ht]
  \centering
  \includegraphics[width=0.425\columnwidth]{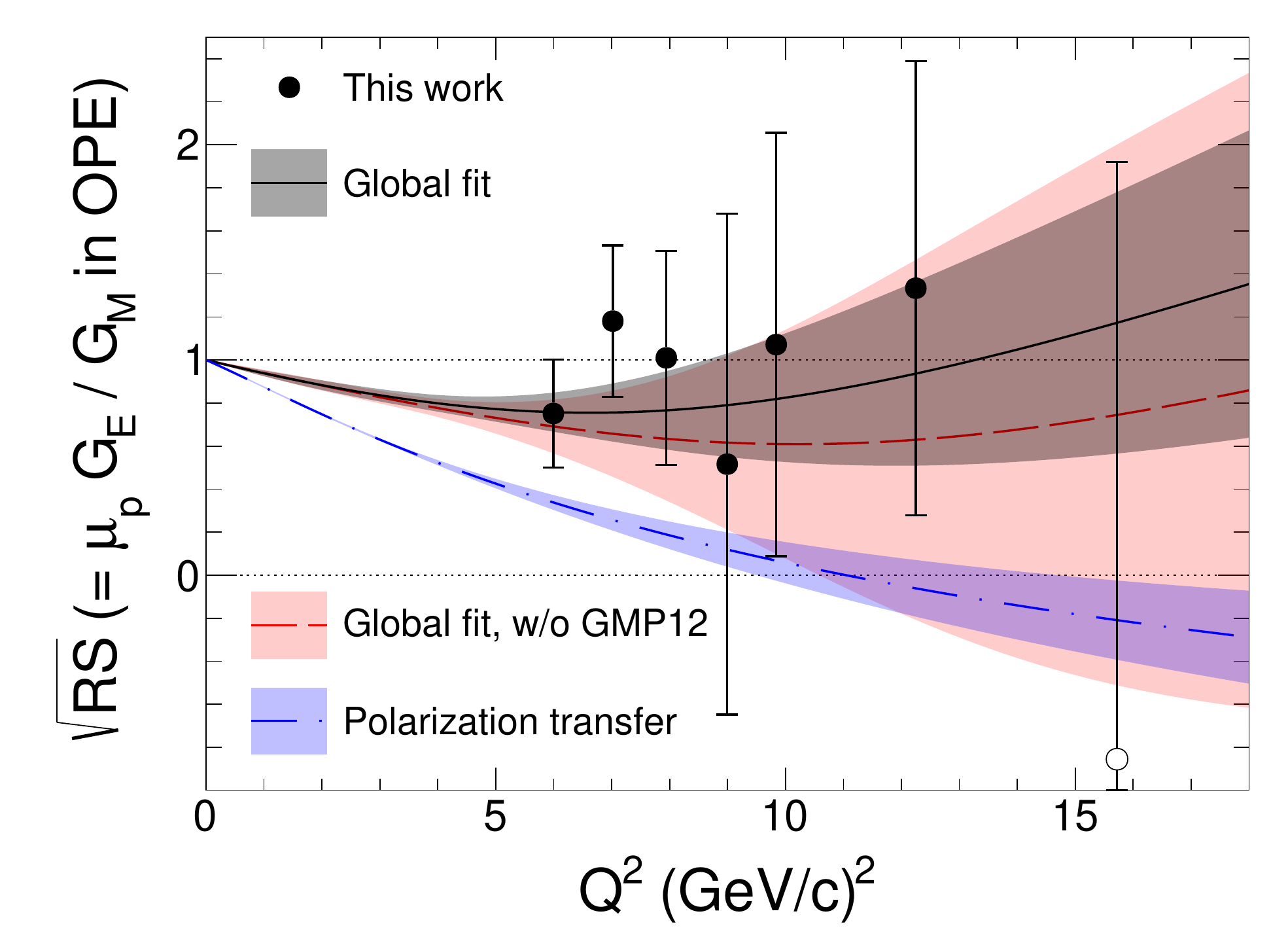}
  \includegraphics[width=0.47\columnwidth]{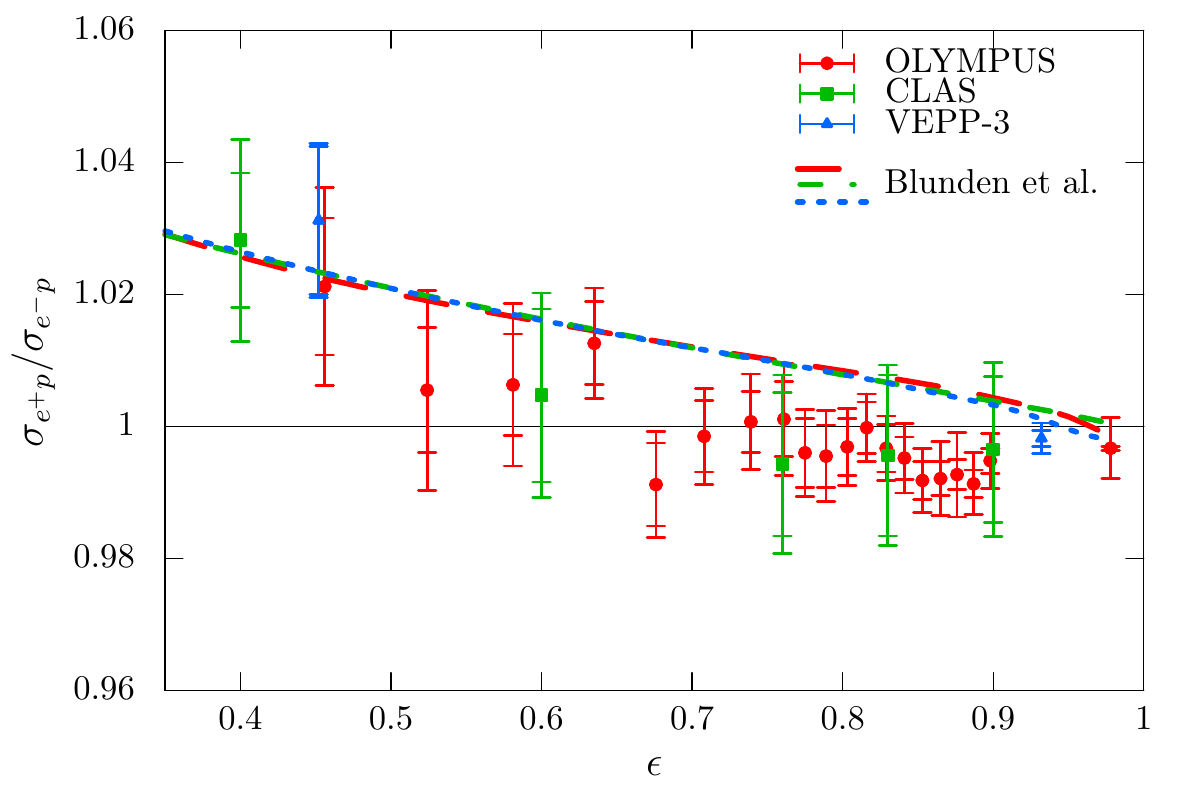}
  \caption{\label{fig:r2g_compare}
(Left) Direct Rosenbluth separation results
for $\sqrt{\text{RS}}$ ( $ = \mu_p$G$_{_E}$/G$_{_M}$ in the one-photon exchange approximation). The black solid (red dashed) curve shows the results of the fit to the cross section data with (without) the new GMp12 data (``This work")~\cite{Christy:2021snt}. The blue dot-dashed curve shows $\mu_p$G$_{_E}$/G$_{_M}$ from a fit to the polarization data. 
(Right) The ratio of positron-proton to electron-proton elastic cross sections as a function
    of $\epsilon$, as measured by OLYMPUS~\cite{OLYMPUS:2016gso}, CLAS~\cite{CLAS:2016fvy}, and at VEPP-3~\cite{Rachek:2014fam}.
    The data are generally closer to unity than 
    the expectation if the difference between the Rosenbluth separation and polarization method is fully attributed to two-photon exchange effects. 
    }
\end{figure}

\vskip 0.3cm
\noindent
{\bf Nucleon form factors at high-$Q^2$ and two-photon exchange physics} Apart from the charge radius determination from the low $Q^2$ measurement, nucleon form factors provide information on the fundamental constituent structure of the nucleon, and at times reveal our lack of understanding in related topics. 
Specifically, the proton electric-to-magnetic form factor ratio  determined from the polarization transfer method had revolutionized the basic understanding of the constituent structure of the proton~\cite{JeffersonLabHallA:1999epl,JeffersonLabHallA:2001qqe}. The discrepancy observed between these measurements and those from the (traditional) Rosenbluth separation method, see left panel of Fig.~\ref{fig:r2g_compare}, has stimulated theoretical investigations into the two-photon exchange (TPE) contribution, currently considered as the leading explanation. 
While a number of recent measurements have shown evidence for sizable TPE in several different observables, the situation is far from resolved. 
For example, several recent experiments were carried out to directly measure 
TPE by looking for a difference in the unpolarized positron-proton and electron-proton elastic cross sections, including the OLYMPUS experiment at DESY~\cite{OLYMPUS:2016gso}, and those utilizing CLAS at JLab~\cite{CLAS:2014xso,CLAS:2016fvy} and the VEPP-3 storage ring at Novosibirsk~\cite{Rachek:2014fam}. The results of these experiments are shown in the right panel of Fig.~\ref{fig:r2g_compare}. The data favor a non-zero slope as a function of the virtual photon polarization parameter, $\epsilon$, which is a sign of TPE. However, these data are limited to the low $Q^2$ region, and are closer to unity than needed to fully explain the proton form factor discrepancy. 
New measurements with greater kinematic reach are needed to fully explain the proton form factor discrepancy and to guide theoretical efforts. 
This is one of the major motivation for the proposed positron beam program at JLab, see Section~\ref{sec:cold_future_positron}, as well as for the proposed TPE Experiment (TPEX) at DESY. 

\subsubsubsection{Nucleon polarizabilities and generalized polarizabilities} 
Nucleon polarizabilities and generalized polarizabilities describe how the charged internal constituents of the nucleon react to external electromagnetic fields and precisely determine the mean-square electromagnetic polarizability radii of the proton. 
Extracting them from the real Compton scattering (RCS),  virtual Compton scattering (VCS) and double virtual Compton scattering (VVCS) processes provides stringent tests of Chiral Effective Field Theory  ($\chi$EFT)~\cite{Drechsel:2002ar,Pasquini:2018wbl,Griesshammer:2012we} and lattice QCD computations~\cite{Howell:2020nob}. They are also essential to extract the hyperfine splitting of muonic hydrogen~\cite{Hagelstein:2015egb}.
Since the 2015 LRP, substantial progress has been made in determining both scalar and spin-dependent static and dynamical polarizabilities of the proton and neutron~\cite{Mornacchi:2022cln,Pasquini:2017ehj,Pasquini:2019nnx,Margaryan:2018opu}, with strong international efforts and synergistic advancements in experiment and theory~\cite{Howell:2020nob}. 
For the proton, key achievements are the first extraction of proton spin polarizabilities from the 
measurements of double polarization observables by the A2 Collaboration at the Mainz microtron (MAMI)~\cite{A2:2014iky,A2:2019bqm}, and new high precision data for unpolarized cross sections and photon beam asymmetry from both MAMI~\cite{A2CollaborationatMAMI:2021vfy} and the High Intensity Gamma-ray Source (HIGS)~\cite{Li:2022vnz}. At HIGS, expertise and techniques have been developed that produce the requisite high-precision RCS cross section measurements on light nuclei~\cite{Li:2019irp, Sikora:2017rfk}, which can be used to determine the neutron polarizabilities. 

\begin{figure}[!ht]
\centering
\includegraphics[width=0.9\textwidth]{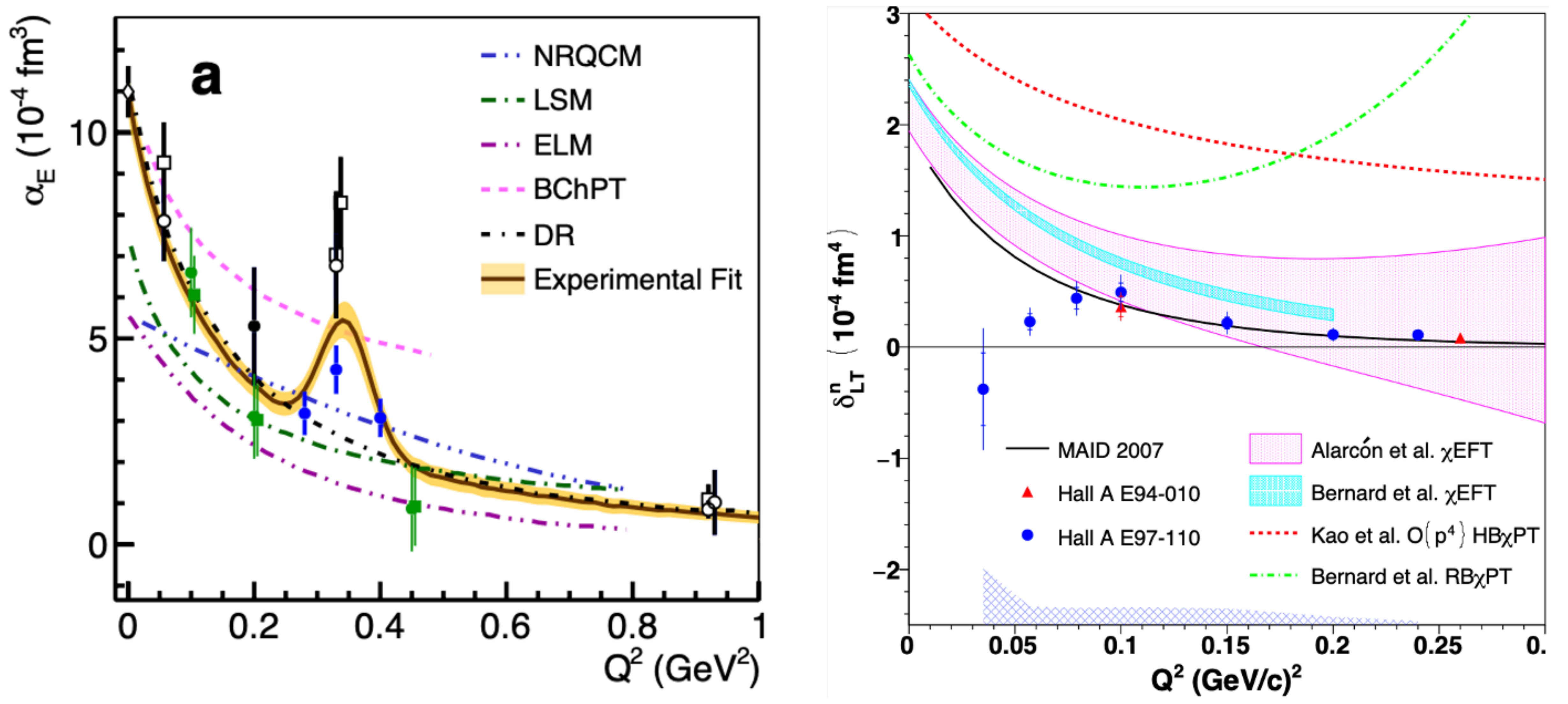}
\caption{Left: world data~\cite{Fonvieille:2019eyf,VCS:2000ldk,A1:2008rjb,A1:2019mrv,A1:2020nof,Blomberg:2019caf,Li:2022sqg,A2CollaborationatMAMI:2021vfy,JeffersonLabHallA:2004vsy,JeffersonLabHallA:2012zwt,Bourgeois:2006js,Bourgeois:2011zz} for the proton electric $\alpha_{\rm E}(Q^2)$ VCS generalized polarizability. The filled symbols mark the recent experiments from MAMI~\cite{A1:2008rjb,A1:2019mrv,A1:2020nof} (green and black solid circles) and from JLab~\cite{Li:2022sqg} (blue solid circles). The dispersion-relations (DR) and low-energy-expansion analysis results are shown with open circles and square symbols, respectively. The theoretical calculations of BChPT~\cite{Lensky:2016nui}, NRQCM~\cite{Pasquini:2000ue}, LSM~\cite{Metz:1996fn}, ELM~\cite{Korchin:1998cx} and DR~\cite{Pasquini:2000pk,Pasquini:2001yy,Drechsel:2002ar}, as well the experimental fit of the electric generalized polarizability that includes all the world data are also shown.
Right: The VVCS generalized longitudinal-transverse spin polarizability $\delta_{\rm LT}^n(Q^2)$ for the neutron measured recently at JLab (blue circles~\cite{E97-110:2021mxm}) and compared with a previous JLab measurement (red triangles~\cite{Amarian:2004yf}), early $\chi$EFT calculations (green~\cite{Bernard:2002pw} and red lines~\cite{Kao:2002cp}), state-of-the-art $\chi$EFT calculations (blue~\cite{Bernard:2012hb} and pink~\cite{Alarcon:2020icz} bands), and the phenomenological MAID~\cite{Drechsel:1998hk} model.}
\label{fig-polar}
\end{figure}

Meanwhile, four high-precision experiments at JLab mapped the very low $Q^2$ behavior of the VVCS generalized forward spin polarizability $\gamma_0(Q^2)$~\cite{Zheng:2021yrn,Adhikari:2017wox, Sulkosky:2021qmh}, and of the generalized longitudinal-transverse spin polarizability $\delta_\mathrm{LT}(Q^2)$~\cite{JeffersonLabHallAg2p:2022qap,Sulkosky:2021qmh}, for both proton and neutron.
A fifth experiment measured the VCS generalized polarizabilities $\alpha_{\rm E}(Q^2)$ and $\beta_{\rm M}(Q^2)$ for the proton~\cite{Li:2022sqg} at intermediate $Q^2$. While some of the results agree with the latest $\chi$EFT calculations, no single calculation describes all of the data well. 
For example, the observed behavior of $\alpha_{\rm E}(Q^2)$ (left panel of Fig.~\ref{fig-polar}) is in sharp contrast with the current theoretical understanding that suggests a monotonic decrease with increasing $Q^2$. Similarly, data on the neutron VVCS spin-dependent generalized polarizability $\delta_{\rm LT}^n(Q^2)$~\cite{E97-110:2021mxm} (right panel of Fig.~\ref{fig-polar}) indicate a small, or even negative, value at very low $Q^2$ and a positive slope, in contrast with predictions from $\chi$EFT~\cite{Bernard:2012hb, Alarcon:2020icz, Bernard:2002pw, Kao:2002cp} and the phenomenological MAID model~\cite{Drechsel:1998hk}. 
These new data pose a challenge to $\chi$EFT and serve as high-precision benchmark data for future non-perturbative QCD calculations. 

\vskip 0.3cm
\noindent
{\bf Precision measurement of the neutral pion lifetime} 
Two fundamental symmetries in QCD are directly involved in both the existence and the lifetime ($\tau$) of the neutral pion ($\pi^0$):  the left-right chiral and axial symmetries. The explicit axial symmetry breaking, due to quantum fluctuations, gives rise to one of the most interesting effects in nature, the chiral (or axial) anomaly. This process is purely responsible for the neutral pion  decay into two photons ($\pi^0\to \gamma\gamma$), defining its unusually short lifetime. The PrimEx collaboration measured the neutral pion decay width $\Gamma(\pi^0\to \gamma\gamma)$ in JLab Hall B with an unprecedented precision~\cite{PrimEx-II:2020jwd}. With its $1.50\%$ total uncertainty, $\tau=8.337\pm 0.056 ({\rm stat.})\pm 0.112 ({\rm syst.})\times 10^{-17}\rm s$, this is the most precise measurement of this critically important quantity, and firmly confirms the prediction of the chiral anomaly in QCD at the percent level. It also played a critical role in the normalization of the neutral pion transition form factor to constrain the hadronic light-by-light scattering contributions to the well-known muon ($g$-2) anomaly in search of new physics.

\subsubsection{One-dimensional Momentum Distributions of the Nucleon} \label{sec:cold_progress_1d}

\medskip
\noindent
{\bf Parton distribution functions} 
Understanding the proton's composition from its underlying constituent quarks is one of the primary goals of all electron-proton scattering experiments. 
Of particular value is the method of deep inelastic lepton-nucleon scattering (DIS), for which data are typically interpreted in terms of the PDFs that describe the momentum distributions of partons in the (one-dimensional) direction parallel to the nucleon momentum.  Tremendous progress in our understanding of the PDFs has been made, most notably in the global effort to determine both quark and gluon PDFs from various high energy experiments, see e.g.~\cite{Hou:2019efy,NNPDF:2017mvq,Alekhin:2017kpj,Cocuzza:2021rfn}. 
Furthermore, the desire to understand PDFs at a more fundamental level is driving experimental programs at both low and high energy facilities. 

\begin{figure} [!ht]
\centering
\includegraphics[width=0.44\columnwidth, angle=0]{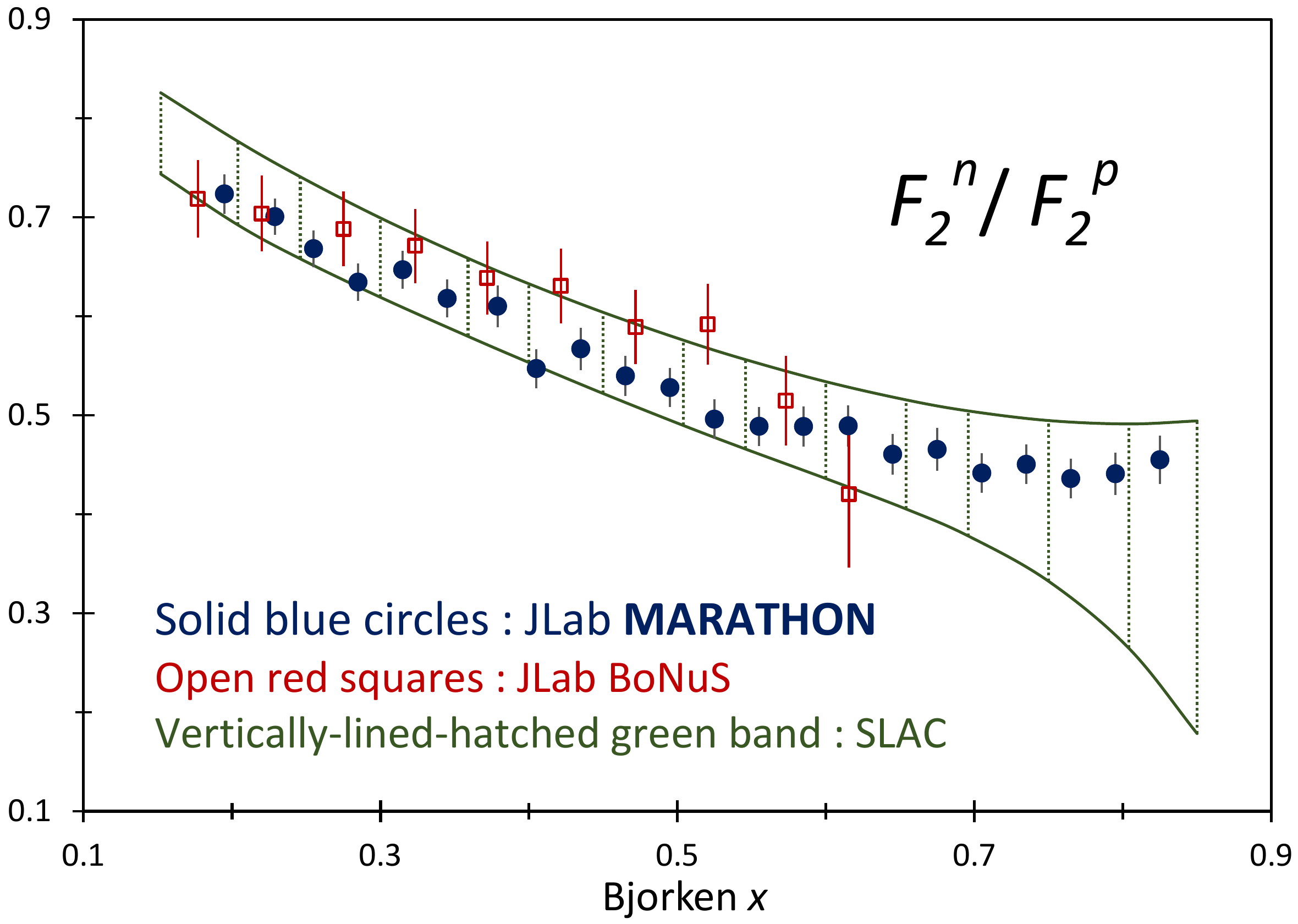}
  \includegraphics[width=0.55\textwidth]{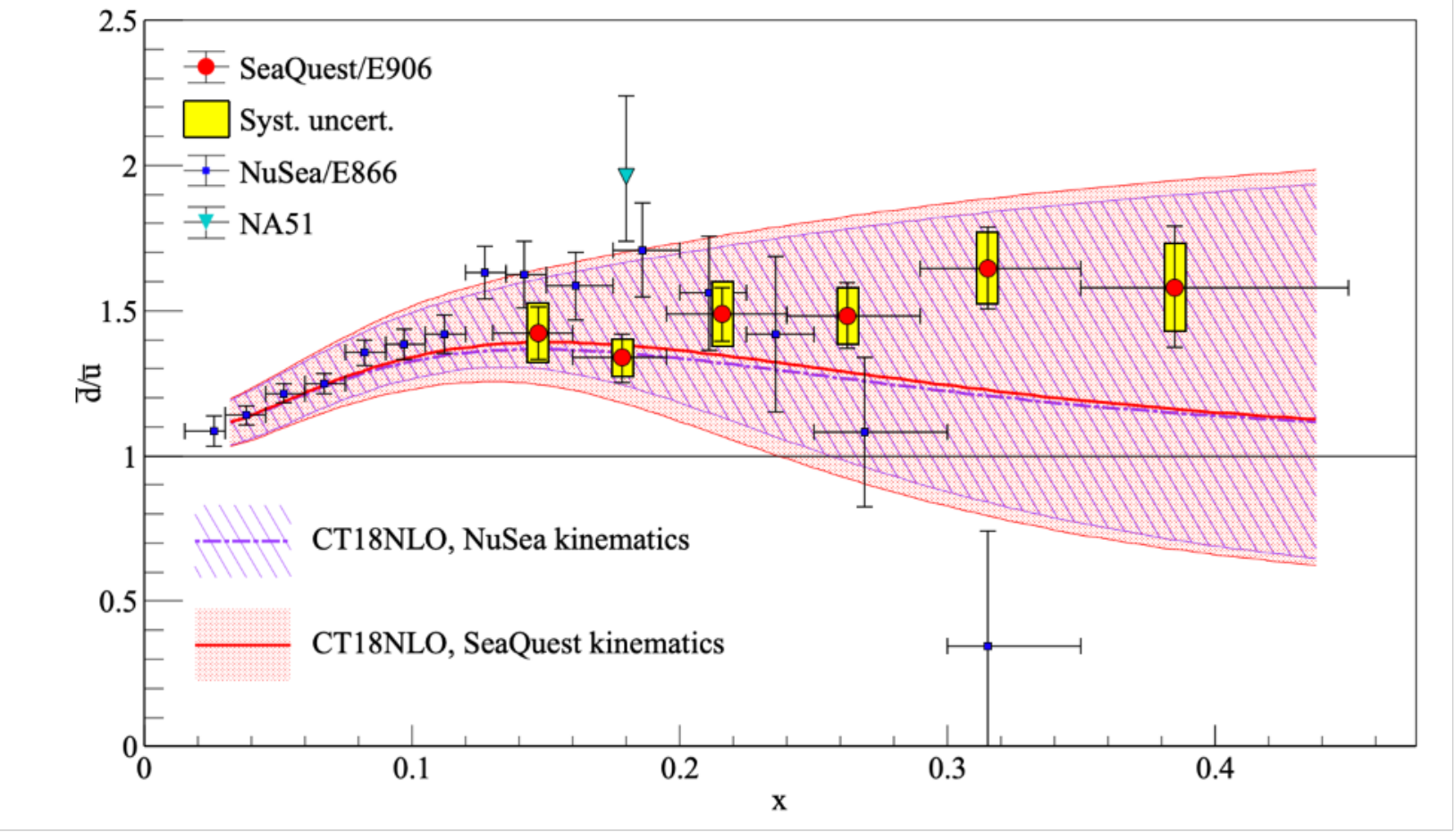}
\caption{\label{fig:quarklargex} Left: The $F_2^n/F_2^p $ ratio plotted versus the Bjorken $x$ from the JLab MARATHON
experiment~\cite{JeffersonLabHallATritium:2021usd}.  
Right: A plot of $\bar d(x)/\bar u(x)$ extracted from the measured SeaQuest $\sigma^{pd} (x)/2\sigma^{pp}(x)$ cross section ratio~\cite{SeaQuest:2021zxb}, compared with data from E866/NuSea~\cite{NuSea:2001idv} and CT18NLO parton distributions.  }
\end{figure}

The asymptotic behavior of the ratio of PDFs in the deep valence quark region $x\to 1$ can test a variety of theoretical predictions. One such ratio is the $d$ over $u$ quark distributions. As featured in LRP15, experiments in Halls A and B at JLab are accessing this ratio with very different approaches. 
The first of these experiments, MARATHON, measured the $^3$H/$^3$He DIS cross sections with the expectation that the effects of nuclear corrections largely cancel between the two ``mirror" nuclei. 
The experiment has been successfully completed and first results on $F_2^n/F_2^p$ have been published~\cite{JeffersonLabHallATritium:2021usd}. 
These data allow for more precise extractions of the underlying $d/u$ ratio~\cite{Cui:2021gzg,Alekhin:2022tip}, while also placing constraints on the isospin-dependence of the nuclear effects~\cite{Cocuzza:2021rfn}. The model dependence of the PDF extraction can be cross checked with the BONuS12 experiment~\cite{JLabPR:BONUS12}, while an extraction of $d/u$, free from the use of any nuclear model, will be made by the Parity-Violating DIS (PVDIS) program of SoLID in JLab Hall A, see Section~\ref{sec:cold_future_solid}.

Meanwhile, the SeaQuest experiment carried out at the Fermilab fixed-target facility has unveiled interesting features of the sea quark distributions~\cite{SeaQuest:2021zxb}. Naively, one expects that the anti-up $\bar u$ and anti-down $\bar d$ quarks should be the same if they both come from the gluon splitting contribution. However, an asymmetry between the two was observed at low $x$ using the Drell-Yan process in the NuSea experiment~\cite{NuSea:2001idv}. 
The recent SeaQuest experiment extended the measurement beyond $x=0.3$ and found that the asymmetry persisted, see the right panel of Fig.~\ref{fig:quarklargex}. Complementary information on $\bar d(x)/\bar u(x)$ has also been studied at RHIC from the ross section ratios of $W$- and $Z$-bosons at mid-rapidity~\cite{STAR:2020vuq}. 

\vskip 0.3cm

\noindent
{\bf Quark and gluon polarizations inside the nucleon} 
DIS measurements with polarized beams and targets and polarized proton-proton collisions probe the polarized (helicity) quark/gluon distribution and the origin of the proton spin. 
Significant progress has been made in assessing the fraction of the proton spin from parton polarizations, see, recent global analyses~\cite{DeFlorian:2019xxt,Nocera:2014gqa,Cocuzza:2022jye}. 

\begin{figure} [!h]
\centering
\begin{minipage}[c]{0.43\textwidth}
\includegraphics[width=\textwidth]{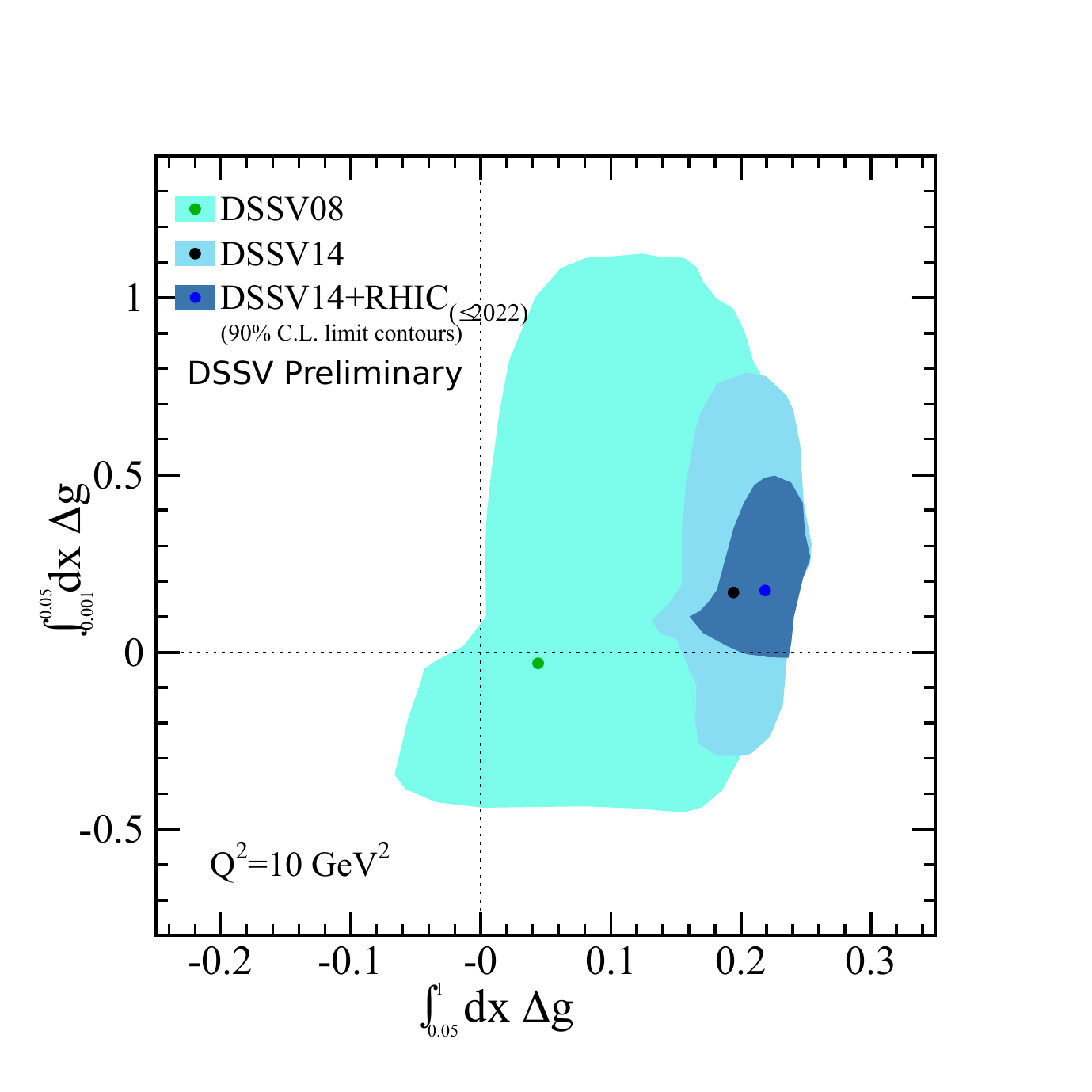}
 \end{minipage}
\begin{minipage}[c]{0.47\textwidth}
\vskip 0.8cm
       \includegraphics[width=\textwidth]{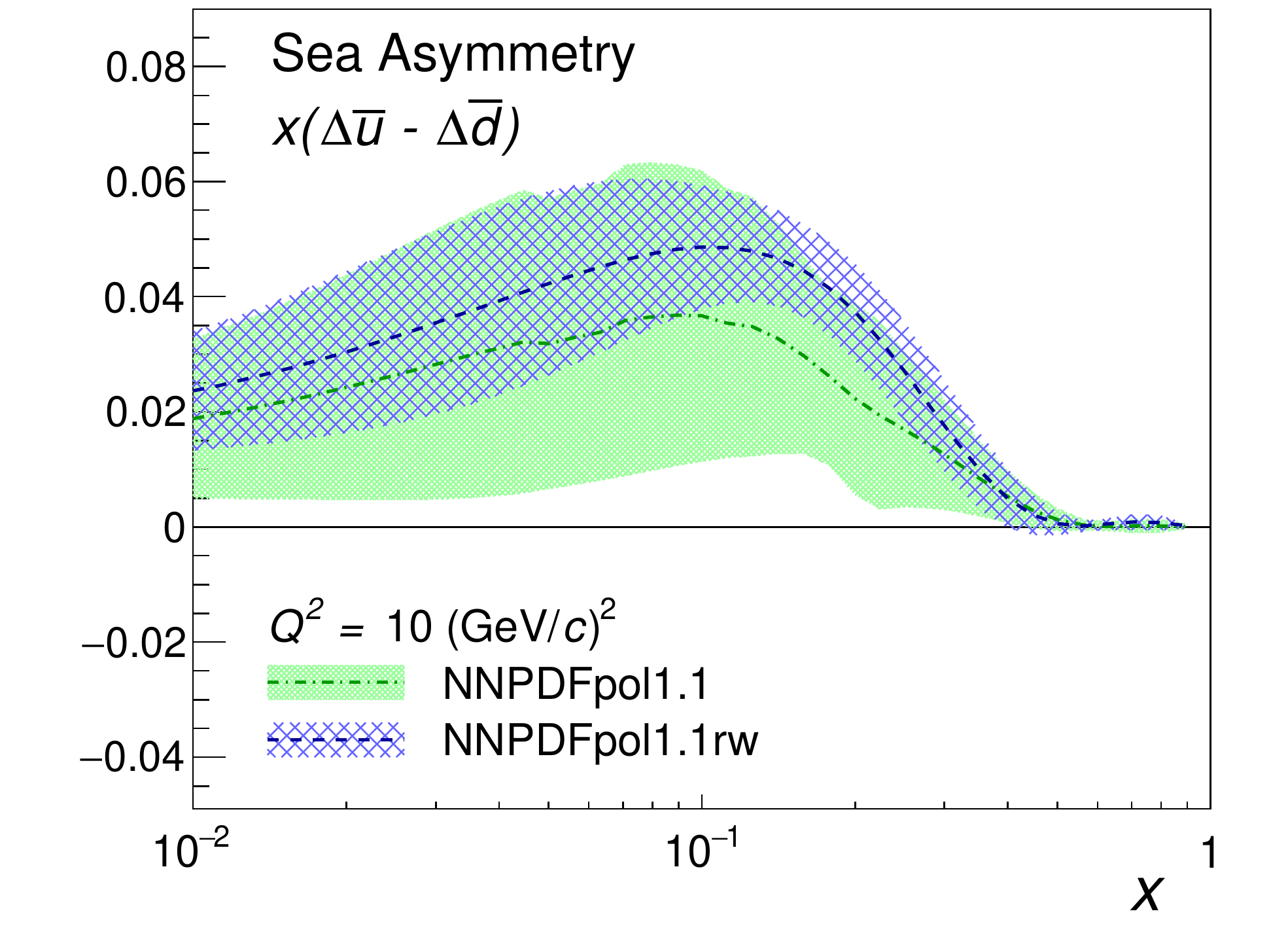}
       \end{minipage} 
\caption{Left: The impact of RHIC data on the truncated moment of the gluon helicity from the new DSSV evaluations \cite{RHIC-Cold-QCD} at $Q^2 = 10\,(\mathrm{GeV}/c)^2$ integrated with the range of $x \in (0.001,0.05)$ versus the range of $x \in (0.05,1)$. The green dot represents the best fit to the data from the DSSV08 evaluations, the black dot shows the DSSV14 results \cite{deFlorian:2014yva}, and the new preliminary fit including the new RHIC data sets is marked with the blue dot. Blue areas represent the 90\% C.L. limit contours. 
Right: The difference of $\bar{u}$ and $\bar{d}$ polarizations as a function of $x$ at a scale of $Q^{2}$= 10 GeV$^{2}$ before and after NNPDFpol1.1 \cite{Nocera:2014gqa} reweighting with STAR 2013 $W$ $A_{L}$~\cite{Adam:2018bam}. The green band shows the NNPDFpol1.1 results \cite{Nocera:2014gqa} and the blue hatched band shows the corresponding distribution after the STAR 2013 $W$ data are included by reweighting.  }\label{fig:rhic-spin-impact}
\end{figure}
The impact from the RHIC spin program with polarized proton-proton collisions has been highlighted in LRP15~\cite{Geesaman:2015fha}. 
Recent STAR results on double-spin asymmetries of inclusive jet and dijet production at center-of-mass energies of 200 GeV and 510 GeV complement and improve the precision of previous measurements~\cite{Adamczyk:2014ozi,Adam:2019aml,STAR:2021mfd,STAR:2021mqa,RHIC-Cold-QCD}, imposing further constraints on the gluon polarization, see the left panel of Fig.~\ref{fig:rhic-spin-impact}. 
Meanwhile, the production of $W$-bosons in longitudinally polarized proton-proton collisions serves as a powerful and elegant tool to access valence and sea quark helicity distributions at a high scale, $Q^2\sim M_W^2$, where $M_W$ the $W$-boson mass. The STAR and PHENIX Collaborations have concluded the measurements of the parity-violating longitudinal single-spin asymmetry in the production of weak bosons and improved the constraints on $\bar u$ and $\bar d$ polarization~\cite{PHENIX:2015ade,PHENIX:2018wuz,Adam:2018bam}. 
The sea quark $\bar u$ helicity, $\Delta \bar u$, is now known to be positive and $\Delta \bar d$ is negative. The opposite sign of the polarized sea-quark distribution with respect to the $\bar d/\bar u$ flavor asymmetry in the unpolarized sea (right panel of Fig.~\ref{fig:quarklargex}) is of special interest due to the differnet predictions in various models of nucleon structure.
The overall impacts of recent jet and dijet, pion, and $W$ data on the quark/gluon helicity distribution based on the global fits are shown in Fig.~\ref{fig:rhic-spin-impact} for $\Delta g$ (left) and $\Delta \bar u-\Delta \bar d$ (right), respectively. 

Besides determining the origin of the proton spin, these data crucially test theories of the strong interaction. Notably, high $Q^2$ studies of the Bjorken sum rule~\cite{Bjorken:1966jh}, defined using the integral of the polarized structure functions of the proton and the neutron $g_1^{p,n}$: $\int_0^{1} dx~g_1^p(x)-g_1^n(x)= \frac{g_A}{6}+${\small (pQCD corrections)} where $g_A$ is the nucleon axial coupling, were the first to show that QCD can describe the strong interaction even when spin degrees of freedom are explicit~\cite{Deur:2018roz}. 
Similarly, low $Q^2$ Bjorken sum data precisely test effective theories that describe the strong interaction at long distances~\cite{Deur:2021klh,Deur:2018roz}. 
The Bjorken sum rule is also used to extract the QCD coupling $\alpha_s(Q^2)$~\cite{Deur:2016tte}, where 
the high-$Q^2$ extractions~\cite{Deur:2014vea} are presently only just competitive with high-energy collider extractions of $\alpha_s$~\cite{dEnterria:2022hzv}. However, they should become more impactful with the EIC, which should provide an accuracy of $\sim$1.5--2\% on $\alpha_s$ (just from the Bjorken sum rule). 

Additionally, quark and gluon polarizations in the nucleon, when measured in specific kinematic regions such as the $x\to 1$ limit, also provide valuable tests of predictions from various quark models, perturbative QCD, and non-perturbative methods. The JLab 6 GeV results~\cite{JeffersonLabHallA:2003joy, JeffersonLabHallA:2004tea} showed that the ratio of the polarized to unpolarized PDF for the down quark, $\Delta d/d$ is negative up to $x=0.61$. That is, the valence down quark spins in the opposite direction of the proton spin. In 2020, the 12 GeV extension of the measurement of the neutron spin was successfully completed in Hall C at JLab, and complementary measurements of the proton and the deuteron are presently underway in Hall B at JLab using CLAS12. A combined analysis of the data from all three targets can assess whether $\Delta d/d$ remains negative up to $x=0.8$, or turns sharply positive at even higher $x$ as predicted by pQCD models~\cite{Avakian:2007xa}. 

\subsubsection{Three-dimensional Tomography of the Nucleon}\label{sec:cold_progress_3d}
To completely understand the proton spin decomposition in terms of quark/gluon spins and their orbital angular momentum contributions, observables and methods that go beyond the one-dimensional PDF discussed above becomes necessary. A new direction that emerged at the time of LRP15 is to pursue three-dimensional (3D) tomography of the nucleon. The first set of tools are focused on the transverse motion of partons: if the nucleon is assumed to move in the $\hat z$-direction, its structure in the transverse direction can be either analysed in coordinate space using generalized parton distributions (GPDs) or in momentum space using transverse momentum dependent parton distributions (TMDs).

\vskip 0.3cm
\begin{figure}[ht!]
\hspace*{0.05\textwidth}
\begin{minipage}[c]{0.45\textwidth}\includegraphics[width=\textwidth]{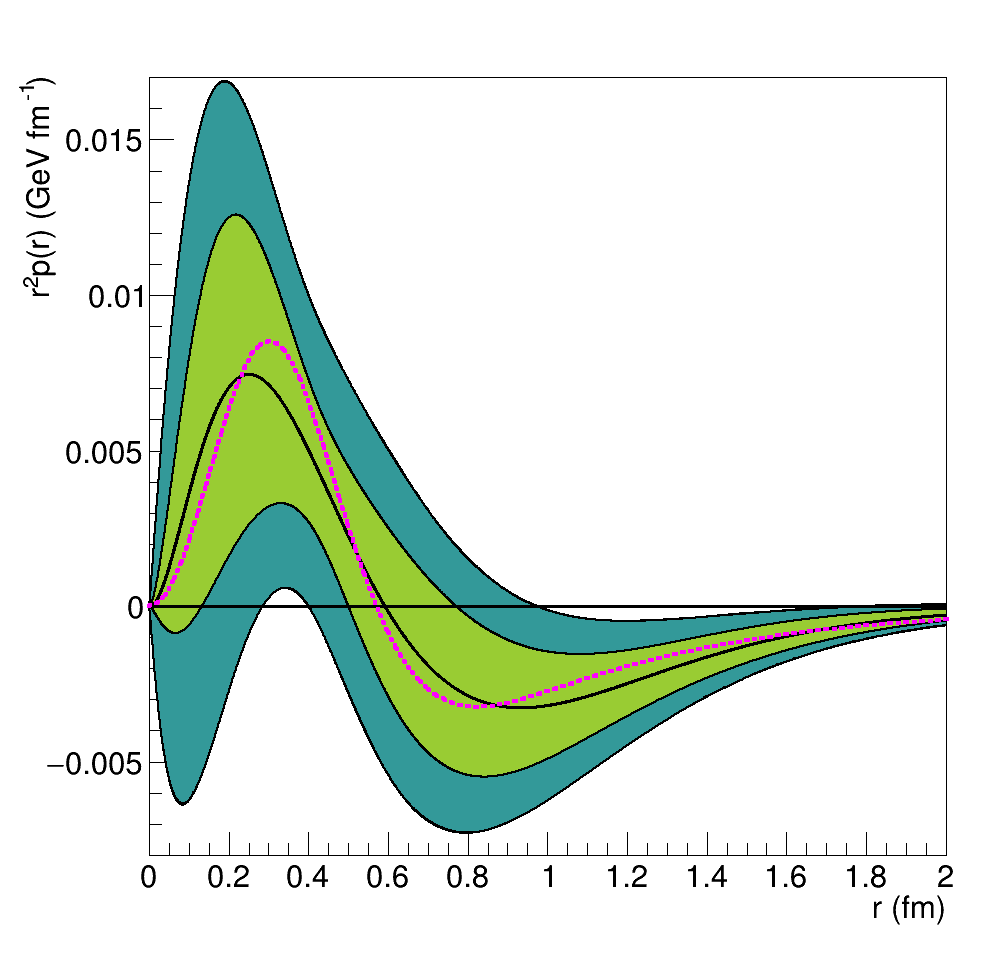}
\end{minipage}
\begin{minipage}[c]{0.41\textwidth}
\vspace*{0.15\textwidth}
\includegraphics[width=\textwidth]
{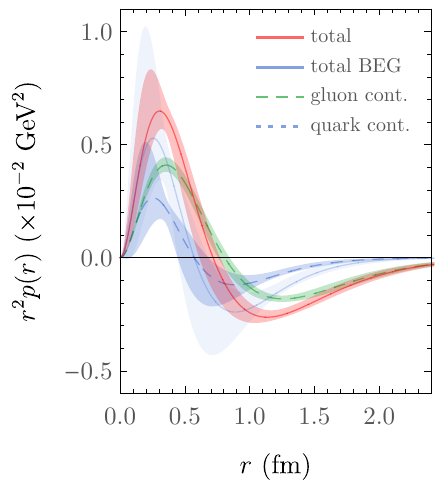}
\end{minipage}
\caption{Pressure distribution in the proton. Left: estimates of the quark contribution from the DVCS measurements~\cite{Burkert:2018bqq}; Right: Lattice QCD calculations of the same quantity, including the gluon contribution~\cite{Shanahan:2018nnv}. 
These results show that the pressure is positive at small distances and negative at large distances. 
}
\label{fig:pressure}
\end{figure}
\noindent
{\bf GPDs and gravitational form factors} 
Deeply Virtual Compton scattering (DVCS) has been identified as a clean process to experimentally access GPDs~\cite{Ji:1996nm} and Compton form factors (CFF) to probe the 3D structure of nucleons and nuclei. 
Data previously collected at various experiments have been used to generate some of the first 3D images of the proton~\cite{Dupre:2017hfs}. %
GPDs can also be used to determine mechanical properties of particles through the gravitational form factors (GFFs)~\cite{Polyakov:2002yz,Polyakov:2018zvc}. 
Using two sets of measurements by CLAS with a 6 GeV polarized electron beam, the beam-spin asymmetry~\cite{CLAS:2007clm} and the differential cross sections~\cite{CLAS:2015bqi}, 
the first data-based estimate has been made on one of the GFFs, the so-called $D(t)$ term~\cite{Burkert:2018bqq}. 
A spherical Bessel function is then employed to Fourier-transform $D(t)$ to the Breit frame and the results are displayed in Fig.~\ref{fig:pressure}. 
Meanwhile, the gluon contributions to the distributions of pressure and shear forces inside the proton were computed using lattice QCD, allowing the first model-independent determination of these fundamental aspects of proton structure~\cite{Shanahan:2018pib}. Combined with the experimental measurements of the quark contribution, this enabled the first complete determination of the pressure and shear distributions of the proton~\cite{Shanahan:2018nnv}. 
More precise determinations of these quantities are a focus of future experiments at JLab and at the EIC. 

\vskip 0.3cm
\begin{figure}[ht]
\centering
\begin{minipage}[c]{0.48\textwidth}
    \includegraphics[width=\textwidth]{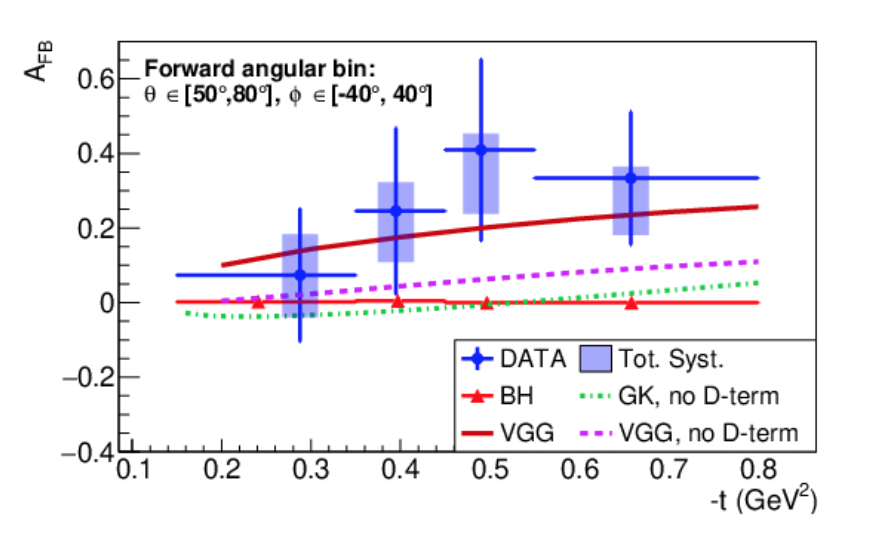}
  \end{minipage}\hfill
  \begin{minipage}[c]{0.48\textwidth}
\caption{Forward-backward asymmetry as a function of the momentum transfer $t$ to the proton. The solid line shows the prediction of a GPD model that describes worldwide DVCS data, including the $D$-term contribution. The red triangles show the asymmetry computed for simulated Bethe-Heitler (BH) events. The dashed and dash-dotted lines are predictions of models without the contribution of the $D$-term. Note that the $D$-term contributes more than 50\% to the asymmetry.}
\label{fig:TCS}
  \end{minipage}
\end{figure}
\noindent
{\bf Time-like Compton scattering}
The time-reversal reaction to DVCS, time-like Compton scattering (TCS), offers unique ways to probe nucleon structure and GPDs. In this case, a quasi-real photon is absorbed by the nucleon which produces a high invariant-mass lepton pair in the final state. While theoretically as clean as DVCS, the experimental measurement of TCS is more challenging due to potential background channels, making the reaction harder to identify. The first measurement of the TCS process was recently performed by the JLab CLAS Collaboration~\cite{CLAS:2021lky}. Phenomenological models that reproduce worldwide data on DVCS satisfactorily describe the photon polarization asymmetry and the forward-backward (FB) asymmetry of TCS, see Fig.~\ref{fig:TCS}. This finding supports the universality of GPDs in describing hard exclusive processes. In addition, TCS is particularly sensitive to the real part of the Compton amplitude and thus to the $D(t)$ term, which can be related to the energy-momentum tensor and pressure distribution inside the nucleon as described above. 

\vskip 0.3cm
\noindent
{\bf 3D momentum tomography of hadrons} 
The TMDs provide not only an intuitive illustration of nucleon tomography, but also the important opportunities to investigate the specific nontrivial QCD dynamics associated with their physics: QCD factorization, universality of the parton distributions and fragmentation functions, and their scale evolutions.
In particular, the quark Sivers functions for semi-inclusive hadron production in DIS (SIDIS) and Drell-Yan lepton pair production differ in sign because of the difference between initial- and final-state interactions.  
This leads to a sign change between the transverse single spin asymmetries (SSAs) in SIDIS and Drell-Yan: ${\rm Sivers~SSA}|_{\rm DY}=-{\rm Sivers~SSA}|_{\rm DIS}$. 
It is important to test this nontrivial QCD prediction by comparing the SSAs in these two processes. The Sivers SSA in SIDIS have been observed by the HERMES, COMPASS, and JLab experiments. There have been significant efforts to measure the Drell-Yan Sivers asymmetry at COMPASS~\cite{COMPASS:2017jbv} and that of $W^{\pm}$ production at RHIC~\cite{STAR:2015vmv}. The analyses of these data provided an indication of a sign change~\cite{Cammarota:2020qcw,Bury:2020vhj,Bury:2021sue,Gamberg:2022kdb,Echevarria:2020hpy,Bacchetta:2020gko}.
More precise measurements are needed to confirm this crucial property.

Precision SIDIS studies in multidimensional space can systematically investigate the production mechanisms and validate the theory assumptions 
in phenomenological TMD studies. 
Recent JLab experiments have studied the contributions from the longitudinally polarized quarks in unpolarized nucleons which are critical for a rigorous TMD interpretation in SIDIS~\cite{CLAS:2021opg,CLAS:2022iqy,CLAS:2020yqf,CLAS:2020igs,Hayward:2021psm,CLAS:2022sqt}. 
The invariant mass dependence of the asymmetries have been observed in two hadron system, indicating an important role of vector meson decay contributions~\cite{CLAS:2020igs,Hayward:2021psm}. 
Finally, the $Q^2$ evolution of the SIDIS structure functions measured at JLab and COMPASS are greatly needed for validation of the current formalism in phenomenology.

\begin{figure}[ht]
    \centering
\includegraphics[width=0.95\columnwidth]{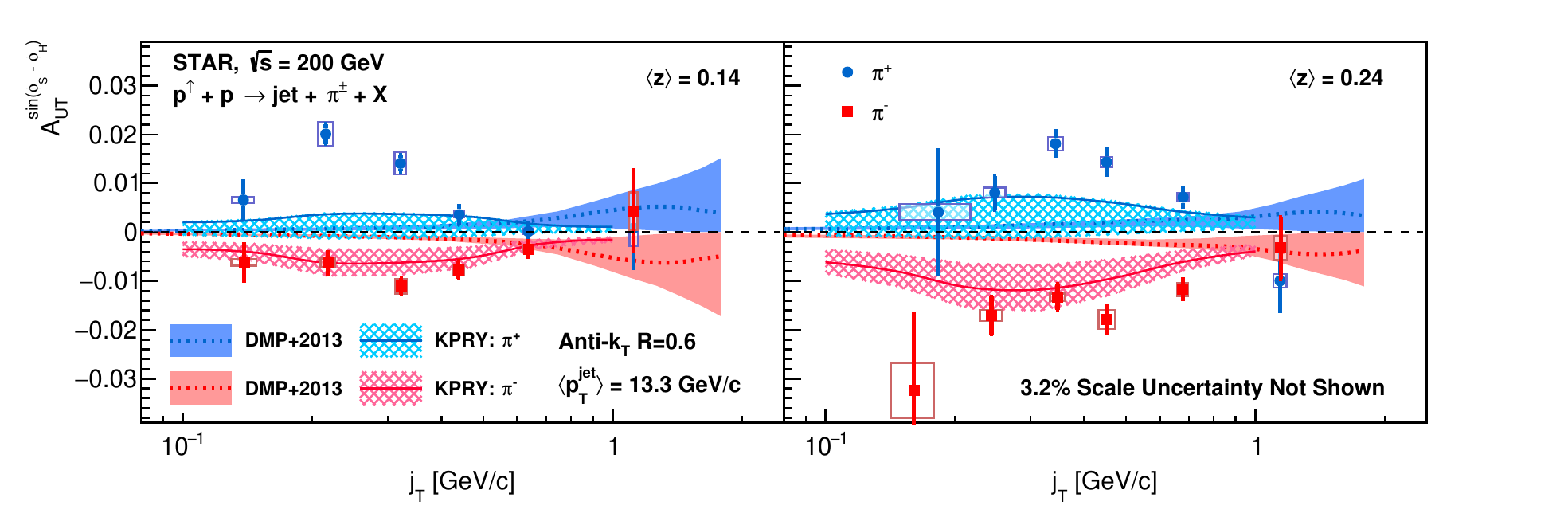}
     \caption{Collins asymmetry plotted for identified $\pi^{+}$ (blue) and $\pi^{-}$ (red) particles as a function of $j_T$ for two different hadron $z$ bins, in jets with $p_T > 9.9$ GeV/$c$ and $0 < \eta < 0.9$ in $p+p$ collisions at 200 GeV \cite{STAR:2022hqg}.  Theoretical evaluations from \cite{Kang:2017btw} and \cite{DAlesio:2017bvu} are also shown.
     }
     \label{fig:Collins_figure_z_jT}
\end{figure}

RHIC experiments have demonstrated the unique ways in which TMDs can be studied at hadron colliders. Azimuthal distributions of identified hadrons in high-energy jets measured at STAR directly probe the \textit{collinear} quark transversity in the proton coupled to the transverse momentum dependent Collins fragmentation function~\cite{Yuan:2007nd,DAlesio:2010sag,Kang:2017glf,Kang:2017btw,DAlesio:2017bvu}. 
Figure~\ref{fig:Collins_figure_z_jT} shows the STAR measurements of Collins asymmetries as functions of $z$, the fraction of jet momentum carried by the hadron, and $j_{T}$, the momentum of the pion transverse to the jet axis, in $p+p$ collisions at 200 GeV~\cite{STAR:2022hqg}. 
The $j_T$ dependence appears to vary with $z$, contrary to the assumptions of most current phenomenological models~\cite{Kang:2017glf,Kang:2017btw,DAlesio:2010sag}. 
STAR has also measured quark transversity through dihadron interference fragmentation functions in 200 and 500 GeV \pp\ collisions~\cite{STAR:2015jkc,STAR:2017wsi}.

Moreover, STAR measurements have shown the persistence of sizeable transverse single-spin asymmetries $A_N$ for forward $\pi^0$ production at RHIC energies up to 510 GeV with a weak energy dependence. 
STAR has explored the SSA in forward electromagnetic jet production as well~\cite{STAR:2020nnl}.
In addition, by utilizing $pA$ collisions at RHIC, both STAR and PHENIX collaborations studied the nuclear modifications of the forward hadron SSAs~\cite{PHENIX:2019ouo,STAR:2020grs}. 
Neither the origin of the nuclear dependence nor the difference between the PHENIX and STAR results is well understood at this time.

\subsubsection{Origin of the Nucleon Mass}

\begin{figure}[!ht]
  \begin{minipage}[c]{0.6\textwidth}
    \includegraphics[width=\textwidth]{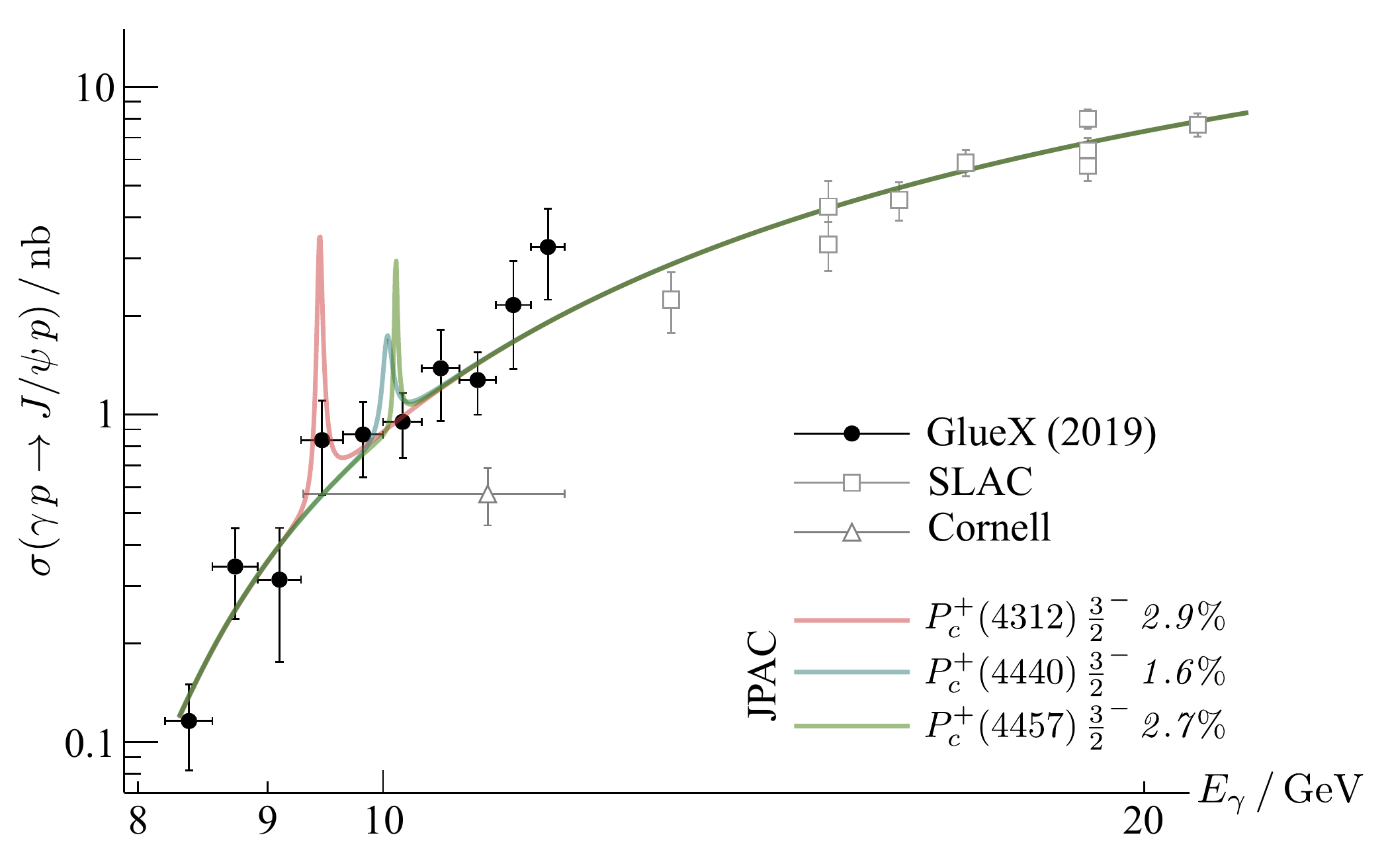}
  \end{minipage}\hfill
  \begin{minipage}[c]{0.4\textwidth}
    \caption{
   Results on $J/\psi$ photoproduction cross section as a function of beam energy from the JLab GlueX experiment~\cite{Ali:2019lzf}, compared to the JPAC model~\cite{HillerBlin:2016odx} including LHCb--motivated pentaquark resonances with hypothetical $J/\psi \, p$ branching ratios in italics. 
    } \label{fig:jpsi.jjd}
  \end{minipage}
\end{figure}

\begin{figure}[!hbt]
  \begin{minipage}[c]{0.55\textwidth}
    \includegraphics[width=\textwidth]{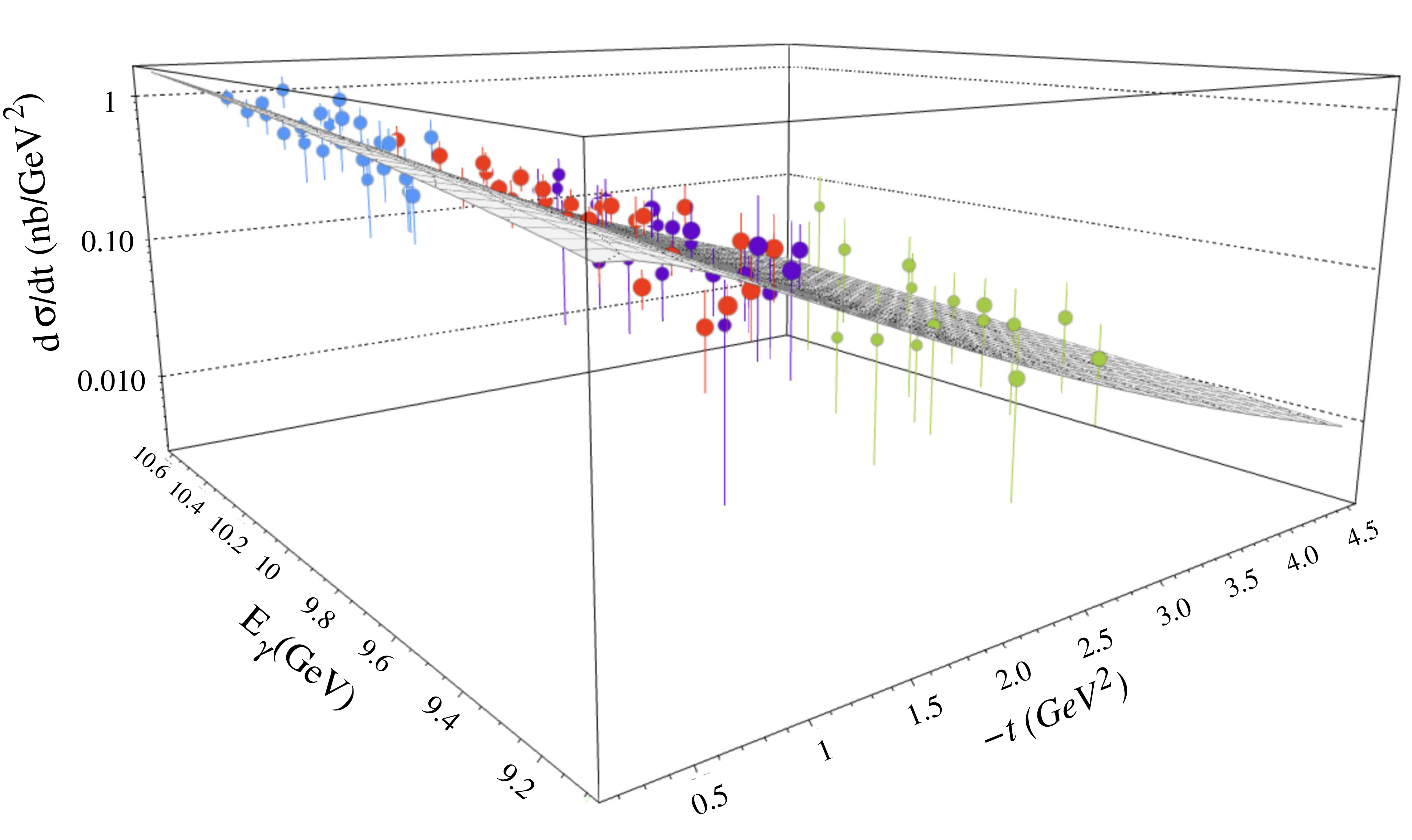}
  \end{minipage}\hfill
  \begin{minipage}[c]{0.35\textwidth}
\caption{Two-dimensional $J/\psi$ photoproduction cross section from the $J/\psi$-007 experiment, as a function of Mandelstam $|t|$ and photon energy $E_\gamma$~\cite{Duran:2022xag}. The grey surface is the result from a fit following a holographic QCD approach.
\label{fig:jpsi-xsec}}
  \end{minipage}
\end{figure}

The origin of the proton mass is one of the central questions in contemporary hadronic physics. The topic, highlighted in LRP15 and the 2018 National Academy of Sciences (NAS) assessment of the EIC, has seen many prominent experimental and theoretical developments in recent years. 
A promising channel to study the emergence of proton mass is quarkonium production near threshold, which is uniquely sensitive to the non-perturbative gluonic structure of the proton. 
In 2019, the GlueX collaboration published the first measurement of $J/\psi$ photoproduction in the near-threshold region~\cite{Ali:2019lzf}, see Fig.~\ref{fig:jpsi.jjd}. Their one-dimensional cross section results trend significantly higher than those previously extrapolated based on older measurements. These results spurred many new theoretical investigations of the gluonic structure of the proton and the origin of its mass~
\cite{Pefkou:2021fni,Kharzeev:2021qkd,Guo:2021ibg,Hatta:2019ocp,Mamo:2019mka,Ji:2021qgo,Sun:2021gmi,Ji:2021mtz,Ji:2021pys,Lorce:2021xku,Mamo:2021krl,Shanahan:2018pib,Wang:2021dis,Wang:2019mza,He:2021bof,Rodini:2020pis,Metz:2020vxd,Lorce:2017xzd,Hatta:2018sqd,Tanaka:2018nae}. In 2022, the JLab Hall C $J/\psi$-007 experiment released the first two-dimensional photoproduction measurement near threshold~\cite{Duran:2022xag}, shown in Fig.~\ref{fig:jpsi-xsec}. These two-dimensional results provide a new window into the gluonic gravitational form factors of the proton. The new data indicate that the proton has a dense, energetic core that contains most of its mass. 
In order to further our understanding of the origin of the proton mass, precision multidimensional measurements of near-threshold quarkonium production are needed, in particular at high momentum transfer. The timely completion of the planned experimental program at JLab, including $J/\psi$ production studies with SoLID, will be crucial. 
More data that may hold the key to a universal understanding of the origin of the proton mass are expected from the EIC, see discussions in Sec.~\ref{sec:eic_science}. 

\subsubsection{Hadron Spectroscopy}
\label{sec:hadronspect}

\vskip 0.3cm
\noindent
{\bf Lightest exotic hybrid meson}
The suggestion that that hybrid mesons, arising from excitations of gluon fields, could exist is as old as QCD. 
A smoking gun signature for such a state would be $J^{PC}$ quantum numbers outside the set allowed for a quark-antiquark pair, with $1^{-+}$ suspected to be the lightest. Experimental data from several facilities suggested \emph{two} low-lying isovector $1^{-+}$ states, a $\pi_1(1400)$ observed in $\eta \pi$, and a $\pi_1(1600)$ observed in $\eta' \pi$ and other channels~\cite{ParticleDataGroup:2022pth}. These were in stark contrast to the results of lattice QCD computations~\cite{Dudek:2013yja} which indicated there should be just a single low-lying state with these quantum numbers. Analyzing recent data from COMPASS on the $\eta \pi$ and $\eta' \pi$ final states~\cite{COMPASS:2014vkj}, the JPAC collaboration utilized a unitary description of coupled-channel amplitudes to show that the enhancements observed could be explained by just \emph{one} resonance, rigorously described by a single pole singularity.  This pole was found deep in the complex energy plane, indicating a broad resonance~\cite{JPAC:2018zyd}. A subsequent lattice QCD calculation~\cite{Woss:2020ayi} using heavier-than-physical quark masses considered the relevant scattering process in which this state appears, finding a single resonance pole in the coupled-channel amplitudes. Upon extrapolation to physical kinematics, relatively small partial widths were found for decay into the observed channels $\eta \pi$ and $\eta' \pi$, but a large coupling to $b_1 \pi$ generated a large total width, in agreement with the JPAC data analysis. GlueX is currently examining the $\eta \pi$ and $\eta' \pi$ final states in photoproduction, and this lattice calculation adds further motivation to the already underway examination of higher multiplicity final states.

\vskip 0.3cm
\noindent
{\bf Heavy quark exotics} While the vast majority of observed hadron states are understood to be composed of three quarks (baryons) or quark-antiquark pairs (mesons), QCD allows for other configurations, including four and five quark states known as tetraquarks and pentaquarks.
Recent observations at several experimental facilities have revealed many candidates for these unconventional states in the charm and bottom sectors~\cite{Esposito:2016noz,Guo:2017jvc,Olsen:2017bmm,Brambilla:2019esw,Chen:2022asf}.  Theoretical models accommodate individual measurements as tightly-bound multiquark states or as hadronic molecules, but no picture can describe all the new observations. Complicating matters further is the fact that many of the states are observed in non-trivial production or decay processes where three-body rescattering effects of essentially kinematic origin could mimic a resonance signature.
Directly producing such states in simple two-particle scattering can eliminate non-resonant interpretations. Utilizing the kinematic reach provided by the 12~GeV CEBAF, near-threshold $J/\psi$ photoproduction was studied for the first time at GlueX~\cite{Ali:2019lzf} and in Hall C~\cite{Duran:2022xag} to search for direct production of the hidden-charm $P_c^+$ pentaquarks observed by LHCb~\cite{LHCb:2015yax,LHCb:2019kea}.  While no resonance signals were observed, as shown in Fig.~\ref{fig:jpsi.jjd}, model-dependent upper limits on the branching ratios provide new constraints on the interpretation of these exotic candidates.  Higher energy photo- and electro-production experiments, such as the EIC and an energy-upgraded CEBAF, can provide new opportunities to directly produce other exotic charmonium-like states and shed light on their nature. 

\vskip 0.3cm
\noindent
{\bf Exotic hadrons in heavy ion collisions}
As described in Sec.~\ref{sec:hot}, the QGP created in heavy ion collisions is an abundant source of deconfined quarks, which can form hadrons by coalescence as the plasma expands and freezes out. Thus 
measurements of exotic states in heavy ion collisions provide new tests of production and transport models~\cite{Wu:2020zbx,Chen:2021akx}
and are potentially sensitive to the structure of the exotic
states themselves~\cite{Zhang:2020dwn}. 
The first measured heavy quark exotic state, X(3872), has recently been measured in $p+p$ (as a function of multiplicity), $p+$Pb and Pb+Pb collisions \cite{LHCb:2020sey,CMS:2021znk}.  While the X(3872) to $\psi$(2S) cross section ratio drops with multiplicity in $p+p$ collisions, there is indication 
of a rise with multiplicity in the larger $p+$Pb and Pb+Pb collision systems.  This varying behavior may indicate that a range of suppression and enhancement effects are coming into play.  Currently these measurements are statistics limited and additional studies with higher statistics data are required to clarify the situation.

In response to these recent measurements, several new theoretical developments have emerged.  Comover 
models have described the multiplicity dependence in $p+p$ in terms of X(3872) breakup via interactions with other particles produced in the event.  The results have been interpreted as providing evidence for the X(3872) as both a compact tetraquark \cite{Esposito:2020ywk} and a hadronic molecule \cite{Braaten:2020iqw}.  Calculations based on quark coalescence at freeze-out using the AMPT model show that, if the X(3872) is a molecular state, it should be greatly enhanced in central $A+A$ collisions relative to a compact tetraquark \cite{Zhang:2020dwn}.  A recent transport calculation comes to the opposite conclusion \cite{Wu:2020zbx}.  While all these models have successfully explained conventional charmonium behavior in both small systems and the QGP, their application to exotic states has provided new challenges.

\subsubsection{QCD in Nuclei}
\label{sec:coldQCD_nuclear}

\vskip 0.3cm
\noindent
{\bf Short range correlated nucleon pairs (SRCs)} SRC pairs are temporal fluctuations of two strongly-interacting nucleons in close proximity. They are characterized by large relative momentum ($p_{\rm rel}>k_F\approx250$ MeV) and smaller total momentum ($p_{\rm tot}\lesssim k_F$)~\cite{Frankfurt:1988nt,Subedi:2008zz,CiofidegliAtti:2015lcu,Ryckebusch:2019oya,Hen:2016kwk,CLAS:2018qpc}. 
At the time of LRP15, it was known that the very high momentum nucleons were almost entirely associated with SRCs, and were strongly dominated by $np$-SRCs for a wide range of nuclei. Since then, there have been significant advances in various aspects of SRCs and their relation to the EMC effect, made possible through extensive investigations of hard exclusive scattering reactions at JLab~\cite{Arrington:2022sov}.

\begin{figure}[!ht]
\centering
\includegraphics[width=0.4\textwidth]{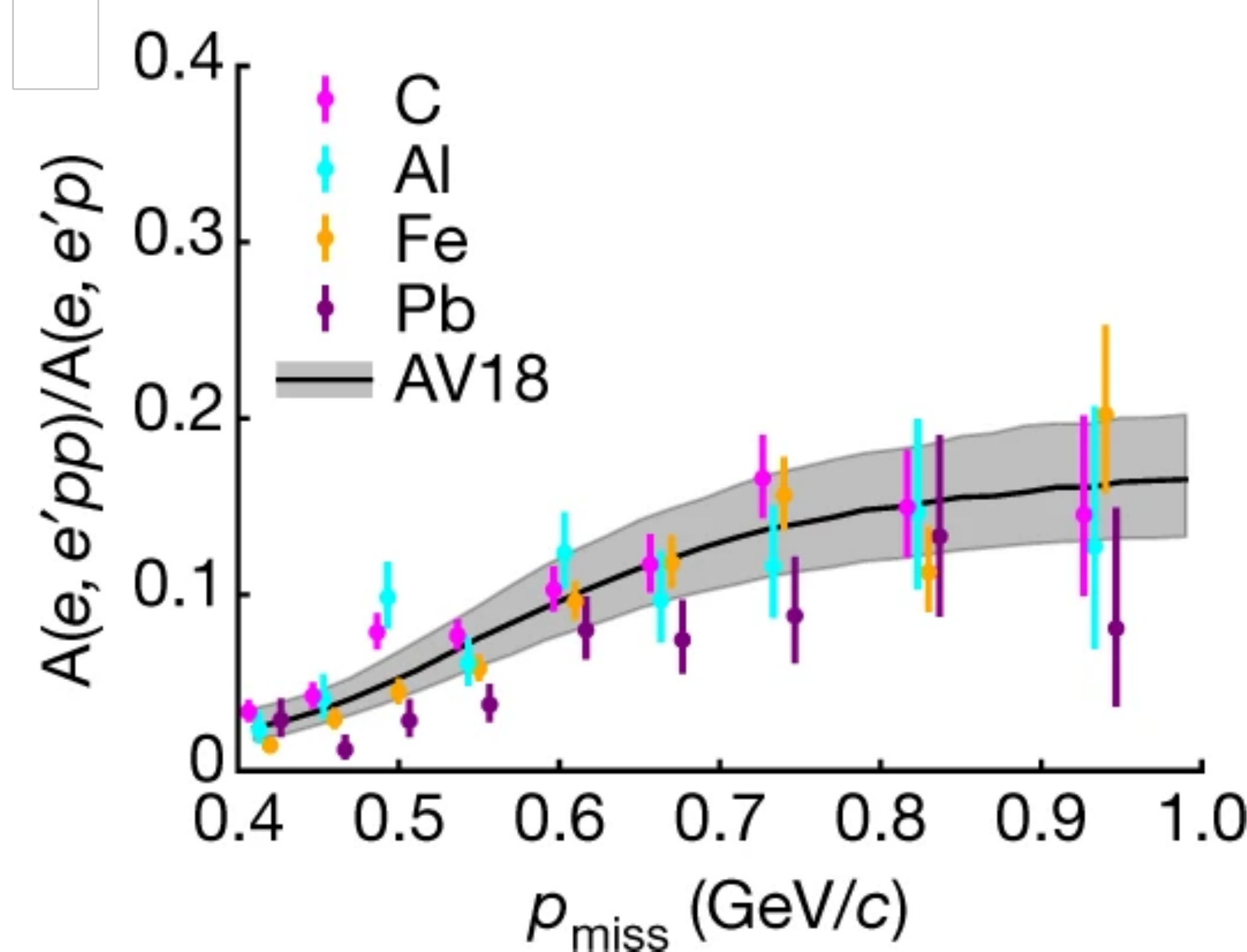}
\hspace{0.3cm}
\includegraphics[width=0.4\textwidth]{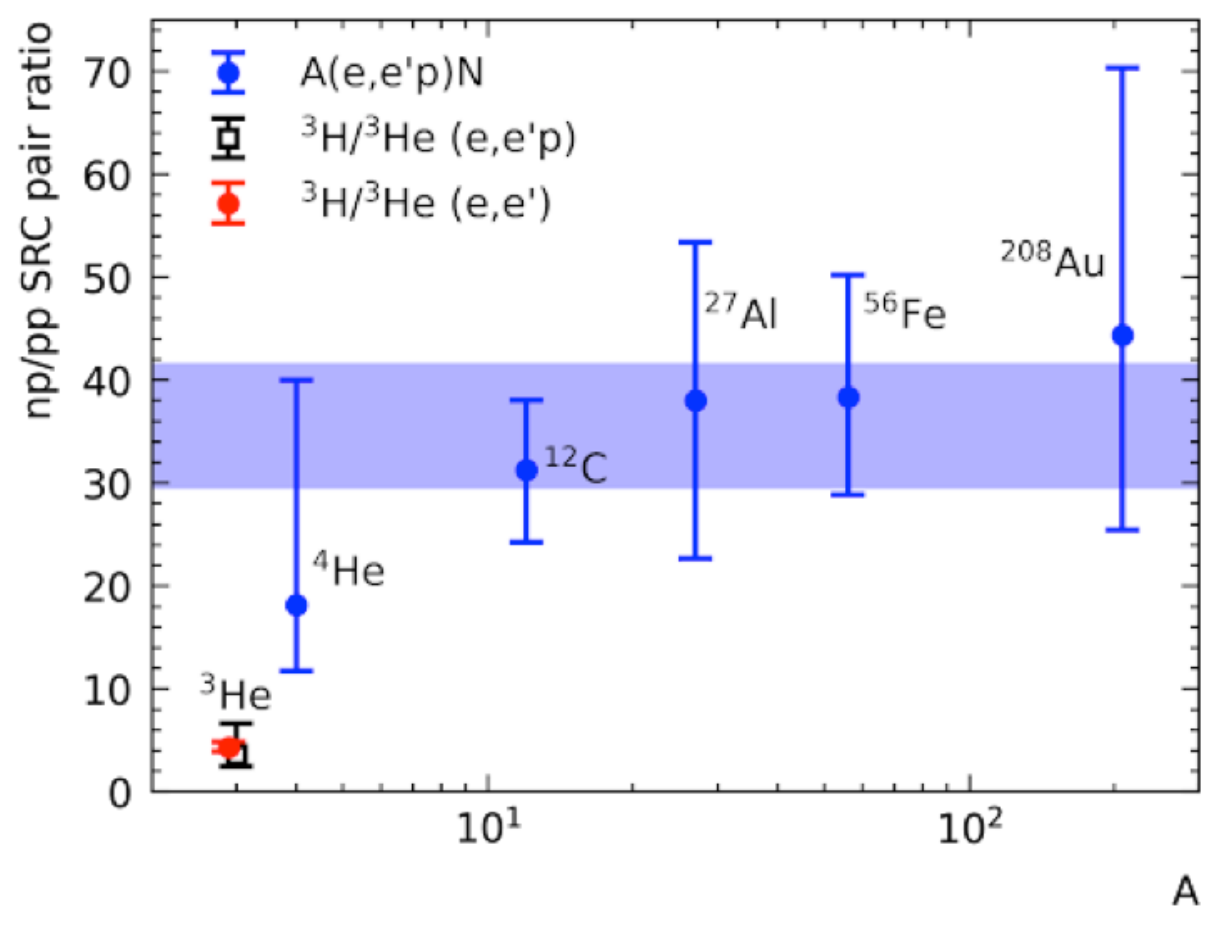}
\caption{{Left:} The ratio of (e,e’pp) to (e,e’p) cross sections for different nuclei as a function of the missing momentum of the first proton~\cite{CLAS:2020mom}.  The gray band shows a factorized calculation for carbon (C)  using the Generalized Contact Formalism (GCF) and the AV18 potential. 
{Right:} 
Extracted $np$-SRC/$pp$-SRC ratio from recent comparisons of $^3$H and $^3$He scattering along with previous world's extractions~\cite{Li:2022fhh}.
}\label{fig:LW}
\end{figure}

First, several new measurements have provided additional confirmation of the universality of the isospin and momentum structure of SRCs. 
Almost all high momentum nucleons in nuclei belong to an SRC pair~\cite{CLAS:2020rue}.  At intermediate relative momenta ($300\leq p_{\rm rel}\leq 600$ MeV), these SRC pairs are predominantly $np$ pairs, due to the tensor part of the nucleon-nucleon ($NN$) interaction~\cite{Hen:2014nza,CLAS:2018yvt,CLAS:2018xvc,Li:2022fhh}. This $np$ dominance can cause momentum inversion, where the minority nucleons (e.g., protons in neutron-rich nuclei) have higher average momentum than the majority nucleons~\cite{CLAS:2018yvt}.  

Second, recent measurements demonstrated that the near total $np$-dominance established above the nuclei Fermi-momentum, where the tensor force dominates, can be modified in special circumstances.
As the momenta probed increases to $700\leq p_{\rm rel}\leq 1000$ MeV, central correlations become more important and the relative numbers of $pp$ and $np$ pairs follow simple spin-state pair counting~\cite{CLAS:2020mom,CLAS:2020rue}, thus observing a new scalar-interaction dominated regime at very short distance scales, see Fig.~\ref{fig:LW} (left panel). 
In addition, recent $^3$H/$^3$He inclusive cross section ratio measurements in the SRC region give a ratio of $0.854\pm 0.010$~\cite{Li:2022fhh}, well below the expected ratio of unity expected from complete $np$-SRC dominance. This ratio, as well as the ratio for $(e,e'p)$ measurements~\cite{JeffersonLabHallATritium:2020mha}, 
can be used to extract the underlying $np$-SRC/$pp$-SRC ratio based on a plane-wave impulse approximation  
picture~\cite{Li:2022fhh}, see right panel of Fig.~\ref{fig:LW}. 
Meanwhile, both data are well described~\cite{CLAS:2020mom,CLAS:2020rue,Weiss:2020bkp} by Generalized Contact Formalism (GCF)~\cite{Weiss:2015mba,Weiss:2016obx,Weiss:2018tbu} calculations using realistic $NN$ interaction models with $np$-dominated SRC contact terms~\cite{Cruz-Torres:2019fum}, and for the inclusive data also by interactions with no tensor-force~\cite{Weiss:2020bkp}. This shows remarkable progress in our understanding of $np$-dominance dynamics and short-distance two-nucleon interactions in all measured nuclei, building connection between scattering data and nuclear theory~\cite{Carlson:2014vla,Pybus:2020itv,Weiss:2020bkp,CLAS:2020mom,CLAS:2020rue,West:2020tyo}. 

The momentum distribution of SRC pairs has also been probed in light nuclei using hard proton knockout from the deuteron~\cite{HallC:2020kdm} and $^3$He and $^3$H~\cite{JeffersonLabHallATritium:2019xlj,JeffersonLabHallATritium:2020mha}. The measured and ab-initio calculated cross-sections show good agreement up to very high momenta.  These measurements therefore  provide new insight to the very high-momentum tails of nucleon distributions in light nuclei, short-distance interactions, and few-nucleon dynamics.
Investigations are currently underway as part of the JLab 12 GeV program to probe beyond the 2N-SRC region and look for 3N-SRCs in nuclei, and test predictions for the universal contribution of 3-body structures at high-momenta.

\begin{figure}[ht]
\centering
\includegraphics[width=0.43\textwidth] {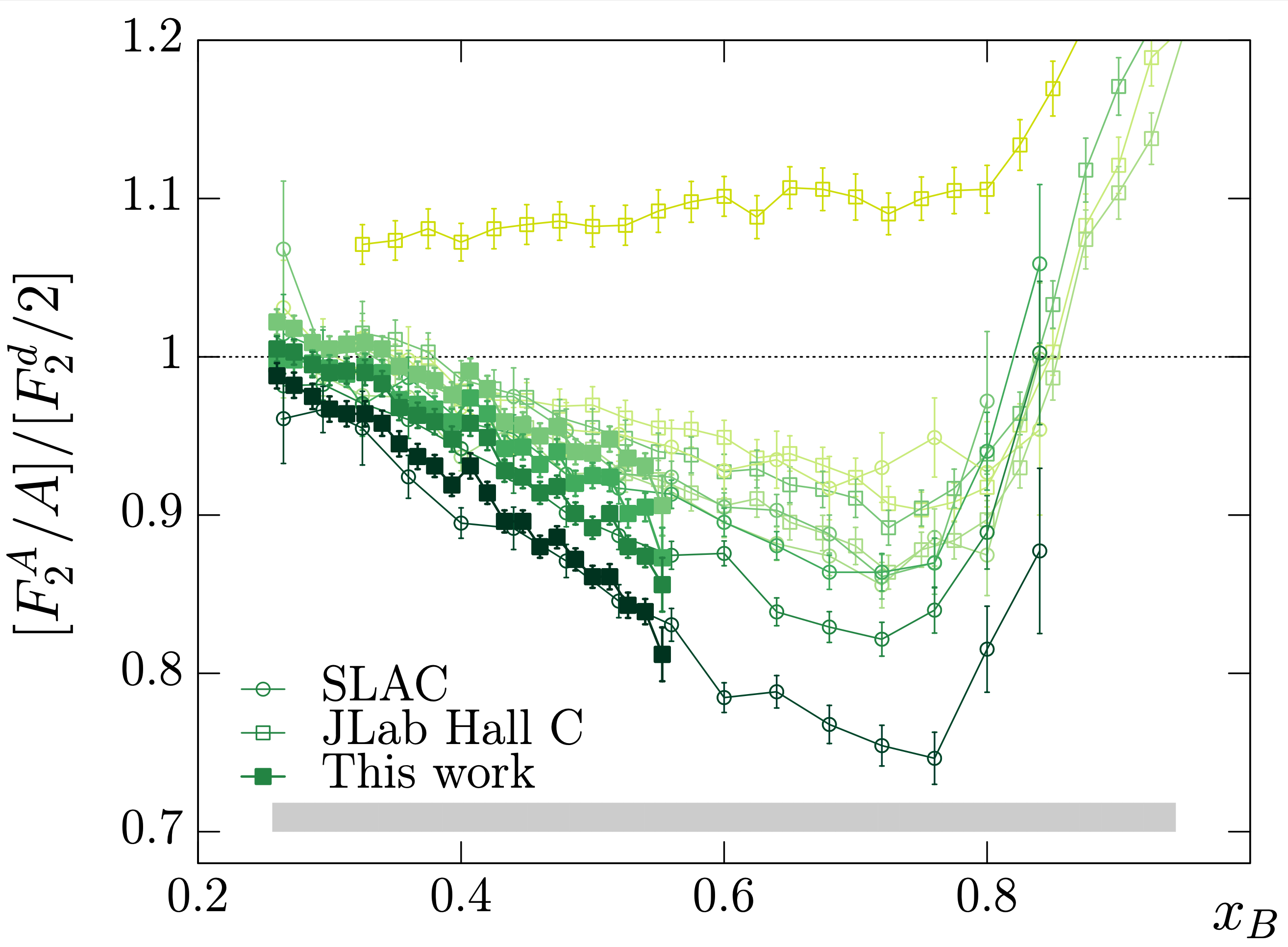}
\includegraphics[width=0.48\textwidth] {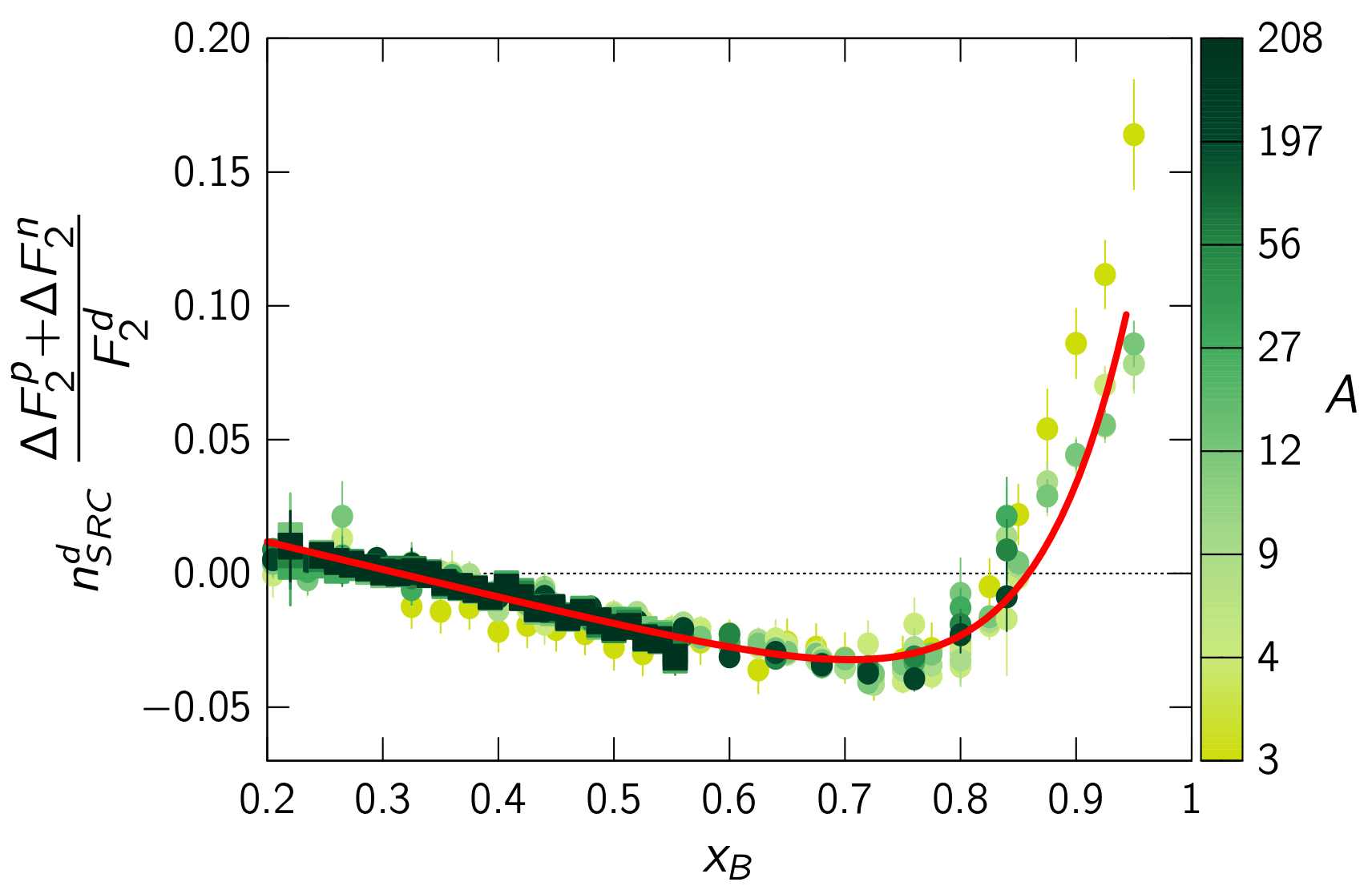}\\
 	     \caption{Left: Measured structure function ratio for different nuclei relative to deuterium~\cite{CLAS:2019vsb}. Right: The extracted universal modification function of nucleons in SRC pairs from a global data analysis~\cite{Segarra:2019gbp}.} 
\label{fig:EMCSRC}
 \end{figure}

\subsubsubsection{Nuclear EMC Effects and SRCs} 
As highlighted in LRP15, the strength of the EMC effect in nuclei, i.e., the nuclear modification of the valance structure functions measured in DIS, is linearly correlated with the abundance of SRC pairs~\cite{Weinstein:2010rt,Hen:2016kwk}. 
This indicates that the short-distance $NN$ interaction could modify nucleon structure.
The measured data could be explained by a universal modification of the structure of nucleons in SRC pairs~\cite{CLAS:2019vsb,Arrington:2019wky,Kim:2022lng}, 
providing empirical corrections of nuclear effects in the extraction of the free-neutron structure function~\cite{Segarra:2019gbp} (see Fig.~\ref{fig:EMCSRC}).

\vskip 0.3cm

At the meantime, results from recent measurements continue to proide insights on the EMC effect in both light and heavy nuclei~\cite{CLAS:2019vsb,Arrington:2021vuu,HallC:2022rzv}. 
In paricular, preliminary results from JLab have shown that the size of the EMC effect is nearly constant for $A=4$ and $A=9$ to $12$, and there is a clear correlation of the EMC effect with the local nuclear density~\cite{HallC:2022rzv}.
More measurements are being carried out to study both the EMC effect and SRCs for all available light nuclei, to study the connection to the detailed nuclear environment, and for heavier nuclei over a range on $N/Z$ to separate $A$-dependence from isospin effects. 

In addition, recent and planned measurements of novel observables such as tagged-DIS (TDIS) will probe the structure function of bound nucleons in specific nuclear states and will provide guidance for constraining off-shell nucleon-modification models that are currently largely unconstrained~\cite{Segarra:2020plg}.
Measurements of the spin structure function EMC effect will also test a complementary set of EMC models~\cite{Cloet:2005rt,Cloet:2006bq,Tronchin:2018mvu} where the bound nucleon modification is driven by the mean-field nuclear potential~\cite{Cloet:2009qs}.
The combination of all these measurements, including those discussed above and in Sec.~\ref{sec:cold_QCD_future_nuclei}, will provide an unprecedented understanding of the impact of the strong nuclear interaction on the internal structure of bound nucleons and thereby the parton structure of nuclei.

\vskip 0.3cm
\noindent
{\bf Nuclear modification of the parton distributions} In addition to the EMC effects discussed above, the nPDFs in the full kinematics, from the shadowing effects at small $x$ to the Fermi motion effects at large $x$, see, Fig.~\ref{fig:nPDF-improve}, provide a framework to study the cold nuclear matter effects. Previously, the nPDFs were extracted through global analysis of the experimental data from fixed-target DIS and Drell-Yan production in $pA$ collisions. In the last few years, proton-lead collisions at the LHC offer a wealth of opportunities to study cold nuclear matter effects, especially by using electroweak bosons~\cite{ATLAS:2015mwq,ALICE:2016rzo,ALICE:2020jff,CMS:2021ynu,LHCb:2022kph,CMS:2019leu,ALICE:2022cxs}.
Combining the LHC data with previous fixed target DIS and Drell-Yan data, the precision of the extracted nPDFs has improved significantly, see, e.g., recent global analyses from several groups~\cite{Duwentaster:2021ioo, Eskola:2021nhw, AbdulKhalek:2022fyi, Helenius:2021tof}. 
Furthermore, recent JLab CLAS data have provided a 3D extension of these studies with the measurement of incoherent and coherent DVCS giving access to both bound proton and nuclear GPDs, respectively~\cite{CLAS:2017udk, CLAS:2018ddh, CLAS:2021ovm}.

\subsubsection{Cold Nuclear Matter Effects in Hadron Production}

\noindent
Apart from the nuclear modification of structure function in DIS (EMC effects) discussed above, cold nuclear matter effects can be studied with semi-inclusive hadron productions in $e+A$ and $p+A$ collisions. The QCD dynamics are much more involved in these processes and the underlying mechanism could come from the parton distribution modifications, hadron formation in a nuclear environment, and small-$x$ gluon saturation in extreme kinematics. Therefore, a comprehensive study of these phenomena requires both theoretical and experimental investigations. Since LRP15, several notable developments have been achieved. Here, we highlight some new results from the LHC, RHIC, and JLab. Future developments are expected, in particular from the planned EIC, see Section~\ref{sec:future_eic}.

\vskip 0.3cm
\begin{figure}[!ht]
\centering
\includegraphics[width=0.9\textwidth] {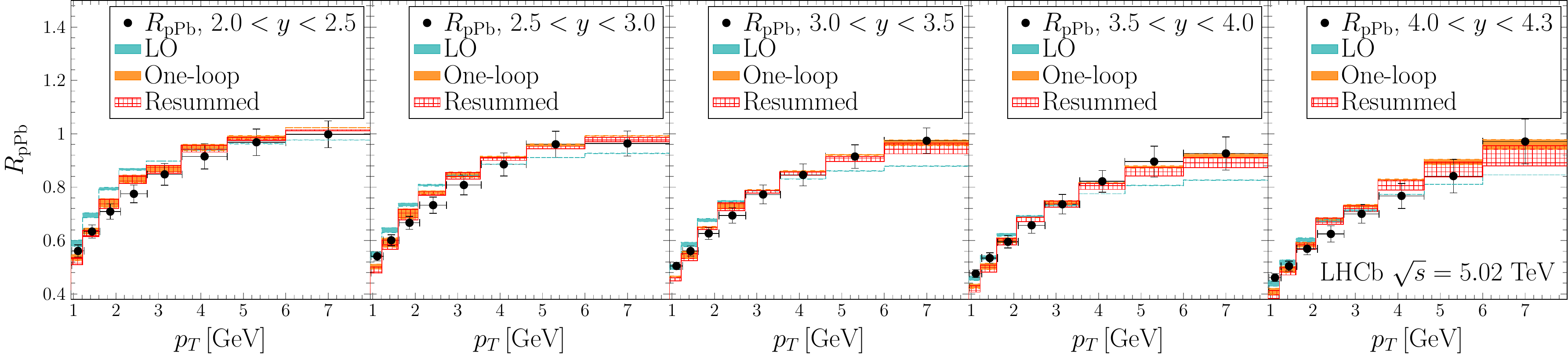}
 	     \caption{ LHCb Collaboration measurement of the nuclear modification factor of charge particles as a function of $p_T$ in different rapidity intervals for the forward region in $p+{\rm Pb}$ collisions at $\sqrt{S_{NN}}=5$~TeV at the LHC~\cite{LHCb:2021vww}, compared to state of the art CGC calculations~\cite{Shi:2021hwx}.}
 \end{figure}
\noindent
{\bf Nuclear modification of forward hadron yields in $p+A$ collisions at LHC and RHIC} Forward hadron production in $p+A$ 
collisions (or ${\rm d}+A$ deuteron-nucleus collisions at RHIC) have attracted a great deal of attention. Experimentally, the evolution of the nuclear modification factor $R_{\rm d{Au}}$ from mid-rapidity to forward-rapidity regions measured at RHIC is considered important evidence of the onset of gluon saturation~\cite{BRAHMS:2004xry,STAR:2006dgg}. Recently, the LHCb collaboration published the measurement of the charged particle spectra in $p+$Pb and $p+p$ collisions at $\sqrt{s_{NN}}=5$~TeV at the LHC, covering a wide range of kinematics, especially the forward pseudorapidity range of $1.6 <\eta < 4.3$~\cite{LHCb:2021vww}. The latest gluon saturation interpretation of all these measurements can be found in~\cite{Shi:2021hwx}, where the CGC calculations are performed at next-to-leading order and supplemented with threshold resummation effect. 

\vskip 0.3cm
\begin{figure}[!ht]
\centering
    \includegraphics[width=0.4\textwidth]{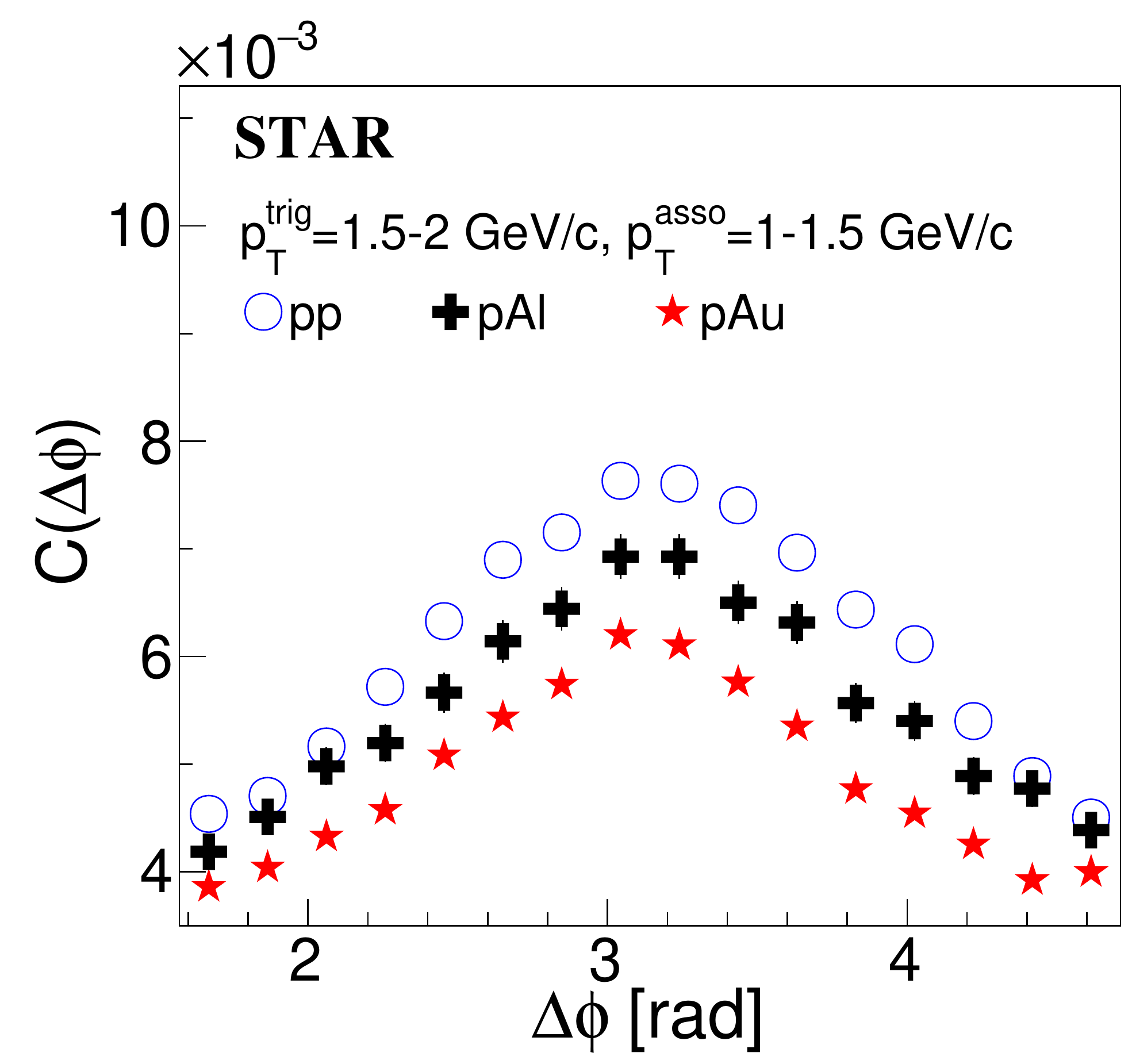}
    \hspace{0.4cm}
    \includegraphics[width=0.38\textwidth]{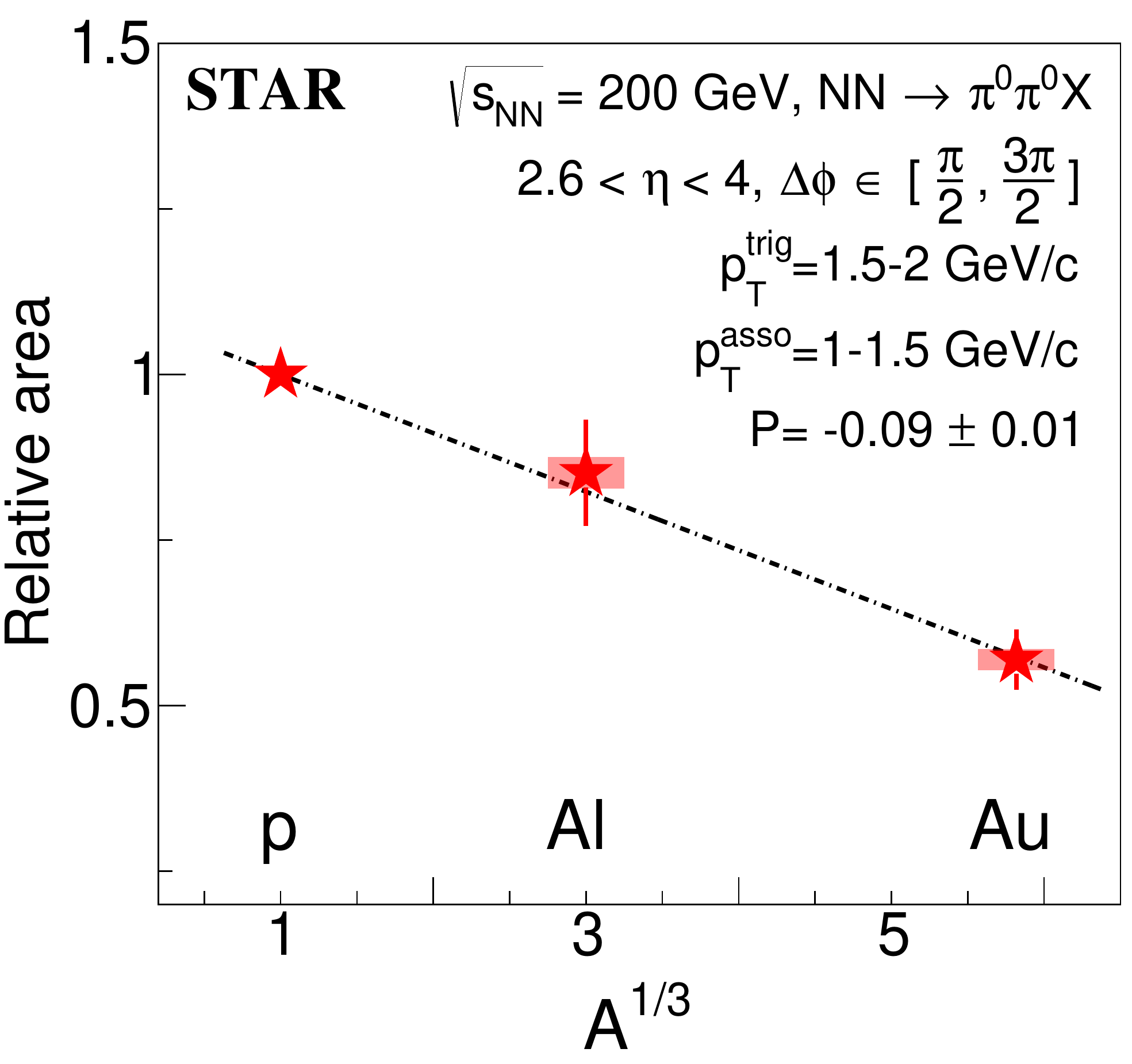}
 	     \caption{ Left: Comparison of 
       correlations functions with azimuthal angle separation between forward ($2.6<\eta<4.0$) $\pi^{0}$s in $p$+$p$, $p+$Al, and $p+$Au collisions at $\sqrt{s_{{_{NN}}}}=200$ GeV. 
Right: Relative area of back-to-back di-$\pi^0$ correlations in $p+$Au and $p+$Al with respect to $p$+$p$ collisions for $p^{\mathrm{trig}}_{T}$ = 1.5$-$2 GeV/$c$ and $p^{\mathrm{asso}}_{T}$ = 1$-$1.5 GeV/$c$. 
    }   \label{fig:Adep}
 \end{figure}
\noindent
{\bf Di-hadron (dijet) correlations in $p+A$ collisions} Understanding the nonlinear behavior of the gluon at small $x$ is one of the most important physics goals for the RHIC/LHC cold QCD program and the future EIC. The back-to-back di-hadron azimuthal angle correlation is one of the most sensitive direct probes the underlying gluon dynamics involved in hard scatterings~\cite{Marquet:2007vb,Dominguez:2011wm}. 
Earlier measurements of di-hadron correlations in d$+$Au collisions have indicated gluon saturation at small-$x$ in large nuclei~\cite{Braidot:2010ig,PHENIX:2011puq,Albacete:2010pg,Stasto:2011ru}.
The STAR Collaboration performed the measurements of back-to-back azimuthal correlations of di-$\pi^0$s produced at forward pseudorapidities ($2.6<\eta<4.0$) in $p + p$, $p + $Al, and $p + $Au collisions at $\sqrt{s_{{_{NN}}}}=200$ GeV. The results have been published~\cite{STAR:2021fgw} recently, showing a clear suppression of the correlated yields of back-to-back $\pi^0$ pairs in $p + $Al and $p + $Au collisions compared to the $p + p$ data. The observed suppression is larger at smaller transverse momentum, which indicates lower $x$ and $Q^2$. The larger suppression found in $p + $Au relative to $p + $Al collisions exhibits a dependence of the suppression on the mass number $A$. 
Higher-precision measurements will be performed with the STAR forward upgrade to further explore the nonlinear gluon dynamics. Those measurements will provide a baseline for searching for gluon saturation at the future EIC. Similarly, dijet correlations in $p+A$ collisions from the LHC have also studied small $x$ physics within a complementary kinematic coverage~\cite{ATLAS:2019jgo}, where theoretical interpretation combines the gluon saturation and high order soft gluon radiation contributions~\cite{vanHameren:2019ysa}.

\vskip 0.3cm
\noindent
{\bf Semi-inclusive hadron production in DIS with nuclear targets} 
Data from previous fixed target HERMES measurement 
have been applied to constrain the cold nuclear matter effects in hadron production in DIS processes.
Several JLab SIDIS experiments on deuterium, carbon, iron, and lead targets have been carried out by the CLAS Collaboration. 
In the first experiment~\cite{CLAS:2021jhm}, the production of single charged pions in SIDIS was measured with first-ever full kinematic coverage. In the second, the azimuthal angle dependence of two pion suppression in nuclear targets was compared to that of nucleon targets~\cite{CLAS:2022asf}, considerably extending the scope of previous studies with HERMES. 
The third studied $\Lambda (uds)$ attenuation and transverse momentum broadening~\cite{CLAS:2022oux}. New data on multi-dimensional attenuation of neutral pions should be released soon.  
All these new studies will explore hadron formation in cold nuclear matter with DIS, helping benchmark future studies at JLab and the EIC.

\vskip 0.3cm
\noindent
{\bf Color transparency} The search for the onset of color transparency (CT) is driven by the need to better understand the connection between the partonic and hadronic degrees of freedom in nuclei. 
Experimental measurements on this topic are not yet conclusive. A recent experiment from Hall C at JLab found no evidence of the onset of CT from proton knockout on a carbon nucleus up to $Q^2=14.2$~(GeV/$c)^2$~\cite{HallC:2020ijh}, whereas previous measurements with exclusive meson electroproduction 
have observed the onset of CT at an order of magnitude lower $Q^2$~\cite{Clasie:2007aa,Qian:2009aa,CLAS:2012tlh}. This is a puzzling situation and future searches for CT at JLab will extend the $Q^2$ range of measurements for pion and rho mesons~\cite{CLAS:2022Elfassi, 2022:Huber}, providing a direct comparison between protons and mesons.

\section{Future Opportunities}\label{sec:future}
\subsection{QCD Theory}

It is widely accepted that QCD is the right theory to describe the strong interaction, which governs all nuclear physics from quarks and gluons to nucleons and nuclei. 
Solving QCD in the non-perturbative region has been a great challenge of modern quantum field theory in the last half century and it will remain so in the foreseeable future. 
In this section, we will lay out the achievements in recent years and then focus on future perspectives related to the QCD theory. 
In particular, the challenges to confront the experimental data have motivated the theory community to form collaborative efforts for all research areas of QCD and examples will be discussed in Sec.~\ref{sec:theory_topical}. These topical collaborations will play crucial roles for future theory developments.
Note that this section is not comprehensive, as many aspects related to theory are discussed in the previous and following sections, along with experimental considerations. This section discusses more formal and purely theoretical aspects of QCD in detail. 

\subsubsection{Lattice QCD}\label{sec:theory_lattice}
Soon after the formulation of QCD, the Euclidean space-time lattice regularization was introduced, paving the way for numerical studies of non-perturbative QCD~\cite{Wilson:1974sk}. Several decades of efforts have demonstrated that lattice QCD is an unmatched tool for understanding strong interaction physics ranging from the partonic structure of nucleons to the QCD phase diagram. 

The structure of the nucleon has been a central component to the development of QCD. Due to their non-perturbative nature, the theoretical determination of many nucleon properties relies crucially on lattice calculations. Since LRP15, there has been tremendous progress in lattice computations of hadron structure, including the axial, scalar and tensor charges and form factors of nucleons, spin and mass decomposition, and various parton distributions. 

\vskip 0.3cm
\noindent
{\bf Nucleon axial, scalar and tensor charges and form factors} The simplest nucleon matrix elements give the nucleon charges and form factors. This includes the axial, scalar, and tensor charges. The axial and tensor charges are related to the longitudinal and transverse spin dependent quark distributions which can be explored in high energy hadronic experiments, such as inclusive DIS and SIDIS. There has been significant progress in lattice calculations of these charges in recent years~\cite{FlavourLatticeAveragingGroupFLAG:2021npn}. 
In particular, the nucleon axial charge $g_A$ served as an important benchmark calculation for lattice QCD applications to nuclear physics. 
The first lattice QCD result that fully addressed all sources of systematic uncertainty appeared in 2016~\cite{Bhattacharya:2016zcn} and results that were also in agreement with the Particle Data Group (PDG) value within one standard deviation appeared in 2017 and thereafter~\cite{Berkowitz:2017gql,Chang:2018uxx,Liang:2018pis,Gupta:2018qil,Shintani:2018ozy,Hasan:2019noy,Harris:2019bih,Alexandrou:2019brg,Walker-Loud:2019cif,Park:2021ypf}.
The precision achieved in the lattice QCD calculations of $g_A$ opens the door for this quantity to be elevated from an important benchmark result to another key quantity needed for precision low-energy tests of the Standard Model.
Meanwhile, at this level of precision, the radiative quantum electrodynamics (QED) corrections must be fully understood, see a recent example of $\mathcal{O}(2\%)$ pion-induced radiative correction to $g_A$~\cite{Cirigliano:2022hob}. 

The lattice calculations of the axial and tensor charges have provided benchmarks for phenomenological extraction of various parton distributions from experimental measurements of spin asymmetries, see, e.g., a combined analysis of nucleon tensor charge from lattice QCD and the quark transversity distributions from experiments~\cite{Lin:2017stx,Gamberg:2022kdb}. Similarly, progress made in the lattice calculations of nucleon form factors~\cite{Shanahan:2018nnv,Shanahan:2018pib,Alexandrou:2018sjm,Shintani:2018ozy,Djukanovic:2021cgp} provides important constraints on the modeling of the quark/gluon GPDs, see e.g.~\cite{Guo:2022upw}. The combination of lattice results with experimental measurements will continue to provide crucial information on nucleon tomography with upcoming programs at JLab and the EIC.

\vskip 0.3cm
\noindent
{\bf Pion and kaon form factors} Pions and kaons are among the most prominent strongly interacting particles next to the nucleon, since they are the
Goldstone bosons of QCD. Thus, it is important to study their internal structure and how it reflects their Goldstone boson nature; a question particularly relevant for understanding the origin of mass generation in QCD \cite{Cui:2020dlm,Roberts:2020udq}. While measuring pion and kaon form factors is one of the goals of the experimental program
at Jlab12 and the EIC \cite{Dudek:2012vr,AbdulKhalek:2021gbh}, current experimental information on the pion and kaon form factors is limited~\cite{Choi:2019nvk}, especially at large momentum transfer, making lattice QCD calculations more relevant. Recent lattice
calculations of the pion form factor have been performed with two flavors ($N_f=2$) of dynamical quarks
\cite{Brommel:2006ww,Frezzotti:2008dr,Aoki:2009qn,Brandt:2013dua,Alexandrou:2017blh},
with physical strange quark and two light quark flavors ($N_f=2+1$) 
\cite{Bonnet:2004fr,Boyle:2008yd,Nguyen:2011ek,Fukaya:2014jka,Aoki:2015pba,Feng:2019geu,Wang:2020nbf,Gao:2021xsm}, 
as well as with dynamical charm quark, strange quark and two flavors of the light 
quarks with nearly physical masses ($N_f=2+1+1$) \cite{Koponen:2015tkr}.
Most of the lattice studies so far focused on the small $Q^2$ behavior of the pion form factor, with the largest
momentum transfer studied so far corresponding to $Q^2 \simeq 1.4~ {\rm GeV}^2$ \cite{Gao:2021xsm}. 
With advanced techniques, such as boosted sources \cite{Bali:2016lva} and increased computational resources
it should be possible to extend the lattice form factor calculations to $Q^2 \simeq 30~{\rm GeV}^2$, i.e.,
the region of interest for the EIC.

\begin{figure}[!ht]
  \centering
 \includegraphics[width=0.42\hsize]{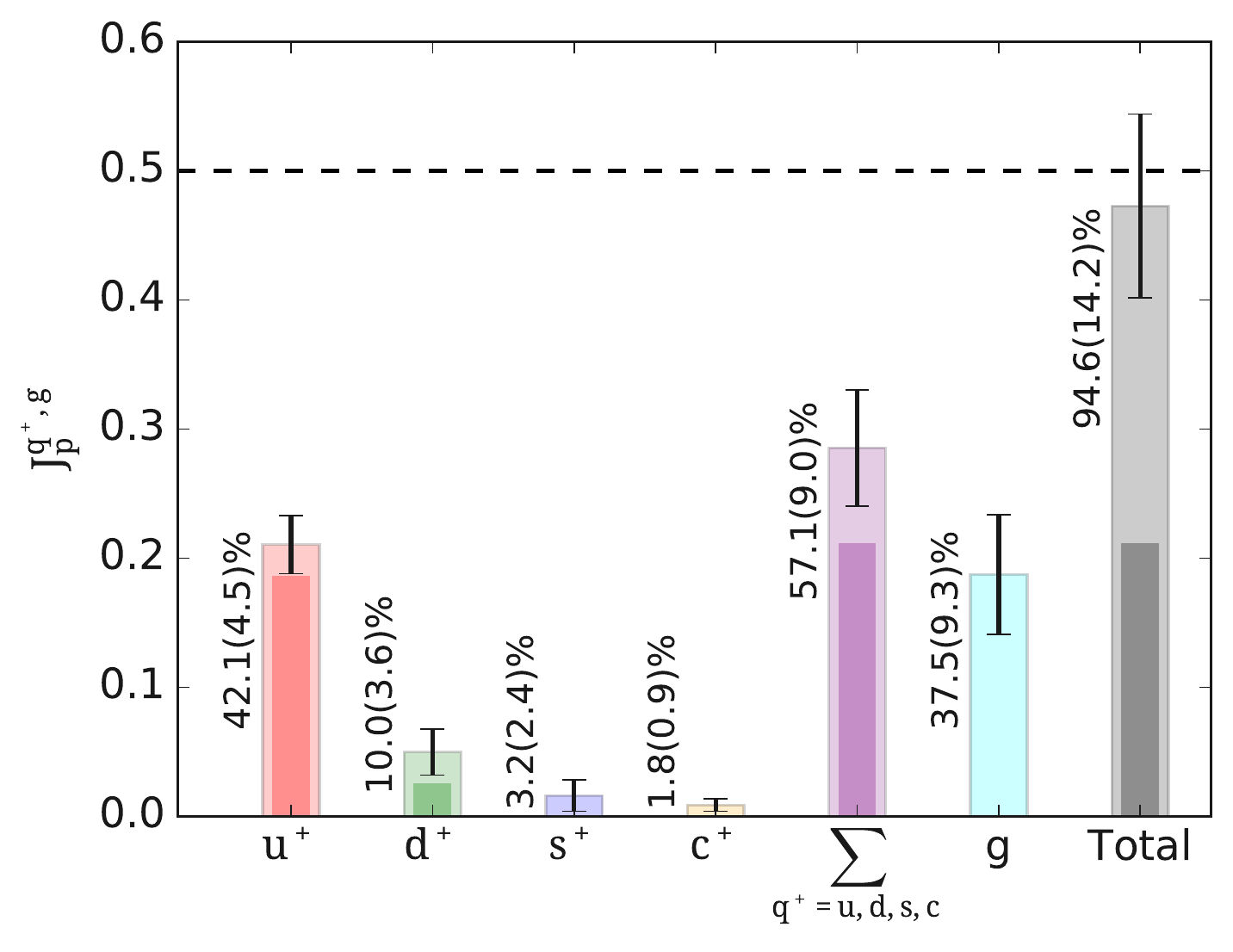}
 \hspace{0.5cm}
  \includegraphics[width=0.43\hsize,angle=0]{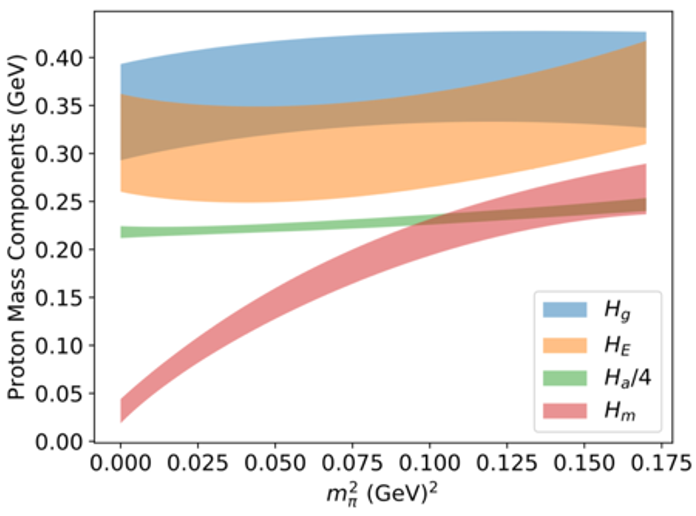}
  \caption{State of art lattice QCD calculations of emerging properties of the nucleon: (left) spin decomposition in terms of the angular momentum $J_q$ for the $u,d$ and $s$ quarks and the gluon angular momentum $J_g$ in the $n_f = 2+1+1$ calculation from the ETMC collaboration~\cite{Alexandrou:2020sml}; (right) mass decomposition in terms of $\langle H_m\rangle$, $\langle H_E\rangle (\mu)$, $ \langle H_g\rangle (\mu)$, and $ \frac{1}{4}\langle H_a\rangle$ at $\mu = 2$ GeV as functions of $m_{\pi}^2$ from $\chi$-QCD collaboration~\cite{Yang:2018nqn}. }
  \label{fig:spinmass}
\end{figure}

\vskip 0.3cm
\noindent 
{\bf Spin and mass decomposition of the nucleon} Lattice QCD has been extensively applied to understand the origin of proton spin and mass. 
Since LRP15, there have been two complete lattice calculations of the nucleon spin decomposition with renormalization, mixing and normalization properly taken into account. In this decomposition, the proton spin is constructed from individual quark and gluon angular momentum contributions, and the quark orbital angular momentum (OAM) contribution can be further derived by subtracting the associated helicity contributions~\cite{Ji:1996ek}. 
One calculation uses the twisted mass fermion on a $N_f = 2+1+1$-flavor lattice with lattice spacing of $a = 0.08$ fm and pion mass of 139 MeV~\cite{Alexandrou:2020sml}. 
The results on the angular momentum fractions $J$ for the $u,d,s,c$ quarks and gluons are shown in the left panel of Fig.~\ref{fig:spinmass}. The summed quark $J_q$ is 57.1(9.0)\% and $J_g$ is 37.5(9.3)\% of the total angular momentum. The quark helicity contribution is also calculated to be $\frac{1}{2} \Delta \Sigma = 0.191(15)$. This leaves the quark OAM with 18.8(10.2)(2)\% of the total spin. Another calculation is based on the valence overlap fermion on a $2+1$-flavor domain wall fermion sea on a $32^3 \times 64$ lattice with $a = 1.43$ fm and pion mass of 171 MeV with a box size of 4.6 fm~\cite{Wang:2021vqy}. 
The results~\cite{Wang:2021vqy} for the percentage contributions are 
$\Delta \Sigma$, $J_g$ and $L_q^{}$ at 40(4)\%, 46(5)\% and 13(5)\%, respectively. 
All these results are matched to the $\overline{\rm MS}$ scheme using the renormalization scale $\mu=2$ GeV. In addition, a complementary approach with direct access to quark OAM, based on Wigner functions, has also been pursued~\cite{Engelhardt:2020qtg,Engelhardt:2021kdo}, yielding compatible results as above and providing further insight on different formalisms for the quark OAM.

The hadron mass and its decomposition can be obtained from the energy-momentum tensor (EMT). 
According to Ref.~\cite{Ji:1994av}, the rest energy has the following expression: $E_0 =   \langle H_m\rangle + \langle H_E\rangle (\mu)+ \langle H_g\rangle (\mu) + \frac{1}{4}\langle H_a\rangle$, where $\langle H_m\rangle$ is the quark condensate, $\langle H_E\rangle (\mu)$ is the quark kinetic and potential energy, and $\langle H_g\rangle (\mu)$ is the glue field energy. Both of them depend on the scale. Finally, $\langle H_a\rangle$ is the trace anomaly and is renormalization group invariant. A lattice calculation of this
decomposition was carried out by the $\chi$QCD Collaboration~\cite{Yang:2018nqn}. This calculation was done with the overlap fermion on four 2+1-flavor domain-wall fermion configuration ensembles for 3 lattice spacings. The largest lattice is at the physical pion point and full non-perturbative renormalization and mixing are performed. 
The right panel of Fig.~\ref{fig:spinmass} shows the fractional decomposition of the rest energy in terms of $\langle H_m\rangle$, $\langle H_E\rangle (\mu)$, $ \langle H_g\rangle (\mu)$, and $ \frac{1}{4}\langle H_a\rangle$ in the $\overline{\rm MS}$ scheme at $\mu = 2$ GeV, as functions of $m_\pi^2$. 
Clearly, except for $\langle H_m\rangle$, the components are fairly independent of the quark mass up to $m_{\pi} = 400$ MeV.

\vskip 0.3cm

\noindent
{\bf Parton distributions} Calculating the partonic structure of bound states from first principles lattice QCD with controlled accuracy remains an important unsolved problem. Previous efforts have been focused on the moments of collinear PDFs, which provide momentum-space ``global” information about partons. 
In recent years, new opportunities emerged for lattice QCD calculations to investigate the partonic structure of hadrons. Novel methods enabled the calculations of the $x$ dependence of PDFs, GPDs, and TMDs, which was previously considered unattainable. The major complication is the inability to obtain light-cone quantities from the Euclidean formulation of lattice QCD. The realization that lattice matrix elements of non-local operators can be related to light-cone distributions has transformed the field of hadron structure~\cite{Monahan:2018euv,Cichy:2018mum,Ji:2020ect,Constantinou:2022yye}, with the U.S. leading several aspects. 

Various methods have been developed~\cite{Liu:1993cv,Liu:1998um,Liu:1999ak,Detmold:2005gg,Detmold:2021uru,Braun:2007wv,Ji:2013dva,Ji:2014gla,Radyushkin:2016hsy,Ma:2014jla,Ma:2014jga,Ma:2017pxb,Chambers:2017dov} and significant progress has been made using two major approaches: large momentum effective theory (LaMET) in which the parton $x$-dependence is calculated directly via a large momentum expansion~\cite{Ji:2013dva,Ji:2014gla}, and short-distance factorization or operator product expansion, which comes under various names such as pseudo distributions~\cite{Radyushkin:2016hsy} or ``good lattice cross sections''~\cite{Ma:2014jla,Ma:2014jga,Ma:2017pxb}, in which the $x$-dependencies are parametrized and fitted with the global analysis method. 
Precision calculations of isovector unpolarized, helicity, and transversity quark PDFs are ongoing, with simulations at the physical quark masses, and with controlled systematic uncertainties, including renormalization of linear divergences and continuum limit, two-loop matching and renormalization group resummation ~\cite{Alexandrou:2018pbm,Lin:2018pvv,Alexandrou:2018eet,Alexandrou:2019lfo,Joo:2020spy,Bhat:2020ktg,HadStruc:2021qdf,Gao:2021dbh,LatticeParton:2022xsd}. While the lattice QCD results for the sea quarks and gluon PDFs still have a long way to go before they reach good accuracy, the field is up-and-coming and more developments are expected to appear shortly~\cite{HadStruc:2021wmh,Fan:2021bcr,Salas-Chavira:2021wui,HadStruc:2022yaw,Egerer:2020hnc}. 

These developments have also been applied to hadron tomography of GPDs and TMDs. 
They are computationally more expensive than the integrated PDFs due to being differential in additional kinematic variables, e.g., the momentum transfer between the initial and final hadronic states for the GPDs, and transverse-momentum degrees of freedom in TMD PDFs and wave functions. 
First exploratory results on the proton unpolarized and helicity GPDs were reported in Refs.~\cite{Alexandrou:2020uyt,Lin:2020rxa,Alexandrou:2021bbo,Lin:2021brq} and more detailed investigations are ongoing. 
On the TMD side, substantial insight into TMD spin physics has been obtained through lattice calculations of TMD ratio observables~\cite{Musch:2011er,Engelhardt:2015xja,Yoon:2017qzo,Engelhardt:2023aem}. Meanwhile, the theoretical developments in the last few years have paved a way to compute the TMDs from lattice based on the LaMET formalism~\cite{Ji:2014hxa,  Ji:2018hvs,Ji:2019sxk,Ji:2019ewn,Ji:2020jeb,Ebert:2019okf,Ebert:2019tvc,Ebert:2020gxr,Shanahan:2019zcq,Shanahan:2020zxr,Shanahan:2021tst,Li:2021wvl,LPC:2022ibr,Ebert:2022fmh,Schindler:2022eva,Zhang:2022xuw}. The associated evolution kernel and soft factors have been computed~\cite{LatticeParton:2020uhz,Shanahan:2020zxr,Li:2021wvl,Ebert:2018gzl,Schlemmer:2021aij,Shanahan:2021tst,Zhang:2020dkn} and a preliminary result for the isovector quark TMDs has recently appeared~\cite{LPC:2022zci}. 

\vskip 0.3cm

\noindent
{\bf Hadron spectroscopy } Hadron spectroscopy as a field is undergoing a rapid development, where the emergent phenomenon of the hadron resonance spectrum as measured in experiments can be directly connected to non-perturbative QCD by means of rigorous amplitude analysis and lattice QCD computation. In the area of lattice computations of meson resonances, researchers in the US are world-leading, in particular the \emph{hadspec} collaboration [{\tt www.hadspec.org}] has made pioneering contributions to the calculations.

The formalism that relates the discrete spectrum of states in a finite volume to elastic two-body scalar particle scattering amplitudes has been in place for over thirty years~\cite{Luscher:1986pf}, but it was only recently that lattice QCD technology developed to the level where the calculations were practical. At the time of the last LRP, only a few elastic scattering systems, and a single coupled-channel system ($\pi K, \eta K$) had been studied in explicit lattice QCD calculations, yielding QCD descriptions of the $\rho$ and $K^*$ resonances. In the years since, many more meson-meson and meson-bayon scattering sectors have been explored in lattice calculations with heavier-than-physical light quarks, exposing the variety of ways resonances can manifest themselves in scattering amplitudes~\cite{Dudek:2016cru, Briceno:2017qmb,Woss:2019hse,Bulava:2022vpq} and providing a first-principles QCD approach to studying longstanding mysteries such as the nature of the light scalar mesons and the decays of the lightest exotic hybrid meson~\cite{Woss:2020ayi}. By coupling currents to scattering systems~\cite{Briceno:2014uqa, Briceno:2015dca, Alexandrou:2018jbt, Radhakrishnan:2022ubg}, the internal structure of unstable resonances can be explored in unprecedented ways.

Going beyond two-particle scattering, finite-volume formalisms to describe three-particle scattering have been derived~\cite{Hansen:2019nir,Mai:2021lwb} and these are currently being applied to explicit lattice QCD calculations in cases of maximal isospin, like $\pi^+ \pi^+ \pi^+$, where there are neither three-body resonances, nor any resonances in two-body sub-channels~\cite{Blanton:2019vdk, Hansen:2020otl, Blanton:2021llb, Brett:2021wyd}. The demonstrated success in these trial channels motivates further ongoing studies of channels in which resonances appear, allowing a much larger fraction of the QCD spectrum to be investigated rigorously, including many exotic hadrons of contemporary interest.

\subsubsubsection{QCD phase diagram}
The study of the QCD phase diagram with lattice simulations has experienced tremendous progress in the last few years. The phase transition line is typically determined by extrapolating chiral observables to finite chemical potential $\mu_B$, and finding the temperature at which the chiral condensate has an inflection point, or the chiral susceptibility has a peak.
The transition temperature as a function of $\mu_B$ can be written as ${T_c(\mu_B)}/{ T_c(\mu_B=0)}=1- \kappa_2 \left({\mu_B}/{T_c(\mu_B)}\right)^2- \kappa_4 \left({\mu_B}/{T_c(\mu_B)}\right)^4+\cdots$.
A high-precision result for the crossover temperature $T_c(\mu_B=0)$ has become available, $T_c(\mu_B=0)$=158 $\pm$ 0.6 MeV \cite{Borsanyi:2020fev}, which is in agreement with the previously quoted value $T_c(\mu_B=0)$=156.5 $\pm$ 1.5 MeV \cite{HotQCD:2018pds}. Current extrapolations to finite chemical potential reach out to $\mu_B \approx 300$ MeV, through the precise knowledge of the coefficients $\kappa_2=0.0153\pm0.0018$ and $\kappa_4=0.00032\pm0.00067$ \cite{Borsanyi:2020fev}.
Similar coefficients for the extrapolation of the transition temperature to finite strangeness, electric charge and isospin chemical potentials were obtained in Ref. \cite{HotQCD:2018pds}.
No sign of criticality is observed from lattice QCD simulations up to $\mu_B\simeq300$ MeV \cite{Borsanyi:2020fev,HotQCD:2019xnw}.
Future challenges include the extrapolation of the phase transition line to larger values of chemical potential and more stringent constraints on the location of the critical point.

\subsubsubsection{QCD equation of state}
The QCD EOS has been known at $\mu_B=0$ with high precision for several years \cite{Borsanyi:2010cj,Borsanyi:2013bia,Bazavov:2014pvz}. The sign problem prevents direct simulations at finite chemical potentials. However, different extrapolation methods have been used to obtain the EOS at moderate values of $\mu_B$. Significant progress has been achieved through a Taylor expansion of the thermodynamic quantities \cite{Bazavov:2017dus,Guenther:2017hnx,Bollweg:2022fqq}, currently limited to $\mu_B/T<3$. A new expansion scheme has extended the range of the EOS to $\mu_B/T<3.5$ \cite{Borsanyi:2021sxv}. An alternative approach with a similar range in $\mu_B/T$ has been developed in \cite{Vovchenko:2017gkg}, where the equation of state has been constructed as a relativistic virial expansion in baryon number fugacity.
All of these equations of state are two-dimensional: thermodynamic variables are calculated as functions of the temperature and the baryon chemical potential. However, in QCD there are other two conserved charges: elecric charge $Q$ and strangeness $S$. A choice needs to be made for the respective 
chemical potentials $\mu_Q$ and $\mu_S$. Typical choices are $\mu_Q=\mu_S=0$ or $\mu_Q=\mu_Q(T,~\mu_B)$ and $\mu_S=\mu_S(T,~\mu_B)$ such that the phenomenological conditions $n_Q=0.4n_B$ and $n_S=0$ are satisfied, with $n_i$ number density for charge $i$. These conditions reflect the overall strangeness and electric charge fraction in the colliding nuclei of a heavy-ion collision. An extension of the new expansion scheme to strangeness neutrality and beyond was presented in Ref. \cite{Borsanyi:2022qlh}.
A full four-dimensional equation of state, expanded as Taylor series in $\mu_B/T,~\mu_S/T$ and $\mu_Q/T$, is available in Refs. \cite{Noronha-Hostler:2019ayj,Monnai:2019hkn}.

The BEST collaboration has built an equation of state which reproduces lattice QCD results up to $\mathcal{O}((\mu_B/T)^4)$ and contains a critical point in the 3D-Ising model universality class \cite{Parotto:2018pwx,Karthein:2021nxe,An:2021wof}. This EOS can be used in hydrodynamic simulations of heavy-ion collisions to check the effect of changing the location and strength of the critical point.
Future challenges, both for lattice QCD and the BEST collaboration equations of state, are to extend them to larger coverage of the QCD phase diagram.

\subsubsubsection{Heavy flavor probes of hot matter on the lattice}
The heavy flavor diffusion coefficient characterizes the movement of a heavy quark with a momentum of at most the order of the temperature with respect to the medium rest frame. For this reason, it can  contribute to our understanding of thermalization of heavy quarks in the QGP. Estimates of this quantity in the deconfined phase of QCD were presented in Refs.~\cite{Ding:2012sp,Francis:2015daa}, albeit in the quenched approximation. This quantity was studied on the lattice by means of the gradient flow method~\cite{Altenkort:2020fgs,Mayer-Steudte:2021hei,Banerjee:2022uge}. A wide temperature range has been explored in Ref. \cite{Brambilla:2020siz}, where the multilevel algorithm was used. The results from the gradient flow and multilevel algorithm methods are consistent with each other. Future challenges include the continuum extrapolation of this observable in full QCD with realistic simulation parameters. 
Calculations with physical quark masses will require exascale computing resources, allocated through the ALCC \cite{ALCC} and INCITE \cite{INCITE} programs. To take advantage of exascale resources, lattice QCD codes must be adapted to the computational hardware, requiring funding from programs like SciDAC \cite{SciDAC}. 
Larger $N_{\tau}$ lattice will not only address bottomonium properties at $T\neq 0$, the complex $Q\overline Q$ potential, and the heavy quark diffusion coefficient but also, with minimal additional investment, improve studies of charm fluctuations and charm baryon number correlations.
The lattice can also study a novel chromoelectric field correlator that describes in-medium dynamics of heavy quark-antiquark pairs \cite{Brambilla:2016wgg,Yao:2020eqy}, which has been shown to be different from the heavy quark diffusion coefficient \cite{Eller:2019spw,Scheihing-Hitschfeld:2022xqx}.

\subsubsection{Theory and Phenomenology of Cold QCD}
\label{sec:theory0}

Applying QCD theory to both hot and cold QCD physics is a great challenge, due to the nonperturbative nature of strong interaction phenomena. 
Therefore, approximations have to be made to confront the experimental measurements, either by using QCD factorization with proper power counting, or building a rigorous numerical framework. 

In the QCD factorization formalism, the hadronic cross sections are factorized into the partonic hard scattering cross sections and the associated non-perturbative hadron structure. The central task for QCD theory is to provide precision computations of the various relevant partonic hard-scattering cross sections and splitting functions at high orders of perturbation theory. The past few years have seen tremendous progress in this area. For example, 
for the longitudinal momentum distribution functions (spin-dependent and spin-independent), the associated DGLAP evolution kernels are now fully known to next-to-next-to-leading order (NNLO)~\cite{Moch:2004pa,Vogt:2004mw,Moch:2014sna}. 
Salient examples of computations of partonic cross sections for, e.g., $e+p$ scattering at the EIC, at NNLO and even beyond, include work on inclusive DIS~\cite{Zijlstra:1992qd,Zijlstra:1993sh,Borsa:2022irn} and jet production in DIS~\cite{Currie:2017tpe,Currie:2018fgr,Boughezal:2018azh,Borsa:2020ulb}.  
\label{sec:theory_hf}
Other developments include heavy quark and quarkonium production in various hard scattering processes~\cite{Bodwin:1994jh,Brambilla:2010cs,Ma:2016exq,Cheung:2017loo,Cheung:2017osx,Cheung:2018tvq,Cheung:2018upe,Cheung:2021epq,Vogt:2018oje,Vogt:2019xmm} and the principle of maximum conformality arguments in perturbative calculations~\cite{Wu:2019mky}. 

The theoretical framework for establishing the tomographic structure of hadrons, as encoded in GPDs, has been well established and higher order perturbative QCD corrections have been calculated. The first computation of NNLO corrections for DVCS has also been reported recently~\cite{Braun:2017cih,Braun:2020yib,Braun:2022bpn}. 
Progress has been made toward a global analysis of GPDs, including a wide range of experiments~\cite{Belitsky:2001ns,Berthou:2015oaw,Kumericki:2016ehc,Kriesten:2019jep,Kriesten:2020wcx,Grigsby:2020auv,Kriesten:2021sqc,Guo:2021gru,Shiells:2021xqo,Guo:2022cgq,Guo:2022upw}. 
Meanwhile, recent global analyses have achieved high precision for the unpolarized TMD quark distribution and fragmentation functions from fits to data on semi-inclusive hard processes ~\cite{Bacchetta:2017gcc,Scimemi:2017etj,Scimemi:2019cmh,Bacchetta:2022awv,Bury:2020vhj,Bury:2021sue}. All these computations will impact the extraction of parton distributions and tomographic structure from future experiments, including JLab and EIC.

More generally, global analysis is a powerful tool that has recently been applied to extracting the unpolarized parton distribution functions~\cite{Hou:2019efy,Bailey:2020ooq,NNPDF:2021njg,Cocuzza:2021rfn,Barry:2021osv,Barry:2018ort}, quark/gluon helicity distributions~\cite{DeFlorian:2019xxt,Nocera:2014gqa,Cocuzza:2022jye}, TMDs and GPDs, as mentioned above. The key feature of these developments is to utilize the computational advances and apply theoretical constraints, including the lattice results. With future data from JLab and EIC on the horizon, the role of global analysis will become even more important. 

An important thread of theory developments is the application of the effective degrees of freedom of QCD to derive an effective field theory (EFT). These developments have not only revealed emerging dynamics of strong interaction physics but also provided advanced techniques to apply perturbative methods to deal with complicated hadronic processes. Recent progress along this direction has made it possible to compute various observables in both hot and cold QCD. In the following, we will describe two examples that have significant impact.

\subsubsubsection{Soft-collinear effective theory}
Soft-collinear effective theory (SCET) is an effective field theory which systematically describes the infrared QCD dynamics in hard collisions, including those associated with soft and collinear degrees of freedom~\cite{Bauer:2000yr,Bauer:2001yt,Bauer:2002nz}. It has been widely applied to a large variety of collider processes. This is partly because SCET provides a systematic and convenient method to perform high order perturbative calculations through the universal steps in deriving factorization in terms of independent functions governing the hard, collinear and soft dynamics of a process. SCET is also transparent in carrying out higher order resummation of large logarithms. Moreover, it has the ability to generalize the factorization to more complicated processes and multiscale observables, and the capability to systematically study power corrections. 
SCET continues to have a significant impact on the field of high precision calculations for hard scattering processes at various colliders, including Higgs/$Z$/$W$ boson production, and inclusive jet and multi-jet production at the LHC. 

In connection to hadron physics, SCET played an important role to clarify the QCD factorization for various hard processes where one can extract nucleon structure, such as the TMDs, see, e.g.~\cite{Ebert:2022fmh}. A key development in recent years is the analysis of power corrections to the factorization formalism~\cite{Ebert:2021jhy}, which will have potential impact on future phenomenological applications at JLab and the EIC. 

\subsubsubsection{Color-glass condensate}
There are compelling theoretical arguments and strong experimental hints that suggest that  gluon distributions saturate at small Bjorken-$x$~\cite{Gribov:1984tu,Mueller:1985wy,McLerran:1993ni,McLerran:1993ka,Iancu:2003xm,Kovchegov:2012mbw}. Gluon saturation occurs when the nonlinear terms in the field strength tensor are of the same magnitude as the kinetic terms or, equivalently, when the occupancy of field modes is $O(1/\alpha_s)$. The CGC is a QCD EFT that describes the physics of small-$x$ modes in protons and large nuclei and the underlying dynamics of gluon saturation at collider energies. The evolution of the complex many-body dynamics of partons in this  regime of QCD with energy is described in the CGC EFT by  powerful renormalization group (RG) equations~\cite{Balitsky:1995ub,Kovchegov:1998bi,JalilianMarian:1997gr,JalilianMarian:1997dw,Iancu:2000hn,Ferreiro:2001qy} that underlie the predictive power of this theoretical framework. The state-of-the art of these RG equations is at NLO accuracy with significant ongoing theoretical and computational work.  

An attractive feature of the CGC EFT is that it can be employed to explore the dynamics of small-$x$ modes and gluon saturation across a wide range of high energy experiments, from electron-hadron DIS from HERA to the EIC, to hadron-hadron, hadron-nucleus and nucleus-nucleus collisions at RHIC and the LHC. In DIS, NLO calculations are emerging for an increasing number of processes in electron-nucleus collisions, while a parallel program of precision comparisons of theoretical predictions to data is underway for proton-nucleus and ultra-peripheral collisions at the LHC (see \cite{Morreale:2021pnn} for a review). 

The CGC EFT also provides a compelling model of the initial conditions in heavy-ion collisions, as shown in Sec.~\ref{sec:init}. A significant body of research in this direction enables one to quantitatively assess the impact of this framework on the space-time evolution of matter in such collisions, and has played a key role in the quantitative extraction of transport coefficients of the quark-gluon plasma. 

Similar data-theory comparisons at the EIC will help solidify and quantify these insights into the 3-D tomography of gluons~\cite{Dominguez:2011wm,Hatta:2016dxp}. These will require a global analysis of data from hadron-nucleus and electron-nucleus collisions in analogy to successful global analysis studies in perturbative QCD discussed above.  
An important theoretical development is the emerging quantitative connections of the CGC EFT to the TMD and GPD frameworks in perturbative QCD. These studies can help refine and expand the predictive power of both frameworks.  
Another set of interesting questions is whether studies of overoccupied states in other systems in nature across wide energy scales  can provide deeper insight into universal features of  gluon saturation; a particularly promising approach is  the perspective provided by quantum information science~\cite{Kharzeev:2017qzs,Hagiwara:2017uaz,Hentschinski:2022rsa,Dvali:2021ooc,Dumitru:2022tud,Duan:2021clk}. An intriguing possibility is that of designing cold atom analog quantum computers (discussed further later) to capture dynamical features of such systems~\cite{Berges:2014bba,Prufer:2018hto}.

\vskip 0.3cm
\noindent
{\bf QCD-inspired models of hadron structure} 
Due to its non-perturbative nature, strong interaction physics has inspired a great deal of models, see textbooks~\cite{Bhaduri:1988gc,Thomas:2001kw} for a summary. In recent years, a number of models have helped to unveil the nontrivial feature of hadron structure and stimulated further theoretical developments. This includes the Schwinger-Dyson (Bethe-Salpeter) equations~\cite{Maris:2003vk,Cloet:2013jya} and instanton liquid models~\cite{Schafer:1996wv} where a certain truncation is needed to apply these models; and the light-front holographic model~\cite{Brodsky:2014yha}; the light-front Hamiltonian model~\cite{Lan:2019vui,Xu:2021wwj,Arifi:2022pal,Choi:2015ywa,Ji:2000fy}. 
All these models have captured certain features of non-perturbarive QCD physics and have gained success to describe the hadron structure to some extent. However, because the connections between the model degrees of freedom and the fundamental ones are unknown, the uncertainties from these calculations may not be under control. 

\vskip 0.3cm
\noindent
{\bf Amplitude analysis to unveil the QCD hadron spectroscopy}
In anticipation of CEBAF 12 GeV operations, in 2013 the Joint Physics Analysis Center (JPAC) was formed to develop the necessary theoretical, phenomenological and computational frameworks for analysis and interpretation of data. The quality and complexity of modern spectroscopy-relevant datasets is such that it is only by collaboration between experimentalists and theorists like those in JPAC that robust results on the hadron spectrum can be obtained.
While the search for light exotic hadrons in experiments at Jefferson Lab continues to be one of the main efforts of JPAC, over time the reach of the center has expanded worldwide with its members now affiliated with experiments outside JLab, including BESIII, COMPASS, and LHCb.

The need for sophisticated amplitude analyses is pressing in view of the copiously produced $XYZ$ states, where what is required is a systematic study of reaction mechanisms to isolate genuine   resonances from other effects, {\it e.g.} kinematical singularities which may generate peaking structures without a resonance being present~\cite{Pilloni:2016obd, COMPASS:2020yhb}. In this context direct production using photon beams would provide an independent validation of the resonance nature of the $XYZ$'s by virtue of the absence of the kinematic singularities present in the three-body production through $b$-hadron decays or $e^+e^-$ annihilation. JPAC has studied both exclusive~\cite{Albaladejo:2020tzt} and semi-inclusive~\cite{Winney:2022tky} photo-production of the $XYZ$ states, and made predictions for future measurements at EIC and an energy-upgraded CEBAF.

 By providing a forum for close collaboration between theory and experiments, JPAC has been successful in effecting integration of theoretical developments into experimental analyses, and in educating a new generation of practitioners in the tools of amplitude analysis.

\subsubsection{Theory and Phenomenology of Hot QCD}

\subsubsubsection{Theory of jets in hot QCD matter} 
The discovery of jet quenching at RHIC in the early 2000s \cite{PHENIX:2001hpc,STAR:2002ggv} and confirmation from the study of fully reconstructed jets at the LHC \cite{ATLAS:2012tjt} has spurred much theoretical and experimental research activity in the past decade with the objective of using jets as a multi-dimensional tool to probe the properties of the quark gluon plasma at various length scales (see Sec.~\ref{sec:jetprogress}). 

The current picture of parton energy loss is based on a medium induced gluon cascade that efficiently transports energy from fast color charges down to the plasma temperature scale where energy is dissipated \cite{Jeon:2003gi,Blaizot:2013vha,Blaizot:2013hx,Mehtar-Tani:2018zba,Schlichting:2020lef}. A future prospect is to improve on the accuracy of such gluon cascades by systematically computing higher order corrections to medium-induced gluon splitting including full kinematics \cite{Sievert:2019cwq,Arnold:2021pin,Arnold:2020uzm,Mehtar-Tani:2019tvy,Barata:2021wuf}.  Another important direction of research is the study of the medium response to the passage of a jet which describes how the distributions of low momentum partons are affected  \cite{Tachibana:2015qxa, Casalderrey-Solana:2016jvj, Tachibana:2017syd, KunnawalkamElayavalli:2017hxo, Casalderrey-Solana:2020rsj, Tachibana:2020mtb, Yang:2021qtl, Yang:2022nei}.

Because jets are complex quantum systems, their energy loss in the QGP is sensitive to color decoherence, an emergent QCD phenomenon caused by rapid color precession of entangled color charges \cite{Mehtar-Tani:2010ebp,Mehtar-Tani:2011hma,Casalderrey-Solana:2012evi,Mehtar-Tani:2017ypq}. 
It was recently investigated in the leading logarithm approximation of the inclusive jet spectrum \cite{Mehtar-Tani:2021fud,Caucal:2020zcz} and was shown to yield an excess of soft particles inside the jet in a study of the jet fragmentation function \cite{Mehtar-Tani:2014yea,Caucal:2018dla}. 
Extensive theoretical studies of jet substructure were carried out to diagnose energy loss mechanisms and color decoherence \cite{Caucal:2021cfb,Caucal:2019uvr,Casalderrey-Solana:2020jbx,Ringer:2019rfk,Chien:2018dfn,Mehtar-Tani:2016aco, Pablos:2022mrx}, and will play a crucial role in the future to fully exploit jet quenching observables to probe the resolution power of the hot QCD media. 

Higher order corrections to jet observables in heavy ion collisions are paramount for precision tests of jet quenching and will certainly constitute a major focus of future theoretical approaches to jet quenching. As an example, it was recently shown that some corrections are enhanced by a large double logarithm in the medium size \cite{Liou:2013qya,Blaizot:2013vha,Blaizot:2014bha,Iancu:2014kga,Ghiglieri:2022gyv,Arnold:2021mow} which, when resummed to all orders, results in an anomalous scaling of transverse momentum broadening that reflects super diffusive behavior \cite{Caucal:2022fhc,Caucal:2021lgf}. Higher order corrections to radiative energy loss were also investigated \cite{Arnold:2016kek,Arnold:2016jnq,Arnold:2020uzm,Blaizot:2014bha,Wu:2014nca}. More progress is required, and will rely on help from high performance computational tools, such as Monte Carlo event generators and lattice techniques, in order to achieve precision tests of non-equilibrium QCD dynamics using jet observables.

\subsubsubsection{Effective theory approaches in hot and dense QCD} \label{sec:eff}
There are two main thrusts in using effective theories to compute properties of QCD at nonzero temperature and density. 
By asymptotic freedom, perturbation theory only allows calculations at very high temperature.
Computations at temperatures $\sim 300$~MeV requires resummation of hard thermal loops
\cite{Haque:2013qta,Haque:2013sja,Andersen:2015eoa,Haque:2020eyj}, very technically challenging.
At low temperatures hadronic models can be used. In between these two, in the
region of greatest experimental interest, numerical simulations are the only method of first principles computation. However, as discussed in more detail above, these methods are limited by the existence of the sign problem at finite quark chemical potential, especially for $\mu_q>T$. 
Thus it is well worth developing effective models which can complement results from the lattice. 
Effective models can advance further into the $T$ and $\mu_q$ plane, particularly for $\mu_q > T$, to determine the transport coefficients as a function of $T$ and $\mu_q$ and explore phenomena such as the location of the critical endpoint, moat spectra and color superconductivity. 

The Functional Renormalization Group \cite{Fu:2022gou} has been applied to QCD, including estimates of the critical endpoint, how trajectories flow in the plane of $T$ and $\mu_q$, {\it etc.}
Results for the shear viscosity have been obtained at both zero
\cite{Horak:2021pfr,Horak:2021syv,Horak:2022myj,Lowdon:2021ehf} and nonzero density
\cite{McLaughlin:2021dph,Grefa:2022sav}.
Another approach is to use approximate solutions to the Schwinger-Dyson equations \cite{Fischer:2018sdj}.
Dynamical transport \cite{Weil:2016zrk, Bleicher:2022kcu} and quasiparticle models have been developed to compute transport properties at high \cite{Cao:2018ews,Aichelin:2019tnk,Moreau:2019vhw,Soloveva:2019xph,Soloveva:2020hpr,Rose:2020sjv,Cassing:2021fkc,Fotakis:2021diq,Soloveva:2021quj,Li:2022ozl}
and intermediate \cite{TMEP:2022xjg}
energies.  These models, while approximate, have the real virtue of being able to compute at nonzero density
with similar efficiency as at zero density.
These models have also been used to compute jet transport coefficients \cite{Grishmanovskii:2022tpb}.
While holography obtains results for the most supersymmetric
$SU(N)$ theory at large $N$, it is a useful approach
\cite{DeWolfe:2010he,Critelli:2017oub,Ishii:2019gta,Jokela:2020piw,Jokela:2021vwy,Grefa:2021qvt,Demircik:2021zll}.
Transport properties have also been obtained in holographic models~\cite{Finazzo:2014cna,Hoyos:2020hmq,Hoyos:2021njg,Grefa:2022sav}.
Lastly, matrix models for the semi-QGP have been developed to describe the equilibrium properties of QCD at both zero
\cite{Hidaka:2009hs,Dumitru:2010mj,Dumitru:2012fw,Pisarski:2016ixt,Hidaka:2020vna}
and nonzero chemical potential \cite{Pisarski:2016ixt}. A preliminary attempt to compute the
shear viscosity was made years ago \cite{Hidaka:2008dr,Hidaka:2009ma}, but needs to be improved by including
a complete effective Lagrangian \cite{Hidaka:2020vna}. 

Further development of these effective theory approaches, along with lattice QCD, will be important for understanding QCD and will be crucial for improving heavy ion phenomenology and the extraction of QGP properties in the coming years.

\subsubsubsection{Hydrodynamics and kinetic theory} \label{sec:hydrotheory} 
Since LRP15, new developments concerning the emergence of hydrodynamics under extreme conditions have shed light on the regime of applicability of hydrodynamics in heavy-ion collisions. Hydrodynamization, the process of approaching hydrodynamic behavior, was systematically investigated in a variety of systems at strong and weak coupling \cite{Chesler:2010bi,Heller:2011ju,Heller:2012km,Heller:2013oxa,Chesler:2013lia,Chesler:2015wra,Keegan:2015avk,Spalinski:2017mel,Denicol:2014xca,Denicol:2014tha,Kurkela:2015qoa,Bazow:2015dha,Romatschke:2017vte,Kurkela:2018wud,Kurkela:2018vqr,Kurkela:2019set,Denicol:2019lio,Brewer:2019oha,Almaalol:2020rnu,Mullins:2022fbx,Brewer:2022vkq}. Results demonstrated that the hydrodynamic gradient expansion in rapidly expanding plasmas can become divergent \cite{Heller:2013fn,Buchel:2016cbj,Denicol:2016bjh,Heller:2016rtz}, which naturally led to question how one may systematically define hydrodynamics. 
The prevailing picture is that the onset of hydrodynamics in high-energy heavy-ion collisions may be identified by the presence of a hydrodynamic attractor \cite{Heller:2015dha}, which provides a key new element in the extension of hydrodynamics towards the far-from-equilibrium regime \cite{Florkowski:2017olj,Romatschke:2017ejr,Berges:2020fwq,Chattopadhyay:2021ive, Jaiswal:2021uvv}. Furthermore, it was systematically investigated how one may resum not only gradients but also the viscous stresses themselves, through anisotropic hydrodynamics \cite{Martinez:2010sc,Florkowski:2010cf,Alqahtani:2017mhy, McNelis:2018jho, McNelis:2021zji}. New causal and stable first-order theories of (general-)relativistic viscous hydrodynamics have been formulated \cite{Bemfica:2017wps,Kovtun:2019hdm,Bemfica:2019knx,Hoult:2020eho,Bemfica:2020zjp,Noronha:2021syv}, which opened up new opportunities to systematically investigate hydrodynamic phenomena without the need to evolve extra variables in addition to the standard hydrodynamic fields, as it occurs in 2$^\mathrm{nd}$ order formulations \cite{Israel:1979wp}. 

Many more theoretical developments are needed in order to fully determine the applicability of hydrodynamics in heavy-ion collisions. A systematic investigation of the nonlinear properties of 2$^\mathrm{nd}$ order hydrodynamics, going beyond the first results of \cite{Bemfica:2020xym} by including all the possible 2$^\mathrm{nd}$ order terms as well as the effects of QCD conserved currents 
\cite{Greif:2017byw,Almaalol:2022pjc} and their initial state fluctuations \cite{Martinez:2019jbu,Carzon:2019qja}, is urgently needed, especially as one moves towards low beam energies. 
The question of causality violation in current simulations~\cite{Chiu:2021muk,Plumberg:2021bme} needs to be addressed to 
avoid instabilities and to better constrain the properties of the pre-hydrodynamic phase. Much progress on the latter has been achieved in recent years using QCD effective kinetic theory \cite{Kurkela:2018wud,Kurkela:2018vqr, Giacalone:2019ldn}. 
A better description of the hydrodynamization process in this context requires the inclusion of fermions \cite{Kurkela:2018xxd,Kurkela:2018oqw} (allowing for the investigation of chemical equilibration), and the inclusion of non-conformal effects when matching to hydrodynamics \cite{NunesdaSilva:2020bfs,Nijs:2020roc,daSilva:2022xwu}.

The possibility of using different definitions of hydrodynamic variables (different hydrodynamic frames) opens up a number of questions in the formulation of hydrodynamics \cite{Noronha:2021syv}. 
Work is needed to systematically formulate 1$^{\rm st}$ and 2$^{\rm nd}$ order stochastic hydrodynamics in general hydrodynamic frames, going beyond existing results~\cite{Akamatsu:2017zzl,Akamatsu:2017rdu,Martinez:2019bsn,An:2019osr}, considering also the effects from fluctuations due to a critical point \cite{Stephanov:2017ghc,An:2019csj} or a first order phase transition. 

The question of how quantum mechanical effects related to spin degrees of freedom or Quantum Field Theory (QFT) anomalies become manifest in relativistic fluids has generated a lot of activity in the field throughout the last decade \cite{Kharzeev:2015znc,Becattini:2022zvf}. 
Chiral (or anomalous) relativistic hydrodynamics includes quantum effects driven by anomalies that manifest in the hydrodynamic regime \cite{Son:2009tf} and influence the dynamics of various systems from the QGP to Weyl semimetals \cite{Kharzeev:2015znc,Huang:2015oca,Hosur:2013kxa}. However, very little is known about the properties of the chiral hydrodynamic equations of motion and their solutions, especially in the nonlinear regime probed in hydrodynamic simulations of the QGP. Initial steps were taken in \cite{Speranza:2021bxf} for ideal hydrodynamics. However, nothing is known about such properties when viscous effects are included in the nonlinear regime. 
Following the measurement of global $\Lambda$ polarization by the STAR collaboration \cite{STAR:2017ckg} (see Sec.\,\ref{sec:vorticity}), the development of consistent theories of spin hydrodynamics is underway \cite{Hattori:2019lfp,Montenegro:2020paq,Gallegos:2020otk,Hongo:2021ona,Gallegos:2022jow,Weickgenannt:2020aaf,Weickgenannt:2022zxs, Weickgenannt:2022qvh, Wagner:2022amr}. There is currently no formulation of viscous spin hydrodynamics that is causal, stable, and well-posed, leaving this as an important task for the next decade.

Hadronic transport codes are necessary in their role as afterburners at high energies, and are currently the only means of describing the largely out-of-equilibrium evolution of heavy-ion collisions at low energies, such as those explored in the BES FXT program.
By comparing simulations with experimental data, hadronic transport can be used to extract the EOS and in-medium properties of nuclear matter at finite $T$ and large $n_B$ \cite{Hartnack:1994ce,Li:1998ze,Danielewicz:2002pu,LeFevre:2015paj,Wang:2018hsw,Nara:2021fuu}, as well as constrain the isospin-dependence of the EOS \cite{Li:2000bj,Fuchs:2000kp,Li:2002qx,Xiao:2008vm,Giordano:2010pv,Li:2014oda,Russotto:2011hq,Cozma:2011nr,Xu:2019hqg,Colonna:2020euy,Yong:2022pyb}, important for understanding the structure of neutron stars. 
Precision extractions require further improvements \cite{Sorensen:2023zkk}, including using maximally flexible parametrizations of the density-, momentum-, and isospin-dependence of nucleon interactions \cite{Sorensen:2020ygf, Oliinychenko:2022uvy}, incorporation of the in-medium properties of nuclear matter as constrained by chiral effective field theory, description of light cluster production, and threshold effects \cite{Sorensen:2023zkk}.
Progress can be made through systematic comparisons between various hadronic transport codes, and the Transport Model Evaluation Project (TMEP) Collaboration has provided several benchmark results and recommendations for improvements \cite{TMEP:2022xjg}.

\subsubsection{Quantum Information Science and QCD}

A rapidly growing area of research within the U.S. Nuclear Physics (NP) research portfolio is the application of Quantum Information Science (QIS) in NP. In fact, while the topic was not discussed in LRP15, its rapid emergence in various disciplines within NP over the past five years promoted NSAC to form a sub-committee in 2019 to report on the opportunities and prospects of QIS in NP. The resulting report~\cite{NSAC-QIS-2019-QuantumInformationScience} identified simulation and sensing as the two major research directions: first since many grand-challenge problems in NP require advanced, and potentially quantum-based, simulation and sensing techniques and technologies, and second since the expertise of nuclear physicists in these sub-areas could lead to transferring some of the current and future developments in NP to the QIS community. In cold and hot QCD research, in particular, simulation has proven the prime first-principles approach. In fact, lattice QCD methods, combined with state-of-the-art high-performance computing, are expected to continue to push the frontiers of accurate studies of nucleon structure, properties of light nuclei, and low-density QCD matter at finite temperatures~\cite{Detmold:2019ghl,Cirigliano:2019jig,Kronfeld:2019nfb,Bazavov:2019lgz,Joo:2019byq}. However, it is conceivable that the range of studies facing a sign problem (or equivalently a signal-to-noise problem) remain infeasible with current techniques. Such studies include finite-density systems aimed at full exploration of the phase diagram of QCD, and of real-time dynamics of QCD processes such as those prevalent in heavy-ion collisions and in early universe, which are essential to understand equilibration, thermalization, hydrodynamization, fragmentation, and hadronization in QCD. 
Additionally, non-equal-time QCD correlation functions generally are not directly accessible, making it challenging to compute hadron and nuclear structure functions, dynamical response functions, transport coefficients, and more.

\begin{figure}
\centering
\includegraphics[scale=0.66]{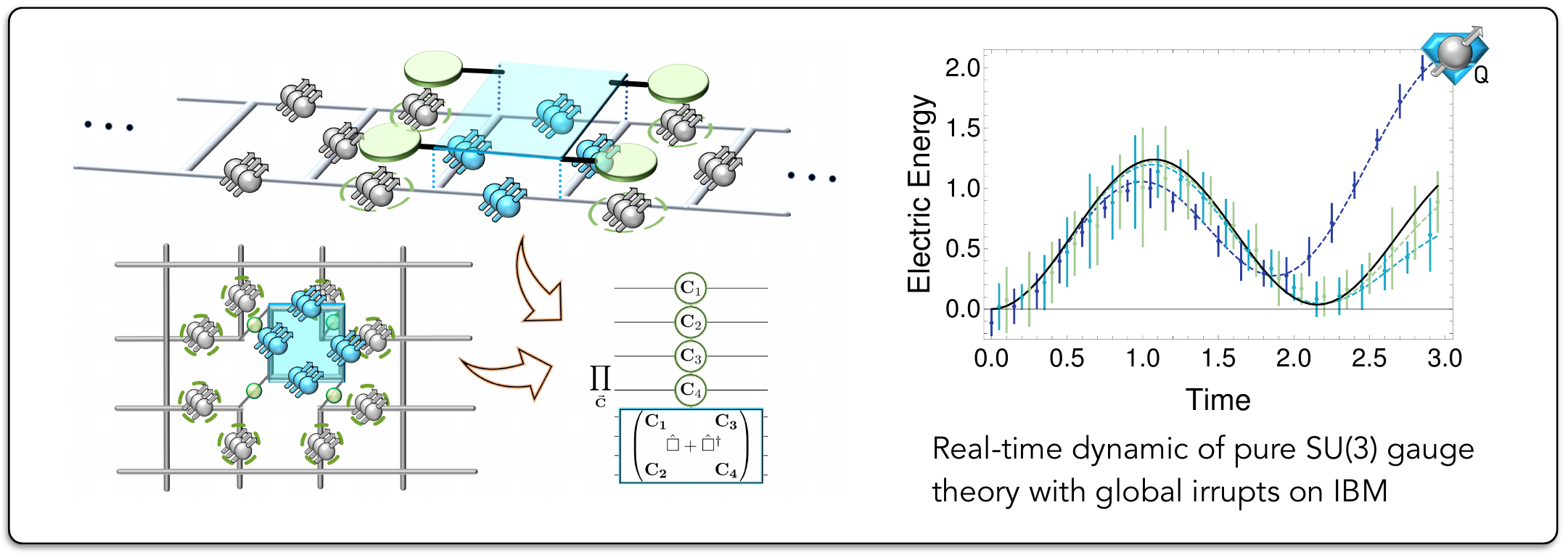}
\caption{The left side depicts the structure of the plaquette operator in the pure SU(3) lattice gauge theory upon integration of local quantum numbers for a one-dimensional string of plaquettes (top) and for a two-dimensional sheet of plaquettes (bottom). The blue squares indicate the active quantum registers while the green circles denote the neighboring controls. The right side shows a real-time simulation of the dynamics of the vacuum in terms of fluctuations in the electric energy of a two-plaquette system in the (truncated) global color parity basis, implemented on the IBM Athens quantum processor. The figure is adopted from Refs.~\cite{Ciavarella:2021nmj,Ciavarella:2022zhe}.}
\end{figure}

Quantum simulation is fundamentally different from classical simulation in that a vast Hilbert space of a quantum many-body system can be encoded exponentially more compactly into quantum units. These quantum units could be two-dimensional spins, or qubit, or higher-dimensional spins called qudits, or even bosons and fermions as in nature. Furthermore, the computations can be parallelized exponentially more efficiently using the principles of superposition and entanglement in quantum mechanics. The time evolution can be efficiently implemented but states need to be initiated and finally measured to access observables, for which many strategies are being developed~\cite{buluta2009quantum,brown2010using,Georgescu:2013oza,altman2021quantum,alexeev2021quantum,Elben:2022jvo}. In the QCD community, the progress has been significant, and while the earlier ground-breaking simulations focused on demonstrating access to non-perturbative real-time phenomena, such as pair production and vacuum fluctuations, in simple models such as the 1+1 dimensional QED~\cite{martinez2016real,klco2018quantum,nguyen2022digital,Mueller:2022xbg,de2021quantum,mil2020scalable,zhou2021thermalization}, a plethora of works in recent years have provided algorithms and strategies for simulating both Abelian and non-Abelian lattice gauge theories in higher dimensions, including three-dimensional SU(3) lattice gauge theory, see e.g., Refs.~\cite{Byrnes:2005qx,Ciavarella:2021nmj,kan2021lattice}. This research has shaped into interconnected theoretical, algorithmic, and hardware implementation and co-design directions: it aims to find the most efficient Hamiltonian formulations of gauge theories of interest in the Standard Model on and off the lattice~\cite{Bauer:2022hpo}, match them to the near- and far-term algorithms that scale increasingly more efficiently with system size~\cite{shaw2020quantum,Byrnes:2005qx,Ciavarella:2021nmj,Lamm:2019bik,kan2021lattice}, and perform small instances of those on current quantum-simulation hardware, in digital or analog modes, to show the potential. See Refs.~\cite{Klco:2019evd,Mueller:2019qqj,Barata:2020jtq,Atas:2021ext,rahman20212,Atas:2022dqm,Illa:2022jqb,Farrell:2022vyh,Bedaque:2022ftd,DeJong:2020riy,Czajka:2021yll,Davoudi:2022uzo,Cohen:2021imf,Lamm:2019uyc,Gong:2021bcp, Yao:2022eqm} for select examples on recent progress on a range of QCD-inspired problems. This had led to the formation of successful co-design efforts among QCD physicists and hardware developers, which may be a critical component of a quantum-simulation program for QCD over the next decade~\cite{Bauer:2022hpo,humble2022snowmass}. The current resource estimates for solving QCD in regimes of interest to phenomenology are far beyond the capabilities of the current quantum hardware. Nonetheless, with the rapid progress in quantum technologies, and the ongoing race toward fault-tolerant quantum computing in academia, industry, and government sectors, it is important for QCD researchers to be ready to take advantage of the new technology, as it is being developed and into the future.


\subsubsection{Topical Collaborations}\label{sec:theory_topical}
As emphasized in the last LRP, many aspects of theoretical nuclear physics can benefit from additional long-term, sustained efforts beyond the base program that bring together the resources of several institutions in a coordinated way to address a well-defined problem or topical area with a clear set of deliverables. When DOE established the first topical collaboration in 2010, the ``Jet" 
Collaboration was selected in the QCD area. In the second round, two topical collaborations from QCD area were selected: ``TMD" and ``BEST".  
In a recent round announced in December 2022, the QCD community received 4 out of 5 awards. Meanwhile, NSF has funded two collaborations: JETSCAPE and MUSES. All these collaborations have been very successful. 

\subsubsubsection{Previous: TMD Collaboration} This collaboration consisted of 3 national laboratories and 11 universities. It addresses the challenge of extracting novel quantitative information about the internal landscape of the nucleon, in particular the three-dimensional (3D) confined partonic motion inside the nucleon, which are encoded in the TMDs. The goal is to develop new theoretical and phenomenological tools that are urgently needed for precision extraction of the 3D tomography of the confined motion of partons inside the nucleon from current and future data. 

\subsubsubsubsection{Bridge position highlight} Prof. Martha Constantinou was hired by Temple University as a bridge position with the TMD Collaboration in 2016. Since then, she has received the DOE early career award and the Sloan foundation research award. The bridge position has enabled her to come to the US from Cyprus. 
She now leads the Quark-Gluon Tomography (QGT) collaboration, funded by DOE in 2023. Hers is a true success story from the DOE topical collaboration program.

\subsubsubsection{Previous: BEST Collaboration} The BEST Collaboration, involving collaborators from two national laboratories and 11 universities, developed a theoretical framework for interpreting the results from the BES program at RHIC. The main goals of this program were to discover, or put constraints on the existence, of a critical point in the QCD phase diagram, and to locate the onset of chiral symmetry restoration by observing correlations related to anomalous hydrodynamic effects in the quark gluon plasma. 

\subsubsubsubsection{Bridge position highlight}
Prof. Chun Shen was hired by Wayne State University in 2018. He received IUPAP Young Scientist Prize in Nuclear Physics in 2019 and a DOE Early Career Award in 2021.
Prof. Vladimir Skokov was hired by North Carolina State University in 2018.

\subsubsubsection{JETSCAPE Collaboration}  
Interpretation of jet measurements requires sophisticated numerical modeling and simulation, and advanced statistical tools for comparison of theory calculations with experimental data.
The JETSCAPE/XScape Collaboration was formed to develop a comprehensive software framework that will provide a systematic, rigorous approach to meet this challenge. It will develop a scalable and portable open source software package to replace a variety of existing codes. 
The collaboration consists of a multi-disciplinary team of physicists, computer scientists and statisticians from 13 institutions, and will create open-source statistical and computational software to help scientists better understand high energy nuclear collisions. 

\subsubsubsection{MUSES Collaboration} This collaboration addresses questions that bridge nuclear physics, heavy-ion physics, and gravitational phenomena such as: What type of matter exists within the core of a neutron star? What temperatures and densities are reached when two neutron stars collide? What can nuclear experiments with heavy-ion collisions teach us about the strongest force in nature and how can we relate heavy-ion collisions to neutron stars? 
The collaboration spans 16+ institutions, hosts annual workshops and biweekly seminars, and supports a number of undergraduates, graduate students, and postdocs. 

\subsubsubsection{QGT Collaboration} The QGT Collaboration brings together a team with broad expertise and leadership across hadron physics theory to drive understanding and discovery in the quark and gluon tomography of hadrons and the origin of their mass and spin. 
This proposal will provide partial support for 11 postdocs, 6 graduate students, and bridge positions in theoretical hadron physics at three institutions: Stony Brook University, Temple University, and University of Washington.

\subsubsubsection{SURGE Collaboration} The Saturated Glue (SURGE) Topical Theory Collaboration aims at the discovery and exploration of the gluon saturation regime in QCD by advancing high precision calculations and developing a comprehensive framework to compare to a wide range of experimental data from hadron/ion colliders and make predictions for the EIC. 
It will provide partial funding for 5 postdocs, 7 graduate students, and 1 undergraduate student at 13 institutions, and will establish a bridge position at the University of Illinois at Urbana Champaign.

\subsubsubsection{HEFTY Collaboration} 
This collaboration combines the capabilities of leading US researchers to develop a rigorous, comprehensive theoretical framework of heavy flavor  particles in QCD matter, from their initial production, their subsequent diffusion through the QGP and hadronization that can be embedded into realistic numerical simulations and compared to data. 
It will provide partial funding for 3 postdocs and 6 graduate students at 7 institutions and establish a bridge position at Kent State University.

\subsubsubsection{ExoHad Collaboration} The Coordinated Theoretical Approach for Exotic Hadron Spectroscopy (ExoHad) Collaboration aims to develop a pathway to study some of the more elusive states formed of quarks and gluons using the the foundational principles of scattering theory and quantum chromodynamics. The funds will support 3 graduate students, 3 postdocs, and two bridge faculty positions at William \& Mary and Indiana University.

\subsection{Future opportunities in Hot QCD}\label{sec:future_hot}

Hot QCD research is addressing questions of fundamental importance that can be summarized in the following main goals:
\begin{itemize}
\item \emph{Determine the phase structure of nuclear matter}
The phase diagram needs to be pinned down as a function of temperature and net-conserved charges, including the determination of a possible QCD critical point, which requires the experimental measurement and theoretical study of collisions with varying collision energy. We need to understand the deconfinement transition and chiral symmetry restoration, and determine the nuclear equation of state for which heavy ion collisions and neutron stars can provide complementary input. 
\item \emph{Understand the mechanisms that lead to the emergence of the fluid behavior of hot and dense nuclear matter}
This requires studying the QGP at short distance scales using hard probes, including jets, heavy flavor hadrons, and quarkonia. Further insight can be gained by pushing the boundaries to e.g.~small collision systems and better constraining the initial state from theory and complementary experiments. Electromagnetic probes carry further information on the time evolution of the system.
\item \emph{Quantify the dynamic properties of the quark gluon plasma}
Transport properties of the QGP, including its shear and bulk viscosity, as well as its interaction with heavy and high momentum probes, need to be determined as functions of temperature and densities of conserved charges, and understood within QCD or effective theories thereof. Further, probing the vortical structure of the fluid flow fields can access spin related transport properties.
\item \emph{Utilize the broad physics reach of heavy ion collisions} Heavy ion collisions provide an incredible amount of information, which, when carefully isolated, allows for physics studies far beyond the QGP and even QCD. Ultra-peripheral collisions can be employed to study photo-nuclear events probing very low $x$, as well as quantum electrodynamic phenomena with some processes sensitive to beyond the standard model physics. Heavy ion collisions also offer a unique opportunity to study quantum anomalies via the chiral magnetic effect. Furthermore, certain observables are highly sensitive to the detailed nuclear structure of the colliding nuclei, and far forward data from heavy ion collisions can provide important information for cosmic ray physics.
\end{itemize}

\subsubsection{Properties of the Quark Gluon Plasma}\label{sec:future_flow}
In the coming years, phenomenological studies of hot many-body QCD systems will focus on obtaining robust constraints on thermodynamic and transport properties of the QGP, exploring the QCD phase structure at large baryon densities, and emerging collectivity in small systems.

\subsubsubsection{Transport properties from Bayesian inference and multi-observable studies} One major goal of studies of the QGP at RHIC and LHC is the determination of transport properties, such as the shear and bulk viscosity to entropy density ratios, $\eta/s$ and $\zeta/s$, as well as relaxation times, electric and heat conductivities, the partonic momentum diffusion coefficient $\hat{q}$, and transport coefficients for single heavy quarks and heavy quark-antiquark pairs. 
The hot QCD community has moved toward determining the temperature dependence of these quantities, as well as their behavior at varying chemical potentials. Observables in heavy ion collisions exhibit varying and complex responses to these QGP properties.
Consequently, systematic and robust phenomenological constraints are best derived from combining multiple measurements via the Bayesian Inference method. Bayesian Inference analyses for the RHIC BES program and including high-statistics observables pose serious numerical challenges to the field. Although using model emulators can effectively reduce the required computational resources, novel techniques are essential to further reduce the required volume of training datasets while keeping good accuracy of the model emulators. To achieve more effective model training, techniques like transfer learning and multi-fidelity training \cite{Heffernan:2022swr,Liyanage:2022byj,Ji:2022xzo} should be employed in full (3+1)D hybrid frameworks. In order to efficiently compute high-statistics observables, additional speed boosts from employing other machine learning tools such as deep neural networks, are needed. The Bayesian model averaging method is crucial for combining different models with their relative statistical weights to systematically fold in theoretical uncertainties. Sophisticated Bayesian Model Mixing techniques are presently being developed by the BAND Collaboration \cite{Phillips:2020dmw}.

A concerted effort will be needed to develop documented and accessible Bayesian inference software frameworks~\cite{Phillips:2020dmw}. Of equal importance will be accessibility to supercomputing infrastructure to perform the large-scale calculations required for Bayesian studies. 
Incorporating more hadronic observables into the Bayesian
analyses should provide stronger constraints than are currently available.
Some work in this direction has been done by including 
normalized symmetric cumulants into a Bayesian analysis and it was
shown that these quantitites are more sensitive to the temperature
dependence of the transport coefficients than \vn~\cite{Parkkila:2021yha}.
Additionally, the transport properties are particularly poorly constrained in the higher temperature range 
accessible only at the LHC (see Fig.\,\ref{fig:jetscape_viscosities}). This could be improved upon for example by measuring the 
\vn\ of dileptons~\cite{Vujanovic:2017psb}. Generally, better constraints on the viscosities could be obtained by using as many observables as possible - combining the low-momentum hadronic observables used in current analyses with electromagnetic probes, as well as jets and heavy flavor probes. 

To maximize information gain on initial state and QGP properties, one should explore new multi-particle correlation observables and their precise experimental measurement. This includes correlations of flow harmonics with mean transverse momentum fluctuations, and higher order versions of those,  normalized symmetric cumulants, mixed harmonic cumulants from 4, 6, 8 or more particle correlations, higher order transverse momentum fluctuations, and non-linear flow mode coefficients. Studying these observables, and analyses of varying collision systems will aid in separating initial state properties from QGP properties and with the extraction of information using Bayesian analyses \cite{Parkkila:2021tqq,Nijs:2021clz}.
Including Hanbury-Brown-Twiss (HBT) observables \cite{PHENIX:2004yan,STAR:2004qya,ALICE:2015hvw,CMS:2017mdg,ATLAS:2017shk} could provide additional constraints as they are more directly sensitive to the spatial size of the emission source.  Measurements of identified particles, in particular the study of strangeness, can elucidate the mechanism of particle production and further constrain the properties of matter created and provide
information on the effect of the hadronic phase in collisions of various size and at varying collision energies.

\begin{figure}
\centering
\includegraphics[width=0.53\textwidth]{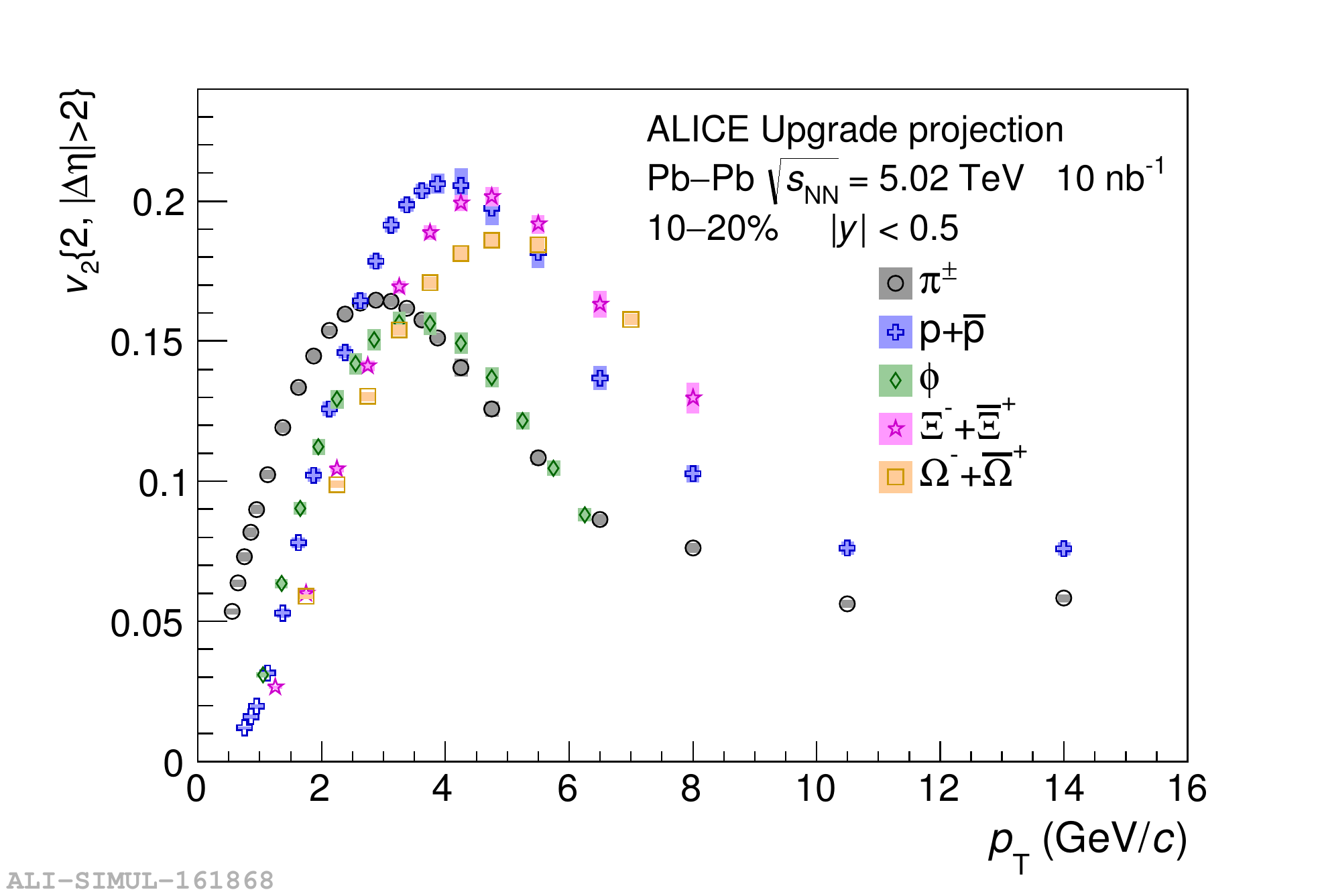}
\includegraphics[width=0.45\textwidth]{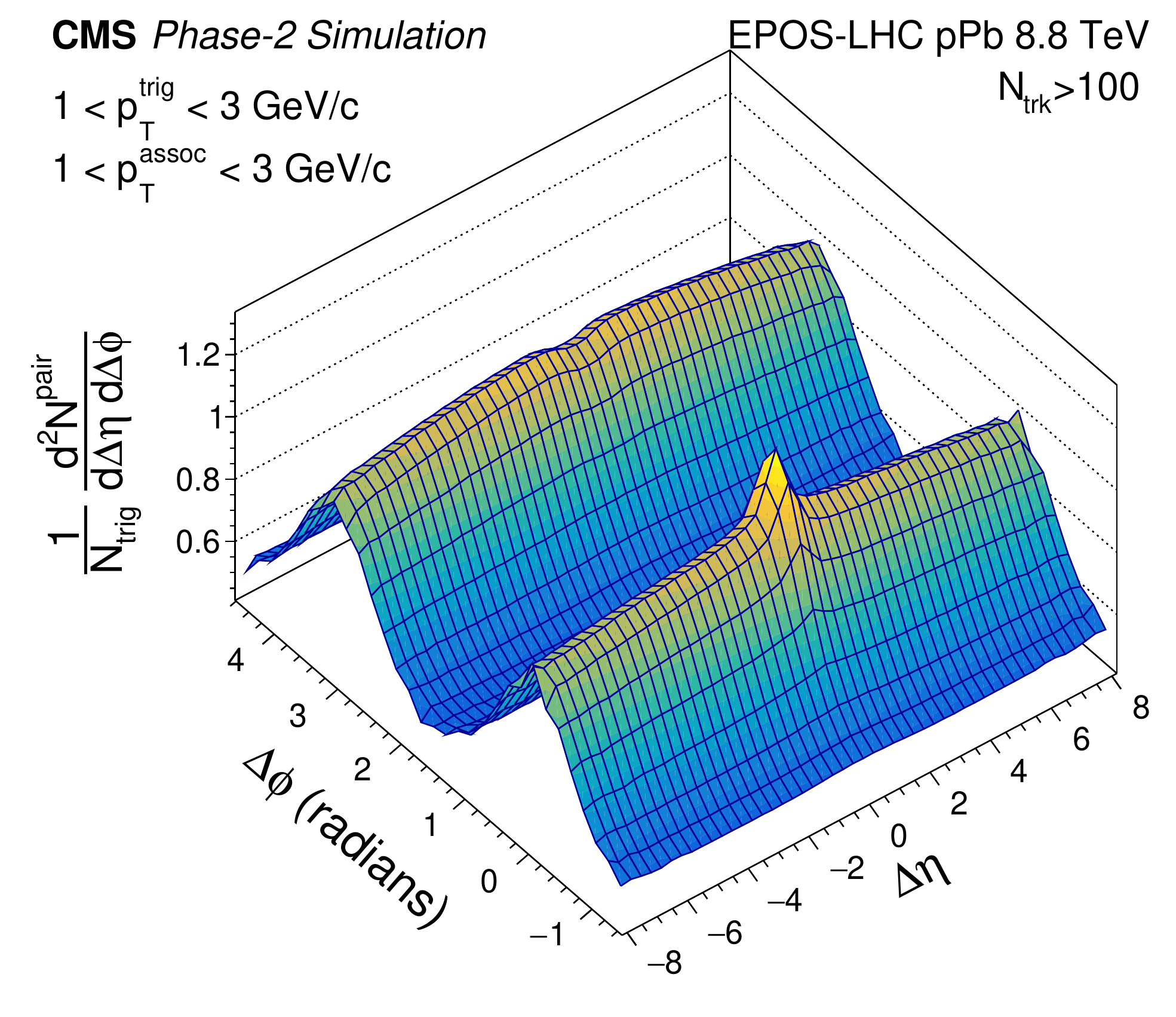}
\caption{Left: ALICE projections for PID $v_2$ in 10--20\% centrality PbPb collisions for an integrated luminosity of 10~nb$^{-1}$ at HL-LHC~\cite{ALICE-PUBLIC-2019-001,Citron:2018lsq}. Right: CMS projection of long-range
two-particle correlations with $\Delta\eta$ up to 8 units from Phase-2 upgrades for the HL-LHC~\cite{CMS-DP-2021-037}.}
\label{fig:longrange_future}
\end{figure}

\subsubsubsection{Detector upgrades} New opportunities to constrain QGP properties in future experimental programs are enabled by not only significantly higher integrated luminosities, but also by detector upgrades. Figure~\ref{fig:longrange_future} (left) shows the projected performance of PID $v_2$ measurements by ALICE for 10--20\% centrality Pb+Pb collisions at the high-luminosity LHC (HL-LHC) with an integrated luminosity of 13~nb$^{-1}$~\cite{ALICE-PUBLIC-2019-001,ALICE:2023udb}. A large variety of baryon and meson $v_2$ (and also higher order $v_n$) including hadrons containing multiple strange quarks will be measured over a wide kinematic range with unprecedented precision, which will impose strong constraints on QGP properties starting from the initial condition, hydrodynamic evolution, to the final hadronization stage. With the Phase-2 upgrades of the CMS and ATLAS experiments, long-range particle correlations and collective behavior of 
the QGP will be explored over 8 units of pseudorapidity $\eta$, as shown  Fig.~\ref{fig:longrange_future} (right) for CMS~\cite{CMS-DP-2021-037}. Moreover, the wide acceptance time-of-flight detector upgrade planned at CMS will bring unique opportunities to study the QGP medium with
identified hadron production and correlations over unprecedented phase space coverage~\cite{CMS-DP-2021-037,CMS-PAS-FTR-22-001,Krintiras:2022ohr}.
LHCb upgrades will increase the centrality range accessible at far forward rapidity and allow new measurements of identified particle and heavy quark collectivity in a unique region of phase space.

\subsubsubsection{Equation of state} It has been proposed that key thermodynamic properties of the QGP can be extracted by the multiplicity dependence of mean $p_{T}$ in ultra-central heavy ion collisions to directly constrain the speed of sound in QGP, and thus the QCD equation of state at high temperatures~\cite{Gardim:2019brr,Nijs:2021clz}. 
The beam-energy dependence of the slope of the directed flow and the elliptic flow have likewise been shown to be highly sensitive to the EOS \cite{Stoecker:1980vf,Ollitrault:1992bk,Rischke:1995pe,Stoecker:2004qu,Brachmann:1999xt,Csernai:1999nf,Ivanov:2014ioa,Hartnack:1994ce,Li:1998ze,Danielewicz:2002pu,LeFevre:2015paj,Wang:2018hsw,Nara:2021fuu}. Extraction of speed of sound via baryon number cumulants has also been proposed in collisions at low energies which probe high baryon densities \cite{Sorensen:2021zme}. These measurements can be explored at both RHIC and the LHC over a wide energy range to obtain key information on the QCD phase diagram. Connections between the grand-canonical susceptibilities of (multiple) QCD conserved charges with the corresponding measurements of (cross-)cumulants in heavy-ion collisions derived in Refs. \cite{Vovchenko:2020tsr,Vovchenko:2020gne} can be utilized to obtain further information about the EOS.

Heavy-ion collisions with $\sqrt{s_{\rm NN}}\sim 10$\,GeV from the current RHIC BES program and future experiments including CBM at the Facility for Antiproton and Ion Research (FAIR) offer a unique opportunity to quantify the QCD phase structure at large baryon densities \cite{Almaalol:2022xwv}. Our phenomenological tools need substantial developments to explore this region of the phase diagram, which is currently inaccessible to first-principles lattice calculations.
As the two incoming nuclei pass through each other, it is crucial to model their interactions dynamically to obtain non-trivial event-by-event distributions of energy density, baryon and electric charge densities for the following macroscopic hydrodynamic evolution. 
Colliding heavy ions with different electric charge to baryon ratios, like isobar pairs, is important to explore the full 4D nature of the QCD phase diagram in $(T, \mu_B, \mu_S, \mu_q)$ \cite{Monnai:2021kgu}. Developments of the parametrized 4D equation of state and the propagation of multiple conserved charge currents and their diffusion in relativistic hydrodynamic frameworks are essential to model the macroscopic evolution of these collision systems and determine how this flavor information leaves its imprint on identified particle production. Constraining the  4D QCD phase diagram in $(T, \mu_B, \mu_S, \mu_q)$ will also make connections with nuclear astrophysics, which focuses on studying the nuclear matter properties in a dense and neutron-rich environment (see sec.\,\ref{sec:neutronstars}) \cite{Lovato:2022vgq}.
In heavy-ion collisions, the out-of-equilibrium propagation of multi-point correlations is crucial to trace the signatures of the QCD critical point and first-order phase transitions. Implementation of dynamical descriptions of the relevant multi-particle correlations in frameworks like hydrodynamics will be an important step in the search for the QCD critical point \cite{An:2021wof}. Another crucial requirement is that descriptions of particlization that retain information on fluctuations and correlations \cite{Pradeep:2022mkf}, are advanced to a level where they can be employed in large scale phenomenological simulations.

Studies utilizing hadronic transport simulations have also been remarkably successful in understanding the dynamics of heavy-ion collisions at low energies from $\sqrt{s_{\rm{NN}}} \approx 1.9$ to $\sqrt{s_{\rm{NN}}} \approx 8.0\ \rm{GeV}$ \cite{Sorensen:2023zkk}. In particular, hadronic transport with mean-field potentials naturally describes the initial state of the collision as well as the interaction between the expanding collision region and the spectators, necessary for understanding the origin of the flow observables including, e.g., the rapidity dependence of the directed flow or the origin of ``squeeze-out'' in the elliptic flow at low energies. 
Currently there are still significant differences between symmetric nuclear matter EOSs extracted from different theoretical fits to heavy-ion collision data \cite{LeFevre:2015paj,Nara:2020ztb,Danielewicz:2002pu,Oliinychenko:2022uvy,Steinheimer:2022gqb}. 
Some of these differences can be assigned to differences in modeling framework. Systematic comparisons between different hadronic transport codes, such as those done within the Transport Model Evaluation Project Collaboration \cite{TMEP:2022xjg}, can provide a common baseline for code development and lead to code and modeling improvements.
Making precise quantitative statements about the properties of dense nuclear matter, including the density-, isospin-, and momentum-dependence of the single-nucleon mean-field potential, will require developing maximally flexible parametrizations of nucleon interactions over large ranges of density and temperature probed in heavy-ion collisions. 
With advances in modeling, the extracted information will achieve unprecedented precision given the forthcoming data, see  Sec.\,\ref{sec:futurephasedigram}. 

\subsubsubsection{Further directions for constraining QGP properties} Electromagnetic probes provide complementary information about the medium properties relative to the hadronic observables, as they provide increased sensitivity to the early stages of the collision. More discussion on the prospects for electromagnetic probes can be found in Sec.\,\ref{sec:em_future}.
Polarized $\Lambda$ hyperons can probe the vortical structure of the fluid flow fields in heavy-ion collisions. The extension and phenomenological applications of recently developed spin-hydrodynamic theories are important to probe the spin related transport properties of the QGP, see Sec.\,\ref{sec:vorticity_future}.

\subsubsection{Hot QCD Studies with Electromagnetic Probes}\label{sec:em_future}

The production of soft photons and dileptons in the little understood early stages of heavy-ion collisions, namely their ``pre-equilibrium  emission'' \cite{Bhattacharya:2015ada,Monnai:2015bca,Linnyk:2015rco,Greif:2016jeb,Vovchenko:2016ijt,Srivastava:2016hwr,Oliva:2017pri,Hauksson:2017udm,Berges:2017eom,Monnai:2019vup,Churchill:2020uvk,Garcia-Montero:2019vju,Gale:2021emg,Khachatryan:2018ori,Coquet:2021gms}, represents one of the most important areas of study for electromagnetic probes. Photons and dileptons can provide critical information on the 
dynamical properties of the early stages, including chemical equilibration. Photons and dileptons will also play a vital role in studying the formation of quark-gluon plasma in collisions of smaller systems, such as proton-gold collisions. Calculations predict a measurable thermal photon signal in  collisions of small systems~\cite{Shen:2016zpp,Shen:2015qba,Gale:2021emg}, and pre-equilibrium photons would likely add to this signal.

Dilepton measurements have already proven valuable in studying lower energy collisions of nuclei~\cite{Endres:2015fna, Galatyuk:2015pkq, Staudenmaier:2017vtq}, providing estimates of the medium temperature~\cite{HADES:2019auv,NA60:2008dcb}. The analysis of high statistics measurements by the STAR BES II program~\cite{Akiba:2015jwa} will provide important new low- and intermediate-mass dilepton measurements that can be used to study the phase diagram of QCD at lower temperatures and higher baryon densities~\cite{Seck:2020qbx}, as well as chiral symmetry restoration. The future experiments NA60+~\cite{Ahdida:2022avx} and CBM~\cite{CBM:2016kpk,Almaalol:2022xwv} will provide high-precision measurements with new detector capabilities.
ALICE will measure thermal dileptons in Runs 3 and 4, which can give access to the system's average temperature. The future experiment ALICE3~\cite{ALICE:2022wwr} will measure low- and intermediate-mass dileptons at higher energies with unprecedented precision to address chiral symmetry restoration through $\rho-a_1$ mixing and to improve the measurement of the plasma temperature (and its time evolution). The novel detector capabilities will also enable differential $v_n$ measurements in various mass ranges, which will provide stringent constraints on medium properties such as shear and bulk viscosity and pre-equilibrium dynamics. The ALICE 3 experiment also aims to study ultrasoft photon emission and the corresponding predictions from Low's theorem~\cite{Low:1958sn}.

Ultimately, the simultaneous systematic study of soft photons and dileptons, along with soft hadrons and other observables, will provide unparalleled constraints on the properties of deconfined nuclear matter.

\subsubsection{QGP Tomography with Hard Probes}
The 2015 LRP cited the importance of measurements of
jets at both RHIC and the LHC in order to
understand the temperature dependence of QGP properties. In the following, new opportunities with jet and heavy flavor probes are discussed. 

\subsubsubsection{Jets}
\label{sec:jetfuture}
Some major open questions in jet physics are listed below.  
These questions are not independent of each other.
Due to the connection between the jet observables and the 
QGP itself, theoretical models which incorporate information
about the soft physics of the QGP, the jet-QGP interactions and
the hadronization process are necessary to compare 
experimental measurements to theory.  
Much recent work in this direction
has been done but more is needed to increase the variety
of measurements and theory compared. 
Further advances in jet substructure measurements will provide further constraints.
Additionally, the upcoming high-luminosity
data from sPHENIX and STAR is necessary to constrain
how the jet-QGP interactions depend on the temperature 
of the QGP.  

\begin{itemize}
\item{\textbf{How does the QGP resolve the color
configuration of the parton shower?}
The parton shower develops from the original hard-scattered
parton to the final observed hadrons in the jet.  The 
structure of the shower varies jet-by-jet with the average
properties dependent on the energy, color-charge and mass
of the parton.  This developing shower interacts with the QGP.  
The question of how the QGP resolves the parton shower is 
key to understanding  how jets are quenched.  Measurements which vary the average jet properties (e.g. photon-tagged jets to enhance the fraction of quark-jets relative to an inclusive jet sample) and measurements which select jet-by-jet on the jet substructure are key to answering this question.
\begin{itemize}
\item{How does the QGP resolve the structure of the 
parton shower?}
\end{itemize}
As discussed previously~\ref{sec:jetprogress}, 
jet quenching has been shown to depend on the structure
of the jet itself.  It is key to understand this quantitatively in terms of whether there is a coherence 
length in the QGP, below which two separate color charges can not be resolved within the QGP~\cite{Casalderrey-Solana:2012evi}.
In order to answer this question measurements of jet quenching as a function of jet substructure and theoretical models that depend on the coherence length are needed at both RHIC and the LHC over a wide kinematic range. 
Applying ML to design new jet observables directly from the data will be helpful in this study as well~\cite{Lai:2021ckt}.
}

\item{\textbf{What is the temperature 
dependence to the QGP opacity?} 
Measurements at sPHENIX, along with improved theoretical models, will provide key constraints on the opacity.}

\item{\textbf{Is there emergent intermediate scale
structure in the QGP?}
Measurements of modifications to the back-to-back jet (hadron) distributions and/or modification of the distance between subjets inside a jet will provide crucial information on this.  }

\item{\textbf{How does jet quenching depend on the 
spacetime evolution of the QGP it travels through? Is there
a minimum time/length of QGP that the jet must interact
with to experience jet quenching?}
Current measurements of the \vtwo\ and \vthree\
of jet/high-\pt\
hadrons show that jet quenching is 
sensitive to the average path length of a class
of jets through the QGP.  However,  a non-zero
value of \vtwo\ is also observed for high-\pt\ hadrons
in \pPb\ collisions \cite{ATLAS:2019vcm}.  Is this \vtwo\ due to some other source?
The most direct way to test this is through measurements
of jet quenching in small symmetric collision systems such
as \oo~\cite{Brewer:2021kiv} where the system
size is near that of central \pA\ collisions, but the 
geometry is more similar to \pbpb\ collisions.  Measurements here 
might suggest that there is a minimum time/length
scale needed for appreciable quenching to occur.
Additionally, since quenching depends on the structure of
the developing parton shower, the path-lenth
dependence of jet quenching could depend on the structure
of the parton shower.  Identifying such a dependence requires
a huge sample of jets to isolate both 
the geometry of the jet trajectory and the structure of the jet.
Jets at RHIC and the LHC also evolve starting 
at very different virtuality scales that influence how they interact with the QGP.
This is a key question for both RHIC and the LHC over the next
several years.}

\item{\textbf{What are the non-equilibrium processes governing
the energy flow from the jet to the QGP over three
orders of magnitude?}
How does the energy lost by the jet become part of the QGP itself?
Are there turbulent processes?  How does the medium
respond to the the energy deposition from the jet?  How does this
process affect the formation of the observed final state hadrons?
Is there an observable vorticity around the jet~\cite{Serenone:2021zef}.
Experimentally isolating the response of the medium
from the quenched jet itself is experimentally challenging.
Recently, it has been proposed that the particle species
mix, the \textit{hadrochemistry}, might be a key
signature of the medium response~\cite{Luo:2021voy}. 
Jet fragmentation
 dominantly produces mesons over baryons.
However,
the enhanced baryon-meson ratio that is characteristic
of coalescence in the soft-sector of the QGP, could be also
seen in medium response.  

}

\end{itemize}

\begin{figure}
\centering
\includegraphics[width=0.95\textwidth]{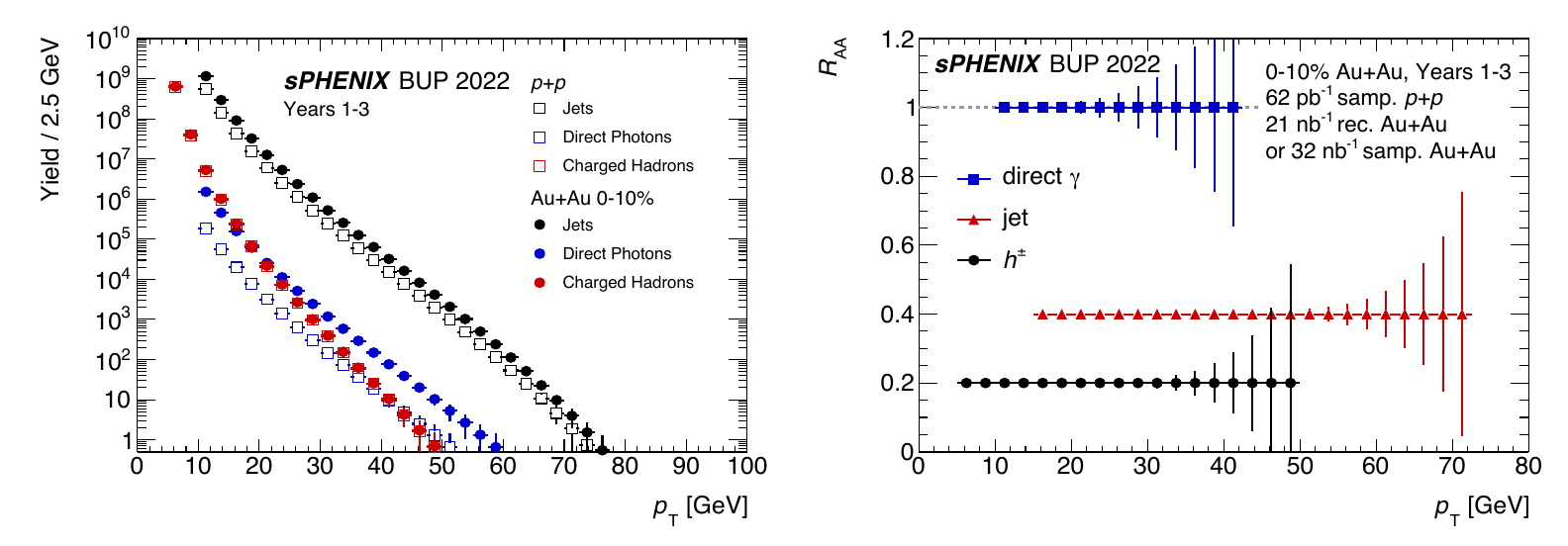}
\caption{Left: Counts of jets, hadrons and direct photons 
projected from sPHENIX operation 
in \pp\ and 0--10\% central
\auau\ collisions. Right: The 
nuclear modification factor \RAA\ as a function of \pt\ for 
0--10\% central \auau\ collisions expected from sPHENIX
operation.  The error bars
show the statistical uncertainties only.  Both plots are from Ref.~\cite{sPHENIXBUP2022}.}
\label{fig:sPHENIXjet}
\end{figure}

Due to the need to measure both the jet structure and geometry
dependence of jet quenching, huge samples of jets
are needed at both RHIC and the LHC to answer the science questions
outlined above. sPHENIX is specifically optimized to measure jets and will provide unbiased samples
of jets over nearly the entire allowed kinematic range at RHIC,
as shown in Fig.~\ref{fig:sPHENIXjet}.  Crucially, this 
will allow measurement of jets at the same \pt\ at both
RHIC and the LHC. STAR and ALICE can also contribute to such comparisons, for example with semi-inclusive gamma+jet and  $h$+jet measurements.
Additionally, along with direct photon data from STAR, sPHENIX will provide a sample of direct photon data sufficient to tag photon-jet pairs for photons with more than $\pt > $~30~GeV.  At the LHC,  the large luminosity
\pbpb\ sample planned for Runs 3 and 4 will provide
much more differential jet measurements than are
currently available~\cite{Citron:2018lsq,CMS-PAS-FTR-22-001}.

There is also interest in running smaller collision systems
at both the LHC and RHIC. The LHC is currently planning on 
a short \oo\ run in 2024. Those data will be important
for understanding the lack of evidence for jet quenching in 
\pA\ collisions.  RHIC has run \oo\ collisions for STAR (though the results are not yet publicly available).  
The sPHENIX Collaboration would like to take data with both \oo\ and \arar\ collisions if the opportunity to run beyond the nominal sPHENIX run plan arises~\cite{sPHENIXBUP2022}.

\begin{figure}
\centering
\includegraphics[width=0.5\textwidth]{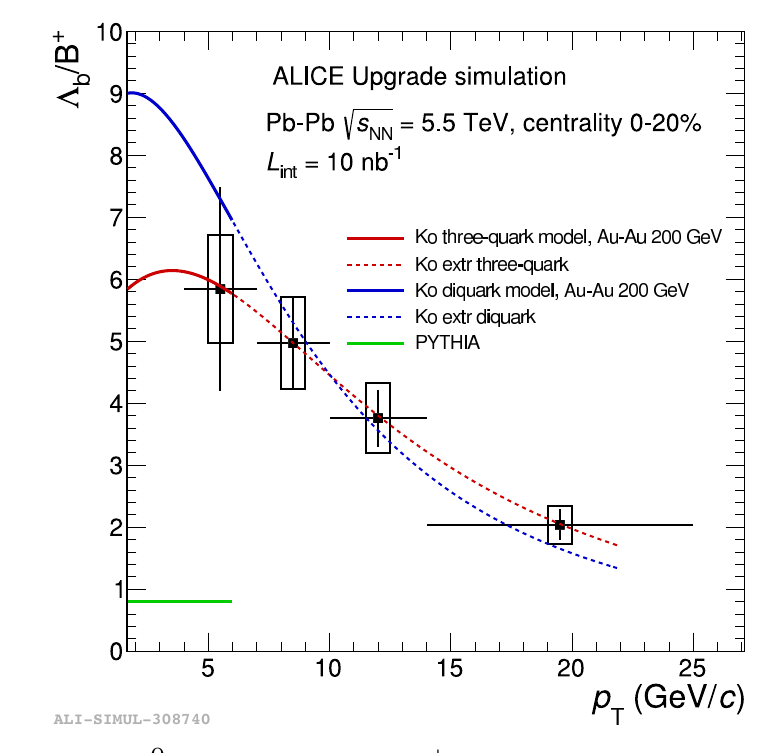}
\caption{Projection for LHC Run 3 and 4 for 
 $\Lambda_b/B^+$ (right) as a function of transverse momentum from ALICE.
 Fig. is from Ref.~\cite{Citron:2018lsq}.}
 \label{fig:hfhadproj}
 \end{figure}

\subsubsubsection{Heavy flavor} 
\label{sec:hf_future} 
 Progress in understanding heavy flavor hadronization requires: an improved space-time picture of coalescence and better constraints on the Wigner functions~\cite{Wang:2019xph}; a more precise determination of the  heavy quark and gluon fragmentation functions~\cite{Chien:2015ctp,Anderle:2017cgl}; and understanding of meson production, absorption and/or dissociation in matter. Here we discuss how the interplay of theoretical developments and new experimental measurements at RHIC and LHC can advance our understanding of heavy flavor. 
Higher statistics $R_{AA}$, $v_2$, and $v_3$ data will significantly reduce the uncertainties on the heavy quark transport coefficients and provide better insights into the initial production, hadronization and evolution of heavy flavor hadrons.
New data expected in LHC Runs 3 and 4 should provide much more
precise constraints than currently available~\cite{Citron:2018lsq,LHCb:2018qbx,CMS:2022cju}.  Figure~\ref{fig:hfhadproj}
shows the projections from ALICE for $\Lambda_b/B^+$, which
has not yet been measured in heavy-ion collisions.  sPHENIX also expects to make very precise measurements of the  $\Lambda_c/D^0$ in \auau\
collisions~\cite{sPHENIXBUP2022}.  In the further future, ALICE3~\cite{ALICE:2022wwr}, the LHCb Upgrade II~\cite{LHCb:2018roe} and the CMS timing detector upgrade~\cite{Butler:2019rpu}
would provide even further improved precision for these observables.

Detector upgrades, together with improved luminosities at both RHIC and LHC, will enable measurements of unprecedented precision of various heavy flavor observables. One targeted measurement is precision $R_{\rm AA}$ and $v_2$ for open bottom hadrons (or their decay daughters) and jets over a broad $p_T$ region. At high $p_T$, combining charm, bottom and light flavor data would allow a systematic investigation of the relative contributions of collisional and radiative energy loss and the transition between them. At low $p_T$, the open bottom $v_2$ (together with charm $v_2$) will address the temperature dependence of the heavy-flavor diffusion coefficient in QCD matter at higher precision. Concurrently, the hadro-chemistry of heavy-flavor hadrons, including charm and bottom baryons, will provide a deeper understanding of the coalescence mechanism and may provide insight into color confinement.

Resummation of in-medium branching processes is necessary to improve predictions of heavy quark tagged jets and their substructure for sPHENIX. Machine learning techniques~\cite{Chien:2018dfn}  can be implemented to analyze high-$p_T$ heavy flavor data from $A+A$ collisions.  The transition from the diffusive elastic to radiative energy loss regimes can be studied theoretically by combining lattice QCD-constrained interactions with effective theories of gluon emission.  Comparison of heavy flavor observables can identify the relevant momentum scales. Finally, the commonalities between heavy flavor production in $A+A$ and $e+A$ collisions~\cite{Li:2020zbk,Li:2021gjw} should be explored.

\begin{figure}[h]
 \begin{minipage} [c] {0.45\linewidth}
 \centering
  \includegraphics[width=0.86\linewidth, angle=-90] {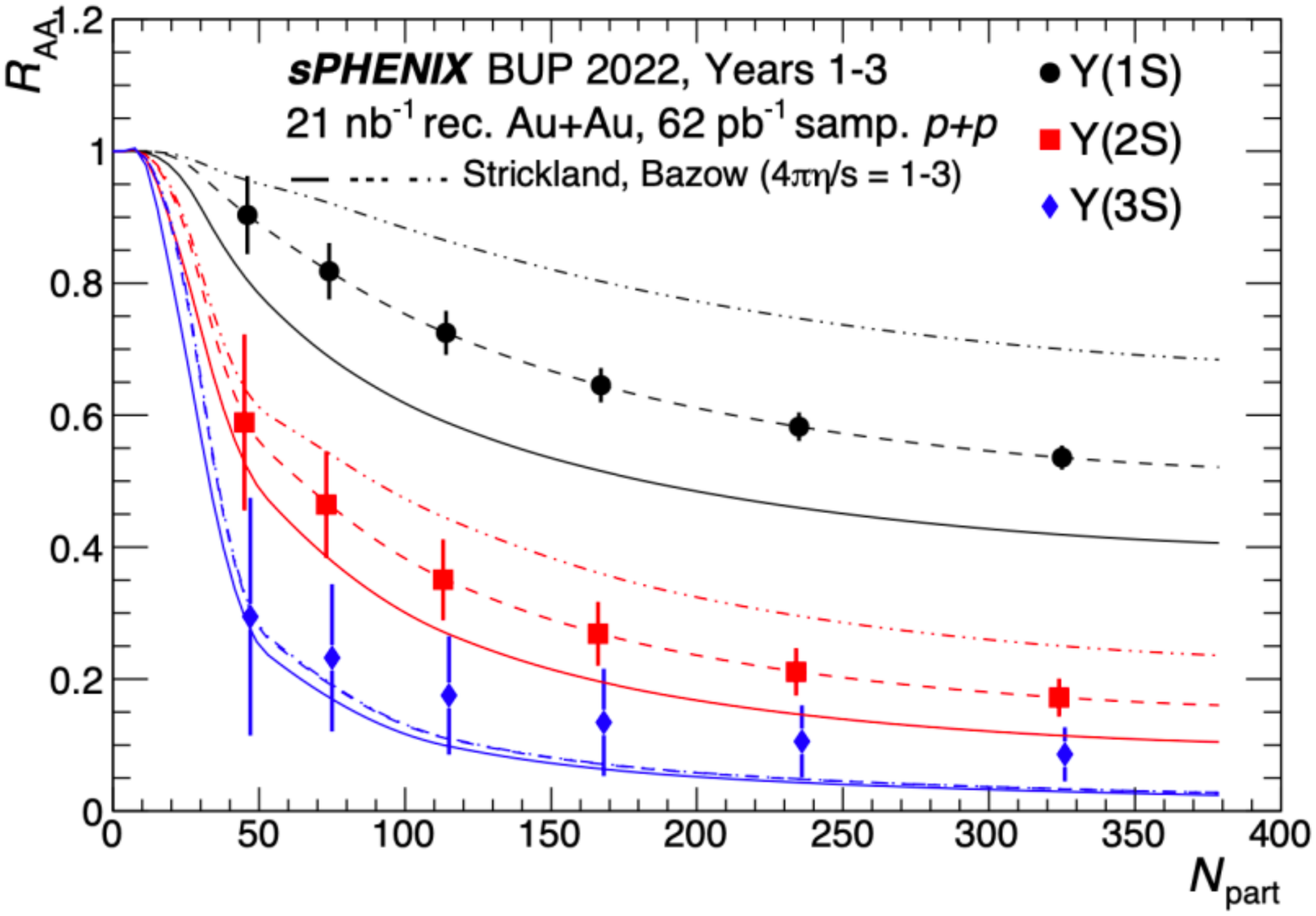}
  \end{minipage} \hfill
\begin{minipage} [c] {0.45\linewidth}
 \centering
 \includegraphics[width=1\linewidth, angle=0] {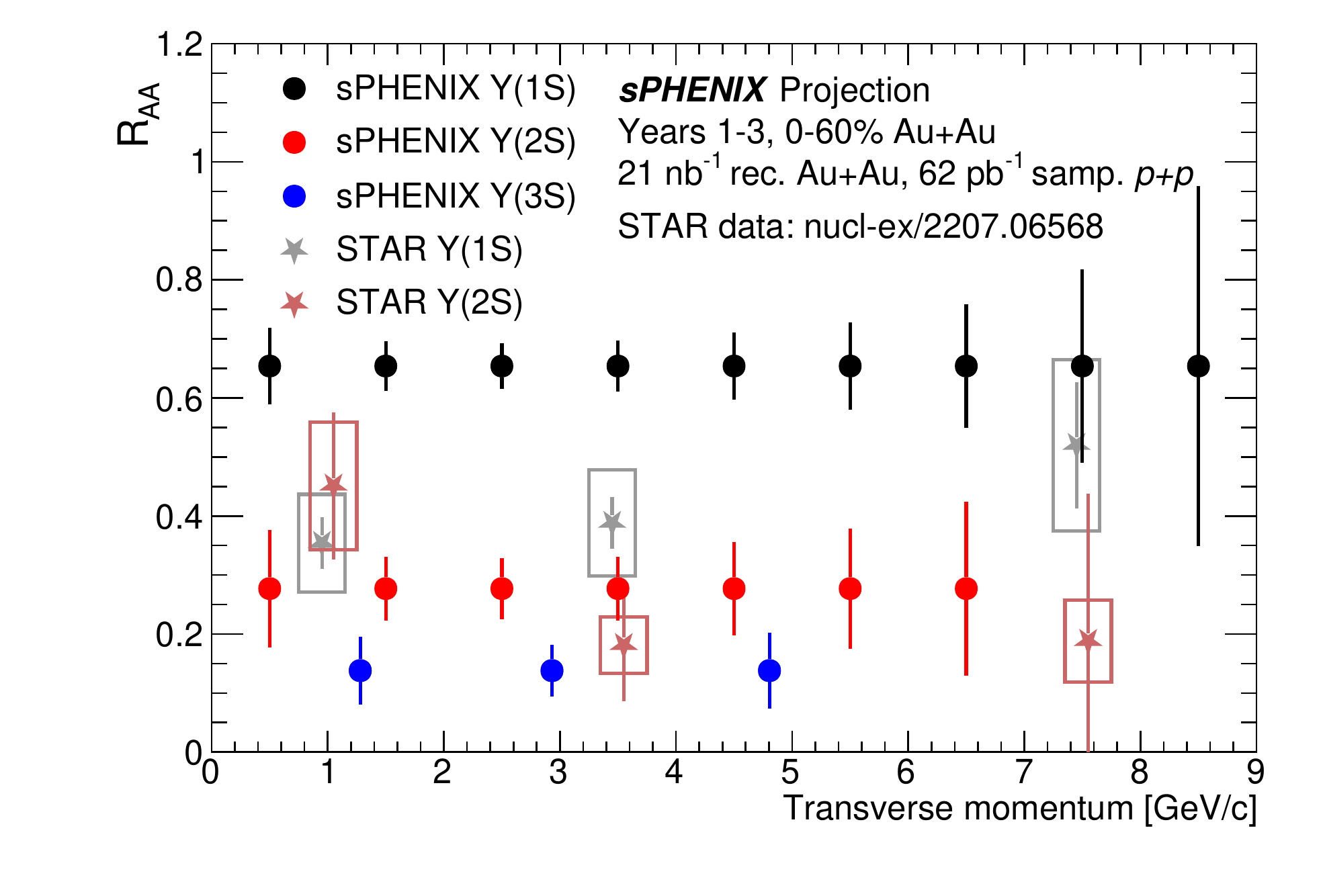}
 \end{minipage} \hfill
\caption{(Left) Simulated precision of the centrality dependence of the Upsilon modification in Au+Au collisions. (Right) The $p_{T}$ dependence of the upsilon \RAA\
for the separated 1S, 2S and 3S states  for 0-60\% central collisions, compared with the STAR measurement
for the  1S separated 2S states~\cite{STAR:2022rpk}. Both figures are from Ref.~\cite{sPHENIXBUP2022}}
 \label{fig:Fig1}
\end{figure}

\subsubsubsection{Quarkonia} The sPHENIX experiment is optimized to measure the nuclear modification of separated Upsilon states to high precision in Au+Au collisions. The design mass resolution of sPHENIX at 10~GeV/$c^2$ is 100 MeV/$c^2$, sufficient to resolve the three states, and Fig.~\ref{fig:Fig1} shows the estimated sPHENIX performance for the $\Upsilon$ measurement over the full three year program~\cite{sPHENIXBUP2022}. 
The $\Upsilon$(3S) was recently observed in \pbpb\ collisions for the first time 
by CMS~\cite{CMS-PAS-HIN-21-007} at the LHC.
In these sPHENIX projections, the modification of the $\Upsilon$(3S) state was assumed to be the same as that observed by CMS; it will be interesting to see
what the behavior of the $\Upsilon$(3S) is at RHIC.  Also shown is the STAR $\Upsilon$(2S) measurement~\cite{STAR:2022rpk}. The sPHENIX $\Upsilon$ measurements, combined with the LHC data, are expected to provide much stronger constraints on bottomonium suppression models in heavy ion collisions. Furthermore, the sPHENIX measurements will provide a unique opportunity to probe the frequency dependence of the chromoelectric field correlator describing quarkonium in-medium dynamics \cite{Yao:2020eqy,Binder:2021otw}.

\noindent

\subsubsubsection{Exotic hadrons} Several key detector upgrades are also currently underway that will directly improve measurements of exotic hadrons in heavy ion collisions.  The entire LHCb tracking system has been replaced with detectors of higher granularity, which will enable measurements in Pb+Pb collisions up to $\sim30\%$ centrality \cite{LHCb:2018qbx}.  An upgrade to the fixed target system at LHCb will enable high-statistics $p+A$ data sets to be collected at lower center of mass energies where coalescence effects are expected to be small \cite{SMOG2}.  The CMS experiment is pursuing the addition of particle ID detectors which will greatly aid in rejecting combinatorial background when reconstructing hadronic decays of exotic hadrons, allowing access to states at lower $p_T$ than currently possible at CMS \cite{CMS_upgrade}.  In the farther future, for Run-5, LHCb will be further upgraded to remove all centrality limitations \cite{LHCb:2018roe}, and the ALICE3 detector, with full particle ID and a fast DAQ, will be well suited to measurements of exotics in heavy ion collisions \cite{ALICE:2022wwr}. 

\subsubsubsection{Intrinsic charm} The existence of intrinsic (non-perturbative) charm in the proton has long been postulated \cite{Brodsky:1980pb,Brodsky:1981se,Hobbs:2013bia} to arise from configurations of the proton such as $|uud c \overline c \rangle$ and manifest at large $x$ when a proton in this state interacts \cite{Brodsky:1980pb,Brodsky:1981se,Brodsky:1991dj}. Experimental measurements \cite{NA3:1983ltt,EuropeanMuon:1981obg,R608:1987dyw} have provided tantalizing hints of intrinsic charm but no firm evidence.  LHCb recently measured $Z + {\rm charm \,\,  jets}$ at large $Z$ rapidity and showed it to be consistent with a 1\% intrinsic charm component \cite{LHCb:2021stx}. In addition, $J/\psi$ distributions from intrinsic charm have been calculated and compared favorably to $p+A$ data~\cite{Vogt:1991qd,Vogt:2021vsc,Vogt:2022glr}.
 The NNPDF Collaboration found evidence for the existence of intrinsic charm to a $3\sigma$ level \cite{Ball:2022qks}, while other global fits~\cite{Pumplin:2007wg,Jimenez-Delgado:2014zga,Hou:2017khm,Guzzi:2022rca} find no such evidence. 
Several experiments, either currently taking data or planned, could help resolve the question of intrinsic charm in the next few years \cite{Vogt:2022glr} including the current System for Measuring Overlap with Gas (SMOG) fixed-target mode at LHCb \cite{SMOG}, at energies of $\sqrt{s_{NN}} \leq 110$~GeV, and future fixed target programs such as NA60+, proposed for the CERN Super Proton Synchrotron (SPS)~\cite{NA60p}. 
Intrinsic charm could also be observed at the EIC, particularly in measurements in the proton-going direction.  
These empirical measurements may also shed light on the necessary theoretical developments to map formulations of nonperturbative charm at the level of the nucleon wave function to QCD factorization-based approaches.

\subsubsection{Initial State and Small-x} \label{sec:init_future}
As details of the initial state become more and more relevant with the increasing experimental precision, all features of the incoming nuclei will have to be considered carefully. This includes nuclear deformation, short-range correlations, alpha-clustering, etc. Close collaboration with nuclear structure experts and research into connecting low and high energy collisions will be important. Furthermore, subnucleonic structure, as measured in electron-ion collisions and quantified with GPDs or even Wigner distributions or generalized transverse momentum dependent parton distribution functions (GTMDs) will play an increasingly important role in the description of $p+p$ and heavy ion collisions. Important measurements will be for example diffractive dijet and vector meson production in UPCs and at the future EIC, which will also require further theory progress, including on fundamental questions concerning the definition of coherent processes \cite{Klein:2023zlf}.

All this information can provide input for a variety of initial state models, of which two major types can be distinguished. On the one hand there will be those models appropriate for low energy collisions, which ideally include some dynamics and are interweaved with the early time evolution. On the other hand, there are more ab-initio models valid in the high energy limit, which should be systematically improved, for example by including quark degrees of freedom and non-conformality, as well as a fully three dimensional spatial distribution. Connecting either model to hydrodynamic simulations might demand an intermediate stage of pre-equilibrium evolution, as further discussed in Sec.\,\ref{sec:hydrotheory}.
Besides the fluctuating spatial distribution of the energy momentum tensor and the conserved charges, computing some observables, such as those sensitive to the chiral magnetic effect, also requires models for the initial electromagnetic fields, which require refinement. 

\subsubsubsection{The case for varying collision systems} 
A way to disentangle initial state from final state properties is to study a wide range of collision systems. 
The nuclear structure and produced initial condition vary in a non-monotonic fashion with $N$ and $Z$, whereas the hydrodynamic response varies smoothly and slowly with the mass number, $N+Z$. Hence, isobar or isobar-like systems with nearly identical hydrodynamic response but large structure differences can be used to separate initial from final state properties and constrain initial state models \cite{Bally:2022vgo}. 

Models can be constrained using collisions of nuclei with well known properties, such as the doubly-magic $^{208}$Pb or $^{132}$Sn. Then, predictions can be made for other species and consistency with low-energy nuclear structure knowledge checked. 
Medium to small systems can expose the role of sub-nucleon fluctuations, initial momentum anisotropy, and the hydrodynamization process. In particular the exploration of isobar or isobar-like collisions in the region from $^{12}$C to $^{48}$Ca with different structures, which are nowadays accessible to cutting-edge {\it ab initio} calculations, will improve our understanding of the emergence of collectivity. 
Exploiting isobar ratios for bulk observables as a function of rapidity and $\sqrt{s_{NN}}$ may further provide access to the $x$-dependence of nPDFs and gluon saturation, complementing the science goals of the EIC.
    
The initial conditions for hard probes are typically modeled by convoluting information from the Glauber model with the nPDF, which contributes to a large uncertainty in the relevant transport properties \cite{Apolinario:2022vzg,Xie:2022ght,JETSCAPE:2022ixz}. By constructing ratios (between collision systems) of selected high-$p_T$ observables at a fixed centrality, jet quenching effects are expected to cancel and deviation of ratios from unity provide access to flavor-dependent nPDFs \cite{Paukkunen:2015bwa,Helenius:2016dsk,Jonas:2021xju}. 
Such measurements would require high-luminosity runs in both collision
systems.  Projections of the feasibility of these measurements have not yet been carried out. 

\subsubsubsection{The role of ultraperipheral collisions} UPCs connect heavy-ion collisions to both cold QCD physics and the EIC.  
In terms of vector meson (VM) and jet photoproduction, other observables, e.g., single jet or high-$p_{\rm T}$ particle production, association of forward neutron production from QED, and light (e.g., $\phi$)  and heavy (e.g., $\Upsilon$) VM threshold production, are of great interest. \textit{Species dependence}, approximately the same as the dipole size dependence and scale dependence, can provide unique insights into the nuclear modification mechanism of parton densities. \textit{Energy dependence}, enabled by experiments at both RHIC and the LHC, e.g., the STAR forward detector and the ALICE's FoCal~\cite{Bylinkin:2022temp}, will provide widest kinematic phase space coverage, which will be complementary to that at the EIC. \textit{New observables}, e.g., combining VMs and jets together in both protons and heavy nuclei, could provide one of the most rigorous experimental tests to nuclear shadowing and gluon saturation models. This is similar to one of the ``day-one'' measurements at the EIC. Looking further ahead, by the mid/late 2030s, the proposed ALICE 3 detector \cite{ALICE:2022wwr} will have acceptance for both charged and neutral particles, over a very wide solid angle, with coverage expected for pseudorapidity. ATLAS and CMS will also cover  $|\eta|<4$ by Run 4. This will offer a very large increase in acceptance for more complex UPC final states.

Furthermore, significant progress has been made in the physics of photon interactions. Experimental measurements and theoretical descriptions have been progressing from the initial observations toward quantitative and precise comparisons. For example, polarized photons have been used and proposed as a tool to test and define the photon Wigner function~\cite{Wang:2021kxm,Klein:2020jom,CMS:2020skx,Sun:2020ygb,Zha:2021jhf}, to probe the properties of the QGP~\cite{STAR:2018ldd,ATLAS:2018pfw,Klein:2018fmp,Wang:2021oqq,An:2021wof,Klusek-Gawenda:2018zfz}, to measure nuclear charge and mass radii~\cite{Wang:2022ihj,STAR:2022wfe,Budker:2021fts}, to study gluon structure inside nuclei~\cite{Hatta:2021jcd,Xing:2020hwh,Bor:2022fga} and to investigate new quantum effects~\cite{STAR:2022wfe,Zha:2018jin,Zha:2020cst,Xing:2020hwh,Dyndal:2020yen,Xu:2022qme}. 

In addition, not only exclusive observables will be measured in the future, recent studies have also indicated important physics implication of inclusive particle photoproduction. The ATLAS measurement of second-order Fourier harmonics of charged particles in $\gamma+$Pb has provided an important experimental input to the origin of collectivity in heavy-ion collisions, a long-standing question to be solved in the next decade. Also, searching for the baryon junction, a fundamental nonperturbative structure connected to color confinement in QCD, has been extensively studied in hadronic collision. Recently, a new idea of searching for baryon junctions has been proposed in $\gamma+$Au UPC events~\cite{Brandenburg:2022hrp}. The physics of baryon stopping is also intimately related to backward photoproduction of mesons \cite{Gayoso:2021rzj}, accessible at the EIC \cite{Cebra:2022avc}.

\subsubsection{Small Size Limit of the QGP}
\label{sec:smallfuture}
There is much work to be done in understanding the small size 
limit of the QGP.  Past measurements have focused on 
\pp\ and \pA\ collisions, particularly
at the LHC.  The large acceptance and high rate 
of sPHENIX will allow for more detailed measurements in \pAu\
collisions at RHIC including multi-particle cumulants, 
open heavy flavor mesons and measurements of
charged particle and jet \vn\ at high transverse momentum.
Additionally, there is great interest in collecting
data with systems which are of similar size to \pA\ collisions but which are symmetric, such as \oo\ \cite{Huang:2019tgz,Brewer:2021kiv} (see Sect~\ref{sec:future_flow}
and Sec.~\ref{sec:jetfuture}) in order to smoothly map the evolution
of small systems to larger ones. The d+Au and O+O data taken by STAR and future p+Au data during the sPHENIX running will enable such mapping, and in particular pin down the role of longitudinal decorrelations and subnucleonic fluctuations. For all small systems, theoretical modeling has to be improved, in particular the initial states and earliest stages of the collision, which are far from equilibrium.

The observation of evidence for collective
flow in \gammaA\ collisions (see Sec.~\ref{sec:smallprogress})
has challenged theoretical modeling of these systems.
Recent work~\cite{Zhao:2022ugy}
suggests that full (3+1)D hydrodynamical modeling is necessary 
to characterize \gammaA\ collisions.  More theoretical
modeling and measurements are needed to see to what
extent this conclusion applies to other asymmetric collision systems.
These can even be applied to high multiplicity events in future electron-ion collisions at the EIC.
Additionally, there is much work to be done to determine if there
are other observables, besides the \vn\ of charged particles,
to further characterize the nature of these collisions.
The next generation of theoretical models needs to include nucleon configurations from \textit{ab initio} nuclear structure physics. Furthermore, integrating with high energy small-$x$ evolution would enable theoretical models to systematically include sub-nucleonic fluctuations and how they evolve with collision energy.

A major outstanding question in small collision systems is the absence of a conclusive observation of jet quenching in any \pA\ or \pp\ measurement.
Experimentally, more precise measurements in \pA\ collisions are necessary to see a potentially small signal.  In order to estimate the size of any jet quenching expectations, light ion collisions are essential.  
This would provide a clean way to bridge between \pA\ and \AA\ collisions because measurements of jet quenching in peripheral \pbpb\ and \auau\ collisions suffer from large uncertainties.  Finally, realistic theoretical modeling
of jet quenching expectations in \pA\ collisions is important
to further constrain the size of any potential effect.  All
of these pieces are necessary to develop a coherent understanding
of the how the QGP works in the small size limit.
Additionally, the modification of heavy flavor production is important for jet quenching in small systems~\cite{Liu:2021izt,Katz:2019qwv,Ke:2022gkq} because the system size changes the relative significance of radiative and collisional energy loss.

\subsubsection{Mapping the QCD Phase Diagram} \label{sec:futurephasedigram}
The goals of the RHIC BES program are to (i) study the QCD phase structure with high-energy nuclear collisions (The BES program covers the widest range in terms of baryon chemical potential, 20 - 780 MeV), and (ii) search for the phase boundary and possible QCD critical point. 
The nuclear matter EOS in the high $\mu_B$ region requires detailed investigations. 
Baryonic interactions including nucleon-nucleon, hyperon-nucleon and hyperon-hyperon interactions are fundamental ingredients to understand QCD and the EOS that governs the properties of nuclear matter and astrophysical objects such as neutron stars~\cite{Lonardoni:2014bwa}.  Precise measurements for a range of observables and collision energies are necessary to understand this physics \cite{ALICE:2020mfd}.

The NA61/SHINE experiment is an ongoing experiment at the CERN SPS, studying the properties of the production of hadrons in collisions of beam particles (pions, and protons, beryllium, argon and xenon) with a variety of fixed nuclear targets. The current program will continue until the end of 2024 and a program beyond that is under discussion. The NA60$^+$ experiment~\cite{NA60:2022sze} is planned as an upgrade to NA60 at the CERN SPS to study dilepton and heavy-quark production in nucleus-nucleus and proton-nucleus collisions with center of mass energies of 6--17.3\,GeV. NA60$^+$ is currently expected to start taking data around 2029. 

\begin{figure}[htb]
\centering
\begin{minipage}{0.45\textwidth}
\includegraphics[width=\columnwidth]{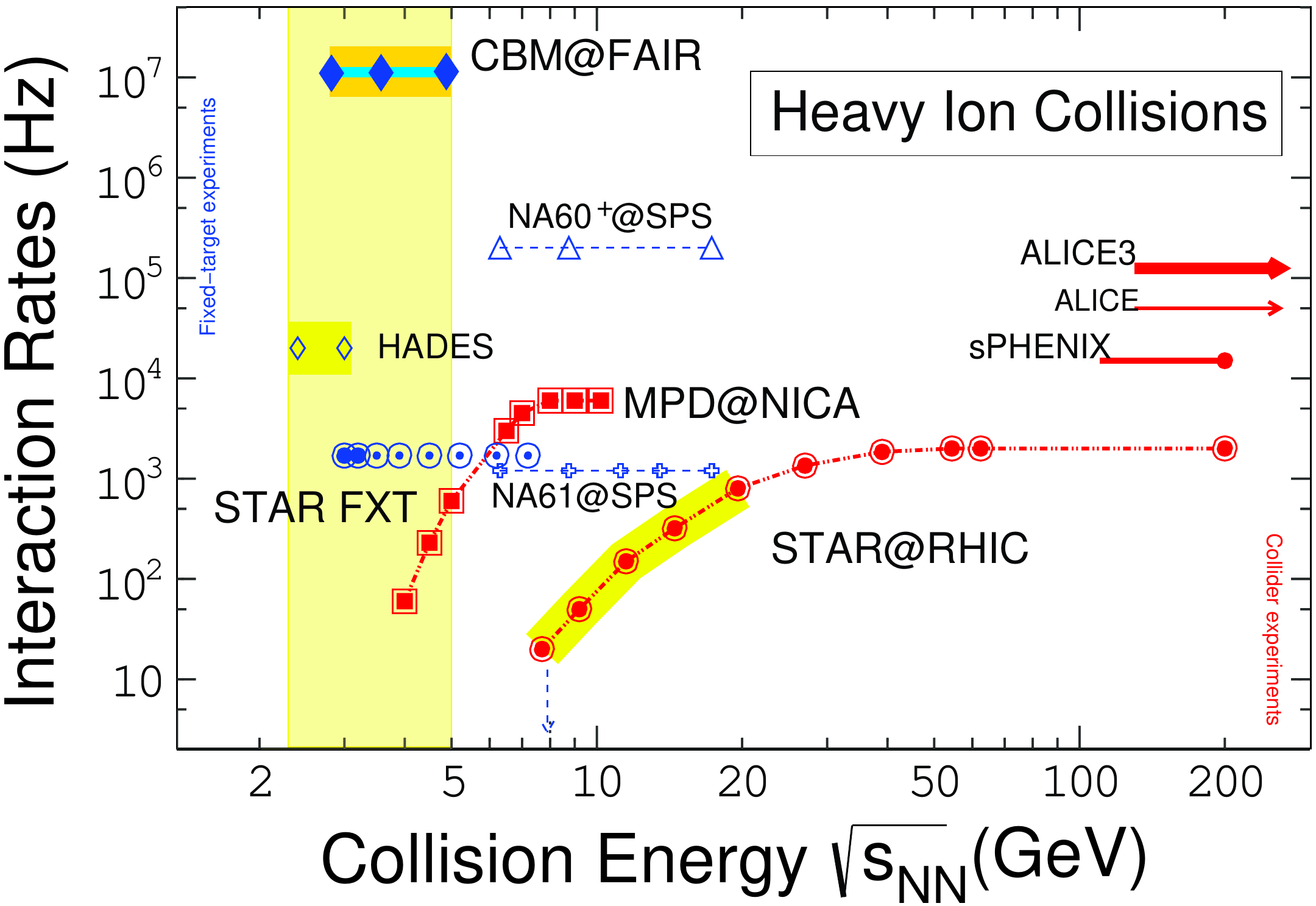} 
\end{minipage}
\hspace{0.3cm}
\begin{minipage}{0.35\textwidth}
\linespread{1.0}\selectfont{} 
\caption{Collision rates as a function of $\snn$  for collider experiments in red, and fixed-target (FXT) experiments in blue. Comparing to the collider experiments, more than four orders of magnitude improvement in collision rates can be achieved with the future CBM experiment at FAIR~\cite{Galatyuk:2019lcf,Fukushima:2020yzx}.}
\label{fig:collisionRates}
\end{minipage}\end{figure}

The CBM experiment at FAIR~\cite{CBM:2016kpk}, will have a uniquely large interaction rate, see Fig.~\ref{fig:collisionRates}. It will determine the EOS (check to be sure that EOS or EoS is used consistently throughout) of QCD matter in the range $\sqrt{s_{NN}}=2.9$--4.9~GeV. FAIR is one of the top-priority facilities for nuclear physics in Europe, according to NuPECC~\cite{NuPECC,NuPECCLRP2017}. The CBM physics program includes net-proton fluctuations, dileptons, multi-strange hyperons and hypernuclei, polarization and spin alignment. These measurements will probe the first-order phase boundary, the QCD critical point, and hypernuclear interactions pertinent also to the inner structure of compact stars. The physics program is currently planned to start later this decade.

To maintain US leadership in the exploration of the QCD phase diagram at high baryon density after the completion of the RHIC BES-II program, opportunities for targeted US participation in international facilities are important to explore. A top priority is to complete the RHIC BES-II data analysis, which will help assess which international experiments present the highest physics potential. One area of interest is the CBM Experiment at FAIR~\cite{Almaalol:2022xwv}.

\subsubsection{Chirality and Vorticity in QCD}
\subsubsubsection{Chiral magnetic effect}
The precision measurements from the RHIC isobar collisions largely constrain the observability of the CME in heavy ion collisions and narrow down future CME searches~\cite{Kharzeev:2022hqz}. A number of possible directions have been identified for the CME in the post-isobar era~\cite{starbur25}. One possible direction is to study lower energy collisions in which the prerequisites of the CME phenomenon are expected to be different due to the enhancement of topological fluctuations~\cite{Ikeda:2020agk} (also see \cite{Cartwright:2021maz} for other effects). 
The recent measurement from the STAR collaboration using the event plane detectors capable of measuring the spectator proton-rich plane has put constraints on the observability of CME in $\sqrt{s_{{NN}}}=27$ GeV Au+Au collisions~\cite{STAR:2022ahj}. This work paves the way to future CME searches with the high statistics data from the RHIC BES-II using novel techniques such as event-shape engineering~\cite{Milton:2021wku}. Another avenue is to revisit the analysis of Au+Au collisions $\sqrt{s_{_{NN}}}=200$ GeV in which signal/background ratio is expected to be larger than that of isobar collisions~\cite{Feng:2021oub}. Estimates from the STAR collaboration indicate that a $5\sigma$ significance on the possible CME signal fraction can be achieved if 20 billion Au+Au events are collected during the remaining RHIC running~\cite{starbur25}.
Besides the CME search, a number of other measurements related to chiral effects will be investigated by the STAR collaboration in the coming years~\cite{STAR:2022hfy,Finch:2017xiz}. 

\subsubsubsection{Vorticity} \label{sec:vorticity_future}
There remain several additional open questions on vorticity 
in heavy-ion collisions. One is the distribution of vorticity in rapidity. Most models that reproduce the falling energy dependence of global polarization at midrapidity predict a rising polarization at forward rapidity, as the fluid vorticity migrates towards the spectator 
region~\cite{Jiang:2016woz,Liang:2019pst,Ivanov:2018eej,Wu:2019eyi,Ivanov:2019ern,Ivanov:2020wak};  others~\cite{Deng:2016gyh,Wei:2018zfb,Xie:2019jun} predict {\it smaller} polarization at forward rapidity.
So far, the data show a rapidity-independent polarization, though the rapidity coverage is quite limited.
Measurements of global polarization at forward rapidity may  discriminate between different physics scenarios that produce similar results at midrapidity.

There are  new vorticity signatures yet to be explored experimentally. At forward rapidity, hydrodynamic and transport simulations at all energies predict~\cite{Baznat:2015eca,Teryaev:2015gxa,Ivanov:2018eej,Xia:2018tes} a circular vorticity pattern superimposed on the global vorticity, due to the interplay between transverse and longitudinal gradients in temperature and flow. The transfer of energy and momentum from a jet to the surrounding medium should produce a toroidal vorticity structure centered about the jet direction~\cite{Betz:2007kg,Tachibana:2012sa,Serenone:2021zef}, which may generate percent-level polarizations observable through $\Lambda$-jet correlations sensitive to the quenching and fluid viscosity~\cite{Serenone:2021zef}. 
Spin alignment of vector mesons is another promising observable sensitive to the large angular momentum of the system and indicative of quark polarization, and can provide new insight into the nature of the vector meson fields \cite{ALICE:2019aid,STAR:2022fan}.
Additionally, it has been suggested~\cite{Voloshin:2017kqp,Lisa:2021zkj} that in \pA\ collisions, a toroidal "vortex tube" may be created at midrapidity, generating something akin to smoke rings centered on the beam direction. This measurement has been proposed by the STAR Collaboration~\cite{StarBur2022}.

\subsubsection{Future Facilities for Hot QCD}

\subsubsubsection{Hot QCD at RHIC}
The RHIC facility began operation in
2000 and over the past two decades has collided nuclei from
protons to uranium at
a wide range of collision energies spanning two orders of 
magnitude. 
The remaining data to be taken at RHIC to 
complete its science mission is the 
sPHENIX program~\cite{sPHENIXBUP2022}.  This program
has three essential components: 
successfully commissioning 
the sPHENIX detector (the first new collider detector to 
be commissioned in over a 
decade); collecting high luminosity $p+p$ and $p+$Au data for nucleon structure studies and as a heavy ion reference system; and taking high luminosity Au+Au data. 
The total recorded \auau\ data is expected to be at least
21~\inb.  The \pp\ luminosity requirement is 62~\inp\
and is driven
by the need to have adequate statistics to use as 
an Au+Au reference measurement. 
These numbers are used in all the projections
in this document.
This data is anticipated to be collected in three years of RHIC running.

Given the versatility of RHIC and the variety of science
questions it can address, the RHIC science potential is in no way
exhausted. 
While the sPHENIX science program is the ``highest priority
for the current RHIC program"~\cite{BNLPAC22}, it is clear there
are additional  high priority
scientific opportunities available at RHIC.
While the currently scheduled RHIC program consistes of \pp,
\pAu, and \auau\ collisions, there are other 
unique opportunities that may become available including:
running \oo\ collisions to understand
the system size dependence of QGP properties, additional
\pA\ running for vorticity measurements, nuclei scans
to measure the initial state, and many others.
Following the completion of the RHIC science program,
the RHIC infrastructure is scheduled to become the basis of the EIC. 

\subsubsubsection{LHC}
Participation in the 
heavy-ion program at the LHC is
a key component of the US heavy-ion program.  All four 
LHC detectors have significant heavy-ion programs and the LHC
heavy-ion program is already planned to go through
the end of Run 4 (currently expected in 2032).  LHC Runs 3 and
4 over the next decade are expected to provide more than approximately 10~\inb\ of \pbpb\ data~\cite{Citron:2018lsq}.  This is approximately
five times the maximum luminosity delivered to the experiments
to date.  Additionally, \pp\ running at the \pbpb\ center 
of mass energy is an essential reference, and $p+$Pb runs are also planned.

In addition, a short run of \oo\ and \po\ collisions is
planned for 2024.  \oo\ collisions would provide key new data
on both soft and hard probes in a system with 
approximately the same number of participants as \pPb\ collisions
but which is, in contrast, symmetric, bridging the gap between peripheral \pbpb\ collisions and \pPb\ collisions.  There is significant
interest in this from the community~\cite{Brewer:2021kiv}.
It is possible that a successful \oo\ run would motivate
further LHC running with light ions.  In addition to the desire to study system size dependence, the total nucleon-nucleon luminosity can be increased in heavy-ion runs by switching from large nuclei (such as lead) to small
nuclei (such as oxygen or argon)~\cite{Citron:2018lsq}.

All experiments have significantly 
upgraded their detectors since the LHC turn on.  With 
the start of LHC Run 3 in 2022, ALICE
began taking data with the "ALICE2" upgrades~\cite{ALICE:2012dtf}
which upgraded the data-taking rate, allowing readout at 50 kHz, and a new Inner Tracking System (ITS)
which will improve the resolution on the distance of 
closest approach to the primary vertex by a factor of 
three.   Together, these upgrades will dramatically 
improve the physics reach of the ALICE detector, especially
for rare probes involving heavy-flavor and other identified particles.
Also prior to Run 3, the LHCb collaboration has completed the first of a series of detector upgrades, Upgrade 1~\cite{LHCb:2012doh}. The entire LHCb tracking system was replaced with higher-granularity detectors, which can reconstruct PbPb collisions up to 30\% centrality (previously the limit was 60\%).  All hardware triggers were removed in favor of an advanced streaming readout system that will sample the full luminosity delivered by the LHC~\cite{LHCbCollaboration:2014vzo}.  
In addition, a dedicated storage cell was installed for the gaseous fixed-target which will greatly increase the rate of beam+gas collisions~\cite{SMOG2}.   This upgraded SMOG2 system is expected to operate concurrently with the collider 
for all beam species, providing large fixed-target data samples with multiple beam and target species at center of mass energies near 100 GeV.

For Run 4, both ATLAS and CMS are planning major upgrades
with direct benefit to the heavy-ion physics program. 
Both ATLAS and CMS will have upgraded trackers which can measure charged particles in
$|\eta| < $~4, compared to $|\eta| < $~2.5 with the 
current detectors.  This increased
acceptance will, among other things, allow for jet
structure and substructure measurements over a wider rapidity
range.  At fixed $p_T$, there is a higher probability for quark jets than gluon jets at forward rapidity, providing a new means of understanding how parton showers develop in the QGP.
Additionally, CMS is planning a new timing 
detector~\cite{Butler:2019rpu} which can make additional measurements with identified hadrons.
Prior to Run 4 LHCb will implement Upgrade 1b, which will include new tracking detectors placed inside the LHCb dipole magnet and a new silicon detector near the beampipe.  The magnet station trackers will allow tracks from soft particles which terminate in the magnet walls to be reconstructed, giving new access to very low \pt\ open heavy flavor and exotic states.  The new silicon detector will provide additional tracking points that will further increase the centrality range accessible by LHCb.
On the same timescale, ALICE is planning to install
a forward calorimeter upgrade, the FOCAL~\cite{ALICE:2020mso,ALICE:2022wwr}.

\subsubsubsection{LHC upgrades for Run 5 and beyond}
For LHC Run 5 (currently planned for 2035-38), ALICE is planning
an entirely new detector, "ALICE3" to further 
improve the most difficult measurements in heavy-ion 
collisions~\cite{ALICE:2022wwr}. 
In the current LHC projections, over Runs 5 and 6, ALICE3 would take 20 times more data than ALICE in Runs 3 and 
4~\cite{Alizadehvandchali:2022ier}.
This extremely large data sample would 
include very detailed identified particle measurements, making qualitative improvements on answers to the questions outlined in the previous section.
Additionally, the very low mass silicon tracking is
expected to make precise dilepton measurements, impossible with any of the current LHC detectors. 
On the same timescale LHCb is planning Upgrade II~\cite{LHCb:2018roe,LHCb:2022ine}.
This upgrade would allow LHCb to make measurements over the 
full centrality range of heavy-ion collisions for the first 
time, providing very precise forward measurements, including access to quarkonia and open heavy flavors at very low \pt\ and forward rapidity in \pA\ and \AA\ collisions.

The physics case for all of these upgrades will continue to develop with the new measurements coming out of the LHC and RHIC.  In order to meet the Run 5 timeline, funding for R\&D for these projects should begin soon. 

It is possible to increase the nucleon-nucleon luminosity
by running with smaller collision systems.  This is offset
by the expectation that QGP effects will be largest in 
the largest collision systems.  Experience with smaller collision systems in the near future will inform whether there is a more optimal collision system to run at the LHC than \pbpb~\cite{ALICE:2022wwr}.

\subsection{Cold QCD in the Next Decade}
\label{sec:cold_future_jlab}

As outlined and described in Sec.~\ref{sec:coldQCD_progress}, the hadron physics community has made tremendous progress in answering the fundamental questions concerning the building blocks of our universe, such as the mass and spin origins of the nucleon, the tomographic imaging of partons inside the hadrons, and nucleon many body interactions encoded in partonic structures in the nucleus. The progress made in these directions since the last LRP has demonstrated the powerful reach of hadron physics facilities to unveil the underlying QCD dynamics and the associated non-perturbative structure of nucleons and nuclei. 
We will continue to deepen our understanding of these questions, focusing on the following aspects:
\begin{itemize}
  \setlength\itemsep{-0.2em}
    \item Nucleon properties including the proton charge radius and (generalized) polarizabilities of the nucleon;
    \item Precision measurements of the polarized and unpolarized quark distributions in the large-$x$ region, in particular when $x\to 1$;
    \item Unprecedented mapping of the 3D tomography of quark distributions inside nucleons;
    \item Unveiling the spin and mass origins of the nucleon, especially for the quark orbital angular momentum contribution to the proton spin and the trace anomaly contribution to the proton mass;
    \item Nucleon-nucleon short range correlations in nuclei and the nuclear modification of the parton distributions in the valence region;
        \item Precision meson and baryon spectroscopy to unravel the spectrum and structure of conventional and exotic hadrons;
    \item Parity-violation measurements and connections to other fields.
\end{itemize}
As this document is written, RHIC will be transitioning to EIC construction within the next 5 years, while CEBAF will continue to operate with fully scheduled programs for at least another decade. We describe below cold QCD research expected from both facilities. 

\begin{figure}[!ht]
\centering
 \begin{minipage}[r]{0.6\textwidth}
  \includegraphics[width=\textwidth]{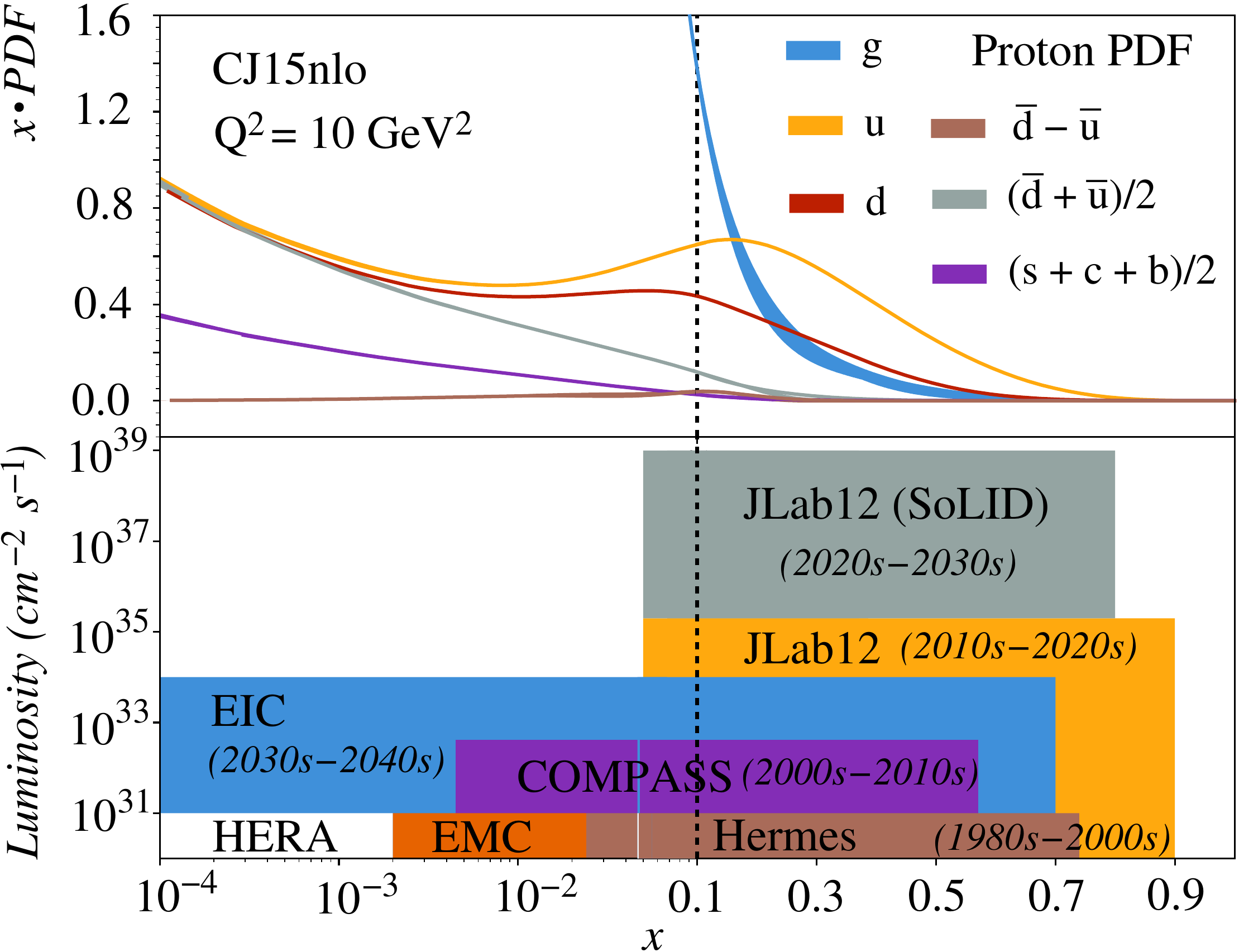}
 \end{minipage}
 \hspace{0.4cm}
\begin{minipage}[c]{0.3\textwidth} 
  \caption{Landscape of the cold QCD program at the DIS facilities.  SoLID expands the luminosity frontier in the large $x$ region whereas the EIC does the same for low $x$. Together, JLab+SoLID and the EIC will, over the next several decades, cover a broad and largely complementary kinematic range, with SoLID probing key physics and providing precision data primarily in the high-$x$ region. Figure 
  from~\cite{Arrington:2021alx, JeffersonLabSoLID:2022iod}.}
\label{fig:Cold_landscape}
 \end{minipage}
\end{figure}

\subsubsection{Cold QCD with CEBAF and the SoLID Physics Program}\label{sec:cold_future_solid}
CEBAF was originally designed to conduct coincidence experiments, but its physics program as well as experimental halls have evolved to meet the ever-changing development and needs of hadronic physics studies over the years.  Most notably, CEBAF was successfully upgraded to double its energy to 12 GeV during the previous LRP period. The higher beam energy, along with the addition of experimental Hall D and upgrades of detectors in other halls, allowed our studies of hadronic physics to expand into new kinematic regions and to search for and study exotic hybrid mesons. 
Moving forward, CEBAF 
will remain in high demand as a QCD facility because of its high luminosity and the mid-scale, high intensity SoLID program, see Fig.~\ref{fig:Cold_landscape}. 

Meanwhile, smaller-scale spectrometers and specialized detectors continue to be built, such as the Super BigBite Spectrometer (SBS), Neutral Particle Spectrometer (NPS), Low Energy Recoil Tracker (ALERT), to name a few.  
The remainder of this section presents an overview of upcoming programs at CEBAF. 
Some of the topics presented could persist into the EIC era.

\begin{figure}[!ht]
  \begin{minipage}[r]{0.7\textwidth}
 \includegraphics[width=\textwidth]{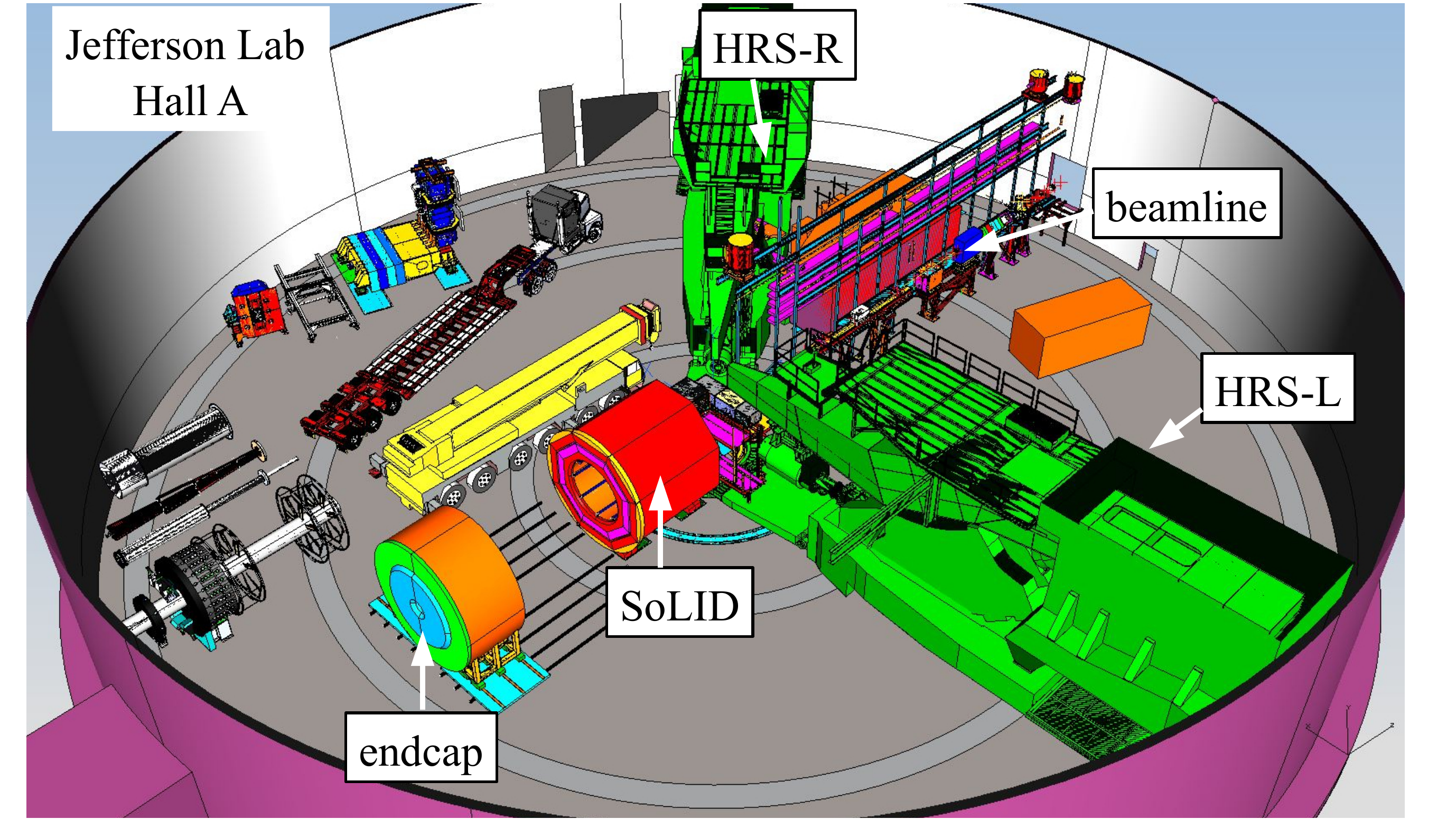}
  \end{minipage}
  \begin{minipage}[c]{0.25\textwidth}
 \caption{Schematic layout of SoLID in Hall A of JLab, with the endcap pulled downstream to allow detector installation and reconfiguration. The two high resolution spectrometers (HRS-L and HRS-R, not in use) are parked at backward angles.}\label{fig:solid_in_halla}
  \end{minipage}
\end{figure}

\subsubsubsection{The SoLID physics program}
With the study of nucleon structure evolving from single- to multi-dimensional measurements that utilize exclusive processes, the quest for understanding the origin of the proton mass based on studies of near-threshold meson production, frontier cold QCD research requires, first and foremost, higher statistics. Similarly, Parity Violating Electron Scattering (PVES) that requires increasing statistical precision to test the Standard Model at low- to medium-energies.  
Such emerging needs from both QCD and fundamental symmetries call for a truly large acceptance, high-intensity device, to fully capitalize on the high-luminosity beam of CEBAF. The Solenoidal Large Intensity Device (SoLID), planned for JLab as an integral part of the CEBAF 12 GeV program, was designed to meet such needs. SoLID will utilize the CLEO-II 1.4-T solenoid magnet and a large-acceptance detector system covering $2\pi$ in azimuth and will be able to operate at luminosities up to $10^{39}$~cm$^{-2}$s$^{-1}$. 
The realization of SoLID in JLab Hall A is shown in Fig.~\ref{fig:solid_in_halla}.

\begin{figure}[hbt!]
\includegraphics[width=0.44\textwidth]{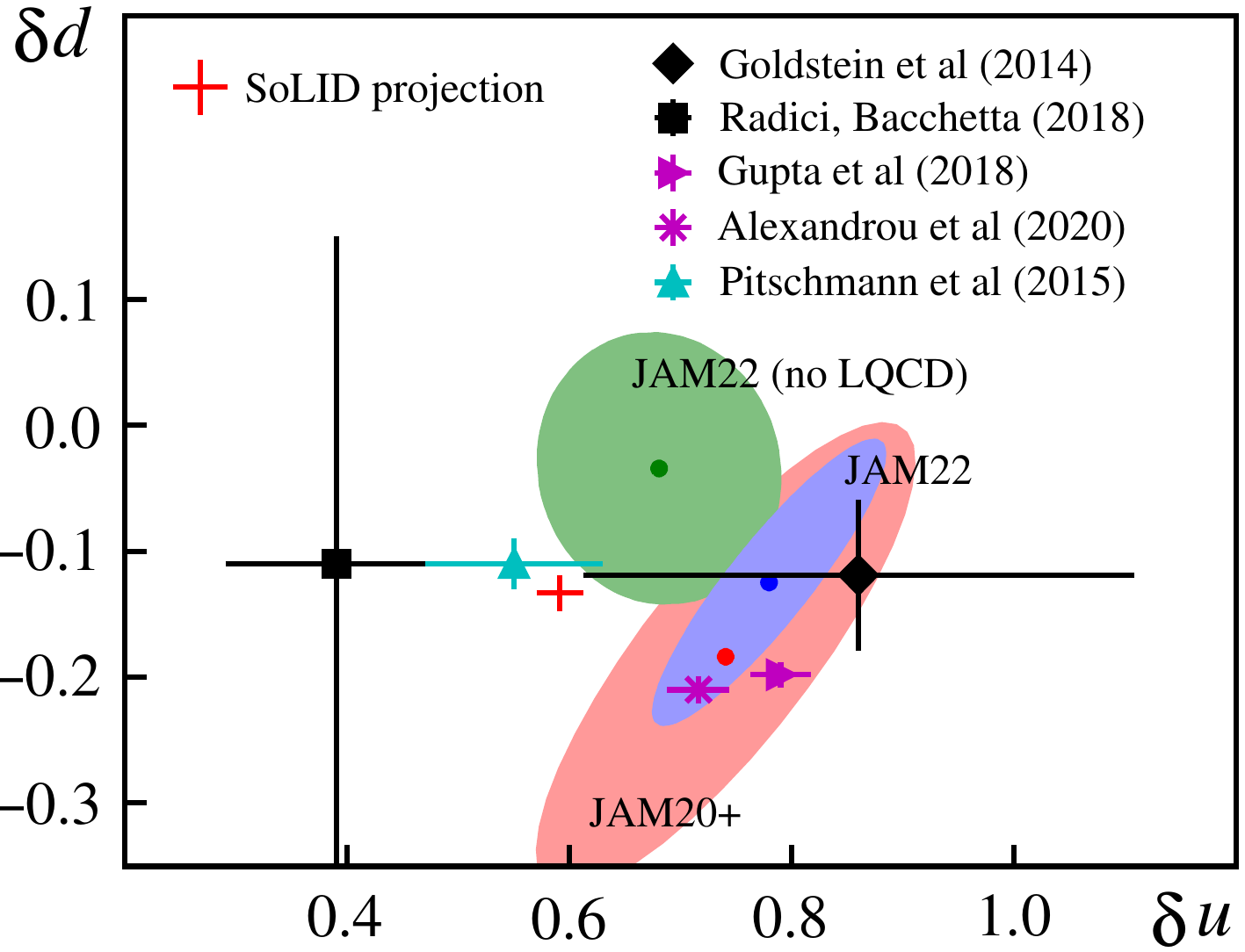}
\hspace*{3mm}
    \includegraphics[width=0.5\textwidth]{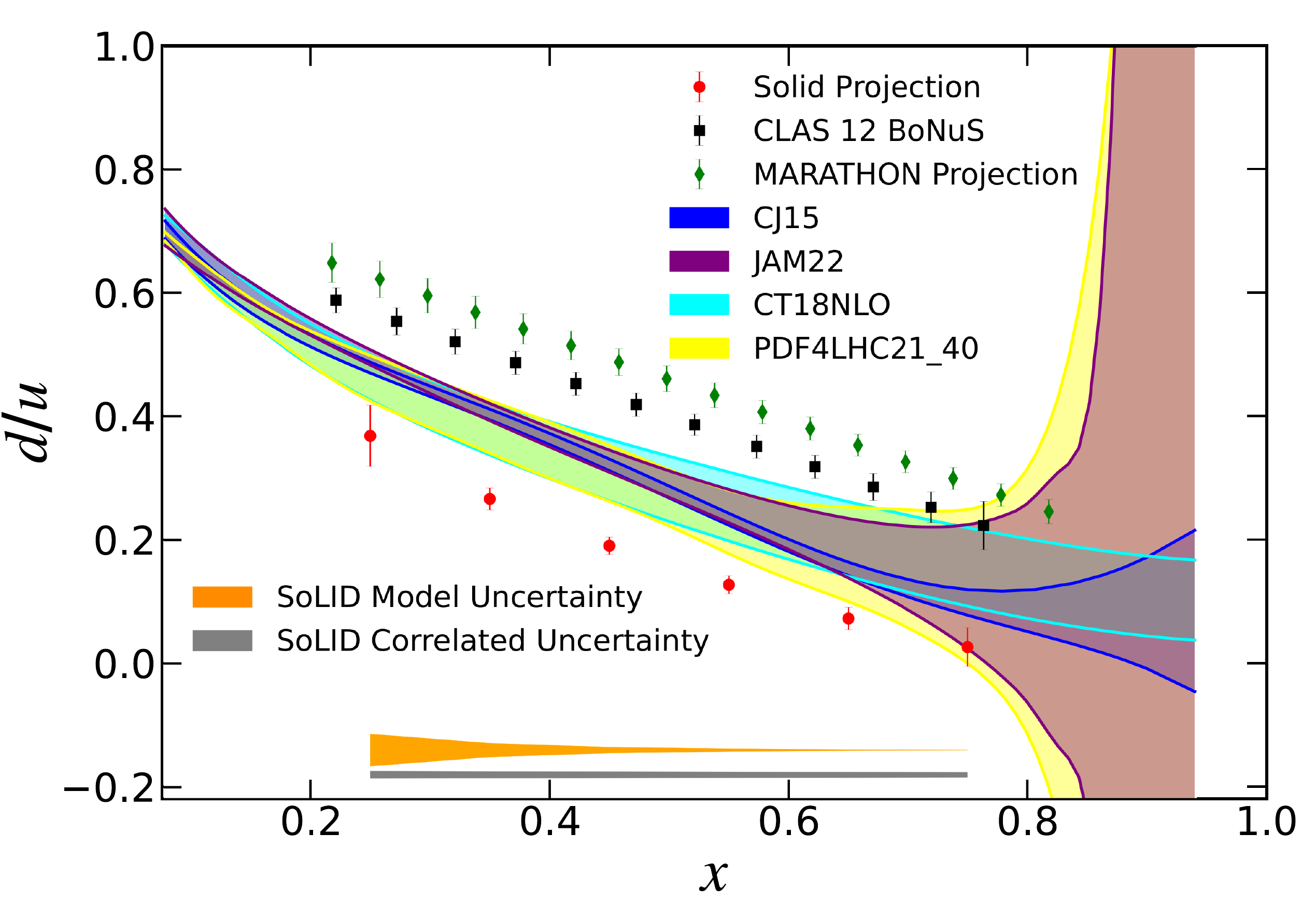}
\caption{Projected impact of the SoLID program on: (left) the $d$ vs. $u$ tensor charge from SIDIS measurements; and (right) the PDF ratio $d/u$ from PVDIS proton measurement. See~\cite{JeffersonLabSoLID:2022iod} for details.
    }
    \label{fig:future_solid_phys} 
\end{figure}

As a multi-purpose device, SoLID currently has seven experimental proposals and several run-group proposals approved by the JLab Program Advisory Committee. Three SIDIS experiments, with transversely and longitudinally polarized $^3$He and transversely polarized protons, will precisely extract TMDs in the valance quark region and determine the $u$ and $d$-quark tensor charge, see Fig.~\ref{fig:future_solid_phys} left.  An experiment studying electro- and photo-production of $J/\psi$ near threshold probes the gluonic field and its contribution to the proton mass, see Section~\ref{sec:cold_future_femtography}. A parity-violating DIS experiment will determine the effective electron-quark couplings of the Standard Model, pushing the limits in phase space to search for new physics (Section~\ref{sec:otherfield_pves}), and will provide the PDF ratio $d/u$ at high $x$, see Fig.~\ref{fig:future_solid_phys} right. 
The two most recently approved experiments include a measurement of TPE with beam SSA in DIS, and a PVES measurement to study isospin dependence of the EMC effect. The run-group experiments include SIDIS with kaon and di-hadron production, transverse inclusive spin structure functions, and exploration of GPDs with deep-exclusive reactions to study the 3D structure of the nucleon in coordinate space. 

\subsubsection{Properties of the Nucleon }

\subsubsubsection{PRad-II} The PRad experiment (see Section~\ref{sec:cold_highlight_longrange}), using innovative methods, provided data on the proton charge radius with high precision, but is in direct conflict with all modern electron scattering experiments. 
The newly approved PRad-II experiment will address this discrepancy with a projected total uncertainty of 0.43\%, a factor of 3.8 smaller than that of the PRad result, and better than the most precise result from ordinary hydrogen spectroscopy measurements~\cite{PRad:2020oor}.  This level of precision has the potential to inform whether there is any difference between $e+p$ scattering and muonic hydrogen results, as well as to evaluate the consistency of systematic uncertainties of muonic hydrogen measurements.

\subsubsubsection{Future polarizability and generalized polarizability measurements} 
In the next seven years, the complementarity of the MAMI and HIGS facilities will be leveraged to access a wide variety of energies and observables for the nucleon polarizability measurements~\cite{Howell:2020nob}, with strong collaboration between experimental and theoretical efforts, see, e.g, ~\cite{Melendez:2020ikd}. 
Exploring a variety of few-nucleon targets is essential for high-accuracy extractions of the neutron polarizabilities and validation of the subtraction of nuclear binding effects. The first values of the neutron spin polarizabilities and improved determinations for the proton will provide insights into the low-energy spin structure of the nucleon, enhancing our understanding of the mechanisms that generate them and complementing the nucleon-structure experiments at JLab, RHIC, and the EIC. 

Plans for future measurements of the proton generalized polarizabilities at JLab are currently underway. 
One major goal is to determine the shape of $\alpha_E(Q^2)$ to high precision.
This will serve as valuable theoretical input for determining the mechanism responsible for the effect. 
Another goal is to accurately describe $\beta_M(Q^2)$ at low $Q^2$ which is currently not well understood due to large uncertainties on the existing data, in particularly at $Q^2 = 0$ where recent results are in conflict~\cite{A2CollaborationatMAMI:2021vfy,Li:2022vnz}. 
A positron beam, proposed to be developed at JLab, can provide an independent cross-check in a different reaction channel~\cite{Pasquini:2021qdi}, particularly in light of the recently reported puzzling results for the proton $\alpha_E (Q^2)$~\cite{Li:2022sqg}. 

\subsubsubsection{Two photon exchange measurements} 
As mentioned in Section~\ref{sec:cold_highlight_longrange}, a full understanding of TPE in $e+p$ elastic scattering is critical for correctly interpreting proton form factor measurements. 
The luminosity and quality of a positron beam would provide dramatically improved direct measurements of TPE in elastic scattering, in particular in the large-angle region where TPE is most important. In addition, several new TPE observables can be measured for the first time. As an example, the asymmetry of electron scattering when only the beam or only the hadron (target) spin is polarized normal to the scattering plane is related to the imaginary part of the TPE amplitude. 
Such SSAs have been measured in PVES experiments~\cite{Esser:2020vjb,Rios:2017vsw,Gou:2020viq,QWeak:2020fih,QWeak:2021jew,PREX:2021uwt} with a polarized beam for elastic scattering. 
 While SSAs on lighter nuclei have confirmed theoretical predictions, the SSA on lead is unexplainably small.
Similar SSAs were measured at HERMES and JLab Hall A with polarized targets~\cite{HERMES:2009hsi,Katich:2013atq,Zhang:2015kna}. 
Each of these observables provide independent constraints on the TPE amplitude, and are valuable for making theoretical progress on the problem of so-called box diagrams which include TPE as well as the $\gamma$Z-box correction relevant to PVES, and the $\gamma$W-box contributing to $\beta$-decay.
An experiment was recently approved to access TPE by measuring transversely-polarized beam SSA in DIS using SoLID, adding a new observable to the TPE study. 

\subsubsubsection{Quark distributions and polarizations at $x\to 1$} As part of a complete three-dimensional mapping of the parton (quark and gluon) distributions in the nucleon, the longitudinal momentum and spin carried by valence quarks at very high Bjorken-$x$ is still of great theoretical and experimental interest. At the same time, quark distributions at large $x$ are also needed as input for cross section calculations at colliders such as the LHC or the Tevatron (see for example the recent results on the $W$ mass~\cite{CDF:2022hxs}). The current and future JLab program studies the large $x$ quark distributions and polarizations in three different experiments. 
The first experiment, MARATHON, was highlighted in Section~\ref{sec:cold_progress_1d} and provided precision data on $F_2^n/F_2^p$. Data have been collected by the second such experiment, BONuS12~\cite{JLabPR:BONUS12}, and results are expected soon. The PVDIS proton program of SoLID will provide $d/u$ at high $x$ without the use of nuclear models, as shown in the right panel of Fig.~\ref{fig:future_solid_phys}.
Additionally, data have been collected on double-polarization asymmetries of both the proton and $^3$He and results on the down quark polarization $\Delta d/d$ are expected to be available concurrent with the release of the 2023 LRP. 

\subsubsection{Nucleon Femtography}\label{sec:cold_future_femtography}

As described in previous sections, the study of the nucleon structure is evolving from 1D structure functions connected to collinear PDFs to also include multi-dimensional tomography in terms of parton GPDs and TMDs. The ultimate goal is to experimentally determine the quantum mechanical
Wigner distribution~\cite{Belitsky:2003nz} in phase space. Semi-inclusive measurements, including spin polarization observables, were provided by the pioneering measurements at HERMES, COMPASS, and the JLab 6 GeV program, among others. Results on GPDs and TMDs are now published over limited ranges of the relevant kinematic variables. The upgraded detectors and CEBAF beam energy and intensity, as well as the potential for polarized positron beams, promise to provide a more detailed three-dimensional spatial mapping of the nucleon. Indeed, this is a major thrust of the JLab 12 GeV facility. Mapping the (2+1)D mixed spatial-momentum images of the nucleon in terms of GPDs has been one of the important goals. 
On the other hand, 3D images in pure momentum space can be made with other generalized distributions: the TMDs. These femto-scale images (or femtography) will provide, among other insights, an intuitive understanding of how the fundamental properties of the nucleon, such as its mass and spin, arise from the underlying quark and gluon degrees of freedom.

\begin{figure}[!ht]
\centering
 \includegraphics[width=0.7\textwidth]{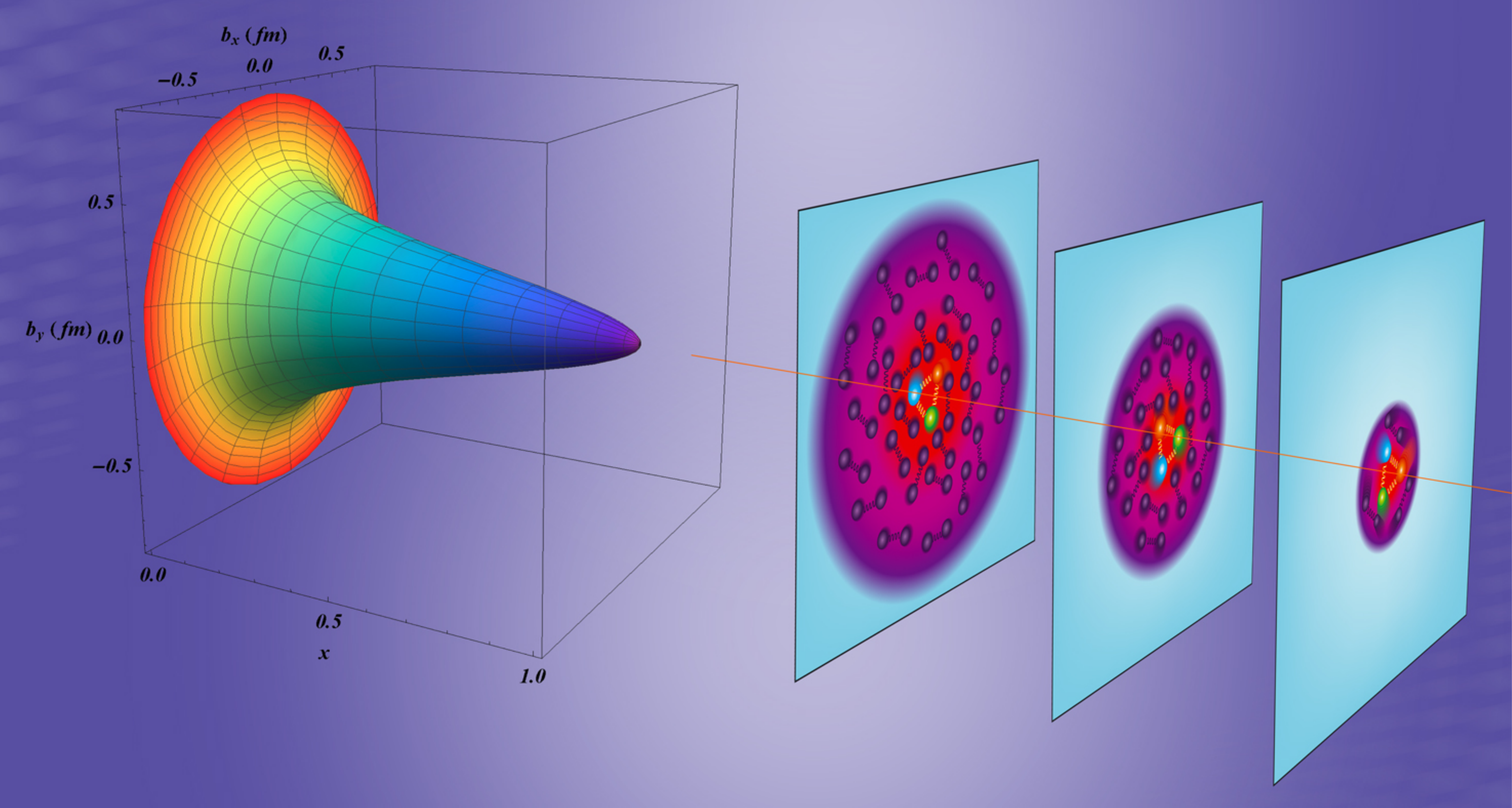}
 \caption{Left: 3-dimensional representation of the $x$-dependence of the proton transverse charge radius. Right: artistic illustration of the corresponding rising quark density and transverse extent as a function of $x$.}\label{fig:coldQCD_Image}
\end{figure}

\subsubsubsection{3D tomography from GPD measurements} 
From the analysis of the DVCS data from HERA and HERMES at DESY, as well as the results of new dedicated experiments at JLab, and at COMPASS at CERN, the experimental constraints on the CFFs and the associated GPDs have been obtained from global extraction fits~\cite{Kumericki:2016ehc,Moutarde:2019tqa}. These data have also been used to generate some of the first 3D images of the proton~\cite{Dupre:2017hfs}, shown in Fig.~\ref{fig:coldQCD_Image}.  However, data covering a sufficiently large kinematic range, and the many different polarization observables, have not been systematically available. Moreover, meson 
production at JLab 6 GeV has not yet shown parton dominance in scattering. The 12 GeV program at JLab will provide comprehensive information on these hard diffractive processes, entering the precision era for GPD studies. Extracting a complete set of CFFs independently in fixed kinematics requires a complete set of experiments. 
In addition, one needs to explore processes that will give both $x$ and skewness parameter $\xi$ information, such as Double DVCS (DDVCS) or similar processes. 

The extensive GPD program from the JLab experiments will provide unprecedented kinematic coverage and precision. Among the approved experiments, Hall A proposed a precision measurement of the helicity dependent and helity independent {\it cross sections} for the $ep \rightarrow ep\gamma$ reaction in DVCS kinematics. This is a follow up to the successful Hall A DVCS run at 5.75 GeV~\cite{JeffersonLabHallA:2015dwe}. 
There are two important DVCS experiments in Hall B using CLAS12 at 11 GeV and at lower energies of 6.6 and 8.8 GeV. 
These measurements cover a large kinematic range, allowing a more comprehensive study of GPDs. 
In addition, to perform the flavor separation of GPDs, the proposed 
experiment in Hall B will measure the beam spin asymmetry for incoherent DVCS scattering on the 
deuteron, detecting the recoil neutron. 
Similar measurements will be made with spectator proton detection in the BONUS and ALERT experiments. 
An experiment has been proposed to measure the target single spin asymmetry on transversely polarized protons. 
The asymmetry has particular sensitivity to the GPD $E(x,\xi)$ which is related to the spin flip nucleon matrix element and hence carries important information on the up and down quark OAM. For nuclear targets, the ALERT detector in tandem with CLAS12 will measure DVCS and deeply virtual meson production (DVMP) off deuterium and $^4$He targets to explore nuclear effects on bound nucleon GPDs. 

Two other Compton-like processes, TCS and DDVCS, as well as hard exclusive meson production, are accessible with JLab 12 GeV and have much to offer. As decsribed in Sec.~\ref{sec:coldQCD_progress}, the preliminary result on the TCS has demonstrated a unique perspective to constrain the quark GPDs. Future experiments of TCS and DDVCS in JLab Halls B and C and SoLID in Hall A will continue to play important roles in comprehensive GPD studies. In addition, experimental data from the 11 GeV beam will provide an important test of the DVMP mechanism. Experiments have been proposed for $\pi^0$ and $\eta$ production with CLAS12 
running contemporaneously with the DVCS experiment, together with 8 GeV beam experiments to separate $\sigma_L$ and $\sigma_T$. Measurement of the $\phi$ meson will provide information on the gluon GPDs 
as well as study intrinsic strangeness. 

\begin{figure}[!ht]
\centering
  \begin{minipage}[r]{0.5\textwidth}
 \includegraphics[width=\textwidth]{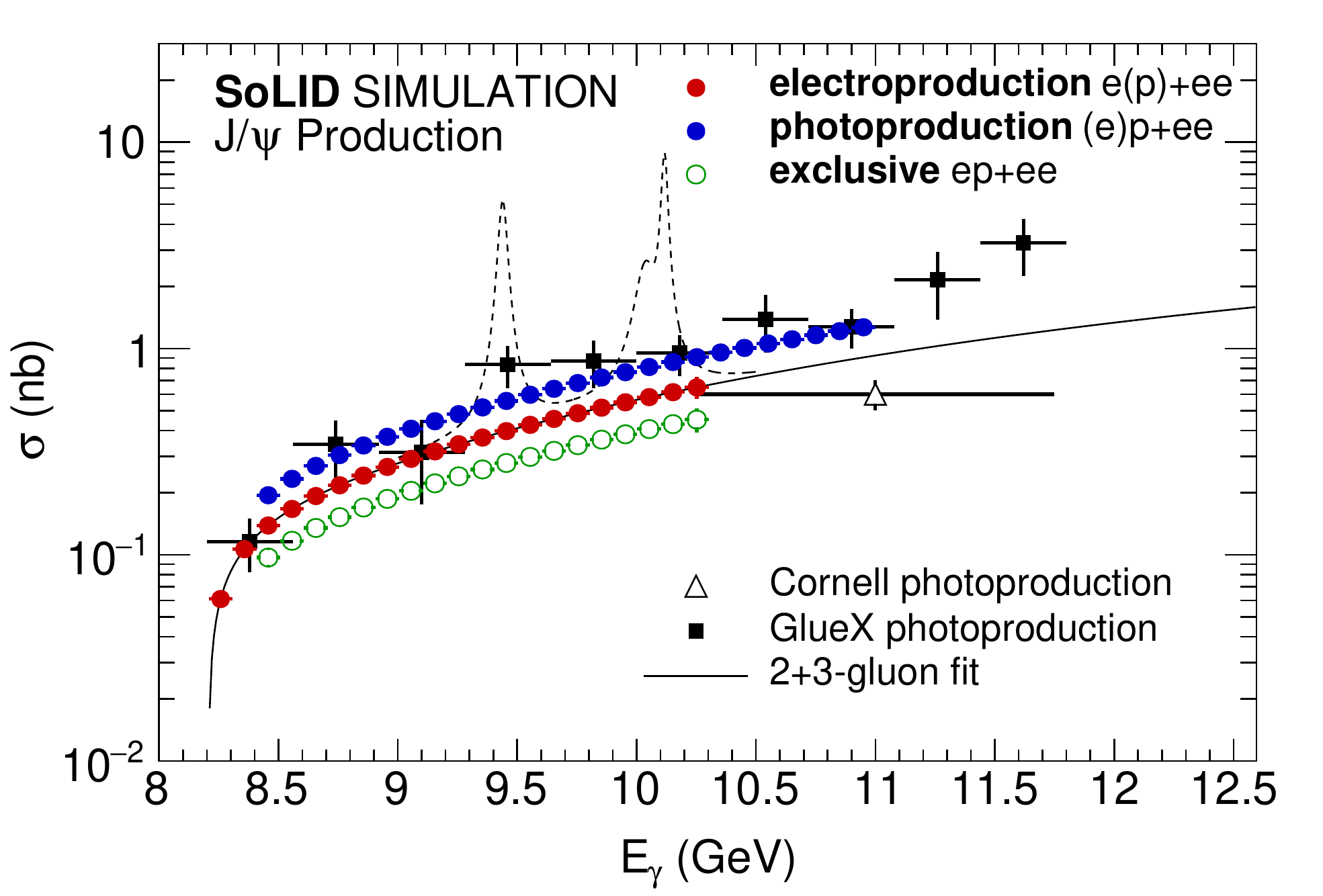}
  \end{minipage}
  \begin{minipage}[c]{0.45\textwidth}
 \caption{Projected 1D $J/\psi$ cross sections as a function of photon energy $E_\gamma$, compared with the available world data. The blue disks show the photoproduction results, while the red disks show the electroproduction results, and the green circles show the results for exclusive electroproduction measurement. Each of the measurements in this figure has a corresponding high-precision measurement of the $t$-dependent differential cross section. Figure from~\cite{JeffersonLabSoLID:2022iod}.}
    \label{fig:ColdQCD_jpsi:1d}
  \end{minipage}
\end{figure}
\vskip 0.3cm 
\noindent
{\bf Threshold $J/\psi$ production and proton mass} Near threshold $J/\psi$ production can provide unique access to the gluon GPD and the form factors of the gluon EMT, providing important information on the mass structure of the nucleon. 
At JLab 12 GeV, the $J/\psi$ can be produced by photon and electron beams on the proton and nuclear targets near threshold. 
Recent experimental results from JLab Halls C and D have been summarized in Sec.~\ref{sec:coldQCD_progress}. There are ongoing experiments in Hall B to measure TCS and $J/\psi$ in photo-production on a hydrogen 
target, with $\sim 10$K events expected after the luminosity upgrade.  Similar statistics are expected for a deuterium target. 
Hall A has an approved experiment using SoLID and can obtain at least another order of magnitude more events ($\sim$800K in photoproduction and $\sim$20K in electroproduction), see Fig.~\ref{fig:ColdQCD_jpsi:1d}. With this large number of  threshold events, one can fit the cross section as a function of $W$ and $t$ to obtain all three gluon EMT form factors, and hence could shed light on the origin of the nucleon mass. 

\begin{figure}[!ht]
\centering
 \includegraphics[width=0.8\textwidth]{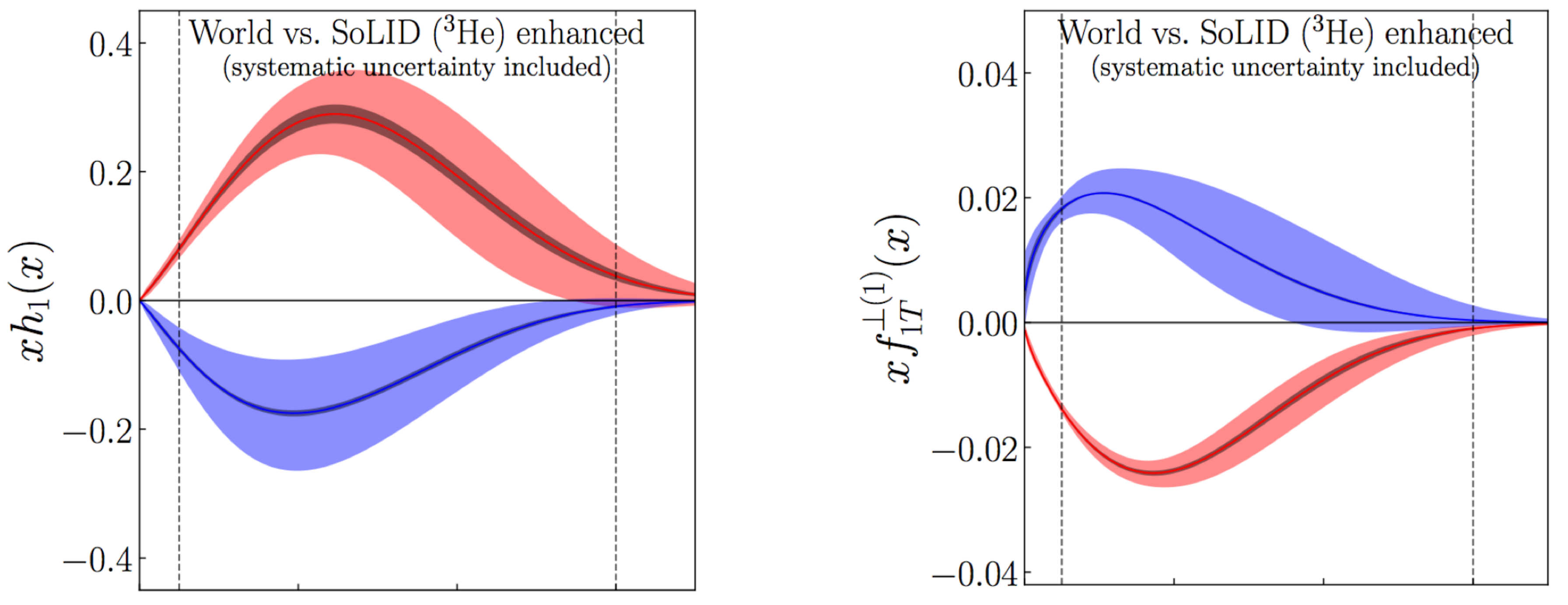}
 \caption{The impact of the SoLID SIDIS program on the $u$ and $d$ quark transversity (left) and Sivers distribution (right).  The wide uncertainty bands show the current results based on a global analysis of world data while the narrower, darker bands show the SoLID projections. Figure from~\cite{JeffersonLabSoLID:2022iod}.
}
    \label{fig:ColdQCD_transversity}
\end{figure}
\vskip 0.3cm
\noindent
{\bf Momentum tomography of the nucleon} One of the most important questions about the 3D structure of the nucleon is the transverse momentum dependence of parton distributions and fragmentation functions. The TMDs, especially those depending on the correlations between the transverse momentum and the polarizations of the partons and hadrons, provide a unique perspective on 3D nucleon tomography. 
At JLab, Halls A, B, and C are all involved in 3D structure studies through measurements of azimuthal modulations in SIDIS for different hadron types, targets, and polarizations over a broad kinematic range. 
The most celebrated SIDIS measurements on TMDs are the surprising non-zero results of the Sivers asymmetries and the Collins asymmetries~\cite{HERMES:2004mhh,COMPASS:2014kcy, JeffersonLabHallA:2011ayy}. These initial explorations established the significance of the SIDIS-TMD experiments and attracted increased efforts in both experimental and theoretical studies of TMDs. The planned SoLID experiments with transversely polarized proton and $^3$He (neutron) targets will provide the most precise measurements of Sivers and Collins asymmetries of charged pion and kaon production in the valance quark (large-$x$) region in 4-dimensions ($x$, $Q^2$, $z$ and $p_T$). 
Figure~\ref{fig:ColdQCD_transversity} shows the projected precision of the extracted transversity $h_{1}(x)$ and Sivers function $f_{1T}^{\perp(1)}(x)$ from the SoLID enhanced configuration for both $u$ and $d$ quarks compared to current results obtained from a global analysis of world data~\cite{Ye:2016prn}. The impact on the nucleon tensor charge from these measurements was  highlighted in Sec.~\ref{sec:cold_future_solid}.

\subsubsection{Meson Structure} 

Several experiments at JLab and planned for the EIC will validate the framework of meson femtography, deepening the understanding of pions and kaons through studies of their form factors, structure functions and masses~\cite{Aguilar:2019teb,Arrington:2021biu}.  Extracting precise meson form factor data requires $L/T$ separated cross sections and control over systematic uncertainties. Over the last decade, JLab measurements have established confidence in the reliability of deep exclusive meson production for probing internal meson structure~\cite{Carmignotto:2018uqj,JeffersonLabFpi-2:2006ysh,Horn:2007ug,JeffersonLab:2008jve,Horn:2012zza,JeffersonLabFpi:2014ykr,JeffersonLabFpi:2014tbe}. 

\begin{figure}[!ht]
\centering
 \includegraphics[width=0.8\textwidth]{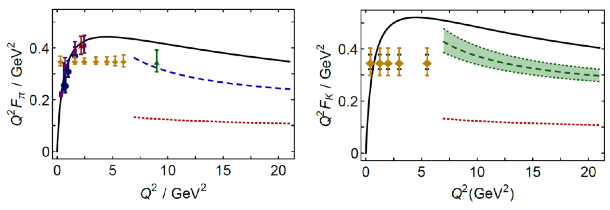}
 \caption{The left (right) panels show calculations, measurements, and projected uncertainties of recent measurements of the pion (kaon) form factors.} 
    \label{fig:ColdQCD_meson}
 \end{figure}

The Super High Momentum Spectrometer (SHMS) in JLab Hall C is a unique facility for making precision 12 GeV measurements~\cite{Benesch:2022xmb}, able to access meson form factors at high momentum transfer and small spatial resolution. 
Two experiments, studying exclusive pion and kaon electroproduction respectively, made precision separations of the $L$ and $T$ cross sections over the last three years.
The projected uncertainties for these experiments are shown in Fig.~\ref{fig:ColdQCD_meson}. In addition, the quark and gluon distributions are expected to differ substantially in pions, kaons and nucleons. 
Planned measurements at JLab using the TDIS technique will provide data that have the potential to settle the issues of quark distributions in the pions at high-$x$ and provide the first data on kaons. 

\subsubsection{Hadron Spectroscopy}

The energies and quantum numbers of the excited state of any physical system provide important clues to the underlying dynamics and relevant degrees of freedom. This is especially true in the case of hadrons, where the spectrum of meson and baryon excitations established the quark model and QCD~\cite{Godfrey:1985xj,Capstick:1986ter} and continues to provide unique information on strong interaction dynamics~\cite{Proceedings:2020fyd,Aznauryan:2011qj}. 
The experimental results from the JLab 6 GeV program on the nucleon resonance electroexcitation amplitudes provided unique information on the structure of the excited states of the nucleon and their evolution with photon virtuality $Q$~\cite{Proceedings:2020fyd,Carman:2020qmb,Mokeev:2022xfo}. 
Extension of these efforts towards high $Q^2$ at JLab 12 GeV 
will explore the transition from the strongly coupled to perturbative QCD regimes is anticipated.

However, the full picture has not been experimentally verified. It is generally argued that the best discovery path is through searching for so-called ``exotic'' meson states, which have quantum numbers that cannot be obtained with only quark--antiquark degrees of freedom. 
QCD and the quark model also predict a number of baryon excitations that have yet to be observed experimentally. A new program at JLab will focus on mapping the spectrum of baryons with strangeness. Excited states in this sector should be less numerous and more narrow than for the nonstrange baryons, which will ameliorate the difficulties associated with overlapping resonances.
There have been a number of narrow charmonium states discovered in recent years~\cite{Guo:2017jvc}, which defy description in terms of the quark model. Their existence points to dynamics of multiquark states that should in principle be predicted by QCD.

The experimental program at JLab is aggressively pursuing the current spectroscopic understanding of QCD dynamics. This includes photoproduction of meson and baryon states in GlueX and CLAS12. 
It also includes new advances in lattice QCD to clarify hadron spectroscopy, in concert with experimental measurements, and to quantify the photoproduction cross sections of hadronic excited states, see Sec.~\ref{sec:theory_lattice}.

\subsubsection{QCD Studies of Nuclei}\label{sec:cold_QCD_future_nuclei}
It has been nearly four decades since the European Muon Collaboration (EMC) published an astonishing finding on how the nucleon PDFs are strongly modified in 
iron nuclei~\cite{EuropeanMuon:1983wih}. 
Although some recent studies suggest a connection to SRCs in nuclei, a full understanding of this phenomenon is still desired.
Indeed, there are several ways in which QCD manifests itself in complex nuclei. CEBAF has contributed to this area of study, see Sec.~\ref{sec:coldQCD_nuclear}, and will continue to provide new experimental data. 

Electron scattering gives access to a range of unique aspects of nuclear structure, providing important data relevant to nuclear interactions at short distances, modifications of nucleon substructure in the nuclear medium, 
and quark/hadron interactions in cold QCD matter. Precision measurements of nuclear elastic, quasielastic, and inelastic scattering, especially those associated with the high-momentum part of the nucleon distributions, provide critical nuclear structure information needed in a range of other areas of nuclear and high-energy physics. Such data are needed as inputs to analyses of neutrino-nucleus scattering, nuclear astrophysics, lepton-nucleus scattering, and heavy-ion collisions, as well as providing important constraints on models of neutron stars. 
Studies of the partonic structure of nuclei provide insights into the impact of the dense nuclear medium on the structure of protons and neutrons and will image the nuclear gluon distribution for the first time. 
In addition, measurements at higher energy will study hadron formation over a wide range of kinematics, as well as quark and hadron interactions with cold, dense nuclear matter, including color transparency, attempting to isolate interactions of small-sized ``pre-hadronic" quark configurations.

Key future programs include measurements which probe nuclear structure at extremes of nucleon momentum, studies of the impact of the dense nuclear environment on the structure of the nucleon, and finally the use of the nucleus to study the interaction of quarks and how they form hadrons in cold nuclear matter. Among these, flavor dependence in the EMC effect should manifest in a number of experiments, e.g., by contrasting structure function measurements in $^{40}$Ca and $^{48}$Ca. A novel method to measure the isovector EMC effect is via PVDIS to obtain the $\gamma$-$Z$ interference structure function $F_2^{\gamma Z}(x)$ and contrast this with the usual DIS structure function to separate the $u$ and $d$ quark PDFs in the same nuclear target. A proposal to perform this experiment on $^{48}$Ca using SoLID was approved for the JLab 12\,GeV program. Interesting opportunities also exist in the comparison of SIDIS on $^3$H and $^3$He with either $\pi^+$ or $\pi^-$ detected in the final state. In addition, as discussed in Sec.~\ref{sec:coldQCD_nuclear}, the spin structure function EMC effect will provide complementary information on the EMC physcis and a first measurement of the polarized EMC ratio in $^7$Li is planned to run at JLab using the CLAS12 spectrometer. 

Last, a growing program of spectator tagging measurements has been recently developed, 
accessing both free and bound nucleon structure, the latter by mapping out the impacts of the nuclear medium and strong nuclear interactions.  First measurements were recently made by the BONUS and BAND experiments, probing the free neutron and deeply-bound proton respectively.  
The Hall B ALERT and Hall C Large Angle Detector (LAD) experiments are approved to extend these studies as part of the JLab 12\,GeV program and anticipated to continue at the EIC using its far-forward fragment spectrometer.

\subsubsection{Cold QCD Program at RHIC}
As the realization of the EIC draws nearer, there is a growing scientific imperative to complete a set of ``must-do” measurements in $p+p$ and $p+A$ collisions in the remaining RHIC runs. The ongoing RHIC cold QCD program of both STAR and the new sPHENIX will build on RHIC’s unique ability to collide a variety of ion beams in addition to polarized protons~\cite{RHIC-Cold-QCD}. 
The STAR forward upgrades, including forward tracking capabilities, will make charged hadron identification and full jet reconstruction possible for the first time in the forward direction.
The sPHENIX detector is optimized for full jet econstruction at mid-rapidity and heavy-flavor measurements~\cite{sPHENIXBUP}.

The new detectors will allow RHIC to extend the full complement of the existing transverse spin program into new kinematic regimes of both lower and higher $x$ domains. This includes measurements of forward single spin asymmetries $A_N$ for charged hadrons $h^{+/-}$, isolated $\pi^0$ and full jet.
The high statistical precision of recently collected data 
at 510 GeV (Run 17 and Run 22) and at 200 GeV at the upcoming Run 24 will enable detailed multi-dimensional binning for the Collins asymmetry measurements for $h^{+/-}$ and $\pi^0$. 
STAR will extend the Collins effect measurements to nuclei and investigate the universality of the Collins effect in hadron production and the spin dependence of the hadronization process in cold nuclear matter. Moreover, the recently collected and future STAR data will further reduce the uncertainties on the single-spin asymmetry of dijet opening angle sensitive to the Sivers TMD parton distribution. 
This will provide a detailed mapping vs $x$ for comparison to results for Sivers functions extracted from SIDIS, Drell-Yan, and vector boson production.

In addition, RHIC will further explore exciting new signatures of gluon saturation and non-linear gluon dynamics. The ratios of forward Drell-Yan and photon-jet yields in $p+p$ and $p+A$/$A+A$ collisions are clean probes of nuclear modifications to initial state parton distributions as well as gluon saturation effects. All of these measurements rely critically on the successful completion of scheduled RHIC operations. Overall, all data will provide valuable information about evolution effects and, with the projected statistical precision, will establish the most precise benchmark for future comparisons with the $ep$ data from the EIC.

\subsubsection{Cold QCD Program at LHC}

The LHC experiments have significantly impacted our understanding of the PDFs in the nucleon and nucleus from various hard scattering processes, including high energy jet and electroweak boson 
production in $p+p$ and $p+A$ collisions, see recent global analyses of the proton and nuclear PDFs~\cite{Hou:2019efy,Bailey:2020ooq,NNPDF:2021njg,Cocuzza:2021rfn,Duwentaster:2021ioo, Eskola:2021nhw, AbdulKhalek:2022fyi,Helenius:2021tof}. These impacts will likely continue with ongoing experimental programs at the LHC with luminosity upgrades. In addition, future measurements at the LHC will impact cold QCD physics in several different ways. Here we highlight two examples.

The unique fixed-target SMOG program at LHCb~\cite{LHCb:2014vhh} took $p+$He, $p+$Ne, $p+$Ar, and Pb+Ar data at $\sqrt{s_{NN}} = $ 69--110~GeV. These measurements~\cite{LHCb:2018jry,LHCb:2018ygc,LHCb:2022lnf,LHCb:2022tum} can constrain nPDFs over a range of nuclei, and in a different kinematic region than that accessible to other experiments, and provide insights into the charmonium production mechanism. The recently installed SMOG II upgrade will allow orders of magnitude higher luminosities and a wider range of possible targets~\cite{Loizides:2020tey}. The LHCSpin project~\cite{Aidala:2019pit,Santimaria:2022tss} would replace SMOG II with a transversely polarized target during the LHC Long Shutdown 3 and start taking data in 2029.  The project has the support of the LHCb Collaboration and the LHC machine, and R\&D is ongoing at LHC Interaction Region 3.  A polarized gas target cell similar to the HERMES polarized target at HERA as well as an alternative jet target are under consideration.  The project would provide singly polarized proton-proton collisions at $\sqrt{s}\approx 115$~GeV, and $p^\uparrow + A$ collisions with a nuclear beam would also be possible.

The ALICE Collaboration at the LHC is proposing an upgrade for LHC Run 4 (2029-32) of a very forward calorimeter, called FoCal, to study the small-$x$ gluon dynamics of hadrons and nuclei~\cite{Bylinkin:2022temp}. The FoCal consists of a highly-granular Si+W electromagnetic calorimeter followed by a conventional sampling hadronic calorimeter, covering the pseudo-rapidity interval $3.4<\eta<5.5$ over the full azimuth. The FoCal design is optimized for the measurement of isolated photons at forward rapidity for $p_T>2$ GeV/c. The FoCal will measure a suite of theoretically well-motivated observables in $p+p$ and $p+$Pb collisions that probe the gluon distribution at small-$x$ (down to approximately 10$^{-6}$) and low to moderate $Q^2$. These observables include isolated photon, neutral meson, and jet inclusive production and correlations in hadronic collisions, and the measurement of vector meson photoproduction in ultra-peripheral collisions. The FoCal scientific program will explore gluon dynamics and non-linear QCD evolution at the lowest values of Bjorken-$x$ that are accessible at any current or planned facility world-wide. FoCal measurements, combined with the comprehensive experimental program at the EIC and other forward measurements at RHIC and the LHC, will enable incisive tests of the universality of linear and non-linear QCD evolution in hadronic matter over an unprecedented kinematic range.

\subsubsection{CEBAF Upgrade Initiatives for Cold QCD}
With the physics program at CEBAF for the next decade outlined as above, one could envision the possibility of CEBAF continuing to operate with a fixed target program at the ``luminosity frontier'', up to $10^{39}~{\rm cm}^{-2}{\rm s}^{-1}$ and with large acceptance detection systems, presenting complementary capabilities in the era of EIC operations. 
One such example is the 3D imaging of the nucleon structure through DDVCS. With a cross section a factor 100 lower than DVCS, DDVCS is not viable at EIC and must be studied at JLab. 
Possible additions and upgrades to the CEBAF facility will further enhance such complementarity. 
In the following, we present two initiatives that were discussed in the QCD Town Meeting. 

\subsubsubsection{Future opportunities with positron beams}\label{sec:cold_future_positron}
Development in many hadronic physics topics calls for additional tools to probe the nucleon structure. In this aspect, the addition of an anti-lepton beam will greatly expand our arsenal of experimental probes and provide data and information that are otherwise unattainable with a lepton beam alone. 
Most prominently, experimental results on the proton form factors and a full mapping of the generalized parton distributions of the nucleon pointed towards the importance of positron beams for the experimental determination of these fundamental quantities of the nucleon structure. Further ideas emerged about testing the electroweak Standard Model and exploring the dark matter sector. A comprehensive research effort was then started both in the physics and the technical areas to assess the potential of an experimental program and to address the possible technological issues of high duty cycle positron beams~\cite{Arrington:2021alx}, as described below. 

\subsubsubsubsection{Two-Photon Exchange Physics with Positrons} 
Our interpretation of data on the proton electromagnetic form factor ratio $G_E^p/G_M^p$ is still clouded by the lack of understanding of TPE effects. 
Investigations of TPE with other observables have produced new questions: First, the GEp-2$\gamma$ experiment at JLab looked for evidence of TPE in the $\epsilon$-dependence of polarization transfer. While no dependence was found in the form factor ratio, an unexplainably large and $\epsilon$-dependent enhancement was found in the longitudinal polarization component~\cite{Puckett:2017flj}. Second, while observables such as SSA (see Section~\ref{sec:cold_future_jlab}) provide information on the imaginary part of TPE, they do not directly address the proton form factor discrepancy. 
A highly desired and probably the most efficient way to study TPE towards a better understanding of the nucleon structure is yet to be provided by high-precision measurement of the lepton-charge difference in $e+p$ elastic scattering. 
Such experiments have been carried out at other facilities than CEBAF, but the beam 
and detectors (VEPP-2, OLYMPUS) suffered from uncorrelated systematic uncertainties in the relative $e^+/e^-$ intensity calibration, either due to lower beam energies or smaller acceptances. 
To this end, the addition of a positron beam to CEBAF, along with its unique large-acceptance detectors already available or under development, will measure the lepton charge difference for all $\epsilon$ points at once and at high $Q^2$ where TPE is expected to be large, and will likely provide an unambiguous explanation of the proton form factor discrepancy. 

\subsubsubsubsection{Nucleon Femtography with Positrons} 
An exciting scientific frontier is the 3D  exploration of nucleon (and nuclear) structure, i.e, nucleon femtography. 
The cleanest reaction to access GPDs is DVCS: $\gamma^*p\to\gamma p$. However, DVCS interferes with the Bethe-Heitler (BH) process, where the lepton scatters elastically off the nucleon and emits a high energy photon before or after the interaction. 
The cleanest way to separate the DVCS and BH amplitudes is to compare electron and positron scattering, as the BH amplitude is lepton-charge even while the DVCS amplitude is lepton-charge odd. 
This method will isolate not only the DVCS amplitude contribution to the cross section, but also the interference term between DVCS and BH amplitudes, with the latter providing direct linear access to DVCS at the amplitude level. 
In short, the use of both positron and electron beams provides direct access to the nucleon structure carried in the DVCS amplitude and indisputable access to the square of the DVCS amplitude, representing a true qualitative shift in the 3D imaging of nucleons and nuclei. 

\begin{figure}[!h]
    \centering
    \includegraphics[width=\linewidth]{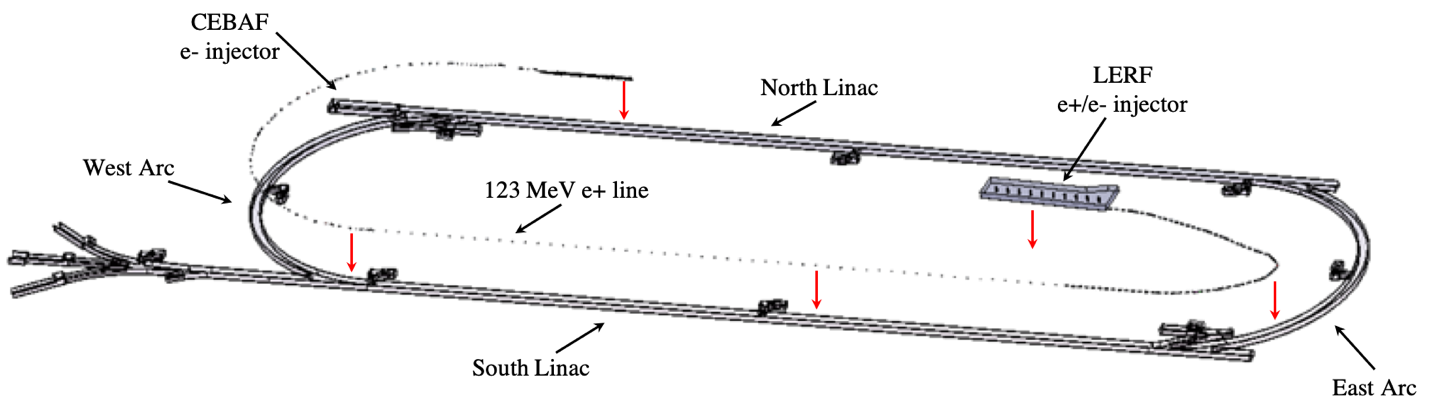}
    \caption{A new tunnel and beam line (shown raised) connects the LERF to CEBAF and transports the 123~MeV $e^+$ beam for injection and acceleration into CEBAF 12~GeV.} 
    \label{fig:fig2}
\end{figure}

\subsubsubsubsection{Positron beams at JLab} 
The challenging creation of positrons with a high degree of polarization at JLab relies on its unique source of polarized $\gamma$ rays produced by Bremsstrahlung radiation from CEBAF's polarized electron beams~\cite{PEPPo:2016saj}. 
Within this framework, a polarized-electron driven positron injector is currently being designed and evaluated~\cite{Habet:2022fch}.
The polarizied electron source needed for such approach is in the range of $>1$~mA. While not routine, such capacity has been demonstrated~\cite{Talman:2018voj} and is assumed.
Additionally, the approach now focuses on utilizing the Low Energy Research Facility (LERF, formerly known as the Free-Electron Laser or FEL) as the site for the new positron beam source. The LERF includes significant existing facilities (cryogenics, low conductivity water, shielding, electronics bays, radio-frequency penetrations, control room) and can 
provide up to 3 superconducting radio-frequency (SRF) cryomodules to support the $e^-$ drive beam and $e^+$ acceleration. 
The selected positron bunch train passes a momentum selection chicane prior to entering a SRF cryomodule for acceleration up to 123~MeV and a bunch compression chicane to match CEBAF acceptance.
Once the 123~MeV $e^+$ beam is produced, it is then a matter of transporting it to CEBAF for acceleration. This can be achieved by a new connector tunnel from the LERF exit to the lower elevation of the CEBAF enclosure, which will allow the 123~MeV $e^+$ beam to be injected at the usual point in front of the north linac for multi-pass acceleration and beam extraction to any of the four Halls at any of the passes, see Fig.~\ref{fig:fig2}. Additionally, the intention is for all of the CEBAF electro-magnets to have a capability for polarity reversal on the scale of a day, for experiments which required both $e^+$ and $e^-$ pair-created beams from the LERF source.
Given the promise of this approach, JLab is continuing expanded follow-on studies of a future positron beam source and its acceleration for CEBAF.

\subsubsubsection{Future opportunties with energy doubling of CEBAF}
Recently, the Cornell Brookhaven Electron Test Accelerator (CBETA) facility has demonstrated eight-pass recirculation of an electron beam with energy recovery~\cite{Bartnik:2020pos}. 
All eight beams -- four accelerating passes and four decelerating passes -- are recirculated by single arcs of fixed field alternating gradient (FFA) magnets. This exciting new technology carries the potential to enable a cost-effective method to double the energy of CEBAF, allowing wider kinematic reach for nucleon femtography studies. 
Furthermore, it will enable new scientific opportunities that include: 
(1) first-time production of various $X$ and $Z$ states in photon(lepton)-hadron collisions; (2) precision studies of near-threshold production of higher mass charmonium states $\chi_c$ and $\psi'$; 
(3) precision measurements of the radiative decay width and transition form factor of $\pi^0$ off an electron for the first time, offering a stringent test of low-energy QCD. 
Ongoing, further investigations and simulations will strengthen the science opportunities introduced here. 
Meanwhile, technical studies of the implementation of FFA technology at CEBAF are in progress and are described in more detail below. 

\begin{figure}[h]
\centering
\includegraphics*[width=0.4\textwidth]{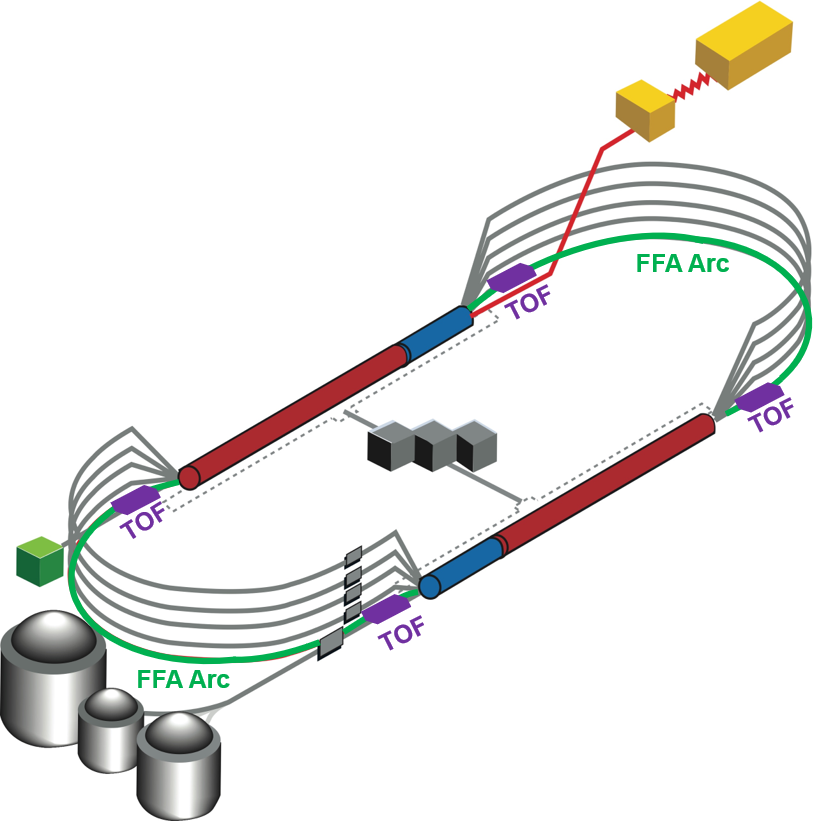}
\caption{Sketch of the CEBAF accelerator with the two highest energy arcs, Arc 9 and Arc A, replaced with a pair of FFA arcs (green). Figure from \cite{Arrington:2021alx}.}
\label{CEBAF}
\end{figure}

\subsubsubsubsection{CEBAF Energy ``Doubling'' -- Accelerator Concept} 
The recent success of the CBETA project demonstrated the possibility to extend the energy reach of CEBAF up to 22 GeV within the existing tunnel footprint. 
Such an increase can be achieved by increasing the number of recirculations through the accelerating cavities, and by replacing the highest-energy arcs with FFA arcs, see Fig.~\ref{CEBAF}. 
The new pair of arcs configured with an FFA lattice would support simultaneous transport of 6 beam passes with energies spanning a factor of two, each beam pass with very small transverse orbit offsets due to the small dispersion function.

Transporting high energy beams (10-22 GeV) while staying within the CEBAF footprint calls for special mitigation of synchrotron radiation effects. One of them is to increase the bend radius at the arc dipoles to suppress adverse effects of the synchrotron radiation on beam quality, including dilution of the transverse and longitudinal emittance due to quantum excitations. Further recirculation beyond 22 GeV is limited by large energy loss due to synchrotron radiation, which depends on energy to the fourth power.
Therefore, using FFA to double CEBAF energy will finally be pushing its energy to a limit set by its footprint. 

\subsubsubsection{Connection to accelerator physics}
Both positron beams and energy doubling will contribute to accelerator physics development, and connect to possible future needs of the EIC and other high energy physics (HEP) and NP facilities which will rely on beam recirculation (e.g. LHeC) or FFA technology. Developments of these two CEBAF upgrades will help to maintain and enhance US leadership in accelerator science and technology.

\subsection{Future QCD Studies at Other Facilities}\label{sec:otherfacility}

\subsubsubsection{Drell-Yan at Fermilab fixed target}
The SpinQuest experiment will investigate whether the sea quarks are orbiting around the center of the nucleon by measuring the Sivers asymmetry with the use of a solid-state target of polarized protons and deuterons.  This measurement provides information on the correlation between the angular distribution of the dimuons in the Drell-Yan process and the nucleon spin at high {$x_B$} with a virtuality of $Q^2\sim10$ GeV$^2$. The observation of a nonzero Sivers Asymmetry would be a strong indication of non-zero sea quark orbital angular momentum. The SpinQuest experiment can also probe the sea quark transversity distribution. Additionally, a proposal has been submitted to upgrade SpinQuest with a specialized RF-modulated target that can be used to separate tensor from vector polarized observables of the deuteron, providing access to additional sea quark and gluon transversity TMDs. The gluon transversity TMD only exists for targets of spin greater or equal to 1 and does not mix with quark distributions at leading twist, thereby providing a particularly clean probe of gluonic degrees of freedom.

\subsubsubsection{HIGS}
The High Intensity Gamma-Ray Source (HIGS), operated by the Triangle Universities Nuclear Laboratory, is capable of providing nearly mono-energetic, polarized gamma-ray beams with energies ranging from 1 to 120 MeV. 
HIGS is the highest flux Compton gamma-ray source ever built and operated. The gamma-ray beam flux delivered to experiments at 100 MeV is approximately $1\times 10^{7} \gamma/s$.
The Compton-scattering program at HIGS, which is carried out by research groups from 13 institutions, is mapping out the energy dependence of the dynamical scalar electromagnetic polarizabilities of the neutron over photon beam energies from 60 to 120 MeV and extending proton measurements from about 100 to 120 MeV. Free-Electron Laser cavity mirror R\&D is underway to increase the maximum gamma-ray beam energy at HIGS from 120 to 150 MeV to enable measurements up to the pion production threshold where the electric dipole polarizability of the nucleon is largest. The work by the HIGS Compton Scattering Collaboration will illuminate differences in the neutron and proton scalar polarizabilities, providing stringent tests of chiral effective theories and a new prediction of the electromagnetic contribution of the proton-neutron mass difference~\cite{Gasser:2020hzn}. 
A program to develop cryogenic polarized target capability at HIGS will enable measurements of spin-dependent nucleon polarizabilities at photon beam energies below the photo-pion production threshold. The Compton-scattering data from HIGS are complementary in both energy and technique to the data measured by the A2 Collaboration at MAMI.

\subsubsubsection{MUSE}
The Muon Scattering Experiment (MUSE) at the PiM1 beam line of the Paul Scherrer Institute (PSI) measures scattering of a mixed beam of electrons, muons, and pions from a liquid hydrogen cryotarget~\cite{MUSE:2017dod,Cline:2021vlw}. The experiment was initially motivated by the proton radius puzzle, and PSI was chosen for its unique capability to provide simultaneous low energy electron and muon beams. MUSE features a large-solid-angle, non-magnetic detector and cryogenic target system, and will test lepton universality through the comparison of cross sections, form factors, and proton radii extracted from electron and muon scattering. Beams of both positive and negative polarity leptons will determine two-photon exchange corrections, testing predictions. A forward-angle calorimeter tests initial-state radiative corrections. The background pions in the beam allow determination of $\pi$-$N$ cross sections of interest to low-energy effective field theories used in the strong QCD regime. The simultaneous measurement of electron and muon scattering, and the measurement of both charge states in the same experiment suppresses many systematic uncertainties. MUSE will be the first experiment to provide elastic muon scattering of sufficient precision to address the puzzle. 

\subsubsubsection{MESA} 
The University of Mainz is currently constructing a new electron accelerator (MESA). 
In 2024, an electron beam is expected to be generated with MESA for the first time. 
It will offer ideal conditions in which scientists will be able to explore the limits of Standard Model physics. 
Several key experiments are currently under development. 
Among them, MAGIX is a multi-purpose spectrometer which allows the precise measurement of proton form factors at the lowest impulse transfer rates. This will contribute decisively to the clarification of the existing contradictions in the experimental determination of the proton radius (the so-called proton radius puzzle) and to dark matter searches. 

\vskip 0.3cm
\noindent
{\bf ULQ2} is a new high-resolution spectrometer facility for low-energy electron scattering at the Research Center for Electron Photon Science (ELPH) at Tohoku University in Sendai, Japan. The facility has been constructed and commissioned from 2017-2022 and will provide precision 
measurements of the proton elastic form factors at very low momentum transfer, the proton charge, magnetic, and Zemach radii, and of low-energy nuclear structure.

\subsubsubsection{J-PARC}
\noindent 
The Japan Proton Accelerator Research Complex, \href{https://www.j-parc.jp}{J-PARC}, is Japan’s leading accelerator facility, which has cascaded proton accelerators including the 400-MeV linear accelerator, the 3-GeV rapid cycling synchrotron (RCS) and the main ring operated at 30 GeV.  There are experimental facilities such as the Materials and Life Science Facility at RCS, and the Neutrino Facility and the Hadron Experimental Facility both at the main ring. In addition to applied physics research, there are two major basic research activities:
(1) {\bf Neutrino Facility} 
Neutrino as well as anti-neutrino beams produced at J-PARC are sent to the Super-Kamiokande located about 295 km west of J-PARC. The research topics at the Neutrino Facility include QCD-related physics such as neutrino-nucleus interactions.
(2) The {\bf Hadron Experimental Facility} is a unique experimental complex which utilizes the secondary beams to perform precision measurements on hyper-nuclear spectroscopy, hyperon-nucleon scattering, and kaonic nuclei, to name a few. Major physics interests of these programs are hadron interactions, including  hyperon-nucleon interactions, hyperon-hyperon interactions, and kaon-nucleon interactions. Upgrades to this experiment, which could measure proton generalized parton distributions and pion distribution amplitudes, are also being discussed.

\subsubsubsection{FAIR}
\href{https://www.gsi.de/en/researchaccelerators/fair}{FAIR} is a European flagship facility~\cite{NuPECC,NuPECCLRP2017}. This worldwide unique accelerator and experimental facility will conduct unprecedented forefront research in physics and applied sciences on both a microscopic and a cosmic scale. 
While the center of mass energies of heavy-ion beams ($\sqrt{s_{NN}}=2.9-4.9$ GeV) are designed for the CBM experiment, the $1.5 - 15$ GeV/$c$ momentum beam of anti-protons will be generated and collected in the high energy storage ring before being sent to the experiment PANDA. There are three major experiments in FAIR designed for fundamental research: 
(1)~The {\bf NUSTAR} experiment is designed together with the Super-FRS and storage cooler rings, and will allow discovery measurements in nuclear structure and nuclear astrophysics;
(2)~The {\bf CBM} experiment is a high-energy nuclear collision experiment with high rate capabilities for determining the location of the QCD critical point, the first-order phase boundary, the equation of state of nuclear matter at high baryon density and the hypernuclear interactions pertinent to the inner structure of compact stars; 
(3)~The {\bf PANDA} experiment, designed at the antiproton storage cooler ring HESR, will provide a unique research environment for an extensive program in hadron spectroscopy, hadron structure and hadronic interactions. In particular, the studies of hadron structure in the relatively large$-x$ region complement the structure measurement at small-$x$ at the EIC in the coming decades.

\subsubsubsection{BELLE II}
High luminosity $e^+e^-$ experiments always played an important and complementary role in the study of QCD, alongside nucleon scattering experiments. While in the latter a spacelike gauge boson is exchanged and the nucleon is used as a QCD laboratory, at $e^+e^-$ machines, the complementary timelike process can be used to study quarks traversing the vacuum and their subsequent hadronization with a precision that cannot be reached in hadronic scattering experiments. 
Belle II~\cite{Belle-II:2018jsg} is taking data at SuperKEKB, a second generation $B$-factory delivering world record luminosities. Over the next decade, Belle II plans to collect 50 ab$^{-1}$ integrated luminosity, about a factor 50 more than Belle. 
The large Belle II dataset will enable the precise determination of complex correlations in the hadronization process, which are necessary for a detailed mapping of the QCD dynamics at play. 
Therefore, support for a robust QCD program at Belle II is essential to make progress in our description of hadronization and precision tests of QCD in hadronization at the pace necessary to analyze data from current and future SIDIS and hadronic scattering programs at JLab, the EIC, and the LHC. 
Recently, the community formulated a broad and widely supported program which is documented in a recent whitepaper~\cite{Accardi:2022oog}. Relevant topics include the spin-orbit correlation in hadronization, measurement of polarized and unpolarized fragmentation functions, hadronization effects in jets, precision tests of perturbative QCD calculations in jet and event shape measurements, and constraining the value of $\alpha_s$.
A focus of the program is on the modeling of hadronization in (polarized) Monte Carlo event generators. Belle II data in conjunction with LEP data will provide the necessary information to test the energy dependence of these models needed for the EIC or the LHC.

\subsubsubsection{AMBER} 
In the context of the physics-beyond-colliders  (PBC) initiative at CERN, the COMPASS++/AMBER (proto-) collaboration proposes to establish a ``New QCD facility at the M2 beam line of the CERN SPS"~\cite{Adams:2018pwt}. 
The proposed measurements cover a wide range of $Q^2$: from lowest values of $Q^2$, where it is planned to measure the proton charge radius by elastic muon-proton scattering, to intermediate $Q^2$ to study the spectroscopy of mesons and baryons by using dedicated meson beams, to high $Q^2$ to study the structure of mesons and baryons via the Drell-Yan process. The whole project is intended to run over the next 10 to 15 years. 
{\bf AMBER} will play a crucial role as it can uniquely provide pion (kaon) Drell-Yan measurements in the center-of-mass energy region $10-20$ GeV. 
These measurements are essential for a global effort aimed at pion structure function measurements (also providing a handle on determination of the so-called ``pion flux" for EIC Sullivan process measurements) and kaon structure function data map.

\section{Electron-Ion Collider}
\label{sec:future_eic}

The scientific foundation for the EIC has been built for over two decades. Throughout, the EIC initiative was driven by maintaining U.S. leadership in both nuclear science and accelerator physics. These dual goals were clear from the outset, starting with the 2002 NSAC LRP~\cite{Symons:2002} where \textit{``R$\&$D over the next three years to address EIC design issues"} was placed at high priority. Support from the community continued with the 2007 LRP~\cite{Tribble:2007}, which recommended \textit{``the allocation of resources to develop accelerator and detector technology necessary to lay the foundation for a polarized Electron-Ion Collider"} and culminated in the 2015 plan, where the EIC was identified as the \textit{``highest priority for new facility construction following the completion of FRIB"}~\cite{Geesaman:2015fha}. 

\begin{figure}[htb]
\begin{center}
\vspace*{2mm}
\includegraphics[width=\textwidth]{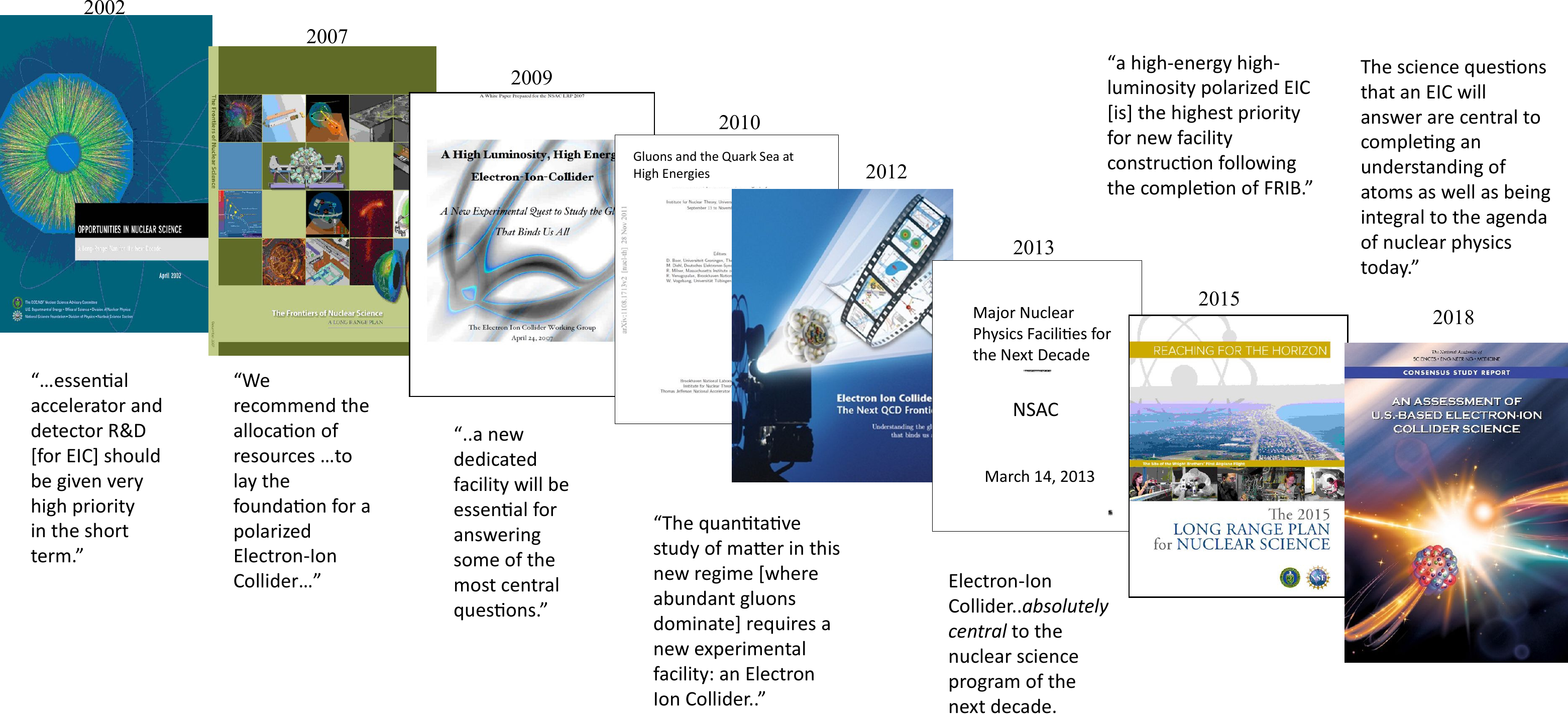}
\caption{A chronological display of the publications that document the development of the EIC science case. From left to right: The 2002~\cite{Symons:2002} and 2007~\cite{Tribble:2007} LRPs, a 2009 report of the EIC Working Group~\cite{EIC-WP-2007LRP}, a report on the joint 2010 BNL/INT/JLab program on EIC~\cite{Boer:2011fh}, the 2012 EIC White Paper~\cite{Accardi:2012qut}, the 2013 NSAC Subcommitte Report on Scientific Facilities, the 2015 LRP~\cite{Geesaman:2015fha}, and the NAS report~\cite{NAP25171}. Figure from~\cite{EIC-WP-2023LRP}. }
\label{EICfoundation}
\end{center}
\end{figure}

During this period the science case underpinning these recommendations was continually developed and documented by the growing EIC community, as illustrated in \fig{EICfoundation}. A series of workshops hosted by the Institute for Nuclear Theory laid the foundation for a White Paper titled "Understanding the glue that binds us all"~\cite{Accardi:2012qut}. The studies developed for the EIC White Paper~\cite{Accardi:2012qut}, combined with continued progress in accelerator $R\&D$, served as input to a critical review in 2018 by the NAS. Their final report, An Assessment of the U.S. Based Electron-Ion Collider Science, concluded that \textit{``the EIC science is compelling, fundamental, and timely."}~\cite{NAP25171}. 

Below we will summarize the flagship components of the EIC science case, consisting of understanding the origin of the proton spin and mass, proton tomography, gluon saturation, cold nuclear phenomena, and fundamental symmetries. We will also briefly discuss the EIC project detector that will be built at the 6 o'clock interaction region by the ~\epic collaboration, along with the plans for a second, complementary detector to be constructed at the 8 o'clock region.

\subsection{The EIC Science}
\label{sec:eic_science}
Decades of scattering experiments and their theoretical interpretation have produced an intriguing picture of the proton and neutron. These particles are held inside the atomic nucleus by the strong force, the same force that generates the dynamic landscape of quarks and gluons that form the substructure of the nucleon. Some quantum numbers of the nucleon, like its electric charge, are easily reproduced by simply summing the properties of the three valence quarks. Yet, the quarks contribute only a third of the total nucleon spin and an even smaller fraction of the total mass. Clearly, many of the fundamental properties of the nucleon must emerge from the gluons, the carriers of the strong force that confine the quarks inside the nucleon, and from the copious $q\bar{q}$ pairs that form the quark sea. Our interest goes beyond reconstructing the fundamental properties of the parent nucleon: our ultimate goal is understanding the dynamics of the dense partonic environment found in nucleons and nuclei. The EIC will be an amazingly versatile machine that will allow experiments to map out the spatial and momentum distributions for quarks and gluons, study how the gluon density evolves with the resolution of the electron probe and with the momentum fraction $x$ carried by the interacting gluon, and observe how transitions from partonic to hadronic degrees of freedom are modified in increasingly dense nuclear matter. These key science questions (and more!) can be summarized by the following lines of inquiry: 
\begin{itemize}
  \setlength\itemsep{-0.2em}
\item How do the nucleonic properties such as mass and spin emerge from partons and their underlying interactions?
\item How are partons inside the nucleon distributed in both momentum and position space?
\item What happens to the gluon density in nucleons and nuclei at small $x$? Does it saturate at high energy, 
giving rise to gluonic matter with universal properties in all nuclei (and perhaps even in nucleons)?  
\item How do color-charged quarks and gluons, and jets, interact with a nuclear medium? How do  confined hadronic states emerge from these quarks and gluons? How do the quark-gluon interactions generate nuclear binding?
\item{Do signals from beyond-the-standard-model physics manifest in electron-proton/ion collisions? If so, what can we learn about the nature of these new particles and forces?}
\end{itemize}

\begin{figure}[htb]
\centering
\begin{minipage}[c]{0.44\textwidth}
\includegraphics[width=\textwidth]{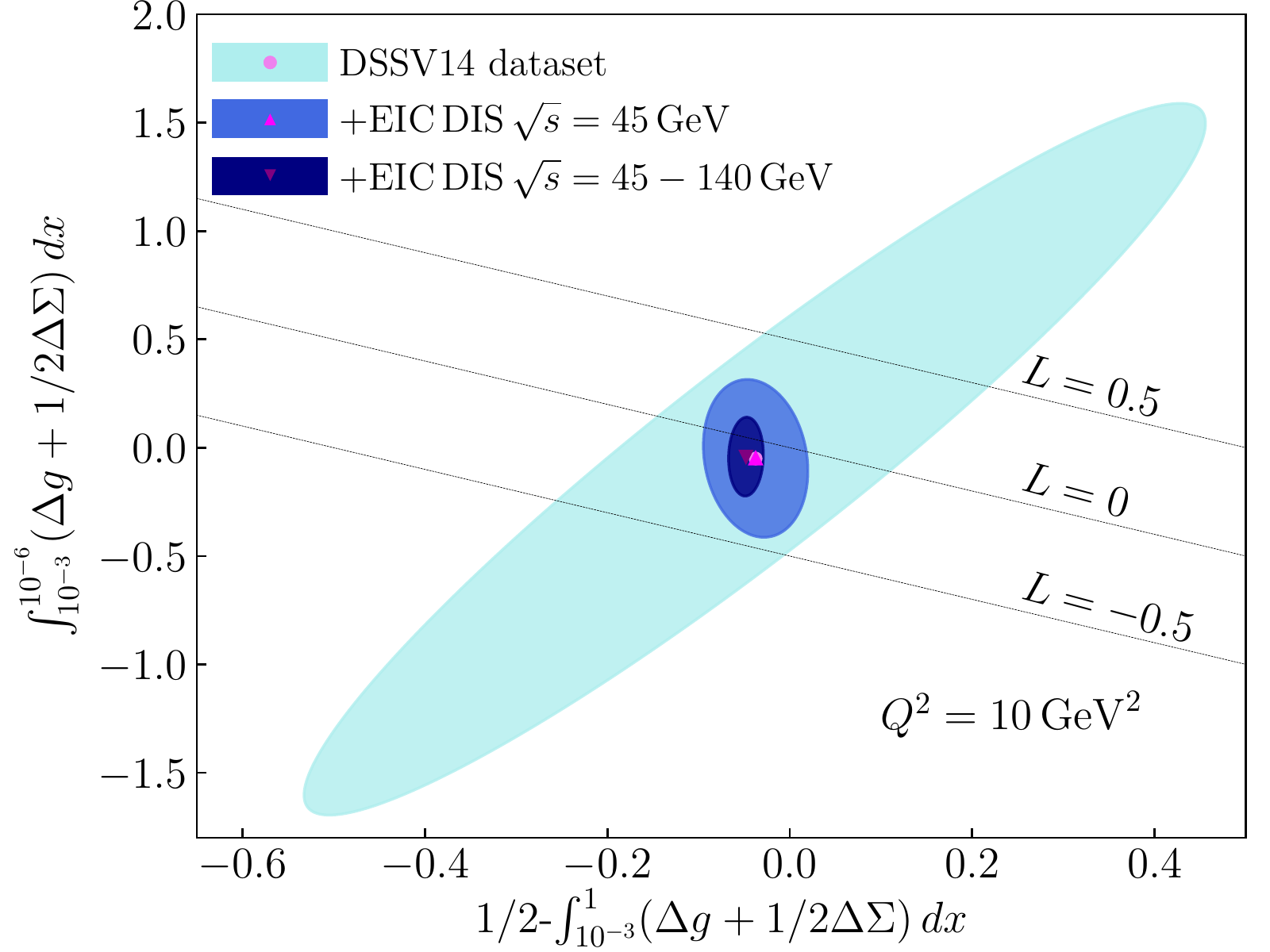}
 \end{minipage}
\begin{minipage}[c]{0.54\textwidth}
       \includegraphics[width=\textwidth]{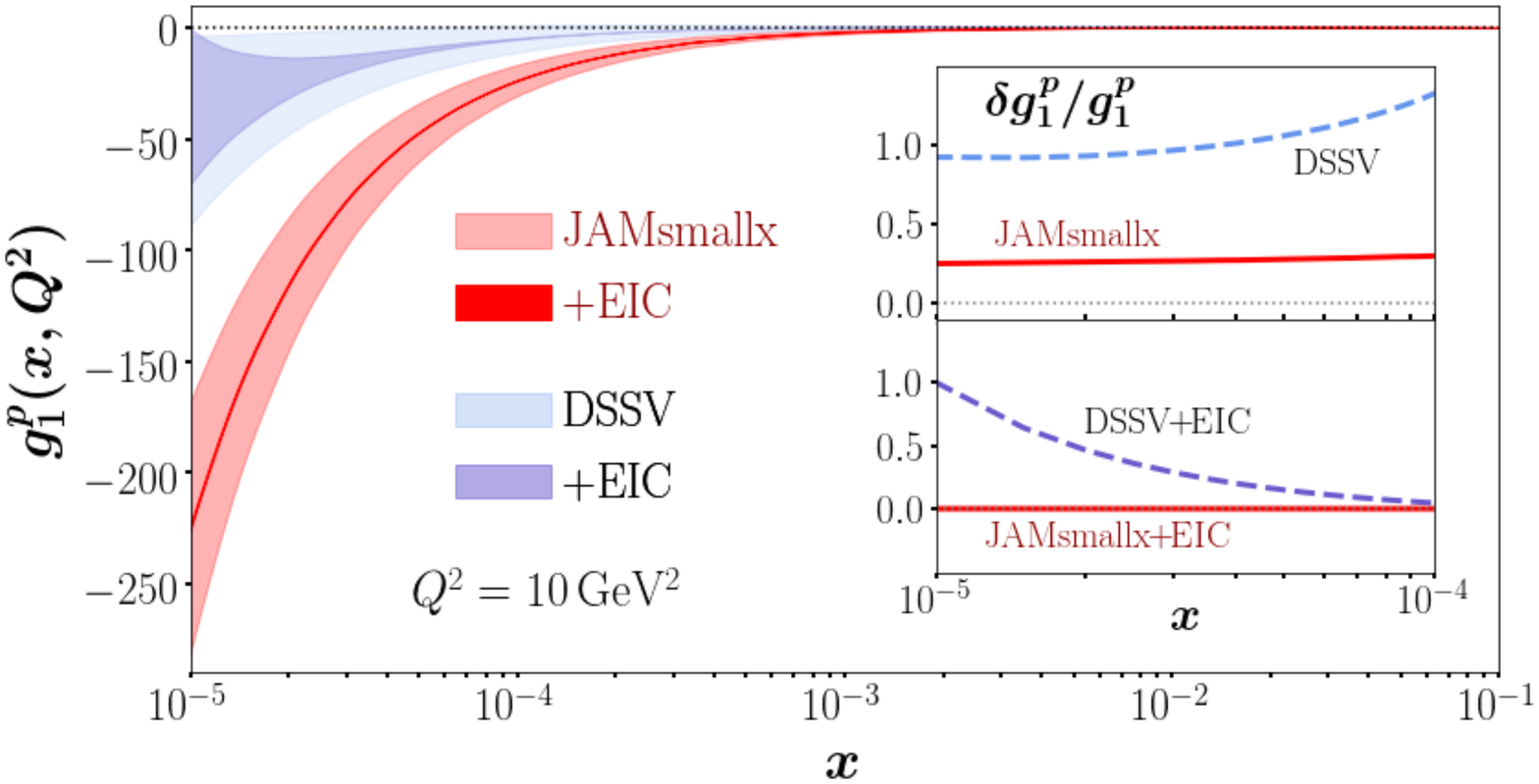}
       \end{minipage}
    \caption{Left: Impact of the projected EIC  pseudodata on the spin decomposition of the proton based on the most recent version of the DSSV14 parametrization~\cite{Borsa:2020lsz,deFlorian:2014yva, deFlorian:2019zkl}; Right: EIC impact on the $g_1$ structure function based on parameterizations with or without the theory-inspired small-$x$ extrapolation~\cite{Adamiak:2021ppq} (see text).} 
    \label{fig:eic_g1}
\end{figure}

In the following, we will highlight the impacts that the EIC will bring to the above fundamental questions of hadron structure and strong interaction physics.

\vskip 0.3cm
\noindent
{\bf Proton spin} Nucleon spin has played a central role in driving hadron physics for over three decades. The EIC will have unprecedented impact on our understanding of this physics. With the unique coverage in both $x$ and $Q^2$ (for polarized DIS), along with very high luminosity, the EIC will provide the most powerful constraints on the quark and gluon helicity contributions to the proton spin yet. The left panel of \fig{fig:eic_g1} depicts the contributions of the large-$x$ ($x \in [10^{-3}, 1]$) quark and gluon helicities (subtracted out of the proton spin of $1/2$) on the horizontal axis and of the small-$x$ ($x \in [10^{-6}, 10^{-3}]$) quark and gluon helicities on the vertical axis, along with the possible lines corresponding to different values of the OAM ($L$) carried by the partons. The EIC data will significantly shrink the error bars of the quark and gluon helicities. 
The precision of the polarized structure functions (and parton helicity distributions) may be further improved by implementing the theoretical predictions for their behavior at small-$x$~\cite{Bartels:1996wc,Kovchegov:2015pbl,Hatta:2016aoc, Boussarie:2019icw, Chirilli:2021lif, Cougoulic:2022gbk}, as shown in the right panel of Fig.~\ref{fig:eic_g1} which compares the (more conventional) DGLAP-based predictions for the proton $g_1$ structure function \cite{Borsa:2020lsz} (in blue) to those based on the small-$x$ evolution~\cite{Adamiak:2021ppq} (in red). 
Apart from the total quark helicity contribution to the proton spin, the sea quark polarization will be determined to higher precision through semi-inclusive hadron production in DIS. 
In addition, a systematic investigation of various hard exclusive processes will provide information on the partonic orbital angular momentum contributions to the proton spin~\cite{Ji:2016jgn,Bhattacharya:2022vvo}.

\vskip 0.3cm
\begin{figure}[htb]
    \centering
\includegraphics[width=0.9\textwidth]{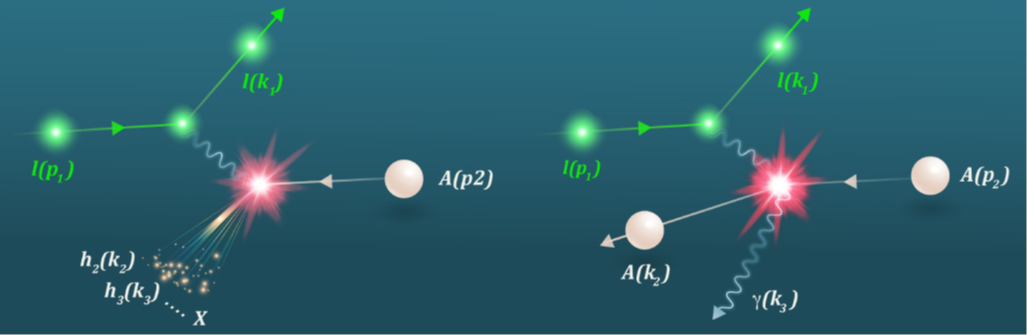}
\vskip 0.2cm
\includegraphics[width=0.9\textwidth]{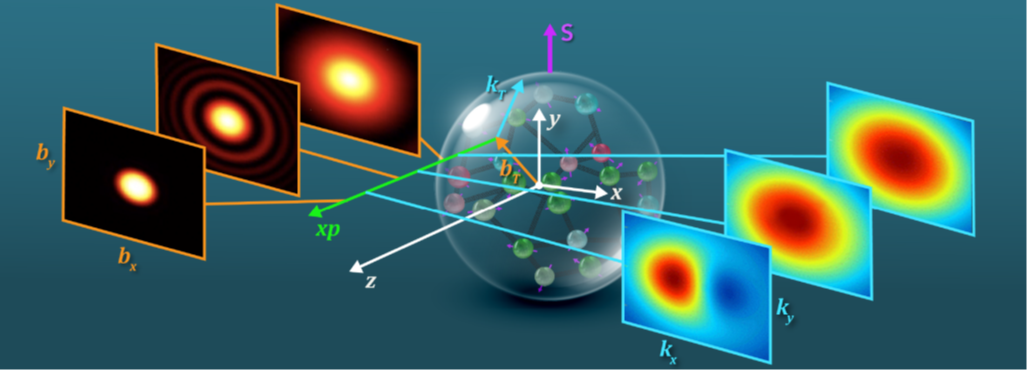}
    \caption{ {\it Upper panel:} Illustration of the two types of processes that occur in lepton-nucleus collisions: a semi-inclusive process where a hadron, hadron pair, jet or dijet is measured and the remnant nucleus is destroyed (left) and an exclusive process where the nucleus remains intact (right). {\it Lower panel:} Tomographic images in slices of $x$ for the quarks and gluons in a nucleus: (transverse) spatial tomography in $\bf{b_T}$-space provided by exclusive processes (left); (transverse) momentum tomography in $\bf{k_T}$-space provided by semi-inclusive processes (right). Figure from~\cite{EIC-WP-2023LRP}.}
    \label{fig:3Dprocesses}
\end{figure}
\noindent
{\bf Nucleon tomography and the origin of mass}
The EIC will significantly extend our knowledge of the distribution of quarks and gluons in nucleons and nuclei, both in position and momentum space. Examples of processes that can provide information beyond the original 1D Feynman parton picture are illustrated in \fig{fig:3Dprocesses}. On one hand, in elastic processes (see the right side of the upper panel in \fig{fig:3Dprocesses}), detecting the full final state of the proton beam provides information about the transverse position of the partons – quarks and gluons – that reside inside nucleons and nuclei. On the other hand, using a related class of inelastic observables gathered from data where the scattered electron is measured in tandem with an electro-produced hadron, or a jet, or a pair of hadrons (see the left side of the upper panel in \fig{fig:3Dprocesses}), the EIC will also measure the transverse motion of partons. 
These measurements will enable parton tomography, a series of 2D images of the nucleon, both in transverse position and momentum space. This is illustrated in the lower panel of \fig{fig:3Dprocesses}, with such snapshots stacked along the Bjorken-$x$ direction. Starting at large $x$, in the domain of valence quarks, and proceeding toward lower $x$, the regime of sea quarks and gluons, these images will reveal where quarks and gluons are located and how their momenta are distributed in the transverse plane. 
The full richness of transverse momentum information is explored when transverse polarization (with the proton spin direction orthogonal to the direction of motion) is added. In this case, orbital motion leads to correlations between spin and transverse momentum, generating an asymmetric transverse momentum distribution, such that the parton tomography provides a series of images of transverse momentum distributions that are fully 2+1-dimensional. 

The tomographic techniques will provide insight into the origin of the proton mass. Studying the processes of elastic $J/\psi$ and $\Upsilon$ production near threshold at the EIC, we will be able to extract the gravitational form factors which shed light on the amount of the proton mass carried by the QCD trace anomaly contribution. The EIC will provide a unique opportunity to better measure the gravitational form factors by providing a lever arm in $Q^2$ for $J/\psi$ or (heavier) $\Upsilon$ elastic production processes.  Understanding the origin of the proton mass is an important and fundamental question, related to our understanding of the origin of mass in the visible universe. 

\vskip 0.3cm

\begin{figure}[!ht]
    \centering
\includegraphics[width=0.45\textwidth]{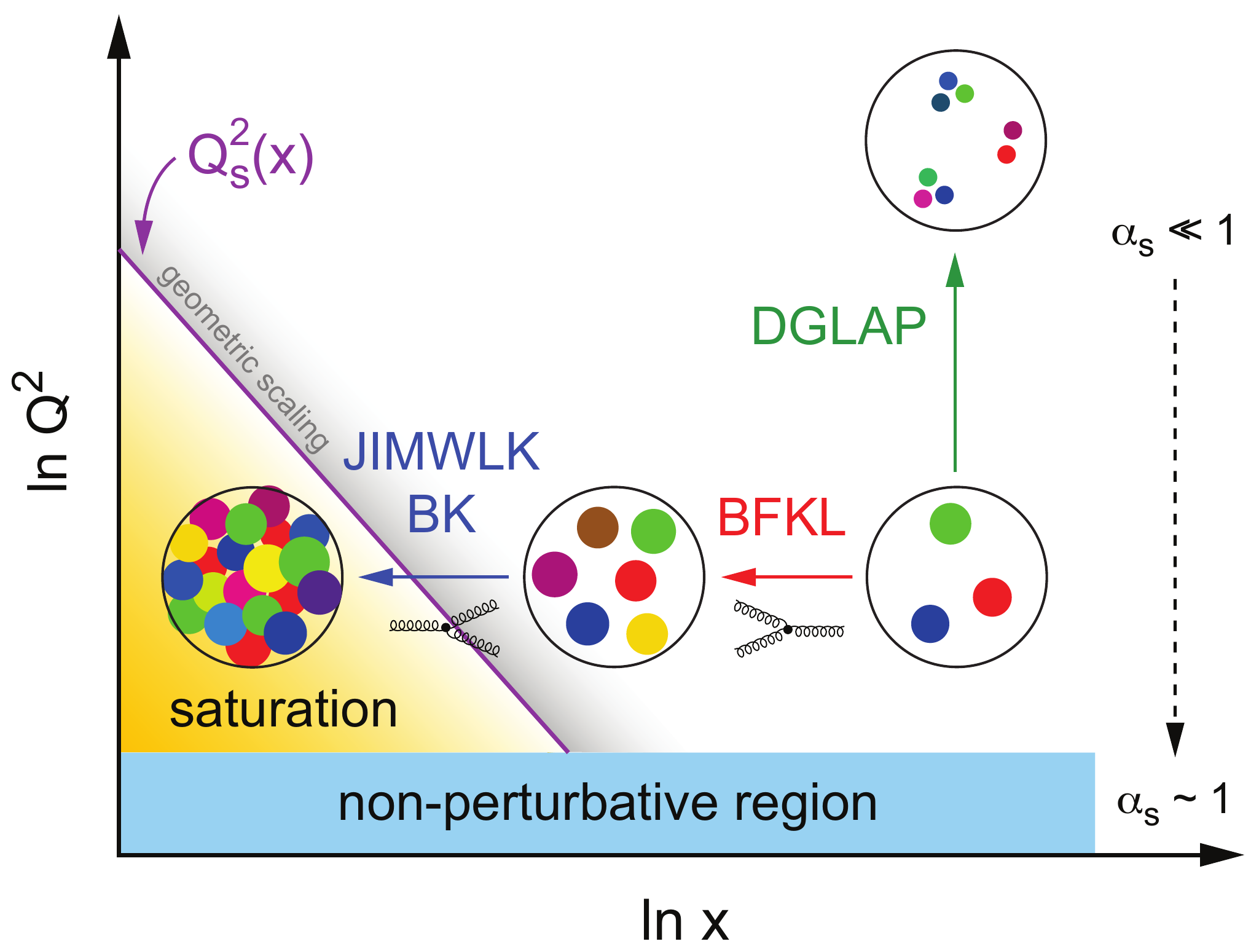}
\includegraphics[width=0.40\textwidth]{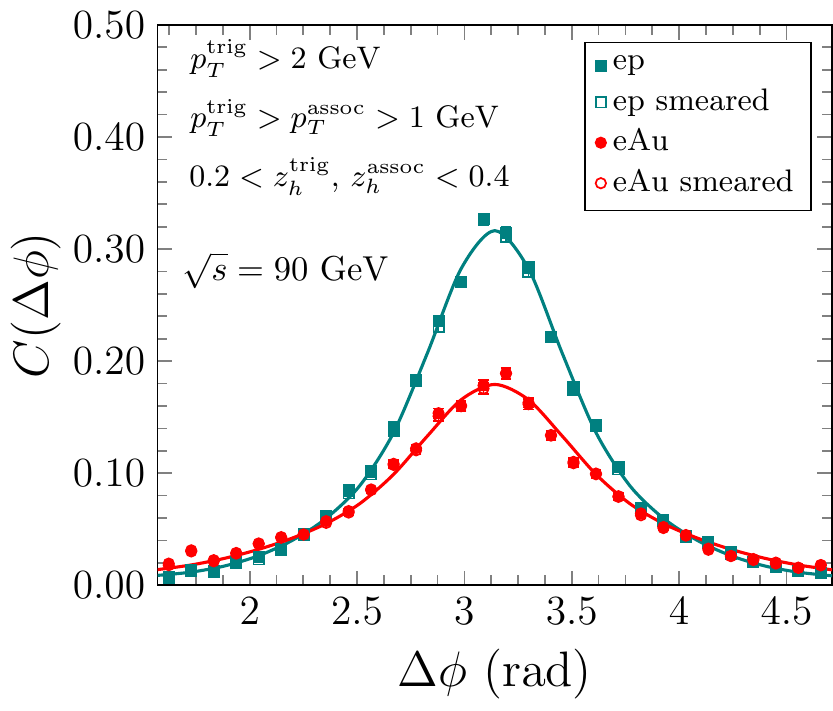}
\caption{Left: Schematic illustration of the
probe resolution, $Q^2$, versus $x$, indicating regions of non-perturbative (band at the bottom) and perturbative QCD (everything above the non-perturbative region), including in the latter, low to high saturated parton density, and the transition region between them \cite{Accardi:2012qut}. 
The saturation region is shown in yellow.
Right: A saturation model prediction of the hadron-hadron correlation function $C (\Delta \phi)$ to be measured in \ep \ and \eA  \ collisions at EIC plotted as a function of the azimuthal angle $\Delta \phi$: the away side peak at $\Delta \phi = \pi$ decreases as one goes from \ep \ to \eA \ due to the increase in the saturation scale with $A$. The ranges of transverse momenta ($p_T$) and longitudinal momentum fractions ($z_h$) of the trigger and associated hadrons are specified on the plot. Figure from~\cite{EIC-WP-2023LRP}.}
\label{fig:eic_gluon}
\end{figure}

\noindent
{\bf Gluon dynamics in a dense medium} 
The gluon and sea quark PDFs in the proton grow rapidly with decreasing Bjorken-$x$. 
The dynamical mechanism responsible for this growth is the
splitting of gluons into pairs of gluons or quark antiquark pairs and the splitting of quarks into quarks and gluons~\cite{Gribov:1972ri,Altarelli:1977zs,Dokshitzer:1977sg,Kuraev:1977fs,Balitsky:1978ic}.  
The large number of gluons and sea quarks at low $x$ confined to the transverse area of the proton results in a high parton density. 
But will high density keep increasing as we probe lower and lower values of $x$? Would the physics change in the high density regime? As was originally conjectured in~\cite{Gribov:1984tu}, the growth of the gluon density should {\sl saturate} at some small value of $x$, leading to the novel regime of {\sl gluon saturation} (see \cite{Iancu:2003xm,Weigert:2005us,Jalilian-Marian:2005ccm,Gelis:2010nm,Albacete:2014fwa,Kovchegov:2012mbw,Morreale:2021pnn} for reviews). The new dynamics in the saturation regime are due to gluon mergers: the mergers compensate for the splittings, leading to saturation of the gluon density. The transition from the splittings-dominated regime to the saturation regime is described by the nonlinear small-$x$ evolution equations~\cite{Balitsky:1995ub,Balitsky:1998ya,Kovchegov:1999yj,Kovchegov:1999ua,Jalilian-Marian:1997jhx,Jalilian-Marian:1997qno,Weigert:2000gi,Iancu:2001ad,Iancu:2000hn,Ferreiro:2001qy}, which are a manifestation of the nonlinear nature of QCD.

A key feature of gluon saturation is the emergence of a momentum scale $Q_{s}$, known as the saturation scale. The scale is predicted by the nonlinear evolution equations  \cite{Balitsky:1995ub,Balitsky:1998ya,Kovchegov:1999yj,Kovchegov:1999ua,Jalilian-Marian:1997jhx,Jalilian-Marian:1997qno,Weigert:2000gi,Iancu:2001ad,Iancu:2000hn,Ferreiro:2001qy} and designates a transition from the low-density regime ($Q>Q_s$) to the high-density saturated regime ($Q<Q_s$) as indicated in the left panel of \fig{fig:eic_gluon}. The saturation scale grows with decreasing $x$, $Q_s^2 \sim 1/x^{0.3}$. 
When this scale significantly exceeds the QCD confinement scale $\Lambda_{\rm QCD}$, the dynamics of strongly correlated gluons can be described by weak coupling QCD methods. The framework that enables such computations is the CGC effective field theory~\cite{Iancu:2003xm,Weigert:2005us,Jalilian-Marian:2005ccm,Gelis:2010nm,Albacete:2014fwa,Kovchegov:2012mbw,Morreale:2021pnn}, see, Sec.~\ref{sec:theory0}. It is expected that the saturation phenomenon grows with the nuclear mass number $A$, $Q_{s}^{2} \propto A^{1/3}$ \cite{Mueller:1989st,McLerran:1993ni,McLerran:1993ka,McLerran:1994vd}; thus, the novel domain of saturated gluon fields can be accessed especially well in large nuclei. 
Unambiguously establishing this novel domain of QCD and its detailed study is one of the most critical goals of the EIC.

Multiple experimental signatures of saturation have been discussed in the literature \cite{Accardi:2012qut}. The EIC program follows a multi-pronged approach taking advantage of the versatility of the EIC facility. Day-one measurements include the proton and nuclear structure functions $F_2$ and $F_L$, which are sensitive to saturation physics. One of the other key signatures concerns the suppression of
di-hadron angular correlations in the process $e+\mathrm{Au} \rightarrow e^{\prime}+h_{1}+h_{2}+X$. The angle between the two hadrons $h_{1}$ and $h_{2}$ in the azimuthal plane, $\Delta \phi$, is sensitive to the transverse momentum of gluons and their self-interaction: the mechanism that leads to saturation. 
The experimental signature of saturation is a progressive suppression of the away-side ($\Delta \phi = \pi$)
correlations of hadrons with increasing atomic number A at a fixed value of $x$, as demonstrated in the right panel of Fig.~\ref{fig:eic_gluon}. Diffraction and diffractive particle production in \eA\ scattering is another promising avenue to establish the existence of saturation and to study 
the underlying dynamics. 
Diffraction entails the exchange of a color-neutral object between the virtual photon and the proton remnant. As a consequence, there
is a rapidity gap between the scattered target and the diffractively produced system. At HERA, these types of diffractive events made up a large fraction of the total \ep\ cross section 
(10--15\%).
Saturation models predict that at the EIC, more than $20\%$ of the cross section will be diffractive. In simplified terms (and at leading order), since diffractive cross sections are proportional to the square of the nuclear gluon distribution, $\sigma \propto [g(x, Q^2)]^2$, they are very sensitive
to the onset of non-linear dynamics in QCD.
An early measurement
of coherent diffraction in \eA\ collisions
at the EIC would provide the first unambiguous
evidence of gluon saturation. Further studies at small $x$ that can provide insight into the spatial and momentum distribution of gluons include coherent and incoherent diffractive vector meson production, deeply virtual Compton scattering (and their dependence on azimuthal angle between the produced particle and the electron plane), as well as inclusive and exclusive dijet production \cite{Morreale:2021pnn}. In particular, access to the gluon Wigner distribution is possible using diffractive dijets \cite{Hatta:2016dxp,CMS:2022lbi}.  

\begin{figure}[!ht]
    \centering
\includegraphics[width=0.42\textwidth]{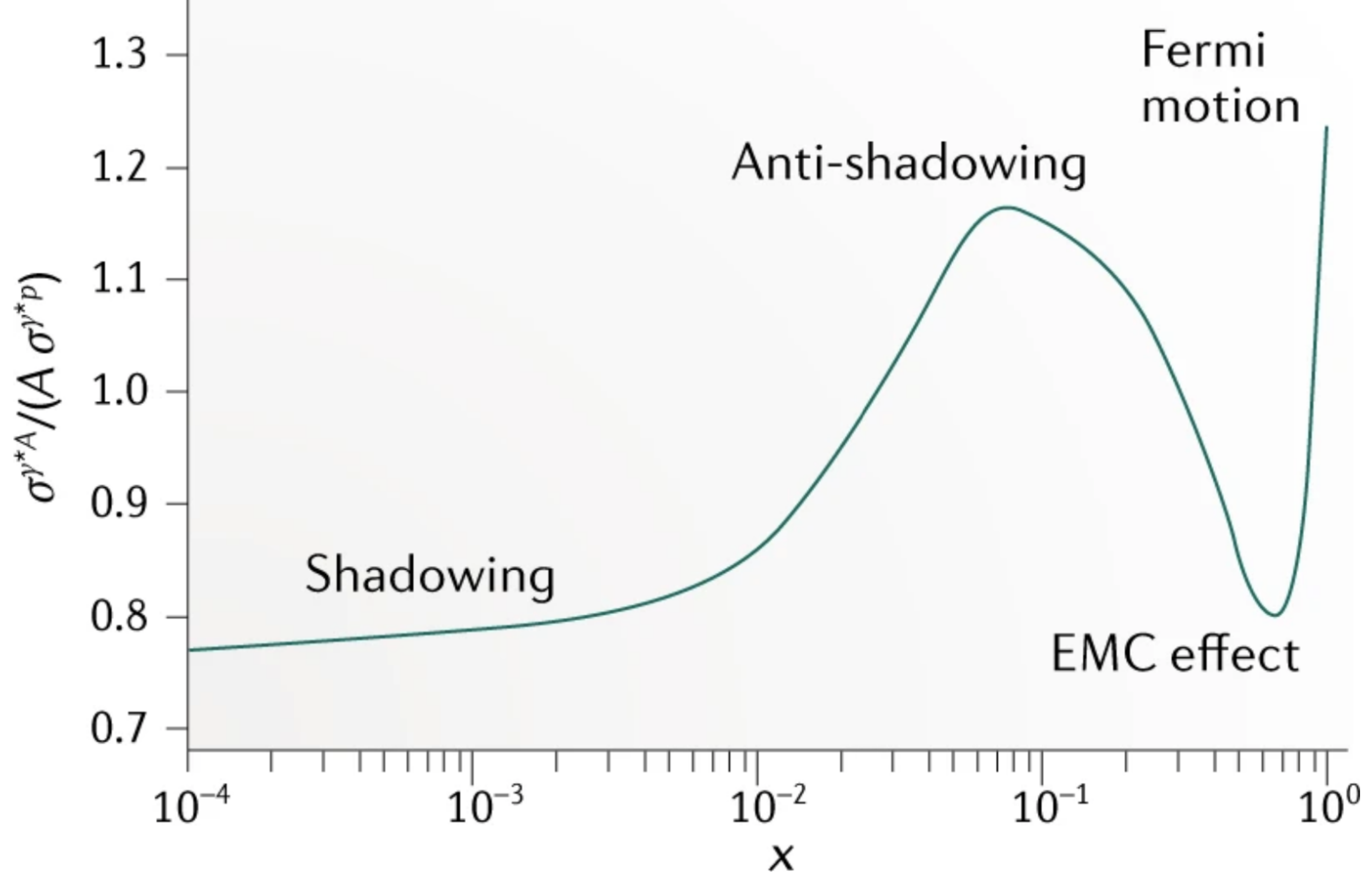}
\includegraphics[width=0.45\textwidth]{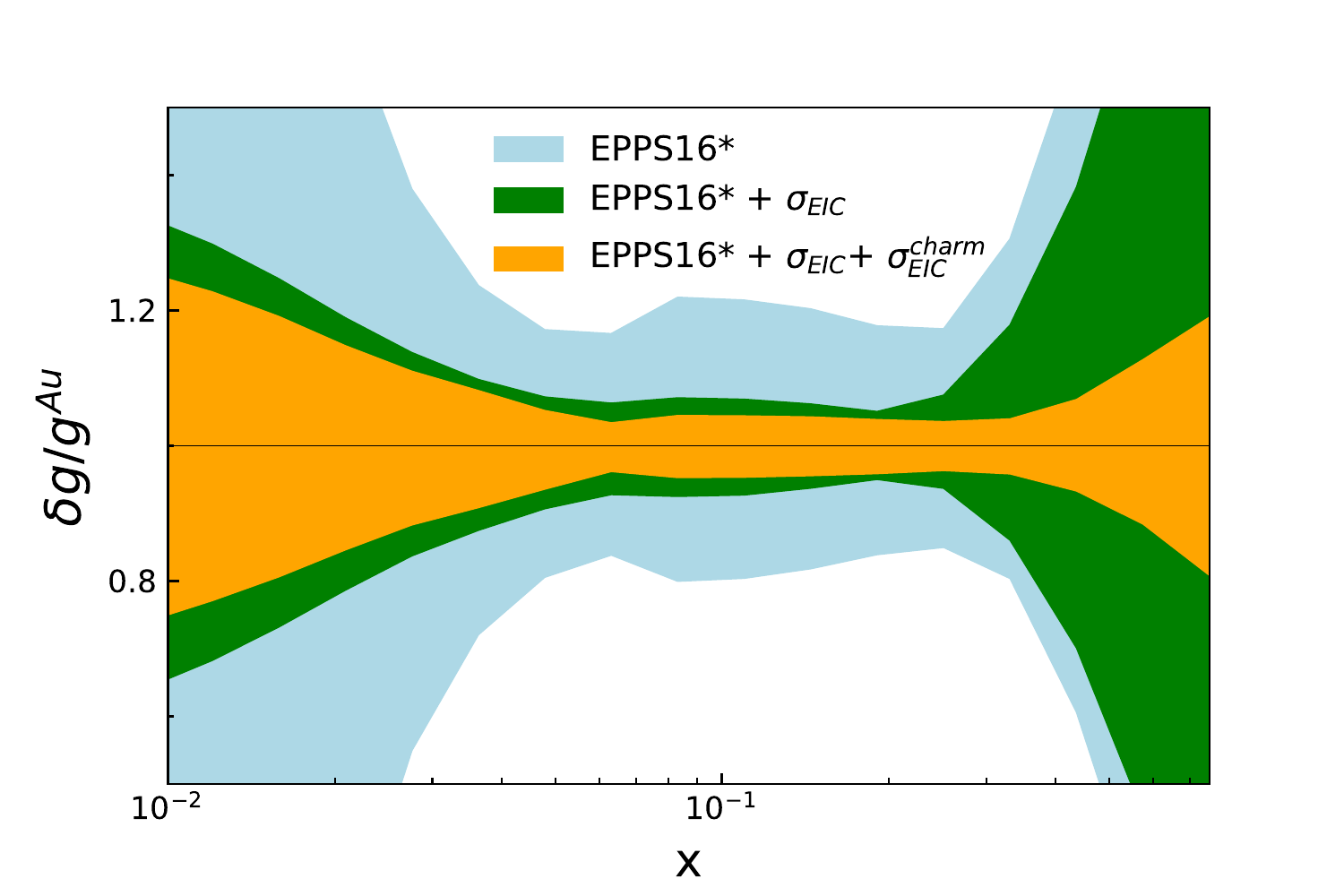}
    \caption{Left: The cross section ratio $\sigma^{\gamma^* A}/(A \, \sigma^{\gamma^* p})$ measures the nuclear modification to the parton distribution functions (figure from \cite{Klein:2019qfb}). Right: Relative uncertainty bands for the gluon distributions in gold nuclei at $Q^{2}=1.69\text{ GeV}^{2}$. The blue band is the original EPPS16* fit, the green band incorporates inclusive 
    cross section pseudodata and the orange band also adds the charm cross section $\sigma^{\rm charm}$ (figure from~\cite{EIC-WP-2023LRP}).}
    \label{fig:nPDF-improve}
\end{figure}

\vskip 0.3cm
\noindent
{\bf Nuclear modifications of parton distributions}
High energy electron-nucleus collisions at the EIC will enable measurements of nuclear modification of the parton distribution functions over a broad and continuous range in $x$ and $Q^{2}$. This will lead to the study of the PDF differences between the bound and unbound nucleons with unprecedented precision. These differences are often quantified via the ratios of the nuclear PDF to the proton PDF divided by the nuclear mass number $A$. Nuclear modifications described by such ratio range from suppression (below unity) in the so-called ``shadowing" domain of small-$x$ to enhancement in the ``anti-shadowing" (moderate-$x$) region and again to suppression in the ``EMC" (large-$x$) regime, as illustrated in the left panel of \fig{fig:nPDF-improve}. For the most part, these modifications are only phenomenologically modeled.

Measurements of nuclear structure functions elucidate to what extent a nucleus could be described by a collection of independent nucleons -- a  fundamental question about nuclear properties in QCD. The effect of the EIC data on our knowledge of the nuclear gluon distribution function is shown in the right panel of Fig.~\ref{fig:nPDF-improve}, where the relative uncertainties clearly shrink as one includes EIC pseudo-data. As can be seen, the EIC provides broad kinematic coverage, mapping the shadowing and anti-shadowing regimes, as well as part of the EMC regime. 

The EIC will also provide novel insights into the physics of SRC in nuclei \cite{Hauenstein:2021zql} and how they relate to the mechanism by which QCD generates the nuclear force~\cite{Tu:2020ymk}, as well as into their possible connections to the nPDF EMC effect. Using far-forward tagging techniques, the EIC will probe the structure of nucleons in varying nuclear states, thereby disentangling the impact of the strong nuclear interaction on the bound nucleon structure. Such 'spectator tagging' techniques can be applied in conjunction with all reactions that are sensitive to nucleon structure, from inclusive DIS, to SIDIS to DVCS/DVMP, and can thus provide unprecedented insight into the impact of strong nuclear interactions and the dense nucleon medium on the structure of bound nucleons. This has been demonstrated in numeric simulations for the deuteron~\cite{Tu:2020ymk,Jentsch:2021qdp} and ${}^3$He~\cite{Friscic:2021oti}. 
This extension of the free-nucleon structure program to bound nucleons via spectator tagging techniques is a novel frontier. It requires performing high-precision measurements over a wide kinematic phase space~\cite{CiofidegliAtti:2010uwl,CiofidegliAtti:2011rm,Strikman:2017koc,Cosyn:2019hem,Cosyn:2020kwu}. 

\vskip 0.3cm
\begin{figure}[tbh]
    \centering
\includegraphics[width=\textwidth]{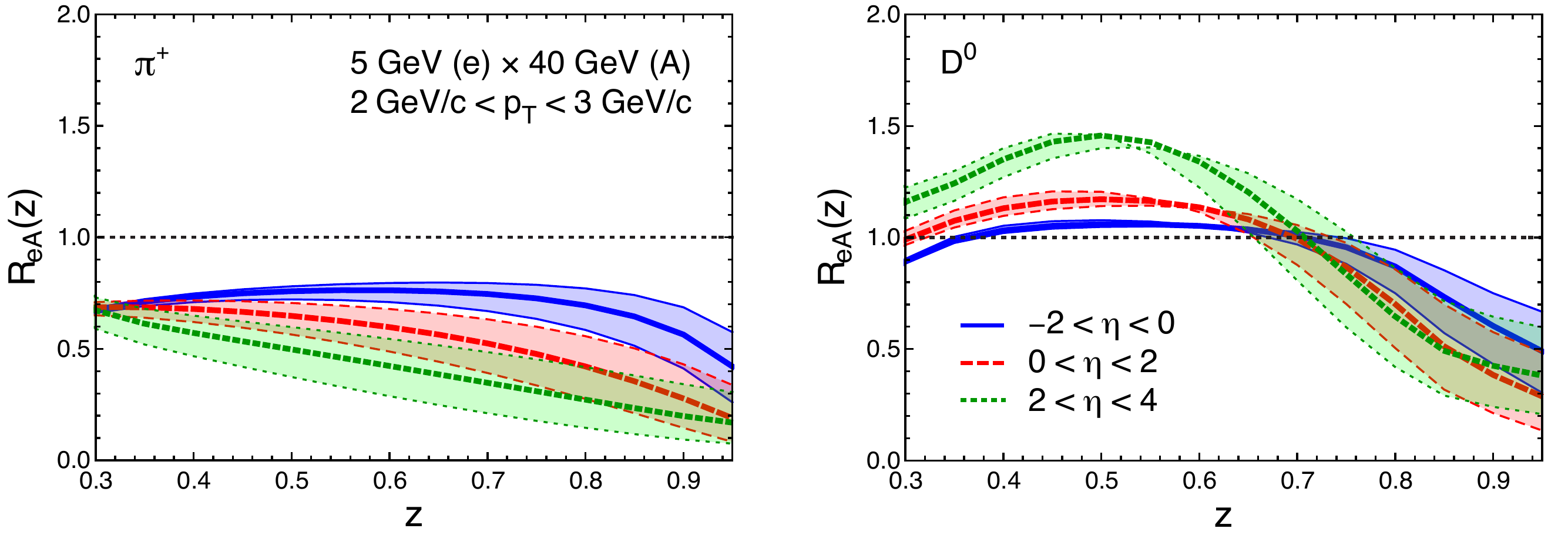}
    \caption{Ratio of relative particle production ($N_h/N_\mathrm{incl}$) in \eA\ over that in \ep\ as a function of $z$, the momentum fraction of the parton carried by the respective hadron.
Light pions (left) show the largest nuclear suppression at the EIC. However, heavy flavor meson ratios (right)  have to be measured to differentiate models of hadronization since they show a substantially different modification in \eA. (Figure is from \cite{Li:2021lbj}.)}
    \label{fig:hadronization_double_ratio}
\end{figure}
\noindent
{\bf Hard probes in cold nuclei} The EIC will make substantial progress in our understanding of hadron \emph{formation}, including inside nuclear matter. Especially, studying hadronization for light and heavy quarks in cold nuclear matter can unravel some of the mysteries surrounding energy loss in a quark-gluon plasma~\cite{Li:2021gjw}. 
At the EIC, the large $Q^2$ range will permit measurements in the perturbative regime with enough leverage to determine nuclear modifications of the fragmentation functions. 
The high luminosity will permit the multidimensional binning necessary for separating the many competing mechanisms. The large photon energy (in the nucleon rest frame), $\nu \approx 10 - 1000$ GeV, will isolate in-medium parton propagation effects (large $\nu$), and to cleanly extract color neutralization and hadron formation times (small $\nu$). Studies of particle production for identified hadron species are required to accomplish these goals (see Fig.~\ref{fig:hadronization_double_ratio}).
The present phenomenological description of in-medium fragmentation describes the observed attenuation of light hadron production through modification of splitting functions in the presence of nuclear matter. Jet substructure studies at the EIC will provide direct experimental input for constraining the evolution of splitting functions in nuclear matter.

\vskip 0.3cm
\noindent
{\bf Fundamental symmetry physics} 
The high luminosity, polarized lepton and polarized hadron beams, and kinematic coverage provided by the \epic\ detector afford unique opportunities for a variety of electroweak (EW) and beyond-the-standard model (BSM) physics topics. Among them, precision measurements of parity violating asymmetries over a wide range of $x$ and $Q^2$, when combined with knowledge of the PDF, can determine the value of the weak mixing angle $\sin^2{\theta_W}$ at an energy scale between fixed-target and high-energy collider facilities, and help narrow down the mass range of possible dark $Z$ bosons ($Z_d$). Additionally, such PVES asymmetries provide nearly orthogonal constraints to Drell-Yan processes measured at the LHC, on new contact interactions when analyzed in the framework of the Standard Model Effective Field Theory (SMEFT). The availability of polarized hadron beams at the EIC will measure new electroweak structure functions, the $g_{1,5}^{\gamma Z}$, for the first time. 
Second, lepton flavor violation observed in neutrino oscillations implies a similar violation in the charged lepton sector, charged lepton flavor violation (CLFV). While CLFV due to Standard Model processes are too suppressed to be observed by current or planned experiments, many BSM scenarios predict much higher 
rates that could be detected by the EIC. 
In particular, electron-to-tau conversion ($e+p\rightarrow \tau + X$), mediated by leptoquarks, is one of the most promising CLFV channels to be studied at the EIC because of its higher luminosity and the exquisite vertex resolution provided by the \epic ~detector. Such limits would 
potentially surpass limits set by the HERA experiments and would be sensitive to the difference between scalar and vector leptoquarks. Another opportunity in $e-\tau$ CLFV is via a possible $e+A\rightarrow\tau+A+a$ process, where $A$ is a high-$Z$ ion and $a$ is an Axion-Like Particle (ALP)~\cite{Davoudiasl:2021mjy}. The polarized beams at the EIC will provide a unique sensitivity to parity violating ALPs. 
Lastly, by measuring the charge-current DIS cross section at different electron beam polarizations, it is possible to set constraints on right-handed $W$ bosons and thus test the chiral structure of the Standard Model. 

\subsection{The EIC Facility}

The EIC will be a new, innovative, large-scale particle accelerator facility capable of colliding high energy beams of polarized electrons with heavy ions and polarized protons and light ions. It is a joint endeavor between BNL and JLab that will be built on the current site of RHIC.  In December 2019, the EIC was launched as an official project of the US government when it was granted Critical Decision Zero (CD-0). Soon after, in June of 2021, the project was awarded CD1 status. Beam operation
is currently expected to commence in the early 2030s.

\begin{figure}[!ht]
\begin{minipage}[c]{0.5\textwidth}
\centering
\vspace*{2mm}
\includegraphics[width=1.0\textwidth]{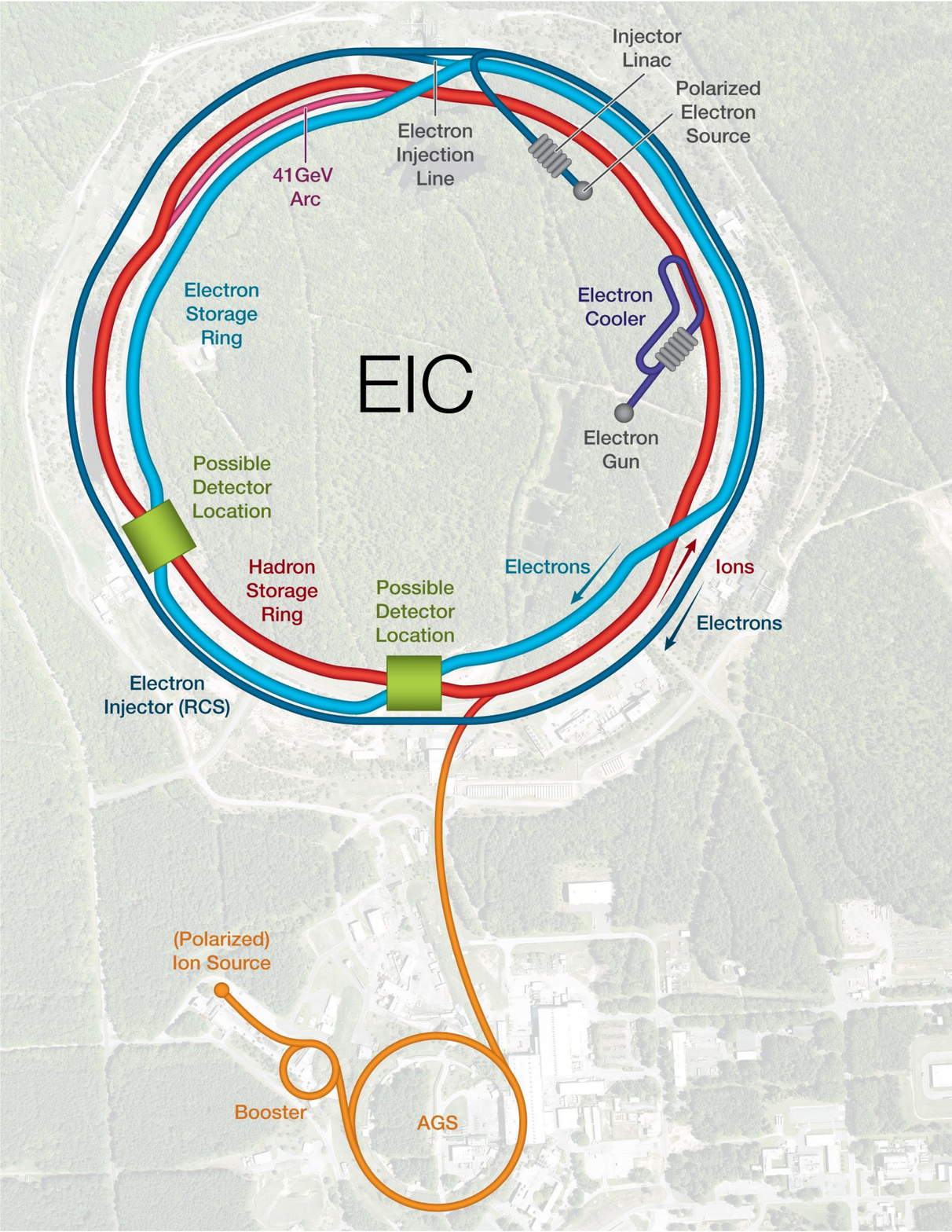}
\caption{Planned EIC accelerator. Figure from~\cite{EIC-WP-2023LRP}.}
\label{EIClayout}
\end{minipage}\hfill
\begin{minipage}[c]{0.46\textwidth}
In order to address the crucial scientific questions discussed in the previous sections, the EIC must provide:
\begin{itemize}
  \setlength\itemsep{-0.2em}
\item Highly polarized electron ($\sim$70\%) and proton ($\sim$70\%) beams;
\item Ion beams from deuterons to heavy nuclei such as gold, lead, or uranium;
\item Variable \ep~center-of-mass energies from 28$-$100 GeV, upgradable to 28$-$140 GeV;
\item High collision electron-nucleon luminosity 10$^{33}-$10$^{34}$  cm$^{-2}$ s$^{-1}$;
\item The possibility of more than one interaction region.
\end{itemize}
\end{minipage}\hfill
\end{figure}

Shown schematically in Fig.~\ref{EIClayout}, the EIC will collide bright, intense counter-circulating beams of electrons and ions at two possible interactions regions, the Interaction Point (IP) 6 and IP8, at 6 and 8 o'clock position in Fig.~\ref{EIClayout}, respectively. The DOE has committed to building a general-purpose, large acceptance detector that is capable of addressing the science case outlined in the NAS report~\cite{NAP25171}. In 2020 the EIC Users Group launched a year-long effort to explore possible detector technologies and codify the detector requirements needed to address the NAS science case. The results of this study have been collected and published as the EIC Yellow Report~\cite{AbdulKhalek:2021gbh}. With the detector requirements defined, BNL and JLab extended a call to the community in March of 2021 for Collaboration Proposals for the reference detector. A Detector Proposal Advisory Panel (DPAP), an international committee of detector experts and theorists, was assembled to review the proposals submitted by the ATHENA, CORE and ECCE proto-collaborations. The outcome of that competitive review process is the \epic\ collaboration, which is in the process of finalizing the technology choices and detector designs for the detector at IP6, starting from and extending the ECCE proposal as its reference design. Details about the current ePIC detector design and plans for the second detector at IP8 will be discussed in Sections~\ref{sec:eic_epic} and \ref{sec:eic_det2}, respectively. 

\begin{figure}[!ht]
\begin{center}
\vspace*{2mm}
\includegraphics[width=1.0\textwidth]{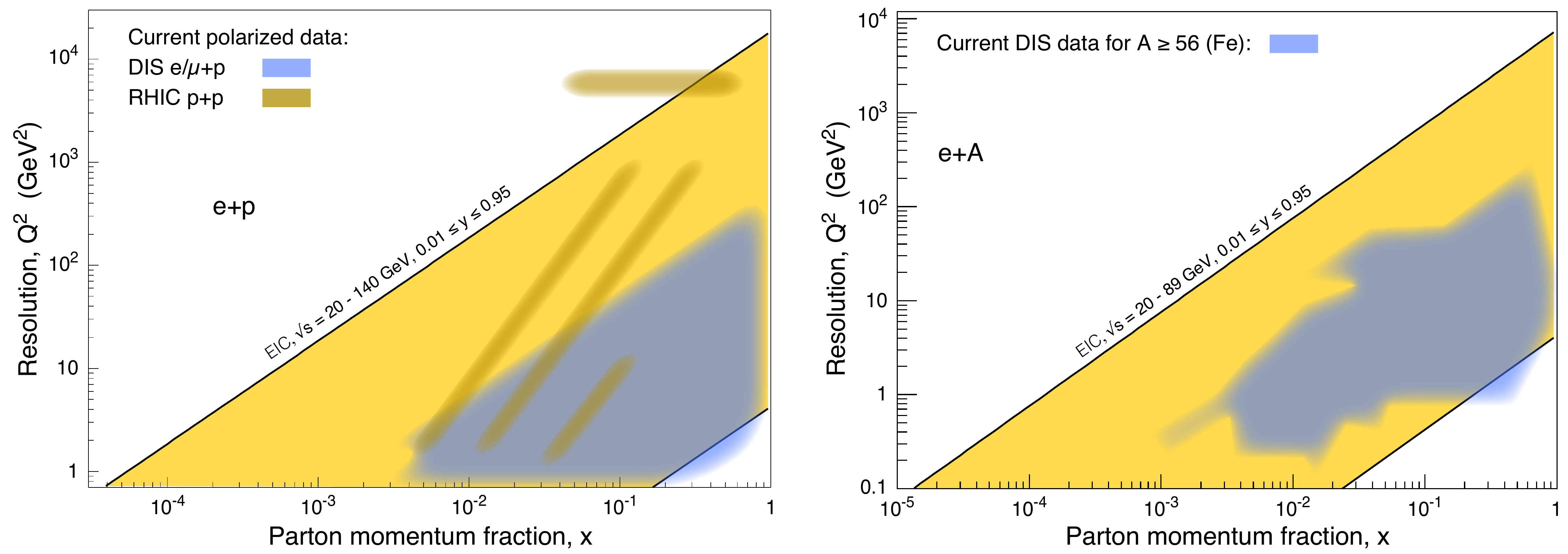}
\caption{\label{fig:xq2}Left: The $x$-$Q^{2}$ range covered by the EIC (yellow) in comparison with past and existing polarized $e/\mu$+$p$ experiments at CERN, DESY, JLab and SLAC, and \pp\ experiments at RHIC. 
Right: The $x$-$Q^{2}$ range for \eA\ collisions for ions heavier than iron (yellow) compared to existing world data. Figure from~\cite{EIC-WP-2023LRP}.} 
\end{center}
\end{figure}

The EIC is being designed to cover a center-of-mass energy range for \ep\ collisions of 28~GeV~$\le \sqrt{s} \le$~140~GeV, which in turn allows for a broad kinematic reach in $x$ and $Q^{2}$ as shown in Fig.~\ref{fig:xq2}. 
The diagonal lines in each plot represent lines of constant inelasticity $y$, which is the ratio of the virtual photon energy to the electron energy in the target rest frame. The quantities $x$, $y$, and $Q^{2}$ are obtained from measurements of energies and angles of final state objects, i.e., the scattered electron, the hadronic final state or a combination of both. The left panel in \fig{fig:xq2} shows the kinematic coverage for \ep\ collisions while the right panel shows the coverage for \eA\ collisions. The EIC will open doors for precision measurement with polarized beams to entirely new regions in both $x$ and $Q^2$ while providing critical overlap with present and past experiments. Access to the low-$x$ region is particularly important as this is the gluon-dominated regime where saturation effects are expected to emerge.

\vskip 0.3cm
\noindent{\bf The accelerator} The EIC must collide electrons with protons and other atomic nuclei (ions) over a range of energies. There must be enough collisions for the experiment to gather adequate data to elucidate or settle the known physics questions, and other questions that may emerge, in a reasonable time. A collider’s ability to squeeze many particles of two beams into a tiny volume where they collide defines its luminosity. The luminosity ultimately required of the EIC is comparable to those of the highest performing colliders built to date, such as the LHC at CERN and the $B$-meson factories at SLAC and KEK. Furthermore, given the crucial role of spin, there must be the capability to polarize both the electron and the proton, neutron or light beams. That is to say, the spins of the individual particles in each beam must be made to line up with each other, overcoming their natural tendency toward randomization.

To achieve these goals, a host of techniques in accelerator physics and technology must be brought to bear. Only a few are mentioned here. State-of-the-art SRF cavities will accelerate high-intensity beams efficiently. Further specialized RF "crab" cavities will rotate the beams as they collide to optimize their overlap. Elaborate interaction region designs must squeeze two very different beams simultaneously into tiny spot sizes using advanced superconducting magnet designs. The hadron beams must be compressed in volume by sophisticated new “beam cooling” techniques that involve subtle interactions with ancillary electron beams. Polarized beams require polarized particle sources, special magnets, and a further level of mastery of beam physics to preserve the polarization through the acceleration process to the collisions. Polarized colliding stored beams have been achieved before only at HERA (polarized positrons or electrons on unpolarized protons) and at RHIC (both colliding proton beams polarized).

EIC accelerator requirements push the current technology and their realization requires significant research and development. Indeed, an important element of the scientific justification for a U.S. electron-ion facility is that it drives advances in accelerator science and technology, which in turn will benefit other fields of accelerator-based science and society. 
The accelerator physics and technology advances required for an EIC will, importantly, have the potential to extend the capabilities of many particle accelerators built for other purposes, from medicine through materials science to elementary particle physics. Construction and future operations of an EIC including an appropriate program of dedicated accelerator test experiments would sustain and develop this precious national asset and help the United States to maintain a leading role in international accelerator-based science.

\subsection{The \epic\ Detector}\label{sec:eic_epic}

The \epic{} detector is a state-of-the-art experimental instrument currently being designed and constructed by a multi-institutional international collaboration including over $600$ scientists. 
To enable the full EIC physics program the \epic{} detector needs to offer complete kinematic coverage for the detection of particles emitted in central ($|\eta| \lesssim 3.7$), far-forward ($\eta \gtrsim 3.7$) and far-backward ($\eta \lesssim -3.7$) directions, where backward and forward refers to the electron and hadron beam directions respectively and the forward acceptance is required to extend down to $10\sigma$ of the beam width away from its central line.
The detected particles should be identified and their momentum measured with high 
precision, over an extensive energy range, $\sim0.1$ -- $50$ GeV~\cite{AbdulKhalek:2021gbh}.
These requirements will ensure all major physics processes: neutral-current and charged-current DIS, SIDIS, and exclusive processes, can be detected, including associated spectator nuclear fragments where relevant.
Special attention was also given to evaluating detector requirements for measurements of processes involving jets, jet substructure, and heavy-flavor hadrons, such as precise vertex resolution, combined precision timing and position measurement, and precision calorimetry. 

Meeting these stringent requirements is a formidable task that is further challenged by the asymmetric nature of EIC collisions and the need to have a non-zero crossing angle between the electron and hadron beams. Therefore, the \epic{} detector requires complete and detailed integration with the EIC interaction point and accelerator beams, a major technical challenge that has been successfully addressed by the EIC community over the last several years.

\begin{figure}[!ht]
\begin{center}
\vspace*{2mm}
\includegraphics[width=0.6\textwidth]{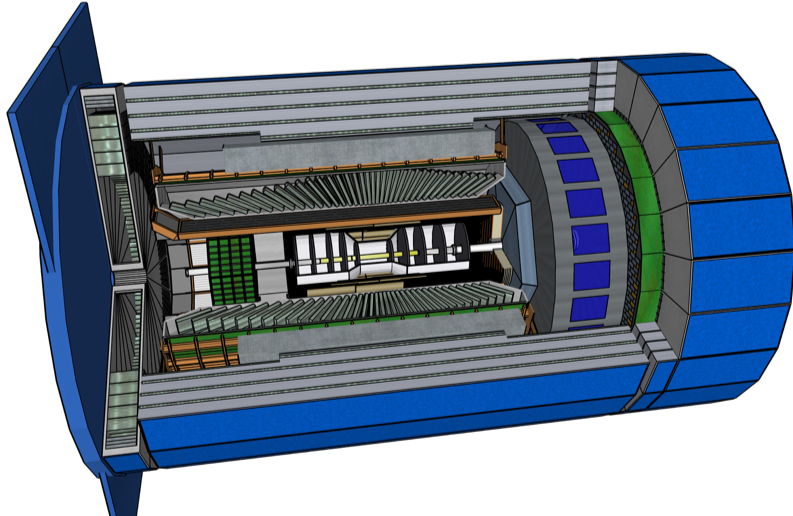}
\caption{\label{fig:epic} Schematic drawing of the ePIC central detector showcasing its high-precision vertexing and tracking detectors, Cerenkov and TOF based PID detectors and electromagnetic and hadronic calorimeters. Figure from~\cite{EIC-WP-2023LRP}.}
\end{center}
\end{figure}

The current layout of the \epic{} central detector is shown in Fig.~\ref{fig:epic}. The central detector is based around a $1.7$~T superconducting solenoid with the same dimensions as the BaBar solenoid used by the sPHENIX experiment. It is divided into a barrel region ($|\eta| \lesssim 1.5$) and forward and backward endcaps ($1.5 \lesssim |\eta| \lesssim 3.7$). All central detector regions follow an overall similar particle detection concept, starting from high-precision vertexing and tracking measurements, continuing with Cherenkov and TOF based PID measurements, followed by electromagnetic and, finally, hadronic calorimetry. 
The tracking system will be based on a set of silicon detectors, using the ITS-2, ITS-3, and AC-LGAD technologies, supplemented by large radii MPGD detectors using $\mu$RWELL and $\mu$Megas technologies. The Cerenkov-based PID uses a high-performance detector of internally reflected Cherenkov light (DIRC) in the barrel, and a dual gas-aerogel RICH and an aerogel-based RICH in the forward and backward endcaps respectively. TOF information for low-momentum PID will be provided by AC-LGAD detectors in the barrel and forward endcap, providing both high-precision tracking and timing information, and in the backward endcap by the LAPPD sensors that will be used to read out the RICH detector. EM calorimetry will be based on PbWO$_4$ crystals in the backward endcap, tungsten SciFi in the forward endcap, and either scintillating-glass crystals or an imaging calorimeter in the barrel. hadronic calorimetry in the barrel will be done by reusing the iron scintillator calorimeter recently built for the sPHENIX experiment, and using a longitudinally segmented iron scintillator and tungsten scintillator sandwich in the forward endcap. The need for a hadronic calorimeter in the backward region is still being investigated and at present an un-instrumented iron scintillator sandwich calorimeter is planed to placed to return the field and allow for a future addition of backward hadron calorimetry.

In addition to the detector suite for the primary interaction region, the \epic{} detector design also includes the far-forward and the far-backward spectrometers that provide beam monitoring, among other functions, that ensure the EIC scientific program can be realized. The far-backward region includes a precision luminosity monitor and two stations of low-$Q^2$ taggers, while the far-forward region includes a ZDC, Roman pots, off-momentum trackers, and a B0 tracker.These detectors use the beam steering magnets themselves as the spectrometer magnet, making their design extremely complex and requiring close integration with the accelerator design.

In addition to the various detector components, a central and novel feature of ePIC is its 'triggerless' readout, 
enabling continuously recording (streaming) and storing all interaction data for off-line analysis. The development of such cutting edge readout system poses a significant challenge that requires the development of advanced zero-suppression and fast hardware-enabled AI algorithms for effective background filtering.

\subsection{Detector II}\label{sec:eic_det2}

A key deliverable of the EIC Project is an accelerator design that can accommodate a second interaction region and detector. The scope of the EIC project includes one detector (the \epic detector). At the same time, it is recognized by all stakeholders that a second, complementary, detector is essential to fully exploit the science potential of the EIC. Historically, projects of similar scientific impact and scope were designed to include two or more complementary detectors and the importance of this model has been demonstrated time and again. Multiple detectors will expand scientific opportunities, draw a more vivid and complete picture of the science, provide independent confirmation for discovery measurements and mitigate potential risks when entering uncharted territory, especially for systematics-limited measurements as the EIC expects to perform. Two detectors will expand the opportunities for a new generation of scientists and encourage technological development and innovation by fostering a natural and healthy competition between the two collaborations.

The timeline for establishing a second experiment at the EIC is crucial. The two experiments should be separated by no more than a few years for scientific validation to be productive. In turn, this delayed time frame can be used to explore new and complementary detector technologies that may not have been utilized by the \epic~detector. The EIC community has emphasized the need for at least two detectors for many years and the Detector Proposal Advisory Panel (DPAP) echoed this support stating in their report that \textit{``A strong case for two complementary general-purpose detectors has been made during the panel review"} and that \textit{``There is significant support in the community and from the panel for a second general-purpose detector system to be installed in IR8 when resources are available."} The DPAP also concluded that \and that \textit{``it is essential to have two detectors with a sufficient degree of complementarity in layout and detector technologies."} In particular, the panel made a convincing case for the significant gain in physics reach achievable with a secondary focus:
\begin{itemize}
  \setlength\itemsep{-0.2em}
\item increased acceptance in the invariant momentum transfer $t$ of the scattered proton in \ep\ collisions, which directly translates into an increased resolution power for imaging partons in the transverse plane;
\item significantly improved abilities to detect nuclear breakup in exclusive and diffractive scattering on light and heavy nuclei. The distinction between coherent and incoherent scattering is essential for the physics interpretation of these processes;
\item prospects for a program of low-background $\gamma$ gamma spectroscopy with rare isotopes in the beam fragments.
\end{itemize}
The panel further pointed out that \textit{``the additional R\&D required for a second detector will bring additional benefits in developing technologies and in training the associated workforce."} The DOE Office of Nuclear Physics has followed up on this and restarted a generic EIC-related detector R\&D program. The EIC Users Group is in the process of refining the science case for a second detector and is actively working to engage additional national and international resources for this effort.

\subsection{EIC-Theory Alliance}

As described above, the EIC will be a unique and versatile facility that will enable us to understand some of the most compelling questions in the physics of the strong nuclear force. 
To fully exploit the potential of the EIC, a focused theory effort will be required. The need for such an effort was already pointed out in the NAS report \cite{NAP25171}. The best way to achieve this goal is the creation of a national EIC Theory Alliance.

The goal of the EIC Theory Alliance is to provide support and stewardship of the theory effort in EIC physics broadly defined for the life span of the facility. It will promote EIC theory and contribute to workforce development through: i) support of graduate students, ii) EIC Theory Fellow positions, iii) bridge positions at universities, and iv) short and long term visitor programs to enhance collaboration between various groups. In addition, the alliance will organize topical schools and workshops. 

The EIC Theory Alliance will be a decentralized organization open to participation by anyone in the community who is interested in EIC physics, i.e., it  will be a membership organization, where members elect an executive board which effectively runs the alliance. The executive board will determine the major scientific thrusts of the theory alliance, make decisions regarding at which universities bridge faculty positions will be created, and serve as a search committee for EIC related positions. Furthermore, the executive board will coordinate the organization of workshops and schools related to the research activities of the alliance. In addition, the EIC Theory Alliance will seek and nurture international cooperation to maximally leverage the available funding. The structure of the EIC Theory Alliance to some extent will resemble the structure of topical collaborations in nuclear theory. However, unlike topical collaborations, it will have a significantly longer life span and involve a large international component.

\section{Connections to Other Sub-fields of Nuclear}\label{sec:other_fields}
All QCD facilities, including CEBAF, RHIC, and the LHC, cover physics programs beyond the subjects of cold and hot QCD. The strong overlap between QCD physics and other nuclear science disciplines has always been a unique feature at these facilities. In previous sections, we have discussed, for example, the nuclear EMC effects and its close relation to nucleon-nucleon short range correlations which play an important role in the nuclear structure studies and are a crucial part of the physics program at FRIB. In the following, we highlight additional connections between QCD studies and other nuclear science fields and beyond. 

\subsection{Probing Novel Regimes of QED in Ultra-Peripheral Heavy-Ion Collisions} \label{sec:UPC_QED} 
The lowest order QED calculation~\cite{Vidovic:1992ik,Baltz:2007kq,Klein:2018cjh,Klein:2018fmp,Brandenburg:2022tna} of lepton pair production via photon-photon fusion in the equivalent photon approximation~\cite{Baltz:2007kq,Klein:2018cjh,Klein:2018fmp,Brandenburg:2022tna} as input for the photon flux can be used to describe the unpolarized cross section in UPCs measured by RHIC and LHC~\cite{Adams:2004rz,Afanasiev:2009hy,Adam:2018tdm,STAR:2019wlg,Abbas:2013oua,ATLAS:2020epq,ATLAS:2022srr,ATLAS:2022ryk,ATLAS:2022vbe,CMS:2020skx,Acharya:2018nxm,Aaboud:2018eph}. This is true also when making selections on various topologies of forward neutron production using ZDCs, which are wellknown to select on the internuclear impact parameter~\cite{Klein:2020fmr,Klein:2016yzr,Suranyi:2021ssd,SantosDiaz:2022hxl}.
Coherent photons are highly linearly polarized with the polarization vector aligned along its transverse momentum direction. A sizable $\cos 4\phi$ azimuthal asymmetry induced by  linearly polarized  coherent photons  was observed in a  STAR measurement~\cite{STAR:2019wlg} in agreement with theoretical predictions~\cite{Li:2019yzy,Li:2019sin}. 
With future high statistics data with larger acceptance in UPCs at RHIC and LHC,  the phase space of photon collisions in transverse momentum, rapidity and momentum-space-spin correlations  can be explored in extreme regions of QED fields~\cite{Li:2019yzy,Klein:2020jom,Brandenburg:2021lnj}. More importantly, these measurements provide a precision calibration necessary for photons as sources of the photonuclear processes~\cite{STAR:2022wfe,ATLAS-CONF-2022-021} (see Sec.\,\ref{sec:init}) and the initial electromagnetic field, necessary for studies of emerging QCD phenomena (see Sec.\,\ref{sec:cme}).

\subsection{Connection to Nuclear Astrophysics} \label{sec:neutronstars}
Astrophysical observations have entered a new era with measurements of neutron star radii and tidal deformabilities that can be used to infer the neutron star equation of state at large baryon densities and vanishing temperature. In 2017 the first gravitational waves from merging binary neutron stars were measured as well as the electromagnetic component of the merger~\cite{TheLIGOScientific:2017qsa}.  Since then other potential mergers of neutron stars (either with other neutron stars or black holes) have been detected. 
 The first radius measurement of a two-solar-mass neutron 
star~\cite{Riley:2021pdl,Miller:2021qha,Fonseca:2021wxt} was done in NASA's NICER mission. Both hot and cold QCD programs cover physics research that are closely related to the EOS of dense QCD matter. The upcoming years are expected to produce many other observations, providing unprecedented constraints on the dense matter equation of state. 

\subsubsection{Heavy Ion Collisions to Explore the QCD EOS}

Low-energy heavy-ion collisions probe densities similar to neutron stars, albeit at much higher temperatures. 
However, significant theoretical development is needed to reliably make direct connections between neutron stars and heavy-ion collisions (see e.g. \cite{An:2021wof,Lovato:2022vgq, Sorensen:2023zkk}), including further development of heavy-ion collisions simulations. One must also keep in mind that heavy-ion collision and neutron stars probe different regions of the phase diagram: while heavy-ion collisions are governed by the EOS of nearly symmetric nuclear matter, neutron stars are extremely neutron rich environments with very few charged hadrons.
Thus, one must have a strong understanding of how properties of QCD are affected when comparing symmetric to asymmetric matter. Fortunately, measurements of mirror nuclei from the future CBM experiment at FAIR or FRIB~\cite{Pineda:2021shy} could provide crucial insight into subtle differences between heavy-ion collisions and neutron stars. 

The description of neutron star mergers can also benefit greatly from theoretical advances in relativistic viscous hydrodynamics, which were driven mostly by applications in heavy ion collisions. In such mergers, rapid changes in $T$ and $\mu_B$ \cite{Alford:2017rxf,Perego:2019adq} can drive fluid elements out of chemical equilibrium, and weakly interacting processes will relax them back to equilibrium. If the corresponding timescale is of the order of milliseconds \cite{Alford:2017rxf,Arras:2018fxj}, this may influence the hydrodynamic evolution and leave imprints in the post-merger gravitational wave emission \cite{Most:2021zvc,Hammond:2022uua,Most:2022yhe}. In this case, the detection of post-merger gravitational waves (using upcoming 3G detectors) could provide information not only about the dense matter EOS but also about its novel transport properties. 
Furthermore, under certain conditions \cite{Gavassino:2020kwo,Camelio:2022ljs}, the chemical imbalance associated with neutrino processes admits a viscous hydrodynamic description in terms of equations of motion similar to those  investigated in heavy-ion collisions \cite{Israel:1979wp,Denicol:2012cn} (though now the transport coefficients are determined by weak-interaction processes). This synergy can foster new collaborations between heavy-ion physicists, nuclear astrophysicists, gravitation wave scientists, and numerical relativity experts (see, e.g, \cite{Most:2021zvc,Most:2022yhe}).

\begin{figure}[!ht]
\includegraphics[width=0.5\textwidth]{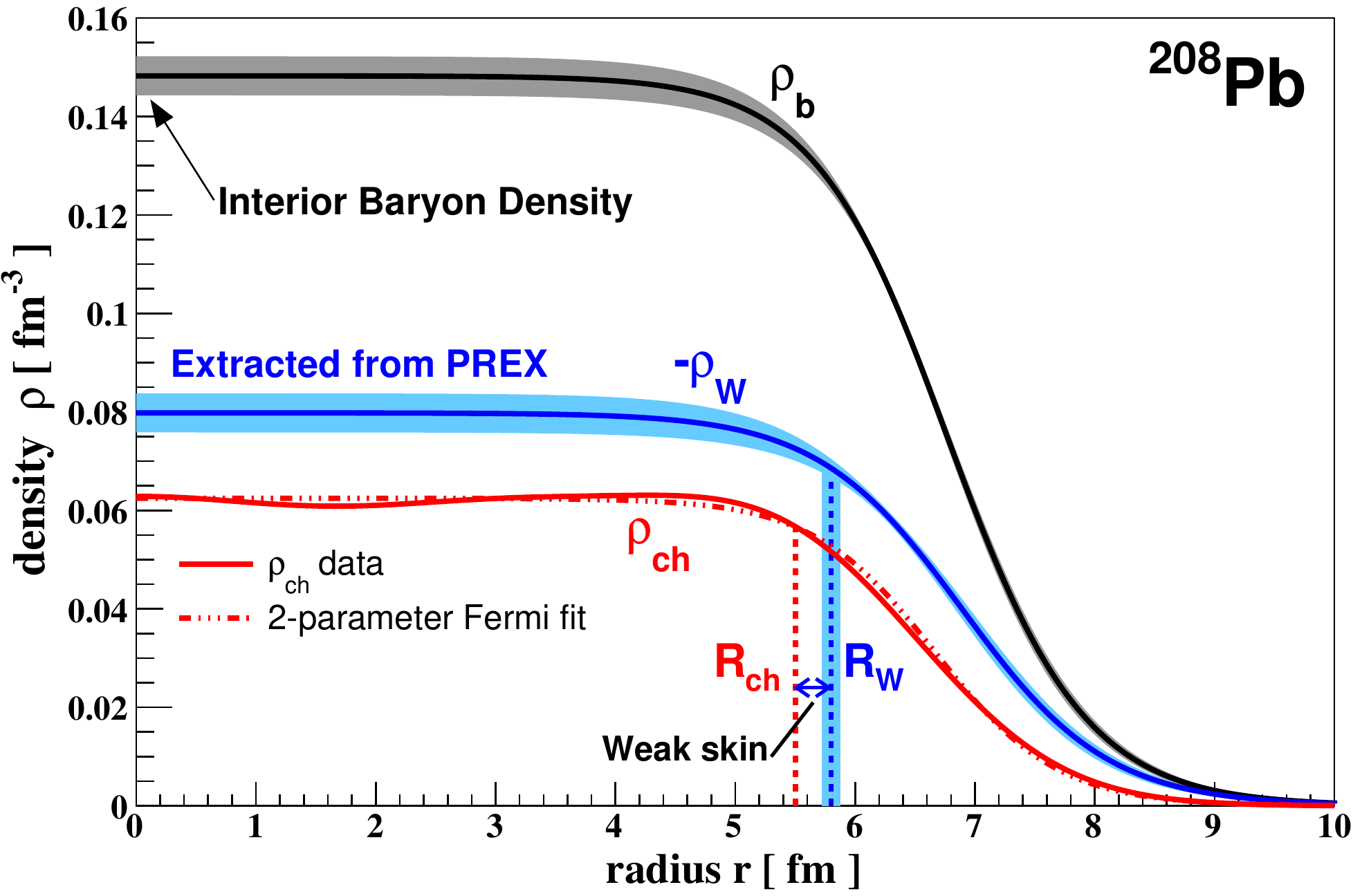}
\includegraphics[width=0.5\textwidth]{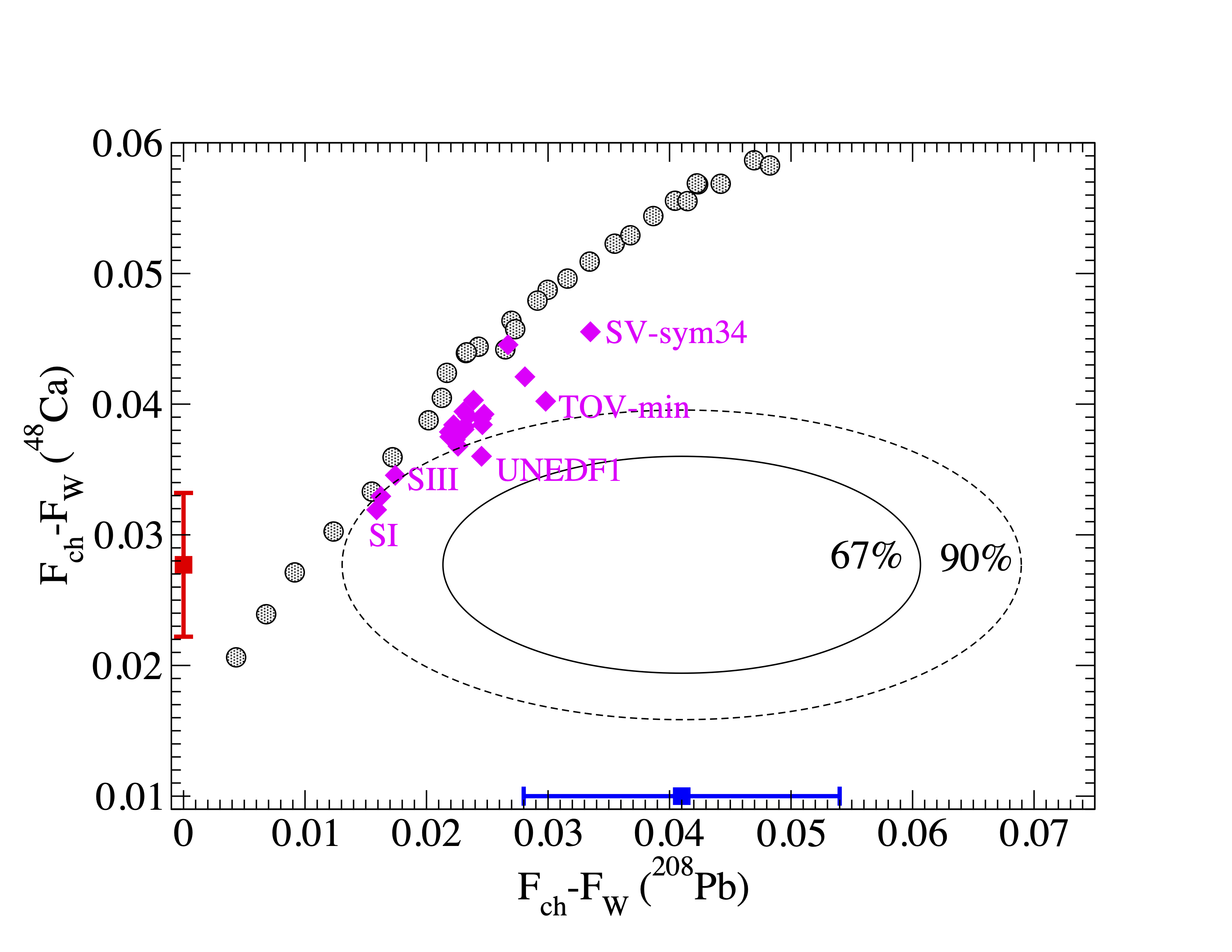}
\caption{ Left: $\epb$ weak and baryon densities from the combined PREX datasets, with uncertainties shaded. The charge density is also shown~\cite{PREX:2021umo}. Right: Difference between the charge and weak form factors of $\eca$ (CREX) versus that of $\epb$ (PREX-2) at their respective momentum transfers. The blue (red) data point shows the PREX-2 (CREX) measurements. The ellipses are joint PREX-2 and CREX 67\% and 90\% probability contours. The gray circles (magenta diamonds) are a range of relativistic (non-relativistic) density functional models~\cite{CREX:2022kgg}.}
\label{fig:prex}
\end{figure}
\subsubsection{Neutron Skin Thickness in Heavy Nuclei and Connection to Neutron Stars}
Nuclei are known to be very dense with nucleons packed against each other. They could in fact be considered a terrestrial laboratory to study the behavior of extremely dense nuclear matter contained within celestial objects of the same nature, e.g., neutron stars. 
In neutron-rich nuclei, neutrons are expected to be distributed over a larger volume than protons, forming a neutron ``skin" around the nucleus.  The thickness of this skin is sensitive to the equation of state for nuclear matter, and specifically to the density dependence of the symmetry energy near saturation density.
A direct measure of the neutron skin thickness has long been an elusive goal. The situation drastically changed with the use of observables involving more than electromagnetic interactions. 
Because the weak charge of the neutron is much larger than that of the proton, PVES provides a highly interpretable, model-independent probe of neutron densities. 
 
Results from two such high-precision measurements at JLab have become available very recently: the PREX-2 experiment on $\epb$ found its neutron skin thickness to be $0.28 \pm 0.07$~fm~\cite{PREX:2021umo}, while the CREX experiment found the neutron skin of $\eca$ to be $0.121 \pm 0.036$~fm~\cite{CREX:2022kgg}. 
Here, the large $\epb$ nucleus provides a close approximation to uniform nuclear matter and the data imply an interior nuclear baryon density of $0.148 \pm 0.038~\rm{fm}^{-3}$, while the $\eca$ system is more sensitive to details of nuclear structure and therefore presents additional tests of models. 
Both PREX-2 and CREX provide a direct, model independent measurement of the difference between the weak and electromagnetic form factors. The difference found by PREX-2 is relatively large, in contrast with the CREX result that shows smaller-than-expected differences. Nuclear model predictions tend to correlate between the two systems, thus the differing results present an important empirical challenge to precise modeling of nuclear structure. 
An experiment called MREX is being planned to measure the $\epb$ neutron radius to $0.03$~fm precision at the new MESA facility in Mainz, Germany, which will help clarify the intriguing observation from PREX-2 of a thick neutron skin.

\subsubsection{Cosmic-rays and Nuclear Physics}

QCD studies have played an important role in one of the most compelling mysteries in astrophysics: 
the nature (nuclear composition) and origin of ultra-high energy cosmic rays (UHECR).   Incident UHECR produce energetic air showers that develop as they propagate through the atmosphere.   
The relationship between observables, such as the size of the electromagnetic shower reaching the ground and the number of muons and the energy and species of the incident particle depends critically on the hadronic physics that is used to model the air shower; different simulation codes predict rather different results~\cite{IceCube:2021ixw}. 
An improved understanding of air shower development is critical in view of several outstanding mysteries in the field: the long-standing unresolved tension between Southern hemisphere observations by the Auger observatory~\cite{PierreAuger:2022atd} and Northern hemisphere measurements by the Telescope Array (TA)~\cite{Hanlon:2018hlv}, and the apparent excess of muons in high-energy air showers seen by multiple experiments~\cite{Albrecht:2021cxw,Pajares:2000sn,Alvarez-Muniz:2006wca}.

Although fixed target RHIC and LHC data have been helpful in tuning Monte Carlo models, there are still significant uncertainties~\cite{Baur:2019cpv}, and predictions are sensitive to parton behavior at low-$x$~\cite{Hentschinski:2022xnd}. Better LHC data is needed in the far-forward region~\cite{Bylinkin:2022temp,CMS:2020ldm,Feng:2022inv}, which is most important for determining the particle fluxes reaching the ground.
This data will also be helpful in better estimating the atmospheric neutrino flux, including the prompt flux, where there are still significant uncertainties \cite {Bhattacharya:2020hfs}. 
Meanwhile, cosmic-rays offer us the opportunity to make nuclear-physics measurements that are not possible with current or planned accelerators~\cite{IceCube:2017roe,IceCube:2018pgc}.  
Future radio-detection experiments should extend the cross-section measurements to energies above $10^{19}$ eV, and thereby probe parton distributions at $x$ values that are lower than are accessible at the LHC ~\cite{Klein:2019nbu}, extending searches for saturation into a new regime. 

\subsection{Electron-nucleus Experiments and Connections to Neutrino Oscillation Measurements} 
\begin{figure}[ht!]
  \begin{minipage}[c]{0.5\textwidth}
    \includegraphics[width=\textwidth]{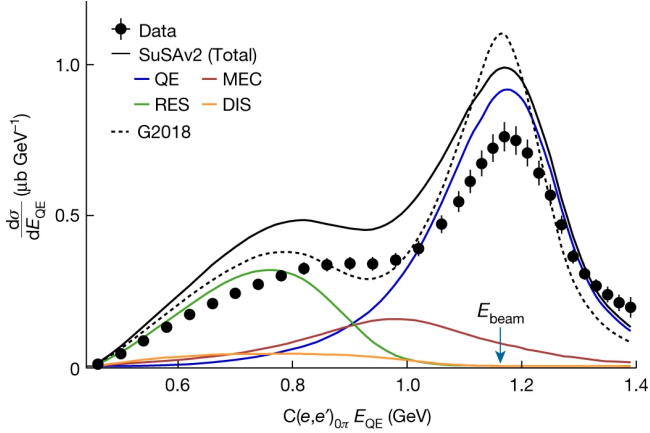}
  \end{minipage}\hfill
  \begin{minipage}[c]{0.45\textwidth}
    \caption{ Quasi-elastic reconstructed energy. The 1.159 GeV $C(e, e')_{0\pi}$ cross section plotted as a function of the reconstructed energy $E_{\rm QE}$ for data (black points), and widely used interaction  models such as GENIE SuSAv2 (solid black curve) and GENIE G2018 (dotted black curve). The colored lines show the contributions of different processes to the GENIE SuSAv2 cross section: quasi-elastic (QE), Meson Exchange Current (MEC), resonances (RES) and DIS. 
    It can be seen that the reconstructed energy distributions based on these models agree only qualitatively with data and the difference can be up to 25\%. Figure from~\cite{CLAS:2021neh}. 
    } \label{fig:e4v}
  \end{minipage}
\end{figure}

The precision of neutrino oscillation experiments depends on the ability to reconstruct the incident neutrino flux as a function of their energy at the detector position. 
As neutrinos are detected following their interaction with atomic nuclei in the detector, this extraction strongly relies on the precise understanding of neutrino-nucleus interaction cross sections.
Current oscillation experiments report significant systematic uncertainties due to these interaction models~\cite{NuSTEC:2017hzk,T2K:2018rhz,T2K:2019bcf,NOvA:2018gge} and simulations show that energy reconstruction errors can lead to significant biases in extracting the CP violating phase in neutrino oscillations at DUNE~\cite{Ankowski:2015kya}. The {\bf e4$\nu$} Collaboration exploits the similarity between electron- and neutrino-nucleus interactions to test and constrain these models. Utilizing the well known energy of the JLab beam and the large acceptance of the CLAS detector, e4$\nu$ performed wide phase-space scattering measurements on relevant nuclear targets and used their data to test energy reconstruction methods and constrain the interaction models for neutrino experiments. In a recent publication~\cite{CLAS:2021neh} they showed a quantitative disagreement (see Fig.~\ref{fig:e4v}) between electron scattering data and interaction models utilizing quasi-elastic-like topology, which is considered to be the simplest interaction one can measure and is used in many oscillations analyses. 
This disagreement grows with energy and nuclear mass 
number, as well as at large transverse momentum. 
Complementary measurements were also done by the JLab E12-14-012 experiment~\cite{Dai:2018xhi,Dai:2018gch,Murphy:2019wed,Gu:2020rcp,JeffersonLabHallA:2022cit,JeffersonLabHallA:2022msz} to improve our understanding of the spectral function of Argon, the target nucleus used by most neutrino detectors in DUNE~Refs.~\cite{Benhar:2010nx,Benhar:2015wva,Ankowski:2016jdd,Rocco:2018mwt}. 

The e4$\nu$ collaboration recently collected data with the CLAS12 detector at various energies and on different targets including argon. The collaboration expects to analyze various interaction channels and use its results to obtain an electron-tuned set of energy-reconstruction models for use by the neutrino oscillations community.  
In parallel, work is underway to unify the neutrino and electron modes in the widely applied GENIE event generators~\cite{electronsforneutrinos:2020tbf} to consistently analyze electron and neutrino data for reliable tests of the standard model in long baseline neutrino oscillation measurements.

\subsection{Connections to Physics Beyond the Standard Model Searches}

\subsubsection{Searches for BSM Physics in Ultra-Peripheral Heavy-Ion Collisions}
Ultra-peripheral heavy ion collisions provide a unique
environment to look for BSM physics in regions of phase space not easily accessible to \pp\ collisions \cite{Bruce:2018yzs,dEnterria:2022sut,Xu:2022qme}.
As discussed in Sec.\,\ref{sec:UPC_QED}, in these collisions the nuclei do not get closer than twice
the nuclear radius to each other and interact only
via QED.  This interaction is very strong (for QED) because
the nuclei have both been stripped of their electrons.
The two photons can interact via \textit{light-by-light scattering}~\cite{dEnterria:2013zqi},
a process which was first measured in these collisions at the
LHC~\cite{ATLAS:2017fur,CMS:2018erd}.  The two-photon
final-state in these collisions could be increased by BSM physics.  Measurements of
the cross section for these collisions~\cite{ATLAS:2019azn,CMS:2018erd} 
are consistent with expectations from the Standard Model.
The limits on new physics from this process, such as the existence of ALPs, are expected
to become stronger with the increased LHC luminosity
in Runs 3 and 4~\cite{Citron:2018lsq,Bruce:2018yzs}.
The anomalous magnetic moment of the $\tau$ lepton ($g_{\tau}$-2)
is also sensitive to new physics beyond the Standard Model~\cite{delAguila:1991rm} (as for $g_{\mu}-2$~\cite{Muong-2:2021ojo})
and can be extracted from the $\gamma\gamma\rightarrow\tau^{+}\tau^{-}$
process in UPCs. Early measurements from ATLAS and CMS~\cite{CMS:2022arf,ATLAS:2022ryk}, along with feasibility studies from ALICE and LHCb \cite{Burmasov:2022gnl}, have already
demonstrated a sensitivity competitive with that from previous LEP measurements~\cite{DELPHI:2003nah}.

\subsubsection{Parity-Violating Electron Scattering and EW/BSM Physics} \label{sec:otherfield_pves}
While CEBAF is considered primarily a QCD facility, the development of high-precision PVES has enabled it to make significant impact on low- and medium-energy tests of the neutral-current (NC) EW sector of the Standard Model and BSM physics. We describe a variety of such PVES measurements below. 

\subsubsubsection{The proton weak charge} 
While electric charges of the proton and the neutron are well known static properties of the nucleon, their counterpart the weak charge, predicted by the theory of electroweak unification, is not as well constrained. The recent QWeak experiment~\cite{Androic:2013rhu,Androic:2018kni} at JLab 
measured the proton weak charge $Q_W^p$ for the first time using the parity-violating asymmetry between right- and left-handed electron elastic scattering off the proton. It was determined to be $Q_w^p=0.0719\pm 0.0045$, which leads to a determination of the weak mixing angle $\sin^2\theta_W=0.2383\pm 0.0011$, both in good agreement with the Standard Model. When combined with atomic parity violation experiments~\cite{Wood:1997zq,Guena:2005uj,Toh:2019iro}, the Qweak experiment provides the best constraint on the NC electron-quark coupling $g_{AV}^{eq}$ to date. The P2 experiment planned at Mainz will improve the uncertainty over Qweak and will determine $Q_W^p$ to $\pm 1.83\%$ and $\sin^2\theta_W$ to $\pm 0.00033$~\cite{Becker:2018ggl}. The chiral counterpart of $g_{AV}^{eq}$, the electron-quark vector-axial coupling $g_{VA}^{eq}$, was measured by the JLab 6 GeV PVDIS experiment~\cite{Wang:2014bba,Wang:2014guo} and is one central focus of the planned SoLID program at JLab, see next paragraph. 

\subsubsubsection{Parity violation DIS and effective electron-quark couplings}
\begin{figure}[!bht]
\begin{minipage}[r]{0.52\textwidth}
\includegraphics[height=\textwidth,angle=270]{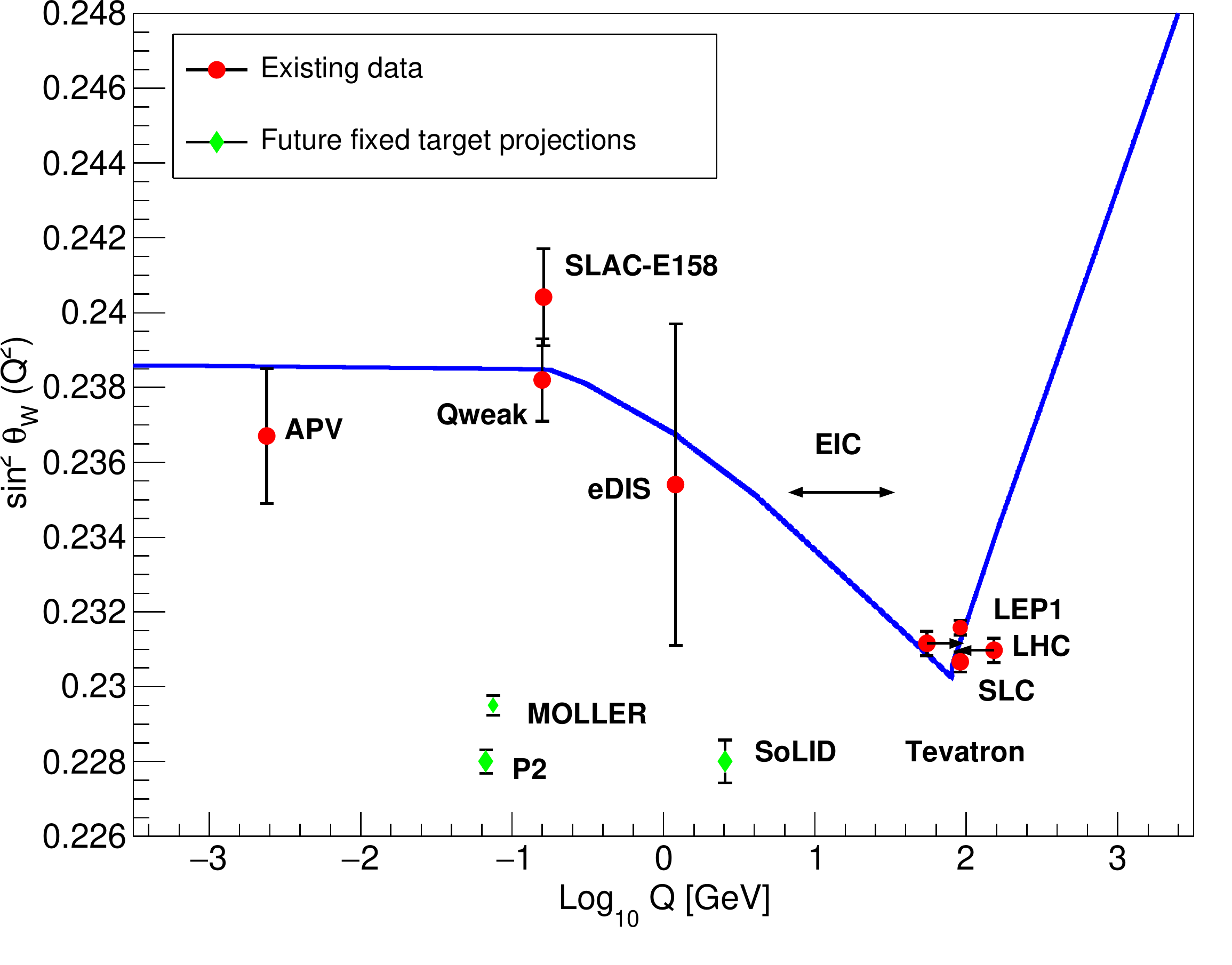}
\end{minipage}
\begin{minipage}[r]{0.45\textwidth}
\includegraphics[width=\textwidth]{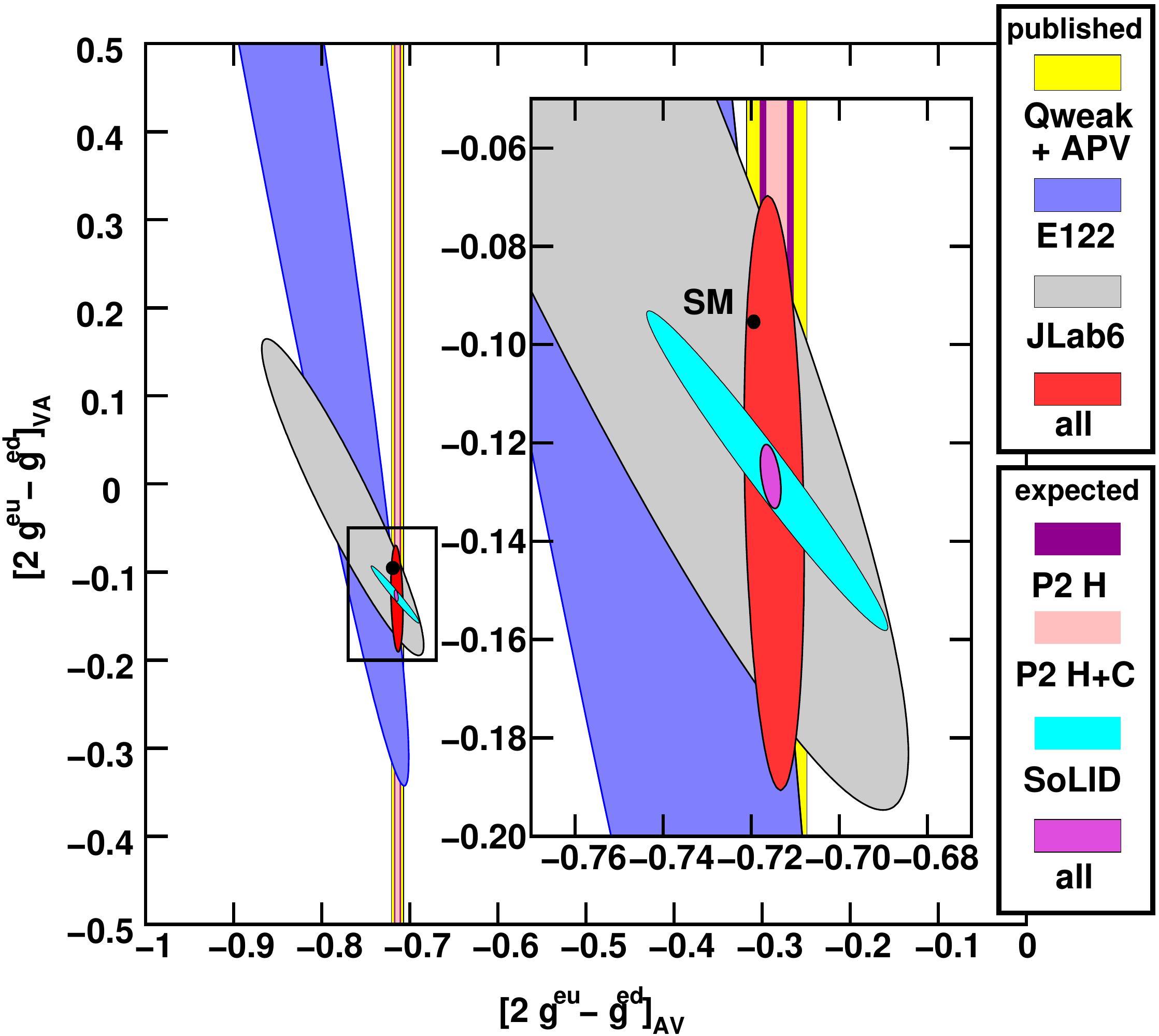}
\end{minipage}
\caption{Left: experimental determination of the weak mixing angle $\sin^2\theta_W$ including expected results from MOLLER and SoLID PVDIS. Data points for the Tevatron and the LHC are shifted horizontally for clarity. Right: Current experimental knowledge of the couplings $g_{VA}^{eq}$ (vertical axis), with the projected SoLID result shown by the cyan ellipse. Also shown are expected results from P2 at Mainz (purple and pink vertical bands) and the combined projection using SoLID, P2, and all existing world data (magenta ellipse), centered at the current best fit values. See ~\cite{JeffersonLabSoLID:2022iod} for details.
}\label{fig:future_pves}
\end{figure}
The aforementioned Qweak experiment provided the first result on the proton weak charge~\cite{Androic:2018kni}, and the 6 GeV PVDIS experiment provided the first evidence that the electron-quark vector-axial effective coupling $g_{VA}^{eq}$ is non-zero at the $2\sigma$ level~\cite{Wang:2014bba,Wang:2014guo}. The future of JLab Hall A will be comprised of two experiments that push the EW/BSM physics further, see Fig.~\ref{fig:future_pves}. The first is the MOLLER experiment, that will measure the electron weak charge and determine the weak mixing angle with a precision comparable to high energy collider experiments. The second is the SoLID PVDIS experiment with a deuterium target, which is the only planned experiment that will improve the precision on $g_{VA}^{eq}$ by an order of magnitude over the 6 GeV JLab result. A new Beam Dump Experiment (BDX) is planned that would run parasitically with MOLLER (or other high luminosity experiments), which will search for dark sector particles produced in the JLab Hall~A beam dump.

\section{Workforce Development and DEI}
\label{sec:workforce}

\begin{figure}[h]
    \centering
    \includegraphics[ width=.8\textwidth]{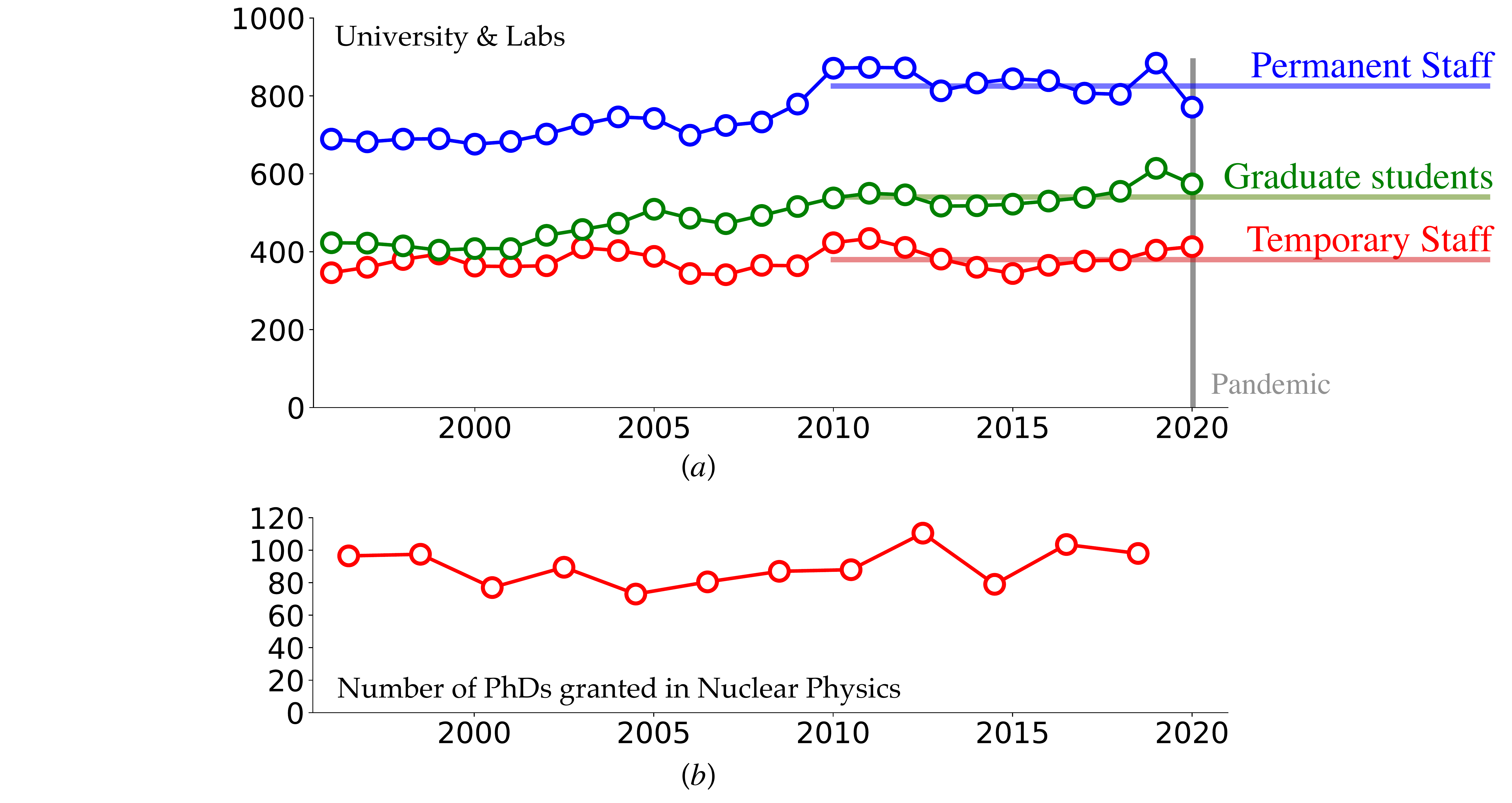}
    \caption{ DOE - FY2020 Nuclear Physics Workforce Survey of USA stats \href{https://science.osti.gov/np/-/media/np/pdf/about/Workforce_Survey.pdf}{[link]}
    }
    \label{fig:Unis_Labs_Phds}
\end{figure}

The success of the long-term future of our science relies on the ability to attract and retain a diverse and talented workforce, as well as a durable pipeline for sustaining it. As articulated in LRP15, “\emph{A highly qualified workforce trained in nuclear science is the most important element in realizing the scientific goals of the field.}”. 
Despite the previous recommendations for the field to grow, it has stagnated in size at best. This is partially reflected in Fig.~\ref{fig:Unis_Labs_Phds}~(a), which shows the numbers of NP graduate students and staff (permanent and temporary) present in American institutions. As can be seen, these numbers have approximately flattened since the 2010s. A similar number is the number of NP PhDs awarded in American institutions, Fig.~\ref{fig:Unis_Labs_Phds}~(b), which shows a similar plateau.  

The reason why the field has stagnated is unclear. Before discussing possible recommendations, it is useful to categorize and enumerate the factors that may be working against the growth of the field. These can generally be categorized as \emph{internal} and \emph{external} to the field. As one may expect, the community may have a better handle addressing the internal ones, but we must also be aware of external ones such that we can work towards countering these factors. 

Among the key \emph{internal} 
factors that may be making the field stagnant is its environment. Ultimately, a community is comprised of a set of individuals, each behaving as they see fit. As a result, the only tool at our disposal is defining guidelines that the community as a whole deems to represent acceptable behavior, in other words, Codes of Conduct. Within \emph{external} factors, there are those that we can effectively address and those that are practically out of our control. These include talent recruitment and retention, an education and outreach.  
In the following, we will discuss these factors that led to the recommendations listed in Sec.~\ref{sec:exe_sum}. We emphasize that all these factors are intertwined.

\subsection{Code of Conduct, Diversity, Equity, and Inclusion}

A Code of Conduct generally includes scientists' duty to behave ethically, respectfully, and inclusively toward one another and to reveal potential conflicts of interest. Together with Diversity, Equity, and Inclusion (DEI) committees, Codes of Conduct have become ubiquitous in the past few years. This is due to the regrettable fact that equity in science is still a mirage, as shown by results obtained from survey after survey. A considerable effort has been devoted by various groups to developing community guidelines, see, for example, Ref.~\cite{Barzi:2022edg}. To make this work, these guidelines have to be accompanied by {\bf accountability} and {\bf enforcement processes}. Their beneficial impact for the current community and the positive investment for future generations of physics make these a critical part of our culture going forward. 

To this end, one of the key recommendations reached within the QCD Town Meeting pertains to Codes of Conducts. In particular, it has been well recognized that part of recruiting and maintaining a diverse workforce requires treating all community members with respect and dignity. The QCD community supports the recent initiatives by the APS and DNP, such as the \href{https://engage.aps.org/dnp/programs/allies-program1}{DNP Allies Program} and \href{https://www.aps.org/meetings/policies/code-conduct.cfm}{Code of Conduct for APS Meetings}, to develop community-wide standards of conduct and recommends that host laboratories and user facilities require the establishment and/or adoption of enforceable conduct standards by all of the experimental and theoretical collaborations they support. The enforcement of such standards is the combined responsibility of all laboratories, theoretical and experimental collaborations, conference organizers, and individual investigators supported by the NP research program.

\bigskip
Meanwhile, DEI is a crucial component in workforce development:
\begin{itemize}
  \setlength\itemsep{-0.2em}
\item Establishing a {\bf diverse} workforce: This includes actively recruiting and hiring individuals from different backgrounds, experiences, and perspectives.

\item Promoting an {\bf inclusive culture}: This includes creating an environment where all individuals feel respected, valued, and heard. This can be done through training, open communication, and active engagement with diverse groups.

\item Providing opportunities for professional development and advancement: This includes providing {\bf equal} opportunities for all members in the QCD community at all career stages to learn, grow, and advance in their careers, regardless of their background.

\item Holding leadership {\bf accountable} and providing support and resources for underrepresented groups, such as employee resource groups, mentoring programs, and counseling services.

\item Continuously evaluating and improving: This includes regularly evaluating all institutional and organizational {DEI efforts, providing feedback, and urging them to make adjustments and improvements as necessary. }
\end{itemize}

It is important to note that promoting {DEI} is not a one-time event or a short-term project, it is a continuous process that requires commitment and effort from all members of the QCD community.

\subsection{Talent Retention for a Diverse Workforce}

The QCD community has experienced and benefitted from the bridge/joint faculty positions in the past, including those associated with BNL/RIKEN center (joint faculty with universities) and JLab bridge positions (with nearby universities). In addition, recent establishment of the DOE topical theoretical collaborations have opened up quite a few bridge faculty positions, see, Sec.~\ref{sec:theory_topical}, for the successful stories. At the QCD Town Meeting, it was strongly suggested by the community to expand such programs of bridge positions, fellowships, traineeships, and other incentives, to continue recruiting and retaining a more diverse group of junior faculty and staff at universities and national laboratories. In particular, that recruitment and retention of certain under-represented groups in NP need to be emphasized. The DEI principle discussed above should play a central role in these programs.

The imbalance of representation in NP almost certainly points to a pipeline issue.  The QCD community has a consensus to support the development and expansion of programs that enable participation in research by students from under-represented communities at national labs and/or research universities, including extended support for researchers from minority-serving and non-PhD granting institutions (MSI). Supporting under-represented communities is essential in realizing DEI. 
This is in line with two newly established funding opportunities by the U.S. DOE Office of Science:  Reaching a New Energy Sciences Workforce (RENEW) initiative and Funding for the Accelerated, Inclusive Research (FAIR) initiative to support research at MSIs and emerging research institutions. 

Great opportunity is also on the horizon, with the planned construction of the EIC. We envision a nationwide (distributed) “EIC Center of Excellence for Science and Diverse Workforce Development” that combines the discovery science, building and supporting research at MSIs, and developing a diverse and talented workforce for the field and beyond. 
Specifically, the Center will establish joint faculty positions between US national laboratories and MSIs, support undergraduate and graduate student fellowships, and postdoctoral fellowships focusing on students from MSIs and early-career researchers from all underrepresented groups. 
This Center will offer a concrete platform and mechanism to support and develop research at MSIs and develop a diverse STEM workforce in partnership with MSIs in a sustainable way. Additional collaborations with other minority serving professional societies will be crucial for success as well. 

\subsection{Education and Outreach}

Among the other external factors that are affecting the field, the first is the ``enrollment cliff'', the shrinking of college population as a result of the Great Recession. The enrollment drop was further accelerated in year 2020 by the Covid-19 pandemic. 
As the number of college students shrinks, the number of faculty tracks will inevitably decrease. Such decrease in faculty tracks could be countered by enhancing the number of available career opportunities in the field, as described in the previous section. Specifically, prestigious fellowships as well as joint and bridge positions will encourage universities to invest more in physics.

Another external factor that we can have an impact on is the perception that society has of physics in general. A main issue that the field of physics encounters in the US is the fact that most high school recommend students to take Biology, Chemistry, and then Physics, in this order, which results in students not taking Physics until the 11th grade. 
At that point, the more ambitious students have likely zeroed in their career path. 
This ordering of courses is partly due to the rigorous 
mathematical requirements for physics, and partly due to physics being interpreted as the hardest science because of the problem solving skills that it requires. The reasoning that one should take the hardest science the latest could be inverted, as the earlier and the longer students learn, the easier it becomes to master the critical thinking skill so important to physics education. 
Short of restructuring the American educational system, as a community we can remedy this obstacle by pushing early-education outreach to local middle and high schools, by encouraging high schoolers to take physics as soon as they finish Algebra II and to take more advanced courses in subsequent years, and by providing enrollment opportunities for introductory physics courses at nearby universities. 

One more issue that the pre-college education system faces is the lack of high quality training in physics education. Among high schoolers who actually took physics courses, some did not have a positive experience because of the teaching quality, making physics an even less desired career path.  
Some high school physics courses stayed with the ``plug and chuck" approach, which does not expose students to the essence of physics that differentiates it from other subjects. 
As a scientific community, we can partly remedy this issue by offering summer training programs for high school teachers such that they can master modern pedagogy and a deep understanding of the beauty of physics, who will in turn instill an interest and passion for physics in their students. 
Similar argument can be made for the teaching quality of college introductory (``gateway") physics courses, as student experience in such courses directly determines whether they would consider physics to be a helpful subject or a viable career. 

Finally, there is the public interpretation of what ``nuclear physics" refers to. For the general public, the term ``nuclear" is tied to nuclear reactors, nuclear 
engineering, and weapons development.  It is important that we provide the public an opportunity to understand that modern nuclear physics is the study of matter at the most fundamental level, and that the technology developed in such research would benefit society, e.g. medical imaging and radiotherapy. In this aspect, holding open houses and public lecture series would be essential, and quite practical to do at national facilities, research labs, and university departments.

\section{Computing}\label{sec:computing}
\subsection{Software and Algorithm Development, Including AI/ML}

Modern NP includes a broad research program at a varied range of collaborative scales, from a few collaborators up to large experiments at scales comparable to those typical of HEP research~\cite{Diefenthaler2022}. Consequently, there is a wide range of scales in the accompanying software efforts, from small pragmatic approaches to substantial organized software and computing activities. The trend for increased software and computational needs is well established~\cite{Geesaman:2015fha} and will continue~\cite{ATLAS_Collaboration_2022,boccali2019computing}. The NP community is conscientiously moving towards the next generation of data processing and analysis workflows that will maximize the science output~\cite{future_trends_in_nuclear_physics_computing}. Programs at NP facilities that include JLab, BNL, and the EIC in particular
\cite{
Accardi:2012qut,NAP25171,AbdulKhalek:2021gbh}, 
will continue to drive computational advances. 

AI and ML have become important tools in NP theory, simulation, and data analysis~\cite{Boehnlein2020}. 
It is anticipated that their role will continue to grow over the coming years. Organizational efforts are being made to develop best practices and common toolkits for AI/ML technology, including for the EIC~\cite{AI4EIC}. Large focused efforts such as SciDAC~\cite{SciDAC} and CSSI~\cite{CSSI}, 
as well as the establishment of AI Institutes~\cite{AI_institutes}, should receive continued support. 
Integrating the technologies and techniques developed within these efforts into larger software ecosystems will require some effort. Particularly integration into mixed heterogeneous computing environments will be a particular challenge, requiring support for career paths with multi-domain expertise over the coming decades. 
In the following, we provide two detailed examples where AI/ML has already made an impact on NP research as well as a glimpse into its possible future development. 

\subsubsection{AI/ML in Data Analysis and Experimental Design}
AI/ML has shown great promise, relative to traditional approaches, in the analysis of the large complex data sets  provided by NP experiments. For example: analysis of charged particle tracks~\cite{Gavalian2020}, reconstruction and identification of electromagnetic showers in calorimeters~\cite{Barsotti2020}, jet tagging~\cite{Bielčíková2021,Lee2022}, particle identification~\cite{Fanelli2022a}, and event-level reconstruction of kinematic observables~\cite{Diefenthaler2022}. ML has been used recently in the unfolding of H1 data~\cite{Andreassen2020,Andreev2020} as well as for fast reconstruction algorithms~\cite{Fanelli2020}. 
Simulation has also benefited through the use of generative models~\cite{Aad2022}. 
Finally, uncertainty quantification~\cite{Nachman2020,Schram2022}, robustness and explainability~\cite{Kitouni2021} are of particular importance to experimental NP, with unique requirements not being addressed by industry. 

AI/ML is also being used to assist in experimental design~\cite{Cisbani2020,Fanelli2022b}. This can help to improve the efficiency of experiments, and has the potential to reduce the cost and time required to carry out the experiment. Modern electronics in streaming readout DAQ systems~\cite{Liu2022,Ameli2022} makes it possible to incorporate high-level AI algorithms directly in the DAQ-analysis pipeline. This will lead to better data quality control and shorter analysis cycles. Autonomous control in detectors~\cite{Jeske2022} will lead to faster calibration and alignment of detectors which will eventually realize self-driven experiments.

\subsubsection{AI/ML Application in Accelerator Science}
In addition to NP and QCD research, AI/ML has wide application in other closely related areas. 
In only the last five years, the application of AI/ML to accelerator facilities has grown exponentially. 
A representative sampling of the research is given in Refs.
\cite{
PhysRevAccelBeams.25.104604,
PhysRevAccelBeams.25.104601,
PhysRevAccelBeams.25.064402,
PhysRevAccelBeams.25.014601,
PhysRevAccelBeams.24.114601,
PhysRevAccelBeams.24.104601,
PhysRevAccelBeams.24.082802,
PhysRevAccelBeams.24.072802,
PhysRevAccelBeams.23.114601,
PhysRevAccelBeams.23.044601,
PhysRevAccelBeams.23.032805,
PhysRevAccelBeams.22.093001,
PhysRevAccelBeams.22.052801,
PhysRevAccelBeams.21.112802,
PhysRevAccelBeams.21.054601}. 
Applications include improved optimization for beam tuning, surrogate models to reduce simulation run times, novel anomaly detection schemes, prognostics, and automation. 
However, we are still far from realizing the full potential of AI/ML.  Present conventional instrumentation, computing architectures and control systems were not designed to support collection of "ML-ready" data from thousands of instruments in km-scale accelerators.
For instance, the increasing need to move large amounts of data around quickly may stress current networks. A recent solicitation for SBIR proposals describes the current situation~\cite{SBIRFY2022PhaseIRelease2}. Next-generation facilities will require the entire data flow cycle -- including infrastructure, data taking, handling, storage and access -- to be revisited.

\subsection{High-Performance and High-Throughput Computing and High-Capacity Data Systems}
The NP experimental program will see increasing detector complexity as well as experiments with higher interaction rates than are common today. 
During the next two decades, 
two activities will be drivers for computing requirements: simulation and data processing. %
Simulation is necessary for hardware systems, such as accelerators and detectors, as well as science data.
Data processing includes data acquisition, calibration, reconstruction, and analysis activities. 

Simulation is well suited to be distributed as workflows across computational facilities, particularly High-Throughput Computing (HTC) facilities. 
Integration of heterogeneous hardware and AI/ML into simulation frameworks such as GEANT4 is already underway. For example, AI/ML has been successfully used to guide EIC detector design~\cite{Fanelli2022b}. These are activities where investment should continue.

The large data sets expected for NP research will require high-capacity data systems and data management tools. 
Despite large projected data rates and detector complexity,  it is anticipated that providing the compute cycles to process data from the experimental program will not be significantly more challenging than it is today. Processor performance per dollar is expected to increase, as is the use of technologies such as heterogeneous hardware and AI/ML. Continued investment in R\&D and deployment of advanced scientific computing technologies will help contain computing cost even as data set sizes grow.
Instead, the main challenges for distribution of large data sets across facilities are data transport, workflow management, and data management. The FAIR Data Principles -- Findable, Accessible, Interoperable, and Reusable -- must be followed to ensure that the data are in place when the computing resources are available. Increased investment in accessibility is needed. Additional challenges exist in cyber-security policies, which at present are not aligned, and federated access for login or service access is a patchwork. DOE is currently considering the development of an Integrated Research Infrastructure with the goal of eliminating many of these challenges. Participation by the NP community will ensure that its needs are considered.

\subsection{Workforce Development and Retention in Computing and AI/ML} 
An increased AI/ML workforce is needed to apply these techniques to NP over the coming decades.  Such an expanded workforce is needed enterprise-wide, across the DOE, not just in NP.
However, development and retention of a diverse, multi-disciplinary workforce in computing and AI/ML face their own unique challenges. 
A recent Secretary of Energy Advisory Board (SEAB) report accurately describes the current environment~\cite{SEABAIMLWorkingGroup2020}, pointing out the extraordinary difficulty for national labs to compete with the private sector in attracting AI/ML talent. A more sustainable strategy would be to provide training for domain experts who have a desire to add AI/ML proficiency to their repertoire of skills.  Conferences, workshops (such as AI for NP~\cite{AIforNuclearPhysicsWorkshop2020} or AI4EIC \cite{AI4EICWorkshop}), schools~\cite{AIforNuclearPhysicsWinterSchool}, hackathons~\cite{AI4EICHackathon,AIHackathonJLab} and other educational/training activities demonstrate the interest from the community. At the same time, effort must be made to incorporate domain experts from Data Science where possible. The careful, systematic analysis of NP data with a strong emphasis on accurate uncertainty quantification also has the potential for NP to feedback to the AI/ML best practices in the field of data Science.

\section{Nuclear Data}\label{sec:nucl_data}

Nuclear data are required for detector development and simulations of detector performance.  One of the most crucial aspects of the design of physics experiments as well as in accelerator development and medical applications is the transport and interactions of particles in a material, be it a detector for physics applications or the human body for medical applications.  The design of any experiment relies on factors such as material budget, how much material is required for each detector component; energy loss (stopping power), how far a particle will travel before it is stopped in a given material; energy and position resolution; and radiation tolerances.  Once a detector is built and being placed in operation, further simulations are necessary to understand the systematic uncertainties on the data including effects on particle tracking such as multiple scattering in the material, affecting the momentum resolution, energy loss, and particle conversion.  These transport models are also needed to determine the detector efficiency.

In high energy experiments, the code packages most commonly used are Geant4 \cite{GEANT4:2002zbu} and FLUKA \cite{Ahdida:2022gjl}.  For example, the data used in Geant4 for photon evaporation, radioactive decay, and nuclide properties are taken directly from the Evaluated Nuclear Structure Data File (ENSDF) \cite{Tepel:1984}, maintained at the National Nuclear Data Center at BNL~\cite{NNDC}.  Neutron cross sections and final states are based on nuclear data libraries such as JEFF-3.3 \cite{Plompen:2020due} and ENDF/B-VII.1 \cite{Chadwick:2011xwu} while the TENDL library \cite{Koning:2019qbo} is used for interactions of incident protons with matter.  The SAID database is used for proton, neutron and pion inelastic, elastic and and charge exchange reaction cross sections for interactions with nucleons below 3 GeV \cite{Arndt:2007qn}.  Nuclear shell effects are based on the liquid drop model of the nucleus, including ground state deformations.  Nuclear data are also required for the nuclear density profiles, photoelectric interactions, impact ionization, and optical reflectance, see Ref.~\cite{GEANT4:2002zbu} for more references and details.

Nuclear data have played a direct role in data analysis by the ALICE Collaboration \cite{ALICE:2022zuz}.  The collaboration was able to make the first determination of the $^3\overline{\rm He}$ (anti $^3$He) absorption cross section in matter by its interactions in different components of the detector made up of different materials with different average nuclear mass values.  This result has cosmological implications for $^3\overline{\rm He}$ production in the galaxy by cosmic ray interactions and dark matter annihilation \cite{ALICE:2022zuz}.

Recently, it has been suggested that a scan of collision species at colliders could complement low energy nuclear structure studies \cite{Bally:2022vgo}.  In particular, the ground state deformations of nuclei plays an important role in the initial conditions of the quark-gluon plasma, leading to very different predictions of the transverse flow patterns for collisions of nuclei with the same mass number but different nuclear shapes, as was shown for collisions of cylindrically-shaped $^{96}{\rm Ru} + ^{96}{\rm Ru}$  compared to those of more irregularly-shaped $^{96}{\rm Zr} + ^{96}{\rm Zr}$ \cite{Bally:2022vgo}.  This is similar to the motivation for the earlier ${\rm U} + {\rm U}$ run at RHIC: collisions of strongly deformed $^{238}$U nuclei could lead to very different initial densities and temperatures depending on whether the collisions were tip-to-tip or side-on-side \cite{Nepali:2006ep}.

Nuclear data also play an important role in applications.  Space exploration is one such application where high energy nuclear data in particular are critical, primarily due to the harmful effects of the space radiation environment.  The wide range of energies, up to the TeV scale, and species, $1 < Z < 28$, of galactic cosmic rays (GCRs) \cite{Badhwar:1992} make it challenging to determine all their potential effects on spacecraft and astronauts.  While the Earth's atmosphere has a protective effect, cosmic ray showers reach the ground all over the Earth and, in fact, have been studied using collider detectors.  In particular, muons from cosmic rays pass all the way through these detectors, producing tracks perpendicular to those from beam-beam collisions and are present even when the beam is not on, see Refs.~\cite{Grupen:2003ih,Ridky:2005mx,L3:2004sed,ALICE:2015wfa}.  The ALICE detector at the LHC includes the dedicated cosmic ray detector ACORDE \cite{FernandezTellez:2007yxa}, used in analysis of Ref.~\cite{ALICE:2015wfa}.  Collisions of GCRs with nuclei in the Earth's atmosphere or a spacecraft in orbit can generate showers of particles, including pions, muons, neutrinos, electrons, and photons as well as protons and neutrons. 

The penetrating power of the initial GCRs and the secondaries generated by their interaction with matter, can have a serious impact on the safety and viability of space exploration.  The 1\% of GCR primaries heavier than He nuclei can be especially serious because the damage they inflict scales as $Z^2$.  The secondary particles generated from GCR interactions with spacecraft material \cite{Finckenor:2018} such as aluminum, polyethylene, and composites can harm astronauts and disrupt or disable electronic systems. The spacecraft shielding designed to reduce the GCR flux is also a target that can increase the secondary flux. Because of the wide variety of possible shielding materials and thicknesses, modeling is essential to determine the sensitivity of the secondaries (both in flux and composition) to different shielding configurations, as well as to determine the subsequent harmful impact of those secondaries on electronic systems \cite{Hoeffgen:2020} and humans \cite{Durante:2011}.

Understanding the effects of the highest energy cosmic rays requires high energy (GeV range) nuclear data and modeling.  However, there are no measurements for incident projectile energies greater than 3~GeV/nucleon.  There is a possibility to fill part of these critical gaps in nuclear data employing fixed-target collisions at RHIC. A proposal \cite{45} was recently made to bombard C, Al, and Fe targets with C, Al, and Fe ions at energies from 5 to 50 GeV, and measure the produced secondaries using the STAR detector. This measurement, however, would have to be completed before RHIC is shut down and EIC construction has begun.

Due to the lack of data at the appropriate energies, simulations of space radiation effects have large uncertainties.  The space research community has generally relied on phenomenological nuclear reaction models such as the Double Differential Fragmentation model (DDFRG) \cite{Norbury:2021coa}.  Many of the models rely on abrasion-ablation models \cite{Hufner:1975zz,Werneth:2021} or semi-empirical parameterizations, see Ref.~\cite{Luoni:2021bne}.  Researchers modeling these interactions could benefit from codes developed to study data from RHIC.  The use of hadronic cascade models such as the UrQMD code \cite{Bleicher:1999xi}, which was shown to be able to predict proton and deuteron yields from the BNL Alternating Gradient Synchrotron studies of 15~GeV protons on Be and Au targets \cite{Sombun:2018yqh,E-802:1991unu}, could significantly advance simulations of collisions relevant for space exploration.  For further information about nuclear data needs for space applications, see Refs.~\cite{Kolos:2022stv,Smith:2022xyz}.

\vspace*{1cm}
\centerline{\Large\bf {Acknowledgement}}

\vspace*{0.4cm}
We would like to thank the Laboratory for Nuclear Science (LNS) and the Physics Department of MIT for hosting the QCD Town Meeting. We would like to thank LNS staff: Alisa Cabral, Anna Maria Convertino, Iling Hong, Elsye Luc, Laura Pingcuoluomu, Lauren Saragosa, Caitlin Sulham, and Lily Xu for their support.

\appendix
\section{Agenda of the Hot \& Cold QCD Town Meeting}
\begin{center} \includegraphics[width=\textwidth]{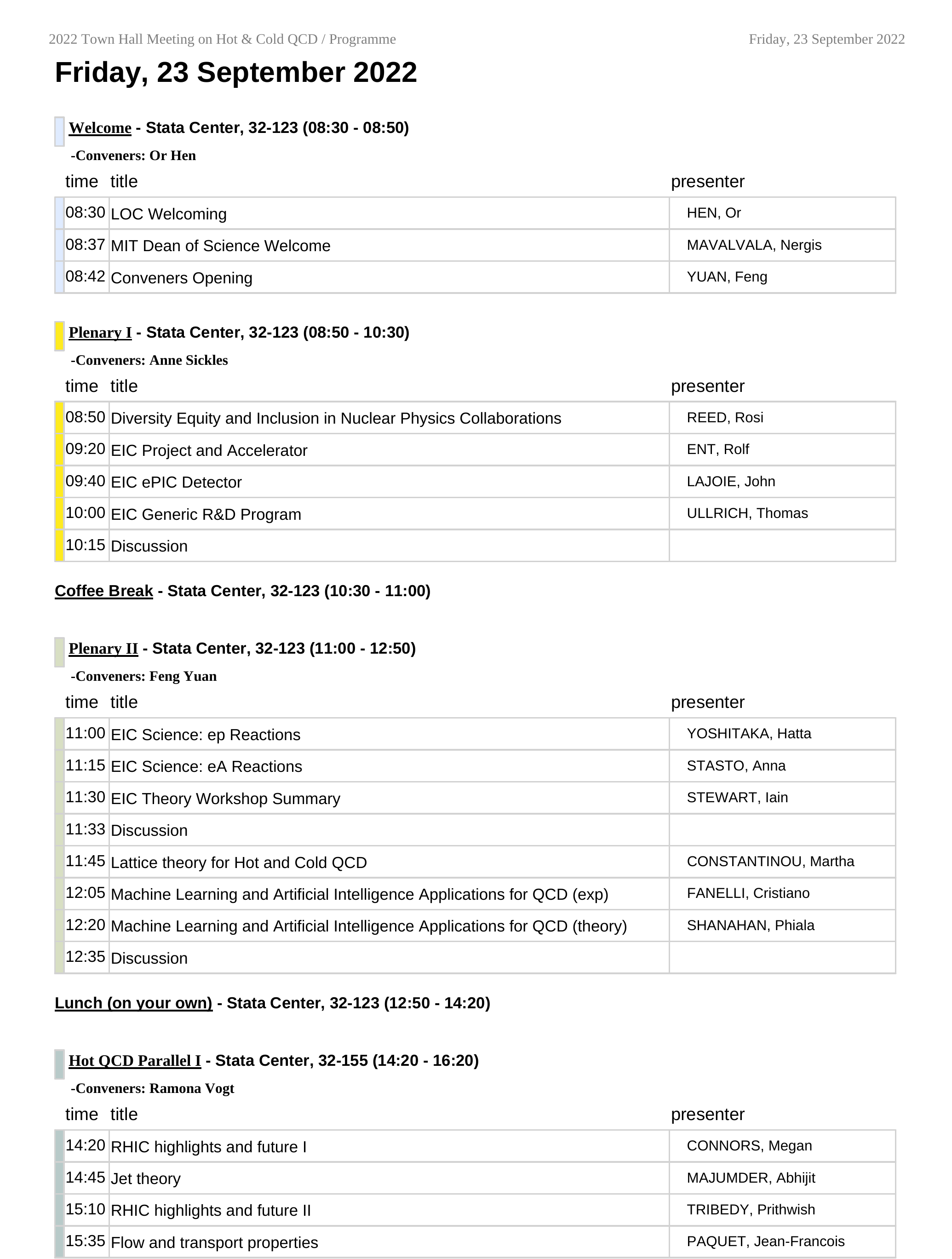} \end{center}
\newpage
\begin{center} \includegraphics[page=2,width=\textwidth]{files_app/timetable-nopage.pdf} \end{center}
\newpage
\begin{center} \includegraphics[page=3,width=\textwidth]{files_app/timetable-nopage.pdf} \end{center}
\newpage
\begin{center} \includegraphics[page=4,width=\textwidth]{files_app/timetable-nopage.pdf} \end{center}
\newpage
\begin{center} \includegraphics[page=5,width=\textwidth]{files_app/timetable-nopage.pdf} \end{center}
\newpage

\begin{center} 
\includegraphics[width=\textwidth]{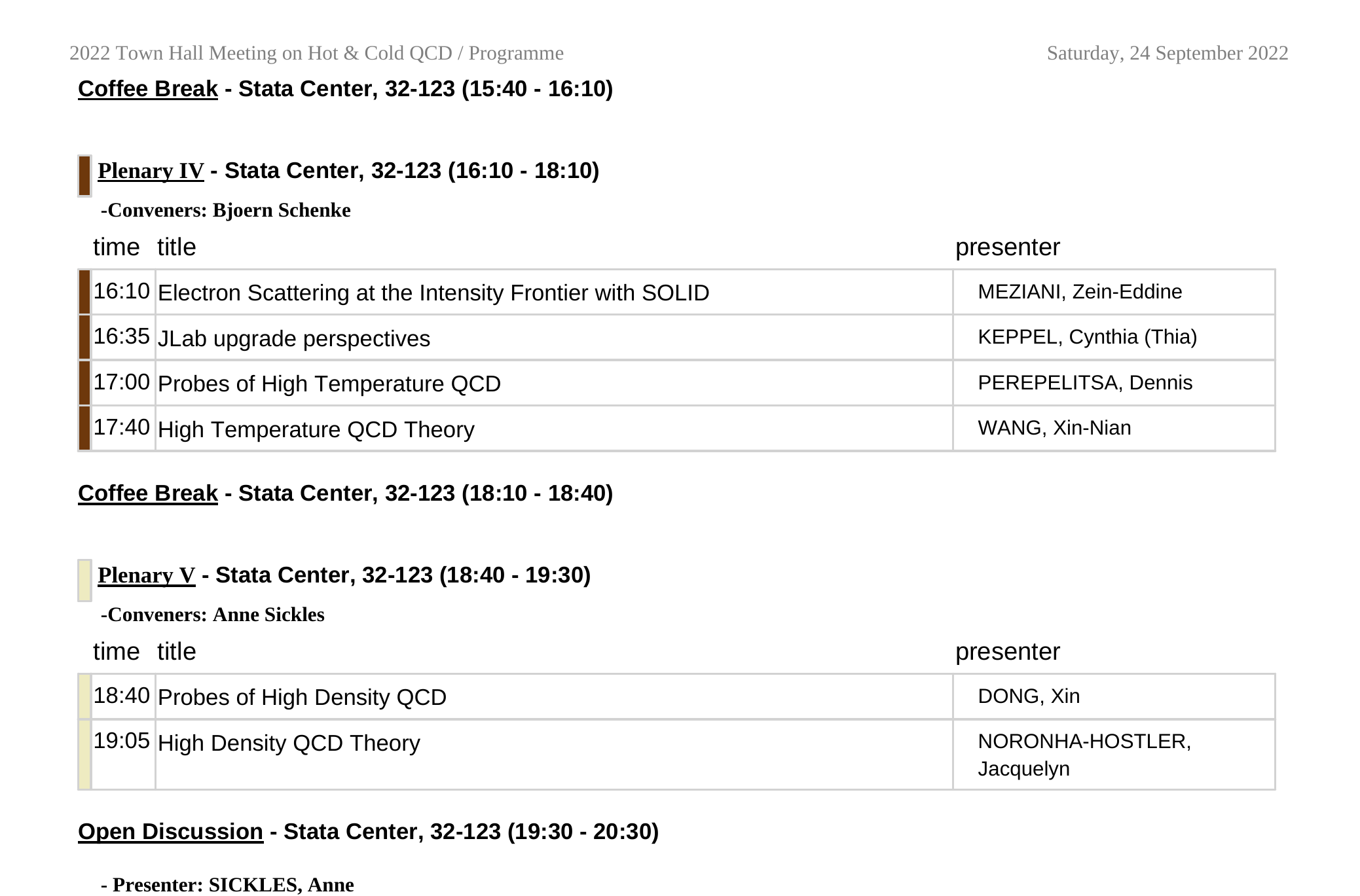}
\end{center}

\vspace*{1cm}
\begin{center} 
\includegraphics[width=\textwidth]{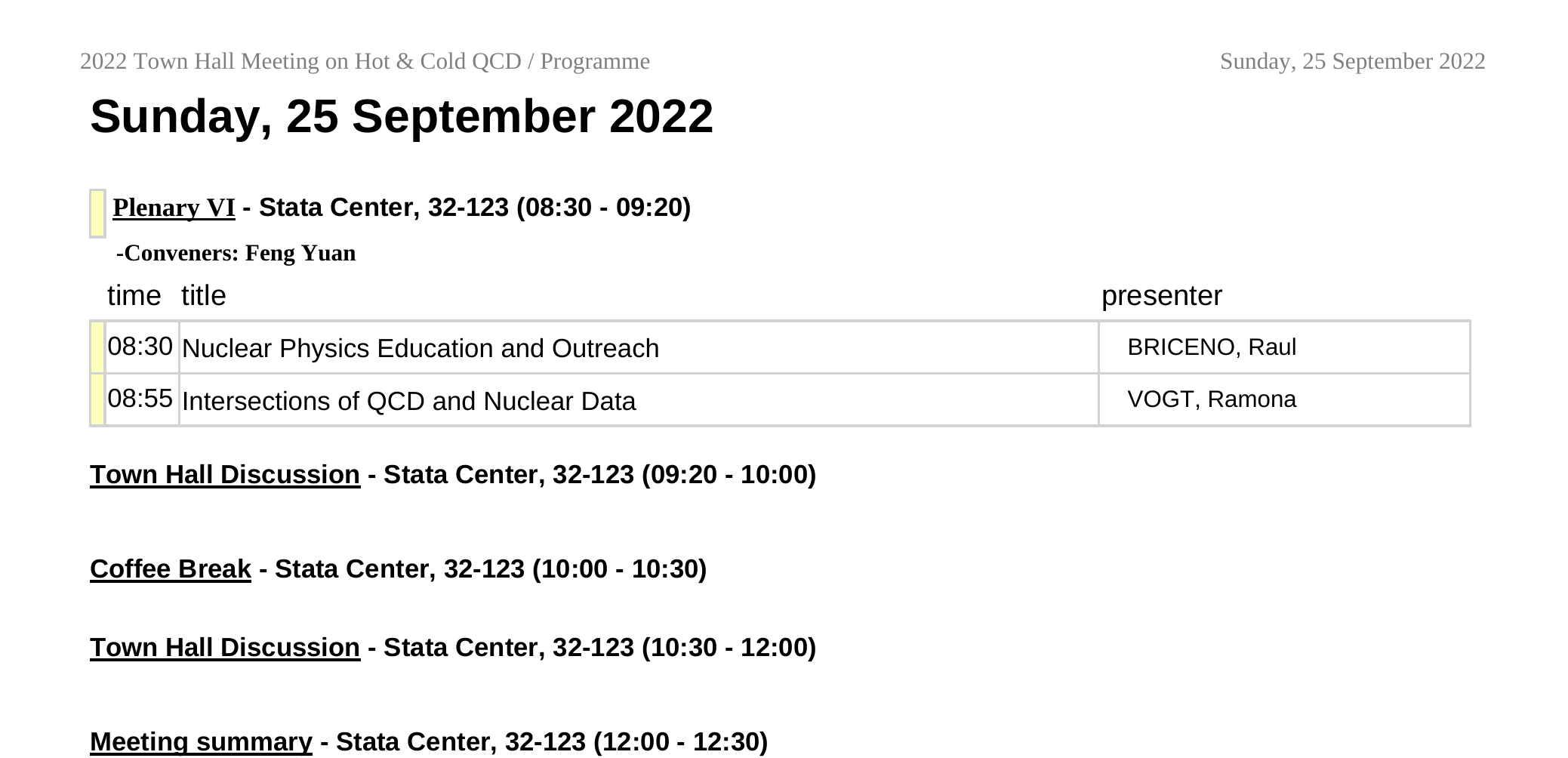}
\end{center}

\newpage


\newpage
\section{References}
\bibliographystyle{unsrturl}
\bibliography{
files_bib/hotQCD,
files_bib/hotQCD_no-inspire,
files_bib/coldQCD, 
files_bib/computing,
files_bib/nucleardata,
files_bib/eic,
files_bib/other_fields,
files_bib/future_others,
files_bib/theory,
files_bib/heavyflavor_QIS
}

\begin{thebibliography}{1000}

\bibitem{MITTownMeeting}
2022 {T}own {H}all {M}eeting on {H}ot and {C}old {QCD}.
\newblock URL: \url{https://indico.mit.edu/event/538/}.

\bibitem{Geesaman:2015fha}
Ani Aprahamian et~al.
\newblock {Reaching for the horizon: The 2015 long range plan for nuclear
  science}.
\newblock 10 2015.

\bibitem{Sorensen:2009cz}
Paul Sorensen.
\newblock {\em {Elliptic Flow: A Study of Space-Momentum Correlations In
  Relativistic Nuclear Collisions}}, pages 323--374.
\newblock 2010.
\newblock \href {http://arxiv.org/abs/0905.0174} {\path{arXiv:0905.0174}},
  \href {https://doi.org/10.1142/9789814293297_0006}
  {\path{doi:10.1142/9789814293297_0006}}.

\bibitem{Shen:2015msa}
Chun Shen and Ulrich Heinz.
\newblock {The road to precision: Extraction of the specific shear viscosity of
  the quark-gluon plasma}.
\newblock {\em Nucl. Phys. News}, 25(2):6--11, 2015.
\newblock \href {http://arxiv.org/abs/1507.01558} {\path{arXiv:1507.01558}},
  \href {https://doi.org/10.1080/10619127.2015.1006502}
  {\path{doi:10.1080/10619127.2015.1006502}}.

\bibitem{ALICE:2016ccg}
Jaroslav Adam et~al.
\newblock {Anisotropic flow of charged particles in Pb-Pb collisions at
  $\sqrt{s_{\rm NN}}=5.02$ TeV}.
\newblock {\em Phys. Rev. Lett.}, 116(13):132302, 2016.
\newblock \href {http://arxiv.org/abs/1602.01119} {\path{arXiv:1602.01119}},
  \href {https://doi.org/10.1103/PhysRevLett.116.132302}
  {\path{doi:10.1103/PhysRevLett.116.132302}}.

\bibitem{CMS:2017xgk}
A.~M. Sirunyan et~al.
\newblock {Azimuthal anisotropy of charged particles with transverse momentum
  up to 100 GeV/ c in PbPb collisions at $\sqrt {s}_{{NN}}$=5.02 TeV}.
\newblock {\em Phys. Lett. B}, 776:195--216, 2018.
\newblock \href {http://arxiv.org/abs/1702.00630} {\path{arXiv:1702.00630}},
  \href {https://doi.org/10.1016/j.physletb.2017.11.041}
  {\path{doi:10.1016/j.physletb.2017.11.041}}.

\bibitem{ALICE:2018lao}
S.~Acharya et~al.
\newblock {Anisotropic flow in Xe-Xe collisions at $\mathbf{\sqrt{s_{\rm{NN}}}
  = 5.44}$ TeV}.
\newblock {\em Phys. Lett. B}, 784:82--95, 2018.
\newblock \href {http://arxiv.org/abs/1805.01832} {\path{arXiv:1805.01832}},
  \href {https://doi.org/10.1016/j.physletb.2018.06.059}
  {\path{doi:10.1016/j.physletb.2018.06.059}}.

\bibitem{ATLAS:2018ezv}
Morad Aaboud et~al.
\newblock {Measurement of the azimuthal anisotropy of charged particles
  produced in $\sqrt{s_{_\text {NN}}}$ = 5.02 TeV Pb+Pb collisions with the
  ATLAS detector}.
\newblock {\em Eur. Phys. J. C}, 78(12):997, 2018.
\newblock \href {http://arxiv.org/abs/1808.03951} {\path{arXiv:1808.03951}},
  \href {https://doi.org/10.1140/epjc/s10052-018-6468-7}
  {\path{doi:10.1140/epjc/s10052-018-6468-7}}.

\bibitem{PHENIX:2018lia}
C.~Aidala et~al.
\newblock {Creation of quark\textendash{}gluon plasma droplets with three
  distinct geometries}.
\newblock {\em Nature Phys.}, 15(3):214--220, 2019.
\newblock \href {http://arxiv.org/abs/1805.02973} {\path{arXiv:1805.02973}},
  \href {https://doi.org/10.1038/s41567-018-0360-0}
  {\path{doi:10.1038/s41567-018-0360-0}}.

\bibitem{CMS:2019cyz}
Albert~M Sirunyan et~al.
\newblock {Charged-particle angular correlations in XeXe collisions at
  $\sqrt{s_{_\mathrm{NN}}}=$ 5.44 TeV}.
\newblock {\em Phys. Rev. C}, 100(4):044902, 2019.
\newblock \href {http://arxiv.org/abs/1901.07997} {\path{arXiv:1901.07997}},
  \href {https://doi.org/10.1103/PhysRevC.100.044902}
  {\path{doi:10.1103/PhysRevC.100.044902}}.

\bibitem{ATLAS:2019dct}
Georges Aad et~al.
\newblock {Measurement of the azimuthal anisotropy of charged-particle
  production in $Xe+Xe$ collisions at $\sqrt{s_{\mathrm{NN}}}=5.44$ TeV with
  the ATLAS detector}.
\newblock {\em Phys. Rev. C}, 101(2):024906, 2020.
\newblock \href {http://arxiv.org/abs/1911.04812} {\path{arXiv:1911.04812}},
  \href {https://doi.org/10.1103/PhysRevC.101.024906}
  {\path{doi:10.1103/PhysRevC.101.024906}}.

\bibitem{STAR:2019zaf}
Jaroslav Adam et~al.
\newblock {Azimuthal Harmonics in Small and Large Collision Systems at RHIC Top
  Energies}.
\newblock {\em Phys. Rev. Lett.}, 122(17):172301, 2019.
\newblock \href {http://arxiv.org/abs/1901.08155} {\path{arXiv:1901.08155}},
  \href {https://doi.org/10.1103/PhysRevLett.122.172301}
  {\path{doi:10.1103/PhysRevLett.122.172301}}.

\bibitem{STAR:2022gki}
Mohamed Abdallah et~al.
\newblock {Collision-System and Beam-Energy Dependence of Anisotropic Flow
  Fluctuations}.
\newblock {\em Phys. Rev. Lett.}, 129(25):252301, 2022.
\newblock \href {http://arxiv.org/abs/2201.10365} {\path{arXiv:2201.10365}},
  \href {https://doi.org/10.1103/PhysRevLett.129.252301}
  {\path{doi:10.1103/PhysRevLett.129.252301}}.

\bibitem{ALICE:2022wpn}
{The ALICE experiment -- A journey through QCD}.
\newblock 11 2022.
\newblock \href {http://arxiv.org/abs/2211.04384} {\path{arXiv:2211.04384}}.

\bibitem{ALICE:2018yph}
S.~Acharya et~al.
\newblock {Anisotropic flow of identified particles in Pb-Pb collisions at $
  {\sqrt{s}}_{\mathrm{NN}}=5.02 $ TeV}.
\newblock {\em JHEP}, 09:006, 2018.
\newblock \href {http://arxiv.org/abs/1805.04390} {\path{arXiv:1805.04390}},
  \href {https://doi.org/10.1007/JHEP09(2018)006}
  {\path{doi:10.1007/JHEP09(2018)006}}.

\bibitem{STAR:2022ncy}
Mohamed Abdallah et~al.
\newblock {Centrality and transverse momentum dependence of higher-order flow
  harmonics of identified hadrons in Au+Au collisions at $\sqrt{s_{NN}}$ = 200
  GeV}.
\newblock {\em Phys. Rev. C}, 105(6):064911, 2022.
\newblock \href {http://arxiv.org/abs/2203.07204} {\path{arXiv:2203.07204}},
  \href {https://doi.org/10.1103/PhysRevC.105.064911}
  {\path{doi:10.1103/PhysRevC.105.064911}}.

\bibitem{CMS:2022bmk}
{Strange hadron collectivity in pPb and PbPb collisions}.
\newblock 4 2022.
\newblock \href {http://arxiv.org/abs/2205.00080} {\path{arXiv:2205.00080}}.

\bibitem{ATLAS:2014qxy}
Georges Aad et~al.
\newblock {Measurement of flow harmonics with multi-particle cumulants in Pb+Pb
  collisions at $\sqrt{s_{\mathrm {NN}}}=2.76$ TeV with the ATLAS detector}.
\newblock {\em Eur. Phys. J. C}, 74(11):3157, 2014.
\newblock \href {http://arxiv.org/abs/1408.4342} {\path{arXiv:1408.4342}},
  \href {https://doi.org/10.1140/epjc/s10052-014-3157-z}
  {\path{doi:10.1140/epjc/s10052-014-3157-z}}.

\bibitem{ALICE:2014dwt}
Betty~Bezverkhny Abelev et~al.
\newblock {Multiparticle azimuthal correlations in p -Pb and Pb-Pb collisions
  at the CERN Large Hadron Collider}.
\newblock {\em Phys. Rev. C}, 90(5):054901, 2014.
\newblock \href {http://arxiv.org/abs/1406.2474} {\path{arXiv:1406.2474}},
  \href {https://doi.org/10.1103/PhysRevC.90.054901}
  {\path{doi:10.1103/PhysRevC.90.054901}}.

\bibitem{CMS:2013jlh}
Serguei Chatrchyan et~al.
\newblock {Multiplicity and Transverse Momentum Dependence of Two- and
  Four-Particle Correlations in pPb and PbPb Collisions}.
\newblock {\em Phys. Lett. B}, 724:213--240, 2013.
\newblock \href {http://arxiv.org/abs/1305.0609} {\path{arXiv:1305.0609}},
  \href {https://doi.org/10.1016/j.physletb.2013.06.028}
  {\path{doi:10.1016/j.physletb.2013.06.028}}.

\bibitem{STAR:2014ofx}
N.~M. Abdelwahab et~al.
\newblock {Isolation of Flow and Nonflow Correlations by Two- and Four-Particle
  Cumulant Measurements of Azimuthal Harmonics in $\sqrt{s_{_{\rm NN}}} =$ 200
  GeV Au+Au Collisions}.
\newblock {\em Phys. Lett. B}, 745:40--47, 2015.
\newblock \href {http://arxiv.org/abs/1409.2043} {\path{arXiv:1409.2043}},
  \href {https://doi.org/10.1016/j.physletb.2015.04.033}
  {\path{doi:10.1016/j.physletb.2015.04.033}}.

\bibitem{STAR:2015mki}
L.~Adamczyk et~al.
\newblock {Azimuthal anisotropy in U$+$U and Au$+$Au collisions at RHIC}.
\newblock {\em Phys. Rev. Lett.}, 115(22):222301, 2015.
\newblock \href {http://arxiv.org/abs/1505.07812} {\path{arXiv:1505.07812}},
  \href {https://doi.org/10.1103/PhysRevLett.115.222301}
  {\path{doi:10.1103/PhysRevLett.115.222301}}.

\bibitem{ATLAS:2017rtr}
Morad Aaboud et~al.
\newblock {Measurement of long-range multiparticle azimuthal correlations with
  the subevent cumulant method in $pp$ and $p + Pb$ collisions with the ATLAS
  detector at the CERN Large Hadron Collider}.
\newblock {\em Phys. Rev. C}, 97(2):024904, 2018.
\newblock \href {http://arxiv.org/abs/1708.03559} {\path{arXiv:1708.03559}},
  \href {https://doi.org/10.1103/PhysRevC.97.024904}
  {\path{doi:10.1103/PhysRevC.97.024904}}.

\bibitem{CMS:2017glf}
Albert~M Sirunyan et~al.
\newblock {Non-Gaussian elliptic-flow fluctuations in PbPb collisions at
  $\sqrt{\smash[b]{s_{_\text{NN}}}} = 5.02$ TeV}.
\newblock {\em Phys. Lett. B}, 789:643--665, 2019.
\newblock \href {http://arxiv.org/abs/1711.05594} {\path{arXiv:1711.05594}},
  \href {https://doi.org/10.1016/j.physletb.2018.11.063}
  {\path{doi:10.1016/j.physletb.2018.11.063}}.

\bibitem{ALICE:2019zfl}
Shreyasi Acharya et~al.
\newblock {Investigations of Anisotropic Flow Using Multiparticle Azimuthal
  Correlations in pp, p-Pb, Xe-Xe, and Pb-Pb Collisions at the LHC}.
\newblock {\em Phys. Rev. Lett.}, 123(14):142301, 2019.
\newblock \href {http://arxiv.org/abs/1903.01790} {\path{arXiv:1903.01790}},
  \href {https://doi.org/10.1103/PhysRevLett.123.142301}
  {\path{doi:10.1103/PhysRevLett.123.142301}}.

\bibitem{ATLAS:2019peb}
Morad Aaboud et~al.
\newblock {Fluctuations of anisotropic flow in Pb+Pb collisions at $
  \sqrt{{\mathrm{s}}_{\mathrm{NN}}} $ = 5.02 TeV with the ATLAS detector}.
\newblock {\em JHEP}, 01:051, 2020.
\newblock \href {http://arxiv.org/abs/1904.04808} {\path{arXiv:1904.04808}},
  \href {https://doi.org/10.1007/JHEP01(2020)051}
  {\path{doi:10.1007/JHEP01(2020)051}}.

\bibitem{ALICE:2016kpq}
Jaroslav Adam et~al.
\newblock {Correlated event-by-event fluctuations of flow harmonics in Pb-Pb
  collisions at $\sqrt{s_{_{\rm NN}}}=2.76$ TeV}.
\newblock {\em Phys. Rev. Lett.}, 117:182301, 2016.
\newblock \href {http://arxiv.org/abs/1604.07663} {\path{arXiv:1604.07663}},
  \href {https://doi.org/10.1103/PhysRevLett.117.182301}
  {\path{doi:10.1103/PhysRevLett.117.182301}}.

\bibitem{CMS:2017kcs}
Albert~M Sirunyan et~al.
\newblock {Observation of Correlated Azimuthal Anisotropy Fourier Harmonics in
  $pp$ and $p+Pb$ Collisions at the LHC}.
\newblock {\em Phys. Rev. Lett.}, 120(9):092301, 2018.
\newblock \href {http://arxiv.org/abs/1709.09189} {\path{arXiv:1709.09189}},
  \href {https://doi.org/10.1103/PhysRevLett.120.092301}
  {\path{doi:10.1103/PhysRevLett.120.092301}}.

\bibitem{ATLAS:2018ngv}
Morad Aaboud et~al.
\newblock {Correlated long-range mixed-harmonic fluctuations measured in $pp$,
  $p$+Pb and low-multiplicity Pb+Pb collisions with the ATLAS detector}.
\newblock {\em Phys. Lett. B}, 789:444--471, 2019.
\newblock \href {http://arxiv.org/abs/1807.02012} {\path{arXiv:1807.02012}},
  \href {https://doi.org/10.1016/j.physletb.2018.11.065}
  {\path{doi:10.1016/j.physletb.2018.11.065}}.

\bibitem{CMS:2019nct}
Albert~M Sirunyan et~al.
\newblock {Mixed higher-order anisotropic flow and nonlinear response
  coefficients of charged particles in $\mathrm {PbPb}$ collisions at
  $\sqrt{\smash [b]{s_{_{\mathrm {NN}}}}} = 2.76$ and 5.02$\,\text {TeV}$}.
\newblock {\em Eur. Phys. J. C}, 80(6):534, 2020.
\newblock \href {http://arxiv.org/abs/1910.08789} {\path{arXiv:1910.08789}},
  \href {https://doi.org/10.1140/epjc/s10052-020-7834-9}
  {\path{doi:10.1140/epjc/s10052-020-7834-9}}.

\bibitem{ALICE:2021adw}
Shreyasi Acharya et~al.
\newblock {Measurements of mixed harmonic cumulants in Pb\textendash{}Pb
  collisions at $\sqrt {s_{NN}}$ = 5.02 TeV}.
\newblock {\em Phys. Lett. B}, 818:136354, 2021.
\newblock \href {http://arxiv.org/abs/2102.12180} {\path{arXiv:2102.12180}},
  \href {https://doi.org/10.1016/j.physletb.2021.136354}
  {\path{doi:10.1016/j.physletb.2021.136354}}.

\bibitem{ALICE:2021gxt}
Shreyasi Acharya et~al.
\newblock {Characterizing the initial conditions of heavy-ion collisions at the
  LHC with mean transverse momentum and anisotropic flow correlations}.
\newblock {\em Phys. Lett. B}, 834:137393, 2022.
\newblock \href {http://arxiv.org/abs/2111.06106} {\path{arXiv:2111.06106}},
  \href {https://doi.org/10.1016/j.physletb.2022.137393}
  {\path{doi:10.1016/j.physletb.2022.137393}}.

\bibitem{ATLAS:2022dov}
{Correlations between flow and transverse momentum in Xe+Xe and Pb+Pb
  collisions at the LHC with the ATLAS detector: a probe of the heavy-ion
  initial state and nuclear deformation}.
\newblock 4 2022.
\newblock \href {http://arxiv.org/abs/2205.00039} {\path{arXiv:2205.00039}}.

\bibitem{Giacalone:2020byk}
Giuliano Giacalone, Bj\"orn Schenke, and Chun Shen.
\newblock {Observable signatures of initial state momentum anisotropies in
  nuclear collisions}.
\newblock {\em Phys. Rev. Lett.}, 125(19):192301, 2020.
\newblock \href {http://arxiv.org/abs/2006.15721} {\path{arXiv:2006.15721}},
  \href {https://doi.org/10.1103/PhysRevLett.125.192301}
  {\path{doi:10.1103/PhysRevLett.125.192301}}.

\bibitem{CMS-PAS-HIN-21-012}
{Correlations between multiparticle cumulants and mean transverse momentum in
  small collision systems with the CMS detector}.
\newblock Technical report, CERN, Geneva, 2022.
\newblock URL: \url{https://cds.cern.ch/record/2805932}.

\bibitem{PHENIX:2004yan}
S.~S. Adler et~al.
\newblock {Bose-Einstein correlations of charged pion pairs in Au + Au
  collisions at s(NN)**(1/2) = 200-GeV}.
\newblock {\em Phys. Rev. Lett.}, 93:152302, 2004.
\newblock \href {http://arxiv.org/abs/nucl-ex/0401003}
  {\path{arXiv:nucl-ex/0401003}}, \href
  {https://doi.org/10.1103/PhysRevLett.93.152302}
  {\path{doi:10.1103/PhysRevLett.93.152302}}.

\bibitem{STAR:2004qya}
J.~Adams et~al.
\newblock {Pion interferometry in Au+Au collisions at S(NN)**(1/2) = 200-GeV}.
\newblock {\em Phys. Rev. C}, 71:044906, 2005.
\newblock \href {http://arxiv.org/abs/nucl-ex/0411036}
  {\path{arXiv:nucl-ex/0411036}}, \href
  {https://doi.org/10.1103/PhysRevC.71.044906}
  {\path{doi:10.1103/PhysRevC.71.044906}}.

\bibitem{ALICE:2015hvw}
Jaroslav Adam et~al.
\newblock {One-dimensional pion, kaon, and proton femtoscopy in Pb-Pb
  collisions at $\sqrt{s_{\rm {NN}}}$ =2.76 TeV}.
\newblock {\em Phys. Rev. C}, 92(5):054908, 2015.
\newblock \href {http://arxiv.org/abs/1506.07884} {\path{arXiv:1506.07884}},
  \href {https://doi.org/10.1103/PhysRevC.92.054908}
  {\path{doi:10.1103/PhysRevC.92.054908}}.

\bibitem{CMS:2017mdg}
Albert~M Sirunyan et~al.
\newblock {Bose-Einstein correlations in $pp, p\mathrm{Pb}$, and PbPb
  collisions at $\sqrt{{s}_{NN}}=0.9-7$ TeV}.
\newblock {\em Phys. Rev. C}, 97(6):064912, 2018.
\newblock \href {http://arxiv.org/abs/1712.07198} {\path{arXiv:1712.07198}},
  \href {https://doi.org/10.1103/PhysRevC.97.064912}
  {\path{doi:10.1103/PhysRevC.97.064912}}.

\bibitem{ATLAS:2017shk}
Morad Aaboud et~al.
\newblock {Femtoscopy with identified charged pions in proton-lead collisions
  at $\sqrt{s_{\mathrm{NN}}}=5.02$ TeV with ATLAS}.
\newblock {\em Phys. Rev. C}, 96(6):064908, 2017.
\newblock \href {http://arxiv.org/abs/1704.01621} {\path{arXiv:1704.01621}},
  \href {https://doi.org/10.1103/PhysRevC.96.064908}
  {\path{doi:10.1103/PhysRevC.96.064908}}.

\bibitem{CMS:2023jjt}
{K$^0_\mathrm{S}$ and $\Lambda(\overline\Lambda)$ two-particle femtoscopic
  correlations in PbPb collisions at $\sqrt{s_\mathrm{NN}}$ = 5.02 TeV}.
\newblock 1 2023.
\newblock \href {http://arxiv.org/abs/2301.05290} {\path{arXiv:2301.05290}}.

\bibitem{Adamczyk:2015obl}
L.~Adamczyk et~al.
\newblock {Azimuthal anisotropy in U$+$U and Au$+$Au collisions at RHIC}.
\newblock {\em Phys. Rev. Lett.}, 115(22):222301, 2015.
\newblock \href {http://arxiv.org/abs/1505.07812} {\path{arXiv:1505.07812}},
  \href {https://doi.org/10.1103/PhysRevLett.115.222301}
  {\path{doi:10.1103/PhysRevLett.115.222301}}.

\bibitem{Adam:2019woz}
Jaroslav Adam et~al.
\newblock {Azimuthal Harmonics in Small and Large Collision Systems at RHIC Top
  Energies}.
\newblock {\em Phys. Rev. Lett.}, 122(17):172301, 2019.
\newblock \href {http://arxiv.org/abs/1901.08155} {\path{arXiv:1901.08155}},
  \href {https://doi.org/10.1103/PhysRevLett.122.172301}
  {\path{doi:10.1103/PhysRevLett.122.172301}}.

\bibitem{Aidala:2017ajz}
C.~Aidala et~al.
\newblock {Measurements of Multiparticle Correlations in $d+\mathrm{Au}$
  Collisions at 200, 62.4, 39, and 19.6 GeV and $p+\mathrm{Au}$ Collisions at
  200 GeV and Implications for Collective Behavior}.
\newblock {\em Phys. Rev. Lett.}, 120(6):062302, 2018.
\newblock \href {http://arxiv.org/abs/1707.06108} {\path{arXiv:1707.06108}},
  \href {https://doi.org/10.1103/PhysRevLett.120.062302}
  {\path{doi:10.1103/PhysRevLett.120.062302}}.

\bibitem{Schenke:2020mbo}
Bjoern Schenke, Chun Shen, and Prithwish Tribedy.
\newblock {Running the gamut of high energy nuclear collisions}.
\newblock {\em Phys. Rev. C}, 102(4):044905, 2020.
\newblock \href {http://arxiv.org/abs/2005.14682} {\path{arXiv:2005.14682}},
  \href {https://doi.org/10.1103/PhysRevC.102.044905}
  {\path{doi:10.1103/PhysRevC.102.044905}}.

\bibitem{Shen:2022oyg}
Chun Shen and Bj\"orn Schenke.
\newblock {Longitudinal dynamics and particle production in relativistic
  nuclear collisions}.
\newblock {\em Phys. Rev. C}, 105(6):064905, 2022.
\newblock \href {http://arxiv.org/abs/2203.04685} {\path{arXiv:2203.04685}},
  \href {https://doi.org/10.1103/PhysRevC.105.064905}
  {\path{doi:10.1103/PhysRevC.105.064905}}.

\bibitem{Carzon:2019qja}
Patrick Carzon, Mauricio Martinez, Matthew~D. Sievert, Douglas~E. Wertepny, and
  Jacquelyn Noronha-Hostler.
\newblock {Monte Carlo event generator for initial conditions of conserved
  charges in nuclear geometry}.
\newblock {\em Phys. Rev. C}, 105(3):034908, 2022.
\newblock \href {http://arxiv.org/abs/1911.12454} {\path{arXiv:1911.12454}},
  \href {https://doi.org/10.1103/PhysRevC.105.034908}
  {\path{doi:10.1103/PhysRevC.105.034908}}.

\bibitem{Schenke:2016ksl}
Bjoern Schenke and Soeren Schlichting.
\newblock {3D glasma initial state for relativistic heavy ion collisions}.
\newblock {\em Phys. Rev. C}, 94(4):044907, 2016.
\newblock \href {http://arxiv.org/abs/1605.07158} {\path{arXiv:1605.07158}},
  \href {https://doi.org/10.1103/PhysRevC.94.044907}
  {\path{doi:10.1103/PhysRevC.94.044907}}.

\bibitem{Schenke:2022mjv}
Bjoern Schenke, Soeren Schlichting, and Pragya Singh.
\newblock {Rapidity dependence of initial state geometry and momentum
  correlations in p+Pb collisions}.
\newblock {\em Phys. Rev. D}, 105(9):094023, 2022.
\newblock \href {http://arxiv.org/abs/2201.08864} {\path{arXiv:2201.08864}},
  \href {https://doi.org/10.1103/PhysRevD.105.094023}
  {\path{doi:10.1103/PhysRevD.105.094023}}.

\bibitem{Kurkela:2018wud}
Aleksi Kurkela, Aleksas Mazeliauskas, Jean-Fran\c{c}ois Paquet, S\"oren
  Schlichting, and Derek Teaney.
\newblock {Matching the Nonequilibrium Initial Stage of Heavy Ion Collisions to
  Hydrodynamics with QCD Kinetic Theory}.
\newblock {\em Phys. Rev. Lett.}, 122(12):122302, 2019.
\newblock \href {http://arxiv.org/abs/1805.01604} {\path{arXiv:1805.01604}},
  \href {https://doi.org/10.1103/PhysRevLett.122.122302}
  {\path{doi:10.1103/PhysRevLett.122.122302}}.

\bibitem{Alqahtani:2017mhy}
Mubarak Alqahtani, Mohammad Nopoush, and Michael Strickland.
\newblock {Relativistic anisotropic hydrodynamics}.
\newblock {\em Prog. Part. Nucl. Phys.}, 101:204--248, 2018.
\newblock \href {http://arxiv.org/abs/1712.03282} {\path{arXiv:1712.03282}},
  \href {https://doi.org/10.1016/j.ppnp.2018.05.004}
  {\path{doi:10.1016/j.ppnp.2018.05.004}}.

\bibitem{McNelis:2018jho}
M.~McNelis, D.~Bazow, and U.~Heinz.
\newblock {(3+1)-dimensional anisotropic fluid dynamics with a lattice QCD
  equation of state}.
\newblock {\em Phys. Rev. C}, 97(5):054912, 2018.
\newblock \href {http://arxiv.org/abs/1803.01810} {\path{arXiv:1803.01810}},
  \href {https://doi.org/10.1103/PhysRevC.97.054912}
  {\path{doi:10.1103/PhysRevC.97.054912}}.

\bibitem{Bhadury:2020cop}
Samapan Bhadury, Wojciech Florkowski, Amaresh Jaiswal, Avdhesh Kumar, and
  Radoslaw Ryblewski.
\newblock {Dissipative Spin Dynamics in Relativistic Matter}.
\newblock {\em Phys. Rev. D}, 103(1):014030, 2021.
\newblock \href {http://arxiv.org/abs/2008.10976} {\path{arXiv:2008.10976}},
  \href {https://doi.org/10.1103/PhysRevD.103.014030}
  {\path{doi:10.1103/PhysRevD.103.014030}}.

\bibitem{Shi:2019wzi}
Shuzhe Shi, Hui Zhang, Defu Hou, and Jinfeng Liao.
\newblock {Signatures of Chiral Magnetic Effect in the Collisions of Isobars}.
\newblock {\em Phys. Rev. Lett.}, 125:242301, 2020.
\newblock \href {http://arxiv.org/abs/1910.14010} {\path{arXiv:1910.14010}},
  \href {https://doi.org/10.1103/PhysRevLett.125.242301}
  {\path{doi:10.1103/PhysRevLett.125.242301}}.

\bibitem{Ammon:2020rvg}
Martin Ammon, Sebastian Grieninger, Juan Hernandez, Matthias Kaminski, Roshan
  Koirala, Julian Leiber, and Jackson Wu.
\newblock {Chiral hydrodynamics in strong external magnetic fields}.
\newblock {\em JHEP}, 04:078, 2021.
\newblock \href {http://arxiv.org/abs/2012.09183} {\path{arXiv:2012.09183}},
  \href {https://doi.org/10.1007/JHEP04(2021)078}
  {\path{doi:10.1007/JHEP04(2021)078}}.

\bibitem{Parotto:2018pwx}
Paolo Parotto, Marcus Bluhm, Debora Mroczek, Marlene Nahrgang, Jacquelyn
  Noronha-Hostler, Krishna Rajagopal, Claudia Ratti, Thomas Sch\"afer, and
  Mikhail Stephanov.
\newblock {QCD equation of state matched to lattice data and exhibiting a
  critical point singularity}.
\newblock {\em Phys. Rev. C}, 101(3):034901, 2020.
\newblock \href {http://arxiv.org/abs/1805.05249} {\path{arXiv:1805.05249}},
  \href {https://doi.org/10.1103/PhysRevC.101.034901}
  {\path{doi:10.1103/PhysRevC.101.034901}}.

\bibitem{Monnai:2019hkn}
Akihiko Monnai, Bj\"orn Schenke, and Chun Shen.
\newblock {Equation of state at finite densities for QCD matter in nuclear
  collisions}.
\newblock {\em Phys. Rev. C}, 100(2):024907, 2019.
\newblock \href {http://arxiv.org/abs/1902.05095} {\path{arXiv:1902.05095}},
  \href {https://doi.org/10.1103/PhysRevC.100.024907}
  {\path{doi:10.1103/PhysRevC.100.024907}}.

\bibitem{Noronha-Hostler:2019ayj}
J.~Noronha-Hostler, P.~Parotto, C.~Ratti, and J.~M. Stafford.
\newblock {Lattice-based equation of state at finite baryon number, electric
  charge and strangeness chemical potentials}.
\newblock {\em Phys. Rev. C}, 100(6):064910, 2019.
\newblock \href {http://arxiv.org/abs/1902.06723} {\path{arXiv:1902.06723}},
  \href {https://doi.org/10.1103/PhysRevC.100.064910}
  {\path{doi:10.1103/PhysRevC.100.064910}}.

\bibitem{Monnai:2021kgu}
Akihiko Monnai, Bj\"orn Schenke, and Chun Shen.
\newblock {QCD Equation of State at Finite Chemical Potentials for Relativistic
  Nuclear Collisions}.
\newblock {\em Int. J. Mod. Phys. A}, 36(07):2130007, 2021.
\newblock \href {http://arxiv.org/abs/2101.11591} {\path{arXiv:2101.11591}},
  \href {https://doi.org/10.1142/S0217751X21300076}
  {\path{doi:10.1142/S0217751X21300076}}.

\bibitem{An:2021wof}
Xin An et~al.
\newblock {The BEST framework for the search for the QCD critical point and the
  chiral magnetic effect}.
\newblock {\em Nucl. Phys. A}, 1017:122343, 2022.
\newblock \href {http://arxiv.org/abs/2108.13867} {\path{arXiv:2108.13867}},
  \href {https://doi.org/10.1016/j.nuclphysa.2021.122343}
  {\path{doi:10.1016/j.nuclphysa.2021.122343}}.

\bibitem{Casalderrey-Solana:2011dxg}
Jorge Casalderrey-Solana, Hong Liu, David Mateos, Krishna Rajagopal, and
  Urs~Achim Wiedemann.
\newblock {\em {Gauge/String Duality, Hot QCD and Heavy Ion Collisions}}.
\newblock Cambridge University Press, 2014.
\newblock \href {http://arxiv.org/abs/1101.0618} {\path{arXiv:1101.0618}},
  \href {https://doi.org/10.1017/CBO9781139136747}
  {\path{doi:10.1017/CBO9781139136747}}.

\bibitem{Nahrgang:2018afz}
Marlene Nahrgang, Marcus Bluhm, Thomas Schaefer, and Steffen~A. Bass.
\newblock {Diffusive dynamics of critical fluctuations near the QCD critical
  point}.
\newblock {\em Phys. Rev. D}, 99(11):116015, 2019.
\newblock \href {http://arxiv.org/abs/1804.05728} {\path{arXiv:1804.05728}},
  \href {https://doi.org/10.1103/PhysRevD.99.116015}
  {\path{doi:10.1103/PhysRevD.99.116015}}.

\bibitem{Nahrgang:2020yxm}
Marlene Nahrgang and Marcus Bluhm.
\newblock {Modeling the diffusive dynamics of critical fluctuations near the
  QCD critical point}.
\newblock {\em Phys. Rev. D}, 102(9):094017, 2020.
\newblock \href {http://arxiv.org/abs/2007.10371} {\path{arXiv:2007.10371}},
  \href {https://doi.org/10.1103/PhysRevD.102.094017}
  {\path{doi:10.1103/PhysRevD.102.094017}}.

\bibitem{Akamatsu:2016llw}
Yukinao Akamatsu, Aleksas Mazeliauskas, and Derek Teaney.
\newblock {A kinetic regime of hydrodynamic fluctuations and long time tails
  for a Bjorken expansion}.
\newblock {\em Phys. Rev. C}, 95(1):014909, 2017.
\newblock \href {http://arxiv.org/abs/1606.07742} {\path{arXiv:1606.07742}},
  \href {https://doi.org/10.1103/PhysRevC.95.014909}
  {\path{doi:10.1103/PhysRevC.95.014909}}.

\bibitem{Stephanov:2017ghc}
M.~Stephanov and Y.~Yin.
\newblock {Hydrodynamics with parametric slowing down and fluctuations near the
  critical point}.
\newblock {\em Phys. Rev. D}, 98(3):036006, 2018.
\newblock \href {http://arxiv.org/abs/1712.10305} {\path{arXiv:1712.10305}},
  \href {https://doi.org/10.1103/PhysRevD.98.036006}
  {\path{doi:10.1103/PhysRevD.98.036006}}.

\bibitem{An:2019osr}
Xin An, Gokce Basar, Mikhail Stephanov, and Ho-Ung Yee.
\newblock {Relativistic Hydrodynamic Fluctuations}.
\newblock {\em Phys. Rev. C}, 100(2):024910, 2019.
\newblock \href {http://arxiv.org/abs/1902.09517} {\path{arXiv:1902.09517}},
  \href {https://doi.org/10.1103/PhysRevC.100.024910}
  {\path{doi:10.1103/PhysRevC.100.024910}}.

\bibitem{Rajagopal:2019xwg}
Krishna Rajagopal, Gregory Ridgway, Ryan Weller, and Yi~Yin.
\newblock {Understanding the out-of-equilibrium dynamics near a critical point
  in the QCD phase diagram}.
\newblock {\em Phys. Rev. D}, 102(9):094025, 2020.
\newblock \href {http://arxiv.org/abs/1908.08539} {\path{arXiv:1908.08539}},
  \href {https://doi.org/10.1103/PhysRevD.102.094025}
  {\path{doi:10.1103/PhysRevD.102.094025}}.

\bibitem{An:2019csj}
Xin An, G\"ok\c{c}e Ba\c{s}ar, Mikhail Stephanov, and Ho-Ung Yee.
\newblock {Fluctuation dynamics in a relativistic fluid with a critical point}.
\newblock {\em Phys. Rev. C}, 102(3):034901, 2020.
\newblock \href {http://arxiv.org/abs/1912.13456} {\path{arXiv:1912.13456}},
  \href {https://doi.org/10.1103/PhysRevC.102.034901}
  {\path{doi:10.1103/PhysRevC.102.034901}}.

\bibitem{An:2020vri}
Xin An, G\"ok\c{c}e Ba\c{s}ar, Mikhail Stephanov, and Ho-Ung Yee.
\newblock {Evolution of Non-Gaussian Hydrodynamic Fluctuations}.
\newblock {\em Phys. Rev. Lett.}, 127(7):072301, 2021.
\newblock \href {http://arxiv.org/abs/2009.10742} {\path{arXiv:2009.10742}},
  \href {https://doi.org/10.1103/PhysRevLett.127.072301}
  {\path{doi:10.1103/PhysRevLett.127.072301}}.

\bibitem{Du:2020bxp}
Lipei Du, Ulrich Heinz, Krishna Rajagopal, and Yi~Yin.
\newblock {Fluctuation dynamics near the QCD critical point}.
\newblock {\em Phys. Rev. C}, 102(5):054911, 2020.
\newblock \href {http://arxiv.org/abs/2004.02719} {\path{arXiv:2004.02719}},
  \href {https://doi.org/10.1103/PhysRevC.102.054911}
  {\path{doi:10.1103/PhysRevC.102.054911}}.

\bibitem{De:2022tkb}
Aritra De, Chun Shen, and Joseph~I. Kapusta.
\newblock {Stochastic hydrodynamics and hydro-kinetics: Similarities and
  differences}.
\newblock {\em Phys. Rev. C}, 106(5):054903, 2022.
\newblock \href {http://arxiv.org/abs/2203.02134} {\path{arXiv:2203.02134}},
  \href {https://doi.org/10.1103/PhysRevC.106.054903}
  {\path{doi:10.1103/PhysRevC.106.054903}}.

\bibitem{Oliinychenko:2019zfk}
Dmytro Oliinychenko and Volker Koch.
\newblock {Microcanonical Particlization with Local Conservation Laws}.
\newblock {\em Phys. Rev. Lett.}, 123(18):182302, 2019.
\newblock \href {http://arxiv.org/abs/1902.09775} {\path{arXiv:1902.09775}},
  \href {https://doi.org/10.1103/PhysRevLett.123.182302}
  {\path{doi:10.1103/PhysRevLett.123.182302}}.

\bibitem{Pradeep:2022mkf}
Maneesha Pradeep, Krishna Rajagopal, Mikhail Stephanov, and Yi~Yin.
\newblock {Freezing out fluctuations in Hydro+ near the QCD critical point}.
\newblock {\em Phys. Rev. D}, 106(3):036017, 2022.
\newblock \href {http://arxiv.org/abs/2204.00639} {\path{arXiv:2204.00639}},
  \href {https://doi.org/10.1103/PhysRevD.106.036017}
  {\path{doi:10.1103/PhysRevD.106.036017}}.

\bibitem{Kanakubo:2021qcw}
Yuuka Kanakubo, Yasuki Tachibana, and Tetsufumi Hirano.
\newblock {Interplay between core and corona components in high-energy nuclear
  collisions}.
\newblock {\em Phys. Rev. C}, 105(2):024905, 2022.
\newblock \href {http://arxiv.org/abs/2108.07943} {\path{arXiv:2108.07943}},
  \href {https://doi.org/10.1103/PhysRevC.105.024905}
  {\path{doi:10.1103/PhysRevC.105.024905}}.

\bibitem{Danielewicz:2002pu}
Pawel Danielewicz, Roy Lacey, and William~G. Lynch.
\newblock {Determination of the equation of state of dense matter}.
\newblock {\em Science}, 298:1592--1596, 2002.
\newblock \href {http://arxiv.org/abs/nucl-th/0208016}
  {\path{arXiv:nucl-th/0208016}}, \href
  {https://doi.org/10.1126/science.1078070}
  {\path{doi:10.1126/science.1078070}}.

\bibitem{LeFevre:2015paj}
A.~Le~F\`evre, Y.~Leifels, W.~Reisdorf, J.~Aichelin, and Ch. Hartnack.
\newblock {Constraining the nuclear matter equation of state around twice
  saturation density}.
\newblock {\em Nucl. Phys. A}, 945:112--133, 2016.
\newblock \href {http://arxiv.org/abs/1501.05246} {\path{arXiv:1501.05246}},
  \href {https://doi.org/10.1016/j.nuclphysa.2015.09.015}
  {\path{doi:10.1016/j.nuclphysa.2015.09.015}}.

\bibitem{Oliinychenko:2022uvy}
Dmytro Oliinychenko, Agnieszka Sorensen, Volker Koch, and Larry McLerran.
\newblock {Sensitivity of Au+Au collisions to the symmetric nuclear matter
  equation of state at 2 -- 5 nuclear saturation densities}.
\newblock 8 2022.
\newblock \href {http://arxiv.org/abs/2208.11996} {\path{arXiv:2208.11996}}.

\bibitem{Steinheimer:2022gqb}
Jan Steinheimer, Anton Motornenko, Agnieszka Sorensen, Yasushi Nara, Volker
  Koch, and Marcus Bleicher.
\newblock {The high-density equation of state in heavy-ion collisions:
  constraints from proton flow}.
\newblock {\em Eur. Phys. J. C}, 82(10):911, 2022.
\newblock \href {http://arxiv.org/abs/2208.12091} {\path{arXiv:2208.12091}},
  \href {https://doi.org/10.1140/epjc/s10052-022-10894-w}
  {\path{doi:10.1140/epjc/s10052-022-10894-w}}.

\bibitem{Colonna:2020euy}
Maria Colonna.
\newblock {Collision dynamics at medium and relativistic energies}.
\newblock 3 2020.
\newblock \href {http://arxiv.org/abs/2003.02500} {\path{arXiv:2003.02500}},
  \href {https://doi.org/10.1016/j.ppnp.2020.103775}
  {\path{doi:10.1016/j.ppnp.2020.103775}}.

\bibitem{Fuchs:2000kp}
C.~Fuchs, Amand Faessler, E.~Zabrodin, and Yu-Ming Zheng.
\newblock {Probing the nuclear equation of state by K+ production in heavy ion
  collisions}.
\newblock {\em Phys. Rev. Lett.}, 86:1974--1977, 2001.
\newblock \href {http://arxiv.org/abs/nucl-th/0011102}
  {\path{arXiv:nucl-th/0011102}}, \href
  {https://doi.org/10.1103/PhysRevLett.86.1974}
  {\path{doi:10.1103/PhysRevLett.86.1974}}.

\bibitem{Li:2002qx}
Bao-An Li.
\newblock {Probing the high density behavior of nuclear symmetry energy with
  high-energy heavy ion collisions}.
\newblock {\em Phys. Rev. Lett.}, 88:192701, 2002.
\newblock \href {http://arxiv.org/abs/nucl-th/0205002}
  {\path{arXiv:nucl-th/0205002}}, \href
  {https://doi.org/10.1103/PhysRevLett.88.192701}
  {\path{doi:10.1103/PhysRevLett.88.192701}}.

\bibitem{Xiao:2008vm}
Zhigang Xiao, Bao-An Li, Lie-Wen Chen, Gao-Chan Yong, and Ming Zhang.
\newblock {Circumstantial Evidence for a Soft Nuclear Symmetry Energy at
  Suprasaturation Densities}.
\newblock {\em Phys. Rev. Lett.}, 102:062502, 2009.
\newblock \href {http://arxiv.org/abs/0808.0186} {\path{arXiv:0808.0186}},
  \href {https://doi.org/10.1103/PhysRevLett.102.062502}
  {\path{doi:10.1103/PhysRevLett.102.062502}}.

\bibitem{Yong:2022pyb}
Gao-Chan Yong, Bao-An Li, Zhi-Gang Xiao, and Zi-Wei Lin.
\newblock {Probing the high-density nuclear symmetry energy with the
  \ensuremath{\Xi}\ensuremath{-}/\ensuremath{\Xi}0 ratio in heavy-ion
  collisions at sNN\ensuremath{\approx}3 GeV}.
\newblock {\em Phys. Rev. C}, 106(2):024902, 2022.
\newblock \href {http://arxiv.org/abs/2206.10766} {\path{arXiv:2206.10766}},
  \href {https://doi.org/10.1103/PhysRevC.106.024902}
  {\path{doi:10.1103/PhysRevC.106.024902}}.

\bibitem{SRIT:2021gcy}
J.~Estee et~al.
\newblock {Probing the Symmetry Energy with the Spectral Pion Ratio}.
\newblock {\em Phys. Rev. Lett.}, 126(16):162701, 2021.
\newblock \href {http://arxiv.org/abs/2103.06861} {\path{arXiv:2103.06861}},
  \href {https://doi.org/10.1103/PhysRevLett.126.162701}
  {\path{doi:10.1103/PhysRevLett.126.162701}}.

\bibitem{Li:2000bj}
Bao-An Li.
\newblock {Neutron proton differential flow as a probe of isospin dependence of
  nuclear equation of state}.
\newblock {\em Phys. Rev. Lett.}, 85:4221--4224, 2000.
\newblock \href {http://arxiv.org/abs/nucl-th/0009069}
  {\path{arXiv:nucl-th/0009069}}, \href
  {https://doi.org/10.1103/PhysRevLett.85.4221}
  {\path{doi:10.1103/PhysRevLett.85.4221}}.

\bibitem{Li:2014oda}
Bao-An Li, Angels Ramos, Giuseppe Verde, and Isaac Vidana.
\newblock {Topical issue on nuclear symmetry energy}.
\newblock {\em Eur. Phys. J. A}, 50:9, 2014.
\newblock \href {https://doi.org/10.1140/epja/i2014-14009-x}
  {\path{doi:10.1140/epja/i2014-14009-x}}.

\bibitem{Xu:2019hqg}
Jun Xu.
\newblock {Transport approaches for the description of intermediate-energy
  heavy-ion collisions}.
\newblock {\em Prog. Part. Nucl. Phys.}, 106:312--359, 2019.
\newblock \href {http://arxiv.org/abs/1904.00131} {\path{arXiv:1904.00131}},
  \href {https://doi.org/10.1016/j.ppnp.2019.02.009}
  {\path{doi:10.1016/j.ppnp.2019.02.009}}.

\bibitem{Russotto:2011hq}
P.~Russotto et~al.
\newblock {Symmetry energy from elliptic flow in $^{197}$Au $+ ^{197}$Au}.
\newblock {\em Phys. Lett. B}, 697:471--476, 2011.
\newblock \href {http://arxiv.org/abs/1101.2361} {\path{arXiv:1101.2361}},
  \href {https://doi.org/10.1016/j.physletb.2011.02.033}
  {\path{doi:10.1016/j.physletb.2011.02.033}}.

\bibitem{Cozma:2011nr}
M.~D. Cozma.
\newblock {Neutron-proton elliptic flow difference as a probe for the high
  density dependence of the symmetry energy}.
\newblock {\em Phys. Lett. B}, 700:139--144, 2011.
\newblock \href {http://arxiv.org/abs/1102.2728} {\path{arXiv:1102.2728}},
  \href {https://doi.org/10.1016/j.physletb.2011.05.002}
  {\path{doi:10.1016/j.physletb.2011.05.002}}.

\bibitem{Giordano:2010pv}
V.~Giordano, M.~Colonna, M.~Di~Toro, V.~Greco, and J.~Rizzo.
\newblock {Isospin emission and flows at high baryon density: a test of the
  symmetry potential}.
\newblock {\em Phys. Rev. C}, 81:044611, 2010.
\newblock \href {http://arxiv.org/abs/1001.4961} {\path{arXiv:1001.4961}},
  \href {https://doi.org/10.1103/PhysRevC.81.044611}
  {\path{doi:10.1103/PhysRevC.81.044611}}.

\bibitem{Liu:2019ags}
He~Liu, Feng-Tao Wang, Kai-Jia Sun, Jun Xu, and Che~Ming Ko.
\newblock {Isospin splitting of pion elliptic flow in relativistic heavy-ion
  collisions}.
\newblock {\em Phys. Lett. B}, 798:135002, 2019.
\newblock \href {http://arxiv.org/abs/1908.01156} {\path{arXiv:1908.01156}},
  \href {https://doi.org/10.1016/j.physletb.2019.135002}
  {\path{doi:10.1016/j.physletb.2019.135002}}.

\bibitem{Li:1996ju}
Guo-Qiang Li and C.~M. Ko.
\newblock {Lambda flow in heavy ion collisions: The Role of final state
  interactions}.
\newblock {\em Phys. Rev. C}, 54:1897--1902, 1996.
\newblock \href {http://arxiv.org/abs/nucl-th/9608049}
  {\path{arXiv:nucl-th/9608049}}, \href
  {https://doi.org/10.1103/PhysRevC.54.1897}
  {\path{doi:10.1103/PhysRevC.54.1897}}.

\bibitem{Wang:1998ew}
Z.~S. Wang, Amand Faessler, C.~Fuchs, and T.~Waindzoch.
\newblock {Lambda collective flow in heavy ion reactions}.
\newblock {\em Nucl. Phys. A}, 645:177--188, 1999.
\newblock [Erratum: Nucl.Phys.A 648, 281--281 (1999)].
\newblock \href {http://arxiv.org/abs/nucl-th/9811090}
  {\path{arXiv:nucl-th/9811090}}, \href
  {https://doi.org/10.1016/S0375-9474(98)00605-8}
  {\path{doi:10.1016/S0375-9474(98)00605-8}}.

\bibitem{Ko:2000cd}
Che~Ming Ko.
\newblock {Medium effects on the flow of strange particles in heavy ion
  collisions}.
\newblock {\em J. Phys. G}, 27:327--336, 2001.
\newblock \href {http://arxiv.org/abs/nucl-th/0009040}
  {\path{arXiv:nucl-th/0009040}}, \href
  {https://doi.org/10.1088/0954-3899/27/3/310}
  {\path{doi:10.1088/0954-3899/27/3/310}}.

\bibitem{Sorensen:2020ygf}
Agnieszka Sorensen and Volker Koch.
\newblock {Phase transitions and critical behavior in hadronic transport with a
  relativistic density functional equation of state}.
\newblock {\em Phys. Rev. C}, 104(3):034904, 2021.
\newblock \href {http://arxiv.org/abs/2011.06635} {\path{arXiv:2011.06635}},
  \href {https://doi.org/10.1103/PhysRevC.104.034904}
  {\path{doi:10.1103/PhysRevC.104.034904}}.

\bibitem{JETSCAPE:2020shq}
D.~Everett et~al.
\newblock {Phenomenological constraints on the transport properties of QCD
  matter with data-driven model averaging}.
\newblock {\em Phys. Rev. Lett.}, 126(24):242301, 2021.
\newblock \href {http://arxiv.org/abs/2010.03928} {\path{arXiv:2010.03928}},
  \href {https://doi.org/10.1103/PhysRevLett.126.242301}
  {\path{doi:10.1103/PhysRevLett.126.242301}}.

\bibitem{Petersen:2010zt}
Hannah Petersen, Christopher Coleman-Smith, Steffen~A. Bass, and Robert
  Wolpert.
\newblock {Constraining the initial state granularity with bulk observables in
  Au+Au collisions at $\sqrt{s_{\rm NN}}=200$ GeV}.
\newblock {\em J. Phys. G}, 38:045102, 2011.
\newblock \href {http://arxiv.org/abs/1012.4629} {\path{arXiv:1012.4629}},
  \href {https://doi.org/10.1088/0954-3899/38/4/045102}
  {\path{doi:10.1088/0954-3899/38/4/045102}}.

\bibitem{Novak:2013bqa}
John Novak, Kevin Novak, Scott Pratt, Joshua Vredevoogd, Chris Coleman-Smith,
  and Robert Wolpert.
\newblock {Determining Fundamental Properties of Matter Created in
  Ultrarelativistic Heavy-Ion Collisions}.
\newblock {\em Phys. Rev. C}, 89(3):034917, 2014.
\newblock \href {http://arxiv.org/abs/1303.5769} {\path{arXiv:1303.5769}},
  \href {https://doi.org/10.1103/PhysRevC.89.034917}
  {\path{doi:10.1103/PhysRevC.89.034917}}.

\bibitem{Sangaline:2015isa}
Evan Sangaline and Scott Pratt.
\newblock {Toward a deeper understanding of how experiments constrain the
  underlying physics of heavy-ion collisions}.
\newblock {\em Phys. Rev. C}, 93(2):024908, 2016.
\newblock \href {http://arxiv.org/abs/1508.07017} {\path{arXiv:1508.07017}},
  \href {https://doi.org/10.1103/PhysRevC.93.024908}
  {\path{doi:10.1103/PhysRevC.93.024908}}.

\bibitem{Bernhard:2015hxa}
Jonah~E. Bernhard, Peter~W. Marcy, Christopher~E. Coleman-Smith, Snehalata
  Huzurbazar, Robert~L. Wolpert, and Steffen~A. Bass.
\newblock {Quantifying properties of hot and dense QCD matter through
  systematic model-to-data comparison}.
\newblock {\em Phys. Rev. C}, 91(5):054910, 2015.
\newblock \href {http://arxiv.org/abs/1502.00339} {\path{arXiv:1502.00339}},
  \href {https://doi.org/10.1103/PhysRevC.91.054910}
  {\path{doi:10.1103/PhysRevC.91.054910}}.

\bibitem{Bernhard:2016tnd}
Jonah~E. Bernhard, J.~Scott Moreland, Steffen~A. Bass, Jia Liu, and Ulrich
  Heinz.
\newblock {Applying Bayesian parameter estimation to relativistic heavy-ion
  collisions: simultaneous characterization of the initial state and
  quark-gluon plasma medium}.
\newblock {\em Phys. Rev. C}, 94(2):024907, 2016.
\newblock \href {http://arxiv.org/abs/1605.03954} {\path{arXiv:1605.03954}},
  \href {https://doi.org/10.1103/PhysRevC.94.024907}
  {\path{doi:10.1103/PhysRevC.94.024907}}.

\bibitem{Moreland:2018gsh}
J.~Scott Moreland, Jonah~E. Bernhard, and Steffen~A. Bass.
\newblock {Bayesian calibration of a hybrid nuclear collision model using p-Pb
  and Pb-Pb data at energies available at the CERN Large Hadron Collider}.
\newblock {\em Phys. Rev. C}, 101(2):024911, 2020.
\newblock \href {http://arxiv.org/abs/1808.02106} {\path{arXiv:1808.02106}},
  \href {https://doi.org/10.1103/PhysRevC.101.024911}
  {\path{doi:10.1103/PhysRevC.101.024911}}.

\bibitem{Bernhard:2019bmu}
Jonah~E. Bernhard, J.~Scott Moreland, and Steffen~A. Bass.
\newblock {Bayesian estimation of the specific shear and bulk viscosity of
  quark\textendash{}gluon plasma}.
\newblock {\em Nature Phys.}, 15(11):1113--1117, 2019.
\newblock \href {https://doi.org/10.1038/s41567-019-0611-8}
  {\path{doi:10.1038/s41567-019-0611-8}}.

\bibitem{JETSCAPE:2020mzn}
D.~Everett et~al.
\newblock {Multisystem Bayesian constraints on the transport coefficients of
  QCD matter}.
\newblock {\em Phys. Rev. C}, 103(5):054904, 2021.
\newblock \href {http://arxiv.org/abs/2011.01430} {\path{arXiv:2011.01430}},
  \href {https://doi.org/10.1103/PhysRevC.103.054904}
  {\path{doi:10.1103/PhysRevC.103.054904}}.

\bibitem{Nijs:2020ors}
Govert Nijs, Wilke van~der Schee, Umut G\"ursoy, and Raimond Snellings.
\newblock {Transverse Momentum Differential Global Analysis of Heavy-Ion
  Collisions}.
\newblock {\em Phys. Rev. Lett.}, 126(20):202301, 2021.
\newblock \href {http://arxiv.org/abs/2010.15130} {\path{arXiv:2010.15130}},
  \href {https://doi.org/10.1103/PhysRevLett.126.202301}
  {\path{doi:10.1103/PhysRevLett.126.202301}}.

\bibitem{Nijs:2020roc}
Govert Nijs, Wilke van~der Schee, Umut G\"ursoy, and Raimond Snellings.
\newblock {Bayesian analysis of heavy ion collisions with the heavy ion
  computational framework Trajectum}.
\newblock {\em Phys. Rev. C}, 103(5):054909, 2021.
\newblock \href {http://arxiv.org/abs/2010.15134} {\path{arXiv:2010.15134}},
  \href {https://doi.org/10.1103/PhysRevC.103.054909}
  {\path{doi:10.1103/PhysRevC.103.054909}}.

\bibitem{Nijs:2021clz}
Govert Nijs and Wilke van~der Schee.
\newblock {Predictions and postdictions for relativistic lead and oxygen
  collisions with the computational simulation code Trajectum}.
\newblock {\em Phys. Rev. C}, 106(4):044903, 2022.
\newblock \href {http://arxiv.org/abs/2110.13153} {\path{arXiv:2110.13153}},
  \href {https://doi.org/10.1103/PhysRevC.106.044903}
  {\path{doi:10.1103/PhysRevC.106.044903}}.

\bibitem{Parkkila:2021tqq}
J.~E. Parkkila, A.~Onnerstad, and D.~J. Kim.
\newblock {Bayesian estimation of the specific shear and bulk viscosity of the
  quark-gluon plasma with additional flow harmonic observables}.
\newblock {\em Phys. Rev. C}, 104(5):054904, 2021.
\newblock \href {http://arxiv.org/abs/2106.05019} {\path{arXiv:2106.05019}},
  \href {https://doi.org/10.1103/PhysRevC.104.054904}
  {\path{doi:10.1103/PhysRevC.104.054904}}.

\bibitem{JETSCAPE:2020avt}
Jean-Fran\c{c}ois Paquet et~al.
\newblock {Revisiting Bayesian constraints on the transport coefficients of
  QCD}.
\newblock {\em Nucl. Phys. A}, 1005:121749, 2021.
\newblock \href {http://arxiv.org/abs/2002.05337} {\path{arXiv:2002.05337}},
  \href {https://doi.org/10.1016/j.nuclphysa.2020.121749}
  {\path{doi:10.1016/j.nuclphysa.2020.121749}}.

\bibitem{Xie:2022ght}
Man Xie, Weiyao Ke, Hanzhong Zhang, and Xin-Nian Wang.
\newblock {Information field based global Bayesian inference of the jet
  transport coefficient}.
\newblock 6 2022.
\newblock \href {http://arxiv.org/abs/2206.01340} {\path{arXiv:2206.01340}}.

\bibitem{Xie:2022fak}
Man Xie, Weiyao Ke, Hanzhong Zhang, and Xin-Nian Wang.
\newblock {Global constraint on the jet transport coefficient from single
  hadron, dihadron and $\gamma$-hadron spectra in high-energy heavy-ion
  collisions}.
\newblock 8 2022.
\newblock \href {http://arxiv.org/abs/2208.14419} {\path{arXiv:2208.14419}}.

\bibitem{Soltz:2020xvw}
R.~A. Soltz.
\newblock {A Comprehensive Monte Carlo Framework for Jet-Quenching}.
\newblock {\em Nucl. Phys. A}, 1005:122040, 2021.
\newblock \href {http://arxiv.org/abs/2003.11728} {\path{arXiv:2003.11728}},
  \href {https://doi.org/10.1016/j.nuclphysa.2020.122040}
  {\path{doi:10.1016/j.nuclphysa.2020.122040}}.

\bibitem{JETSCAPE:2021ehl}
S.~Cao et~al.
\newblock {Determining the jet transport coefficient $\hat q$ from inclusive
  hadron suppression measurements using Bayesian parameter estimation}.
\newblock {\em Phys. Rev. C}, 104(2):024905, 2021.
\newblock \href {http://arxiv.org/abs/2102.11337} {\path{arXiv:2102.11337}},
  \href {https://doi.org/10.1103/PhysRevC.104.024905}
  {\path{doi:10.1103/PhysRevC.104.024905}}.

\bibitem{Bernhard:2018hnz}
Jonah~E. Bernhard.
\newblock {\em {Bayesian parameter estimation for relativistic heavy-ion
  collisions}}.
\newblock PhD thesis, Duke U., 4 2018.
\newblock \href {http://arxiv.org/abs/1804.06469} {\path{arXiv:1804.06469}}.

\bibitem{Geurts:2022xmk}
Frank Geurts and Ralf-Arno Tripolt.
\newblock {Electromagnetic probes: Theory and experiment}.
\newblock {\em Prog. Part. Nucl. Phys.}, 128:104004, 2023.
\newblock \href {http://arxiv.org/abs/2210.01622} {\path{arXiv:2210.01622}},
  \href {https://doi.org/10.1016/j.ppnp.2022.104004}
  {\path{doi:10.1016/j.ppnp.2022.104004}}.

\bibitem{STAR:2016use}
L.~Adamczyk et~al.
\newblock {Direct virtual photon production in Au+Au collisions at
  $\sqrt{s_{NN}}$ = 200 GeV}.
\newblock {\em Phys. Lett. B}, 770:451--458, 2017.
\newblock \href {http://arxiv.org/abs/1607.01447} {\path{arXiv:1607.01447}},
  \href {https://doi.org/10.1016/j.physletb.2017.04.050}
  {\path{doi:10.1016/j.physletb.2017.04.050}}.

\bibitem{PHENIX:2018for}
A.~Adare et~al.
\newblock {Beam Energy and Centrality Dependence of Direct-Photon Emission from
  Ultrarelativistic Heavy-Ion Collisions}.
\newblock {\em Phys. Rev. Lett.}, 123(2):022301, 2019.
\newblock \href {http://arxiv.org/abs/1805.04084} {\path{arXiv:1805.04084}},
  \href {https://doi.org/10.1103/PhysRevLett.123.022301}
  {\path{doi:10.1103/PhysRevLett.123.022301}}.

\bibitem{PHENIX:2022qfp}
U.~A. Acharya et~al.
\newblock {Low-$p_T$ direct-photon production in Au$+$Au collisions at
  $\sqrt{s_{_{NN}}}=39$ and 62.4 GeV}.
\newblock 3 2022.
\newblock \href {http://arxiv.org/abs/2203.12354} {\path{arXiv:2203.12354}}.

\bibitem{PHENIX:2022rsx}
U.~A. Acharya et~al.
\newblock {Nonprompt direct-photon production in Au$+$Au collisions at
  $\sqrt{s_{_{NN}}}=200$ GeV}.
\newblock 3 2022.
\newblock \href {http://arxiv.org/abs/2203.17187} {\path{arXiv:2203.17187}}.

\bibitem{ALICE:2015xmh}
Jaroslav Adam et~al.
\newblock {Direct photon production in Pb-Pb collisions at $\sqrt{s_{NN}} =$
  2.76 TeV}.
\newblock {\em Phys. Lett. B}, 754:235--248, 2016.
\newblock \href {http://arxiv.org/abs/1509.07324} {\path{arXiv:1509.07324}},
  \href {https://doi.org/10.1016/j.physletb.2016.01.020}
  {\path{doi:10.1016/j.physletb.2016.01.020}}.

\bibitem{ALICE:2018dti}
Shreyasi Acharya et~al.
\newblock {Direct photon elliptic flow in Pb-Pb collisions at $\sqrt{s_{\rm
  NN}}$ = 2.76 TeV}.
\newblock {\em Phys. Lett. B}, 789:308--322, 2019.
\newblock \href {http://arxiv.org/abs/1805.04403} {\path{arXiv:1805.04403}},
  \href {https://doi.org/10.1016/j.physletb.2018.11.039}
  {\path{doi:10.1016/j.physletb.2018.11.039}}.

\bibitem{PHENIX:2015igl}
A.~Adare et~al.
\newblock {Azimuthally anisotropic emission of low-momentum direct photons in
  Au$+$Au collisions at $\sqrt{s_{_{NN}}}=200$ GeV}.
\newblock {\em Phys. Rev. C}, 94(6):064901, 2016.
\newblock \href {http://arxiv.org/abs/1509.07758} {\path{arXiv:1509.07758}},
  \href {https://doi.org/10.1103/PhysRevC.94.064901}
  {\path{doi:10.1103/PhysRevC.94.064901}}.

\bibitem{Gale:2021emg}
Charles Gale, Jean-Fran\c{c}ois Paquet, Bj\"orn Schenke, and Chun Shen.
\newblock {Multimessenger heavy-ion collision physics}.
\newblock {\em Phys. Rev. C}, 105(1):014909, 2022.
\newblock \href {http://arxiv.org/abs/2106.11216} {\path{arXiv:2106.11216}},
  \href {https://doi.org/10.1103/PhysRevC.105.014909}
  {\path{doi:10.1103/PhysRevC.105.014909}}.

\bibitem{ALICE:2018ael}
Shreyasi Acharya et~al.
\newblock {Measurement of dielectron production in central Pb-Pb collisions at
  $\sqrt{{\textit{s}}_{\mathrm{NN}}}$ = 2.76 TeV}.
\newblock {\em Phys. Rev. C}, 99(2):024002, 2019.
\newblock \href {http://arxiv.org/abs/1807.00923} {\path{arXiv:1807.00923}},
  \href {https://doi.org/10.1103/PhysRevC.99.024002}
  {\path{doi:10.1103/PhysRevC.99.024002}}.

\bibitem{STAR:2015zal}
L.~Adamczyk et~al.
\newblock {Energy dependence of acceptance-corrected dielectron excess mass
  spectrum at mid-rapidity in Au$+$Au collisions at $\sqrt{s_{NN}} =$ 19.6 and
  200 GeV}.
\newblock {\em Phys. Lett. B}, 750:64--71, 2015.
\newblock \href {http://arxiv.org/abs/1501.05341} {\path{arXiv:1501.05341}},
  \href {https://doi.org/10.1016/j.physletb.2015.08.044}
  {\path{doi:10.1016/j.physletb.2015.08.044}}.

\bibitem{STAR:2015tnn}
L.~Adamczyk et~al.
\newblock {Measurements of Dielectron Production in Au$+$Au Collisions at
  $\sqrt{s_{\rm NN}}$ = 200 GeV from the STAR Experiment}.
\newblock {\em Phys. Rev. C}, 92(2):024912, 2015.
\newblock \href {http://arxiv.org/abs/1504.01317} {\path{arXiv:1504.01317}},
  \href {https://doi.org/10.1103/PhysRevC.92.024912}
  {\path{doi:10.1103/PhysRevC.92.024912}}.

\bibitem{STAR:2018xaj}
Jaroslav Adam et~al.
\newblock {Measurements of Dielectron Production in Au$+$Au Collisions at
  $\sqrt{s_{NN}}$= 27, 39, and 62.4 GeV from the STAR Experiment}.
\newblock 10 2018.
\newblock \href {http://arxiv.org/abs/1810.10159} {\path{arXiv:1810.10159}}.

\bibitem{STAR:2018ldd}
Jaroslav Adam et~al.
\newblock {Low-$p_T$ $e^{+}e^{-}$ pair production in Au$+$Au collisions at
  $\sqrt{s_{NN}}$ = 200 GeV and U$+$U collisions at $\sqrt{s_{NN}}$ = 193 GeV
  at STAR}.
\newblock {\em Phys. Rev. Lett.}, 121(13):132301, 2018.
\newblock \href {http://arxiv.org/abs/1806.02295} {\path{arXiv:1806.02295}},
  \href {https://doi.org/10.1103/PhysRevLett.121.132301}
  {\path{doi:10.1103/PhysRevLett.121.132301}}.

\bibitem{Rapp:2014hha}
Ralf Rapp and Hendrik van Hees.
\newblock {Thermal Dileptons as Fireball Thermometer and Chronometer}.
\newblock {\em Phys. Lett. B}, 753:586--590, 2016.
\newblock \href {http://arxiv.org/abs/1411.4612} {\path{arXiv:1411.4612}},
  \href {https://doi.org/10.1016/j.physletb.2015.12.065}
  {\path{doi:10.1016/j.physletb.2015.12.065}}.

\bibitem{Paquet:2015lta}
Jean-Fran\c{c}ois Paquet, Chun Shen, Gabriel~S. Denicol, Matthew Luzum, Bj\"orn
  Schenke, Sangyong Jeon, and Charles Gale.
\newblock {Production of photons in relativistic heavy-ion collisions}.
\newblock {\em Phys. Rev. C}, 93(4):044906, 2016.
\newblock \href {http://arxiv.org/abs/1509.06738} {\path{arXiv:1509.06738}},
  \href {https://doi.org/10.1103/PhysRevC.93.044906}
  {\path{doi:10.1103/PhysRevC.93.044906}}.

\bibitem{Kim:2016ylr}
Young-Min Kim, Chang-Hwan Lee, Derek Teaney, and Ismail Zahed.
\newblock {Direct photon elliptic flow at energies available at the BNL
  Relativistic Heavy Ion Collider and the CERN Large Hadron Collider}.
\newblock {\em Phys. Rev. C}, 96(1):015201, 2017.
\newblock \href {http://arxiv.org/abs/1610.06213} {\path{arXiv:1610.06213}},
  \href {https://doi.org/10.1103/PhysRevC.96.015201}
  {\path{doi:10.1103/PhysRevC.96.015201}}.

\bibitem{Dasgupta:2018pjm}
Pingal Dasgupta, Somnath De, Rupa Chatterjee, and Dinesh~K. Srivastava.
\newblock {Photon production from Pb+Pb collisions at $\sqrt{s_{\rm {NN}}}$ =
  5.02 TeV at LHC and at $\sqrt{s_{\rm {NN}}}$ = 39 TeV at FCC}.
\newblock {\em Phys. Rev. C}, 98(2):024911, 2018.
\newblock \href {http://arxiv.org/abs/1804.02828} {\path{arXiv:1804.02828}},
  \href {https://doi.org/10.1103/PhysRevC.98.024911}
  {\path{doi:10.1103/PhysRevC.98.024911}}.

\bibitem{Garcia-Montero:2019kjk}
Oscar Garcia-Montero, Nicole L\"oher, Aleksas Mazeliauskas, J\"urgen Berges,
  and Klaus Reygers.
\newblock {Probing the evolution of heavy-ion collisions using direct photon
  interferometry}.
\newblock {\em Phys. Rev. C}, 102(2):024915, 2020.
\newblock \href {http://arxiv.org/abs/1909.12246} {\path{arXiv:1909.12246}},
  \href {https://doi.org/10.1103/PhysRevC.102.024915}
  {\path{doi:10.1103/PhysRevC.102.024915}}.

\bibitem{Monnai:2022hfs}
Akihiko Monnai.
\newblock {Direct photons in hydrodynamic modeling of relativistic nuclear
  collisions}.
\newblock {\em Int. J. Mod. Phys. A}, 37(11n12):2230006, 2022.
\newblock \href {http://arxiv.org/abs/2203.13208} {\path{arXiv:2203.13208}},
  \href {https://doi.org/10.1142/S0217751X2230006X}
  {\path{doi:10.1142/S0217751X2230006X}}.

\bibitem{Chatterjee:2017akg}
Rupa Chatterjee, Pingal Dasgupta, and Dinesh~K. Srivastava.
\newblock {Anisotropic flow of thermal photons at energies available at the BNL
  Relativistic Heavy Ion Collider and at the CERN Large Hadron Collider}.
\newblock {\em Phys. Rev. C}, 96(1):014911, 2017.
\newblock \href {http://arxiv.org/abs/1702.02378} {\path{arXiv:1702.02378}},
  \href {https://doi.org/10.1103/PhysRevC.96.014911}
  {\path{doi:10.1103/PhysRevC.96.014911}}.

\bibitem{Shen:2013vja}
Chun Shen, Ulrich~W Heinz, Jean-Francois Paquet, and Charles Gale.
\newblock {Thermal photons as a quark-gluon plasma thermometer reexamined}.
\newblock {\em Phys. Rev. C}, 89(4):044910, 2014.
\newblock \href {http://arxiv.org/abs/1308.2440} {\path{arXiv:1308.2440}},
  \href {https://doi.org/10.1103/PhysRevC.89.044910}
  {\path{doi:10.1103/PhysRevC.89.044910}}.

\bibitem{Shen:2013cca}
Chun Shen, Ulrich~W. Heinz, Jean-Francois Paquet, Igor Kozlov, and Charles
  Gale.
\newblock {Anisotropic flow of thermal photons as a quark-gluon plasma
  viscometer}.
\newblock {\em Phys. Rev. C}, 91(2):024908, 2015.
\newblock \href {http://arxiv.org/abs/1308.2111} {\path{arXiv:1308.2111}},
  \href {https://doi.org/10.1103/PhysRevC.91.024908}
  {\path{doi:10.1103/PhysRevC.91.024908}}.

\bibitem{Vujanovic:2017psb}
Gojko Vujanovic, Gabriel~S. Denicol, Matthew Luzum, Sangyong Jeon, and Charles
  Gale.
\newblock {Investigating the temperature dependence of the specific shear
  viscosity of QCD matter with dilepton radiation}.
\newblock {\em Phys. Rev. C}, 98(1):014902, 2018.
\newblock \href {http://arxiv.org/abs/1702.02941} {\path{arXiv:1702.02941}},
  \href {https://doi.org/10.1103/PhysRevC.98.014902}
  {\path{doi:10.1103/PhysRevC.98.014902}}.

\bibitem{Vujanovic:2019yih}
Gojko Vujanovic, Jean-Fran\c{c}ois Paquet, Chun Shen, Gabriel~S. Denicol,
  Sangyong Jeon, Charles Gale, and Ulrich Heinz.
\newblock {Exploring the influence of bulk viscosity of QCD on dilepton
  tomography}.
\newblock {\em Phys. Rev. C}, 101:044904, 2020.
\newblock \href {http://arxiv.org/abs/1903.05078} {\path{arXiv:1903.05078}},
  \href {https://doi.org/10.1103/PhysRevC.101.044904}
  {\path{doi:10.1103/PhysRevC.101.044904}}.

\bibitem{Vujanovic:2016anq}
Gojko Vujanovic, Jean-Fran\c{c}ois Paquet, Gabriel~S. Denicol, Matthew Luzum,
  Sangyong Jeon, and Charles Gale.
\newblock {Electromagnetic radiation as a probe of the initial state and of
  viscous dynamics in relativistic nuclear collisions}.
\newblock {\em Phys. Rev. C}, 94(1):014904, 2016.
\newblock \href {http://arxiv.org/abs/1602.01455} {\path{arXiv:1602.01455}},
  \href {https://doi.org/10.1103/PhysRevC.94.014904}
  {\path{doi:10.1103/PhysRevC.94.014904}}.

\bibitem{Liu:2017fib}
Yizhuang Liu and Ismail Zahed.
\newblock {Viscous corrections to electromagnetic emissivities in QCD}.
\newblock {\em Phys. Rev. D}, 96(11):116021, 2017.
\newblock \href {http://arxiv.org/abs/1707.08523} {\path{arXiv:1707.08523}},
  \href {https://doi.org/10.1103/PhysRevD.96.116021}
  {\path{doi:10.1103/PhysRevD.96.116021}}.

\bibitem{Hauksson:2017udm}
Sigtryggur Hauksson, Sangyong Jeon, and Charles Gale.
\newblock {Photon emission from quark-gluon plasma out of equilibrium}.
\newblock {\em Phys. Rev. C}, 97(1):014901, 2018.
\newblock \href {http://arxiv.org/abs/1709.03598} {\path{arXiv:1709.03598}},
  \href {https://doi.org/10.1103/PhysRevC.97.014901}
  {\path{doi:10.1103/PhysRevC.97.014901}}.

\bibitem{Schafer:2021slz}
Anna Sch\"afer, Oscar Garcia-Montero, Jean-Fran\c{c}ois Paquet, Hannah Elfner,
  and Charles Gale.
\newblock {Out-of-equilibrium photon production in the late stages of
  relativistic heavy-ion collisions}.
\newblock {\em Phys. Rev. C}, 105(4):044910, 2022.
\newblock \href {http://arxiv.org/abs/2111.13603} {\path{arXiv:2111.13603}},
  \href {https://doi.org/10.1103/PhysRevC.105.044910}
  {\path{doi:10.1103/PhysRevC.105.044910}}.

\bibitem{Schafer:2019edr}
Anna Sch\"afer, Juan~M. Torres-Rincon, Jonas Rothermel, Niklas Ehlert, Charles
  Gale, and Hannah Elfner.
\newblock {Benchmarking a nonequilibrium approach to photon emission in
  relativistic heavy-ion collisions}.
\newblock {\em Phys. Rev. D}, 99(11):114021, 2019.
\newblock \href {http://arxiv.org/abs/1902.07564} {\path{arXiv:1902.07564}},
  \href {https://doi.org/10.1103/PhysRevD.99.114021}
  {\path{doi:10.1103/PhysRevD.99.114021}}.

\bibitem{Holt:2015cda}
Nathan P.~M. Holt, Paul~M. Hohler, and Ralf Rapp.
\newblock {Thermal photon emission from the
  \ensuremath{\pi}\ensuremath{\rho}\ensuremath{\omega} system}.
\newblock {\em Nucl. Phys. A}, 945:1--20, 2016.
\newblock \href {http://arxiv.org/abs/1506.09205} {\path{arXiv:1506.09205}},
  \href {https://doi.org/10.1016/j.nuclphysa.2015.09.008}
  {\path{doi:10.1016/j.nuclphysa.2015.09.008}}.

\bibitem{Holt:2020mwf}
Nathan P.~M. Holt and Ralf Rapp.
\newblock {Baryonic Sources of Thermal Photons}.
\newblock {\em Eur. Phys. J. A}, 56(11):292, 2020.
\newblock \href {http://arxiv.org/abs/2008.00116} {\path{arXiv:2008.00116}},
  \href {https://doi.org/10.1140/epja/s10050-020-00301-x}
  {\path{doi:10.1140/epja/s10050-020-00301-x}}.

\bibitem{Hidaka:2015ima}
Yoshimasa Hidaka, Shu Lin, Robert~D. Pisarski, and Daisuke Satow.
\newblock {Dilepton and photon production in the presence of a nontrivial
  Polyakov loop}.
\newblock {\em JHEP}, 10:005, 2015.
\newblock \href {http://arxiv.org/abs/1504.01770} {\path{arXiv:1504.01770}},
  \href {https://doi.org/10.1007/JHEP10(2015)005}
  {\path{doi:10.1007/JHEP10(2015)005}}.

\bibitem{Zakharov:2017cul}
B.~G. Zakharov.
\newblock {Phenomenology of collinear photon emission from
  quark\textendash{}gluon plasma in $AA$ collisions}.
\newblock {\em JETP Lett.}, 106(5):283--289, 2017.
\newblock \href {http://arxiv.org/abs/1707.08602} {\path{arXiv:1707.08602}},
  \href {https://doi.org/10.1134/S0021364017170027}
  {\path{doi:10.1134/S0021364017170027}}.

\bibitem{Bandyopadhyay:2015wua}
Aritra Bandyopadhyay, Najmul Haque, Munshi~G. Mustafa, and Michael Strickland.
\newblock {Dilepton rate and quark number susceptibility with the Gribov
  action}.
\newblock {\em Phys. Rev. D}, 93(6):065004, 2016.
\newblock \href {http://arxiv.org/abs/1508.06249} {\path{arXiv:1508.06249}},
  \href {https://doi.org/10.1103/PhysRevD.93.065004}
  {\path{doi:10.1103/PhysRevD.93.065004}}.

\bibitem{Iatrakis:2016ugz}
Ioannis Iatrakis, Elias Kiritsis, Chun Shen, and Di-Lun Yang.
\newblock {Holographic Photon Production in Heavy Ion Collisions}.
\newblock {\em JHEP}, 04:035, 2017.
\newblock \href {http://arxiv.org/abs/1609.07208} {\path{arXiv:1609.07208}},
  \href {https://doi.org/10.1007/JHEP04(2017)035}
  {\path{doi:10.1007/JHEP04(2017)035}}.

\bibitem{Ding:2016hua}
Heng-Tong Ding, Olaf Kaczmarek, and Florian Meyer.
\newblock {Thermal dilepton rates and electrical conductivity of the QGP from
  the lattice}.
\newblock {\em Phys. Rev. D}, 94(3):034504, 2016.
\newblock \href {http://arxiv.org/abs/1604.06712} {\path{arXiv:1604.06712}},
  \href {https://doi.org/10.1103/PhysRevD.94.034504}
  {\path{doi:10.1103/PhysRevD.94.034504}}.

\bibitem{Ghiglieri:2016tvj}
J.~Ghiglieri, O.~Kaczmarek, M.~Laine, and F.~Meyer.
\newblock {Lattice constraints on the thermal photon rate}.
\newblock {\em Phys. Rev. D}, 94(1):016005, 2016.
\newblock \href {http://arxiv.org/abs/1604.07544} {\path{arXiv:1604.07544}},
  \href {https://doi.org/10.1103/PhysRevD.94.016005}
  {\path{doi:10.1103/PhysRevD.94.016005}}.

\bibitem{Jackson:2019yao}
G.~Jackson and M.~Laine.
\newblock {Testing thermal photon and dilepton rates}.
\newblock {\em JHEP}, 11:144, 2019.
\newblock \href {http://arxiv.org/abs/1910.09567} {\path{arXiv:1910.09567}},
  \href {https://doi.org/10.1007/JHEP11(2019)144}
  {\path{doi:10.1007/JHEP11(2019)144}}.

\bibitem{Ce:2020tmx}
Marco C\`e, Tim Harris, Harvey~B. Meyer, Aman Steinberg, and Arianna Toniato.
\newblock {Rate of photon production in the quark-gluon plasma from lattice
  QCD}.
\newblock {\em Phys. Rev. D}, 102(9):091501, 2020.
\newblock \href {http://arxiv.org/abs/2001.03368} {\path{arXiv:2001.03368}},
  \href {https://doi.org/10.1103/PhysRevD.102.091501}
  {\path{doi:10.1103/PhysRevD.102.091501}}.

\bibitem{Ce:2022fot}
Marco C\`e, Tim Harris, Ardit Krasniqi, Harvey~B. Meyer, and Csaba T\"or\"ok.
\newblock {Photon emissivity of the quark-gluon plasma: A lattice QCD analysis
  of the transverse channel}.
\newblock {\em Phys. Rev. D}, 106(5):054501, 2022.
\newblock \href {http://arxiv.org/abs/2205.02821} {\path{arXiv:2205.02821}},
  \href {https://doi.org/10.1103/PhysRevD.106.054501}
  {\path{doi:10.1103/PhysRevD.106.054501}}.

\bibitem{Bhattacharya:2015ada}
Lusaka Bhattacharya, Radoslaw Ryblewski, and Michael Strickland.
\newblock {Photon production from a nonequilibrium quark-gluon plasma}.
\newblock {\em Phys. Rev. D}, 93(6):065005, 2016.
\newblock \href {http://arxiv.org/abs/1507.06605} {\path{arXiv:1507.06605}},
  \href {https://doi.org/10.1103/PhysRevD.93.065005}
  {\path{doi:10.1103/PhysRevD.93.065005}}.

\bibitem{Kasmaei:2018oag}
Babak~S. Kasmaei and Michael Strickland.
\newblock {Dilepton production and elliptic flow from an anisotropic
  quark-gluon plasma}.
\newblock {\em Phys. Rev. D}, 99(3):034015, 2019.
\newblock \href {http://arxiv.org/abs/1811.07486} {\path{arXiv:1811.07486}},
  \href {https://doi.org/10.1103/PhysRevD.99.034015}
  {\path{doi:10.1103/PhysRevD.99.034015}}.

\bibitem{Kasmaei:2019ofu}
Babak~Salehi Kasmaei and Michael Strickland.
\newblock {Photon production and elliptic flow from a momentum-anisotropic
  quark-gluon plasma}.
\newblock {\em Phys. Rev. D}, 102(1):014037, 2020.
\newblock \href {http://arxiv.org/abs/1911.03370} {\path{arXiv:1911.03370}},
  \href {https://doi.org/10.1103/PhysRevD.102.014037}
  {\path{doi:10.1103/PhysRevD.102.014037}}.

\bibitem{Fujii:2022hxa}
Hirotsugu Fujii, Kazunori Itakura, Katsunori Miyachi, and Chiho Nonaka.
\newblock {Radiative hadronization: Photon emission at hadronization from
  quark-gluon plasma}.
\newblock {\em Phys. Rev. C}, 106(3):034906, 2022.
\newblock \href {http://arxiv.org/abs/2204.03116} {\path{arXiv:2204.03116}},
  \href {https://doi.org/10.1103/PhysRevC.106.034906}
  {\path{doi:10.1103/PhysRevC.106.034906}}.

\bibitem{Tuchin:2019jxd}
Kirill Tuchin.
\newblock {Photon radiation in hot nuclear matter by means of chiral
  anomalies}.
\newblock {\em Phys. Rev. C}, 99(6):064907, 2019.
\newblock \href {http://arxiv.org/abs/1903.02629} {\path{arXiv:1903.02629}},
  \href {https://doi.org/10.1103/PhysRevC.99.064907}
  {\path{doi:10.1103/PhysRevC.99.064907}}.

\bibitem{Ayala:2017vex}
Alejandro Ayala, Jorge~David Castano-Yepes, Cesareo~A. Dominguez, Luis~A.
  Hernandez, Saul Hernandez-Ortiz, and Maria~Elena Tejeda-Yeomans.
\newblock {Prompt photon yield and elliptic flow from gluon fusion induced by
  magnetic fields in relativistic heavy-ion collisions}.
\newblock {\em Phys. Rev. D}, 96(1):014023, 2017.
\newblock [Erratum: Phys.Rev.D 96, 119901 (2017)].
\newblock \href {http://arxiv.org/abs/1704.02433} {\path{arXiv:1704.02433}},
  \href {https://doi.org/10.1103/PhysRevD.96.014023}
  {\path{doi:10.1103/PhysRevD.96.014023}}.

\bibitem{Linnyk:2015tha}
O.~Linnyk, V.~Konchakovski, T.~Steinert, W.~Cassing, and E.~L. Bratkovskaya.
\newblock {Hadronic and partonic sources of direct photons in relativistic
  heavy-ion collisions}.
\newblock {\em Phys. Rev. C}, 92(5):054914, 2015.
\newblock \href {http://arxiv.org/abs/1504.05699} {\path{arXiv:1504.05699}},
  \href {https://doi.org/10.1103/PhysRevC.92.054914}
  {\path{doi:10.1103/PhysRevC.92.054914}}.

\bibitem{NA60:2008ctj}
R.~Arnaldi et~al.
\newblock {NA60 results on thermal dimuons}.
\newblock {\em Eur. Phys. J. C}, 61:711--720, 2009.
\newblock \href {http://arxiv.org/abs/0812.3053} {\path{arXiv:0812.3053}},
  \href {https://doi.org/10.1140/epjc/s10052-009-0878-5}
  {\path{doi:10.1140/epjc/s10052-009-0878-5}}.

\bibitem{Geurts:2012rv}
Frank Geurts.
\newblock {The STAR Dilepton Physics Program}.
\newblock {\em Nucl. Phys. A}, 904-905:217c--224c, 2013.
\newblock \href {http://arxiv.org/abs/1210.5549} {\path{arXiv:1210.5549}},
  \href {https://doi.org/10.1016/j.nuclphysa.2013.01.062}
  {\path{doi:10.1016/j.nuclphysa.2013.01.062}}.

\bibitem{Hohler:2013eba}
Paul~M. Hohler and Ralf Rapp.
\newblock {Is $\rho$-Meson Melting Compatible with Chiral Restoration?}
\newblock {\em Phys. Lett. B}, 731:103--109, 2014.
\newblock \href {http://arxiv.org/abs/1311.2921} {\path{arXiv:1311.2921}},
  \href {https://doi.org/10.1016/j.physletb.2014.02.021}
  {\path{doi:10.1016/j.physletb.2014.02.021}}.

\bibitem{CMS:2016jip}
Vardan Khachatryan et~al.
\newblock {Measurement of the double-differential inclusive jet cross section
  in proton\textendash{}proton collisions at $\sqrt{s} = 13\,\text {TeV} $}.
\newblock {\em Eur. Phys. J. C}, 76(8):451, 2016.
\newblock \href {http://arxiv.org/abs/1605.04436} {\path{arXiv:1605.04436}},
  \href {https://doi.org/10.1140/epjc/s10052-016-4286-3}
  {\path{doi:10.1140/epjc/s10052-016-4286-3}}.

\bibitem{ATLAS:2017ble}
M.~Aaboud et~al.
\newblock {Measurement of inclusive jet and dijet cross-sections in
  proton-proton collisions at $\sqrt{s}=13$ TeV with the ATLAS detector}.
\newblock {\em JHEP}, 05:195, 2018.
\newblock \href {http://arxiv.org/abs/1711.02692} {\path{arXiv:1711.02692}},
  \href {https://doi.org/10.1007/JHEP05(2018)195}
  {\path{doi:10.1007/JHEP05(2018)195}}.

\bibitem{PHENIX:2001hpc}
K.~Adcox et~al.
\newblock {Suppression of hadrons with large transverse momentum in central
  Au+Au collisions at $\sqrt{s_{NN}}$ = 130-GeV}.
\newblock {\em Phys. Rev. Lett.}, 88:022301, 2002.
\newblock \href {http://arxiv.org/abs/nucl-ex/0109003}
  {\path{arXiv:nucl-ex/0109003}}, \href
  {https://doi.org/10.1103/PhysRevLett.88.022301}
  {\path{doi:10.1103/PhysRevLett.88.022301}}.

\bibitem{STAR:2002svs}
C.~Adler et~al.
\newblock {Disappearance of back-to-back high $p_{T}$ hadron correlations in
  central Au+Au collisions at $\sqrt{s_{NN}}$ = 200-GeV}.
\newblock {\em Phys. Rev. Lett.}, 90:082302, 2003.
\newblock \href {http://arxiv.org/abs/nucl-ex/0210033}
  {\path{arXiv:nucl-ex/0210033}}, \href
  {https://doi.org/10.1103/PhysRevLett.90.082302}
  {\path{doi:10.1103/PhysRevLett.90.082302}}.

\bibitem{ATLAS:2010isq}
Georges Aad et~al.
\newblock {Observation of a Centrality-Dependent Dijet Asymmetry in Lead-Lead
  Collisions at $\sqrt{s_{NN}}=2.77$ TeV with the ATLAS Detector at the LHC}.
\newblock {\em Phys. Rev. Lett.}, 105:252303, 2010.
\newblock \href {http://arxiv.org/abs/1011.6182} {\path{arXiv:1011.6182}},
  \href {https://doi.org/10.1103/PhysRevLett.105.252303}
  {\path{doi:10.1103/PhysRevLett.105.252303}}.

\bibitem{CMS:2011iwn}
Serguei Chatrchyan et~al.
\newblock {Observation and studies of jet quenching in PbPb collisions at
  nucleon-nucleon center-of-mass energy = 2.76 TeV}.
\newblock {\em Phys. Rev. C}, 84:024906, 2011.
\newblock \href {http://arxiv.org/abs/1102.1957} {\path{arXiv:1102.1957}},
  \href {https://doi.org/10.1103/PhysRevC.84.024906}
  {\path{doi:10.1103/PhysRevC.84.024906}}.

\bibitem{CMS:2016uxf}
Vardan Khachatryan et~al.
\newblock {Measurement of inclusive jet cross sections in $pp$ and PbPb
  collisions at $\sqrt{s_{NN}}=$ 2.76 TeV}.
\newblock {\em Phys. Rev. C}, 96(1):015202, 2017.
\newblock \href {http://arxiv.org/abs/1609.05383} {\path{arXiv:1609.05383}},
  \href {https://doi.org/10.1103/PhysRevC.96.015202}
  {\path{doi:10.1103/PhysRevC.96.015202}}.

\bibitem{CMS:2014jjt}
Serguei Chatrchyan et~al.
\newblock {Measurement of Jet Fragmentation in PbPb and pp Collisions at
  $\sqrt{s_{NN}}= 2.76$ TeV}.
\newblock {\em Phys. Rev. C}, 90(2):024908, 2014.
\newblock \href {http://arxiv.org/abs/1406.0932} {\path{arXiv:1406.0932}},
  \href {https://doi.org/10.1103/PhysRevC.90.024908}
  {\path{doi:10.1103/PhysRevC.90.024908}}.

\bibitem{ATLAS:2014dtd}
Georges Aad et~al.
\newblock {Measurement of inclusive jet charged-particle fragmentation
  functions in Pb+Pb collisions at $\sqrt{s_{NN}}=2.76$ TeV with the ATLAS
  detector}.
\newblock {\em Phys. Lett. B}, 739:320--342, 2014.
\newblock \href {http://arxiv.org/abs/1406.2979} {\path{arXiv:1406.2979}},
  \href {https://doi.org/10.1016/j.physletb.2014.10.065}
  {\path{doi:10.1016/j.physletb.2014.10.065}}.

\bibitem{ATLAS:2018bvp}
Morad Aaboud et~al.
\newblock {Measurement of jet fragmentation in Pb+Pb and $pp$ collisions at
  $\sqrt{s_{NN}} = 5.02$ TeV with the ATLAS detector}.
\newblock {\em Phys. Rev. C}, 98(2):024908, 2018.
\newblock \href {http://arxiv.org/abs/1805.05424} {\path{arXiv:1805.05424}},
  \href {https://doi.org/10.1103/PhysRevC.98.024908}
  {\path{doi:10.1103/PhysRevC.98.024908}}.

\bibitem{CMS:2018fof}
Albert~M Sirunyan et~al.
\newblock {Measurement of the groomed jet mass in PbPb and pp collisions at $
  \sqrt{s_{\mathrm{NN}}}=5.02 $ TeV}.
\newblock {\em JHEP}, 10:161, 2018.
\newblock \href {http://arxiv.org/abs/1805.05145} {\path{arXiv:1805.05145}},
  \href {https://doi.org/10.1007/JHEP10(2018)161}
  {\path{doi:10.1007/JHEP10(2018)161}}.

\bibitem{CMS:2017qlm}
Albert~M Sirunyan et~al.
\newblock {Measurement of the Splitting Function in $pp$ and Pb-Pb Collisions
  at $\sqrt{s_{_{\mathrm{NN}}}} =$ 5.02 TeV}.
\newblock {\em Phys. Rev. Lett.}, 120(14):142302, 2018.
\newblock \href {http://arxiv.org/abs/1708.09429} {\path{arXiv:1708.09429}},
  \href {https://doi.org/10.1103/PhysRevLett.120.142302}
  {\path{doi:10.1103/PhysRevLett.120.142302}}.

\bibitem{ALargeIonColliderExperiment:2021mqf}
Shreyasi Acharya et~al.
\newblock {Measurement of the groomed jet radius and momentum splitting
  fraction in pp and Pb$-$Pb collisions at $\sqrt{s_{NN}} = 5.02$ TeV}.
\newblock {\em Phys. Rev. Lett.}, 128(10):102001, 2022.
\newblock \href {http://arxiv.org/abs/2107.12984} {\path{arXiv:2107.12984}},
  \href {https://doi.org/10.1103/PhysRevLett.128.102001}
  {\path{doi:10.1103/PhysRevLett.128.102001}}.

\bibitem{Hulcher:2017cpt}
Zachary Hulcher, Daniel Pablos, and Krishna Rajagopal.
\newblock {Resolution Effects in the Hybrid Strong/Weak Coupling Model}.
\newblock {\em JHEP}, 03:010, 2018.
\newblock \href {http://arxiv.org/abs/1707.05245} {\path{arXiv:1707.05245}},
  \href {https://doi.org/10.1007/JHEP03(2018)010}
  {\path{doi:10.1007/JHEP03(2018)010}}.

\bibitem{Caucal:2019uvr}
P.~Caucal, E.~Iancu, and G.~Soyez.
\newblock {Deciphering the $z_g$ distribution in ultrarelativistic heavy ion
  collisions}.
\newblock {\em JHEP}, 10:273, 2019.
\newblock \href {http://arxiv.org/abs/1907.04866} {\path{arXiv:1907.04866}},
  \href {https://doi.org/10.1007/JHEP10(2019)273}
  {\path{doi:10.1007/JHEP10(2019)273}}.

\bibitem{Casalderrey-Solana:2019ubu}
J.~Casalderrey-Solana, G.~Milhano, D.~Pablos, and K.~Rajagopal.
\newblock {Modification of Jet Substructure in Heavy Ion Collisions as a Probe
  of the Resolution Length of Quark-Gluon Plasma}.
\newblock {\em JHEP}, 01:044, 2020.
\newblock \href {http://arxiv.org/abs/1907.11248} {\path{arXiv:1907.11248}},
  \href {https://doi.org/10.1007/JHEP01(2020)044}
  {\path{doi:10.1007/JHEP01(2020)044}}.

\bibitem{Caucal:2020uic}
P.~Caucal, E.~Iancu, and G.~Soyez.
\newblock {Jet radiation in a longitudinally expanding medium}.
\newblock {\em JHEP}, 04:209, 2021.
\newblock \href {http://arxiv.org/abs/2012.01457} {\path{arXiv:2012.01457}},
  \href {https://doi.org/10.1007/JHEP04(2021)209}
  {\path{doi:10.1007/JHEP04(2021)209}}.

\bibitem{CMS:2017ehl}
Albert~M Sirunyan et~al.
\newblock {Study of jet quenching with isolated-photon+jet correlations in PbPb
  and pp collisions at $\sqrt{s_{_{\mathrm{NN}}}} =$ 5.02 TeV}.
\newblock {\em Phys. Lett. B}, 785:14--39, 2018.
\newblock \href {http://arxiv.org/abs/1711.09738} {\path{arXiv:1711.09738}},
  \href {https://doi.org/10.1016/j.physletb.2018.07.061}
  {\path{doi:10.1016/j.physletb.2018.07.061}}.

\bibitem{ATLAS:2018gwx}
Morad Aaboud et~al.
\newblock {Measurement of the nuclear modification factor for inclusive jets in
  Pb+Pb collisions at $\sqrt{s_\mathrm{NN}}=5.02$ TeV with the ATLAS detector}.
\newblock {\em Phys. Lett. B}, 790:108--128, 2019.
\newblock \href {http://arxiv.org/abs/1805.05635} {\path{arXiv:1805.05635}},
  \href {https://doi.org/10.1016/j.physletb.2018.10.076}
  {\path{doi:10.1016/j.physletb.2018.10.076}}.

\bibitem{STAR:2017hhs}
L.~Adamczyk et~al.
\newblock {Measurements of jet quenching with semi-inclusive hadron+jet
  distributions in Au+Au collisions at $\sqrt{s_{NN}}$ = 200 GeV}.
\newblock {\em Phys. Rev. C}, 96(2):024905, 2017.
\newblock \href {http://arxiv.org/abs/1702.01108} {\path{arXiv:1702.01108}},
  \href {https://doi.org/10.1103/PhysRevC.96.024905}
  {\path{doi:10.1103/PhysRevC.96.024905}}.

\bibitem{DEramo:2018eoy}
Francesco D'Eramo, Krishna Rajagopal, and Yi~Yin.
\newblock {Moli\`ere scattering in quark-gluon plasma: finding point-like
  scatterers in a liquid}.
\newblock {\em JHEP}, 01:172, 2019.
\newblock \href {http://arxiv.org/abs/1808.03250} {\path{arXiv:1808.03250}},
  \href {https://doi.org/10.1007/JHEP01(2019)172}
  {\path{doi:10.1007/JHEP01(2019)172}}.

\bibitem{DEramo:2018ydo}
Francesco D'Eramo, Krishna Rajagopal, and Yi~Yin.
\newblock {Finding the Scatterers in Hot Quark Soup}.
\newblock {\em PoS}, HardProbes2018:066, 2018.
\newblock \href {http://arxiv.org/abs/1812.06878} {\path{arXiv:1812.06878}},
  \href {https://doi.org/10.22323/1.345.0066} {\path{doi:10.22323/1.345.0066}}.

\bibitem{Barata:2020rdn}
Jo\~ao Barata, Yacine Mehtar-Tani, Alba Soto-Ontoso, and Konrad Tywoniuk.
\newblock {Revisiting transverse momentum broadening in dense QCD media}.
\newblock {\em Phys. Rev. D}, 104(5):054047, 2021.
\newblock \href {http://arxiv.org/abs/2009.13667} {\path{arXiv:2009.13667}},
  \href {https://doi.org/10.1103/PhysRevD.104.054047}
  {\path{doi:10.1103/PhysRevD.104.054047}}.

\bibitem{Ce:2020wgg}
Marco C\`e, Tim Harris, Harvey~B. Meyer, and Arianna Toniato.
\newblock {Deep inelastic scattering on the quark-gluon plasma}.
\newblock {\em JHEP}, 03:035, 2021.
\newblock \href {http://arxiv.org/abs/2012.07522} {\path{arXiv:2012.07522}},
  \href {https://doi.org/10.1007/JHEP03(2021)035}
  {\path{doi:10.1007/JHEP03(2021)035}}.

\bibitem{Hulcher:2022kmn}
Z.~Hulcher, D.~Pablos, and K.~Rajagopal.
\newblock {Sensitivity of jet observables to the presence of quasi-particles in
  QGP}.
\newblock In {\em {29th International Conference on Ultra-relativistic
  Nucleus-Nucleus Collisions}}, 8 2022.
\newblock \href {http://arxiv.org/abs/2208.13593} {\path{arXiv:2208.13593}}.

\bibitem{CMS:2013qak}
Serguei Chatrchyan et~al.
\newblock {Evidence of b-Jet Quenching in PbPb Collisions at
  $\sqrt{s_{NN}}=2.76$ TeV}.
\newblock {\em Phys. Rev. Lett.}, 113(13):132301, 2014.
\newblock [Erratum: Phys.Rev.Lett. 115, 029903 (2015)].
\newblock \href {http://arxiv.org/abs/1312.4198} {\path{arXiv:1312.4198}},
  \href {https://doi.org/10.1103/PhysRevLett.113.132301}
  {\path{doi:10.1103/PhysRevLett.113.132301}}.

\bibitem{ATLAS:2022fgb}
{Measurement of the nuclear modification factor of $b$-jets in 5.02 TeV Pb+Pb
  collisions with the ATLAS detector}.
\newblock 4 2022.
\newblock \href {http://arxiv.org/abs/2204.13530} {\path{arXiv:2204.13530}}.

\bibitem{CMS:2018dqf}
Albert~M Sirunyan et~al.
\newblock {Comparing transverse momentum balance of b jet pairs in pp and PbPb
  collisions at $ \sqrt{s_{\mathrm{NN}}}=5.02 $ TeV}.
\newblock {\em JHEP}, 03:181, 2018.
\newblock \href {http://arxiv.org/abs/1802.00707} {\path{arXiv:1802.00707}},
  \href {https://doi.org/10.1007/JHEP03(2018)181}
  {\path{doi:10.1007/JHEP03(2018)181}}.

\bibitem{Brewer:2020och}
Jasmine Brewer, Jesse Thaler, and Andrew~P. Turner.
\newblock {Data-driven quark and gluon jet modification in heavy-ion
  collisions}.
\newblock {\em Phys. Rev. C}, 103(2):L021901, 2021.
\newblock \href {http://arxiv.org/abs/2008.08596} {\path{arXiv:2008.08596}},
  \href {https://doi.org/10.1103/PhysRevC.103.L021901}
  {\path{doi:10.1103/PhysRevC.103.L021901}}.

\bibitem{Ying:2022jvy}
Yueyang Ying, Jasmine Brewer, Yi~Chen, and Yen-Jie Lee.
\newblock {Data-driven extraction of the substructure of quark and gluon jets
  in proton-proton and heavy-ion collisions}.
\newblock 4 2022.
\newblock \href {http://arxiv.org/abs/2204.00641} {\path{arXiv:2204.00641}}.

\bibitem{ALICE:2015mdb}
Jaroslav Adam et~al.
\newblock {Measurement of jet quenching with semi-inclusive hadron-jet
  distributions in central Pb-Pb collisions at $ \sqrt{s_{\mathrm{NN}}}=2.76 $
  TeV}.
\newblock {\em JHEP}, 09:170, 2015.
\newblock \href {http://arxiv.org/abs/1506.03984} {\path{arXiv:1506.03984}},
  \href {https://doi.org/10.1007/JHEP09(2015)170}
  {\path{doi:10.1007/JHEP09(2015)170}}.

\bibitem{Putschke:2019yrg}
J.~H. Putschke et~al.
\newblock {The JETSCAPE framework}.
\newblock 3 2019.
\newblock \href {http://arxiv.org/abs/1903.07706} {\path{arXiv:1903.07706}}.

\bibitem{Zapp:2012ak}
Korinna~C. Zapp, Frank Krauss, and Urs~A. Wiedemann.
\newblock {A perturbative framework for jet quenching}.
\newblock {\em JHEP}, 03:080, 2013.
\newblock \href {http://arxiv.org/abs/1212.1599} {\path{arXiv:1212.1599}},
  \href {https://doi.org/10.1007/JHEP03(2013)080}
  {\path{doi:10.1007/JHEP03(2013)080}}.

\bibitem{Zapp:2013vla}
Korinna~C. Zapp.
\newblock {JEWEL 2.0.0: directions for use}.
\newblock {\em Eur. Phys. J. C}, 74(2):2762, 2014.
\newblock \href {http://arxiv.org/abs/1311.0048} {\path{arXiv:1311.0048}},
  \href {https://doi.org/10.1140/epjc/s10052-014-2762-1}
  {\path{doi:10.1140/epjc/s10052-014-2762-1}}.

\bibitem{Caucal:2018dla}
P.~Caucal, E.~Iancu, A.~H. Mueller, and G.~Soyez.
\newblock {Vacuum-like jet fragmentation in a dense QCD medium}.
\newblock {\em Phys. Rev. Lett.}, 120:232001, 2018.
\newblock \href {http://arxiv.org/abs/1801.09703} {\path{arXiv:1801.09703}},
  \href {https://doi.org/10.1103/PhysRevLett.120.232001}
  {\path{doi:10.1103/PhysRevLett.120.232001}}.

\bibitem{Chien:2016led}
Yang-Ting Chien and Ivan Vitev.
\newblock {Probing the Hardest Branching within Jets in Heavy-Ion Collisions}.
\newblock {\em Phys. Rev. Lett.}, 119(11):112301, 2017.
\newblock \href {http://arxiv.org/abs/1608.07283} {\path{arXiv:1608.07283}},
  \href {https://doi.org/10.1103/PhysRevLett.119.112301}
  {\path{doi:10.1103/PhysRevLett.119.112301}}.

\bibitem{Chang:2019nrx}
Ning-Bo Chang.
\newblock {Probing medium-induced jet splitting in heavy-ion collisions}.
\newblock {\em PoS}, HardProbes2018:076, 2019.
\newblock \href {https://doi.org/10.22323/1.345.0076}
  {\path{doi:10.22323/1.345.0076}}.

\bibitem{Casalderrey-Solana:2014bpa}
Jorge Casalderrey-Solana, Doga~Can Gulhan, Jos\'e~Guilherme Milhano, Daniel
  Pablos, and Krishna Rajagopal.
\newblock {A Hybrid Strong/Weak Coupling Approach to Jet Quenching}.
\newblock {\em JHEP}, 10:019, 2014.
\newblock [Erratum: JHEP 09, 175 (2015)].
\newblock \href {http://arxiv.org/abs/1405.3864} {\path{arXiv:1405.3864}},
  \href {https://doi.org/10.1007/JHEP09(2015)175}
  {\path{doi:10.1007/JHEP09(2015)175}}.

\bibitem{ATLAS:2022vii}
{Measurement of substructure-dependent jet suppression in Pb+Pb collisions at
  5.02 TeV with the ATLAS detector}.
\newblock 11 2022.
\newblock \href {http://arxiv.org/abs/2211.11470} {\path{arXiv:2211.11470}}.

\bibitem{CMS:2016qnj}
Vardan Khachatryan et~al.
\newblock {Correlations between jets and charged particles in PbPb and pp
  collisions at $ \sqrt{s_{\mathrm{NN}}}=2.76 $ TeV}.
\newblock {\em JHEP}, 02:156, 2016.
\newblock \href {http://arxiv.org/abs/1601.00079} {\path{arXiv:1601.00079}},
  \href {https://doi.org/10.1007/JHEP02(2016)156}
  {\path{doi:10.1007/JHEP02(2016)156}}.

\bibitem{CMS:2016cvr}
Vardan Khachatryan et~al.
\newblock {Decomposing transverse momentum balance contributions for quenched
  jets in PbPb collisions at $ \sqrt{s_{\mathrm{N}\;\mathrm{N}}}=2.76 $ TeV}.
\newblock {\em JHEP}, 11:055, 2016.
\newblock \href {http://arxiv.org/abs/1609.02466} {\path{arXiv:1609.02466}},
  \href {https://doi.org/10.1007/JHEP11(2016)055}
  {\path{doi:10.1007/JHEP11(2016)055}}.

\bibitem{CMS:2018zze}
Albert~M Sirunyan et~al.
\newblock {Jet properties in PbPb and pp collisions at $
  \sqrt{s_{\mathrm{N}\;\mathrm{N}}}=5.02 $ TeV}.
\newblock {\em JHEP}, 05:006, 2018.
\newblock \href {http://arxiv.org/abs/1803.00042} {\path{arXiv:1803.00042}},
  \href {https://doi.org/10.1007/JHEP05(2018)006}
  {\path{doi:10.1007/JHEP05(2018)006}}.

\bibitem{ATLAS:2019pid}
Georges Aad et~al.
\newblock {Measurement of angular and momentum distributions of charged
  particles within and around jets in Pb+Pb and $pp$ collisions at
  $\sqrt{s_{\mathrm{NN}}} = 5.02$ TeV with the ATLAS detector}.
\newblock {\em Phys. Rev. C}, 100(6):064901, 2019.
\newblock [Erratum: Phys.Rev.C 101, 059903 (2020)].
\newblock \href {http://arxiv.org/abs/1908.05264} {\path{arXiv:1908.05264}},
  \href {https://doi.org/10.1103/PhysRevC.100.064901}
  {\path{doi:10.1103/PhysRevC.100.064901}}.

\bibitem{CMS:2012nro}
Serguei Chatrchyan et~al.
\newblock {Measurement of jet fragmentation into charged particles in $pp$ and
  PbPb collisions at $\sqrt{s_{NN}}=2.76$ TeV}.
\newblock {\em JHEP}, 10:087, 2012.
\newblock \href {http://arxiv.org/abs/1205.5872} {\path{arXiv:1205.5872}},
  \href {https://doi.org/10.1007/JHEP10(2012)087}
  {\path{doi:10.1007/JHEP10(2012)087}}.

\bibitem{STAR:2020xiv}
Jaroslav Adam et~al.
\newblock {Measurement of inclusive charged-particle jet production in Au + Au
  collisions at $\sqrt{s_{NN}}=$200 GeV}.
\newblock {\em Phys. Rev. C}, 102(5):054913, 2020.
\newblock \href {http://arxiv.org/abs/2006.00582} {\path{arXiv:2006.00582}},
  \href {https://doi.org/10.1103/PhysRevC.102.054913}
  {\path{doi:10.1103/PhysRevC.102.054913}}.

\bibitem{ATLAS:2012tjt}
Georges Aad et~al.
\newblock {Measurement of the jet radius and transverse momentum dependence of
  inclusive jet suppression in lead-lead collisions at $\sqrt{s_{NN}}$= 2.76
  TeV with the ATLAS detector}.
\newblock {\em Phys. Lett. B}, 719:220--241, 2013.
\newblock \href {http://arxiv.org/abs/1208.1967} {\path{arXiv:1208.1967}},
  \href {https://doi.org/10.1016/j.physletb.2013.01.024}
  {\path{doi:10.1016/j.physletb.2013.01.024}}.

\bibitem{ALICE:2019qyj}
Shreyasi Acharya et~al.
\newblock {Measurements of inclusive jet spectra in pp and central Pb-Pb
  collisions at $\sqrt{s_{\rm{NN}}}$ = 5.02 TeV}.
\newblock {\em Phys. Rev. C}, 101(3):034911, 2020.
\newblock \href {http://arxiv.org/abs/1909.09718} {\path{arXiv:1909.09718}},
  \href {https://doi.org/10.1103/PhysRevC.101.034911}
  {\path{doi:10.1103/PhysRevC.101.034911}}.

\bibitem{CMS:2021vui}
Albert~M Sirunyan et~al.
\newblock {First measurement of large area jet transverse momentum spectra in
  heavy-ion collisions}.
\newblock {\em JHEP}, 05:284, 2021.
\newblock \href {http://arxiv.org/abs/2102.13080} {\path{arXiv:2102.13080}},
  \href {https://doi.org/10.1007/JHEP05(2021)284}
  {\path{doi:10.1007/JHEP05(2021)284}}.

\bibitem{CMS:2021otx}
Albert~M Sirunyan et~al.
\newblock {Using Z Boson Events to Study Parton-Medium Interactions in Pb-Pb
  Collisions}.
\newblock {\em Phys. Rev. Lett.}, 128(12):122301, 2022.
\newblock \href {http://arxiv.org/abs/2103.04377} {\path{arXiv:2103.04377}},
  \href {https://doi.org/10.1103/PhysRevLett.128.122301}
  {\path{doi:10.1103/PhysRevLett.128.122301}}.

\bibitem{Schenke:2009gb}
Bjoern Schenke, Charles Gale, and Sangyong Jeon.
\newblock {MARTINI: An Event generator for relativistic heavy-ion collisions}.
\newblock {\em Phys. Rev. C}, 80:054913, 2009.
\newblock \href {http://arxiv.org/abs/0909.2037} {\path{arXiv:0909.2037}},
  \href {https://doi.org/10.1103/PhysRevC.80.054913}
  {\path{doi:10.1103/PhysRevC.80.054913}}.

\bibitem{He:2018xjv}
Yayun He, Shanshan Cao, Wei Chen, Tan Luo, Long-Gang Pang, and Xin-Nian Wang.
\newblock {Interplaying mechanisms behind single inclusive jet suppression in
  heavy-ion collisions}.
\newblock {\em Phys. Rev. C}, 99(5):054911, 2019.
\newblock \href {http://arxiv.org/abs/1809.02525} {\path{arXiv:1809.02525}},
  \href {https://doi.org/10.1103/PhysRevC.99.054911}
  {\path{doi:10.1103/PhysRevC.99.054911}}.

\bibitem{Ke:2020clc}
Weiyao Ke and Xin-Nian Wang.
\newblock {QGP modification to single inclusive jets in a calibrated transport
  model}.
\newblock {\em JHEP}, 05:041, 2021.
\newblock \href {http://arxiv.org/abs/2010.13680} {\path{arXiv:2010.13680}},
  \href {https://doi.org/10.1007/JHEP05(2021)041}
  {\path{doi:10.1007/JHEP05(2021)041}}.

\bibitem{JET:2013cls}
Karen~M. Burke et~al.
\newblock {Extracting the jet transport coefficient from jet quenching in
  high-energy heavy-ion collisions}.
\newblock {\em Phys. Rev. C}, 90(1):014909, 2014.
\newblock \href {http://arxiv.org/abs/1312.5003} {\path{arXiv:1312.5003}},
  \href {https://doi.org/10.1103/PhysRevC.90.014909}
  {\path{doi:10.1103/PhysRevC.90.014909}}.

\bibitem{ALICE:2015efi}
Jaroslav Adam et~al.
\newblock {Azimuthal anisotropy of charged jet production in $\sqrt{s_{\rm
  NN}}$ = 2.76 TeV Pb-Pb collisions}.
\newblock {\em Phys. Lett. B}, 753:511--525, 2016.
\newblock \href {http://arxiv.org/abs/1509.07334} {\path{arXiv:1509.07334}},
  \href {https://doi.org/10.1016/j.physletb.2015.12.047}
  {\path{doi:10.1016/j.physletb.2015.12.047}}.

\bibitem{ATLAS:2021ktw}
Georges Aad et~al.
\newblock {Measurements of azimuthal anisotropies of jet production in Pb+Pb
  collisions at $\sqrt{s_{NN}} =$ 5.02 TeV with the ATLAS detector}.
\newblock {\em Phys. Rev. C}, 105(6):064903, 2022.
\newblock \href {http://arxiv.org/abs/2111.06606} {\path{arXiv:2111.06606}},
  \href {https://doi.org/10.1103/PhysRevC.105.064903}
  {\path{doi:10.1103/PhysRevC.105.064903}}.

\bibitem{CMS:2022nsv}
{Azimuthal anisotropy of dijet events in PbPb collisions at
  $\sqrt{s_\mathrm{NN}}$ = 5.02 TeV}.
\newblock 10 2022.
\newblock \href {http://arxiv.org/abs/2210.08325} {\path{arXiv:2210.08325}}.

\bibitem{Dong:2019byy}
Xin Dong, Yen-Jie Lee, and Ralf Rapp.
\newblock {Open Heavy-Flavor Production in Heavy-Ion Collisions}.
\newblock {\em Ann. Rev. Nucl. Part. Sci.}, 69:417--445, 2019.
\newblock \href {http://arxiv.org/abs/1903.07709} {\path{arXiv:1903.07709}},
  \href {https://doi.org/10.1146/annurev-nucl-101918-023806}
  {\path{doi:10.1146/annurev-nucl-101918-023806}}.

\bibitem{He:2022ywp}
Min He, Hendrik van Hees, and Ralf Rapp.
\newblock {Heavy-Quark Diffusion in the Quark-Gluon Plasma}.
\newblock 4 2022.
\newblock \href {http://arxiv.org/abs/2204.09299} {\path{arXiv:2204.09299}}.

\bibitem{Akiba:2015jwa}
Yasuyuki Akiba et~al.
\newblock {The Hot QCD White Paper: Exploring the Phases of QCD at RHIC and the
  LHC}.
\newblock 2 2015.
\newblock \href {http://arxiv.org/abs/1502.02730} {\path{arXiv:1502.02730}}.

\bibitem{Matsui:1986dk}
T.~Matsui and H.~Satz.
\newblock {$J/\psi$ Suppression by Quark-Gluon Plasma Formation}.
\newblock {\em Phys. Lett. B}, 178:416--422, 1986.
\newblock \href {https://doi.org/10.1016/0370-2693(86)91404-8}
  {\path{doi:10.1016/0370-2693(86)91404-8}}.

\bibitem{Rothkopf:2019ipj}
Alexander Rothkopf.
\newblock {Heavy Quarkonium in Extreme Conditions}.
\newblock {\em Phys. Rept.}, 858:1--117, 2020.
\newblock \href {http://arxiv.org/abs/1912.02253} {\path{arXiv:1912.02253}},
  \href {https://doi.org/10.1016/j.physrep.2020.02.006}
  {\path{doi:10.1016/j.physrep.2020.02.006}}.

\bibitem{Akamatsu:2020ypb}
Yukinao Akamatsu.
\newblock {Quarkonium in quark\textendash{}gluon plasma: Open quantum system
  approaches re-examined}.
\newblock {\em Prog. Part. Nucl. Phys.}, 123:103932, 2022.
\newblock \href {http://arxiv.org/abs/2009.10559} {\path{arXiv:2009.10559}},
  \href {https://doi.org/10.1016/j.ppnp.2021.103932}
  {\path{doi:10.1016/j.ppnp.2021.103932}}.

\bibitem{Sharma:2021vvu}
Rishi Sharma.
\newblock {Quarkonium propagation in the quark\textendash{}gluon plasma}.
\newblock {\em Eur. Phys. J. ST}, 230(3):697--718, 2021.
\newblock \href {http://arxiv.org/abs/2101.04268} {\path{arXiv:2101.04268}},
  \href {https://doi.org/10.1140/epjs/s11734-021-00025-z}
  {\path{doi:10.1140/epjs/s11734-021-00025-z}}.

\bibitem{Yao:2021lus}
Xiaojun Yao.
\newblock {Open quantum systems for quarkonia}.
\newblock {\em Int. J. Mod. Phys. A}, 36(20):2130010, 2021.
\newblock \href {http://arxiv.org/abs/2102.01736} {\path{arXiv:2102.01736}},
  \href {https://doi.org/10.1142/S0217751X21300106}
  {\path{doi:10.1142/S0217751X21300106}}.

\bibitem{Bouttefeux:2020ycy}
A.~Bouttefeux and M.~Laine.
\newblock {Mass-suppressed effects in heavy quark diffusion}.
\newblock {\em JHEP}, 12:150, 2020.
\newblock \href {http://arxiv.org/abs/2010.07316} {\path{arXiv:2010.07316}},
  \href {https://doi.org/10.1007/JHEP12(2020)150}
  {\path{doi:10.1007/JHEP12(2020)150}}.

\bibitem{Altenkort:2020fgs}
Luis Altenkort, Alexander~M. Eller, Olaf Kaczmarek, Lukas Mazur, Guy~D. Moore,
  and Hai-Tao Shu.
\newblock {Heavy quark momentum diffusion from the lattice using gradient
  flow}.
\newblock {\em Phys. Rev. D}, 103(1):014511, 2021.
\newblock \href {http://arxiv.org/abs/2009.13553} {\path{arXiv:2009.13553}},
  \href {https://doi.org/10.1103/PhysRevD.103.014511}
  {\path{doi:10.1103/PhysRevD.103.014511}}.

\bibitem{Mayer-Steudte:2021hei}
Julian Mayer-Steudte, Nora Brambilla, Viljami Leino, and Peter Petreczky.
\newblock {Chromoelectric and chromomagnetic correlators at high temperature
  from gradient flow}.
\newblock {\em PoS}, LATTICE2021:318, 2022.
\newblock \href {http://arxiv.org/abs/2111.10340} {\path{arXiv:2111.10340}},
  \href {https://doi.org/10.22323/1.396.0318} {\path{doi:10.22323/1.396.0318}}.

\bibitem{Rapp:2018qla}
A.~Beraudo et~al.
\newblock {Extraction of Heavy-Flavor Transport Coefficients in QCD Matter}.
\newblock {\em Nucl. Phys. A}, 979:21--86, 2018.
\newblock \href {http://arxiv.org/abs/1803.03824} {\path{arXiv:1803.03824}},
  \href {https://doi.org/10.1016/j.nuclphysa.2018.09.002}
  {\path{doi:10.1016/j.nuclphysa.2018.09.002}}.

\bibitem{Cao:2018ews}
Shanshan Cao et~al.
\newblock {Toward the determination of heavy-quark transport coefficients in
  quark-gluon plasma}.
\newblock {\em Phys. Rev. C}, 99(5):054907, 2019.
\newblock \href {http://arxiv.org/abs/1809.07894} {\path{arXiv:1809.07894}},
  \href {https://doi.org/10.1103/PhysRevC.99.054907}
  {\path{doi:10.1103/PhysRevC.99.054907}}.

\bibitem{STAR:2019ank}
Jaroslav Adam et~al.
\newblock {First measurement of $\Lambda_c$ baryon production in Au+Au
  collisions at $\sqrt{s_{\rm NN}}$ = 200 GeV}.
\newblock {\em Phys. Rev. Lett.}, 124(17):172301, 2020.
\newblock \href {http://arxiv.org/abs/1910.14628} {\path{arXiv:1910.14628}},
  \href {https://doi.org/10.1103/PhysRevLett.124.172301}
  {\path{doi:10.1103/PhysRevLett.124.172301}}.

\bibitem{CMS:2019uws}
Albert~M Sirunyan et~al.
\newblock {Production of $\Lambda_\mathrm{c}^+$ baryons in proton-proton and
  lead-lead collisions at $\sqrt{s_\mathrm{NN}}=$ 5.02 TeV}.
\newblock {\em Phys. Lett. B}, 803:135328, 2020.
\newblock \href {http://arxiv.org/abs/1906.03322} {\path{arXiv:1906.03322}},
  \href {https://doi.org/10.1016/j.physletb.2020.135328}
  {\path{doi:10.1016/j.physletb.2020.135328}}.

\bibitem{ALICE:2021bib}
Shreyasi Acharya et~al.
\newblock {Constraining hadronization mechanisms with $\rm \Lambda_{\rm
  c}^{+}$/D$^0$ production ratios in Pb-Pb collisions at $\sqrt{s_{\rm NN}} =
  5.02$ TeV}.
\newblock 12 2021.
\newblock \href {http://arxiv.org/abs/2112.08156} {\path{arXiv:2112.08156}}.

\bibitem{LHCb:2022ddg}
{Measurement of the $\Lambda_c^+$ to $D^0$ production cross-section ratio in
  peripheral PbPb collisions}.
\newblock 10 2022.
\newblock \href {http://arxiv.org/abs/2210.06939} {\path{arXiv:2210.06939}}.

\bibitem{He:2019vgs}
Min He and Ralf Rapp.
\newblock {Hadronization and Charm-Hadron Ratios in Heavy-Ion Collisions}.
\newblock {\em Phys. Rev. Lett.}, 124(4):042301, 2020.
\newblock \href {http://arxiv.org/abs/1905.09216} {\path{arXiv:1905.09216}},
  \href {https://doi.org/10.1103/PhysRevLett.124.042301}
  {\path{doi:10.1103/PhysRevLett.124.042301}}.

\bibitem{Plumari:2017ntm}
Salvatore Plumari, Vincenzo Minissale, Santosh~K. Das, G.~Coci, and V.~Greco.
\newblock {Charmed Hadrons from Coalescence plus Fragmentation in relativistic
  nucleus-nucleus collisions at RHIC and LHC}.
\newblock {\em Eur. Phys. J. C}, 78(4):348, 2018.
\newblock \href {http://arxiv.org/abs/1712.00730} {\path{arXiv:1712.00730}},
  \href {https://doi.org/10.1140/epjc/s10052-018-5828-7}
  {\path{doi:10.1140/epjc/s10052-018-5828-7}}.

\bibitem{Andronic:2021erx}
Anton Andronic, Peter Braun-Munzinger, Markus~K. K\"ohler, Aleksas
  Mazeliauskas, Krzysztof Redlich, Johanna Stachel, and Vytautas Vislavicius.
\newblock {The multiple-charm hierarchy in the statistical hadronization
  model}.
\newblock {\em JHEP}, 07:035, 2021.
\newblock \href {http://arxiv.org/abs/2104.12754} {\path{arXiv:2104.12754}},
  \href {https://doi.org/10.1007/JHEP07(2021)035}
  {\path{doi:10.1007/JHEP07(2021)035}}.

\bibitem{Zhao:2018jlw}
Jiaxing Zhao, Shuzhe Shi, Nu~Xu, and Pengfei Zhuang.
\newblock {Sequential Coalescence with Charm Conservation in High Energy
  Nuclear Collisions}.
\newblock 5 2018.
\newblock \href {http://arxiv.org/abs/1805.10858} {\path{arXiv:1805.10858}}.

\bibitem{Cho:2019lxb}
Sungtae Cho, Kai-Jia Sun, Che~Ming Ko, Su~Houng Lee, and Yongseok Oh.
\newblock {Charmed hadron production in an improved quark coalescence model}.
\newblock {\em Phys. Rev. C}, 101(2):024909, 2020.
\newblock \href {http://arxiv.org/abs/1905.09774} {\path{arXiv:1905.09774}},
  \href {https://doi.org/10.1103/PhysRevC.101.024909}
  {\path{doi:10.1103/PhysRevC.101.024909}}.

\bibitem{Cao:2019iqs}
Shanshan Cao, Kai-Jia Sun, Shu-Qing Li, Shuai Y.~F. Liu, Wen-Jing Xing,
  Guang-You Qin, and Che~Ming Ko.
\newblock {Charmed hadron chemistry in relativistic heavy-ion collisions}.
\newblock {\em Phys. Lett. B}, 807:135561, 2020.
\newblock \href {http://arxiv.org/abs/1911.00456} {\path{arXiv:1911.00456}},
  \href {https://doi.org/10.1016/j.physletb.2020.135561}
  {\path{doi:10.1016/j.physletb.2020.135561}}.

\bibitem{STAR:2021tte}
J.~Adam et~al.
\newblock {Observation of $D_{s}^{\pm}/D^0$ enhancement in Au+Au collisions at
  $\sqrt{s_{_{NN}}}$ = 200 GeV}.
\newblock {\em Phys. Rev. Lett.}, 127:092301, 2021.
\newblock \href {http://arxiv.org/abs/2101.11793} {\path{arXiv:2101.11793}},
  \href {https://doi.org/10.1103/PhysRevLett.127.092301}
  {\path{doi:10.1103/PhysRevLett.127.092301}}.

\bibitem{ALICE:2021kfc}
Shreyasi Acharya et~al.
\newblock {Measurement of prompt $D_s^+$-meson production and azimuthal
  anisotropy in Pb\textendash{}Pb collisions at $\sqrt {s_{NN}}$=5.02TeV}.
\newblock {\em Phys. Lett. B}, 827:136986, 2022.
\newblock \href {http://arxiv.org/abs/2110.10006} {\path{arXiv:2110.10006}},
  \href {https://doi.org/10.1016/j.physletb.2022.136986}
  {\path{doi:10.1016/j.physletb.2022.136986}}.

\bibitem{CMS:2021mzx}
Armen Tumasyan et~al.
\newblock {Observation of Bs0 mesons and measurement of the Bs0/B+ yield ratio
  in PbPb collisions at Image 1 TeV}.
\newblock {\em Phys. Lett. B}, 829:137062, 2022.
\newblock \href {http://arxiv.org/abs/2109.01908} {\path{arXiv:2109.01908}},
  \href {https://doi.org/10.1016/j.physletb.2022.137062}
  {\path{doi:10.1016/j.physletb.2022.137062}}.

\bibitem{Kang:2016ofv}
Zhong-Bo Kang, Felix Ringer, and Ivan Vitev.
\newblock {Effective field theory approach to open heavy flavor production in
  heavy-ion collisions}.
\newblock {\em JHEP}, 03:146, 2017.
\newblock \href {http://arxiv.org/abs/1610.02043} {\path{arXiv:1610.02043}},
  \href {https://doi.org/10.1007/JHEP03(2017)146}
  {\path{doi:10.1007/JHEP03(2017)146}}.

\bibitem{Li:2018xuv}
Hai~Tao Li and Ivan Vitev.
\newblock {Inclusive heavy flavor jet production with semi-inclusive jet
  functions: from proton to heavy-ion collisions}.
\newblock {\em JHEP}, 07:148, 2019.
\newblock \href {http://arxiv.org/abs/1811.07905} {\path{arXiv:1811.07905}},
  \href {https://doi.org/10.1007/JHEP07(2019)148}
  {\path{doi:10.1007/JHEP07(2019)148}}.

\bibitem{Buzzatti:2011vt}
Alessandro Buzzatti and Miklos Gyulassy.
\newblock {Jet Flavor Tomography of Quark Gluon Plasmas at RHIC and LHC}.
\newblock {\em Phys. Rev. Lett.}, 108:022301, 2012.
\newblock \href {http://arxiv.org/abs/1106.3061} {\path{arXiv:1106.3061}},
  \href {https://doi.org/10.1103/PhysRevLett.108.022301}
  {\path{doi:10.1103/PhysRevLett.108.022301}}.

\bibitem{Dokshitzer:2001zm}
Yuri~L. Dokshitzer and D.~E. Kharzeev.
\newblock {Heavy quark colorimetry of QCD matter}.
\newblock {\em Phys. Lett. B}, 519:199--206, 2001.
\newblock \href {http://arxiv.org/abs/hep-ph/0106202}
  {\path{arXiv:hep-ph/0106202}}, \href
  {https://doi.org/10.1016/S0370-2693(01)01130-3}
  {\path{doi:10.1016/S0370-2693(01)01130-3}}.

\bibitem{STAR:2021uzu}
M.~S. Abdallah et~al.
\newblock {Evidence of Mass Ordering of Charm and Bottom Quark Energy Loss in
  Au+Au Collisions at RHIC}.
\newblock {\em Eur. Phys. J. C}, 82(12):1150, 2022.
\newblock \href {http://arxiv.org/abs/2111.14615} {\path{arXiv:2111.14615}},
  \href {https://doi.org/10.1140/epjc/s10052-022-11003-7}
  {\path{doi:10.1140/epjc/s10052-022-11003-7}}.

\bibitem{PHENIX:2022wim}
U.~A. Acharya et~al.
\newblock {Charm- and Bottom-Quark Production in Au$+$Au Collisions at
  $\sqrt{s_{_{NN}}}$ = 200 GeV}.
\newblock 3 2022.
\newblock \href {http://arxiv.org/abs/2203.17058} {\path{arXiv:2203.17058}}.

\bibitem{CMS:2018bwt}
Albert~M Sirunyan et~al.
\newblock {Studies of Beauty Suppression via Nonprompt $D^0$ Mesons in Pb-Pb
  Collisions at $Q^2 = 4$ $\rm GeV^2$}.
\newblock {\em Phys. Rev. Lett.}, 123(2):022001, 2019.
\newblock \href {http://arxiv.org/abs/1810.11102} {\path{arXiv:1810.11102}},
  \href {https://doi.org/10.1103/PhysRevLett.123.022001}
  {\path{doi:10.1103/PhysRevLett.123.022001}}.

\bibitem{CMS:2017uuv}
Albert~M Sirunyan et~al.
\newblock {Measurement of prompt and nonprompt charmonium suppression in $\text
  {PbPb}$ collisions at 5.02 $\,\text {Te}\text {V}$}.
\newblock {\em Eur. Phys. J. C}, 78(6):509, 2018.
\newblock \href {http://arxiv.org/abs/1712.08959} {\path{arXiv:1712.08959}},
  \href {https://doi.org/10.1140/epjc/s10052-018-5950-6}
  {\path{doi:10.1140/epjc/s10052-018-5950-6}}.

\bibitem{CMS:2017qjw}
Albert~M Sirunyan et~al.
\newblock {Nuclear modification factor of D$^0$ mesons in PbPb collisions at
  $\sqrt{s_\mathrm{NN}} = 5.02$ TeV}.
\newblock {\em Phys. Lett. B}, 782:474--496, 2018.
\newblock \href {http://arxiv.org/abs/1708.04962} {\path{arXiv:1708.04962}},
  \href {https://doi.org/10.1016/j.physletb.2018.05.074}
  {\path{doi:10.1016/j.physletb.2018.05.074}}.

\bibitem{ALICE:2019nuy}
Shreyasi Acharya et~al.
\newblock {Measurement of electrons from semileptonic heavy-flavour hadron
  decays at midrapidity in pp and Pb-Pb collisions at $\sqrt{s_{\rm{NN}}}$ =
  5.02 TeV}.
\newblock {\em Phys. Lett. B}, 804:135377, 2020.
\newblock \href {http://arxiv.org/abs/1910.09110} {\path{arXiv:1910.09110}},
  \href {https://doi.org/10.1016/j.physletb.2020.135377}
  {\path{doi:10.1016/j.physletb.2020.135377}}.

\bibitem{CMS:2022sxl}
Armen Tumasyan et~al.
\newblock {Observation of the $B_c^+$ Meson in Pb-Pb and pp Collisions at
  $\sqrt{s_{NN}}$=5.02\,\,TeV and Measurement of its Nuclear Modification
  Factor}.
\newblock {\em Phys. Rev. Lett.}, 128(25):252301, 2022.
\newblock \href {http://arxiv.org/abs/2201.02659} {\path{arXiv:2201.02659}},
  \href {https://doi.org/10.1103/PhysRevLett.128.252301}
  {\path{doi:10.1103/PhysRevLett.128.252301}}.

\bibitem{Li:2017wwc}
Hai~Tao Li and Ivan Vitev.
\newblock {Inverting the mass hierarchy of jet quenching effects with prompt
  $b$-jet substructure}.
\newblock {\em Phys. Lett. B}, 793:259--264, 2019.
\newblock \href {http://arxiv.org/abs/1801.00008} {\path{arXiv:1801.00008}},
  \href {https://doi.org/10.1016/j.physletb.2019.04.052}
  {\path{doi:10.1016/j.physletb.2019.04.052}}.

\bibitem{Bala:2021fkm}
Dibyendu Bala, Olaf Kaczmarek, Rasmus Larsen, Swagato Mukherjee, Gaurang
  Parkar, Peter Petreczky, Alexander Rothkopf, and Johannes~Heinrich Weber.
\newblock {Static quark-antiquark interactions at nonzero temperature from
  lattice QCD}.
\newblock {\em Phys. Rev. D}, 105(5):054513, 2022.
\newblock \href {http://arxiv.org/abs/2110.11659} {\path{arXiv:2110.11659}},
  \href {https://doi.org/10.1103/PhysRevD.105.054513}
  {\path{doi:10.1103/PhysRevD.105.054513}}.

\bibitem{Mukherjee:2015mxc}
Swagato Mukherjee, Peter Petreczky, and Sayantan Sharma.
\newblock {Charm degrees of freedom in the quark gluon plasma}.
\newblock {\em Phys. Rev. D}, 93(1):014502, 2016.
\newblock \href {http://arxiv.org/abs/1509.08887} {\path{arXiv:1509.08887}},
  \href {https://doi.org/10.1103/PhysRevD.93.014502}
  {\path{doi:10.1103/PhysRevD.93.014502}}.

\bibitem{Petreczky:2021zmz}
Peter Petreczky, Sayantan Sharma, and Johannes~Heinrich Weber.
\newblock {Bottomonium melting from screening correlators at high temperature}.
\newblock {\em Phys. Rev. D}, 104(5):054511, 2021.
\newblock \href {http://arxiv.org/abs/2107.11368} {\path{arXiv:2107.11368}},
  \href {https://doi.org/10.1103/PhysRevD.104.054511}
  {\path{doi:10.1103/PhysRevD.104.054511}}.

\bibitem{Adamczyk:2017xur}
L.~Adamczyk et~al.
\newblock {Measurement of $D^0$ Azimuthal Anisotropy at Midrapidity in Au+Au
  Collisions at $\sqrt{s_{NN}}$=200 GeV}.
\newblock {\em Phys. Rev. Lett.}, 118(21):212301, 2017.
\newblock \href {http://arxiv.org/abs/1701.06060} {\path{arXiv:1701.06060}},
  \href {https://doi.org/10.1103/PhysRevLett.118.212301}
  {\path{doi:10.1103/PhysRevLett.118.212301}}.

\bibitem{Adam:2018inb}
Jaroslav Adam et~al.
\newblock {Centrality and transverse momentum dependence of $D^0$-meson
  production at mid-rapidity in Au+Au collisions at ${\sqrt{s_{\rm NN}} =
  \rm{200\,GeV}}$}.
\newblock {\em Phys. Rev. C}, 99(3):034908, 2019.
\newblock \href {http://arxiv.org/abs/1812.10224} {\path{arXiv:1812.10224}},
  \href {https://doi.org/10.1103/PhysRevC.99.034908}
  {\path{doi:10.1103/PhysRevC.99.034908}}.

\bibitem{ALICE:2018lyv}
S.~Acharya et~al.
\newblock {Measurement of D$^{0}$, D$^{+}$, D$^{*+}$ and D$_{s}^{+}$ production
  in Pb-Pb collisions at $ \sqrt{{\mathrm{s}}_{\mathrm{NN}}}=5.02 $ TeV}.
\newblock {\em JHEP}, 10:174, 2018.
\newblock \href {http://arxiv.org/abs/1804.09083} {\path{arXiv:1804.09083}},
  \href {https://doi.org/10.1007/JHEP10(2018)174}
  {\path{doi:10.1007/JHEP10(2018)174}}.

\bibitem{CMS:2017vhp}
Albert~M Sirunyan et~al.
\newblock {Measurement of prompt $D^0$ meson azimuthal anisotropy in Pb-Pb
  collisions at $\sqrt{{s}_{NN}}$ = 5.02 TeV}.
\newblock {\em Phys. Rev. Lett.}, 120(20):202301, 2018.
\newblock \href {http://arxiv.org/abs/1708.03497} {\path{arXiv:1708.03497}},
  \href {https://doi.org/10.1103/PhysRevLett.120.202301}
  {\path{doi:10.1103/PhysRevLett.120.202301}}.

\bibitem{CMS:2020bnz}
Albert~M Sirunyan et~al.
\newblock {Measurement of prompt ${\mathrm{D^0}}$ and
  ${\mathrm{\overline{D}}{}^0}$ meson azimuthal anisotropy and search for
  strong electric fields in PbPb collisions at $\sqrt{s_\mathrm{NN}} =$ 5.02
  TeV}.
\newblock {\em Phys. Lett. B}, 816:136253, 2021.
\newblock \href {http://arxiv.org/abs/2009.12628} {\path{arXiv:2009.12628}},
  \href {https://doi.org/10.1016/j.physletb.2021.136253}
  {\path{doi:10.1016/j.physletb.2021.136253}}.

\bibitem{ALICE:2021rxa}
Shreyasi Acharya et~al.
\newblock {Prompt D$^{0}$, D$^{+}$, and D$^{*+}$ production in
  Pb\textendash{}Pb collisions at $ \sqrt{s_{\mathrm{NN}}} $ = 5.02 TeV}.
\newblock {\em JHEP}, 01:174, 2022.
\newblock \href {http://arxiv.org/abs/2110.09420} {\path{arXiv:2110.09420}},
  \href {https://doi.org/10.1007/JHEP01(2022)174}
  {\path{doi:10.1007/JHEP01(2022)174}}.

\bibitem{PHENIX:2011img}
A.~Adare et~al.
\newblock {$J/\psi$ suppression at forward rapidity in Au+Au collisions at
  $\sqrt{s_{NN}}=200$ GeV}.
\newblock {\em Phys. Rev. C}, 84:054912, 2011.
\newblock \href {http://arxiv.org/abs/1103.6269} {\path{arXiv:1103.6269}},
  \href {https://doi.org/10.1103/PhysRevC.84.054912}
  {\path{doi:10.1103/PhysRevC.84.054912}}.

\bibitem{PHENIX:2012czk}
A.~Adare et~al.
\newblock {Transverse-Momentum Dependence of the $J/\psi$ Nuclear Modification
  in $d+$Au Collisions at $\sqrt{s_{NN}}=200$ GeV}.
\newblock {\em Phys. Rev. C}, 87(3):034904, 2013.
\newblock \href {http://arxiv.org/abs/1204.0777} {\path{arXiv:1204.0777}},
  \href {https://doi.org/10.1103/PhysRevC.87.034904}
  {\path{doi:10.1103/PhysRevC.87.034904}}.

\bibitem{PHENIX:2019brm}
U.~Acharya et~al.
\newblock {Measurement of $J/\psi$ at forward and backward rapidity in $p+p$,
  $p+A$l, $p+A$u, and $^3$He$+$Au collisions at $\sqrt{s_{_{NN}}}=200~{\rm
  GeV}$}.
\newblock {\em Phys. Rev. C}, 102(1):014902, 2020.
\newblock \href {http://arxiv.org/abs/1910.14487} {\path{arXiv:1910.14487}},
  \href {https://doi.org/10.1103/PhysRevC.102.014902}
  {\path{doi:10.1103/PhysRevC.102.014902}}.

\bibitem{PHENIX:2012xtg}
A.~Adare et~al.
\newblock {$J/\psi$ suppression at forward rapidity in Au+Au collisions at
  $\sqrt{s_{NN}}=39$ and 62.4 GeV}.
\newblock {\em Phys. Rev. C}, 86:064901, 2012.
\newblock \href {http://arxiv.org/abs/1208.2251} {\path{arXiv:1208.2251}},
  \href {https://doi.org/10.1103/PhysRevC.86.064901}
  {\path{doi:10.1103/PhysRevC.86.064901}}.

\bibitem{STAR:2016utm}
L.~Adamczyk et~al.
\newblock {Energy dependence of $J/\psi$ production in Au+Au collisions at
  $\sqrt{s_{NN}} =$ 39, 62.4 and 200 GeV}.
\newblock {\em Phys. Lett. B}, 771:13--20, 2017.
\newblock \href {http://arxiv.org/abs/1607.07517} {\path{arXiv:1607.07517}},
  \href {https://doi.org/10.1016/j.physletb.2017.04.078}
  {\path{doi:10.1016/j.physletb.2017.04.078}}.

\bibitem{ALICE:2019lga}
Shreyasi Acharya et~al.
\newblock {Studies of J/$\psi$ production at forward rapidity in Pb-Pb
  collisions at $\sqrt{s_{\rm{NN}}}$ = 5.02 TeV}.
\newblock {\em JHEP}, 02:041, 2020.
\newblock \href {http://arxiv.org/abs/1909.03158} {\path{arXiv:1909.03158}},
  \href {https://doi.org/10.1007/JHEP02(2020)041}
  {\path{doi:10.1007/JHEP02(2020)041}}.

\bibitem{STAR:2012jzy}
L.~Adamczyk et~al.
\newblock {Measurement of $J/\psi$ Azimuthal Anisotropy in Au+Au Collisions at
  $\sqrt{s_{NN}}$ = 200 GeV}.
\newblock {\em Phys. Rev. Lett.}, 111(5):052301, 2013.
\newblock \href {http://arxiv.org/abs/1212.3304} {\path{arXiv:1212.3304}},
  \href {https://doi.org/10.1103/PhysRevLett.111.052301}
  {\path{doi:10.1103/PhysRevLett.111.052301}}.

\bibitem{ALICE:2020pvw}
Shreyasi Acharya et~al.
\newblock {J/$\psi$ elliptic and triangular flow in Pb-Pb collisions at
  $\sqrt{s_{\rm NN}}$ = 5.02 TeV}.
\newblock {\em JHEP}, 10:141, 2020.
\newblock \href {http://arxiv.org/abs/2005.14518} {\path{arXiv:2005.14518}},
  \href {https://doi.org/10.1007/JHEP10(2020)141}
  {\path{doi:10.1007/JHEP10(2020)141}}.

\bibitem{PHENIX:2013pmn}
A.~Adare et~al.
\newblock {Nuclear Modification of $\psi',\chi_c,J/\psi$ Production in d+Au
  Collisions at $\sqrt{s_{NN}}$=200 GeV}.
\newblock {\em Phys. Rev. Lett.}, 111(20):202301, 2013.
\newblock \href {http://arxiv.org/abs/1305.5516} {\path{arXiv:1305.5516}},
  \href {https://doi.org/10.1103/PhysRevLett.111.202301}
  {\path{doi:10.1103/PhysRevLett.111.202301}}.

\bibitem{PHENIX:2016vmz}
A.~Adare et~al.
\newblock {Measurement of the relative yields of $\psi(2S)$ to $\psi(1S)$
  mesons produced at forward and backward rapidity in $p+p$, $p+$Al, $p+$Au,
  and $^{3}$He$+$Au collisions at $\sqrt{s_{_{NN}}}=200$ GeV}.
\newblock {\em Phys. Rev. C}, 95(3):034904, 2017.
\newblock \href {http://arxiv.org/abs/1609.06550} {\path{arXiv:1609.06550}},
  \href {https://doi.org/10.1103/PhysRevC.95.034904}
  {\path{doi:10.1103/PhysRevC.95.034904}}.

\bibitem{PHENIX:2022nrm}
U.~A. Acharya et~al.
\newblock {Measurement of $\psi(2S)$ nuclear modification at backward and
  forward rapidity in $p$ $+$ $p$, $p$ $+$ Al, and $p$ $+$ Au collisions at
  $\sqrt{s_{_{NN}}}=200$ GeV}.
\newblock {\em Phys. Rev. C}, 105(6):064912, 2022.
\newblock \href {http://arxiv.org/abs/2202.03863} {\path{arXiv:2202.03863}},
  \href {https://doi.org/10.1103/PhysRevC.105.064912}
  {\path{doi:10.1103/PhysRevC.105.064912}}.

\bibitem{CMS:2016wgo}
Albert~M Sirunyan et~al.
\newblock {Relative Modification of Prompt \ensuremath{\psi}(2S) and
  J/\ensuremath{\psi} Yields from pp to PbPb Collisions at $\sqrt{s_{NN}}=5.02$
  TeV}.
\newblock {\em Phys. Rev. Lett.}, 118(16):162301, 2017.
\newblock \href {http://arxiv.org/abs/1611.01438} {\path{arXiv:1611.01438}},
  \href {https://doi.org/10.1103/PhysRevLett.118.162301}
  {\path{doi:10.1103/PhysRevLett.118.162301}}.

\bibitem{LHCb:2016vqr}
Roel Aaij et~al.
\newblock {Study of $\psi(2S)$ production and cold nuclear matter effects in
  pPb collisions at $\sqrt{s_{NN}}=5~\mathrm{TeV}$}.
\newblock {\em JHEP}, 03:133, 2016.
\newblock \href {http://arxiv.org/abs/1601.07878} {\path{arXiv:1601.07878}},
  \href {https://doi.org/10.1007/JHEP03(2016)133}
  {\path{doi:10.1007/JHEP03(2016)133}}.

\bibitem{ATLAS:2017prf}
Morad Aaboud et~al.
\newblock {Measurement of quarkonium production in proton\textendash{}lead and
  proton\textendash{}proton collisions at $5.02~\mathrm {TeV}$ with the ATLAS
  detector}.
\newblock {\em Eur. Phys. J. C}, 78(3):171, 2018.
\newblock \href {http://arxiv.org/abs/1709.03089} {\path{arXiv:1709.03089}},
  \href {https://doi.org/10.1140/epjc/s10052-018-5624-4}
  {\path{doi:10.1140/epjc/s10052-018-5624-4}}.

\bibitem{CMS:2018gbb}
Albert~M Sirunyan et~al.
\newblock {Measurement of prompt $\psi$(2S) production cross sections in
  proton-lead and proton-proton collisions at $\sqrt{s_{_\mathrm{NN}}}=$ 5.02
  TeV}.
\newblock {\em Phys. Lett. B}, 790:509--532, 2019.
\newblock \href {http://arxiv.org/abs/1805.02248} {\path{arXiv:1805.02248}},
  \href {https://doi.org/10.1016/j.physletb.2019.01.058}
  {\path{doi:10.1016/j.physletb.2019.01.058}}.

\bibitem{ALICE:2020tsj}
Shreyasi Acharya et~al.
\newblock {Centrality dependence of J/$\psi$ and $\psi$(2S) production and
  nuclear modification in p-Pb collisions at $\sqrt{s_{\rm NN}} =$ 8.16 TeV}.
\newblock {\em JHEP}, 02:002, 2021.
\newblock \href {http://arxiv.org/abs/2008.04806} {\path{arXiv:2008.04806}},
  \href {https://doi.org/10.1007/JHEP02(2021)002}
  {\path{doi:10.1007/JHEP02(2021)002}}.

\bibitem{CMS:2022wfi}
Armen Tumasyan et~al.
\newblock {Nuclear modification of $\Upsilon$ states in pPb collisions at
  $\sqrt{s_\mathrm{NN}}$ = 5.02 TeV}.
\newblock {\em Phys. Lett. B}, 835:137397, 2022.
\newblock \href {http://arxiv.org/abs/2202.11807} {\path{arXiv:2202.11807}},
  \href {https://doi.org/10.1016/j.physletb.2022.137397}
  {\path{doi:10.1016/j.physletb.2022.137397}}.

\bibitem{ALICE:2022jeh}
{$\psi(2S)$ suppression in Pb-Pb collisions at the LHC}.
\newblock 10 2022.
\newblock \href {http://arxiv.org/abs/2210.08893} {\path{arXiv:2210.08893}}.

\bibitem{CMS:2018zza}
Albert~M Sirunyan et~al.
\newblock {Measurement of nuclear modification factors of $\Upsilon$(1S),
  $\Upsilon$(2S), and $\Upsilon$(3S) mesons in PbPb collisions at
  $\sqrt{s_{_\mathrm{NN}}} =$ 5.02 TeV}.
\newblock {\em Phys. Lett. B}, 790:270--293, 2019.
\newblock \href {http://arxiv.org/abs/1805.09215} {\path{arXiv:1805.09215}},
  \href {https://doi.org/10.1016/j.physletb.2019.01.006}
  {\path{doi:10.1016/j.physletb.2019.01.006}}.

\bibitem{PHENIX:2014tbe}
A.~Adare et~al.
\newblock {Measurement of $\Upsilon(1S+2S+3S)$ production in $p+p$ and Au$+$Au
  collisions at $\sqrt{s_{_{NN}}}=200$ GeV}.
\newblock {\em Phys. Rev. C}, 91(2):024913, 2015.
\newblock \href {http://arxiv.org/abs/1404.2246} {\path{arXiv:1404.2246}},
  \href {https://doi.org/10.1103/PhysRevC.91.024913}
  {\path{doi:10.1103/PhysRevC.91.024913}}.

\bibitem{STAR:2016pof}
L.~Adamczyk et~al.
\newblock {$\Upsilon$ production in U + U collisions at $\sqrt{{s}_{NN}}=$ 193
  GeV measured with the STAR experiment}.
\newblock {\em Phys. Rev. C}, 94(6):064904, 2016.
\newblock \href {http://arxiv.org/abs/1608.06487} {\path{arXiv:1608.06487}},
  \href {https://doi.org/10.1103/PhysRevC.94.064904}
  {\path{doi:10.1103/PhysRevC.94.064904}}.

\bibitem{STAR:2022rpk}
{Observation of sequential $\Upsilon$ suppression in Au+Au collisions at
  $\sqrt{s_{_\mathrm{NN}}}$ = 200 GeV with the STAR experiment}.
\newblock 7 2022.
\newblock \href {http://arxiv.org/abs/2207.06568} {\path{arXiv:2207.06568}}.

\bibitem{Schenke:2021mxx}
Bj\"orn Schenke.
\newblock {The smallest fluid on Earth}.
\newblock {\em Rept. Prog. Phys.}, 84(8):082301, 2021.
\newblock \href {http://arxiv.org/abs/2102.11189} {\path{arXiv:2102.11189}},
  \href {https://doi.org/10.1088/1361-6633/ac14c9}
  {\path{doi:10.1088/1361-6633/ac14c9}}.

\bibitem{Mantysaari:2016ykx}
Heikki M\"antysaari and Bj\"orn Schenke.
\newblock {Evidence of strong proton shape fluctuations from incoherent
  diffraction}.
\newblock {\em Phys. Rev. Lett.}, 117(5):052301, 2016.
\newblock \href {http://arxiv.org/abs/1603.04349} {\path{arXiv:1603.04349}},
  \href {https://doi.org/10.1103/PhysRevLett.117.052301}
  {\path{doi:10.1103/PhysRevLett.117.052301}}.

\bibitem{Mantysaari:2020lhf}
Heikki M\"antysaari, Kaushik Roy, Farid Salazar, and Bj\"orn Schenke.
\newblock {Gluon imaging using azimuthal correlations in diffractive scattering
  at the Electron-Ion Collider}.
\newblock {\em Phys. Rev. D}, 103(9):094026, 2021.
\newblock \href {http://arxiv.org/abs/2011.02464} {\path{arXiv:2011.02464}},
  \href {https://doi.org/10.1103/PhysRevD.103.094026}
  {\path{doi:10.1103/PhysRevD.103.094026}}.

\bibitem{Mantysaari:2019csc}
Heikki M\"antysaari, Niklas Mueller, and Bj\"orn Schenke.
\newblock {Diffractive Dijet Production and Wigner Distributions from the Color
  Glass Condensate}.
\newblock {\em Phys. Rev. D}, 99(7):074004, 2019.
\newblock \href {http://arxiv.org/abs/1902.05087} {\path{arXiv:1902.05087}},
  \href {https://doi.org/10.1103/PhysRevD.99.074004}
  {\path{doi:10.1103/PhysRevD.99.074004}}.

\bibitem{Mantysaari:2019hkq}
Heikki M\"antysaari, Niklas Mueller, Farid Salazar, and Bj\"orn Schenke.
\newblock {Multigluon Correlations and Evidence of Saturation from Dijet
  Measurements at an Electron-Ion Collider}.
\newblock {\em Phys. Rev. Lett.}, 124(11):112301, 2020.
\newblock \href {http://arxiv.org/abs/1912.05586} {\path{arXiv:1912.05586}},
  \href {https://doi.org/10.1103/PhysRevLett.124.112301}
  {\path{doi:10.1103/PhysRevLett.124.112301}}.

\bibitem{Mantysaari:2022sux}
Heikki M\"antysaari, Farid Salazar, and Bj\"orn Schenke.
\newblock {Nuclear geometry at high energy from exclusive vector meson
  production}.
\newblock {\em Phys. Rev. D}, 106(7):074019, 2022.
\newblock \href {http://arxiv.org/abs/2207.03712} {\path{arXiv:2207.03712}},
  \href {https://doi.org/10.1103/PhysRevD.106.074019}
  {\path{doi:10.1103/PhysRevD.106.074019}}.

\bibitem{Krasnitz:1998ns}
Alex Krasnitz and Raju Venugopalan.
\newblock {Nonperturbative computation of gluon minijet production in nuclear
  collisions at very high-energies}.
\newblock {\em Nucl. Phys. B}, 557:237, 1999.
\newblock \href {http://arxiv.org/abs/hep-ph/9809433}
  {\path{arXiv:hep-ph/9809433}}, \href
  {https://doi.org/10.1016/S0550-3213(99)00366-1}
  {\path{doi:10.1016/S0550-3213(99)00366-1}}.

\bibitem{Krasnitz:1999wc}
Alex Krasnitz and Raju Venugopalan.
\newblock {The Initial energy density of gluons produced in very high-energy
  nuclear collisions}.
\newblock {\em Phys. Rev. Lett.}, 84:4309--4312, 2000.
\newblock \href {http://arxiv.org/abs/hep-ph/9909203}
  {\path{arXiv:hep-ph/9909203}}, \href
  {https://doi.org/10.1103/PhysRevLett.84.4309}
  {\path{doi:10.1103/PhysRevLett.84.4309}}.

\bibitem{Krasnitz:2000gz}
Alex Krasnitz and Raju Venugopalan.
\newblock {The Initial gluon multiplicity in heavy ion collisions}.
\newblock {\em Phys. Rev. Lett.}, 86:1717--1720, 2001.
\newblock \href {http://arxiv.org/abs/hep-ph/0007108}
  {\path{arXiv:hep-ph/0007108}}, \href
  {https://doi.org/10.1103/PhysRevLett.86.1717}
  {\path{doi:10.1103/PhysRevLett.86.1717}}.

\bibitem{Schenke:2012wb}
Bjoern Schenke, Prithwish Tribedy, and Raju Venugopalan.
\newblock {Fluctuating Glasma initial conditions and flow in heavy ion
  collisions}.
\newblock {\em Phys. Rev. Lett.}, 108:252301, 2012.
\newblock \href {http://arxiv.org/abs/1202.6646} {\path{arXiv:1202.6646}},
  \href {https://doi.org/10.1103/PhysRevLett.108.252301}
  {\path{doi:10.1103/PhysRevLett.108.252301}}.

\bibitem{Schenke:2012hg}
Bjoern Schenke, Prithwish Tribedy, and Raju Venugopalan.
\newblock {Event-by-event gluon multiplicity, energy density, and
  eccentricities in ultrarelativistic heavy-ion collisions}.
\newblock {\em Phys. Rev. C}, 86:034908, 2012.
\newblock \href {http://arxiv.org/abs/1206.6805} {\path{arXiv:1206.6805}},
  \href {https://doi.org/10.1103/PhysRevC.86.034908}
  {\path{doi:10.1103/PhysRevC.86.034908}}.

\bibitem{Giacalone:2021clp}
Giuliano Giacalone, Bj\"orn Schenke, and Chun Shen.
\newblock {Constraining the Nucleon Size with Relativistic Nuclear Collisions}.
\newblock {\em Phys. Rev. Lett.}, 128(4):042301, 2022.
\newblock \href {http://arxiv.org/abs/2111.02908} {\path{arXiv:2111.02908}},
  \href {https://doi.org/10.1103/PhysRevLett.128.042301}
  {\path{doi:10.1103/PhysRevLett.128.042301}}.

\bibitem{Bally:2022vgo}
Benjamin Bally et~al.
\newblock {Imaging the initial condition of heavy-ion collisions and nuclear
  structure across the nuclide chart}.
\newblock 9 2022.
\newblock \href {http://arxiv.org/abs/2209.11042} {\path{arXiv:2209.11042}}.

\bibitem{ALICE:2022xir}
{ALICE luminosity determination for Pb$-$Pb collisions at
  $\sqrt{s_{\mathrm{NN}}} = 5.02$ TeV}.
\newblock 4 2022.
\newblock \href {http://arxiv.org/abs/2204.10148} {\path{arXiv:2204.10148}}.

\bibitem{Nijs:2022rme}
Govert Nijs and Wilke van~der Schee.
\newblock {Hadronic Nucleus-Nucleus Cross Section and the Nucleon Size}.
\newblock {\em Phys. Rev. Lett.}, 129(23):232301, 2022.
\newblock \href {http://arxiv.org/abs/2206.13522} {\path{arXiv:2206.13522}},
  \href {https://doi.org/10.1103/PhysRevLett.129.232301}
  {\path{doi:10.1103/PhysRevLett.129.232301}}.

\bibitem{STAR:2021mii}
Mohamed Abdallah et~al.
\newblock {Search for the chiral magnetic effect with isobar collisions at
  $\sqrt {s_{NN}}$=200 GeV by the STAR Collaboration at the BNL Relativistic
  Heavy Ion Collider}.
\newblock {\em Phys. Rev. C}, 105(1):014901, 2022.
\newblock \href {http://arxiv.org/abs/2109.00131} {\path{arXiv:2109.00131}},
  \href {https://doi.org/10.1103/PhysRevC.105.014901}
  {\path{doi:10.1103/PhysRevC.105.014901}}.

\bibitem{Jia:2021oyt}
Jiangyong Jia and Chun-Jian Zhang.
\newblock {Scaling approach to nuclear structure in high-energy heavy-ion
  collisions}.
\newblock 11 2021.
\newblock \href {http://arxiv.org/abs/2111.15559} {\path{arXiv:2111.15559}}.

\bibitem{Jia:2022qgl}
Jiangyong Jia, Giuliano Giacalone, and Chunjian Zhang.
\newblock {Separating the impact of nuclear skin and nuclear deformation on
  elliptic flow and its fluctuations in high-energy isobar collisions}.
\newblock 6 2022.
\newblock \href {http://arxiv.org/abs/2206.10449} {\path{arXiv:2206.10449}}.

\bibitem{Xu:2021vpn}
Hao-jie Xu, Hanlin Li, Xiaobao Wang, Caiwan Shen, and Fuqiang Wang.
\newblock {Determine the neutron skin type by relativistic isobaric
  collisions}.
\newblock {\em Phys. Lett. B}, 819:136453, 2021.
\newblock \href {http://arxiv.org/abs/2103.05595} {\path{arXiv:2103.05595}},
  \href {https://doi.org/10.1016/j.physletb.2021.136453}
  {\path{doi:10.1016/j.physletb.2021.136453}}.

\bibitem{Xu:2021uar}
Hao-jie Xu, Wenbin Zhao, Hanlin Li, Ying Zhou, Lie-Wen Chen, and Fuqiang Wang.
\newblock {Probing nuclear structure with mean transverse momentum in
  relativistic isobar collisions}.
\newblock 11 2021.
\newblock \href {http://arxiv.org/abs/2111.14812} {\path{arXiv:2111.14812}}.

\bibitem{Cao:2020rgr}
Yuchen Cao, Sylvester~E. Agbemava, Anatoli~V. Afanasjev, Witold Nazarewicz, and
  Erik Olsen.
\newblock {Landscape of pear-shaped even-even nuclei}.
\newblock {\em Phys. Rev. C}, 102(2):024311, 2020.
\newblock \href {http://arxiv.org/abs/2004.01319} {\path{arXiv:2004.01319}},
  \href {https://doi.org/10.1103/PhysRevC.102.024311}
  {\path{doi:10.1103/PhysRevC.102.024311}}.

\bibitem{Bertulani:2005ru}
Carlos~A. Bertulani, Spencer~R. Klein, and Joakim Nystrand.
\newblock {Physics of ultra-peripheral nuclear collisions}.
\newblock {\em Ann. Rev. Nucl. Part. Sci.}, 55:271--310, 2005.
\newblock \href {http://arxiv.org/abs/nucl-ex/0502005}
  {\path{arXiv:nucl-ex/0502005}}, \href
  {https://doi.org/10.1146/annurev.nucl.55.090704.151526}
  {\path{doi:10.1146/annurev.nucl.55.090704.151526}}.

\bibitem{Baltz:2007kq}
A.~J. Baltz.
\newblock {The Physics of Ultraperipheral Collisions at the LHC}.
\newblock {\em Phys. Rept.}, 458:1--171, 2008.
\newblock \href {http://arxiv.org/abs/0706.3356} {\path{arXiv:0706.3356}},
  \href {https://doi.org/10.1016/j.physrep.2007.12.001}
  {\path{doi:10.1016/j.physrep.2007.12.001}}.

\bibitem{Contreras:2015dqa}
J.~G. Contreras and J.~D. Tapia~Takaki.
\newblock {Ultra-peripheral heavy-ion collisions at the LHC}.
\newblock {\em Int. J. Mod. Phys. A}, 30:1542012, 2015.
\newblock \href {https://doi.org/10.1142/S0217751X15420129}
  {\path{doi:10.1142/S0217751X15420129}}.

\bibitem{Klein:2017nqo}
Spencer Klein and Joakim Nystrand.
\newblock {Ultraperipheral nuclear collisions}.
\newblock {\em Phys. Today}, 70(10):40--47, 2017.
\newblock \href {https://doi.org/10.1063/PT.3.3727}
  {\path{doi:10.1063/PT.3.3727}}.

\bibitem{Klein:2019qfb}
Spencer~R. Klein and Heikki M\"antysaari.
\newblock {Imaging the nucleus with high-energy photons}.
\newblock {\em Nature Rev. Phys.}, 1(11):662--674, 2019.
\newblock \href {http://arxiv.org/abs/1910.10858} {\path{arXiv:1910.10858}},
  \href {https://doi.org/10.1038/s42254-019-0107-6}
  {\path{doi:10.1038/s42254-019-0107-6}}.

\bibitem{Klein:2020fmr}
Spencer Klein and Peter Steinberg.
\newblock {Photonuclear and Two-photon Interactions at High-Energy Nuclear
  Colliders}.
\newblock {\em Ann. Rev. Nucl. Part. Sci.}, 70:323--354, 2020.
\newblock \href {http://arxiv.org/abs/2005.01872} {\path{arXiv:2005.01872}},
  \href {https://doi.org/10.1146/annurev-nucl-030320-033923}
  {\path{doi:10.1146/annurev-nucl-030320-033923}}.

\bibitem{Eskola:2022vaf}
Kari~J. Eskola, Christopher~A. Flett, Vadim Guzey, Topi L\"oyt\"ainen, and
  Hannu Paukkunen.
\newblock {Next-to-leading order perturbative QCD predictions for exclusive
  $J/\psi$ photoproduction in oxygen-oxygen and lead-lead collisions at the
  LHC}.
\newblock 10 2022.
\newblock \href {http://arxiv.org/abs/2210.16048} {\path{arXiv:2210.16048}}.

\bibitem{STAR:2002caw}
C.~Adler et~al.
\newblock {Coherent rho0 production in ultraperipheral heavy ion collisions}.
\newblock {\em Phys. Rev. Lett.}, 89:272302, 2002.
\newblock \href {http://arxiv.org/abs/nucl-ex/0206004}
  {\path{arXiv:nucl-ex/0206004}}, \href
  {https://doi.org/10.1103/PhysRevLett.89.272302}
  {\path{doi:10.1103/PhysRevLett.89.272302}}.

\bibitem{STAR:2007elq}
B.~I. Abelev et~al.
\newblock {$\rho^0$ photoproduction in ultraperipheral relativistic heavy ion
  collisions at $\sqrt{s_{NN}}$ = 200 GeV}.
\newblock {\em Phys. Rev. C}, 77:034910, 2008.
\newblock \href {http://arxiv.org/abs/0712.3320} {\path{arXiv:0712.3320}},
  \href {https://doi.org/10.1103/PhysRevC.77.034910}
  {\path{doi:10.1103/PhysRevC.77.034910}}.

\bibitem{STAR:2011wtm}
G.~Agakishiev et~al.
\newblock {$\rho^{0}$ Photoproduction in AuAu Collisions at
  $\sqrt{s_{NN}}$=62.4 GeV with STAR}.
\newblock {\em Phys. Rev. C}, 85:014910, 2012.
\newblock \href {http://arxiv.org/abs/1107.4630} {\path{arXiv:1107.4630}},
  \href {https://doi.org/10.1103/PhysRevC.85.014910}
  {\path{doi:10.1103/PhysRevC.85.014910}}.

\bibitem{STAR:2017enh}
L.~Adamczyk et~al.
\newblock {Coherent diffractive photoproduction of $\rho^{0}$mesons on gold
  nuclei at 200 GeV/nucleon-pair at the Relativistic Heavy Ion Collider}.
\newblock {\em Phys. Rev. C}, 96(5):054904, 2017.
\newblock \href {http://arxiv.org/abs/1702.07705} {\path{arXiv:1702.07705}},
  \href {https://doi.org/10.1103/PhysRevC.96.054904}
  {\path{doi:10.1103/PhysRevC.96.054904}}.

\bibitem{CMS:2016itn}
Vardan Khachatryan et~al.
\newblock {Coherent $J/\psi$ photoproduction in ultra-peripheral PbPb
  collisions at $\sqrt {s_{NN}} =$ 2.76 TeV with the CMS experiment}.
\newblock {\em Phys. Lett. B}, 772:489--511, 2017.
\newblock \href {http://arxiv.org/abs/1605.06966} {\path{arXiv:1605.06966}},
  \href {https://doi.org/10.1016/j.physletb.2017.07.001}
  {\path{doi:10.1016/j.physletb.2017.07.001}}.

\bibitem{ALICE:2013wjo}
E.~Abbas et~al.
\newblock {Charmonium and $e^+e^-$ pair photoproduction at mid-rapidity in
  ultra-peripheral Pb-Pb collisions at $\sqrt{s_{\rm NN}}$=2.76 TeV}.
\newblock {\em Eur. Phys. J. C}, 73(11):2617, 2013.
\newblock \href {http://arxiv.org/abs/1305.1467} {\path{arXiv:1305.1467}},
  \href {https://doi.org/10.1140/epjc/s10052-013-2617-1}
  {\path{doi:10.1140/epjc/s10052-013-2617-1}}.

\bibitem{ALICE:2019tqa}
Shreyasi Acharya et~al.
\newblock {Coherent J/$\psi$ photoproduction at forward rapidity in
  ultra-peripheral Pb-Pb collisions at $\sqrt{s_{\rm{NN}}}=5.02$ TeV}.
\newblock {\em Phys. Lett. B}, 798:134926, 2019.
\newblock \href {http://arxiv.org/abs/1904.06272} {\path{arXiv:1904.06272}},
  \href {https://doi.org/10.1016/j.physletb.2019.134926}
  {\path{doi:10.1016/j.physletb.2019.134926}}.

\bibitem{ALICE:2015nmy}
Jaroslav Adam et~al.
\newblock {Coherent $\psi$(2S) photo-production in ultra-peripheral Pb Pb
  collisions at $\sqrt{s}_{\rm NN}$ = 2.76 TeV}.
\newblock {\em Phys. Lett. B}, 751:358--370, 2015.
\newblock \href {http://arxiv.org/abs/1508.05076} {\path{arXiv:1508.05076}},
  \href {https://doi.org/10.1016/j.physletb.2015.10.040}
  {\path{doi:10.1016/j.physletb.2015.10.040}}.

\bibitem{ALICE:2021gpt}
Shreyasi Acharya et~al.
\newblock {Coherent $J/\psi$ and $\psi'$ photoproduction at midrapidity in
  ultra-peripheral Pb-Pb collisions at $\sqrt{s_{\mathrm{NN}}}~=~5.02$ TeV}.
\newblock {\em Eur. Phys. J. C}, 81(8):712, 2021.
\newblock \href {http://arxiv.org/abs/2101.04577} {\path{arXiv:2101.04577}},
  \href {https://doi.org/10.1140/epjc/s10052-021-09437-6}
  {\path{doi:10.1140/epjc/s10052-021-09437-6}}.

\bibitem{CMS:2019awk}
Albert~M Sirunyan et~al.
\newblock {Measurement of exclusive $\rho(770)^0$ photoproduction in
  ultraperipheral pPb collisions at $\sqrt{s_\mathrm{NN}} =$ 5.02 TeV}.
\newblock {\em Eur. Phys. J. C}, 79(8):702, 2019.
\newblock \href {http://arxiv.org/abs/1902.01339} {\path{arXiv:1902.01339}},
  \href {https://doi.org/10.1140/epjc/s10052-019-7202-9}
  {\path{doi:10.1140/epjc/s10052-019-7202-9}}.

\bibitem{ALICE:2021tyx}
Shreyasi Acharya et~al.
\newblock {First measurement of the |$t$|-dependence of coherent $J/\psi$
  photonuclear production}.
\newblock {\em Phys. Lett. B}, 817:136280, 2021.
\newblock \href {http://arxiv.org/abs/2101.04623} {\path{arXiv:2101.04623}},
  \href {https://doi.org/10.1016/j.physletb.2021.136280}
  {\path{doi:10.1016/j.physletb.2021.136280}}.

\bibitem{LHCb:2021bfl}
Roel Aaij et~al.
\newblock {Study of coherent $J/\psi$ production in lead-lead collisions at $
  \sqrt{{\mathrm{s}}_{\mathrm{NN}}} $ = 5 TeV}.
\newblock {\em JHEP}, 07:117, 2022.
\newblock \href {http://arxiv.org/abs/2107.03223} {\path{arXiv:2107.03223}},
  \href {https://doi.org/10.1007/JHEP07(2022)117}
  {\path{doi:10.1007/JHEP07(2022)117}}.

\bibitem{LHCb:2022ahs}
{Study of coherent charmonium production in ultra-peripheral lead-lead
  collisions}.
\newblock 6 2022.
\newblock \href {http://arxiv.org/abs/2206.08221} {\path{arXiv:2206.08221}}.

\bibitem{STAR:2022wfe}
Mohamed Abdallah et~al.
\newblock {Tomography of ultrarelativistic nuclei with polarized photon-gluon
  collisions}.
\newblock {\em Sci. Adv.}, 9(1):eabq3903, 2023.
\newblock \href {http://arxiv.org/abs/2204.01625} {\path{arXiv:2204.01625}},
  \href {https://doi.org/10.1126/sciadv.abq3903}
  {\path{doi:10.1126/sciadv.abq3903}}.

\bibitem{ALICE:2015mzu}
Jaroslav Adam et~al.
\newblock {Measurement of an excess in the yield of $J/\psi$ at very low
  $p_{\rm T}$ in Pb-Pb collisions at $\sqrt{s_{\rm NN}}$ = 2.76 TeV}.
\newblock {\em Phys. Rev. Lett.}, 116(22):222301, 2016.
\newblock \href {http://arxiv.org/abs/1509.08802} {\path{arXiv:1509.08802}},
  \href {https://doi.org/10.1103/PhysRevLett.116.222301}
  {\path{doi:10.1103/PhysRevLett.116.222301}}.

\bibitem{Zha:2017jch}
W.~Zha, S.~R. Klein, R.~Ma, L.~Ruan, T.~Todoroki, Z.~Tang, Z.~Xu, C.~Yang,
  Q.~Yang, and S.~Yang.
\newblock {Coherent J/$\psi$ photoproduction in hadronic heavy-ion collisions}.
\newblock {\em Phys. Rev. C}, 97(4):044910, 2018.
\newblock \href {http://arxiv.org/abs/1705.01460} {\path{arXiv:1705.01460}},
  \href {https://doi.org/10.1103/PhysRevC.97.044910}
  {\path{doi:10.1103/PhysRevC.97.044910}}.

\bibitem{STAR:2019yox}
J.~Adam et~al.
\newblock {Observation of excess J/$\psi$ yield at very low transverse momenta
  in Au+Au collisions at $\sqrt{s_{\rm{NN}}} =$ 200 GeV and U+U collisions at
  $\sqrt{s_{\rm{NN}}} =$ 193 GeV}.
\newblock {\em Phys. Rev. Lett.}, 123(13):132302, 2019.
\newblock \href {http://arxiv.org/abs/1904.11658} {\path{arXiv:1904.11658}},
  \href {https://doi.org/10.1103/PhysRevLett.123.132302}
  {\path{doi:10.1103/PhysRevLett.123.132302}}.

\bibitem{STAR:2021wwq}
Mohamed Abdallah et~al.
\newblock {Probing the Gluonic Structure of the Deuteron with $J/\psi$
  Photoproduction in d+Au Ultraperipheral Collisions}.
\newblock {\em Phys. Rev. Lett.}, 128(12):122303, 2022.
\newblock \href {http://arxiv.org/abs/2109.07625} {\path{arXiv:2109.07625}},
  \href {https://doi.org/10.1103/PhysRevLett.128.122303}
  {\path{doi:10.1103/PhysRevLett.128.122303}}.

\bibitem{CMS:2022lbi}
{Azimuthal correlations within exclusive dijets with large momentum transfer in
  photon-lead collisions}.
\newblock 4 2022.
\newblock \href {http://arxiv.org/abs/2205.00045} {\path{arXiv:2205.00045}}.

\bibitem{Guzey:2016tek}
V.~Guzey and M.~Klasen.
\newblock {Diffractive dijet photoproduction in ultraperipheral collisions at
  the LHC in next-to-leading order QCD}.
\newblock {\em JHEP}, 04:158, 2016.
\newblock \href {http://arxiv.org/abs/1603.06055} {\path{arXiv:1603.06055}},
  \href {https://doi.org/10.1007/JHEP04(2016)158}
  {\path{doi:10.1007/JHEP04(2016)158}}.

\bibitem{Dumitru:2018kuw}
Adrian Dumitru, Vladimir Skokov, and Thomas Ullrich.
\newblock {Measuring the Weizs\"acker-Williams distribution of linearly
  polarized gluons at an electron-ion collider through dijet azimuthal
  asymmetries}.
\newblock {\em Phys. Rev. C}, 99(1):015204, 2019.
\newblock \href {http://arxiv.org/abs/1809.02615} {\path{arXiv:1809.02615}},
  \href {https://doi.org/10.1103/PhysRevC.99.015204}
  {\path{doi:10.1103/PhysRevC.99.015204}}.

\bibitem{Jung:1993gf}
Hannes Jung.
\newblock {Hard diffractive scattering in high-energy e p collisions and the
  Monte Carlo generator RAPGAP}.
\newblock {\em Comput. Phys. Commun.}, 86:147--161, 1995.
\newblock \href {https://doi.org/10.1016/0010-4655(94)00150-Z}
  {\path{doi:10.1016/0010-4655(94)00150-Z}}.

\bibitem{Hatta:2020bgy}
Yoshitaka Hatta, Bo-Wen Xiao, Feng Yuan, and Jian Zhou.
\newblock {Anisotropy in Dijet Production in Exclusive and Inclusive
  Processes}.
\newblock {\em Phys. Rev. Lett.}, 126(14):142001, 2021.
\newblock \href {http://arxiv.org/abs/2010.10774} {\path{arXiv:2010.10774}},
  \href {https://doi.org/10.1103/PhysRevLett.126.142001}
  {\path{doi:10.1103/PhysRevLett.126.142001}}.

\bibitem{Hatta:2021jcd}
Yoshitaka Hatta, Bo-Wen Xiao, Feng Yuan, and Jian Zhou.
\newblock {Azimuthal angular asymmetry of soft gluon radiation in jet
  production}.
\newblock {\em Phys. Rev. D}, 104(5):054037, 2021.
\newblock \href {http://arxiv.org/abs/2106.05307} {\path{arXiv:2106.05307}},
  \href {https://doi.org/10.1103/PhysRevD.104.054037}
  {\path{doi:10.1103/PhysRevD.104.054037}}.

\bibitem{CMS:2010ifv}
Vardan Khachatryan et~al.
\newblock {Observation of Long-Range Near-Side Angular Correlations in
  Proton-Proton Collisions at the LHC}.
\newblock {\em JHEP}, 09:091, 2010.
\newblock \href {http://arxiv.org/abs/1009.4122} {\path{arXiv:1009.4122}},
  \href {https://doi.org/10.1007/JHEP09(2010)091}
  {\path{doi:10.1007/JHEP09(2010)091}}.

\bibitem{CMS:2012qk}
Serguei Chatrchyan et~al.
\newblock {Observation of Long-Range Near-Side Angular Correlations in
  Proton-Lead Collisions at the LHC}.
\newblock {\em Phys. Lett. B}, 718:795--814, 2013.
\newblock \href {http://arxiv.org/abs/1210.5482} {\path{arXiv:1210.5482}},
  \href {https://doi.org/10.1016/j.physletb.2012.11.025}
  {\path{doi:10.1016/j.physletb.2012.11.025}}.

\bibitem{ALICE:2012eyl}
Betty Abelev et~al.
\newblock {Long-range angular correlations on the near and away side in $p$-Pb
  collisions at $\sqrt{s_{NN}}=5.02$ TeV}.
\newblock {\em Phys. Lett. B}, 719:29--41, 2013.
\newblock \href {http://arxiv.org/abs/1212.2001} {\path{arXiv:1212.2001}},
  \href {https://doi.org/10.1016/j.physletb.2013.01.012}
  {\path{doi:10.1016/j.physletb.2013.01.012}}.

\bibitem{ATLAS:2012cix}
Georges Aad et~al.
\newblock {Observation of Associated Near-Side and Away-Side Long-Range
  Correlations in $\sqrt{s_{NN}}$=5.02 TeV Proton-Lead Collisions with the
  ATLAS Detector}.
\newblock {\em Phys. Rev. Lett.}, 110(18):182302, 2013.
\newblock \href {http://arxiv.org/abs/1212.5198} {\path{arXiv:1212.5198}},
  \href {https://doi.org/10.1103/PhysRevLett.110.182302}
  {\path{doi:10.1103/PhysRevLett.110.182302}}.

\bibitem{Dusling:2015gta}
Kevin Dusling, Wei Li, and Bj\"orn Schenke.
\newblock {Novel collective phenomena in high-energy proton\textendash{}proton
  and proton\textendash{}nucleus collisions}.
\newblock {\em Int. J. Mod. Phys. E}, 25(01):1630002, 2016.
\newblock \href {http://arxiv.org/abs/1509.07939} {\path{arXiv:1509.07939}},
  \href {https://doi.org/10.1142/S0218301316300022}
  {\path{doi:10.1142/S0218301316300022}}.

\bibitem{Nagle:2018nvi}
James~L. Nagle and William~A. Zajc.
\newblock {Small System Collectivity in Relativistic Hadronic and Nuclear
  Collisions}.
\newblock {\em Ann. Rev. Nucl. Part. Sci.}, 68:211--235, 2018.
\newblock \href {http://arxiv.org/abs/1801.03477} {\path{arXiv:1801.03477}},
  \href {https://doi.org/10.1146/annurev-nucl-101916-123209}
  {\path{doi:10.1146/annurev-nucl-101916-123209}}.

\bibitem{CMS:2014und}
Vardan Khachatryan et~al.
\newblock {Long-range two-particle correlations of strange hadrons with charged
  particles in pPb and PbPb collisions at LHC energies}.
\newblock {\em Phys. Lett. B}, 742:200--224, 2015.
\newblock \href {http://arxiv.org/abs/1409.3392} {\path{arXiv:1409.3392}},
  \href {https://doi.org/10.1016/j.physletb.2015.01.034}
  {\path{doi:10.1016/j.physletb.2015.01.034}}.

\bibitem{ATLAS:2015hzw}
Georges Aad et~al.
\newblock {Observation of Long-Range Elliptic Azimuthal Anisotropies in
  $\sqrt{s}=$13 and 2.76 TeV $pp$ Collisions with the ATLAS Detector}.
\newblock {\em Phys. Rev. Lett.}, 116(17):172301, 2016.
\newblock \href {http://arxiv.org/abs/1509.04776} {\path{arXiv:1509.04776}},
  \href {https://doi.org/10.1103/PhysRevLett.116.172301}
  {\path{doi:10.1103/PhysRevLett.116.172301}}.

\bibitem{CMS:2015yux}
Vardan Khachatryan et~al.
\newblock {Evidence for Collective Multiparticle Correlations in p-Pb
  Collisions}.
\newblock {\em Phys. Rev. Lett.}, 115(1):012301, 2015.
\newblock \href {http://arxiv.org/abs/1502.05382} {\path{arXiv:1502.05382}},
  \href {https://doi.org/10.1103/PhysRevLett.115.012301}
  {\path{doi:10.1103/PhysRevLett.115.012301}}.

\bibitem{CMS:2016fnw}
Vardan Khachatryan et~al.
\newblock {Evidence for collectivity in pp collisions at the LHC}.
\newblock {\em Phys. Lett. B}, 765:193--220, 2017.
\newblock \href {http://arxiv.org/abs/1606.06198} {\path{arXiv:1606.06198}},
  \href {https://doi.org/10.1016/j.physletb.2016.12.009}
  {\path{doi:10.1016/j.physletb.2016.12.009}}.

\bibitem{CMS:2019lin}
Albert~M Sirunyan et~al.
\newblock {Correlations of azimuthal anisotropy Fourier harmonics with subevent
  cumulants in $pPb$ collisions at $\sqrt{s_{NN}}=$8.16TeV}.
\newblock {\em Phys. Rev. C}, 103(1):014902, 2021.
\newblock \href {http://arxiv.org/abs/1905.09935} {\path{arXiv:1905.09935}},
  \href {https://doi.org/10.1103/PhysRevC.103.014902}
  {\path{doi:10.1103/PhysRevC.103.014902}}.

\bibitem{CMS:2019wiy}
Albert~M Sirunyan et~al.
\newblock {Multiparticle correlation studies in pPb collisions at
  $\sqrt{s_\mathrm{NN}} =$ 8.16 TeV}.
\newblock {\em Phys. Rev. C}, 101(1):014912, 2020.
\newblock \href {http://arxiv.org/abs/1904.11519} {\path{arXiv:1904.11519}},
  \href {https://doi.org/10.1103/PhysRevC.101.014912}
  {\path{doi:10.1103/PhysRevC.101.014912}}.

\bibitem{CMS:2017xnj}
Albert~M Sirunyan et~al.
\newblock {Pseudorapidity and transverse momentum dependence of flow harmonics
  in pPb and PbPb collisions}.
\newblock {\em Phys. Rev. C}, 98(4):044902, 2018.
\newblock \href {http://arxiv.org/abs/1710.07864} {\path{arXiv:1710.07864}},
  \href {https://doi.org/10.1103/PhysRevC.98.044902}
  {\path{doi:10.1103/PhysRevC.98.044902}}.

\bibitem{Habich:2014jna}
M.~Habich, J.~L. Nagle, and P.~Romatschke.
\newblock {Particle spectra and HBT radii for simulated central nuclear
  collisions of C + C, Al + Al, Cu + Cu, Au + Au, and Pb + Pb from
  $\sqrt{s}=62.4$ - $2760$ GeV}.
\newblock {\em Eur. Phys. J. C}, 75(1):15, 2015.
\newblock \href {http://arxiv.org/abs/1409.0040} {\path{arXiv:1409.0040}},
  \href {https://doi.org/10.1140/epjc/s10052-014-3206-7}
  {\path{doi:10.1140/epjc/s10052-014-3206-7}}.

\bibitem{Shen:2016zpp}
Chun Shen, Jean-Fran\c{c}ois Paquet, Gabriel~S. Denicol, Sangyong Jeon, and
  Charles Gale.
\newblock {Collectivity and electromagnetic radiation in small systems}.
\newblock {\em Phys. Rev. C}, 95(1):014906, 2017.
\newblock \href {http://arxiv.org/abs/1609.02590} {\path{arXiv:1609.02590}},
  \href {https://doi.org/10.1103/PhysRevC.95.014906}
  {\path{doi:10.1103/PhysRevC.95.014906}}.

\bibitem{Mace:2018vwq}
Mark Mace, Vladimir~V. Skokov, Prithwish Tribedy, and Raju Venugopalan.
\newblock {Hierarchy of Azimuthal Anisotropy Harmonics in Collisions of Small
  Systems from the Color Glass Condensate}.
\newblock {\em Phys. Rev. Lett.}, 121(5):052301, 2018.
\newblock [Erratum: Phys.Rev.Lett. 123, 039901 (2019)].
\newblock \href {http://arxiv.org/abs/1805.09342} {\path{arXiv:1805.09342}},
  \href {https://doi.org/10.1103/PhysRevLett.121.052301}
  {\path{doi:10.1103/PhysRevLett.121.052301}}.

\bibitem{STAR:2022pfn}
{Measurements of the elliptic and triangular azimuthal anisotropies in central
  3He+Au, d+Au and p+Au collisions at \ensuremath{\sqrt{}} sNN = 200 GeV}.
\newblock 10 2022.
\newblock \href {http://arxiv.org/abs/2210.11352} {\path{arXiv:2210.11352}}.

\bibitem{ALICE:2016fzo}
Jaroslav Adam et~al.
\newblock {Enhanced production of multi-strange hadrons in high-multiplicity
  proton-proton collisions}.
\newblock {\em Nature Phys.}, 13:535--539, 2017.
\newblock \href {http://arxiv.org/abs/1606.07424} {\path{arXiv:1606.07424}},
  \href {https://doi.org/10.1038/nphys4111} {\path{doi:10.1038/nphys4111}}.

\bibitem{ATLAS:2021jhn}
Georges Aad et~al.
\newblock {Two-particle azimuthal correlations in photonuclear ultraperipheral
  Pb+Pb collisions at 5.02 TeV with ATLAS}.
\newblock {\em Phys. Rev. C}, 104(1):014903, 2021.
\newblock \href {http://arxiv.org/abs/2101.10771} {\path{arXiv:2101.10771}},
  \href {https://doi.org/10.1103/PhysRevC.104.014903}
  {\path{doi:10.1103/PhysRevC.104.014903}}.

\bibitem{Badea:2019vey}
Anthony Badea, Austin Baty, Paoti Chang, Gian~Michele Innocenti, Marcello
  Maggi, Christopher Mcginn, Michael Peters, Tzu-An Sheng, Jesse Thaler, and
  Yen-Jie Lee.
\newblock {Measurements of two-particle correlations in $e^+e^-$ collisions at
  91 GeV with ALEPH archived data}.
\newblock {\em Phys. Rev. Lett.}, 123(21):212002, 2019.
\newblock \href {http://arxiv.org/abs/1906.00489} {\path{arXiv:1906.00489}},
  \href {https://doi.org/10.1103/PhysRevLett.123.212002}
  {\path{doi:10.1103/PhysRevLett.123.212002}}.

\bibitem{ZEUS:2019jya}
I.~Abt et~al.
\newblock {Two-particle azimuthal correlations as a probe of collective
  behaviour in deep inelastic $ep$ scattering at HERA}.
\newblock {\em JHEP}, 04:070, 2020.
\newblock \href {http://arxiv.org/abs/1912.07431} {\path{arXiv:1912.07431}},
  \href {https://doi.org/10.1007/JHEP04(2020)070}
  {\path{doi:10.1007/JHEP04(2020)070}}.

\bibitem{ZEUS:2021qzg}
I.~Abt et~al.
\newblock {Azimuthal correlations in photoproduction and deep inelastic $ep$
  scattering at HERA}.
\newblock {\em JHEP}, 12:102, 2021.
\newblock \href {http://arxiv.org/abs/2106.12377} {\path{arXiv:2106.12377}},
  \href {https://doi.org/10.1007/JHEP12(2021)102}
  {\path{doi:10.1007/JHEP12(2021)102}}.

\bibitem{CMS:2022doq}
{Two-particle azimuthal correlations in $\gamma$p interactions using pPb
  collisions at $\sqrt{s_\mathrm{NN}}$ = 8.16 TeV}.
\newblock 4 2022.
\newblock \href {http://arxiv.org/abs/2204.13486} {\path{arXiv:2204.13486}}.

\bibitem{Zhao:2022ayk}
Wenbin Zhao, Chun Shen, and Bj\"orn Schenke.
\newblock {Collectivity in Ultraperipheral Pb+Pb Collisions at the Large Hadron
  Collider}.
\newblock {\em Phys. Rev. Lett.}, 129(25):252302, 2022.
\newblock \href {http://arxiv.org/abs/2203.06094} {\path{arXiv:2203.06094}},
  \href {https://doi.org/10.1103/PhysRevLett.129.252302}
  {\path{doi:10.1103/PhysRevLett.129.252302}}.

\bibitem{LHCb:2022syj}
{Evidence for modification of $b$ quark hadronization in high-multiplicity $pp$
  collisions at $\sqrt{s} = 13$ TeV}.
\newblock 4 2022.
\newblock \href {http://arxiv.org/abs/2204.13042} {\path{arXiv:2204.13042}}.

\bibitem{CMS:2016xef}
Vardan Khachatryan et~al.
\newblock {Charged-particle nuclear modification factors in PbPb and pPb
  collisions at $ \sqrt{s_{\mathrm{N}\;\mathrm{N}}}=5.02 $ TeV}.
\newblock {\em JHEP}, 04:039, 2017.
\newblock \href {http://arxiv.org/abs/1611.01664} {\path{arXiv:1611.01664}},
  \href {https://doi.org/10.1007/JHEP04(2017)039}
  {\path{doi:10.1007/JHEP04(2017)039}}.

\bibitem{CMS:2016svx}
Vardan Khachatryan et~al.
\newblock {Measurement of inclusive jet production and nuclear modifications in
  pPb collisions at $\sqrt{s_{_\mathrm {NN}}} =$ 5.02 TeV}.
\newblock {\em Eur. Phys. J. C}, 76(7):372, 2016.
\newblock \href {http://arxiv.org/abs/1601.02001} {\path{arXiv:1601.02001}},
  \href {https://doi.org/10.1140/epjc/s10052-016-4205-7}
  {\path{doi:10.1140/epjc/s10052-016-4205-7}}.

\bibitem{PHENIX:2015fgy}
A.~Adare et~al.
\newblock {Centrality-dependent modification of jet-production rates in
  deuteron-gold collisions at $\sqrt{s_{NN}}$=200 GeV}.
\newblock {\em Phys. Rev. Lett.}, 116(12):122301, 2016.
\newblock \href {http://arxiv.org/abs/1509.04657} {\path{arXiv:1509.04657}},
  \href {https://doi.org/10.1103/PhysRevLett.116.122301}
  {\path{doi:10.1103/PhysRevLett.116.122301}}.

\bibitem{ALICE:2015umm}
Jaroslav Adam et~al.
\newblock {Measurement of charged jet production cross sections and nuclear
  modification in p-Pb collisions at $\sqrt{s_{NN}} = 5.02$ TeV}.
\newblock {\em Phys. Lett. B}, 749:68--81, 2015.
\newblock \href {http://arxiv.org/abs/1503.00681} {\path{arXiv:1503.00681}},
  \href {https://doi.org/10.1016/j.physletb.2015.07.054}
  {\path{doi:10.1016/j.physletb.2015.07.054}}.

\bibitem{ATLAS:2014cpa}
Georges Aad et~al.
\newblock {Centrality and rapidity dependence of inclusive jet production in
  $\sqrt{s_\mathrm{NN}} = 5.02$ TeV proton-lead collisions with the ATLAS
  detector}.
\newblock {\em Phys. Lett. B}, 748:392--413, 2015.
\newblock \href {http://arxiv.org/abs/1412.4092} {\path{arXiv:1412.4092}},
  \href {https://doi.org/10.1016/j.physletb.2015.07.023}
  {\path{doi:10.1016/j.physletb.2015.07.023}}.

\bibitem{CMS:2016wma}
Albert~M Sirunyan et~al.
\newblock {Measurements of the charm jet cross section and nuclear modification
  factor in pPb collisions at $\sqrt{{s}_{NN}}$ = 5.02 TeV}.
\newblock {\em Phys. Lett. B}, 772:306--329, 2017.
\newblock \href {http://arxiv.org/abs/1612.08972} {\path{arXiv:1612.08972}},
  \href {https://doi.org/10.1016/j.physletb.2017.06.053}
  {\path{doi:10.1016/j.physletb.2017.06.053}}.

\bibitem{STAR:2003pjh}
J.~Adams et~al.
\newblock {Evidence from d + Au measurements for final state suppression of
  high p(T) hadrons in Au+Au collisions at RHIC}.
\newblock {\em Phys. Rev. Lett.}, 91:072304, 2003.
\newblock \href {http://arxiv.org/abs/nucl-ex/0306024}
  {\path{arXiv:nucl-ex/0306024}}, \href
  {https://doi.org/10.1103/PhysRevLett.91.072304}
  {\path{doi:10.1103/PhysRevLett.91.072304}}.

\bibitem{PHENIX:2003qdw}
S.~S. Adler et~al.
\newblock {Absence of suppression in particle production at large transverse
  momentum in S(NN)**(1/2) = 200-GeV d + Au collisions}.
\newblock {\em Phys. Rev. Lett.}, 91:072303, 2003.
\newblock \href {http://arxiv.org/abs/nucl-ex/0306021}
  {\path{arXiv:nucl-ex/0306021}}, \href
  {https://doi.org/10.1103/PhysRevLett.91.072303}
  {\path{doi:10.1103/PhysRevLett.91.072303}}.

\bibitem{CMS:2019isl}
Albert~M Sirunyan et~al.
\newblock {Strange hadron production in pp and pPb collisions at
  $\sqrt{s_\mathrm{NN}}= $ 5.02 TeV}.
\newblock {\em Phys. Rev. C}, 101(6):064906, 2020.
\newblock \href {http://arxiv.org/abs/1910.04812} {\path{arXiv:1910.04812}},
  \href {https://doi.org/10.1103/PhysRevC.101.064906}
  {\path{doi:10.1103/PhysRevC.101.064906}}.

\bibitem{ALICE:2017svf}
Shreyasi Acharya et~al.
\newblock {Constraints on jet quenching in p-Pb collisions at
  $\mathbf{\sqrt{s_{NN}}}$ = 5.02 TeV measured by the event-activity dependence
  of semi-inclusive hadron-jet distributions}.
\newblock {\em Phys. Lett. B}, 783:95--113, 2018.
\newblock \href {http://arxiv.org/abs/1712.05603} {\path{arXiv:1712.05603}},
  \href {https://doi.org/10.1016/j.physletb.2018.05.059}
  {\path{doi:10.1016/j.physletb.2018.05.059}}.

\bibitem{ATLAS:2019vcm}
Georges Aad et~al.
\newblock {Transverse momentum and process dependent azimuthal anisotropies in
  $\sqrt{s_{\mathrm{NN}}}=8.16$ TeV $p$+Pb collisions with the ATLAS detector}.
\newblock {\em Eur. Phys. J. C}, 80(1):73, 2020.
\newblock \href {http://arxiv.org/abs/1910.13978} {\path{arXiv:1910.13978}},
  \href {https://doi.org/10.1140/epjc/s10052-020-7624-4}
  {\path{doi:10.1140/epjc/s10052-020-7624-4}}.

\bibitem{ALICE:2017smo}
S.~Acharya et~al.
\newblock {Search for collectivity with azimuthal J/$\psi$-hadron correlations
  in high multiplicity p-Pb collisions at $\sqrt{s_{\rm NN}}$ = 5.02 and 8.16
  TeV}.
\newblock {\em Phys. Lett. B}, 780:7--20, 2018.
\newblock \href {http://arxiv.org/abs/1709.06807} {\path{arXiv:1709.06807}},
  \href {https://doi.org/10.1016/j.physletb.2018.02.039}
  {\path{doi:10.1016/j.physletb.2018.02.039}}.

\bibitem{CMS:2018loe}
A.~M. Sirunyan et~al.
\newblock {Elliptic flow of charm and strange hadrons in high-multiplicity pPb
  collisions at $\sqrt{s_{_\mathrm{NN}}} =$ 8.16 TeV}.
\newblock {\em Phys. Rev. Lett.}, 121(8):082301, 2018.
\newblock \href {http://arxiv.org/abs/1804.09767} {\path{arXiv:1804.09767}},
  \href {https://doi.org/10.1103/PhysRevLett.121.082301}
  {\path{doi:10.1103/PhysRevLett.121.082301}}.

\bibitem{CMS:2018duw}
Albert~M Sirunyan et~al.
\newblock {Observation of prompt J/$\psi$ meson elliptic flow in
  high-multiplicity pPb collisions at $\sqrt{s_\mathrm{NN}} =$ 8.16 TeV}.
\newblock {\em Phys. Lett. B}, 791:172--194, 2019.
\newblock \href {http://arxiv.org/abs/1810.01473} {\path{arXiv:1810.01473}},
  \href {https://doi.org/10.1016/j.physletb.2019.02.018}
  {\path{doi:10.1016/j.physletb.2019.02.018}}.

\bibitem{ATLAS:2019xqc}
Georges Aad et~al.
\newblock {Measurement of azimuthal anisotropy of muons from charm and bottom
  hadrons in $pp$ collisions at $\sqrt{s}=13$ TeV with the ATLAS detector}.
\newblock {\em Phys. Rev. Lett.}, 124(8):082301, 2020.
\newblock \href {http://arxiv.org/abs/1909.01650} {\path{arXiv:1909.01650}},
  \href {https://doi.org/10.1103/PhysRevLett.124.082301}
  {\path{doi:10.1103/PhysRevLett.124.082301}}.

\bibitem{CMS:2020qul}
Albert~M Sirunyan et~al.
\newblock {Studies of charm and beauty hadron long-range correlations in pp and
  pPb collisions at LHC energies}.
\newblock {\em Phys. Lett. B}, 813:136036, 2021.
\newblock \href {http://arxiv.org/abs/2009.07065} {\path{arXiv:2009.07065}},
  \href {https://doi.org/10.1016/j.physletb.2020.136036}
  {\path{doi:10.1016/j.physletb.2020.136036}}.

\bibitem{Aoki:2006we}
Y.~Aoki, G.~Endrodi, Z.~Fodor, S.~D. Katz, and K.~K. Szabo.
\newblock {The Order of the quantum chromodynamics transition predicted by the
  standard model of particle physics}.
\newblock {\em Nature}, 443:675--678, 2006.
\newblock \href {http://arxiv.org/abs/hep-lat/0611014}
  {\path{arXiv:hep-lat/0611014}}, \href {https://doi.org/10.1038/nature05120}
  {\path{doi:10.1038/nature05120}}.

\bibitem{Fukushima:2013rx}
Kenji Fukushima and Chihiro Sasaki.
\newblock {The phase diagram of nuclear and quark matter at high baryon
  density}.
\newblock {\em Prog. Part. Nucl. Phys.}, 72:99--154, 2013.
\newblock \href {http://arxiv.org/abs/1301.6377} {\path{arXiv:1301.6377}},
  \href {https://doi.org/10.1016/j.ppnp.2013.05.003}
  {\path{doi:10.1016/j.ppnp.2013.05.003}}.

\bibitem{Stephanov:1998dy}
Misha~A. Stephanov, K.~Rajagopal, and Edward~V. Shuryak.
\newblock {Signatures of the tricritical point in QCD}.
\newblock {\em Phys. Rev. Lett.}, 81:4816--4819, 1998.
\newblock \href {http://arxiv.org/abs/hep-ph/9806219}
  {\path{arXiv:hep-ph/9806219}}, \href
  {https://doi.org/10.1103/PhysRevLett.81.4816}
  {\path{doi:10.1103/PhysRevLett.81.4816}}.

\bibitem{Stephanov:1999zu}
Misha~A. Stephanov, K.~Rajagopal, and Edward~V. Shuryak.
\newblock {Event-by-event fluctuations in heavy ion collisions and the QCD
  critical point}.
\newblock {\em Phys. Rev. D}, 60:114028, 1999.
\newblock \href {http://arxiv.org/abs/hep-ph/9903292}
  {\path{arXiv:hep-ph/9903292}}, \href
  {https://doi.org/10.1103/PhysRevD.60.114028}
  {\path{doi:10.1103/PhysRevD.60.114028}}.

\bibitem{Bzdak:2019pkr}
Adam Bzdak, Shinichi Esumi, Volker Koch, Jinfeng Liao, Mikhail Stephanov, and
  Nu~Xu.
\newblock {Mapping the Phases of Quantum Chromodynamics with Beam Energy Scan}.
\newblock {\em Phys. Rept.}, 853:1--87, 2020.
\newblock \href {http://arxiv.org/abs/1906.00936} {\path{arXiv:1906.00936}},
  \href {https://doi.org/10.1016/j.physrep.2020.01.005}
  {\path{doi:10.1016/j.physrep.2020.01.005}}.

\bibitem{Bazavov:2020bjn}
A.~Bazavov et~al.
\newblock {Skewness, kurtosis, and the fifth and sixth order cumulants of net
  baryon-number distributions from lattice QCD confront high-statistics STAR
  data}.
\newblock {\em Phys. Rev. D}, 101(7):074502, 2020.
\newblock \href {http://arxiv.org/abs/2001.08530} {\path{arXiv:2001.08530}},
  \href {https://doi.org/10.1103/PhysRevD.101.074502}
  {\path{doi:10.1103/PhysRevD.101.074502}}.

\bibitem{STAR:2021iop}
Mohamed Abdallah et~al.
\newblock {Cumulants and correlation functions of net-proton, proton, and
  antiproton multiplicity distributions in Au+Au collisions at energies
  available at the BNL Relativistic Heavy Ion Collider}.
\newblock {\em Phys. Rev. C}, 104(2):024902, 2021.
\newblock \href {http://arxiv.org/abs/2101.12413} {\path{arXiv:2101.12413}},
  \href {https://doi.org/10.1103/PhysRevC.104.024902}
  {\path{doi:10.1103/PhysRevC.104.024902}}.

\bibitem{Adamczewski-Musch:2020slf}
J.~Adamczewski-Musch et~al.
\newblock {Proton-number fluctuations in $\sqrt {s_{NN}}$ =2.4 GeV Au + Au
  collisions studied with the High-Acceptance DiElectron Spectrometer (HADES)}.
\newblock {\em Phys. Rev. C}, 102(2):024914, 2020.
\newblock \href {http://arxiv.org/abs/2002.08701} {\path{arXiv:2002.08701}},
  \href {https://doi.org/10.1103/PhysRevC.102.024914}
  {\path{doi:10.1103/PhysRevC.102.024914}}.

\bibitem{STAR:2021fge}
M.~S. Abdallah et~al.
\newblock {Measurements of Proton High Order Cumulants in
  $\sqrt{s_{_{\mathrm{NN}}}}$ = 3 GeV Au+Au Collisions and Implications for the
  QCD Critical Point}.
\newblock {\em Phys. Rev. Lett.}, 128(20):202303, 2022.
\newblock \href {http://arxiv.org/abs/2112.00240} {\path{arXiv:2112.00240}},
  \href {https://doi.org/10.1103/PhysRevLett.128.202303}
  {\path{doi:10.1103/PhysRevLett.128.202303}}.

\bibitem{Braun-Munzinger:2020jbk}
P.~Braun-Munzinger, B.~Friman, K.~Redlich, A.~Rustamov, and J.~Stachel.
\newblock {Relativistic nuclear collisions: Establishing a non-critical
  baseline for fluctuation measurements}.
\newblock {\em Nucl. Phys. A}, 1008:122141, 2021.
\newblock \href {http://arxiv.org/abs/2007.02463} {\path{arXiv:2007.02463}},
  \href {https://doi.org/10.1016/j.nuclphysa.2021.122141}
  {\path{doi:10.1016/j.nuclphysa.2021.122141}}.

\bibitem{Bleicher:1999xi}
M.~Bleicher et~al.
\newblock {Relativistic hadron hadron collisions in the ultrarelativistic
  quantum molecular dynamics model}.
\newblock {\em J. Phys. G}, 25:1859--1896, 1999.
\newblock \href {http://arxiv.org/abs/hep-ph/9909407}
  {\path{arXiv:hep-ph/9909407}}, \href
  {https://doi.org/10.1088/0954-3899/25/9/308}
  {\path{doi:10.1088/0954-3899/25/9/308}}.

\bibitem{Bass:1998ca}
S.~A. Bass et~al.
\newblock {Microscopic models for ultrarelativistic heavy ion collisions}.
\newblock {\em Prog. Part. Nucl. Phys.}, 41:255--369, 1998.
\newblock \href {http://arxiv.org/abs/nucl-th/9803035}
  {\path{arXiv:nucl-th/9803035}}, \href
  {https://doi.org/10.1016/S0146-6410(98)00058-1}
  {\path{doi:10.1016/S0146-6410(98)00058-1}}.

\bibitem{Stephanov:2011pb}
M.~A. Stephanov.
\newblock {On the sign of kurtosis near the QCD critical point}.
\newblock {\em Phys. Rev. Lett.}, 107:052301, 2011.
\newblock \href {http://arxiv.org/abs/1104.1627} {\path{arXiv:1104.1627}},
  \href {https://doi.org/10.1103/PhysRevLett.107.052301}
  {\path{doi:10.1103/PhysRevLett.107.052301}}.

\bibitem{STAR:2020dav}
J.~Adam et~al.
\newblock {Flow and interferometry results from Au+Au collisions at
  $\sqrt{s_{NN}} = 4.5$ GeV}.
\newblock {\em Phys. Rev. C}, 103(3):034908, 2021.
\newblock \href {http://arxiv.org/abs/2007.14005} {\path{arXiv:2007.14005}},
  \href {https://doi.org/10.1103/PhysRevC.103.034908}
  {\path{doi:10.1103/PhysRevC.103.034908}}.

\bibitem{STAR:2017ieb}
L.~Adamczyk et~al.
\newblock {Beam Energy Dependence of Jet-Quenching Effects in Au+Au Collisions
  at $\sqrt{s_{_{ \mathrm{NN}}}}$ = 7.7, 11.5, 14.5, 19.6, 27, 39, and 62.4
  GeV}.
\newblock {\em Phys. Rev. Lett.}, 121(3):032301, 2018.
\newblock \href {http://arxiv.org/abs/1707.01988} {\path{arXiv:1707.01988}},
  \href {https://doi.org/10.1103/PhysRevLett.121.032301}
  {\path{doi:10.1103/PhysRevLett.121.032301}}.

\bibitem{STAR:2013cow}
L.~Adamczyk et~al.
\newblock {Observation of an Energy-Dependent Difference in Elliptic Flow
  between Particles and Antiparticles in Relativistic Heavy Ion Collisions}.
\newblock {\em Phys. Rev. Lett.}, 110(14):142301, 2013.
\newblock \href {http://arxiv.org/abs/1301.2347} {\path{arXiv:1301.2347}},
  \href {https://doi.org/10.1103/PhysRevLett.110.142301}
  {\path{doi:10.1103/PhysRevLett.110.142301}}.

\bibitem{STAR:2020tga}
J.~Adam et~al.
\newblock {Nonmonotonic Energy Dependence of Net-Proton Number Fluctuations}.
\newblock {\em Phys. Rev. Lett.}, 126(9):092301, 2021.
\newblock \href {http://arxiv.org/abs/2001.02852} {\path{arXiv:2001.02852}},
  \href {https://doi.org/10.1103/PhysRevLett.126.092301}
  {\path{doi:10.1103/PhysRevLett.126.092301}}.

\bibitem{STAR:2021rls}
Mohamed Abdallah et~al.
\newblock {Measurement of the Sixth-Order Cumulant of Net-Proton Multiplicity
  Distributions in Au+Au Collisions at $\sqrt{s_{NN}}=$ 27, 54.4, and 200 GeV
  at RHIC}.
\newblock {\em Phys. Rev. Lett.}, 127(26):262301, 2021.
\newblock \href {http://arxiv.org/abs/2105.14698} {\path{arXiv:2105.14698}},
  \href {https://doi.org/10.1103/PhysRevLett.127.262301}
  {\path{doi:10.1103/PhysRevLett.127.262301}}.

\bibitem{STAR:2022vlo}
{Beam Energy Dependence of Fifth and Sixth-Order Net-proton Number Fluctuations
  in Au+Au Collisions at RHIC}.
\newblock 7 2022.
\newblock \href {http://arxiv.org/abs/2207.09837} {\path{arXiv:2207.09837}}.

\bibitem{Vovchenko:2021kxx}
Volodymyr Vovchenko, Volker Koch, and Chun Shen.
\newblock {Proton number cumulants and correlation functions in Au-Au
  collisions at sNN=7.7\textendash{}200 GeV from hydrodynamics}.
\newblock {\em Phys. Rev. C}, 105(1):014904, 2022.
\newblock \href {http://arxiv.org/abs/2107.00163} {\path{arXiv:2107.00163}},
  \href {https://doi.org/10.1103/PhysRevC.105.014904}
  {\path{doi:10.1103/PhysRevC.105.014904}}.

\bibitem{STAR:2021yiu}
M.~S. Abdallah et~al.
\newblock {Disappearance of partonic collectivity in sNN=3GeV Au+Au collisions
  at RHIC}.
\newblock {\em Phys. Lett. B}, 827:137003, 2022.
\newblock \href {http://arxiv.org/abs/2108.00908} {\path{arXiv:2108.00908}},
  \href {https://doi.org/10.1016/j.physletb.2022.137003}
  {\path{doi:10.1016/j.physletb.2022.137003}}.

\bibitem{STAR:2021hyx}
M.~S. Abdallah et~al.
\newblock {Probing strangeness canonical ensemble with K\ensuremath{-},
  \ensuremath{\phi}(1020) and \ensuremath{\Xi}\ensuremath{-} production in
  Au+Au collisions at sNN=3 GeV}.
\newblock {\em Phys. Lett. B}, 831:137152, 2022.
\newblock \href {http://arxiv.org/abs/2108.00924} {\path{arXiv:2108.00924}},
  \href {https://doi.org/10.1016/j.physletb.2022.137152}
  {\path{doi:10.1016/j.physletb.2022.137152}}.

\bibitem{HADES:2020wpc}
J.~Adamczewski-Musch et~al.
\newblock {Proton-number fluctuations in $\sqrt {s_{NN}}$ =2.4 GeV Au + Au
  collisions studied with the High-Acceptance DiElectron Spectrometer (HADES)}.
\newblock {\em Phys. Rev. C}, 102(2):024914, 2020.
\newblock \href {http://arxiv.org/abs/2002.08701} {\path{arXiv:2002.08701}},
  \href {https://doi.org/10.1103/PhysRevC.102.024914}
  {\path{doi:10.1103/PhysRevC.102.024914}}.

\bibitem{Kharzeev:1999cz}
Dmitri Kharzeev and Robert~D. Pisarski.
\newblock {Pionic measures of parity and CP violation in high-energy nuclear
  collisions}.
\newblock {\em Phys. Rev. D}, 61:111901, 2000.
\newblock \href {http://arxiv.org/abs/hep-ph/9906401}
  {\path{arXiv:hep-ph/9906401}}, \href
  {https://doi.org/10.1103/PhysRevD.61.111901}
  {\path{doi:10.1103/PhysRevD.61.111901}}.

\bibitem{Kharzeev:2004ey}
Dmitri Kharzeev.
\newblock {Parity violation in hot QCD: Why it can happen, and how to look for
  it}.
\newblock {\em Phys. Lett. B}, 633:260--264, 2006.
\newblock \href {http://arxiv.org/abs/hep-ph/0406125}
  {\path{arXiv:hep-ph/0406125}}, \href
  {https://doi.org/10.1016/j.physletb.2005.11.075}
  {\path{doi:10.1016/j.physletb.2005.11.075}}.

\bibitem{Gursoy:2014aka}
Umut Gursoy, Dmitri Kharzeev, and Krishna Rajagopal.
\newblock {Magnetohydrodynamics, charged currents and directed flow in heavy
  ion collisions}.
\newblock {\em Phys. Rev. C}, 89(5):054905, 2014.
\newblock \href {http://arxiv.org/abs/1401.3805} {\path{arXiv:1401.3805}},
  \href {https://doi.org/10.1103/PhysRevC.89.054905}
  {\path{doi:10.1103/PhysRevC.89.054905}}.

\bibitem{Das:2016cwd}
Santosh~K. Das, Salvatore Plumari, Sandeep Chatterjee, Jane Alam, Francesco
  Scardina, and Vincenzo Greco.
\newblock {Directed Flow of Charm Quarks as a Witness of the Initial Strong
  Magnetic Field in Ultra-Relativistic Heavy Ion Collisions}.
\newblock {\em Phys. Lett. B}, 768:260--264, 2017.
\newblock \href {http://arxiv.org/abs/1608.02231} {\path{arXiv:1608.02231}},
  \href {https://doi.org/10.1016/j.physletb.2017.02.046}
  {\path{doi:10.1016/j.physletb.2017.02.046}}.

\bibitem{Gursoy:2018yai}
Umut G\"ursoy, Dmitri Kharzeev, Eric Marcus, Krishna Rajagopal, and Chun Shen.
\newblock {Charge-dependent Flow Induced by Magnetic and Electric Fields in
  Heavy Ion Collisions}.
\newblock {\em Phys. Rev. C}, 98(5):055201, 2018.
\newblock \href {http://arxiv.org/abs/1806.05288} {\path{arXiv:1806.05288}},
  \href {https://doi.org/10.1103/PhysRevC.98.055201}
  {\path{doi:10.1103/PhysRevC.98.055201}}.

\bibitem{Sun:2020wkg}
Yifeng Sun, Salvatore Plumari, and Vincenzo Greco.
\newblock {Probing the electromagnetic fields in ultrarelativistic collisions
  with leptons from $Z^0$ decay and charmed mesons}.
\newblock {\em Phys. Lett. B}, 816:136271, 2021.
\newblock \href {http://arxiv.org/abs/2004.09880} {\path{arXiv:2004.09880}},
  \href {https://doi.org/10.1016/j.physletb.2021.136271}
  {\path{doi:10.1016/j.physletb.2021.136271}}.

\bibitem{Oliva:2020doe}
Lucia Oliva, Salvatore Plumari, and Vincenzo Greco.
\newblock {Directed flow of D mesons at RHIC and LHC: non-perturbative
  dynamics, longitudinal bulk matter asymmetry and electromagnetic fields}.
\newblock {\em JHEP}, 05:034, 2021.
\newblock \href {http://arxiv.org/abs/2009.11066} {\path{arXiv:2009.11066}},
  \href {https://doi.org/10.1007/JHEP05(2021)034}
  {\path{doi:10.1007/JHEP05(2021)034}}.

\bibitem{Feng:2022yus}
Yicheng Feng.
\newblock {Estimate of a nonflow baseline for the chiral magnetic effect in
  isobar collisions at RHIC}.
\newblock In {\em {20th International Conference on Strangeness in Quark Matter
  2022}}, 9 2022.
\newblock \href {http://arxiv.org/abs/2209.13078} {\path{arXiv:2209.13078}}.

\bibitem{Voloshin:2004vk}
Sergei~A. Voloshin.
\newblock {Parity violation in hot QCD: How to detect it}.
\newblock {\em Phys. Rev. C}, 70:057901, 2004.
\newblock \href {http://arxiv.org/abs/hep-ph/0406311}
  {\path{arXiv:hep-ph/0406311}}, \href
  {https://doi.org/10.1103/PhysRevC.70.057901}
  {\path{doi:10.1103/PhysRevC.70.057901}}.

\bibitem{Abelev:2009ac}
B.~I. Abelev et~al.
\newblock {Azimuthal Charged-Particle Correlations and Possible Local Strong
  Parity Violation}.
\newblock {\em Phys. Rev. Lett.}, 103:251601, 2009.
\newblock \href {http://arxiv.org/abs/0909.1739} {\path{arXiv:0909.1739}},
  \href {https://doi.org/10.1103/PhysRevLett.103.251601}
  {\path{doi:10.1103/PhysRevLett.103.251601}}.

\bibitem{Wang:2009kd}
Fuqiang Wang.
\newblock {Effects of Cluster Particle Correlations on Local Parity Violation
  Observables}.
\newblock {\em Phys. Rev. C}, 81:064902, 2010.
\newblock \href {http://arxiv.org/abs/0911.1482} {\path{arXiv:0911.1482}},
  \href {https://doi.org/10.1103/PhysRevC.81.064902}
  {\path{doi:10.1103/PhysRevC.81.064902}}.

\bibitem{Bzdak:2009fc}
Adam Bzdak, Volker Koch, and Jinfeng Liao.
\newblock {Remarks on possible local parity violation in heavy ion collisions}.
\newblock {\em Phys. Rev. C}, 81:031901, 2010.
\newblock \href {http://arxiv.org/abs/0912.5050} {\path{arXiv:0912.5050}},
  \href {https://doi.org/10.1103/PhysRevC.81.031901}
  {\path{doi:10.1103/PhysRevC.81.031901}}.

\bibitem{Bzdak:2010fd}
Adam Bzdak, Volker Koch, and Jinfeng Liao.
\newblock {Azimuthal correlations from transverse momentum conservation and
  possible local parity violation}.
\newblock {\em Phys. Rev. C}, 83:014905, 2011.
\newblock \href {http://arxiv.org/abs/1008.4919} {\path{arXiv:1008.4919}},
  \href {https://doi.org/10.1103/PhysRevC.83.014905}
  {\path{doi:10.1103/PhysRevC.83.014905}}.

\bibitem{Schlichting:2010qia}
Soren Schlichting and Scott Pratt.
\newblock {Charge conservation at energies available at the BNL Relativistic
  Heavy Ion Collider and contributions to local parity violation observables}.
\newblock {\em Phys. Rev. C}, 83:014913, 2011.
\newblock \href {http://arxiv.org/abs/1009.4283} {\path{arXiv:1009.4283}},
  \href {https://doi.org/10.1103/PhysRevC.83.014913}
  {\path{doi:10.1103/PhysRevC.83.014913}}.

\bibitem{Pratt:2010zn}
Scott Pratt, Soeren Schlichting, and Sean Gavin.
\newblock {Effects of Momentum Conservation and Flow on Angular Correlations at
  RHIC}.
\newblock {\em Phys. Rev. C}, 84:024909, 2011.
\newblock \href {http://arxiv.org/abs/1011.6053} {\path{arXiv:1011.6053}},
  \href {https://doi.org/10.1103/PhysRevC.84.024909}
  {\path{doi:10.1103/PhysRevC.84.024909}}.

\bibitem{Abelev:2009ad}
B.~I. Abelev et~al.
\newblock {Observation of charge-dependent azimuthal correlations and possible
  local strong parity violation in heavy ion collisions}.
\newblock {\em Phys. Rev. C}, 81:054908, 2010.
\newblock \href {http://arxiv.org/abs/0909.1717} {\path{arXiv:0909.1717}},
  \href {https://doi.org/10.1103/PhysRevC.81.054908}
  {\path{doi:10.1103/PhysRevC.81.054908}}.

\bibitem{Abelev:2012pa}
Betty Abelev et~al.
\newblock {Charge separation relative to the reaction plane in Pb-Pb collisions
  at $\sqrt{s_{NN}}= 2.76$ TeV}.
\newblock {\em Phys. Rev. Lett.}, 110(1):012301, 2013.
\newblock \href {http://arxiv.org/abs/1207.0900} {\path{arXiv:1207.0900}},
  \href {https://doi.org/10.1103/PhysRevLett.110.012301}
  {\path{doi:10.1103/PhysRevLett.110.012301}}.

\bibitem{Adamczyk:2013hsi}
L.~Adamczyk et~al.
\newblock {Fluctuations of charge separation perpendicular to the event plane
  and local parity violation in $\sqrt{s_{NN}}=200$ GeV Au+Au collisions at the
  BNL Relativistic Heavy Ion Collider}.
\newblock {\em Phys. Rev. C}, 88(6):064911, 2013.
\newblock \href {http://arxiv.org/abs/1302.3802} {\path{arXiv:1302.3802}},
  \href {https://doi.org/10.1103/PhysRevC.88.064911}
  {\path{doi:10.1103/PhysRevC.88.064911}}.

\bibitem{Adamczyk:2013kcb}
L.~Adamczyk et~al.
\newblock {Measurement of charge multiplicity asymmetry correlations in
  high-energy nucleus-nucleus collisions at $\sqrt{{s}_{NN}} =$ 200 GeV}.
\newblock {\em Phys. Rev. C}, 89(4):044908, 2014.
\newblock \href {http://arxiv.org/abs/1303.0901} {\path{arXiv:1303.0901}},
  \href {https://doi.org/10.1103/PhysRevC.89.044908}
  {\path{doi:10.1103/PhysRevC.89.044908}}.

\bibitem{Adamczyk:2014mzf}
L.~Adamczyk et~al.
\newblock {Beam-energy dependence of charge separation along the magnetic field
  in Au+Au collisions at RHIC}.
\newblock {\em Phys. Rev. Lett.}, 113:052302, 2014.
\newblock \href {http://arxiv.org/abs/1404.1433} {\path{arXiv:1404.1433}},
  \href {https://doi.org/10.1103/PhysRevLett.113.052302}
  {\path{doi:10.1103/PhysRevLett.113.052302}}.

\bibitem{Adam:2015vje}
Jaroslav Adam et~al.
\newblock {Charge-dependent flow and the search for the chiral magnetic wave in
  Pb-Pb collisions at $\sqrt{s_{\rm NN}} =$ 2.76 TeV}.
\newblock {\em Phys. Rev. C}, 93(4):044903, 2016.
\newblock \href {http://arxiv.org/abs/1512.05739} {\path{arXiv:1512.05739}},
  \href {https://doi.org/10.1103/PhysRevC.93.044903}
  {\path{doi:10.1103/PhysRevC.93.044903}}.

\bibitem{Khachatryan:2016got}
Vardan Khachatryan et~al.
\newblock {Observation of charge-dependent azimuthal correlations in $p$-Pb
  collisions and its implication for the search for the chiral magnetic
  effect}.
\newblock {\em Phys. Rev. Lett.}, 118(12):122301, 2017.
\newblock \href {http://arxiv.org/abs/1610.00263} {\path{arXiv:1610.00263}},
  \href {https://doi.org/10.1103/PhysRevLett.118.122301}
  {\path{doi:10.1103/PhysRevLett.118.122301}}.

\bibitem{Acharya:2017fau}
Shreyasi Acharya et~al.
\newblock {Constraining the magnitude of the Chiral Magnetic Effect with Event
  Shape Engineering in Pb-Pb collisions at $\sqrt{s_\mathrm{NN}}$ = 2.76 TeV}.
\newblock {\em Phys. Lett. B}, 777:151--162, 2018.
\newblock \href {http://arxiv.org/abs/1709.04723} {\path{arXiv:1709.04723}},
  \href {https://doi.org/10.1016/j.physletb.2017.12.021}
  {\path{doi:10.1016/j.physletb.2017.12.021}}.

\bibitem{Sirunyan:2017quh}
Albert~M Sirunyan et~al.
\newblock {Constraints on the chiral magnetic effect using charge-dependent
  azimuthal correlations in $p\mathrm{Pb}$ and PbPb collisions at the CERN
  Large Hadron Collider}.
\newblock {\em Phys. Rev. C}, 97(4):044912, 2018.
\newblock \href {http://arxiv.org/abs/1708.01602} {\path{arXiv:1708.01602}},
  \href {https://doi.org/10.1103/PhysRevC.97.044912}
  {\path{doi:10.1103/PhysRevC.97.044912}}.

\bibitem{STAR:2019xzd}
J.~Adam et~al.
\newblock {Charge-dependent pair correlations relative to a third particle in
  $p$ + Au and $d$+ Au collisions at RHIC}.
\newblock {\em Phys. Lett. B}, 798:134975, 2019.
\newblock \href {http://arxiv.org/abs/1906.03373} {\path{arXiv:1906.03373}},
  \href {https://doi.org/10.1016/j.physletb.2019.134975}
  {\path{doi:10.1016/j.physletb.2019.134975}}.

\bibitem{STAR:2020gky}
M.~S. Abdallah et~al.
\newblock {Pair invariant mass to isolate background in the search for the
  chiral magnetic effect in Au~+~Au collisions at sNN=200~GeV}.
\newblock {\em Phys. Rev. C}, 106(3):034908, 2022.
\newblock \href {http://arxiv.org/abs/2006.05035} {\path{arXiv:2006.05035}},
  \href {https://doi.org/10.1103/PhysRevC.106.034908}
  {\path{doi:10.1103/PhysRevC.106.034908}}.

\bibitem{STAR:2021pwb}
M.~S. Abdallah et~al.
\newblock {Search for the Chiral Magnetic Effect via Charge-Dependent Azimuthal
  Correlations Relative to Spectator and Participant Planes in Au+Au Collisions
  at $\sqrt{s_{NN}}$ =\, 200\,GeV}.
\newblock {\em Phys. Rev. Lett.}, 128(9):092301, 2022.
\newblock \href {http://arxiv.org/abs/2106.09243} {\path{arXiv:2106.09243}},
  \href {https://doi.org/10.1103/PhysRevLett.128.092301}
  {\path{doi:10.1103/PhysRevLett.128.092301}}.

\bibitem{ALICE:2020siw}
Shreyasi Acharya et~al.
\newblock {Constraining the Chiral Magnetic Effect with charge-dependent
  azimuthal correlations in Pb-Pb collisions at $ \sqrt{s_{\mathrm{NN}}} $ =
  2.76 and 5.02 TeV}.
\newblock {\em JHEP}, 09:160, 2020.
\newblock \href {http://arxiv.org/abs/2005.14640} {\path{arXiv:2005.14640}},
  \href {https://doi.org/10.1007/JHEP09(2020)160}
  {\path{doi:10.1007/JHEP09(2020)160}}.

\bibitem{Collaboration:2022flo}
{Search for the Chiral Magnetic Effect with charge-dependent azimuthal
  correlations in Xe-Xe collisions at $\sqrt{s_{\mathrm{NN}}} = 5.44$ TeV}.
\newblock 10 2022.
\newblock \href {http://arxiv.org/abs/2210.15383} {\path{arXiv:2210.15383}}.

\bibitem{Liang:2004ph}
Zuo-Tang Liang and Xin-Nian Wang.
\newblock {Globally polarized quark-gluon plasma in non-central A+A
  collisions}.
\newblock {\em Phys. Rev. Lett.}, 94:102301, 2005.
\newblock [Erratum: Phys.Rev.Lett. 96, 039901 (2006)].
\newblock \href {http://arxiv.org/abs/nucl-th/0410079}
  {\path{arXiv:nucl-th/0410079}}, \href
  {https://doi.org/10.1103/PhysRevLett.94.102301}
  {\path{doi:10.1103/PhysRevLett.94.102301}}.

\bibitem{Becattini:2007sr}
F.~Becattini, F.~Piccinini, and J.~Rizzo.
\newblock {Angular momentum conservation in heavy ion collisions at very high
  energy}.
\newblock {\em Phys. Rev. C}, 77:024906, 2008.
\newblock \href {http://arxiv.org/abs/0711.1253} {\path{arXiv:0711.1253}},
  \href {https://doi.org/10.1103/PhysRevC.77.024906}
  {\path{doi:10.1103/PhysRevC.77.024906}}.

\bibitem{STAR:2017ckg}
L.~Adamczyk et~al.
\newblock {Global $\Lambda$ hyperon polarization in nuclear collisions:
  evidence for the most vortical fluid}.
\newblock {\em Nature}, 548:62--65, 2017.
\newblock \href {http://arxiv.org/abs/1701.06657} {\path{arXiv:1701.06657}},
  \href {https://doi.org/10.1038/nature23004} {\path{doi:10.1038/nature23004}}.

\bibitem{Becattini:2016gvu}
F.~Becattini, I.~Karpenko, M.~Lisa, I.~Upsal, and S.~Voloshin.
\newblock {Global hyperon polarization at local thermodynamic equilibrium with
  vorticity, magnetic field and feed-down}.
\newblock {\em Phys. Rev. C}, 95(5):054902, 2017.
\newblock \href {http://arxiv.org/abs/1610.02506} {\path{arXiv:1610.02506}},
  \href {https://doi.org/10.1103/PhysRevC.95.054902}
  {\path{doi:10.1103/PhysRevC.95.054902}}.

\bibitem{STAR:2020xbm}
J.~Adam et~al.
\newblock {Global Polarization of $\Xi$ and $\Omega$ Hyperons in Au+Au
  Collisions at $\sqrt {s_{NN}}$ = 200 GeV}.
\newblock {\em Phys. Rev. Lett.}, 126(16):162301, 2021.
\newblock \href {http://arxiv.org/abs/2012.13601} {\path{arXiv:2012.13601}},
  \href {https://doi.org/10.1103/PhysRevLett.126.162301}
  {\path{doi:10.1103/PhysRevLett.126.162301}}.

\bibitem{STAR:2021beb}
M.~S. Abdallah et~al.
\newblock {Global $\Lambda$-hyperon polarization in Au+Au collisions at $\sqrt
  {s_{NN}}$=3~GeV}.
\newblock {\em Phys. Rev. C}, 104(6):L061901, 2021.
\newblock \href {http://arxiv.org/abs/2108.00044} {\path{arXiv:2108.00044}},
  \href {https://doi.org/10.1103/PhysRevC.104.L061901}
  {\path{doi:10.1103/PhysRevC.104.L061901}}.

\bibitem{STAR:2007ccu}
B.~I. Abelev et~al.
\newblock {Global polarization measurement in Au+Au collisions}.
\newblock {\em Phys. Rev. C}, 76:024915, 2007.
\newblock [Erratum: Phys.Rev.C 95, 039906 (2017)].
\newblock \href {http://arxiv.org/abs/0705.1691} {\path{arXiv:0705.1691}},
  \href {https://doi.org/10.1103/PhysRevC.76.024915}
  {\path{doi:10.1103/PhysRevC.76.024915}}.

\bibitem{STAR:2018gyt}
Jaroslav Adam et~al.
\newblock {Global polarization of $\Lambda$ hyperons in Au+Au collisions at
  $\sqrt{s_{_{NN}}}$ = 200 GeV}.
\newblock {\em Phys. Rev. C}, 98:014910, 2018.
\newblock \href {http://arxiv.org/abs/1805.04400} {\path{arXiv:1805.04400}},
  \href {https://doi.org/10.1103/PhysRevC.98.014910}
  {\path{doi:10.1103/PhysRevC.98.014910}}.

\bibitem{ALICE:2019onw}
Shreyasi Acharya et~al.
\newblock {Global polarization of $\Lambda \bar \Lambda$ hyperons in Pb-Pb
  collisions at $\sqrt {s_{NN}}$ = 2.76 and 5.02 TeV}.
\newblock {\em Phys. Rev. C}, 101(4):044611, 2020.
\newblock [Erratum: Phys.Rev.C 105, 029902 (2022)].
\newblock \href {http://arxiv.org/abs/1909.01281} {\path{arXiv:1909.01281}},
  \href {https://doi.org/10.1103/PhysRevC.101.044611}
  {\path{doi:10.1103/PhysRevC.101.044611}}.

\bibitem{HADES:2022enx}
R.~Abou~Yassine et~al.
\newblock {Measurement of global polarization of \ensuremath{\Lambda} hyperons
  in few-GeV heavy-ion collisions}.
\newblock {\em Phys. Lett. B}, 835:137506, 2022.
\newblock \href {http://arxiv.org/abs/2207.05160} {\path{arXiv:2207.05160}},
  \href {https://doi.org/10.1016/j.physletb.2022.137506}
  {\path{doi:10.1016/j.physletb.2022.137506}}.

\bibitem{Becattini:2020ngo}
Francesco Becattini and Michael~A. Lisa.
\newblock {Polarization and Vorticity in the Quark\textendash{}Gluon Plasma}.
\newblock {\em Ann. Rev. Nucl. Part. Sci.}, 70:395--423, 2020.
\newblock \href {http://arxiv.org/abs/2003.03640} {\path{arXiv:2003.03640}},
  \href {https://doi.org/10.1146/annurev-nucl-021920-095245}
  {\path{doi:10.1146/annurev-nucl-021920-095245}}.

\bibitem{ParticleDataGroup:2020ssz}
P.~A. Zyla et~al.
\newblock {Review of Particle Physics}.
\newblock {\em PTEP}, 2020(8):083C01, 2020.
\newblock \href {https://doi.org/10.1093/ptep/ptaa104}
  {\path{doi:10.1093/ptep/ptaa104}}.

\bibitem{Karpenko:2016jyx}
I.~Karpenko and F.~Becattini.
\newblock {Study of $\Lambda $ polarization in relativistic nuclear collisions
  at $\sqrt{s_\mathrm {NN}}=7.7$ \textendash{}200 GeV}.
\newblock {\em Eur. Phys. J. C}, 77(4):213, 2017.
\newblock \href {http://arxiv.org/abs/1610.04717} {\path{arXiv:1610.04717}},
  \href {https://doi.org/10.1140/epjc/s10052-017-4765-1}
  {\path{doi:10.1140/epjc/s10052-017-4765-1}}.

\bibitem{Sun:2017xhx}
Yifeng Sun and Che~Ming Ko.
\newblock {$\Lambda$ hyperon polarization in relativistic heavy ion collisions
  from a chiral kinetic approach}.
\newblock {\em Phys. Rev. C}, 96(2):024906, 2017.
\newblock \href {http://arxiv.org/abs/1706.09467} {\path{arXiv:1706.09467}},
  \href {https://doi.org/10.1103/PhysRevC.96.024906}
  {\path{doi:10.1103/PhysRevC.96.024906}}.

\bibitem{Guo:2021udq}
Yu~Guo, Jinfeng Liao, Enke Wang, Hongxi Xing, and Hui Zhang.
\newblock {Hyperon polarization from the vortical fluid in low-energy nuclear
  collisions}.
\newblock {\em Phys. Rev. C}, 104(4):L041902, 2021.
\newblock \href {http://arxiv.org/abs/2105.13481} {\path{arXiv:2105.13481}},
  \href {https://doi.org/10.1103/PhysRevC.104.L041902}
  {\path{doi:10.1103/PhysRevC.104.L041902}}.

\bibitem{Ivanov:2020udj}
Yu~B. Ivanov.
\newblock {Global $\Lambda$ polarization in moderately relativistic nuclear
  collisions}.
\newblock {\em Phys. Rev. C}, 103(3):L031903, 2021.
\newblock \href {http://arxiv.org/abs/2012.07597} {\path{arXiv:2012.07597}},
  \href {https://doi.org/10.1103/PhysRevC.103.L031903}
  {\path{doi:10.1103/PhysRevC.103.L031903}}.

\bibitem{Adams:2019fpo}
Joseph Adams et~al.
\newblock {The STAR Event Plane Detector}.
\newblock {\em Nucl. Instrum. Meth. A}, 968:163970, 2020.
\newblock \href {http://arxiv.org/abs/1912.05243} {\path{arXiv:1912.05243}},
  \href {https://doi.org/10.1016/j.nima.2020.163970}
  {\path{doi:10.1016/j.nima.2020.163970}}.

\bibitem{Muller:2018ibh}
Berndt M\"uller and Andreas Sch\"afer.
\newblock {Chiral magnetic effect and an experimental bound on the late time
  magnetic field strength}.
\newblock {\em Phys. Rev. D}, 98(7):071902, 2018.
\newblock \href {http://arxiv.org/abs/1806.10907} {\path{arXiv:1806.10907}},
  \href {https://doi.org/10.1103/PhysRevD.98.071902}
  {\path{doi:10.1103/PhysRevD.98.071902}}.

\bibitem{Becattini:2017gcx}
F.~Becattini and Iu. Karpenko.
\newblock {Collective Longitudinal Polarization in Relativistic Heavy-Ion
  Collisions at Very High Energy}.
\newblock {\em Phys. Rev. Lett.}, 120(1):012302, 2018.
\newblock \href {http://arxiv.org/abs/1707.07984} {\path{arXiv:1707.07984}},
  \href {https://doi.org/10.1103/PhysRevLett.120.012302}
  {\path{doi:10.1103/PhysRevLett.120.012302}}.

\bibitem{Voloshin:2017kqp}
Sergei~A. Voloshin.
\newblock {Vorticity and particle polarization in heavy ion collisions
  (experimental perspective)}.
\newblock {\em EPJ Web Conf.}, 171:07002, 2018.
\newblock \href {http://arxiv.org/abs/1710.08934} {\path{arXiv:1710.08934}},
  \href {https://doi.org/10.1051/epjconf/201817107002}
  {\path{doi:10.1051/epjconf/201817107002}}.

\bibitem{STAR:2019erd}
Jaroslav Adam et~al.
\newblock {Polarization of $\Lambda$ ($\bar{\Lambda}$) hyperons along the beam
  direction in Au+Au collisions at $\sqrt{s_{_{NN}}}$ = 200 GeV}.
\newblock {\em Phys. Rev. Lett.}, 123(13):132301, 2019.
\newblock \href {http://arxiv.org/abs/1905.11917} {\path{arXiv:1905.11917}},
  \href {https://doi.org/10.1103/PhysRevLett.123.132301}
  {\path{doi:10.1103/PhysRevLett.123.132301}}.

\bibitem{ALICE:2021pzu}
Shreyasi Acharya et~al.
\newblock {Polarization of $\Lambda$ and $\bar \Lambda$ Hyperons along the Beam
  Direction in Pb-Pb Collisions at $\sqrt {s_{NN}}$=5.02\,\,TeV}.
\newblock {\em Phys. Rev. Lett.}, 128(17):172005, 2022.
\newblock \href {http://arxiv.org/abs/2107.11183} {\path{arXiv:2107.11183}},
  \href {https://doi.org/10.1103/PhysRevLett.128.172005}
  {\path{doi:10.1103/PhysRevLett.128.172005}}.

\bibitem{Becattini:2021iol}
F.~Becattini, M.~Buzzegoli, G.~Inghirami, I.~Karpenko, and A.~Palermo.
\newblock {Local Polarization and Isothermal Local Equilibrium in Relativistic
  Heavy Ion Collisions}.
\newblock {\em Phys. Rev. Lett.}, 127(27):272302, 2021.
\newblock \href {http://arxiv.org/abs/2103.14621} {\path{arXiv:2103.14621}},
  \href {https://doi.org/10.1103/PhysRevLett.127.272302}
  {\path{doi:10.1103/PhysRevLett.127.272302}}.

\bibitem{Liu:2021uhn}
Shuai Y.~F. Liu and Yi~Yin.
\newblock {Spin polarization induced by the hydrodynamic gradients}.
\newblock {\em JHEP}, 07:188, 2021.
\newblock \href {http://arxiv.org/abs/2103.09200} {\path{arXiv:2103.09200}},
  \href {https://doi.org/10.1007/JHEP07(2021)188}
  {\path{doi:10.1007/JHEP07(2021)188}}.

\bibitem{Fu:2021pok}
Baochi Fu, Shuai Y.~F. Liu, Longgang Pang, Huichao Song, and Yi~Yin.
\newblock {Shear-Induced Spin Polarization in Heavy-Ion Collisions}.
\newblock {\em Phys. Rev. Lett.}, 127(14):142301, 2021.
\newblock \href {http://arxiv.org/abs/2103.10403} {\path{arXiv:2103.10403}},
  \href {https://doi.org/10.1103/PhysRevLett.127.142301}
  {\path{doi:10.1103/PhysRevLett.127.142301}}.

\bibitem{Alzhrani:2022dpi}
Sahr Alzhrani, Sangwook Ryu, and Chun Shen.
\newblock {\ensuremath{\Lambda} spin polarization in event-by-event
  relativistic heavy-ion collisions}.
\newblock {\em Phys. Rev. C}, 106(1):014905, 2022.
\newblock \href {http://arxiv.org/abs/2203.15718} {\path{arXiv:2203.15718}},
  \href {https://doi.org/10.1103/PhysRevC.106.014905}
  {\path{doi:10.1103/PhysRevC.106.014905}}.

\bibitem{Pohl:2010zza}
Randolf Pohl et~al.
\newblock {The size of the proton}.
\newblock {\em Nature}, 466:213--216, 2010.
\newblock \href {https://doi.org/10.1038/nature09250}
  {\path{doi:10.1038/nature09250}}.

\bibitem{Xiong:2019umf}
W.~Xiong et~al.
\newblock {A small proton charge radius from an electron\textendash{}proton
  scattering experiment}.
\newblock {\em Nature}, 575(7781):147--150, 2019.
\newblock \href {https://doi.org/10.1038/s41586-019-1721-2}
  {\path{doi:10.1038/s41586-019-1721-2}}.

\bibitem{Gao:2021sml}
Haiyan Gao and Marc Vanderhaeghen.
\newblock {The proton charge radius}.
\newblock {\em Rev. Mod. Phys.}, 94(1):015002, 2022.
\newblock \href {http://arxiv.org/abs/2105.00571} {\path{arXiv:2105.00571}},
  \href {https://doi.org/10.1103/RevModPhys.94.015002}
  {\path{doi:10.1103/RevModPhys.94.015002}}.

\bibitem{Christy:2021snt}
M.~E. Christy et~al.
\newblock {Form Factors and Two-Photon Exchange in High-Energy Elastic
  Electron-Proton Scattering}.
\newblock {\em Phys. Rev. Lett.}, 128(10):102002, 2022.
\newblock \href {http://arxiv.org/abs/2103.01842} {\path{arXiv:2103.01842}},
  \href {https://doi.org/10.1103/PhysRevLett.128.102002}
  {\path{doi:10.1103/PhysRevLett.128.102002}}.

\bibitem{OLYMPUS:2016gso}
B.~S. Henderson et~al.
\newblock {Hard Two-Photon Contribution to Elastic Lepton-Proton Scattering:
  Determined by the OLYMPUS Experiment}.
\newblock {\em Phys. Rev. Lett.}, 118(9):092501, 2017.
\newblock \href {http://arxiv.org/abs/1611.04685} {\path{arXiv:1611.04685}},
  \href {https://doi.org/10.1103/PhysRevLett.118.092501}
  {\path{doi:10.1103/PhysRevLett.118.092501}}.

\bibitem{CLAS:2016fvy}
D.~Rimal et~al.
\newblock {Measurement of two-photon exchange effect by comparing elastic
  $e^\pm p$ cross sections}.
\newblock {\em Phys. Rev. C}, 95(6):065201, 2017.
\newblock \href {http://arxiv.org/abs/1603.00315} {\path{arXiv:1603.00315}},
  \href {https://doi.org/10.1103/PhysRevC.95.065201}
  {\path{doi:10.1103/PhysRevC.95.065201}}.

\bibitem{Rachek:2014fam}
I.~A. Rachek et~al.
\newblock {Measurement of the two-photon exchange contribution to the elastic
  $e^{\pm}p$ scattering cross sections at the VEPP-3 storage ring}.
\newblock {\em Phys. Rev. Lett.}, 114(6):062005, 2015.
\newblock \href {http://arxiv.org/abs/1411.7372} {\path{arXiv:1411.7372}},
  \href {https://doi.org/10.1103/PhysRevLett.114.062005}
  {\path{doi:10.1103/PhysRevLett.114.062005}}.

\bibitem{JeffersonLabHallA:1999epl}
M.~K. Jones et~al.
\newblock {G(E(p)) / G(M(p)) ratio by polarization transfer in polarized e p
  ---\ensuremath{>} e polarized p}.
\newblock {\em Phys. Rev. Lett.}, 84:1398--1402, 2000.
\newblock \href {http://arxiv.org/abs/nucl-ex/9910005}
  {\path{arXiv:nucl-ex/9910005}}, \href
  {https://doi.org/10.1103/PhysRevLett.84.1398}
  {\path{doi:10.1103/PhysRevLett.84.1398}}.

\bibitem{JeffersonLabHallA:2001qqe}
O.~Gayou et~al.
\newblock {Measurement of G(Ep) / G(Mp) in polarized-e p ---\ensuremath{>} e
  polarized-p to Q**2 = 5.6-GeV**2}.
\newblock {\em Phys. Rev. Lett.}, 88:092301, 2002.
\newblock \href {http://arxiv.org/abs/nucl-ex/0111010}
  {\path{arXiv:nucl-ex/0111010}}, \href
  {https://doi.org/10.1103/PhysRevLett.88.092301}
  {\path{doi:10.1103/PhysRevLett.88.092301}}.

\bibitem{CLAS:2014xso}
D.~Adikaram et~al.
\newblock {Towards a resolution of the proton form factor problem: new electron
  and positron scattering data}.
\newblock {\em Phys. Rev. Lett.}, 114:062003, 2015.
\newblock \href {http://arxiv.org/abs/1411.6908} {\path{arXiv:1411.6908}},
  \href {https://doi.org/10.1103/PhysRevLett.114.062003}
  {\path{doi:10.1103/PhysRevLett.114.062003}}.

\bibitem{Drechsel:2002ar}
D.~Drechsel, B.~Pasquini, and M.~Vanderhaeghen.
\newblock {Dispersion relations in real and virtual Compton scattering}.
\newblock {\em Phys. Rept.}, 378:99--205, 2003.
\newblock \href {http://arxiv.org/abs/hep-ph/0212124}
  {\path{arXiv:hep-ph/0212124}}, \href
  {https://doi.org/10.1016/S0370-1573(02)00636-1}
  {\path{doi:10.1016/S0370-1573(02)00636-1}}.

\bibitem{Pasquini:2018wbl}
Barbara Pasquini and Marc Vanderhaeghen.
\newblock {Dispersion Theory in Electromagnetic Interactions}.
\newblock {\em Ann. Rev. Nucl. Part. Sci.}, 68:75--103, 2018.
\newblock \href {http://arxiv.org/abs/1805.10482} {\path{arXiv:1805.10482}},
  \href {https://doi.org/10.1146/annurev-nucl-101917-020843}
  {\path{doi:10.1146/annurev-nucl-101917-020843}}.

\bibitem{Griesshammer:2012we}
H.~W. Griesshammer, J.~A. McGovern, D.~R. Phillips, and G.~Feldman.
\newblock {Using effective field theory to analyse low-energy Compton
  scattering data from protons and light nuclei}.
\newblock {\em Prog. Part. Nucl. Phys.}, 67:841--897, 2012.
\newblock \href {http://arxiv.org/abs/1203.6834} {\path{arXiv:1203.6834}},
  \href {https://doi.org/10.1016/j.ppnp.2012.04.003}
  {\path{doi:10.1016/j.ppnp.2012.04.003}}.

\bibitem{Howell:2020nob}
C.~R. Howell et~al.
\newblock {International workshop on next generation gamma-ray source}.
\newblock {\em J. Phys. G}, 49(1):010502, 2022.
\newblock \href {http://arxiv.org/abs/2012.10843} {\path{arXiv:2012.10843}},
  \href {https://doi.org/10.1088/1361-6471/ac2827}
  {\path{doi:10.1088/1361-6471/ac2827}}.

\bibitem{Hagelstein:2015egb}
Franziska Hagelstein, Rory Miskimen, and Vladimir Pascalutsa.
\newblock {Nucleon Polarizabilities: from Compton Scattering to Hydrogen Atom}.
\newblock {\em Prog. Part. Nucl. Phys.}, 88:29--97, 2016.
\newblock \href {http://arxiv.org/abs/1512.03765} {\path{arXiv:1512.03765}},
  \href {https://doi.org/10.1016/j.ppnp.2015.12.001}
  {\path{doi:10.1016/j.ppnp.2015.12.001}}.

\bibitem{Mornacchi:2022cln}
E.~Mornacchi, S.~Rodini, B.~Pasquini, and P.~Pedroni.
\newblock {First Concurrent Extraction of the Leading-Order Scalar and Spin
  Proton Polarizabilities}.
\newblock {\em Phys. Rev. Lett.}, 129(10):102501, 2022.
\newblock \href {http://arxiv.org/abs/2204.13491} {\path{arXiv:2204.13491}},
  \href {https://doi.org/10.1103/PhysRevLett.129.102501}
  {\path{doi:10.1103/PhysRevLett.129.102501}}.

\bibitem{Pasquini:2017ehj}
B.~Pasquini, P.~Pedroni, and S.~Sconfietti.
\newblock {First extraction of the scalar proton dynamical polarizabilities
  from real Compton scattering data}.
\newblock {\em Phys. Rev. C}, 98(1):015204, 2018.
\newblock \href {http://arxiv.org/abs/1711.07401} {\path{arXiv:1711.07401}},
  \href {https://doi.org/10.1103/PhysRevC.98.015204}
  {\path{doi:10.1103/PhysRevC.98.015204}}.

\bibitem{Pasquini:2019nnx}
B.~Pasquini, P.~Pedroni, and S.~Sconfietti.
\newblock {Proton scalar dipole polarizabilities from real Compton scattering
  data, using fixed-t subtracted dispersion relations and the bootstrap
  method}.
\newblock {\em J. Phys. G}, 46(10):104001, 2019.
\newblock \href {http://arxiv.org/abs/1903.07952} {\path{arXiv:1903.07952}},
  \href {https://doi.org/10.1088/1361-6471/ab323a}
  {\path{doi:10.1088/1361-6471/ab323a}}.

\bibitem{Margaryan:2018opu}
Arman Margaryan, Bruno Strandberg, Harald~W. Griesshammer, Judith~A. Mcgovern,
  Daniel~R. Phillips, and Deepshikha Shukla.
\newblock {Elastic Compton scattering from$^{3}$He and the role of the Delta}.
\newblock {\em Eur. Phys. J. A}, 54(7):125, 2018.
\newblock \href {http://arxiv.org/abs/1804.00956} {\path{arXiv:1804.00956}},
  \href {https://doi.org/10.1140/epja/i2018-12554-x}
  {\path{doi:10.1140/epja/i2018-12554-x}}.

\bibitem{A2:2014iky}
P.~P. Martel et~al.
\newblock {Measurements of Double-Polarized Compton Scattering Asymmetries and
  Extraction of the Proton Spin Polarizabilities}.
\newblock {\em Phys. Rev. Lett.}, 114(11):112501, 2015.
\newblock \href {http://arxiv.org/abs/1408.1576} {\path{arXiv:1408.1576}},
  \href {https://doi.org/10.1103/PhysRevLett.114.112501}
  {\path{doi:10.1103/PhysRevLett.114.112501}}.

\bibitem{A2:2019bqm}
D.~Paudyal et~al.
\newblock {Extracting the spin polarizabilities of the proton by measurement of
  Compton double-polarization observables}.
\newblock {\em Phys. Rev. C}, 102(3):035205, 2020.
\newblock \href {http://arxiv.org/abs/1909.02032} {\path{arXiv:1909.02032}},
  \href {https://doi.org/10.1103/PhysRevC.102.035205}
  {\path{doi:10.1103/PhysRevC.102.035205}}.

\bibitem{A2CollaborationatMAMI:2021vfy}
E.~Mornacchi et~al.
\newblock {Measurement of Compton Scattering at MAMI for the Extraction of the
  Electric and Magnetic Polarizabilities of the Proton}.
\newblock {\em Phys. Rev. Lett.}, 128(13):132503, 2022.
\newblock \href {http://arxiv.org/abs/2110.15691} {\path{arXiv:2110.15691}},
  \href {https://doi.org/10.1103/PhysRevLett.128.132503}
  {\path{doi:10.1103/PhysRevLett.128.132503}}.

\bibitem{Li:2022vnz}
X.~Li et~al.
\newblock {Proton Compton Scattering from Linearly Polarized Gamma Rays}.
\newblock {\em Phys. Rev. Lett.}, 128:132502, 2022.
\newblock \href {http://arxiv.org/abs/2205.10533} {\path{arXiv:2205.10533}},
  \href {https://doi.org/10.1103/PhysRevLett.128.132502}
  {\path{doi:10.1103/PhysRevLett.128.132502}}.

\bibitem{Li:2019irp}
X.~Li et~al.
\newblock {Compton scattering from $^4$He at the TUNL HI$\gamma$S facility}.
\newblock {\em Phys. Rev. C}, 101(3):034618, 2020.
\newblock \href {http://arxiv.org/abs/1912.06915} {\path{arXiv:1912.06915}},
  \href {https://doi.org/10.1103/PhysRevC.101.034618}
  {\path{doi:10.1103/PhysRevC.101.034618}}.

\bibitem{Sikora:2017rfk}
M.~H. Sikora et~al.
\newblock {Compton scattering from $^4$He at 61 MeV}.
\newblock {\em Phys. Rev. C}, 96(5):055209, 2017.
\newblock \href {https://doi.org/10.1103/PhysRevC.96.055209}
  {\path{doi:10.1103/PhysRevC.96.055209}}.

\bibitem{Fonvieille:2019eyf}
H.~Fonvieille, B.~Pasquini, and N.~Sparveris.
\newblock {Virtual Compton Scattering and Nucleon Generalized
  Polarizabilities}.
\newblock {\em Prog. Part. Nucl. Phys.}, 113:103754, 2020.
\newblock \href {http://arxiv.org/abs/1910.11071} {\path{arXiv:1910.11071}},
  \href {https://doi.org/10.1016/j.ppnp.2020.103754}
  {\path{doi:10.1016/j.ppnp.2020.103754}}.

\bibitem{VCS:2000ldk}
J.~Roche et~al.
\newblock {The First determination of generalized polarizabilities of the
  proton by a virtual Compton scattering experiment}.
\newblock {\em Phys. Rev. Lett.}, 85:708, 2000.
\newblock \href {http://arxiv.org/abs/hep-ex/0007053}
  {\path{arXiv:hep-ex/0007053}}, \href
  {https://doi.org/10.1103/PhysRevLett.85.708}
  {\path{doi:10.1103/PhysRevLett.85.708}}.

\bibitem{A1:2008rjb}
P.~Janssens et~al.
\newblock {A New measurement of the structure functions P(LL) - P(TT)/ epsilon
  and P(LT) in virtual Compton scattering at Q**2 = 0.33 (GeV/c)**2}.
\newblock {\em Eur. Phys. J. A}, 37:1, 2008.
\newblock \href {http://arxiv.org/abs/0803.0911} {\path{arXiv:0803.0911}},
  \href {https://doi.org/10.1140/epja/i2008-10609-3}
  {\path{doi:10.1140/epja/i2008-10609-3}}.

\bibitem{A1:2019mrv}
J.~Beri\v{c}i\v{c} et~al.
\newblock {New Insight in the $Q^2$-Dependence of Proton Generalized
  Polarizabilities}.
\newblock {\em Phys. Rev. Lett.}, 123(19):192302, 2019.
\newblock \href {http://arxiv.org/abs/1907.09954} {\path{arXiv:1907.09954}},
  \href {https://doi.org/10.1103/PhysRevLett.123.192302}
  {\path{doi:10.1103/PhysRevLett.123.192302}}.

\bibitem{A1:2020nof}
H.~Fonvieille et~al.
\newblock {Measurement of the generalized polarizabilities of the proton at
  intermediate $Q^2$}.
\newblock {\em Phys. Rev. C}, 103(2):025205, 2021.
\newblock \href {http://arxiv.org/abs/2008.08958} {\path{arXiv:2008.08958}},
  \href {https://doi.org/10.1103/PhysRevC.103.025205}
  {\path{doi:10.1103/PhysRevC.103.025205}}.

\bibitem{Blomberg:2019caf}
A.~Blomberg et~al.
\newblock {Virtual Compton Scattering measurements in the nucleon resonance
  region}.
\newblock {\em Eur. Phys. J. A}, 55(10):182, 2019.
\newblock \href {http://arxiv.org/abs/1901.08951} {\path{arXiv:1901.08951}},
  \href {https://doi.org/10.1140/epja/i2019-12877-0}
  {\path{doi:10.1140/epja/i2019-12877-0}}.

\bibitem{Li:2022sqg}
R.~Li et~al.
\newblock {Measured proton electromagnetic structure deviates from theoretical
  predictions}.
\newblock {\em Nature}, 611(7935):265--270, 2022.
\newblock \href {http://arxiv.org/abs/2210.11461} {\path{arXiv:2210.11461}},
  \href {https://doi.org/10.1038/s41586-022-05248-1}
  {\path{doi:10.1038/s41586-022-05248-1}}.

\bibitem{JeffersonLabHallA:2004vsy}
G.~Laveissiere et~al.
\newblock {Measurement of the generalized polarizabilities of the proton in
  virtual Compton scattering at Q**2 = 0.92-GeV**2 and 1.76-GeV**2}.
\newblock {\em Phys. Rev. Lett.}, 93:122001, 2004.
\newblock \href {http://arxiv.org/abs/hep-ph/0404243}
  {\path{arXiv:hep-ph/0404243}}, \href
  {https://doi.org/10.1103/PhysRevLett.93.122001}
  {\path{doi:10.1103/PhysRevLett.93.122001}}.

\bibitem{JeffersonLabHallA:2012zwt}
H.~Fonvieille et~al.
\newblock {Virtual Compton Scattering and the Generalized Polarizabilities of
  the Proton at $Q^2=0.92$ and 1.76 GeV$^2$}.
\newblock {\em Phys. Rev. C}, 86:015210, 2012.
\newblock \href {http://arxiv.org/abs/1205.3387} {\path{arXiv:1205.3387}},
  \href {https://doi.org/10.1103/PhysRevC.86.015210}
  {\path{doi:10.1103/PhysRevC.86.015210}}.

\bibitem{Bourgeois:2006js}
P.~Bourgeois et~al.
\newblock {Measurements of the generalized electric and magnetic
  polarizabilities of the proton at low Q**2 using the VCS reaction}.
\newblock {\em Phys. Rev. Lett.}, 97:212001, 2006.
\newblock \href {http://arxiv.org/abs/nucl-ex/0605009}
  {\path{arXiv:nucl-ex/0605009}}, \href
  {https://doi.org/10.1103/PhysRevLett.97.212001}
  {\path{doi:10.1103/PhysRevLett.97.212001}}.

\bibitem{Bourgeois:2011zz}
P.~Bourgeois et~al.
\newblock {Measurements of the generalized electric and magnetic
  polarizabilities of the proton at low Q-2 using the virtual Compton
  scattering reaction}.
\newblock {\em Phys. Rev. C}, 84:035206, 2011.
\newblock \href {https://doi.org/10.1103/PhysRevC.84.035206}
  {\path{doi:10.1103/PhysRevC.84.035206}}.

\bibitem{Lensky:2016nui}
Vadim Lensky, Vladimir Pascalutsa, and Marc Vanderhaeghen.
\newblock {Generalized polarizabilities of the nucleon in baryon chiral
  perturbation theory}.
\newblock {\em Eur. Phys. J. C}, 77(2):119, 2017.
\newblock \href {http://arxiv.org/abs/1612.08626} {\path{arXiv:1612.08626}},
  \href {https://doi.org/10.1140/epjc/s10052-017-4652-9}
  {\path{doi:10.1140/epjc/s10052-017-4652-9}}.

\bibitem{Pasquini:2000ue}
B.~Pasquini, S.~Scherer, and D.~Drechsel.
\newblock {Generalized polarizabilities of the proton in a constituent quark
  model revisited}.
\newblock {\em Phys. Rev. C}, 63:025205, 2001.
\newblock \href {http://arxiv.org/abs/nucl-th/0008046}
  {\path{arXiv:nucl-th/0008046}}, \href
  {https://doi.org/10.1103/PhysRevC.63.025205}
  {\path{doi:10.1103/PhysRevC.63.025205}}.

\bibitem{Metz:1996fn}
A.~Metz and D.~Drechsel.
\newblock {Generalized polarizabilities of the nucleon studied in the linear
  sigma model}.
\newblock {\em Z. Phys. A}, 356:351--357, 1996.
\newblock \href {https://doi.org/10.1007/s002180050188}
  {\path{doi:10.1007/s002180050188}}.

\bibitem{Korchin:1998cx}
A.~Yu. Korchin and O.~Scholten.
\newblock {Nucleon polarizabilities in virtual Compton scattering}.
\newblock {\em Phys. Rev. C}, 58:1098--1100, 1998.
\newblock \href {https://doi.org/10.1103/PhysRevC.58.1098}
  {\path{doi:10.1103/PhysRevC.58.1098}}.

\bibitem{Pasquini:2000pk}
B.~Pasquini, D.~Drechsel, M.~Gorchtein, A.~Metz, and M.~Vanderhaeghen.
\newblock {Dispersion relation formalism for virtual Compton scattering and the
  generalized polarizabilities of the nucleon}.
\newblock {\em Phys. Rev. C}, 62:052201, 2000.
\newblock \href {http://arxiv.org/abs/hep-ph/0007144}
  {\path{arXiv:hep-ph/0007144}}, \href
  {https://doi.org/10.1103/PhysRevC.62.052201}
  {\path{doi:10.1103/PhysRevC.62.052201}}.

\bibitem{Pasquini:2001yy}
B.~Pasquini, M.~Gorchtein, D.~Drechsel, A.~Metz, and M.~Vanderhaeghen.
\newblock {Dispersion relation formalism for virtual Compton scattering off the
  proton}.
\newblock {\em Eur. Phys. J. A}, 11:185--208, 2001.
\newblock \href {http://arxiv.org/abs/hep-ph/0102335}
  {\path{arXiv:hep-ph/0102335}}, \href {https://doi.org/10.1007/s100500170084}
  {\path{doi:10.1007/s100500170084}}.

\bibitem{E97-110:2021mxm}
Vincent Sulkosky et~al.
\newblock {Measurement of the generalized spin polarizabilities of the neutron
  in the low-$Q^2$ region}.
\newblock {\em Nature Phys.}, 17(6):687--692, 2021.
\newblock [Erratum: Nature Phys. 18, (2022)].
\newblock \href {http://arxiv.org/abs/2103.03333} {\path{arXiv:2103.03333}},
  \href {https://doi.org/10.1038/s41567-021-01245-9}
  {\path{doi:10.1038/s41567-021-01245-9}}.

\bibitem{Amarian:2004yf}
M.~Amarian et~al.
\newblock {Measurement of the generalized forward spin polarizabilities of the
  neutron}.
\newblock {\em Phys. Rev. Lett.}, 93:152301, 2004.
\newblock \href {http://arxiv.org/abs/nucl-ex/0406005}
  {\path{arXiv:nucl-ex/0406005}}, \href
  {https://doi.org/10.1103/PhysRevLett.93.152301}
  {\path{doi:10.1103/PhysRevLett.93.152301}}.

\bibitem{Bernard:2002pw}
Veronique Bernard, Thomas~R. Hemmert, and Ulf-G. Meissner.
\newblock {Spin structure of the nucleon at low-energies}.
\newblock {\em Phys. Rev. D}, 67:076008, 2003.
\newblock \href {http://arxiv.org/abs/hep-ph/0212033}
  {\path{arXiv:hep-ph/0212033}}, \href
  {https://doi.org/10.1103/PhysRevD.67.076008}
  {\path{doi:10.1103/PhysRevD.67.076008}}.

\bibitem{Kao:2002cp}
Chung~Wen Kao, Thomas Spitzenberg, and Marc Vanderhaeghen.
\newblock {Burkhardt-Cottingham sum rule and forward spin polarizabilities in
  heavy baryon chiral perturbation theory}.
\newblock {\em Phys. Rev. D}, 67:016001, 2003.
\newblock \href {http://arxiv.org/abs/hep-ph/0209241}
  {\path{arXiv:hep-ph/0209241}}, \href
  {https://doi.org/10.1103/PhysRevD.67.016001}
  {\path{doi:10.1103/PhysRevD.67.016001}}.

\bibitem{Bernard:2012hb}
Veronique Bernard, Evgeny Epelbaum, Hermann Krebs, and Ulf-G. Meissner.
\newblock {New insights into the spin structure of the nucleon}.
\newblock {\em Phys. Rev. D}, 87(5):054032, 2013.
\newblock \href {http://arxiv.org/abs/1209.2523} {\path{arXiv:1209.2523}},
  \href {https://doi.org/10.1103/PhysRevD.87.054032}
  {\path{doi:10.1103/PhysRevD.87.054032}}.

\bibitem{Alarcon:2020icz}
Jose~Manuel Alarc\'on, Franziska Hagelstein, Vadim Lensky, and Vladimir
  Pascalutsa.
\newblock {Forward doubly-virtual Compton scattering off the nucleon in chiral
  perturbation theory: II. Spin polarizabilities and moments of polarized
  structure functions}.
\newblock {\em Phys. Rev. D}, 102(11):114026, 2020.
\newblock \href {http://arxiv.org/abs/2006.08626} {\path{arXiv:2006.08626}},
  \href {https://doi.org/10.1103/PhysRevD.102.114026}
  {\path{doi:10.1103/PhysRevD.102.114026}}.

\bibitem{Drechsel:1998hk}
D.~Drechsel, O.~Hanstein, S.~S. Kamalov, and L.~Tiator.
\newblock {A Unitary isobar model for pion photoproduction and
  electroproduction on the proton up to 1-GeV}.
\newblock {\em Nucl. Phys. A}, 645:145--174, 1999.
\newblock \href {http://arxiv.org/abs/nucl-th/9807001}
  {\path{arXiv:nucl-th/9807001}}, \href
  {https://doi.org/10.1016/S0375-9474(98)00572-7}
  {\path{doi:10.1016/S0375-9474(98)00572-7}}.

\bibitem{Zheng:2021yrn}
X.~Zheng et~al.
\newblock {Measurement of the proton spin structure at long distances}.
\newblock {\em Nature Phys.}, 17(6):736--741, 2021.
\newblock \href {http://arxiv.org/abs/2102.02658} {\path{arXiv:2102.02658}},
  \href {https://doi.org/10.1038/s41567-021-01198-z}
  {\path{doi:10.1038/s41567-021-01198-z}}.

\bibitem{Adhikari:2017wox}
K.~P. Adhikari et~al.
\newblock {Measurement of the ${Q}^{2}$ Dependence of the Deuteron Spin
  Structure Function ${g}_{1}$ and its Moments at Low ${Q}^{2}$ with CLAS}.
\newblock {\em Phys. Rev. Lett.}, 120(6):062501, 2018.
\newblock \href {http://arxiv.org/abs/1711.01974} {\path{arXiv:1711.01974}},
  \href {https://doi.org/10.1103/PhysRevLett.120.062501}
  {\path{doi:10.1103/PhysRevLett.120.062501}}.

\bibitem{Sulkosky:2021qmh}
Vincent Sulkosky et~al.
\newblock {Measurement of the generalized spin polarizabilities of the neutron
  in the low-$Q^2$ region}.
\newblock {\em Nature Phys.}, 17(6):687--692, 2021.
\newblock [Erratum: Nature Phys. 18, (2022)].
\newblock \href {http://arxiv.org/abs/2103.03333} {\path{arXiv:2103.03333}},
  \href {https://doi.org/10.1038/s41567-021-01245-9}
  {\path{doi:10.1038/s41567-021-01245-9}}.

\bibitem{JeffersonLabHallAg2p:2022qap}
D.~Ruth et~al.
\newblock {Proton spin structure and generalized polarizabilities in the strong
  quantum chromodynamics regime}.
\newblock {\em Nature Phys.}, 18(12):1441--1446, 2022.
\newblock \href {http://arxiv.org/abs/2204.10224} {\path{arXiv:2204.10224}},
  \href {https://doi.org/10.1038/s41567-022-01781-y}
  {\path{doi:10.1038/s41567-022-01781-y}}.

\bibitem{PrimEx-II:2020jwd}
I.~Larin et~al.
\newblock {Precision measurement of the neutral pion lifetime}.
\newblock {\em Science}, 368(6490):506--509, 2020.
\newblock \href {https://doi.org/10.1126/science.aay6641}
  {\path{doi:10.1126/science.aay6641}}.

\bibitem{Hou:2019efy}
Tie-Jiun Hou et~al.
\newblock {New CTEQ global analysis of quantum chromodynamics with
  high-precision data from the LHC}.
\newblock {\em Phys. Rev. D}, 103(1):014013, 2021.
\newblock \href {http://arxiv.org/abs/1912.10053} {\path{arXiv:1912.10053}},
  \href {https://doi.org/10.1103/PhysRevD.103.014013}
  {\path{doi:10.1103/PhysRevD.103.014013}}.

\bibitem{NNPDF:2017mvq}
Richard~D. Ball et~al.
\newblock {Parton distributions from high-precision collider data}.
\newblock {\em Eur. Phys. J. C}, 77(10):663, 2017.
\newblock \href {http://arxiv.org/abs/1706.00428} {\path{arXiv:1706.00428}},
  \href {https://doi.org/10.1140/epjc/s10052-017-5199-5}
  {\path{doi:10.1140/epjc/s10052-017-5199-5}}.

\bibitem{Alekhin:2017kpj}
S.~Alekhin, J.~Bl\"umlein, S.~Moch, and R.~Placakyte.
\newblock {Parton distribution functions, $\alpha_s$, and heavy-quark masses
  for LHC Run II}.
\newblock {\em Phys. Rev. D}, 96(1):014011, 2017.
\newblock \href {http://arxiv.org/abs/1701.05838} {\path{arXiv:1701.05838}},
  \href {https://doi.org/10.1103/PhysRevD.96.014011}
  {\path{doi:10.1103/PhysRevD.96.014011}}.

\bibitem{Cocuzza:2021rfn}
C.~Cocuzza, C.~E. Keppel, H.~Liu, W.~Melnitchouk, A.~Metz, N.~Sato, and A.~W.
  Thomas.
\newblock {Isovector EMC Effect from Global QCD Analysis with MARATHON Data}.
\newblock {\em Phys. Rev. Lett.}, 127(24):242001, 2021.
\newblock \href {http://arxiv.org/abs/2104.06946} {\path{arXiv:2104.06946}},
  \href {https://doi.org/10.1103/PhysRevLett.127.242001}
  {\path{doi:10.1103/PhysRevLett.127.242001}}.

\bibitem{JeffersonLabHallATritium:2021usd}
D.~Abrams et~al.
\newblock {Measurement of the Nucleon $F^n_2/F^p_2$ Structure Function Ratio by
  the Jefferson Lab MARATHON Tritium/Helium-3 Deep Inelastic Scattering
  Experiment}.
\newblock {\em Phys. Rev. Lett.}, 128(13):132003, 2022.
\newblock \href {http://arxiv.org/abs/2104.05850} {\path{arXiv:2104.05850}},
  \href {https://doi.org/10.1103/PhysRevLett.128.132003}
  {\path{doi:10.1103/PhysRevLett.128.132003}}.

\bibitem{SeaQuest:2021zxb}
J.~Dove et~al.
\newblock {The asymmetry of antimatter in the proton}.
\newblock {\em Nature}, 590(7847):561--565, 2021.
\newblock [Erratum: Nature 604, E26 (2022)].
\newblock \href {http://arxiv.org/abs/2103.04024} {\path{arXiv:2103.04024}},
  \href {https://doi.org/10.1038/s41586-022-04707-z}
  {\path{doi:10.1038/s41586-022-04707-z}}.

\bibitem{NuSea:2001idv}
R.~S. Towell et~al.
\newblock {Improved measurement of the anti-d / anti-u asymmetry in the nucleon
  sea}.
\newblock {\em Phys. Rev. D}, 64:052002, 2001.
\newblock \href {http://arxiv.org/abs/hep-ex/0103030}
  {\path{arXiv:hep-ex/0103030}}, \href
  {https://doi.org/10.1103/PhysRevD.64.052002}
  {\path{doi:10.1103/PhysRevD.64.052002}}.

\bibitem{Cui:2021gzg}
Zhu-Fang Cui, Fei Gao, Daniele Binosi, Lei Chang, Craig~D. Roberts, and
  Sebastian~M. Schmidt.
\newblock {Valence Quark Ratio in the Proton}.
\newblock {\em Chin. Phys. Lett.}, 39(4):041401, 2022.
\newblock \href {http://arxiv.org/abs/2108.11493} {\path{arXiv:2108.11493}},
  \href {https://doi.org/10.1088/0256-307X/39/4/041401}
  {\path{doi:10.1088/0256-307X/39/4/041401}}.

\bibitem{Alekhin:2022tip}
S.~I. Alekhin, S.~A. Kulagin, and R.~Petti.
\newblock {Nuclear effects in the deuteron and global QCD analyses}.
\newblock {\em Phys. Rev. D}, 105(11):114037, 2022.
\newblock \href {http://arxiv.org/abs/2203.07333} {\path{arXiv:2203.07333}},
  \href {https://doi.org/10.1103/PhysRevD.105.114037}
  {\path{doi:10.1103/PhysRevD.105.114037}}.

\bibitem{JLabPR:BONUS12}
S.~Kuhn~(contact), S.~Bueltmann, M.~Christy, K.~Griffioen, M.~Hattawy,
  C.~Keppel, W.~Melnitchouk, et~al.
\newblock {The Structure of the Free Neutron at Large x-Bjorken}.
\newblock Jefferson Lab Experiment E12-06-113, 2006 with ? update.

\bibitem{STAR:2020vuq}
Jaroslav Adam et~al.
\newblock {Measurements of $W$ and $Z/\gamma^*$ cross sections and their ratios
  in p+p collisions at RHIC}.
\newblock {\em Phys. Rev. D}, 103(1):012001, 2021.
\newblock \href {http://arxiv.org/abs/2011.04708} {\path{arXiv:2011.04708}},
  \href {https://doi.org/10.1103/PhysRevD.103.012001}
  {\path{doi:10.1103/PhysRevD.103.012001}}.

\bibitem{DeFlorian:2019xxt}
Daniel De~Florian, Gonzalo~Agust\'\i{}n Lucero, Rodolfo Sassot, Marco
  Stratmann, and Werner Vogelsang.
\newblock {Monte Carlo sampling variant of the DSSV14 set of helicity parton
  densities}.
\newblock {\em Phys. Rev. D}, 100(11):114027, 2019.
\newblock \href {http://arxiv.org/abs/1902.10548} {\path{arXiv:1902.10548}},
  \href {https://doi.org/10.1103/PhysRevD.100.114027}
  {\path{doi:10.1103/PhysRevD.100.114027}}.

\bibitem{Nocera:2014gqa}
Emanuele~R. Nocera, Richard~D. Ball, Stefano Forte, Giovanni Ridolfi, and Juan
  Rojo.
\newblock {A first unbiased global determination of polarized PDFs and their
  uncertainties}.
\newblock {\em Nucl. Phys. B}, 887:276--308, 2014.
\newblock \href {http://arxiv.org/abs/1406.5539} {\path{arXiv:1406.5539}},
  \href {https://doi.org/10.1016/j.nuclphysb.2014.08.008}
  {\path{doi:10.1016/j.nuclphysb.2014.08.008}}.

\bibitem{Cocuzza:2022jye}
C.~Cocuzza, W.~Melnitchouk, A.~Metz, and N.~Sato.
\newblock {Polarized antimatter in the proton from a global QCD analysis}.
\newblock {\em Phys. Rev. D}, 106(3):L031502, 2022.
\newblock \href {http://arxiv.org/abs/2202.03372} {\path{arXiv:2202.03372}},
  \href {https://doi.org/10.1103/PhysRevD.106.L031502}
  {\path{doi:10.1103/PhysRevD.106.L031502}}.

\bibitem{RHIC-Cold-QCD}


\bibitem{deFlorian:2014yva}
Daniel de~Florian, Rodolfo Sassot, Marco Stratmann, and Werner Vogelsang.
\newblock {Evidence for polarization of gluons in the proton}.
\newblock {\em Phys. Rev. Lett.}, 113(1):012001, 2014.
\newblock \href {http://arxiv.org/abs/1404.4293} {\path{arXiv:1404.4293}},
  \href {https://doi.org/10.1103/PhysRevLett.113.012001}
  {\path{doi:10.1103/PhysRevLett.113.012001}}.

\bibitem{Adam:2018bam}
Jaroslav Adam et~al.
\newblock {Measurement of the longitudinal spin asymmetries for weak boson
  production in proton-proton collisions at $\sqrt{s}$ = 510 GeV}.
\newblock {\em Phys. Rev. D}, 99(5):051102, 2019.
\newblock \href {http://arxiv.org/abs/1812.04817} {\path{arXiv:1812.04817}},
  \href {https://doi.org/10.1103/PhysRevD.99.051102}
  {\path{doi:10.1103/PhysRevD.99.051102}}.

\bibitem{Adamczyk:2014ozi}
L.~Adamczyk et~al.
\newblock {Precision Measurement of the Longitudinal Double-spin Asymmetry for
  Inclusive Jet Production in Polarized Proton Collisions at $\sqrt{s}=200$
  GeV}.
\newblock {\em Phys. Rev. Lett.}, 115(9):092002, 2015.
\newblock \href {http://arxiv.org/abs/1405.5134} {\path{arXiv:1405.5134}},
  \href {https://doi.org/10.1103/PhysRevLett.115.092002}
  {\path{doi:10.1103/PhysRevLett.115.092002}}.

\bibitem{Adam:2019aml}
J.~Adam et~al.
\newblock {Longitudinal double-spin asymmetry for inclusive jet and dijet
  production in pp collisions at $\sqrt{s} = 510$ GeV}.
\newblock {\em Phys. Rev. D}, 100(5):052005, 2019.
\newblock \href {http://arxiv.org/abs/1906.02740} {\path{arXiv:1906.02740}},
  \href {https://doi.org/10.1103/PhysRevD.100.052005}
  {\path{doi:10.1103/PhysRevD.100.052005}}.

\bibitem{STAR:2021mfd}
M.~S. Abdallah et~al.
\newblock {Longitudinal double-spin asymmetry for inclusive jet and dijet
  production in polarized proton collisions at $\sqrt{s}=200$ GeV}.
\newblock {\em Phys. Rev. D}, 103(9):L091103, 2021.
\newblock \href {http://arxiv.org/abs/2103.05571} {\path{arXiv:2103.05571}},
  \href {https://doi.org/10.1103/PhysRevD.103.L091103}
  {\path{doi:10.1103/PhysRevD.103.L091103}}.

\bibitem{STAR:2021mqa}
M.~S. Abdallah et~al.
\newblock {Longitudinal double-spin asymmetry for inclusive jet and dijet
  production in polarized proton collisions at $\sqrt{s}=510$ GeV}.
\newblock {\em Phys. Rev. D}, 105(9):092011, 2022.
\newblock \href {http://arxiv.org/abs/2110.11020} {\path{arXiv:2110.11020}},
  \href {https://doi.org/10.1103/PhysRevD.105.092011}
  {\path{doi:10.1103/PhysRevD.105.092011}}.

\bibitem{PHENIX:2015ade}
A.~Adare et~al.
\newblock {Measurement of parity-violating spin asymmetries in W$^{\pm}$
  production at midrapidity in longitudinally polarized $p+p$ collisions}.
\newblock {\em Phys. Rev. D}, 93(5):051103, 2016.
\newblock \href {http://arxiv.org/abs/1504.07451} {\path{arXiv:1504.07451}},
  \href {https://doi.org/10.1103/PhysRevD.93.051103}
  {\path{doi:10.1103/PhysRevD.93.051103}}.

\bibitem{PHENIX:2018wuz}
A.~Adare et~al.
\newblock {Cross section and longitudinal single-spin asymmetry $A_L$ for
  forward $W^{\pm}\rightarrow\mu^{\pm}\nu$ production in polarized $p+p$
  collisions at $\sqrt{s}=510$ GeV}.
\newblock {\em Phys. Rev. D}, 98(3):032007, 2018.
\newblock \href {http://arxiv.org/abs/1804.04181} {\path{arXiv:1804.04181}},
  \href {https://doi.org/10.1103/PhysRevD.98.032007}
  {\path{doi:10.1103/PhysRevD.98.032007}}.

\bibitem{Bjorken:1966jh}
J.~D. Bjorken.
\newblock {Applications of the Chiral U(6) x (6) Algebra of Current Densities}.
\newblock {\em Phys. Rev.}, 148:1467--1478, 1966.
\newblock \href {https://doi.org/10.1103/PhysRev.148.1467}
  {\path{doi:10.1103/PhysRev.148.1467}}.

\bibitem{Deur:2018roz}
Alexandre Deur, Stanley~J. Brodsky, and Guy~F. De~T\'eramond.
\newblock {The Spin Structure of the Nucleon}.
\newblock 7 2018.
\newblock \href {http://arxiv.org/abs/1807.05250} {\path{arXiv:1807.05250}},
  \href {https://doi.org/10.1088/1361-6633/ab0b8f}
  {\path{doi:10.1088/1361-6633/ab0b8f}}.

\bibitem{Deur:2021klh}
A.~Deur et~al.
\newblock {Experimental study of the behavior of the Bjorken sum at very low
  Q2}.
\newblock {\em Phys. Lett. B}, 825:136878, 2022.
\newblock \href {http://arxiv.org/abs/2107.08133} {\path{arXiv:2107.08133}},
  \href {https://doi.org/10.1016/j.physletb.2022.136878}
  {\path{doi:10.1016/j.physletb.2022.136878}}.

\bibitem{Deur:2016tte}
Alexandre Deur, Stanley~J. Brodsky, and Guy~F. de~Teramond.
\newblock {The QCD Running Coupling}.
\newblock {\em Nucl. Phys.}, 90:1, 2016.
\newblock \href {http://arxiv.org/abs/1604.08082} {\path{arXiv:1604.08082}},
  \href {https://doi.org/10.1016/j.ppnp.2016.04.003}
  {\path{doi:10.1016/j.ppnp.2016.04.003}}.

\bibitem{Deur:2014vea}
A.~Deur, Y.~Prok, V.~Burkert, D.~Crabb, F.~X. Girod, K.~A. Griffioen, N.~Guler,
  S.~E. Kuhn, and N.~Kvaltine.
\newblock {High precision determination of the $Q^2$ evolution of the Bjorken
  Sum}.
\newblock {\em Phys. Rev. D}, 90(1):012009, 2014.
\newblock \href {http://arxiv.org/abs/1405.7854} {\path{arXiv:1405.7854}},
  \href {https://doi.org/10.1103/PhysRevD.90.012009}
  {\path{doi:10.1103/PhysRevD.90.012009}}.

\bibitem{dEnterria:2022hzv}
D.~d'Enterria et~al.
\newblock {The strong coupling constant: State of the art and the decade
  ahead}.
\newblock 3 2022.
\newblock \href {http://arxiv.org/abs/2203.08271} {\path{arXiv:2203.08271}}.

\bibitem{JeffersonLabHallA:2003joy}
X.~Zheng et~al.
\newblock {Precision measurement of the neutron spin asymmetry A1**N and spin
  flavor decomposition in the valence quark region}.
\newblock {\em Phys. Rev. Lett.}, 92:012004, 2004.
\newblock \href {http://arxiv.org/abs/nucl-ex/0308011}
  {\path{arXiv:nucl-ex/0308011}}, \href
  {https://doi.org/10.1103/PhysRevLett.92.012004}
  {\path{doi:10.1103/PhysRevLett.92.012004}}.

\bibitem{JeffersonLabHallA:2004tea}
X.~Zheng et~al.
\newblock {Precision measurement of the neutron spin asymmetries and
  spin-dependent structure functions in the valence quark region}.
\newblock {\em Phys. Rev. C}, 70:065207, 2004.
\newblock \href {http://arxiv.org/abs/nucl-ex/0405006}
  {\path{arXiv:nucl-ex/0405006}}, \href
  {https://doi.org/10.1103/PhysRevC.70.065207}
  {\path{doi:10.1103/PhysRevC.70.065207}}.

\bibitem{Avakian:2007xa}
Harut Avakian, Stanley~J. Brodsky, Alexandre Deur, and Feng Yuan.
\newblock {Effect of Orbital Angular Momentum on Valence-Quark Helicity
  Distributions}.
\newblock {\em Phys. Rev. Lett.}, 99:082001, 2007.
\newblock \href {http://arxiv.org/abs/0705.1553} {\path{arXiv:0705.1553}},
  \href {https://doi.org/10.1103/PhysRevLett.99.082001}
  {\path{doi:10.1103/PhysRevLett.99.082001}}.

\bibitem{Burkert:2018bqq}
V.~D. Burkert, L.~Elouadrhiri, and F.~X. Girod.
\newblock {The pressure distribution inside the proton}.
\newblock {\em Nature}, 557(7705):396--399, 2018.
\newblock \href {https://doi.org/10.1038/s41586-018-0060-z}
  {\path{doi:10.1038/s41586-018-0060-z}}.

\bibitem{Shanahan:2018nnv}
P.~E. Shanahan and W.~Detmold.
\newblock {Pressure Distribution and Shear Forces inside the Proton}.
\newblock {\em Phys. Rev. Lett.}, 122(7):072003, 2019.
\newblock \href {http://arxiv.org/abs/1810.07589} {\path{arXiv:1810.07589}},
  \href {https://doi.org/10.1103/PhysRevLett.122.072003}
  {\path{doi:10.1103/PhysRevLett.122.072003}}.

\bibitem{Ji:1996nm}
Xiang-Dong Ji.
\newblock {Deeply virtual Compton scattering}.
\newblock {\em Phys. Rev. D}, 55:7114--7125, 1997.
\newblock \href {http://arxiv.org/abs/hep-ph/9609381}
  {\path{arXiv:hep-ph/9609381}}, \href
  {https://doi.org/10.1103/PhysRevD.55.7114}
  {\path{doi:10.1103/PhysRevD.55.7114}}.

\bibitem{Dupre:2017hfs}
Rapha\"el Dupr\'e, Michel Guidal, Silvia Niccolai, and Marc Vanderhaeghen.
\newblock {Analysis of Deeply Virtual Compton Scattering Data at Jefferson Lab
  and Proton Tomography}.
\newblock {\em Eur. Phys. J. A}, 53(8):171, 2017.
\newblock \href {http://arxiv.org/abs/1704.07330} {\path{arXiv:1704.07330}},
  \href {https://doi.org/10.1140/epja/i2017-12356-8}
  {\path{doi:10.1140/epja/i2017-12356-8}}.

\bibitem{Polyakov:2002yz}
M.~V. Polyakov.
\newblock {Generalized parton distributions and strong forces inside nucleons
  and nuclei}.
\newblock {\em Phys. Lett. B}, 555:57--62, 2003.
\newblock \href {http://arxiv.org/abs/hep-ph/0210165}
  {\path{arXiv:hep-ph/0210165}}, \href
  {https://doi.org/10.1016/S0370-2693(03)00036-4}
  {\path{doi:10.1016/S0370-2693(03)00036-4}}.

\bibitem{Polyakov:2018zvc}
Maxim~V. Polyakov and Peter Schweitzer.
\newblock {Forces inside hadrons: pressure, surface tension, mechanical radius,
  and all that}.
\newblock {\em Int. J. Mod. Phys. A}, 33(26):1830025, 2018.
\newblock \href {http://arxiv.org/abs/1805.06596} {\path{arXiv:1805.06596}},
  \href {https://doi.org/10.1142/S0217751X18300259}
  {\path{doi:10.1142/S0217751X18300259}}.

\bibitem{CLAS:2007clm}
F.~X. Girod et~al.
\newblock {Measurement of Deeply virtual Compton scattering beam-spin
  asymmetries}.
\newblock {\em Phys. Rev. Lett.}, 100:162002, 2008.
\newblock \href {http://arxiv.org/abs/0711.4805} {\path{arXiv:0711.4805}},
  \href {https://doi.org/10.1103/PhysRevLett.100.162002}
  {\path{doi:10.1103/PhysRevLett.100.162002}}.

\bibitem{CLAS:2015bqi}
S.~Pisano et~al.
\newblock {Single and double spin asymmetries for deeply virtual Compton
  scattering measured with CLAS and a longitudinally polarized proton target}.
\newblock {\em Phys. Rev. D}, 91(5):052014, 2015.
\newblock \href {http://arxiv.org/abs/1501.07052} {\path{arXiv:1501.07052}},
  \href {https://doi.org/10.1103/PhysRevD.91.052014}
  {\path{doi:10.1103/PhysRevD.91.052014}}.

\bibitem{Shanahan:2018pib}
P.~E. Shanahan and W.~Detmold.
\newblock {Gluon gravitational form factors of the nucleon and the pion from
  lattice QCD}.
\newblock {\em Phys. Rev. D}, 99(1):014511, 2019.
\newblock \href {http://arxiv.org/abs/1810.04626} {\path{arXiv:1810.04626}},
  \href {https://doi.org/10.1103/PhysRevD.99.014511}
  {\path{doi:10.1103/PhysRevD.99.014511}}.

\bibitem{CLAS:2021lky}
P.~Chatagnon et~al.
\newblock {First Measurement of Timelike Compton Scattering}.
\newblock {\em Phys. Rev. Lett.}, 127(26):262501, 2021.
\newblock \href {http://arxiv.org/abs/2108.11746} {\path{arXiv:2108.11746}},
  \href {https://doi.org/10.1103/PhysRevLett.127.262501}
  {\path{doi:10.1103/PhysRevLett.127.262501}}.

\bibitem{COMPASS:2017jbv}
M.~Aghasyan et~al.
\newblock {First measurement of transverse-spin-dependent azimuthal asymmetries
  in the Drell-Yan process}.
\newblock {\em Phys. Rev. Lett.}, 119(11):112002, 2017.
\newblock \href {http://arxiv.org/abs/1704.00488} {\path{arXiv:1704.00488}},
  \href {https://doi.org/10.1103/PhysRevLett.119.112002}
  {\path{doi:10.1103/PhysRevLett.119.112002}}.

\bibitem{STAR:2015vmv}
L.~Adamczyk et~al.
\newblock {Measurement of the transverse single-spin asymmetry in $p^\uparrow+p
  \to W^{\pm}/Z^0$ at RHIC}.
\newblock {\em Phys. Rev. Lett.}, 116(13):132301, 2016.
\newblock \href {http://arxiv.org/abs/1511.06003} {\path{arXiv:1511.06003}},
  \href {https://doi.org/10.1103/PhysRevLett.116.132301}
  {\path{doi:10.1103/PhysRevLett.116.132301}}.

\bibitem{Cammarota:2020qcw}
Justin Cammarota, Leonard Gamberg, Zhong-Bo Kang, Joshua~A. Miller, Daniel
  Pitonyak, Alexei Prokudin, Ted~C. Rogers, and Nobuo Sato.
\newblock {Origin of single transverse-spin asymmetries in high-energy
  collisions}.
\newblock {\em Phys. Rev. D}, 102(5):054002, 2020.
\newblock \href {http://arxiv.org/abs/2002.08384} {\path{arXiv:2002.08384}},
  \href {https://doi.org/10.1103/PhysRevD.102.054002}
  {\path{doi:10.1103/PhysRevD.102.054002}}.

\bibitem{Bury:2020vhj}
Marcin Bury, Alexei Prokudin, and Alexey Vladimirov.
\newblock {Extraction of the Sivers Function from SIDIS, Drell-Yan, and
  $W^{\pm}/Z$ Data at Next-to-Next-to-Next-to Leading Order}.
\newblock {\em Phys. Rev. Lett.}, 126(11):112002, 2021.
\newblock \href {http://arxiv.org/abs/2012.05135} {\path{arXiv:2012.05135}},
  \href {https://doi.org/10.1103/PhysRevLett.126.112002}
  {\path{doi:10.1103/PhysRevLett.126.112002}}.

\bibitem{Bury:2021sue}
Marcin Bury, Alexei Prokudin, and Alexey Vladimirov.
\newblock {Extraction of the Sivers function from SIDIS, Drell-Yan, and
  $W^\pm/Z$ boson production data with TMD evolution}.
\newblock {\em JHEP}, 05:151, 2021.
\newblock \href {http://arxiv.org/abs/2103.03270} {\path{arXiv:2103.03270}},
  \href {https://doi.org/10.1007/JHEP05(2021)151}
  {\path{doi:10.1007/JHEP05(2021)151}}.

\bibitem{Gamberg:2022kdb}
Leonard Gamberg, Michel Malda, Joshua~A. Miller, Daniel Pitonyak, Alexei
  Prokudin, and Nobuo Sato.
\newblock {Updated QCD global analysis of single transverse-spin asymmetries:
  Extracting H\textasciitilde{}, and the role of the Soffer bound and lattice
  QCD}.
\newblock {\em Phys. Rev. D}, 106(3):034014, 2022.
\newblock \href {http://arxiv.org/abs/2205.00999} {\path{arXiv:2205.00999}},
  \href {https://doi.org/10.1103/PhysRevD.106.034014}
  {\path{doi:10.1103/PhysRevD.106.034014}}.

\bibitem{Echevarria:2020hpy}
Miguel~G. Echevarria, Zhong-Bo Kang, and John Terry.
\newblock {Global analysis of the Sivers functions at NLO+NNLL in QCD}.
\newblock {\em JHEP}, 01:126, 2021.
\newblock \href {http://arxiv.org/abs/2009.10710} {\path{arXiv:2009.10710}},
  \href {https://doi.org/10.1007/JHEP01(2021)126}
  {\path{doi:10.1007/JHEP01(2021)126}}.

\bibitem{Bacchetta:2020gko}
Alessandro Bacchetta, Filippo Delcarro, Cristian Pisano, and Marco Radici.
\newblock {The 3-dimensional distribution of quarks in momentum space}.
\newblock {\em Phys. Lett. B}, 827:136961, 2022.
\newblock \href {http://arxiv.org/abs/2004.14278} {\path{arXiv:2004.14278}},
  \href {https://doi.org/10.1016/j.physletb.2022.136961}
  {\path{doi:10.1016/j.physletb.2022.136961}}.

\bibitem{CLAS:2021opg}
S.~Diehl et~al.
\newblock {Multidimensional, High Precision Measurements of Beam Single Spin
  Asymmetries in Semi-inclusive $\pi^{+}$~Electroproduction off Protons in the
  Valence Region}.
\newblock {\em Phys. Rev. Lett.}, 128(6):062005, 2022.
\newblock \href {http://arxiv.org/abs/2101.03544} {\path{arXiv:2101.03544}},
  \href {https://doi.org/10.1103/PhysRevLett.128.062005}
  {\path{doi:10.1103/PhysRevLett.128.062005}}.

\bibitem{CLAS:2022iqy}
S.~Diehl et~al.
\newblock {A multidimensional study of the structure function ratio
  $\sigma_{LT}$/$\sigma_0$ from hard exclusive $\pi^+$ electro-production off
  protons in the GPD regime}.
\newblock {\em Phys. Lett. B}, 839:137761, 2023.
\newblock \href {http://arxiv.org/abs/2210.14557} {\path{arXiv:2210.14557}},
  \href {https://doi.org/10.1016/j.physletb.2023.137761}
  {\path{doi:10.1016/j.physletb.2023.137761}}.

\bibitem{CLAS:2020yqf}
S.~Diehl et~al.
\newblock {Extraction of Beam-Spin Asymmetries from the Hard Exclusive $\pi^+$
  Channel off Protons in a Wide Range of Kinematics}.
\newblock {\em Phys. Rev. Lett.}, 125(18):182001, 2020.
\newblock \href {http://arxiv.org/abs/2007.15677} {\path{arXiv:2007.15677}},
  \href {https://doi.org/10.1103/PhysRevLett.125.182001}
  {\path{doi:10.1103/PhysRevLett.125.182001}}.

\bibitem{CLAS:2020igs}
M.~Mirazita et~al.
\newblock {Beam Spin Asymmetry in Semi-Inclusive Electroproduction of Hadron
  Pairs}.
\newblock {\em Phys. Rev. Lett.}, 126(6):062002, 2021.
\newblock \href {http://arxiv.org/abs/2010.09544} {\path{arXiv:2010.09544}},
  \href {https://doi.org/10.1103/PhysRevLett.126.062002}
  {\path{doi:10.1103/PhysRevLett.126.062002}}.

\bibitem{Hayward:2021psm}
T.~B. Hayward et~al.
\newblock {Observation of Beam Spin Asymmetries in the Process
  $ep\rightarrow{e}^{'}{\pi}^{+}{\pi}^{-}X$ with CLAS12}.
\newblock {\em Phys. Rev. Lett.}, 126:152501, 2021.
\newblock \href {http://arxiv.org/abs/2101.04842} {\path{arXiv:2101.04842}},
  \href {https://doi.org/10.1103/PhysRevLett.126.152501}
  {\path{doi:10.1103/PhysRevLett.126.152501}}.

\bibitem{CLAS:2022sqt}
H.~Avakian et~al.
\newblock {Observation of Correlations between Spin and Transverse Momenta in
  Back-to-Back Dihadron Production at CLAS12}.
\newblock {\em Phys. Rev. Lett.}, 130(2):022501, 2023.
\newblock \href {http://arxiv.org/abs/2208.05086} {\path{arXiv:2208.05086}},
  \href {https://doi.org/10.1103/PhysRevLett.130.022501}
  {\path{doi:10.1103/PhysRevLett.130.022501}}.

\bibitem{STAR:2022hqg}
Mohamed Abdallah et~al.
\newblock {Azimuthal transverse single-spin asymmetries of inclusive jets and
  identified hadrons within jets from polarized $pp$ collisions at $\sqrt{s}$ =
  200 GeV}.
\newblock {\em Phys. Rev. D}, 106(7):072010, 2022.
\newblock \href {http://arxiv.org/abs/2205.11800} {\path{arXiv:2205.11800}},
  \href {https://doi.org/10.1103/PhysRevD.106.072010}
  {\path{doi:10.1103/PhysRevD.106.072010}}.

\bibitem{Kang:2017btw}
Zhong-Bo Kang, Alexei Prokudin, Felix Ringer, and Feng Yuan.
\newblock {Collins azimuthal asymmetries of hadron production inside jets}.
\newblock {\em Phys. Lett. B}, 774:635--642, 2017.
\newblock \href {http://arxiv.org/abs/1707.00913} {\path{arXiv:1707.00913}},
  \href {https://doi.org/10.1016/j.physletb.2017.10.031}
  {\path{doi:10.1016/j.physletb.2017.10.031}}.

\bibitem{DAlesio:2017bvu}
Umberto D'Alesio, Francesco Murgia, and Cristian Pisano.
\newblock {Testing the universality of the Collins function in pion-jet
  production at RHIC}.
\newblock {\em Phys. Lett. B}, 773:300--306, 2017.
\newblock \href {http://arxiv.org/abs/1707.00914} {\path{arXiv:1707.00914}},
  \href {https://doi.org/10.1016/j.physletb.2017.08.023}
  {\path{doi:10.1016/j.physletb.2017.08.023}}.

\bibitem{Yuan:2007nd}
Feng Yuan.
\newblock {Azimuthal asymmetric distribution of hadrons inside a jet at hadron
  collider}.
\newblock {\em Phys. Rev. Lett.}, 100:032003, 2008.
\newblock \href {http://arxiv.org/abs/0709.3272} {\path{arXiv:0709.3272}},
  \href {https://doi.org/10.1103/PhysRevLett.100.032003}
  {\path{doi:10.1103/PhysRevLett.100.032003}}.

\bibitem{DAlesio:2010sag}
Umberto D'Alesio, Francesco Murgia, and Cristian Pisano.
\newblock {Azimuthal asymmetries for hadron distributions inside a jet in
  hadronic collisions}.
\newblock {\em Phys. Rev. D}, 83:034021, 2011.
\newblock \href {http://arxiv.org/abs/1011.2692} {\path{arXiv:1011.2692}},
  \href {https://doi.org/10.1103/PhysRevD.83.034021}
  {\path{doi:10.1103/PhysRevD.83.034021}}.

\bibitem{Kang:2017glf}
Zhong-Bo Kang, Xiaohui Liu, Felix Ringer, and Hongxi Xing.
\newblock {The transverse momentum distribution of hadrons within jets}.
\newblock {\em JHEP}, 11:068, 2017.
\newblock \href {http://arxiv.org/abs/1705.08443} {\path{arXiv:1705.08443}},
  \href {https://doi.org/10.1007/JHEP11(2017)068}
  {\path{doi:10.1007/JHEP11(2017)068}}.

\bibitem{STAR:2015jkc}
L.~Adamczyk et~al.
\newblock {Observation of Transverse Spin-Dependent Azimuthal Correlations of
  Charged Pion Pairs in $p^\uparrow+p$ at $\sqrt{s}=200$ GeV}.
\newblock {\em Phys. Rev. Lett.}, 115:242501, 2015.
\newblock \href {http://arxiv.org/abs/1504.00415} {\path{arXiv:1504.00415}},
  \href {https://doi.org/10.1103/PhysRevLett.115.242501}
  {\path{doi:10.1103/PhysRevLett.115.242501}}.

\bibitem{STAR:2017wsi}
L.~Adamczyk et~al.
\newblock {Transverse spin-dependent azimuthal correlations of charged pion
  pairs measured in p$^\uparrow$+p collisions at $\sqrt{s}$ = 500 GeV}.
\newblock {\em Phys. Lett. B}, 780:332--339, 2018.
\newblock \href {http://arxiv.org/abs/1710.10215} {\path{arXiv:1710.10215}},
  \href {https://doi.org/10.1016/j.physletb.2018.02.069}
  {\path{doi:10.1016/j.physletb.2018.02.069}}.

\bibitem{STAR:2020nnl}
Jaroslav Adam et~al.
\newblock {Measurement of transverse single-spin asymmetries of $\pi^0$ and
  electromagnetic jets at forward rapidity in 200 and 500 GeV transversely
  polarized proton-proton collisions}.
\newblock {\em Phys. Rev. D}, 103(9):092009, 2021.
\newblock \href {http://arxiv.org/abs/2012.11428} {\path{arXiv:2012.11428}},
  \href {https://doi.org/10.1103/PhysRevD.103.092009}
  {\path{doi:10.1103/PhysRevD.103.092009}}.

\bibitem{PHENIX:2019ouo}
C.~Aidala et~al.
\newblock {Nuclear Dependence of the Transverse Single-Spin Asymmetry in the
  Production of Charged Hadrons at Forward Rapidity in Polarized $p+p$, $p+$Al,
  and $p+$Au Collisions at $\sqrt{s_{_{NN}}}=200$ GeV}.
\newblock {\em Phys. Rev. Lett.}, 123(12):122001, 2019.
\newblock \href {http://arxiv.org/abs/1903.07422} {\path{arXiv:1903.07422}},
  \href {https://doi.org/10.1103/PhysRevLett.123.122001}
  {\path{doi:10.1103/PhysRevLett.123.122001}}.

\bibitem{STAR:2020grs}
Jaroslav Adam et~al.
\newblock {Comparison of transverse single-spin asymmetries for forward
  $\pi^{0}$ production in polarized $pp$, $p\rm{Al}$ and $p\rm{Au}$ collisions
  at nucleon pair c.m. energy $\sqrt{s_{\mathrm{NN}}}= 200$ GeV}.
\newblock {\em Phys. Rev. D}, 103(7):072005, 2021.
\newblock \href {http://arxiv.org/abs/2012.07146} {\path{arXiv:2012.07146}},
  \href {https://doi.org/10.1103/PhysRevD.103.072005}
  {\path{doi:10.1103/PhysRevD.103.072005}}.

\bibitem{Ali:2019lzf}
A.~Ali et~al.
\newblock {First Measurement of Near-Threshold J/\ensuremath{\psi} Exclusive
  Photoproduction off the Proton}.
\newblock {\em Phys. Rev. Lett.}, 123(7):072001, 2019.
\newblock \href {http://arxiv.org/abs/1905.10811} {\path{arXiv:1905.10811}},
  \href {https://doi.org/10.1103/PhysRevLett.123.072001}
  {\path{doi:10.1103/PhysRevLett.123.072001}}.

\bibitem{Pefkou:2021fni}
Dimitra~A. Pefkou, Daniel~C. Hackett, and Phiala~E. Shanahan.
\newblock {Gluon gravitational structure of hadrons of different spin}.
\newblock {\em Phys. Rev. D}, 105(5):054509, 2022.
\newblock \href {http://arxiv.org/abs/2107.10368} {\path{arXiv:2107.10368}},
  \href {https://doi.org/10.1103/PhysRevD.105.054509}
  {\path{doi:10.1103/PhysRevD.105.054509}}.

\bibitem{Kharzeev:2021qkd}
Dmitri~E. Kharzeev.
\newblock {Mass radius of the proton}.
\newblock {\em Phys. Rev. D}, 104(5):054015, 2021.
\newblock \href {http://arxiv.org/abs/2102.00110} {\path{arXiv:2102.00110}},
  \href {https://doi.org/10.1103/PhysRevD.104.054015}
  {\path{doi:10.1103/PhysRevD.104.054015}}.

\bibitem{Guo:2021ibg}
Yuxun Guo, Xiangdong Ji, and Yizhuang Liu.
\newblock {QCD Analysis of Near-Threshold Photon-Proton Production of Heavy
  Quarkonium}.
\newblock {\em Phys. Rev. D}, 103(9):096010, 2021.
\newblock \href {http://arxiv.org/abs/2103.11506} {\path{arXiv:2103.11506}},
  \href {https://doi.org/10.1103/PhysRevD.103.096010}
  {\path{doi:10.1103/PhysRevD.103.096010}}.

\bibitem{Hatta:2019ocp}
Yoshitaka Hatta, Mark Strikman, Ji~Xu, and Feng Yuan.
\newblock {Sub-threshold $J/\psi$ and $\Upsilon$ production in $\gamma A$
  collisions}.
\newblock {\em Phys. Lett. B}, 803:135321, 2020.
\newblock \href {http://arxiv.org/abs/1911.11706} {\path{arXiv:1911.11706}},
  \href {https://doi.org/10.1016/j.physletb.2020.135321}
  {\path{doi:10.1016/j.physletb.2020.135321}}.

\bibitem{Mamo:2019mka}
Kiminad~A. Mamo and Ismail Zahed.
\newblock {Diffractive photoproduction of $J/\psi$ and $\Upsilon$ using
  holographic QCD: gravitational form factors and GPD of gluons in the proton}.
\newblock {\em Phys. Rev. D}, 101(8):086003, 2020.
\newblock \href {http://arxiv.org/abs/1910.04707} {\path{arXiv:1910.04707}},
  \href {https://doi.org/10.1103/PhysRevD.101.086003}
  {\path{doi:10.1103/PhysRevD.101.086003}}.

\bibitem{Ji:2021qgo}
Xiangdong Ji, Yizhuang Liu, and Andreas Sch\"afer.
\newblock {Scale symmetry breaking, quantum anomalous energy and proton mass
  decomposition}.
\newblock {\em Nucl. Phys. B}, 971:115537, 2021.
\newblock \href {http://arxiv.org/abs/2105.03974} {\path{arXiv:2105.03974}},
  \href {https://doi.org/10.1016/j.nuclphysb.2021.115537}
  {\path{doi:10.1016/j.nuclphysb.2021.115537}}.

\bibitem{Sun:2021gmi}
Peng Sun, Xuan-Bo Tong, and Feng Yuan.
\newblock {Perturbative QCD analysis of near threshold heavy quarkonium
  photoproduction at large momentum transfer}.
\newblock {\em Phys. Lett. B}, 822:136655, 2021.
\newblock \href {http://arxiv.org/abs/2103.12047} {\path{arXiv:2103.12047}},
  \href {https://doi.org/10.1016/j.physletb.2021.136655}
  {\path{doi:10.1016/j.physletb.2021.136655}}.

\bibitem{Ji:2021mtz}
Xiangdong Ji.
\newblock {Proton mass decomposition: naturalness and interpretations}.
\newblock {\em Front. Phys. (Beijing)}, 16(6):64601, 2021.
\newblock \href {http://arxiv.org/abs/2102.07830} {\path{arXiv:2102.07830}},
  \href {https://doi.org/10.1007/s11467-021-1065-x}
  {\path{doi:10.1007/s11467-021-1065-x}}.

\bibitem{Ji:2021pys}
Xiangdong Ji and Yizhuang Liu.
\newblock {Quantum anomalous energy effects on the nucleon mass}.
\newblock {\em Sci. China Phys. Mech. Astron.}, 64(8):281012, 2021.
\newblock \href {http://arxiv.org/abs/2101.04483} {\path{arXiv:2101.04483}},
  \href {https://doi.org/10.1007/s11433-021-1723-2}
  {\path{doi:10.1007/s11433-021-1723-2}}.

\bibitem{Lorce:2021xku}
C\'edric Lorc\'e, Andreas Metz, Barbara Pasquini, and Simone Rodini.
\newblock {Energy-momentum tensor in QCD: nucleon mass decomposition and
  mechanical equilibrium}.
\newblock {\em JHEP}, 11:121, 2021.
\newblock \href {http://arxiv.org/abs/2109.11785} {\path{arXiv:2109.11785}},
  \href {https://doi.org/10.1007/JHEP11(2021)121}
  {\path{doi:10.1007/JHEP11(2021)121}}.

\bibitem{Mamo:2021krl}
Kiminad~A. Mamo and Ismail Zahed.
\newblock {Nucleon mass radii and distribution: Holographic QCD, Lattice QCD
  and GlueX data}.
\newblock {\em Phys. Rev. D}, 103(9):094010, 2021.
\newblock \href {http://arxiv.org/abs/2103.03186} {\path{arXiv:2103.03186}},
  \href {https://doi.org/10.1103/PhysRevD.103.094010}
  {\path{doi:10.1103/PhysRevD.103.094010}}.

\bibitem{Wang:2021dis}
Rong Wang, Wei Kou, Ya-Ping Xie, and Xurong Chen.
\newblock {Extraction of the proton mass radius from the vector meson
  photoproductions near thresholds}.
\newblock {\em Phys. Rev. D}, 103(9):L091501, 2021.
\newblock \href {http://arxiv.org/abs/2102.01610} {\path{arXiv:2102.01610}},
  \href {https://doi.org/10.1103/PhysRevD.103.L091501}
  {\path{doi:10.1103/PhysRevD.103.L091501}}.

\bibitem{Wang:2019mza}
Rong Wang, Jarah Evslin, and Xurong Chen.
\newblock {The origin of proton mass from J/${\Psi }$ photo-production data}.
\newblock {\em Eur. Phys. J. C}, 80(6):507, 2020.
\newblock \href {http://arxiv.org/abs/1912.12040} {\path{arXiv:1912.12040}},
  \href {https://doi.org/10.1140/epjc/s10052-020-8057-9}
  {\path{doi:10.1140/epjc/s10052-020-8057-9}}.

\bibitem{He:2021bof}
Fangcheng He, Peng Sun, and Yi-Bo Yang.
\newblock {Demonstration of the hadron mass origin from the QCD trace anomaly}.
\newblock {\em Phys. Rev. D}, 104(7):074507, 2021.
\newblock \href {http://arxiv.org/abs/2101.04942} {\path{arXiv:2101.04942}},
  \href {https://doi.org/10.1103/PhysRevD.104.074507}
  {\path{doi:10.1103/PhysRevD.104.074507}}.

\bibitem{Rodini:2020pis}
S.~Rodini, A.~Metz, and B.~Pasquini.
\newblock {Mass sum rules of the electron in quantum electrodynamics}.
\newblock {\em JHEP}, 09:067, 2020.
\newblock \href {http://arxiv.org/abs/2004.03704} {\path{arXiv:2004.03704}},
  \href {https://doi.org/10.1007/JHEP09(2020)067}
  {\path{doi:10.1007/JHEP09(2020)067}}.

\bibitem{Metz:2020vxd}
Andreas Metz, Barbara Pasquini, and Simone Rodini.
\newblock {Revisiting the proton mass decomposition}.
\newblock {\em Phys. Rev. D}, 102:114042, 2020.
\newblock \href {http://arxiv.org/abs/2006.11171} {\path{arXiv:2006.11171}},
  \href {https://doi.org/10.1103/PhysRevD.102.114042}
  {\path{doi:10.1103/PhysRevD.102.114042}}.

\bibitem{Lorce:2017xzd}
C\'edric Lorc\'e.
\newblock {On the hadron mass decomposition}.
\newblock {\em Eur. Phys. J. C}, 78(2):120, 2018.
\newblock \href {http://arxiv.org/abs/1706.05853} {\path{arXiv:1706.05853}},
  \href {https://doi.org/10.1140/epjc/s10052-018-5561-2}
  {\path{doi:10.1140/epjc/s10052-018-5561-2}}.

\bibitem{Hatta:2018sqd}
Yoshitaka Hatta, Abha Rajan, and Kazuhiro Tanaka.
\newblock {Quark and gluon contributions to the QCD trace anomaly}.
\newblock {\em JHEP}, 12:008, 2018.
\newblock \href {http://arxiv.org/abs/1810.05116} {\path{arXiv:1810.05116}},
  \href {https://doi.org/10.1007/JHEP12(2018)008}
  {\path{doi:10.1007/JHEP12(2018)008}}.

\bibitem{Tanaka:2018nae}
Kazuhiro Tanaka.
\newblock {Three-loop formula for quark and gluon contributions to the QCD
  trace anomaly}.
\newblock {\em JHEP}, 01:120, 2019.
\newblock \href {http://arxiv.org/abs/1811.07879} {\path{arXiv:1811.07879}},
  \href {https://doi.org/10.1007/JHEP01(2019)120}
  {\path{doi:10.1007/JHEP01(2019)120}}.

\bibitem{Duran:2022xag}
B.~Duran et~al.
\newblock {When Color meets Gravity; Near-Threshold Exclusive $J/\psi$
  Photoproduction on the Proton}.
\newblock 7 2022.
\newblock \href {http://arxiv.org/abs/2207.05212} {\path{arXiv:2207.05212}}.

\bibitem{HillerBlin:2016odx}
A.~N. Hiller~Blin, C.~Fern\'andez-Ram\'\i{}rez, A.~Jackura, V.~Mathieu, V.~I.
  Mokeev, A.~Pilloni, and A.~P. Szczepaniak.
\newblock {Studying the P$_c$(4450) resonance in J/$\psi$ photoproduction off
  protons}.
\newblock {\em Phys. Rev. D}, 94(3):034002, 2016.
\newblock \href {http://arxiv.org/abs/1606.08912} {\path{arXiv:1606.08912}},
  \href {https://doi.org/10.1103/PhysRevD.94.034002}
  {\path{doi:10.1103/PhysRevD.94.034002}}.

\bibitem{ParticleDataGroup:2022pth}
R.~L. Workman et~al.
\newblock {Review of Particle Physics}.
\newblock {\em PTEP}, 2022:083C01, 2022.
\newblock \href {https://doi.org/10.1093/ptep/ptac097}
  {\path{doi:10.1093/ptep/ptac097}}.

\bibitem{Dudek:2013yja}
Jozef~J. Dudek, Robert~G. Edwards, Peng Guo, and Christopher~E. Thomas.
\newblock {Toward the excited isoscalar meson spectrum from lattice QCD}.
\newblock {\em Phys. Rev. D}, 88(9):094505, 2013.
\newblock \href {http://arxiv.org/abs/1309.2608} {\path{arXiv:1309.2608}},
  \href {https://doi.org/10.1103/PhysRevD.88.094505}
  {\path{doi:10.1103/PhysRevD.88.094505}}.

\bibitem{COMPASS:2014vkj}
C.~Adolph et~al.
\newblock {Odd and even partial waves of $\eta\pi^-$ and $\eta'\pi^-$ in
  $\pi^-p\to\eta^{(\prime)}\pi^-p$ at $191\,\textrm{GeV}/c$}.
\newblock {\em Phys. Lett. B}, 740:303--311, 2015.
\newblock [Erratum: Phys.Lett.B 811, 135913 (2020)].
\newblock \href {http://arxiv.org/abs/1408.4286} {\path{arXiv:1408.4286}},
  \href {https://doi.org/10.1016/j.physletb.2014.11.058}
  {\path{doi:10.1016/j.physletb.2014.11.058}}.

\bibitem{JPAC:2018zyd}
A.~Rodas et~al.
\newblock {Determination of the pole position of the lightest hybrid meson
  candidate}.
\newblock {\em Phys. Rev. Lett.}, 122(4):042002, 2019.
\newblock \href {http://arxiv.org/abs/1810.04171} {\path{arXiv:1810.04171}},
  \href {https://doi.org/10.1103/PhysRevLett.122.042002}
  {\path{doi:10.1103/PhysRevLett.122.042002}}.

\bibitem{Woss:2020ayi}
Antoni~J. Woss, Jozef~J. Dudek, Robert~G. Edwards, Christopher~E. Thomas, and
  David~J. Wilson.
\newblock {Decays of an exotic $1{-+}$ hybrid meson resonance in QCD}.
\newblock {\em Phys. Rev. D}, 103(5):054502, 2021.
\newblock \href {http://arxiv.org/abs/2009.10034} {\path{arXiv:2009.10034}},
  \href {https://doi.org/10.1103/PhysRevD.103.054502}
  {\path{doi:10.1103/PhysRevD.103.054502}}.

\bibitem{Esposito:2016noz}
A.~Esposito, A.~Pilloni, and A.~D. Polosa.
\newblock {Multiquark Resonances}.
\newblock {\em Phys. Rept.}, 668:1--97, 2017.
\newblock \href {http://arxiv.org/abs/1611.07920} {\path{arXiv:1611.07920}},
  \href {https://doi.org/10.1016/j.physrep.2016.11.002}
  {\path{doi:10.1016/j.physrep.2016.11.002}}.

\bibitem{Guo:2017jvc}
Feng-Kun Guo, Christoph Hanhart, Ulf-G. Mei\ss{}ner, Qian Wang, Qiang Zhao, and
  Bing-Song Zou.
\newblock {Hadronic molecules}.
\newblock {\em Rev. Mod. Phys.}, 90(1):015004, 2018.
\newblock [Erratum: Rev.Mod.Phys. 94, 029901 (2022)].
\newblock \href {http://arxiv.org/abs/1705.00141} {\path{arXiv:1705.00141}},
  \href {https://doi.org/10.1103/RevModPhys.90.015004}
  {\path{doi:10.1103/RevModPhys.90.015004}}.

\bibitem{Olsen:2017bmm}
Stephen~Lars Olsen, Tomasz Skwarnicki, and Daria Zieminska.
\newblock {Nonstandard heavy mesons and baryons: Experimental evidence}.
\newblock {\em Rev. Mod. Phys.}, 90(1):015003, 2018.
\newblock \href {http://arxiv.org/abs/1708.04012} {\path{arXiv:1708.04012}},
  \href {https://doi.org/10.1103/RevModPhys.90.015003}
  {\path{doi:10.1103/RevModPhys.90.015003}}.

\bibitem{Brambilla:2019esw}
Nora Brambilla, Simon Eidelman, Christoph Hanhart, Alexey Nefediev, Cheng-Ping
  Shen, Christopher~E. Thomas, Antonio Vairo, and Chang-Zheng Yuan.
\newblock {The $XYZ$ states: experimental and theoretical status and
  perspectives}.
\newblock {\em Phys. Rept.}, 873:1--154, 2020.
\newblock \href {http://arxiv.org/abs/1907.07583} {\path{arXiv:1907.07583}},
  \href {https://doi.org/10.1016/j.physrep.2020.05.001}
  {\path{doi:10.1016/j.physrep.2020.05.001}}.

\bibitem{Chen:2022asf}
Hua-Xing Chen, Wei Chen, Xiang Liu, Yan-Rui Liu, and Shi-Lin Zhu.
\newblock {An updated review of the new hadron states}.
\newblock {\em Rept. Prog. Phys.}, 86(2):026201, 2023.
\newblock \href {http://arxiv.org/abs/2204.02649} {\path{arXiv:2204.02649}},
  \href {https://doi.org/10.1088/1361-6633/aca3b6}
  {\path{doi:10.1088/1361-6633/aca3b6}}.

\bibitem{LHCb:2015yax}
Roel Aaij et~al.
\newblock {Observation of $J/\psi p$ Resonances Consistent with Pentaquark
  States in $\Lambda_b^0 \to J/\psi K^- p$ Decays}.
\newblock {\em Phys. Rev. Lett.}, 115:072001, 2015.
\newblock \href {http://arxiv.org/abs/1507.03414} {\path{arXiv:1507.03414}},
  \href {https://doi.org/10.1103/PhysRevLett.115.072001}
  {\path{doi:10.1103/PhysRevLett.115.072001}}.

\bibitem{LHCb:2019kea}
Roel Aaij et~al.
\newblock {Observation of a narrow pentaquark state, $P_c(4312)^+$, and of
  two-peak structure of the $P_c(4450)^+$}.
\newblock {\em Phys. Rev. Lett.}, 122(22):222001, 2019.
\newblock \href {http://arxiv.org/abs/1904.03947} {\path{arXiv:1904.03947}},
  \href {https://doi.org/10.1103/PhysRevLett.122.222001}
  {\path{doi:10.1103/PhysRevLett.122.222001}}.

\bibitem{Wu:2020zbx}
Biaogang Wu, Xiaojian Du, Matthew Sibila, and Ralf Rapp.
\newblock {$X(3872)$transport in heavy-ion collisions}.
\newblock {\em Eur. Phys. J. A}, 57(4):122, 2021.
\newblock [Erratum: Eur.Phys.J.A 57, 314 (2021)].
\newblock \href {http://arxiv.org/abs/2006.09945} {\path{arXiv:2006.09945}},
  \href {https://doi.org/10.1140/epja/s10050-021-00623-4}
  {\path{doi:10.1140/epja/s10050-021-00623-4}}.

\bibitem{Chen:2021akx}
Baoyi Chen, Liu Jiang, Xiao-Hai Liu, Yunpeng Liu, and Jiaxing Zhao.
\newblock {X(3872) production in relativistic heavy-ion collisions}.
\newblock {\em Phys. Rev. C}, 105(5):054901, 2022.
\newblock \href {http://arxiv.org/abs/2107.00969} {\path{arXiv:2107.00969}},
  \href {https://doi.org/10.1103/PhysRevC.105.054901}
  {\path{doi:10.1103/PhysRevC.105.054901}}.

\bibitem{Zhang:2020dwn}
Hui Zhang, Jinfeng Liao, Enke Wang, Qian Wang, and Hongxi Xing.
\newblock {Deciphering the Nature of X(3872) in Heavy Ion Collisions}.
\newblock {\em Phys. Rev. Lett.}, 126(1):012301, 2021.
\newblock \href {http://arxiv.org/abs/2004.00024} {\path{arXiv:2004.00024}},
  \href {https://doi.org/10.1103/PhysRevLett.126.012301}
  {\path{doi:10.1103/PhysRevLett.126.012301}}.

\bibitem{LHCb:2020sey}
Roel Aaij et~al.
\newblock {Observation of Multiplicity Dependent Prompt $\chi_{c1}(3872)$ and
  $\psi(2S)$ Production in $pp$ Collisions}.
\newblock {\em Phys. Rev. Lett.}, 126(9):092001, 2021.
\newblock \href {http://arxiv.org/abs/2009.06619} {\path{arXiv:2009.06619}},
  \href {https://doi.org/10.1103/PhysRevLett.126.092001}
  {\path{doi:10.1103/PhysRevLett.126.092001}}.

\bibitem{CMS:2021znk}
Albert~M. Sirunyan et~al.
\newblock {Evidence for X(3872) in Pb-Pb Collisions and Studies of its Prompt
  Production at $\sqrt {s_{NN}}$=5.02\,\,TeV}.
\newblock {\em Phys. Rev. Lett.}, 128(3):032001, 2022.
\newblock \href {http://arxiv.org/abs/2102.13048} {\path{arXiv:2102.13048}},
  \href {https://doi.org/10.1103/PhysRevLett.128.032001}
  {\path{doi:10.1103/PhysRevLett.128.032001}}.

\bibitem{Esposito:2020ywk}
Angelo Esposito, Elena~G. Ferreiro, Alessandro Pilloni, Antonio~D. Polosa, and
  Carlos~A. Salgado.
\newblock {The nature of X(3872) from high-multiplicity pp collisions}.
\newblock {\em Eur. Phys. J. C}, 81(7):669, 2021.
\newblock \href {http://arxiv.org/abs/2006.15044} {\path{arXiv:2006.15044}},
  \href {https://doi.org/10.1140/epjc/s10052-021-09425-w}
  {\path{doi:10.1140/epjc/s10052-021-09425-w}}.

\bibitem{Braaten:2020iqw}
Eric Braaten, Li-Ping He, Kevin Ingles, and Jun Jiang.
\newblock {Production of $X(3872)$ at High Multiplicity}.
\newblock {\em Phys. Rev. D}, 103(7):L071901, 2021.
\newblock \href {http://arxiv.org/abs/2012.13499} {\path{arXiv:2012.13499}},
  \href {https://doi.org/10.1103/PhysRevD.103.L071901}
  {\path{doi:10.1103/PhysRevD.103.L071901}}.

\bibitem{Frankfurt:1988nt}
L.~L. Frankfurt and M.~I. Strikman.
\newblock {Hard Nuclear Processes and Microscopic Nuclear Structure}.
\newblock {\em Phys. Rept.}, 160:235--427, 1988.
\newblock \href {https://doi.org/10.1016/0370-1573(88)90179-2}
  {\path{doi:10.1016/0370-1573(88)90179-2}}.

\bibitem{Subedi:2008zz}
R.~Subedi et~al.
\newblock {Probing Cold Dense Nuclear Matter}.
\newblock {\em Science}, 320:1476--1478, 2008.
\newblock \href {http://arxiv.org/abs/0908.1514} {\path{arXiv:0908.1514}},
  \href {https://doi.org/10.1126/science.1156675}
  {\path{doi:10.1126/science.1156675}}.

\bibitem{CiofidegliAtti:2015lcu}
Claudio Ciofi~degli Atti.
\newblock {In-medium short-range dynamics of nucleons: Recent theoretical and
  experimental advances}.
\newblock {\em Phys. Rept.}, 590:1--85, 2015.
\newblock \href {https://doi.org/10.1016/j.physrep.2015.06.002}
  {\path{doi:10.1016/j.physrep.2015.06.002}}.

\bibitem{Ryckebusch:2019oya}
Jan Ryckebusch, Wim Cosyn, Tom Vieijra, and Corneel Casert.
\newblock {Isospin composition of the high-momentum fluctuations in nuclei from
  asymptotic momentum distributions}.
\newblock {\em Phys. Rev. C}, 100(5):054620, 2019.
\newblock \href {http://arxiv.org/abs/1907.07259} {\path{arXiv:1907.07259}},
  \href {https://doi.org/10.1103/PhysRevC.100.054620}
  {\path{doi:10.1103/PhysRevC.100.054620}}.

\bibitem{Hen:2016kwk}
O.~Hen, G.~A. Miller, E.~Piasetzky, and L.~B. Weinstein.
\newblock {Nucleon-Nucleon Correlations, Short-lived Excitations, and the
  Quarks Within}.
\newblock {\em Rev. Mod. Phys.}, 89(4):045002, 2017.
\newblock \href {http://arxiv.org/abs/1611.09748} {\path{arXiv:1611.09748}},
  \href {https://doi.org/10.1103/RevModPhys.89.045002}
  {\path{doi:10.1103/RevModPhys.89.045002}}.

\bibitem{CLAS:2018qpc}
E.~O. Cohen et~al.
\newblock {Center of Mass Motion of Short-Range Correlated Nucleon Pairs
  studied via the $A(e,e'pp)$ Reaction}.
\newblock {\em Phys. Rev. Lett.}, 121(9):092501, 2018.
\newblock \href {http://arxiv.org/abs/1805.01981} {\path{arXiv:1805.01981}},
  \href {https://doi.org/10.1103/PhysRevLett.121.092501}
  {\path{doi:10.1103/PhysRevLett.121.092501}}.

\bibitem{Arrington:2022sov}
John Arrington, Nadia Fomin, and Axel Schmidt.
\newblock {Progress in understanding short-range structure in nuclei: an
  experimental perspective}.
\newblock 3 2022.
\newblock \href {http://arxiv.org/abs/2203.02608} {\path{arXiv:2203.02608}},
  \href {https://doi.org/10.1146/annurev-nucl-102020-022253}
  {\path{doi:10.1146/annurev-nucl-102020-022253}}.

\bibitem{CLAS:2020mom}
A.~Schmidt et~al.
\newblock {Probing the core of the strong nuclear interaction}.
\newblock {\em Nature}, 578(7796):540--544, 2020.
\newblock \href {http://arxiv.org/abs/2004.11221} {\path{arXiv:2004.11221}},
  \href {https://doi.org/10.1038/s41586-020-2021-6}
  {\path{doi:10.1038/s41586-020-2021-6}}.

\bibitem{Li:2022fhh}
S.~Li et~al.
\newblock {Revealing the short-range structure of the mirror nuclei $^{3}$H and
  $^{3}$He}.
\newblock {\em Nature}, 609(7925):41--45, 2022.
\newblock \href {http://arxiv.org/abs/2210.04189} {\path{arXiv:2210.04189}},
  \href {https://doi.org/10.1038/s41586-022-05007-2}
  {\path{doi:10.1038/s41586-022-05007-2}}.

\bibitem{CLAS:2020rue}
I.~Korover et~al.
\newblock {12C(e,e'pN) measurements of short range correlations in the
  tensor-to-scalar interaction transition region}.
\newblock {\em Phys. Lett. B}, 820:136523, 2021.
\newblock \href {http://arxiv.org/abs/2004.07304} {\path{arXiv:2004.07304}},
  \href {https://doi.org/10.1016/j.physletb.2021.136523}
  {\path{doi:10.1016/j.physletb.2021.136523}}.

\bibitem{Hen:2014nza}
O.~Hen et~al.
\newblock {Momentum sharing in imbalanced Fermi systems}.
\newblock {\em Science}, 346:614--617, 2014.
\newblock \href {http://arxiv.org/abs/1412.0138} {\path{arXiv:1412.0138}},
  \href {https://doi.org/10.1126/science.1256785}
  {\path{doi:10.1126/science.1256785}}.

\bibitem{CLAS:2018yvt}
M.~Duer et~al.
\newblock {Probing high-momentum protons and neutrons in neutron-rich nuclei}.
\newblock {\em Nature}, 560(7720):617--621, 2018.
\newblock \href {https://doi.org/10.1038/s41586-018-0400-z}
  {\path{doi:10.1038/s41586-018-0400-z}}.

\bibitem{CLAS:2018xvc}
M.~Duer et~al.
\newblock {Direct Observation of Proton-Neutron Short-Range Correlation
  Dominance in Heavy Nuclei}.
\newblock {\em Phys. Rev. Lett.}, 122(17):172502, 2019.
\newblock \href {http://arxiv.org/abs/1810.05343} {\path{arXiv:1810.05343}},
  \href {https://doi.org/10.1103/PhysRevLett.122.172502}
  {\path{doi:10.1103/PhysRevLett.122.172502}}.

\bibitem{JeffersonLabHallATritium:2020mha}
R.~Cruz-Torres et~al.
\newblock {Probing Few-Body Nuclear Dynamics via $^3$H and $^3$He ($e,e'p$)pn
  Cross-Section Measurements}.
\newblock {\em Phys. Rev. Lett.}, 124(21):212501, 2020.
\newblock \href {http://arxiv.org/abs/2001.07230} {\path{arXiv:2001.07230}},
  \href {https://doi.org/10.1103/PhysRevLett.124.212501}
  {\path{doi:10.1103/PhysRevLett.124.212501}}.

\bibitem{Weiss:2020bkp}
R.~Weiss, A.~W. Denniston, J.~R. Pybus, O.~Hen, E.~Piasetzky, A.~Schmidt, L.~B.
  Weinstein, and N.~Barnea.
\newblock {Extracting the number of short-range correlated nucleon pairs from
  inclusive electron scattering data}.
\newblock {\em Phys. Rev. C}, 103(3):L031301, 2021.
\newblock \href {http://arxiv.org/abs/2005.01621} {\path{arXiv:2005.01621}},
  \href {https://doi.org/10.1103/PhysRevC.103.L031301}
  {\path{doi:10.1103/PhysRevC.103.L031301}}.

\bibitem{Weiss:2015mba}
Ronen Weiss, Betzalel Bazak, and Nir Barnea.
\newblock {Generalized nuclear contacts and momentum distributions}.
\newblock {\em Phys. Rev. C}, 92(5):054311, 2015.
\newblock \href {http://arxiv.org/abs/1503.07047} {\path{arXiv:1503.07047}},
  \href {https://doi.org/10.1103/PhysRevC.92.054311}
  {\path{doi:10.1103/PhysRevC.92.054311}}.

\bibitem{Weiss:2016obx}
R.~Weiss, R.~Cruz-Torres, N.~Barnea, E.~Piasetzky, and O.~Hen.
\newblock {The nuclear contacts and short range correlations in nuclei}.
\newblock {\em Phys. Lett. B}, 780:211--215, 2018.
\newblock \href {http://arxiv.org/abs/1612.00923} {\path{arXiv:1612.00923}},
  \href {https://doi.org/10.1016/j.physletb.2018.01.061}
  {\path{doi:10.1016/j.physletb.2018.01.061}}.

\bibitem{Weiss:2018tbu}
Ronen Weiss, Igor Korover, Eliezer Piasetzky, Or~Hen, and Nir Barnea.
\newblock {Energy and momentum dependence of nuclear short-range correlations -
  Spectral function, exclusive scattering experiments and the contact
  formalism}.
\newblock {\em Phys. Lett. B}, 791:242--248, 2019.
\newblock \href {http://arxiv.org/abs/1806.10217} {\path{arXiv:1806.10217}},
  \href {https://doi.org/10.1016/j.physletb.2019.02.019}
  {\path{doi:10.1016/j.physletb.2019.02.019}}.

\bibitem{Cruz-Torres:2019fum}
R.~Cruz-Torres, D.~Lonardoni, R.~Weiss, N.~Barnea, D.~W. Higinbotham,
  E.~Piasetzky, A.~Schmidt, L.~B. Weinstein, R.~B. Wiringa, and O.~Hen.
\newblock {Many-body factorization and position\textendash{}momentum
  equivalence of nuclear short-range correlations}.
\newblock {\em Nature Phys.}, 17(3):306--310, 2021.
\newblock \href {http://arxiv.org/abs/1907.03658} {\path{arXiv:1907.03658}},
  \href {https://doi.org/10.1038/s41567-020-01053-7}
  {\path{doi:10.1038/s41567-020-01053-7}}.

\bibitem{Carlson:2014vla}
J.~Carlson, S.~Gandolfi, F.~Pederiva, Steven~C. Pieper, R.~Schiavilla, K.~E.
  Schmidt, and R.~B. Wiringa.
\newblock {Quantum Monte Carlo methods for nuclear physics}.
\newblock {\em Rev. Mod. Phys.}, 87:1067, 2015.
\newblock \href {http://arxiv.org/abs/1412.3081} {\path{arXiv:1412.3081}},
  \href {https://doi.org/10.1103/RevModPhys.87.1067}
  {\path{doi:10.1103/RevModPhys.87.1067}}.

\bibitem{Pybus:2020itv}
J.~R. Pybus, I.~Korover, R.~Weiss, A.~Schmidt, N.~Barnea, D.~W. Higinbotham,
  E.~Piasetzky, M.~Strikman, L.~B. Weinstein, and O.~Hen.
\newblock {Generalized contact formalism analysis of the 4He(e,e'pN) reaction}.
\newblock {\em Phys. Lett. B}, 805:135429, 2020.
\newblock \href {http://arxiv.org/abs/2003.02318} {\path{arXiv:2003.02318}},
  \href {https://doi.org/10.1016/j.physletb.2020.135429}
  {\path{doi:10.1016/j.physletb.2020.135429}}.

\bibitem{West:2020tyo}
Jennifer~Rittenhouse West.
\newblock {Diquark induced short-range nucleon-nucleon correlations \& the EMC
  effect}.
\newblock {\em Nucl. Phys. A}, 1029:122563, 2023.
\newblock \href {http://arxiv.org/abs/2009.06968} {\path{arXiv:2009.06968}},
  \href {https://doi.org/10.1016/j.nuclphysa.2022.122563}
  {\path{doi:10.1016/j.nuclphysa.2022.122563}}.

\bibitem{HallC:2020kdm}
Carlos Yero et~al.
\newblock {Probing the Deuteron at Very Large Internal Momenta}.
\newblock {\em Phys. Rev. Lett.}, 125(26):262501, 2020.
\newblock \href {http://arxiv.org/abs/2008.08058} {\path{arXiv:2008.08058}},
  \href {https://doi.org/10.1103/PhysRevLett.125.262501}
  {\path{doi:10.1103/PhysRevLett.125.262501}}.

\bibitem{JeffersonLabHallATritium:2019xlj}
R.~Cruz-Torres et~al.
\newblock {Comparing proton momentum distributions in $A=2$ and 3 nuclei via
  $^2$H $^3$H and $^3$He $(e, e'p)$ measurements}.
\newblock {\em Phys. Lett. B}, 797:134890, 2019.
\newblock \href {http://arxiv.org/abs/1902.06358} {\path{arXiv:1902.06358}},
  \href {https://doi.org/10.1016/j.physletb.2019.134890}
  {\path{doi:10.1016/j.physletb.2019.134890}}.

\bibitem{Weinstein:2010rt}
L.~B. Weinstein, E.~Piasetzky, D.~W. Higinbotham, J.~Gomez, O.~Hen, and
  R.~Shneor.
\newblock {Short Range Correlations and the EMC Effect}.
\newblock {\em Phys. Rev. Lett.}, 106:052301, 2011.
\newblock \href {http://arxiv.org/abs/1009.5666} {\path{arXiv:1009.5666}},
  \href {https://doi.org/10.1103/PhysRevLett.106.052301}
  {\path{doi:10.1103/PhysRevLett.106.052301}}.

\bibitem{CLAS:2019vsb}
B.~Schmookler et~al.
\newblock {Modified structure of protons and neutrons in correlated pairs}.
\newblock {\em Nature}, 566(7744):354--358, 2019.
\newblock \href {http://arxiv.org/abs/2004.12065} {\path{arXiv:2004.12065}},
  \href {https://doi.org/10.1038/s41586-019-0925-9}
  {\path{doi:10.1038/s41586-019-0925-9}}.

\bibitem{Arrington:2019wky}
J.~Arrington and N.~Fomin.
\newblock {Searching for flavor dependence in nuclear quark behavior}.
\newblock {\em Phys. Rev. Lett.}, 123(4):042501, 2019.
\newblock \href {http://arxiv.org/abs/1903.12535} {\path{arXiv:1903.12535}},
  \href {https://doi.org/10.1103/PhysRevLett.123.042501}
  {\path{doi:10.1103/PhysRevLett.123.042501}}.

\bibitem{Kim:2022lng}
Dmitriy~N. Kim and Gerald~A. Miller.
\newblock {Light-front holography model of the EMC effect}.
\newblock {\em Phys. Rev. C}, 106(5):055202, 2022.
\newblock \href {http://arxiv.org/abs/2209.13753} {\path{arXiv:2209.13753}},
  \href {https://doi.org/10.1103/PhysRevC.106.055202}
  {\path{doi:10.1103/PhysRevC.106.055202}}.

\bibitem{Segarra:2019gbp}
E.~P. Segarra, A.~Schmidt, T.~Kutz, D.~W. Higinbotham, E.~Piasetzky,
  M.~Strikman, L.~B. Weinstein, and O.~Hen.
\newblock {Neutron Valence Structure from Nuclear Deep Inelastic Scattering}.
\newblock {\em Phys. Rev. Lett.}, 124(9):092002, 2020.
\newblock \href {http://arxiv.org/abs/1908.02223} {\path{arXiv:1908.02223}},
  \href {https://doi.org/10.1103/PhysRevLett.124.092002}
  {\path{doi:10.1103/PhysRevLett.124.092002}}.

\bibitem{Arrington:2021vuu}
J.~Arrington et~al.
\newblock {Measurement of the EMC effect in light and heavy nuclei}.
\newblock {\em Phys. Rev. C}, 104(6):065203, 2021.
\newblock \href {http://arxiv.org/abs/2110.08399} {\path{arXiv:2110.08399}},
  \href {https://doi.org/10.1103/PhysRevC.104.065203}
  {\path{doi:10.1103/PhysRevC.104.065203}}.

\bibitem{HallC:2022rzv}
A.~Karki et~al.
\newblock {First Measurement of the EMC Effect in $^{10}$B and $^{11}$B}.
\newblock 7 2022.
\newblock \href {http://arxiv.org/abs/2207.03850} {\path{arXiv:2207.03850}}.

\bibitem{Segarra:2020plg}
E.~P. Segarra, J.~R. Pybus, F.~Hauenstein, D.~W. Higinbotham, G.~A. Miller,
  E.~Piasetzky, A.~Schmidt, M.~Strikman, L.~B. Weinstein, and O.~Hen.
\newblock {Short-range correlations and the nuclear EMC effect in deuterium and
  helium-3}.
\newblock {\em Phys. Rev. Res.}, 3(2):023240, 2021.
\newblock \href {http://arxiv.org/abs/2006.10249} {\path{arXiv:2006.10249}},
  \href {https://doi.org/10.1103/PhysRevResearch.3.023240}
  {\path{doi:10.1103/PhysRevResearch.3.023240}}.

\bibitem{Cloet:2005rt}
I.~C. Cloet, Wolfgang Bentz, and Anthony~William Thomas.
\newblock {Spin-dependent structure functions in nuclear matter and the
  polarized EMC effect}.
\newblock {\em Phys. Rev. Lett.}, 95:052302, 2005.
\newblock \href {http://arxiv.org/abs/nucl-th/0504019}
  {\path{arXiv:nucl-th/0504019}}, \href
  {https://doi.org/10.1103/PhysRevLett.95.052302}
  {\path{doi:10.1103/PhysRevLett.95.052302}}.

\bibitem{Cloet:2006bq}
I.~C. Cloet, Wolfgang Bentz, and Anthony~William Thomas.
\newblock {EMC and polarized EMC effects in nuclei}.
\newblock {\em Phys. Lett. B}, 642:210--217, 2006.
\newblock \href {http://arxiv.org/abs/nucl-th/0605061}
  {\path{arXiv:nucl-th/0605061}}, \href
  {https://doi.org/10.1016/j.physletb.2006.08.076}
  {\path{doi:10.1016/j.physletb.2006.08.076}}.

\bibitem{Tronchin:2018mvu}
Stephen Tronchin, Hrayr~H. Matevosyan, and Anthony~W. Thomas.
\newblock {Polarized EMC Effect in the QMC Model}.
\newblock {\em Phys. Lett. B}, 783:247--252, 2018.
\newblock \href {http://arxiv.org/abs/1806.00481} {\path{arXiv:1806.00481}},
  \href {https://doi.org/10.1016/j.physletb.2018.06.065}
  {\path{doi:10.1016/j.physletb.2018.06.065}}.

\bibitem{Cloet:2009qs}
I.~C. Cloet, W.~Bentz, and A.~W. Thomas.
\newblock {Isovector EMC effect explains the NuTeV anomaly}.
\newblock {\em Phys. Rev. Lett.}, 102:252301, 2009.
\newblock \href {http://arxiv.org/abs/0901.3559} {\path{arXiv:0901.3559}},
  \href {https://doi.org/10.1103/PhysRevLett.102.252301}
  {\path{doi:10.1103/PhysRevLett.102.252301}}.

\bibitem{ATLAS:2015mwq}
Georges Aad et~al.
\newblock {$Z$ boson production in $p+$Pb collisions at $\sqrt{s_{NN}}=5.02$
  TeV measured with the ATLAS detector}.
\newblock {\em Phys. Rev. C}, 92(4):044915, 2015.
\newblock \href {http://arxiv.org/abs/1507.06232} {\path{arXiv:1507.06232}},
  \href {https://doi.org/10.1103/PhysRevC.92.044915}
  {\path{doi:10.1103/PhysRevC.92.044915}}.

\bibitem{ALICE:2016rzo}
Jaroslav Adam et~al.
\newblock {W and Z boson production in p-Pb collisions at $\sqrt{s_{\rm NN}}$ =
  5.02 TeV}.
\newblock {\em JHEP}, 02:077, 2017.
\newblock \href {http://arxiv.org/abs/1611.03002} {\path{arXiv:1611.03002}},
  \href {https://doi.org/10.1007/JHEP02(2017)077}
  {\path{doi:10.1007/JHEP02(2017)077}}.

\bibitem{ALICE:2020jff}
Shreyasi Acharya et~al.
\newblock {Z-boson production in p-Pb collisions at
  $\sqrt{s_{\mathrm{NN}}}=8.16$ TeV and Pb-Pb collisions at
  $\sqrt{s_{\mathrm{NN}}}=5.02$ TeV}.
\newblock {\em JHEP}, 09:076, 2020.
\newblock \href {http://arxiv.org/abs/2005.11126} {\path{arXiv:2005.11126}},
  \href {https://doi.org/10.1007/JHEP09(2020)076}
  {\path{doi:10.1007/JHEP09(2020)076}}.

\bibitem{CMS:2021ynu}
Albert~M Sirunyan et~al.
\newblock {Study of Drell-Yan dimuon production in proton-lead collisions at
  $\sqrt{s_\mathrm{NN}} =$ 8.16 TeV}.
\newblock {\em JHEP}, 05:182, 2021.
\newblock \href {http://arxiv.org/abs/2102.13648} {\path{arXiv:2102.13648}},
  \href {https://doi.org/10.1007/JHEP05(2021)182}
  {\path{doi:10.1007/JHEP05(2021)182}}.

\bibitem{LHCb:2022kph}
{Measurement of the $Z$ boson production cross-section in proton-lead
  collisions at $\sqrt{s_\mathrm{NN}}=8.16\,\mathrm{TeV}$}.
\newblock 5 2022.
\newblock \href {http://arxiv.org/abs/2205.10213} {\path{arXiv:2205.10213}}.

\bibitem{CMS:2019leu}
Albert~M Sirunyan et~al.
\newblock {Observation of nuclear modifications in W$^\pm$ boson production in
  pPb collisions at $\sqrt{s_\mathrm{NN}} =$ 8.16 TeV}.
\newblock {\em Phys. Lett. B}, 800:135048, 2020.
\newblock \href {http://arxiv.org/abs/1905.01486} {\path{arXiv:1905.01486}},
  \href {https://doi.org/10.1016/j.physletb.2019.135048}
  {\path{doi:10.1016/j.physletb.2019.135048}}.

\bibitem{ALICE:2022cxs}
{W$^\pm$-boson production in p$-$Pb collisions at $\sqrt{s_{NN}} = 8.16$ TeV
  and PbPb collisions at $\sqrt{s_{NN}} = 5.02$ TeV}.
\newblock 4 2022.
\newblock \href {http://arxiv.org/abs/2204.10640} {\path{arXiv:2204.10640}}.

\bibitem{Duwentaster:2021ioo}
P.~Duwent\"aster, L.~A. Husov\'a, T.~Je\v{z}o, M.~Klasen, K.~Kova\v{r}\'\i{}k,
  A.~Kusina, K.~F. Muzakka, F.~I. Olness, I.~Schienbein, and J.~Y. Yu.
\newblock {Impact of inclusive hadron production data on nuclear gluon PDFs}.
\newblock {\em Phys. Rev. D}, 104:094005, 2021.
\newblock \href {http://arxiv.org/abs/2105.09873} {\path{arXiv:2105.09873}},
  \href {https://doi.org/10.1103/PhysRevD.104.094005}
  {\path{doi:10.1103/PhysRevD.104.094005}}.

\bibitem{Eskola:2021nhw}
Kari~J. Eskola, Petja Paakkinen, Hannu Paukkunen, and Carlos~A. Salgado.
\newblock {EPPS21: a global QCD analysis of nuclear PDFs}.
\newblock {\em Eur. Phys. J. C}, 82(5):413, 2022.
\newblock \href {http://arxiv.org/abs/2112.12462} {\path{arXiv:2112.12462}},
  \href {https://doi.org/10.1140/epjc/s10052-022-10359-0}
  {\path{doi:10.1140/epjc/s10052-022-10359-0}}.

\bibitem{AbdulKhalek:2022fyi}
Rabah Abdul~Khalek, Rhorry Gauld, Tommaso Giani, Emanuele~R. Nocera, Tanjona~R.
  Rabemananjara, and Juan Rojo.
\newblock {nNNPDF3.0: evidence for a modified partonic structure in heavy
  nuclei}.
\newblock {\em Eur. Phys. J. C}, 82(6):507, 2022.
\newblock \href {http://arxiv.org/abs/2201.12363} {\path{arXiv:2201.12363}},
  \href {https://doi.org/10.1140/epjc/s10052-022-10417-7}
  {\path{doi:10.1140/epjc/s10052-022-10417-7}}.

\bibitem{Helenius:2021tof}
Ilkka Helenius, Marina Walt, and Werner Vogelsang.
\newblock {NNLO nuclear parton distribution functions with electroweak-boson
  production data from the LHC}.
\newblock {\em Phys. Rev. D}, 105(9):094031, 2022.
\newblock \href {http://arxiv.org/abs/2112.11904} {\path{arXiv:2112.11904}},
  \href {https://doi.org/10.1103/PhysRevD.105.094031}
  {\path{doi:10.1103/PhysRevD.105.094031}}.

\bibitem{CLAS:2017udk}
M.~Hattawy et~al.
\newblock {First Exclusive Measurement of Deeply Virtual Compton Scattering off
  $^4$He: Toward the 3D Tomography of Nuclei}.
\newblock {\em Phys. Rev. Lett.}, 119(20):202004, 2017.
\newblock \href {http://arxiv.org/abs/1707.03361} {\path{arXiv:1707.03361}},
  \href {https://doi.org/10.1103/PhysRevLett.119.202004}
  {\path{doi:10.1103/PhysRevLett.119.202004}}.

\bibitem{CLAS:2018ddh}
M.~Hattawy et~al.
\newblock {Exploring the Structure of the Bound Proton with Deeply Virtual
  Compton Scattering}.
\newblock {\em Phys. Rev. Lett.}, 123(3):032502, 2019.
\newblock \href {http://arxiv.org/abs/1812.07628} {\path{arXiv:1812.07628}},
  \href {https://doi.org/10.1103/PhysRevLett.123.032502}
  {\path{doi:10.1103/PhysRevLett.123.032502}}.

\bibitem{CLAS:2021ovm}
R.~Dupr\'e et~al.
\newblock {Measurement of deeply virtual Compton scattering off
  $^{4}\mathrm{He}$ with the CEBAF Large Acceptance Spectrometer at Jefferson
  Lab}.
\newblock {\em Phys. Rev. C}, 104(2):025203, 2021.
\newblock \href {http://arxiv.org/abs/2102.07419} {\path{arXiv:2102.07419}},
  \href {https://doi.org/10.1103/PhysRevC.104.025203}
  {\path{doi:10.1103/PhysRevC.104.025203}}.

\bibitem{LHCb:2021vww}
Roel Aaij et~al.
\newblock {Measurement of the Nuclear Modification Factor and Prompt Charged
  Particle Production in $p-Pb$ and $pp$ Collisions at $\sqrt
  {s_{NN}}$=5\,\,TeV}.
\newblock {\em Phys. Rev. Lett.}, 128(14):142004, 2022.
\newblock \href {http://arxiv.org/abs/2108.13115} {\path{arXiv:2108.13115}},
  \href {https://doi.org/10.1103/PhysRevLett.128.142004}
  {\path{doi:10.1103/PhysRevLett.128.142004}}.

\bibitem{Shi:2021hwx}
Yu~Shi, Lei Wang, Shu-Yi Wei, and Bo-Wen Xiao.
\newblock {Pursuing the Precision Study for Color Glass Condensate in Forward
  Hadron Productions}.
\newblock {\em Phys. Rev. Lett.}, 128(20):202302, 2022.
\newblock \href {http://arxiv.org/abs/2112.06975} {\path{arXiv:2112.06975}},
  \href {https://doi.org/10.1103/PhysRevLett.128.202302}
  {\path{doi:10.1103/PhysRevLett.128.202302}}.

\bibitem{BRAHMS:2004xry}
I.~Arsene et~al.
\newblock {On the evolution of the nuclear modification factors with rapidity
  and centrality in d + Au collisions at s(NN)**(1/2) = 200-GeV}.
\newblock {\em Phys. Rev. Lett.}, 93:242303, 2004.
\newblock \href {http://arxiv.org/abs/nucl-ex/0403005}
  {\path{arXiv:nucl-ex/0403005}}, \href
  {https://doi.org/10.1103/PhysRevLett.93.242303}
  {\path{doi:10.1103/PhysRevLett.93.242303}}.

\bibitem{STAR:2006dgg}
J.~Adams et~al.
\newblock {Forward neutral pion production in p+p and d+Au collisions at
  s(NN)**(1/2) = 200-GeV}.
\newblock {\em Phys. Rev. Lett.}, 97:152302, 2006.
\newblock \href {http://arxiv.org/abs/nucl-ex/0602011}
  {\path{arXiv:nucl-ex/0602011}}, \href
  {https://doi.org/10.1103/PhysRevLett.97.152302}
  {\path{doi:10.1103/PhysRevLett.97.152302}}.

\bibitem{Marquet:2007vb}
Cyrille Marquet.
\newblock {Forward inclusive dijet production and azimuthal correlations in
  p(A) collisions}.
\newblock {\em Nucl. Phys. A}, 796:41--60, 2007.
\newblock \href {http://arxiv.org/abs/0708.0231} {\path{arXiv:0708.0231}},
  \href {https://doi.org/10.1016/j.nuclphysa.2007.09.001}
  {\path{doi:10.1016/j.nuclphysa.2007.09.001}}.

\bibitem{Dominguez:2011wm}
Fabio Dominguez, Cyrille Marquet, Bo-Wen Xiao, and Feng Yuan.
\newblock {Universality of Unintegrated Gluon Distributions at small x}.
\newblock {\em Phys. Rev. D}, 83:105005, 2011.
\newblock \href {http://arxiv.org/abs/1101.0715} {\path{arXiv:1101.0715}},
  \href {https://doi.org/10.1103/PhysRevD.83.105005}
  {\path{doi:10.1103/PhysRevD.83.105005}}.

\bibitem{Braidot:2010ig}
Ermes Braidot.
\newblock {Two Particle Correlations at Forward Rapidity in STAR}.
\newblock {\em Nucl. Phys. A}, 854:168--174, 2011.
\newblock \href {http://arxiv.org/abs/1008.3989} {\path{arXiv:1008.3989}},
  \href {https://doi.org/10.1016/j.nuclphysa.2011.01.016}
  {\path{doi:10.1016/j.nuclphysa.2011.01.016}}.

\bibitem{PHENIX:2011puq}
A.~Adare et~al.
\newblock {Suppression of back-to-back hadron pairs at forward rapidity in
  $d+$Au Collisions at $\sqrt{s_{NN}}=200$ GeV}.
\newblock {\em Phys. Rev. Lett.}, 107:172301, 2011.
\newblock \href {http://arxiv.org/abs/1105.5112} {\path{arXiv:1105.5112}},
  \href {https://doi.org/10.1103/PhysRevLett.107.172301}
  {\path{doi:10.1103/PhysRevLett.107.172301}}.

\bibitem{Albacete:2010pg}
Javier~L. Albacete and Cyrille Marquet.
\newblock {Azimuthal correlations of forward di-hadrons in d+Au collisions at
  RHIC in the Color Glass Condensate}.
\newblock {\em Phys. Rev. Lett.}, 105:162301, 2010.
\newblock \href {http://arxiv.org/abs/1005.4065} {\path{arXiv:1005.4065}},
  \href {https://doi.org/10.1103/PhysRevLett.105.162301}
  {\path{doi:10.1103/PhysRevLett.105.162301}}.

\bibitem{Stasto:2011ru}
Anna Stasto, Bo-Wen Xiao, and Feng Yuan.
\newblock {Back-to-Back Correlations of Di-hadrons in dAu Collisions at RHIC}.
\newblock {\em Phys. Lett. B}, 716:430--434, 2012.
\newblock \href {http://arxiv.org/abs/1109.1817} {\path{arXiv:1109.1817}},
  \href {https://doi.org/10.1016/j.physletb.2012.08.044}
  {\path{doi:10.1016/j.physletb.2012.08.044}}.

\bibitem{STAR:2021fgw}
M.~S. Abdallah et~al.
\newblock {Evidence for Nonlinear Gluon Effects in QCD and Their Mass Number
  Dependence at STAR}.
\newblock {\em Phys. Rev. Lett.}, 129(9):092501, 2022.
\newblock \href {http://arxiv.org/abs/2111.10396} {\path{arXiv:2111.10396}},
  \href {https://doi.org/10.1103/PhysRevLett.129.092501}
  {\path{doi:10.1103/PhysRevLett.129.092501}}.

\bibitem{ATLAS:2019jgo}
Morad Aaboud et~al.
\newblock {Dijet azimuthal correlations and conditional yields in pp and p+Pb
  collisions at sNN=5.02TeV with the ATLAS detector}.
\newblock {\em Phys. Rev. C}, 100(3):034903, 2019.
\newblock \href {http://arxiv.org/abs/1901.10440} {\path{arXiv:1901.10440}},
  \href {https://doi.org/10.1103/PhysRevC.100.034903}
  {\path{doi:10.1103/PhysRevC.100.034903}}.

\bibitem{vanHameren:2019ysa}
Andreas van Hameren, Piotr Kotko, Krzysztof Kutak, and Sebastian Sapeta.
\newblock {Broadening and saturation effects in dijet azimuthal correlations in
  p-p and p-Pb collisions at $\mathbf{\sqrt{s}} = $ 5.02 TeV}.
\newblock {\em Phys. Lett. B}, 795:511--515, 2019.
\newblock \href {http://arxiv.org/abs/1903.01361} {\path{arXiv:1903.01361}},
  \href {https://doi.org/10.1016/j.physletb.2019.06.055}
  {\path{doi:10.1016/j.physletb.2019.06.055}}.

\bibitem{CLAS:2021jhm}
S.~Moran et~al.
\newblock {Measurement of charged-pion production in deep-inelastic scattering
  off nuclei with the CLAS detector}.
\newblock {\em Phys. Rev. C}, 105(1):015201, 2022.
\newblock \href {http://arxiv.org/abs/2109.09951} {\path{arXiv:2109.09951}},
  \href {https://doi.org/10.1103/PhysRevC.105.015201}
  {\path{doi:10.1103/PhysRevC.105.015201}}.

\bibitem{CLAS:2022asf}
S.~J. Paul et~al.
\newblock {Observation of Azimuth-Dependent Suppression of Hadron Pairs in
  Electron Scattering off Nuclei}.
\newblock {\em Phys. Rev. Lett.}, 129(18):182501, 2022.
\newblock \href {http://arxiv.org/abs/2207.06682} {\path{arXiv:2207.06682}},
  \href {https://doi.org/10.1103/PhysRevLett.129.182501}
  {\path{doi:10.1103/PhysRevLett.129.182501}}.

\bibitem{CLAS:2022oux}
T.~Chetry et~al.
\newblock {First Measurement of $\Lambda$ Electroproduction off Nuclei in the
  Current and Target Fragmentation Regions}.
\newblock 10 2022.
\newblock \href {http://arxiv.org/abs/2210.13691} {\path{arXiv:2210.13691}}.

\bibitem{HallC:2020ijh}
D.~Bhetuwal et~al.
\newblock {Ruling out Color Transparency in Quasielastic $^{12}$C(e,e'p) up to
  $Q^2$ of 14.2 (GeV/c)$^2$}.
\newblock {\em Phys. Rev. Lett.}, 126(8):082301, 2021.
\newblock \href {http://arxiv.org/abs/2011.00703} {\path{arXiv:2011.00703}},
  \href {https://doi.org/10.1103/PhysRevLett.126.082301}
  {\path{doi:10.1103/PhysRevLett.126.082301}}.

\bibitem{Clasie:2007aa}
B.~Clasie et~al.
\newblock {Measurement of nuclear transparency for the A(e, e-prime' pi+)
  reaction}.
\newblock {\em Phys. Rev. Lett.}, 99:242502, 2007.
\newblock \href {http://arxiv.org/abs/0707.1481} {\path{arXiv:0707.1481}},
  \href {https://doi.org/10.1103/PhysRevLett.99.242502}
  {\path{doi:10.1103/PhysRevLett.99.242502}}.

\bibitem{Qian:2009aa}
X.~Qian et~al.
\newblock {Experimental tudy of the A(e,e'$\pi^+$) Reaction on $^1$H, $^2$H,
  $^{12}$C, $^{27}$Al, $^{63}$Cu and $^{197}$Au}.
\newblock {\em Phys. Rev. C}, 81:055209, 2010.
\newblock \href {http://arxiv.org/abs/0908.1616} {\path{arXiv:0908.1616}},
  \href {https://doi.org/10.1103/PhysRevC.81.055209}
  {\path{doi:10.1103/PhysRevC.81.055209}}.

\bibitem{CLAS:2012tlh}
L.~El~Fassi et~al.
\newblock {Evidence for the onset of color transparency in $\rho^0$
  electroproduction off nuclei}.
\newblock {\em Phys. Lett. B}, 712:326--330, 2012.
\newblock \href {http://arxiv.org/abs/1201.2735} {\path{arXiv:1201.2735}},
  \href {https://doi.org/10.1016/j.physletb.2012.05.019}
  {\path{doi:10.1016/j.physletb.2012.05.019}}.

\bibitem{CLAS:2022Elfassi}
Lamiaa {El Fassi}.
\newblock {Chasing QCD Signatures in Nuclei Using Color Coherence Phenomena}.
\newblock {\em Physics}, 4(3):970--980, August 2022.
\newblock \href {https://doi.org/10.3390/physics4030064}
  {\path{doi:10.3390/physics4030064}}.

\bibitem{2022:Huber}
Garth~M. Huber, Wenliang~B. Li, Wim Cosyn, and Bernard Pire.
\newblock u-channel color transparency observables.
\newblock {\em Physics}, 4(2):451--461, apr 2022.
\newblock URL: \url{https://doi.org/10.3390%2Fphysics4020030}, \href
  {https://doi.org/10.3390/physics4020030} {\path{doi:10.3390/physics4020030}}.

\bibitem{Wilson:1974sk}
Kenneth~G. Wilson.
\newblock {Confinement of Quarks}.
\newblock {\em Phys. Rev. D}, 10:2445--2459, 1974.
\newblock \href {https://doi.org/10.1103/PhysRevD.10.2445}
  {\path{doi:10.1103/PhysRevD.10.2445}}.

\bibitem{FlavourLatticeAveragingGroupFLAG:2021npn}
Y.~Aoki et~al.
\newblock {FLAG Review 2021}.
\newblock {\em Eur. Phys. J. C}, 82(10):869, 2022.
\newblock \href {http://arxiv.org/abs/2111.09849} {\path{arXiv:2111.09849}},
  \href {https://doi.org/10.1140/epjc/s10052-022-10536-1}
  {\path{doi:10.1140/epjc/s10052-022-10536-1}}.

\bibitem{Bhattacharya:2016zcn}
Tanmoy Bhattacharya, Vincenzo Cirigliano, Saul Cohen, Rajan Gupta, Huey-Wen
  Lin, and Boram Yoon.
\newblock {Axial, Scalar and Tensor Charges of the Nucleon from 2+1+1-flavor
  Lattice QCD}.
\newblock {\em Phys. Rev. D}, 94(5):054508, 2016.
\newblock \href {http://arxiv.org/abs/1606.07049} {\path{arXiv:1606.07049}},
  \href {https://doi.org/10.1103/PhysRevD.94.054508}
  {\path{doi:10.1103/PhysRevD.94.054508}}.

\bibitem{Berkowitz:2017gql}
Evan Berkowitz et~al.
\newblock {An accurate calculation of the nucleon axial charge with lattice
  QCD}.
\newblock 4 2017.
\newblock \href {http://arxiv.org/abs/1704.01114} {\path{arXiv:1704.01114}}.

\bibitem{Chang:2018uxx}
C.~C. Chang et~al.
\newblock {A per-cent-level determination of the nucleon axial coupling from
  quantum chromodynamics}.
\newblock {\em Nature}, 558(7708):91--94, 2018.
\newblock \href {http://arxiv.org/abs/1805.12130} {\path{arXiv:1805.12130}},
  \href {https://doi.org/10.1038/s41586-018-0161-8}
  {\path{doi:10.1038/s41586-018-0161-8}}.

\bibitem{Liang:2018pis}
Jian Liang, Yi-Bo Yang, Terrence Draper, Ming Gong, and Keh-Fei Liu.
\newblock {Quark spins and Anomalous Ward Identity}.
\newblock {\em Phys. Rev. D}, 98(7):074505, 2018.
\newblock \href {http://arxiv.org/abs/1806.08366} {\path{arXiv:1806.08366}},
  \href {https://doi.org/10.1103/PhysRevD.98.074505}
  {\path{doi:10.1103/PhysRevD.98.074505}}.

\bibitem{Gupta:2018qil}
Rajan Gupta, Yong-Chull Jang, Boram Yoon, Huey-Wen Lin, Vincenzo Cirigliano,
  and Tanmoy Bhattacharya.
\newblock {Isovector Charges of the Nucleon from 2+1+1-flavor Lattice QCD}.
\newblock {\em Phys. Rev. D}, 98:034503, 2018.
\newblock \href {http://arxiv.org/abs/1806.09006} {\path{arXiv:1806.09006}},
  \href {https://doi.org/10.1103/PhysRevD.98.034503}
  {\path{doi:10.1103/PhysRevD.98.034503}}.

\bibitem{Shintani:2018ozy}
Eigo Shintani, Ken-Ichi Ishikawa, Yoshinobu Kuramashi, Shoichi Sasaki, and
  Takeshi Yamazaki.
\newblock {Nucleon form factors and root-mean-square radii on a (10.8 fm)$^4$
  lattice at the physical point}.
\newblock {\em Phys. Rev. D}, 99(1):014510, 2019.
\newblock [Erratum: Phys.Rev.D 102, 019902 (2020)].
\newblock \href {http://arxiv.org/abs/1811.07292} {\path{arXiv:1811.07292}},
  \href {https://doi.org/10.1103/PhysRevD.99.014510}
  {\path{doi:10.1103/PhysRevD.99.014510}}.

\bibitem{Hasan:2019noy}
Nesreen Hasan, Jeremy Green, Stefan Meinel, Michael Engelhardt, Stefan Krieg,
  John Negele, Andrew Pochinsky, and Sergey Syritsyn.
\newblock {Nucleon axial, scalar, and tensor charges using lattice QCD at the
  physical pion mass}.
\newblock {\em Phys. Rev. D}, 99(11):114505, 2019.
\newblock \href {http://arxiv.org/abs/1903.06487} {\path{arXiv:1903.06487}},
  \href {https://doi.org/10.1103/PhysRevD.99.114505}
  {\path{doi:10.1103/PhysRevD.99.114505}}.

\bibitem{Harris:2019bih}
Tim Harris, Georg von Hippel, Parikshit Junnarkar, Harvey~B. Meyer, Konstantin
  Ottnad, Jonas Wilhelm, Hartmut Wittig, and Linus Wrang.
\newblock {Nucleon isovector charges and twist-2 matrix elements with $N_f=2+1$
  dynamical Wilson quarks}.
\newblock {\em Phys. Rev. D}, 100(3):034513, 2019.
\newblock \href {http://arxiv.org/abs/1905.01291} {\path{arXiv:1905.01291}},
  \href {https://doi.org/10.1103/PhysRevD.100.034513}
  {\path{doi:10.1103/PhysRevD.100.034513}}.

\bibitem{Alexandrou:2019brg}
C.~Alexandrou, S.~Bacchio, M.~Constantinou, J.~Finkenrath, K.~Hadjiyiannakou,
  K.~Jansen, G.~Koutsou, and A.~Vaquero Aviles-Casco.
\newblock {Nucleon axial, tensor, and scalar charges and $\sigma$-terms in
  lattice QCD}.
\newblock {\em Phys. Rev. D}, 102(5):054517, 2020.
\newblock \href {http://arxiv.org/abs/1909.00485} {\path{arXiv:1909.00485}},
  \href {https://doi.org/10.1103/PhysRevD.102.054517}
  {\path{doi:10.1103/PhysRevD.102.054517}}.

\bibitem{Walker-Loud:2019cif}
Andr\'e Walker-Loud et~al.
\newblock {Lattice QCD Determination of $g_A$}.
\newblock {\em PoS}, CD2018:020, 2020.
\newblock \href {http://arxiv.org/abs/1912.08321} {\path{arXiv:1912.08321}},
  \href {https://doi.org/10.22323/1.317.0020} {\path{doi:10.22323/1.317.0020}}.

\bibitem{Park:2021ypf}
Sungwoo Park, Rajan Gupta, Boram Yoon, Santanu Mondal, Tanmoy Bhattacharya,
  Yong-Chull Jang, B\'alint Jo\'o, and Frank Winter.
\newblock {Precision nucleon charges and form factors using (2+1)-flavor
  lattice QCD}.
\newblock {\em Phys. Rev. D}, 105(5):054505, 2022.
\newblock \href {http://arxiv.org/abs/2103.05599} {\path{arXiv:2103.05599}},
  \href {https://doi.org/10.1103/PhysRevD.105.054505}
  {\path{doi:10.1103/PhysRevD.105.054505}}.

\bibitem{Cirigliano:2022hob}
Vincenzo Cirigliano, Jordy de~Vries, Leendert Hayen, Emanuele Mereghetti, and
  Andr\'e Walker-Loud.
\newblock {Pion-Induced Radiative Corrections to Neutron \ensuremath{\beta}
  Decay}.
\newblock {\em Phys. Rev. Lett.}, 129(12):121801, 2022.
\newblock \href {http://arxiv.org/abs/2202.10439} {\path{arXiv:2202.10439}},
  \href {https://doi.org/10.1103/PhysRevLett.129.121801}
  {\path{doi:10.1103/PhysRevLett.129.121801}}.

\bibitem{Lin:2017stx}
Huey-Wen Lin, W.~Melnitchouk, Alexei Prokudin, N.~Sato, and H.~Shows.
\newblock {First Monte Carlo Global Analysis of Nucleon Transversity with
  Lattice QCD Constraints}.
\newblock {\em Phys. Rev. Lett.}, 120(15):152502, 2018.
\newblock \href {http://arxiv.org/abs/1710.09858} {\path{arXiv:1710.09858}},
  \href {https://doi.org/10.1103/PhysRevLett.120.152502}
  {\path{doi:10.1103/PhysRevLett.120.152502}}.

\bibitem{Alexandrou:2018sjm}
C.~Alexandrou, S.~Bacchio, M.~Constantinou, J.~Finkenrath, K.~Hadjiyiannakou,
  K.~Jansen, G.~Koutsou, and A.~Vaquero Aviles-Casco.
\newblock {Proton and neutron electromagnetic form factors from lattice QCD}.
\newblock {\em Phys. Rev. D}, 100(1):014509, 2019.
\newblock \href {http://arxiv.org/abs/1812.10311} {\path{arXiv:1812.10311}},
  \href {https://doi.org/10.1103/PhysRevD.100.014509}
  {\path{doi:10.1103/PhysRevD.100.014509}}.

\bibitem{Djukanovic:2021cgp}
D.~Djukanovic, T.~Harris, G.~von Hippel, P.~M. Junnarkar, H.~B. Meyer,
  D.~Mohler, K.~Ottnad, T.~Schulz, J.~Wilhelm, and H.~Wittig.
\newblock {Isovector electromagnetic form factors of the nucleon from lattice
  QCD and the proton radius puzzle}.
\newblock {\em Phys. Rev. D}, 103(9):094522, 2021.
\newblock \href {http://arxiv.org/abs/2102.07460} {\path{arXiv:2102.07460}},
  \href {https://doi.org/10.1103/PhysRevD.103.094522}
  {\path{doi:10.1103/PhysRevD.103.094522}}.

\bibitem{Guo:2022upw}
Yuxun Guo, Xiangdong Ji, and Kyle Shiells.
\newblock {Generalized parton distributions through universal moment
  parameterization: zero skewness case}.
\newblock {\em JHEP}, 09:215, 2022.
\newblock \href {http://arxiv.org/abs/2207.05768} {\path{arXiv:2207.05768}},
  \href {https://doi.org/10.1007/JHEP09(2022)215}
  {\path{doi:10.1007/JHEP09(2022)215}}.

\bibitem{Cui:2020dlm}
Zhu-Fang Cui, Minghui Ding, Fei Gao, Khepani Raya, Daniele Binosi, Lei Chang,
  Craig~D. Roberts, Jose Rodriguez-Quintero, and Sebastian~M. Schmidt.
\newblock {Higgs modulation of emergent mass as revealed in kaon and pion
  parton distributions}.
\newblock {\em Eur. Phys. J. A}, 57(1):5, 2021.
\newblock \href {http://arxiv.org/abs/2006.14075} {\path{arXiv:2006.14075}},
  \href {https://doi.org/10.1140/epja/s10050-020-00318-2}
  {\path{doi:10.1140/epja/s10050-020-00318-2}}.

\bibitem{Roberts:2020udq}
Craig~D. Roberts and Sebastian~M. Schmidt.
\newblock {Reflections upon the emergence of hadronic mass}.
\newblock {\em Eur. Phys. J. ST}, 229(22-23):3319--3340, 2020.
\newblock \href {http://arxiv.org/abs/2006.08782} {\path{arXiv:2006.08782}},
  \href {https://doi.org/10.1140/epjst/e2020-000064-6}
  {\path{doi:10.1140/epjst/e2020-000064-6}}.

\bibitem{Dudek:2012vr}
Jozef Dudek et~al.
\newblock {Physics Opportunities with the 12 GeV Upgrade at Jefferson Lab}.
\newblock {\em Eur. Phys. J. A}, 48:187, 2012.
\newblock \href {http://arxiv.org/abs/1208.1244} {\path{arXiv:1208.1244}},
  \href {https://doi.org/10.1140/epja/i2012-12187-1}
  {\path{doi:10.1140/epja/i2012-12187-1}}.

\bibitem{AbdulKhalek:2021gbh}
R.~Abdul~Khalek et~al.
\newblock {Science Requirements and Detector Concepts for the Electron-Ion
  Collider}: {EIC Yellow Report}.
\newblock {\em Nucl. Phys. A}, 1026:122447, 2022.
\newblock \href {http://arxiv.org/abs/2103.05419} {\path{arXiv:2103.05419}},
  \href {https://doi.org/10.1016/j.nuclphysa.2022.122447}
  {\path{doi:10.1016/j.nuclphysa.2022.122447}}.

\bibitem{Choi:2019nvk}
Ho-Meoyng Choi, T.~Frederico, Chueng-Ryong Ji, and J.~P. B.~C. de~Melo.
\newblock {Pion off-shell electromagnetic form factors: data extraction and
  model analysis}.
\newblock {\em Phys. Rev. D}, 100(11):116020, 2019.
\newblock \href {http://arxiv.org/abs/1908.01185} {\path{arXiv:1908.01185}},
  \href {https://doi.org/10.1103/PhysRevD.100.116020}
  {\path{doi:10.1103/PhysRevD.100.116020}}.

\bibitem{Brommel:2006ww}
D.~Br\"ommel et~al.
\newblock {The Pion form-factor from lattice QCD with two dynamical flavours}.
\newblock {\em Eur. Phys. J. C}, 51:335--345, 2007.
\newblock \href {http://arxiv.org/abs/hep-lat/0608021}
  {\path{arXiv:hep-lat/0608021}}, \href
  {https://doi.org/10.1140/epjc/s10052-007-0295-6}
  {\path{doi:10.1140/epjc/s10052-007-0295-6}}.

\bibitem{Frezzotti:2008dr}
R.~Frezzotti, V.~Lubicz, and S.~Simula.
\newblock {Electromagnetic form factor of the pion from twisted-mass lattice
  QCD at N(f) = 2}.
\newblock {\em Phys. Rev. D}, 79:074506, 2009.
\newblock \href {http://arxiv.org/abs/0812.4042} {\path{arXiv:0812.4042}},
  \href {https://doi.org/10.1103/PhysRevD.79.074506}
  {\path{doi:10.1103/PhysRevD.79.074506}}.

\bibitem{Aoki:2009qn}
S.~Aoki et~al.
\newblock {Pion form factors from two-flavor lattice QCD with exact chiral
  symmetry}.
\newblock {\em Phys. Rev. D}, 80:034508, 2009.
\newblock \href {http://arxiv.org/abs/0905.2465} {\path{arXiv:0905.2465}},
  \href {https://doi.org/10.1103/PhysRevD.80.034508}
  {\path{doi:10.1103/PhysRevD.80.034508}}.

\bibitem{Brandt:2013dua}
Bastian~B. Brandt, Andreas J\"uttner, and Hartmut Wittig.
\newblock {The pion vector form factor from lattice QCD and NNLO chiral
  perturbation theory}.
\newblock {\em JHEP}, 11:034, 2013.
\newblock \href {http://arxiv.org/abs/1306.2916} {\path{arXiv:1306.2916}},
  \href {https://doi.org/10.1007/JHEP11(2013)034}
  {\path{doi:10.1007/JHEP11(2013)034}}.

\bibitem{Alexandrou:2017blh}
C.~Alexandrou et~al.
\newblock {Pion vector form factor from lattice QCD at the physical point}.
\newblock {\em Phys. Rev. D}, 97(1):014508, 2018.
\newblock \href {http://arxiv.org/abs/1710.10401} {\path{arXiv:1710.10401}},
  \href {https://doi.org/10.1103/PhysRevD.97.014508}
  {\path{doi:10.1103/PhysRevD.97.014508}}.

\bibitem{Bonnet:2004fr}
Frederic D.~R. Bonnet, Robert~G. Edwards, George~Tamminga Fleming, Randy Lewis,
  and David~G. Richards.
\newblock {Lattice computations of the pion form-factor}.
\newblock {\em Phys. Rev. D}, 72:054506, 2005.
\newblock \href {http://arxiv.org/abs/hep-lat/0411028}
  {\path{arXiv:hep-lat/0411028}}, \href
  {https://doi.org/10.1103/PhysRevD.72.054506}
  {\path{doi:10.1103/PhysRevD.72.054506}}.

\bibitem{Boyle:2008yd}
P.~A. Boyle, J.~M. Flynn, A.~Juttner, C.~Kelly, H.~Pedroso de~Lima, C.~M.
  Maynard, C.~T. Sachrajda, and J.~M. Zanotti.
\newblock {The Pion's electromagnetic form-factor at small momentum transfer in
  full lattice QCD}.
\newblock {\em JHEP}, 07:112, 2008.
\newblock \href {http://arxiv.org/abs/0804.3971} {\path{arXiv:0804.3971}},
  \href {https://doi.org/10.1088/1126-6708/2008/07/112}
  {\path{doi:10.1088/1126-6708/2008/07/112}}.

\bibitem{Nguyen:2011ek}
Oanh~Hoang Nguyen, Ken-Ichi Ishikawa, Akira Ukawa, and Naoya Ukita.
\newblock {Electromagnetic form factor of pion from $N_f=2+1$ dynamical flavor
  QCD}.
\newblock {\em JHEP}, 04:122, 2011.
\newblock \href {http://arxiv.org/abs/1102.3652} {\path{arXiv:1102.3652}},
  \href {https://doi.org/10.1007/JHEP04(2011)122}
  {\path{doi:10.1007/JHEP04(2011)122}}.

\bibitem{Fukaya:2014jka}
H.~Fukaya, S.~Aoki, S.~Hashimoto, T.~Kaneko, H.~Matsufuru, and J.~Noaki.
\newblock {Computation of the electromagnetic pion form factor from lattice QCD
  in the $\epsilon$ regime}.
\newblock {\em Phys. Rev. D}, 90(3):034506, 2014.
\newblock \href {http://arxiv.org/abs/1405.4077} {\path{arXiv:1405.4077}},
  \href {https://doi.org/10.1103/PhysRevD.90.034506}
  {\path{doi:10.1103/PhysRevD.90.034506}}.

\bibitem{Aoki:2015pba}
S.~Aoki, G.~Cossu, X.~Feng, S.~Hashimoto, T.~Kaneko, J.~Noaki, and T.~Onogi.
\newblock {Light meson electromagnetic form factors from three-flavor lattice
  QCD with exact chiral symmetry}.
\newblock {\em Phys. Rev. D}, 93(3):034504, 2016.
\newblock \href {http://arxiv.org/abs/1510.06470} {\path{arXiv:1510.06470}},
  \href {https://doi.org/10.1103/PhysRevD.93.034504}
  {\path{doi:10.1103/PhysRevD.93.034504}}.

\bibitem{Feng:2019geu}
Xu~Feng, Yang Fu, and Lu-Chang Jin.
\newblock {Lattice QCD calculation of the pion charge radius using a
  model-independent method}.
\newblock {\em Phys. Rev. D}, 101(5):051502, 2020.
\newblock \href {http://arxiv.org/abs/1911.04064} {\path{arXiv:1911.04064}},
  \href {https://doi.org/10.1103/PhysRevD.101.051502}
  {\path{doi:10.1103/PhysRevD.101.051502}}.

\bibitem{Wang:2020nbf}
Gen Wang, Jian Liang, Terrence Draper, Keh-Fei Liu, and Yi-Bo Yang.
\newblock {Lattice Calculation of Pion Form Factor with Overlap Fermions}.
\newblock 6 2020.
\newblock \href {http://arxiv.org/abs/2006.05431} {\path{arXiv:2006.05431}}.

\bibitem{Gao:2021xsm}
Xiang Gao, Nikhil Karthik, Swagato Mukherjee, Peter Petreczky, Sergey Syritsyn,
  and Yong Zhao.
\newblock {Pion form factor and charge radius from lattice QCD at the physical
  point}.
\newblock {\em Phys. Rev. D}, 104(11):114515, 2021.
\newblock \href {http://arxiv.org/abs/2102.06047} {\path{arXiv:2102.06047}},
  \href {https://doi.org/10.1103/PhysRevD.104.114515}
  {\path{doi:10.1103/PhysRevD.104.114515}}.

\bibitem{Koponen:2015tkr}
J.~Koponen, F.~Bursa, C.T.H. Davies, R.J. Dowdall, and G.P. Lepage.
\newblock {Size of the pion from full lattice QCD with physical u , d , s and c
  quarks}.
\newblock {\em Phys. Rev. D}, 93(5):054503, 2016.
\newblock \href {http://arxiv.org/abs/1511.07382} {\path{arXiv:1511.07382}},
  \href {https://doi.org/10.1103/PhysRevD.93.054503}
  {\path{doi:10.1103/PhysRevD.93.054503}}.

\bibitem{Bali:2016lva}
Gunnar~S. Bali, Bernhard Lang, Bernhard~U. Musch, and Andreas Sch\"afer.
\newblock {Novel quark smearing for hadrons with high momenta in lattice QCD}.
\newblock {\em Phys. Rev. D}, 93(9):094515, 2016.
\newblock \href {http://arxiv.org/abs/1602.05525} {\path{arXiv:1602.05525}},
  \href {https://doi.org/10.1103/PhysRevD.93.094515}
  {\path{doi:10.1103/PhysRevD.93.094515}}.

\bibitem{Ji:1996ek}
Xiang-Dong Ji.
\newblock {Gauge-Invariant Decomposition of Nucleon Spin}.
\newblock {\em Phys. Rev. Lett.}, 78:610--613, 1997.
\newblock \href {http://arxiv.org/abs/hep-ph/9603249}
  {\path{arXiv:hep-ph/9603249}}, \href
  {https://doi.org/10.1103/PhysRevLett.78.610}
  {\path{doi:10.1103/PhysRevLett.78.610}}.

\bibitem{Alexandrou:2020sml}
C.~Alexandrou, S.~Bacchio, M.~Constantinou, J.~Finkenrath, K.~Hadjiyiannakou,
  K.~Jansen, G.~Koutsou, H.~Panagopoulos, and G.~Spanoudes.
\newblock {Complete flavor decomposition of the spin and momentum fraction of
  the proton using lattice QCD simulations at physical pion mass}.
\newblock {\em Phys. Rev. D}, 101(9):094513, 2020.
\newblock \href {http://arxiv.org/abs/2003.08486} {\path{arXiv:2003.08486}},
  \href {https://doi.org/10.1103/PhysRevD.101.094513}
  {\path{doi:10.1103/PhysRevD.101.094513}}.

\bibitem{Wang:2021vqy}
Gen Wang, Yi-Bo Yang, Jian Liang, Terrence Draper, and Keh-Fei Liu.
\newblock {Proton momentum and angular momentum decompositions with overlap
  fermions}.
\newblock {\em Phys. Rev. D}, 106(1):014512, 2022.
\newblock \href {http://arxiv.org/abs/2111.09329} {\path{arXiv:2111.09329}},
  \href {https://doi.org/10.1103/PhysRevD.106.014512}
  {\path{doi:10.1103/PhysRevD.106.014512}}.

\bibitem{Engelhardt:2020qtg}
M.~Engelhardt, J.~R. Green, N.~Hasan, S.~Krieg, S.~Meinel, J.~Negele,
  A.~Pochinsky, and S.~Syritsyn.
\newblock {From Ji to Jaffe-Manohar orbital angular momentum in lattice QCD
  using a direct derivative method}.
\newblock {\em Phys. Rev. D}, 102(7):074505, 2020.
\newblock \href {http://arxiv.org/abs/2008.03660} {\path{arXiv:2008.03660}},
  \href {https://doi.org/10.1103/PhysRevD.102.074505}
  {\path{doi:10.1103/PhysRevD.102.074505}}.

\bibitem{Engelhardt:2021kdo}
Michael Engelhardt et~al.
\newblock {Quark spin-orbit correlations in the proton}.
\newblock {\em PoS}, LATTICE2021:413, 2022.
\newblock \href {http://arxiv.org/abs/2112.13464} {\path{arXiv:2112.13464}},
  \href {https://doi.org/10.22323/1.396.0413} {\path{doi:10.22323/1.396.0413}}.

\bibitem{Yang:2018nqn}
Yi-Bo Yang, Jian Liang, Yu-Jiang Bi, Ying Chen, Terrence Draper, Keh-Fei Liu,
  and Zhaofeng Liu.
\newblock {Proton Mass Decomposition from the QCD Energy Momentum Tensor}.
\newblock {\em Phys. Rev. Lett.}, 121(21):212001, 2018.
\newblock \href {http://arxiv.org/abs/1808.08677} {\path{arXiv:1808.08677}},
  \href {https://doi.org/10.1103/PhysRevLett.121.212001}
  {\path{doi:10.1103/PhysRevLett.121.212001}}.

\bibitem{Ji:1994av}
Xiang-Dong Ji.
\newblock {A QCD analysis of the mass structure of the nucleon}.
\newblock {\em Phys. Rev. Lett.}, 74:1071--1074, 1995.
\newblock \href {http://arxiv.org/abs/hep-ph/9410274}
  {\path{arXiv:hep-ph/9410274}}, \href
  {https://doi.org/10.1103/PhysRevLett.74.1071}
  {\path{doi:10.1103/PhysRevLett.74.1071}}.

\bibitem{Monahan:2018euv}
Christopher Monahan.
\newblock {Recent Developments in $x$-dependent Structure Calculations}.
\newblock {\em PoS}, LATTICE2018:018, 2018.
\newblock \href {http://arxiv.org/abs/1811.00678} {\path{arXiv:1811.00678}},
  \href {https://doi.org/10.22323/1.334.0018} {\path{doi:10.22323/1.334.0018}}.

\bibitem{Cichy:2018mum}
Krzysztof Cichy and Martha Constantinou.
\newblock {A guide to light-cone PDFs from Lattice QCD: an overview of
  approaches, techniques and results}.
\newblock {\em Adv. High Energy Phys.}, 2019:3036904, 2019.
\newblock \href {http://arxiv.org/abs/1811.07248} {\path{arXiv:1811.07248}},
  \href {https://doi.org/10.1155/2019/3036904}
  {\path{doi:10.1155/2019/3036904}}.

\bibitem{Ji:2020ect}
Xiangdong Ji, Yu-Sheng Liu, Yizhuang Liu, Jian-Hui Zhang, and Yong Zhao.
\newblock {Large-momentum effective theory}.
\newblock {\em Rev. Mod. Phys.}, 93(3):035005, 2021.
\newblock \href {http://arxiv.org/abs/2004.03543} {\path{arXiv:2004.03543}},
  \href {https://doi.org/10.1103/RevModPhys.93.035005}
  {\path{doi:10.1103/RevModPhys.93.035005}}.

\bibitem{Constantinou:2022yye}
Martha Constantinou et~al.
\newblock {Lattice QCD Calculations of Parton Physics}.
\newblock 2 2022.
\newblock \href {http://arxiv.org/abs/2202.07193} {\path{arXiv:2202.07193}}.

\bibitem{Liu:1993cv}
Keh-Fei Liu and Shao-Jing Dong.
\newblock {Origin of difference between anti-d and anti-u partons in the
  nucleon}.
\newblock {\em Phys. Rev. Lett.}, 72:1790--1793, 1994.
\newblock \href {http://arxiv.org/abs/hep-ph/9306299}
  {\path{arXiv:hep-ph/9306299}}, \href
  {https://doi.org/10.1103/PhysRevLett.72.1790}
  {\path{doi:10.1103/PhysRevLett.72.1790}}.

\bibitem{Liu:1998um}
K.~F. Liu, S.~J. Dong, Terrence Draper, D.~Leinweber, J.~H. Sloan, W.~Wilcox,
  and R.~M. Woloshyn.
\newblock {Valence QCD: Connecting QCD to the quark model}.
\newblock {\em Phys. Rev. D}, 59:112001, 1999.
\newblock \href {http://arxiv.org/abs/hep-ph/9806491}
  {\path{arXiv:hep-ph/9806491}}, \href
  {https://doi.org/10.1103/PhysRevD.59.112001}
  {\path{doi:10.1103/PhysRevD.59.112001}}.

\bibitem{Liu:1999ak}
Keh-Fei Liu.
\newblock {Parton degrees of freedom from the path integral formalism}.
\newblock {\em Phys. Rev. D}, 62:074501, 2000.
\newblock \href {http://arxiv.org/abs/hep-ph/9910306}
  {\path{arXiv:hep-ph/9910306}}, \href
  {https://doi.org/10.1103/PhysRevD.62.074501}
  {\path{doi:10.1103/PhysRevD.62.074501}}.

\bibitem{Detmold:2005gg}
William Detmold and C.~J.~David Lin.
\newblock {Deep-inelastic scattering and the operator product expansion in
  lattice QCD}.
\newblock {\em Phys. Rev. D}, 73:014501, 2006.
\newblock \href {http://arxiv.org/abs/hep-lat/0507007}
  {\path{arXiv:hep-lat/0507007}}, \href
  {https://doi.org/10.1103/PhysRevD.73.014501}
  {\path{doi:10.1103/PhysRevD.73.014501}}.

\bibitem{Detmold:2021uru}
William Detmold, Anthony~V. Grebe, Issaku Kanamori, C.~J.~David Lin, Robert~J.
  Perry, and Yong Zhao.
\newblock {Parton physics from a heavy-quark operator product expansion:
  Formalism and Wilson coefficients}.
\newblock {\em Phys. Rev. D}, 104(7):074511, 2021.
\newblock \href {http://arxiv.org/abs/2103.09529} {\path{arXiv:2103.09529}},
  \href {https://doi.org/10.1103/PhysRevD.104.074511}
  {\path{doi:10.1103/PhysRevD.104.074511}}.

\bibitem{Braun:2007wv}
V.~Braun and Dieter M\"uller.
\newblock {Exclusive processes in position space and the pion distribution
  amplitude}.
\newblock {\em Eur. Phys. J. C}, 55:349--361, 2008.
\newblock \href {http://arxiv.org/abs/0709.1348} {\path{arXiv:0709.1348}},
  \href {https://doi.org/10.1140/epjc/s10052-008-0608-4}
  {\path{doi:10.1140/epjc/s10052-008-0608-4}}.

\bibitem{Ji:2013dva}
Xiangdong Ji.
\newblock {Parton Physics on a Euclidean Lattice}.
\newblock {\em Phys. Rev. Lett.}, 110:262002, 2013.
\newblock \href {http://arxiv.org/abs/1305.1539} {\path{arXiv:1305.1539}},
  \href {https://doi.org/10.1103/PhysRevLett.110.262002}
  {\path{doi:10.1103/PhysRevLett.110.262002}}.

\bibitem{Ji:2014gla}
Xiangdong Ji.
\newblock {Parton Physics from Large-Momentum Effective Field Theory}.
\newblock {\em Sci. China Phys. Mech. Astron.}, 57:1407--1412, 2014.
\newblock \href {http://arxiv.org/abs/1404.6680} {\path{arXiv:1404.6680}},
  \href {https://doi.org/10.1007/s11433-014-5492-3}
  {\path{doi:10.1007/s11433-014-5492-3}}.

\bibitem{Radyushkin:2016hsy}
Anatoly Radyushkin.
\newblock {Nonperturbative Evolution of Parton Quasi-Distributions}.
\newblock {\em Phys. Lett. B}, 767:314--320, 2017.
\newblock \href {http://arxiv.org/abs/1612.05170} {\path{arXiv:1612.05170}},
  \href {https://doi.org/10.1016/j.physletb.2017.02.019}
  {\path{doi:10.1016/j.physletb.2017.02.019}}.

\bibitem{Ma:2014jla}
Yan-Qing Ma and Jian-Wei Qiu.
\newblock {Extracting Parton Distribution Functions from Lattice QCD
  Calculations}.
\newblock {\em Phys. Rev. D}, 98(7):074021, 2018.
\newblock \href {http://arxiv.org/abs/1404.6860} {\path{arXiv:1404.6860}},
  \href {https://doi.org/10.1103/PhysRevD.98.074021}
  {\path{doi:10.1103/PhysRevD.98.074021}}.

\bibitem{Ma:2014jga}
Yan-Qing Ma and Jian-Wei Qiu.
\newblock {QCD Factorization and PDFs from Lattice QCD Calculation}.
\newblock {\em Int. J. Mod. Phys. Conf. Ser.}, 37:1560041, 2015.
\newblock \href {http://arxiv.org/abs/1412.2688} {\path{arXiv:1412.2688}},
  \href {https://doi.org/10.1142/S2010194515600411}
  {\path{doi:10.1142/S2010194515600411}}.

\bibitem{Ma:2017pxb}
Yan-Qing Ma and Jian-Wei Qiu.
\newblock {Exploring Partonic Structure of Hadrons Using ab initio Lattice QCD
  Calculations}.
\newblock {\em Phys. Rev. Lett.}, 120(2):022003, 2018.
\newblock \href {http://arxiv.org/abs/1709.03018} {\path{arXiv:1709.03018}},
  \href {https://doi.org/10.1103/PhysRevLett.120.022003}
  {\path{doi:10.1103/PhysRevLett.120.022003}}.

\bibitem{Chambers:2017dov}
A.~J. Chambers, R.~Horsley, Y.~Nakamura, H.~Perlt, P.~E.~L. Rakow,
  G.~Schierholz, A.~Schiller, K.~Somfleth, R.~D. Young, and J.~M. Zanotti.
\newblock {Nucleon Structure Functions from Operator Product Expansion on the
  Lattice}.
\newblock {\em Phys. Rev. Lett.}, 118(24):242001, 2017.
\newblock \href {http://arxiv.org/abs/1703.01153} {\path{arXiv:1703.01153}},
  \href {https://doi.org/10.1103/PhysRevLett.118.242001}
  {\path{doi:10.1103/PhysRevLett.118.242001}}.

\bibitem{Alexandrou:2018pbm}
Constantia Alexandrou, Krzysztof Cichy, Martha Constantinou, Karl Jansen,
  Aurora Scapellato, and Fernanda Steffens.
\newblock {Light-Cone Parton Distribution Functions from Lattice QCD}.
\newblock {\em Phys. Rev. Lett.}, 121(11):112001, 2018.
\newblock \href {http://arxiv.org/abs/1803.02685} {\path{arXiv:1803.02685}},
  \href {https://doi.org/10.1103/PhysRevLett.121.112001}
  {\path{doi:10.1103/PhysRevLett.121.112001}}.

\bibitem{Lin:2018pvv}
Huey-Wen Lin, Jiunn-Wei Chen, Xiangdong Ji, Luchang Jin, Ruizi Li, Yu-Sheng
  Liu, Yi-Bo Yang, Jian-Hui Zhang, and Yong Zhao.
\newblock {Proton Isovector Helicity Distribution on the Lattice at Physical
  Pion Mass}.
\newblock {\em Phys. Rev. Lett.}, 121(24):242003, 2018.
\newblock \href {http://arxiv.org/abs/1807.07431} {\path{arXiv:1807.07431}},
  \href {https://doi.org/10.1103/PhysRevLett.121.242003}
  {\path{doi:10.1103/PhysRevLett.121.242003}}.

\bibitem{Alexandrou:2018eet}
Constantia Alexandrou, Krzysztof Cichy, Martha Constantinou, Karl Jansen,
  Aurora Scapellato, and Fernanda Steffens.
\newblock {Transversity parton distribution functions from lattice QCD}.
\newblock {\em Phys. Rev. D}, 98(9):091503, 2018.
\newblock \href {http://arxiv.org/abs/1807.00232} {\path{arXiv:1807.00232}},
  \href {https://doi.org/10.1103/PhysRevD.98.091503}
  {\path{doi:10.1103/PhysRevD.98.091503}}.

\bibitem{Alexandrou:2019lfo}
Constantia Alexandrou, Krzysztof Cichy, Martha Constantinou, Kyriakos
  Hadjiyiannakou, Karl Jansen, Aurora Scapellato, and Fernanda Steffens.
\newblock {Systematic uncertainties in parton distribution functions from
  lattice QCD simulations at the physical point}.
\newblock {\em Phys. Rev. D}, 99(11):114504, 2019.
\newblock \href {http://arxiv.org/abs/1902.00587} {\path{arXiv:1902.00587}},
  \href {https://doi.org/10.1103/PhysRevD.99.114504}
  {\path{doi:10.1103/PhysRevD.99.114504}}.

\bibitem{Joo:2020spy}
B\'alint Jo\'o, Joseph Karpie, Kostas Orginos, Anatoly~V. Radyushkin, David~G.
  Richards, and Savvas Zafeiropoulos.
\newblock {Parton Distribution Functions from Ioffe Time Pseudodistributions
  from Lattice Calculations: Approaching the Physical Point}.
\newblock {\em Phys. Rev. Lett.}, 125(23):232003, 2020.
\newblock \href {http://arxiv.org/abs/2004.01687} {\path{arXiv:2004.01687}},
  \href {https://doi.org/10.1103/PhysRevLett.125.232003}
  {\path{doi:10.1103/PhysRevLett.125.232003}}.

\bibitem{Bhat:2020ktg}
Manjunath Bhat, Krzysztof Cichy, Martha Constantinou, and Aurora Scapellato.
\newblock {Flavor nonsinglet parton distribution functions from lattice QCD at
  physical quark masses via the pseudodistribution approach}.
\newblock {\em Phys. Rev. D}, 103(3):034510, 2021.
\newblock \href {http://arxiv.org/abs/2005.02102} {\path{arXiv:2005.02102}},
  \href {https://doi.org/10.1103/PhysRevD.103.034510}
  {\path{doi:10.1103/PhysRevD.103.034510}}.

\bibitem{HadStruc:2021qdf}
Colin Egerer et~al.
\newblock {Transversity parton distribution function of the nucleon using the
  pseudodistribution approach}.
\newblock {\em Phys. Rev. D}, 105(3):034507, 2022.
\newblock \href {http://arxiv.org/abs/2111.01808} {\path{arXiv:2111.01808}},
  \href {https://doi.org/10.1103/PhysRevD.105.034507}
  {\path{doi:10.1103/PhysRevD.105.034507}}.

\bibitem{Gao:2021dbh}
Xiang Gao, Andrew~D. Hanlon, Swagato Mukherjee, Peter Petreczky, Philipp Scior,
  Sergey Syritsyn, and Yong Zhao.
\newblock {Lattice QCD Determination of the Bjorken-x Dependence of Parton
  Distribution Functions at Next-to-Next-to-Leading Order}.
\newblock {\em Phys. Rev. Lett.}, 128(14):142003, 2022.
\newblock \href {http://arxiv.org/abs/2112.02208} {\path{arXiv:2112.02208}},
  \href {https://doi.org/10.1103/PhysRevLett.128.142003}
  {\path{doi:10.1103/PhysRevLett.128.142003}}.

\bibitem{LatticeParton:2022xsd}
Fei Yao et~al.
\newblock {Nucleon Transversity Distribution in the Continuum and Physical Mass
  Limit from Lattice QCD}.
\newblock 8 2022.
\newblock \href {http://arxiv.org/abs/2208.08008} {\path{arXiv:2208.08008}}.

\bibitem{HadStruc:2021wmh}
Tanjib Khan et~al.
\newblock {Unpolarized gluon distribution in the nucleon from lattice quantum
  chromodynamics}.
\newblock {\em Phys. Rev. D}, 104(9):094516, 2021.
\newblock \href {http://arxiv.org/abs/2107.08960} {\path{arXiv:2107.08960}},
  \href {https://doi.org/10.1103/PhysRevD.104.094516}
  {\path{doi:10.1103/PhysRevD.104.094516}}.

\bibitem{Fan:2021bcr}
Zhouyou Fan and Huey-Wen Lin.
\newblock {Gluon parton distribution of the pion from lattice QCD}.
\newblock {\em Phys. Lett. B}, 823:136778, 2021.
\newblock \href {http://arxiv.org/abs/2104.06372} {\path{arXiv:2104.06372}},
  \href {https://doi.org/10.1016/j.physletb.2021.136778}
  {\path{doi:10.1016/j.physletb.2021.136778}}.

\bibitem{Salas-Chavira:2021wui}
Alejandro Salas-Chavira, Zhouyou Fan, and Huey-Wen Lin.
\newblock {First glimpse into the kaon gluon parton distribution using lattice
  QCD}.
\newblock {\em Phys. Rev. D}, 106(9):094510, 2022.
\newblock \href {http://arxiv.org/abs/2112.03124} {\path{arXiv:2112.03124}},
  \href {https://doi.org/10.1103/PhysRevD.106.094510}
  {\path{doi:10.1103/PhysRevD.106.094510}}.

\bibitem{HadStruc:2022yaw}
Colin Egerer et~al.
\newblock {Toward the determination of the gluon helicity distribution in the
  nucleon from lattice quantum chromodynamics}.
\newblock {\em Phys. Rev. D}, 106(9):094511, 2022.
\newblock \href {http://arxiv.org/abs/2207.08733} {\path{arXiv:2207.08733}},
  \href {https://doi.org/10.1103/PhysRevD.106.094511}
  {\path{doi:10.1103/PhysRevD.106.094511}}.

\bibitem{Egerer:2020hnc}
Colin Egerer, Robert~G. Edwards, Kostas Orginos, and David~G. Richards.
\newblock {Distillation at High-Momentum}.
\newblock {\em Phys. Rev. D}, 103(3):034502, 2021.
\newblock \href {http://arxiv.org/abs/2009.10691} {\path{arXiv:2009.10691}},
  \href {https://doi.org/10.1103/PhysRevD.103.034502}
  {\path{doi:10.1103/PhysRevD.103.034502}}.

\bibitem{Alexandrou:2020uyt}
Constantia Alexandrou, Martha Constantinou, Kyriakos Hadjiyiannakou, Karl
  Jansen, and Floriano Manigrasso.
\newblock {Flavor decomposition for the proton helicity parton distribution
  functions}.
\newblock {\em Phys. Rev. Lett.}, 126(10):102003, 2021.
\newblock \href {http://arxiv.org/abs/2009.13061} {\path{arXiv:2009.13061}},
  \href {https://doi.org/10.1103/PhysRevLett.126.102003}
  {\path{doi:10.1103/PhysRevLett.126.102003}}.

\bibitem{Lin:2020rxa}
Huey-Wen Lin.
\newblock {Nucleon Tomography and Generalized Parton Distribution at Physical
  Pion Mass from Lattice QCD}.
\newblock {\em Phys. Rev. Lett.}, 127(18):182001, 2021.
\newblock \href {http://arxiv.org/abs/2008.12474} {\path{arXiv:2008.12474}},
  \href {https://doi.org/10.1103/PhysRevLett.127.182001}
  {\path{doi:10.1103/PhysRevLett.127.182001}}.

\bibitem{Alexandrou:2021bbo}
Constantia Alexandrou, Krzysztof Cichy, Martha Constantinou, Kyriakos
  Hadjiyiannakou, Karl Jansen, Aurora Scapellato, and Fernanda Steffens.
\newblock {Transversity GPDs of the proton from lattice QCD}.
\newblock {\em Phys. Rev. D}, 105(3):034501, 2022.
\newblock \href {http://arxiv.org/abs/2108.10789} {\path{arXiv:2108.10789}},
  \href {https://doi.org/10.1103/PhysRevD.105.034501}
  {\path{doi:10.1103/PhysRevD.105.034501}}.

\bibitem{Lin:2021brq}
Huey-Wen Lin.
\newblock {Nucleon helicity generalized parton distribution at physical pion
  mass from lattice QCD}.
\newblock {\em Phys. Lett. B}, 824:136821, 2022.
\newblock \href {http://arxiv.org/abs/2112.07519} {\path{arXiv:2112.07519}},
  \href {https://doi.org/10.1016/j.physletb.2021.136821}
  {\path{doi:10.1016/j.physletb.2021.136821}}.

\bibitem{Musch:2011er}
B.~U. Musch, Ph. Hagler, M.~Engelhardt, J.~W. Negele, and A.~Schafer.
\newblock {Sivers and Boer-Mulders observables from lattice QCD}.
\newblock {\em Phys. Rev. D}, 85:094510, 2012.
\newblock \href {http://arxiv.org/abs/1111.4249} {\path{arXiv:1111.4249}},
  \href {https://doi.org/10.1103/PhysRevD.85.094510}
  {\path{doi:10.1103/PhysRevD.85.094510}}.

\bibitem{Engelhardt:2015xja}
M.~Engelhardt, P.~H\"agler, B.~Musch, J.~Negele, and A.~Sch\"afer.
\newblock {Lattice QCD study of the Boer-Mulders effect in a pion}.
\newblock {\em Phys. Rev. D}, 93(5):054501, 2016.
\newblock \href {http://arxiv.org/abs/1506.07826} {\path{arXiv:1506.07826}},
  \href {https://doi.org/10.1103/PhysRevD.93.054501}
  {\path{doi:10.1103/PhysRevD.93.054501}}.

\bibitem{Yoon:2017qzo}
Boram Yoon, Michael Engelhardt, Rajan Gupta, Tanmoy Bhattacharya, Jeremy~R.
  Green, Bernhard~U. Musch, John~W. Negele, Andrew~V. Pochinsky, Andreas
  Sch\"afer, and Sergey~N. Syritsyn.
\newblock {Nucleon Transverse Momentum-dependent Parton Distributions in
  Lattice QCD: Renormalization Patterns and Discretization Effects}.
\newblock {\em Phys. Rev. D}, 96(9):094508, 2017.
\newblock \href {http://arxiv.org/abs/1706.03406} {\path{arXiv:1706.03406}},
  \href {https://doi.org/10.1103/PhysRevD.96.094508}
  {\path{doi:10.1103/PhysRevD.96.094508}}.

\bibitem{Engelhardt:2023aem}
Michael Engelhardt, Nesreen Hasan, Taku Izubuchi, Christos Kallidonis, Stefan
  Krieg, Stefan Meinel, John Negele, Andrew Pochinsky, Giorgio Silvi, and
  Sergey Syritsyn.
\newblock {Transverse momentum-dependent parton distributions for
  longitudinally polarized nucleons from domain wall fermion calculations at
  the physical pion mass}.
\newblock {\em PoS}, LATTICE2022:103, 2023.
\newblock \href {http://arxiv.org/abs/2301.06118} {\path{arXiv:2301.06118}},
  \href {https://doi.org/10.22323/1.430.0103} {\path{doi:10.22323/1.430.0103}}.

\bibitem{Ji:2014hxa}
Xiangdong Ji, Peng Sun, Xiaonu Xiong, and Feng Yuan.
\newblock {Soft factor subtraction and transverse momentum dependent parton
  distributions on the lattice}.
\newblock {\em Phys. Rev. D}, 91:074009, 2015.
\newblock \href {http://arxiv.org/abs/1405.7640} {\path{arXiv:1405.7640}},
  \href {https://doi.org/10.1103/PhysRevD.91.074009}
  {\path{doi:10.1103/PhysRevD.91.074009}}.

\bibitem{Ji:2018hvs}
Xiangdong Ji, Lu-Chang Jin, Feng Yuan, Jian-Hui Zhang, and Yong Zhao.
\newblock {Transverse momentum dependent parton quasidistributions}.
\newblock {\em Phys. Rev. D}, 99(11):114006, 2019.
\newblock \href {http://arxiv.org/abs/1801.05930} {\path{arXiv:1801.05930}},
  \href {https://doi.org/10.1103/PhysRevD.99.114006}
  {\path{doi:10.1103/PhysRevD.99.114006}}.

\bibitem{Ji:2019sxk}
Xiangdong Ji, Yizhuang Liu, and Yu-Sheng Liu.
\newblock {TMD soft function from large-momentum effective theory}.
\newblock {\em Nucl. Phys. B}, 955:115054, 2020.
\newblock \href {http://arxiv.org/abs/1910.11415} {\path{arXiv:1910.11415}},
  \href {https://doi.org/10.1016/j.nuclphysb.2020.115054}
  {\path{doi:10.1016/j.nuclphysb.2020.115054}}.

\bibitem{Ji:2019ewn}
Xiangdong Ji, Yizhuang Liu, and Yu-Sheng Liu.
\newblock {Transverse-momentum-dependent parton distribution functions from
  large-momentum effective theory}.
\newblock {\em Phys. Lett. B}, 811:135946, 2020.
\newblock \href {http://arxiv.org/abs/1911.03840} {\path{arXiv:1911.03840}},
  \href {https://doi.org/10.1016/j.physletb.2020.135946}
  {\path{doi:10.1016/j.physletb.2020.135946}}.

\bibitem{Ji:2020jeb}
Xiangdong Ji, Yizhuang Liu, Andreas Sch\"afer, and Feng Yuan.
\newblock {Single Transverse-Spin Asymmetry and Sivers Function in Large
  Momentum Effective Theory}.
\newblock {\em Phys. Rev. D}, 103(7):074005, 2021.
\newblock \href {http://arxiv.org/abs/2011.13397} {\path{arXiv:2011.13397}},
  \href {https://doi.org/10.1103/PhysRevD.103.074005}
  {\path{doi:10.1103/PhysRevD.103.074005}}.

\bibitem{Ebert:2019okf}
Markus~A. Ebert, Iain~W. Stewart, and Yong Zhao.
\newblock {Towards Quasi-Transverse Momentum Dependent PDFs Computable on the
  Lattice}.
\newblock {\em JHEP}, 09:037, 2019.
\newblock \href {http://arxiv.org/abs/1901.03685} {\path{arXiv:1901.03685}},
  \href {https://doi.org/10.1007/JHEP09(2019)037}
  {\path{doi:10.1007/JHEP09(2019)037}}.

\bibitem{Ebert:2019tvc}
Markus~A. Ebert, Iain~W. Stewart, and Yong Zhao.
\newblock {Renormalization and Matching for the Collins-Soper Kernel from
  Lattice QCD}.
\newblock {\em JHEP}, 03:099, 2020.
\newblock \href {http://arxiv.org/abs/1910.08569} {\path{arXiv:1910.08569}},
  \href {https://doi.org/10.1007/JHEP03(2020)099}
  {\path{doi:10.1007/JHEP03(2020)099}}.

\bibitem{Ebert:2020gxr}
Markus~A. Ebert, Stella~T. Schindler, Iain~W. Stewart, and Yong Zhao.
\newblock {One-loop Matching for Spin-Dependent Quasi-TMDs}.
\newblock {\em JHEP}, 09:099, 2020.
\newblock \href {http://arxiv.org/abs/2004.14831} {\path{arXiv:2004.14831}},
  \href {https://doi.org/10.1007/JHEP09(2020)099}
  {\path{doi:10.1007/JHEP09(2020)099}}.

\bibitem{Shanahan:2019zcq}
Phiala Shanahan, Michael~L. Wagman, and Yong Zhao.
\newblock {Nonperturbative renormalization of staple-shaped Wilson line
  operators in lattice QCD}.
\newblock {\em Phys. Rev. D}, 101(7):074505, 2020.
\newblock \href {http://arxiv.org/abs/1911.00800} {\path{arXiv:1911.00800}},
  \href {https://doi.org/10.1103/PhysRevD.101.074505}
  {\path{doi:10.1103/PhysRevD.101.074505}}.

\bibitem{Shanahan:2020zxr}
Phiala Shanahan, Michael Wagman, and Yong Zhao.
\newblock {Collins-Soper kernel for TMD evolution from lattice QCD}.
\newblock {\em Phys. Rev. D}, 102(1):014511, 2020.
\newblock \href {http://arxiv.org/abs/2003.06063} {\path{arXiv:2003.06063}},
  \href {https://doi.org/10.1103/PhysRevD.102.014511}
  {\path{doi:10.1103/PhysRevD.102.014511}}.

\bibitem{Shanahan:2021tst}
Phiala Shanahan, Michael Wagman, and Yong Zhao.
\newblock {Lattice QCD calculation of the Collins-Soper kernel from
  quasi-TMDPDFs}.
\newblock {\em Phys. Rev. D}, 104(11):114502, 2021.
\newblock \href {http://arxiv.org/abs/2107.11930} {\path{arXiv:2107.11930}},
  \href {https://doi.org/10.1103/PhysRevD.104.114502}
  {\path{doi:10.1103/PhysRevD.104.114502}}.

\bibitem{Li:2021wvl}
Yuan Li et~al.
\newblock {Lattice QCD Study of Transverse-Momentum Dependent Soft Function}.
\newblock {\em Phys. Rev. Lett.}, 128(6):062002, 2022.
\newblock \href {http://arxiv.org/abs/2106.13027} {\path{arXiv:2106.13027}},
  \href {https://doi.org/10.1103/PhysRevLett.128.062002}
  {\path{doi:10.1103/PhysRevLett.128.062002}}.

\bibitem{LPC:2022ibr}
Min-Huan Chu et~al.
\newblock {Nonperturbative determination of the Collins-Soper kernel from
  quasitransverse-momentum-dependent wave functions}.
\newblock {\em Phys. Rev. D}, 106(3):034509, 2022.
\newblock \href {http://arxiv.org/abs/2204.00200} {\path{arXiv:2204.00200}},
  \href {https://doi.org/10.1103/PhysRevD.106.034509}
  {\path{doi:10.1103/PhysRevD.106.034509}}.

\bibitem{Ebert:2022fmh}
Markus~A. Ebert, Stella~T. Schindler, Iain~W. Stewart, and Yong Zhao.
\newblock {Factorization connecting continuum \& lattice TMDs}.
\newblock {\em JHEP}, 04:178, 2022.
\newblock \href {http://arxiv.org/abs/2201.08401} {\path{arXiv:2201.08401}},
  \href {https://doi.org/10.1007/JHEP04(2022)178}
  {\path{doi:10.1007/JHEP04(2022)178}}.

\bibitem{Schindler:2022eva}
Stella~T. Schindler, Iain~W. Stewart, and Yong Zhao.
\newblock {One-loop matching for gluon lattice TMDs}.
\newblock {\em JHEP}, 08:084, 2022.
\newblock \href {http://arxiv.org/abs/2205.12369} {\path{arXiv:2205.12369}},
  \href {https://doi.org/10.1007/JHEP08(2022)084}
  {\path{doi:10.1007/JHEP08(2022)084}}.

\bibitem{Zhang:2022xuw}
Kuan Zhang, Xiangdong Ji, Yi-Bo Yang, Fei Yao, and Jian-Hui Zhang.
\newblock {Renormalization of Transverse-Momentum-Dependent Parton Distribution
  on the Lattice}.
\newblock {\em Phys. Rev. Lett.}, 129(8):082002, 2022.
\newblock \href {http://arxiv.org/abs/2205.13402} {\path{arXiv:2205.13402}},
  \href {https://doi.org/10.1103/PhysRevLett.129.082002}
  {\path{doi:10.1103/PhysRevLett.129.082002}}.

\bibitem{LatticeParton:2020uhz}
Qi-An Zhang et~al.
\newblock {Lattice-QCD Calculations of TMD Soft Function Through Large-Momentum
  Effective Theory}.
\newblock {\em Phys. Rev. Lett.}, 125(19):192001, 2020.
\newblock \href {http://arxiv.org/abs/2005.14572} {\path{arXiv:2005.14572}},
  \href {https://doi.org/10.22323/1.396.0477} {\path{doi:10.22323/1.396.0477}}.

\bibitem{Ebert:2018gzl}
Markus~A. Ebert, Iain~W. Stewart, and Yong Zhao.
\newblock {Determining the Nonperturbative Collins-Soper Kernel From Lattice
  QCD}.
\newblock {\em Phys. Rev. D}, 99(3):034505, 2019.
\newblock \href {http://arxiv.org/abs/1811.00026} {\path{arXiv:1811.00026}},
  \href {https://doi.org/10.1103/PhysRevD.99.034505}
  {\path{doi:10.1103/PhysRevD.99.034505}}.

\bibitem{Schlemmer:2021aij}
Maximilian Schlemmer, Alexey Vladimirov, Christian Zimmermann, Michael
  Engelhardt, and Andreas Sch\"afer.
\newblock {Determination of the Collins-Soper Kernel from Lattice QCD}.
\newblock {\em JHEP}, 08:004, 2021.
\newblock \href {http://arxiv.org/abs/2103.16991} {\path{arXiv:2103.16991}},
  \href {https://doi.org/10.1007/JHEP08(2021)004}
  {\path{doi:10.1007/JHEP08(2021)004}}.

\bibitem{Zhang:2020dkn}
Rui Zhang, Huey-Wen Lin, and Boram Yoon.
\newblock {Probing nucleon strange and charm distributions with lattice QCD}.
\newblock {\em Phys. Rev. D}, 104(9):094511, 2021.
\newblock \href {http://arxiv.org/abs/2005.01124} {\path{arXiv:2005.01124}},
  \href {https://doi.org/10.1103/PhysRevD.104.094511}
  {\path{doi:10.1103/PhysRevD.104.094511}}.

\bibitem{LPC:2022zci}
Jin-Chen He, Min-Huan Chu, Jun Hua, Xiangdong Ji, Andreas Sch\"afer, Yushan Su,
  Wei Wang, Yibo Yang, Jian-Hui Zhang, and Qi-An Zhang.
\newblock {Unpolarized Transverse-Momentum-Dependent Parton Distributions of
  the Nucleon from Lattice QCD}.
\newblock 11 2022.
\newblock \href {http://arxiv.org/abs/2211.02340} {\path{arXiv:2211.02340}}.

\bibitem{Luscher:1986pf}
M.~Luscher.
\newblock {Volume Dependence of the Energy Spectrum in Massive Quantum Field
  Theories. 2. Scattering States}.
\newblock {\em Commun. Math. Phys.}, 105:153--188, 1986.
\newblock \href {https://doi.org/10.1007/BF01211097}
  {\path{doi:10.1007/BF01211097}}.

\bibitem{Dudek:2016cru}
Jozef~J. Dudek, Robert~G. Edwards, and David~J. Wilson.
\newblock {An $a_0$ resonance in strongly coupled $\pi \eta$, $K\overline{K}$
  scattering from lattice QCD}.
\newblock {\em Phys. Rev. D}, 93(9):094506, 2016.
\newblock \href {http://arxiv.org/abs/1602.05122} {\path{arXiv:1602.05122}},
  \href {https://doi.org/10.1103/PhysRevD.93.094506}
  {\path{doi:10.1103/PhysRevD.93.094506}}.

\bibitem{Briceno:2017qmb}
Raul~A. Briceno, Jozef~J. Dudek, Robert~G. Edwards, and David~J. Wilson.
\newblock {Isoscalar $\pi\pi, K\overline{K}, \eta\eta$ scattering and the
  $\sigma, f_0, f_2$ mesons from QCD}.
\newblock {\em Phys. Rev. D}, 97(5):054513, 2018.
\newblock \href {http://arxiv.org/abs/1708.06667} {\path{arXiv:1708.06667}},
  \href {https://doi.org/10.1103/PhysRevD.97.054513}
  {\path{doi:10.1103/PhysRevD.97.054513}}.

\bibitem{Woss:2019hse}
Antoni~J. Woss, Christopher~E. Thomas, Jozef~J. Dudek, Robert~G. Edwards, and
  David~J. Wilson.
\newblock {$b_1$ resonance in coupled $\pi\omega$, $\pi\phi$ scattering from
  lattice QCD}.
\newblock {\em Phys. Rev. D}, 100(5):054506, 2019.
\newblock \href {http://arxiv.org/abs/1904.04136} {\path{arXiv:1904.04136}},
  \href {https://doi.org/10.1103/PhysRevD.100.054506}
  {\path{doi:10.1103/PhysRevD.100.054506}}.

\bibitem{Bulava:2022vpq}
John Bulava, Andrew~D. Hanlon, Ben Horz, Colin Morningstar, Amy Nicholson,
  Fernando Romero-Lopez, Sarah Skinner, Pavlos Vranas, and Andre Walker-Loud.
\newblock {Elastic nucleon-pion scattering at $m_\pi=200$MeV from lattice QCD}.
\newblock {\em Nucl. Phys. B}, 987:116105, 2023.
\newblock \href {http://arxiv.org/abs/2208.03867} {\path{arXiv:2208.03867}},
  \href {https://doi.org/10.1016/j.nuclphysb.2023.116105}
  {\path{doi:10.1016/j.nuclphysb.2023.116105}}.

\bibitem{Briceno:2014uqa}
Ra\'ul~A. Brice\~no, Maxwell~T. Hansen, and Andr\'e Walker-Loud.
\newblock {Multichannel 1 $\rightarrow$ 2 transition amplitudes in a finite
  volume}.
\newblock {\em Phys. Rev. D}, 91(3):034501, 2015.
\newblock \href {http://arxiv.org/abs/1406.5965} {\path{arXiv:1406.5965}},
  \href {https://doi.org/10.1103/PhysRevD.91.034501}
  {\path{doi:10.1103/PhysRevD.91.034501}}.

\bibitem{Briceno:2015dca}
Raul~A. Briceno, Jozef~J. Dudek, Robert~G. Edwards, Christian~J. Shultz,
  Christopher~E. Thomas, and David~J. Wilson.
\newblock {The resonant $\pi^+\gamma\to\pi^+\pi^0$ amplitude from Quantum
  Chromodynamics}.
\newblock {\em Phys. Rev. Lett.}, 115:242001, 2015.
\newblock \href {http://arxiv.org/abs/1507.06622} {\path{arXiv:1507.06622}},
  \href {https://doi.org/10.1103/PhysRevLett.115.242001}
  {\path{doi:10.1103/PhysRevLett.115.242001}}.

\bibitem{Alexandrou:2018jbt}
Constantia Alexandrou, Luka Leskovec, Stefan Meinel, John Negele, Srijit Paul,
  Marcus Petschlies, Andrew Pochinsky, Gumaro Rendon, and Sergey Syritsyn.
\newblock {$\pi\gamma \to \pi\pi$ transition and the $\rho$ radiative decay
  width from lattice QCD}.
\newblock {\em Phys. Rev. D}, 98(7):074502, 2018.
\newblock [Erratum: Phys.Rev.D 105, 019902 (2022)].
\newblock \href {http://arxiv.org/abs/1807.08357} {\path{arXiv:1807.08357}},
  \href {https://doi.org/10.1103/PhysRevD.98.074502}
  {\path{doi:10.1103/PhysRevD.98.074502}}.

\bibitem{Radhakrishnan:2022ubg}
Archana Radhakrishnan, Jozef~J. Dudek, and Robert~G. Edwards.
\newblock {Radiative decay of the resonant K* and the
  \ensuremath{\gamma}K\textrightarrow{}K\ensuremath{\pi} amplitude from lattice
  QCD}.
\newblock {\em Phys. Rev. D}, 106(11):114513, 2022.
\newblock \href {http://arxiv.org/abs/2208.13755} {\path{arXiv:2208.13755}},
  \href {https://doi.org/10.1103/PhysRevD.106.114513}
  {\path{doi:10.1103/PhysRevD.106.114513}}.

\bibitem{Hansen:2019nir}
Maxwell~T. Hansen and Stephen~R. Sharpe.
\newblock {Lattice QCD and Three-particle Decays of Resonances}.
\newblock {\em Ann. Rev. Nucl. Part. Sci.}, 69:65--107, 2019.
\newblock \href {http://arxiv.org/abs/1901.00483} {\path{arXiv:1901.00483}},
  \href {https://doi.org/10.1146/annurev-nucl-101918-023723}
  {\path{doi:10.1146/annurev-nucl-101918-023723}}.

\bibitem{Mai:2021lwb}
Maxim Mai, Michael D\"oring, and Akaki Rusetsky.
\newblock {Multi-particle systems on the lattice and chiral extrapolations: a
  brief review}.
\newblock {\em Eur. Phys. J. ST}, 230(6):1623--1643, 2021.
\newblock \href {http://arxiv.org/abs/2103.00577} {\path{arXiv:2103.00577}},
  \href {https://doi.org/10.1140/epjs/s11734-021-00146-5}
  {\path{doi:10.1140/epjs/s11734-021-00146-5}}.

\bibitem{Blanton:2019vdk}
Tyler~D. Blanton, Fernando Romero-L\'opez, and Stephen~R. Sharpe.
\newblock {$I=3$ Three-Pion Scattering Amplitude from Lattice QCD}.
\newblock {\em Phys. Rev. Lett.}, 124(3):032001, 2020.
\newblock \href {http://arxiv.org/abs/1909.02973} {\path{arXiv:1909.02973}},
  \href {https://doi.org/10.1103/PhysRevLett.124.032001}
  {\path{doi:10.1103/PhysRevLett.124.032001}}.

\bibitem{Hansen:2020otl}
Maxwell~T. Hansen, Raul~A. Brice\~no, Robert~G. Edwards, Christopher~E. Thomas,
  and David~J. Wilson.
\newblock {Energy-Dependent $\pi^+ \pi^+ \pi^+$ Scattering Amplitude from QCD}.
\newblock {\em Phys. Rev. Lett.}, 126:012001, 2021.
\newblock \href {http://arxiv.org/abs/2009.04931} {\path{arXiv:2009.04931}},
  \href {https://doi.org/10.1103/PhysRevLett.126.012001}
  {\path{doi:10.1103/PhysRevLett.126.012001}}.

\bibitem{Blanton:2021llb}
Tyler~D. Blanton, Andrew~D. Hanlon, Ben H\"orz, Colin Morningstar, Fernando
  Romero-L\'opez, and Stephen~R. Sharpe.
\newblock {Interactions of two and three mesons including higher partial waves
  from lattice QCD}.
\newblock {\em JHEP}, 10:023, 2021.
\newblock \href {http://arxiv.org/abs/2106.05590} {\path{arXiv:2106.05590}},
  \href {https://doi.org/10.1007/JHEP10(2021)023}
  {\path{doi:10.1007/JHEP10(2021)023}}.

\bibitem{Brett:2021wyd}
Ruair\'\i{} Brett, Chris Culver, Maxim Mai, Andrei Alexandru, Michael D\"oring,
  and Frank~X. Lee.
\newblock {Three-body interactions from the finite-volume QCD spectrum}.
\newblock {\em Phys. Rev. D}, 104(1):014501, 2021.
\newblock \href {http://arxiv.org/abs/2101.06144} {\path{arXiv:2101.06144}},
  \href {https://doi.org/10.1103/PhysRevD.104.014501}
  {\path{doi:10.1103/PhysRevD.104.014501}}.

\bibitem{Borsanyi:2020fev}
Szabolcs Borsanyi, Zoltan Fodor, Jana~N. Guenther, Ruben Kara, Sandor~D. Katz,
  Paolo Parotto, Attila Pasztor, Claudia Ratti, and Kalman~K. Szabo.
\newblock {QCD Crossover at Finite Chemical Potential from Lattice
  Simulations}.
\newblock {\em Phys. Rev. Lett.}, 125(5):052001, 2020.
\newblock \href {http://arxiv.org/abs/2002.02821} {\path{arXiv:2002.02821}},
  \href {https://doi.org/10.1103/PhysRevLett.125.052001}
  {\path{doi:10.1103/PhysRevLett.125.052001}}.

\bibitem{HotQCD:2018pds}
A.~Bazavov et~al.
\newblock {Chiral crossover in QCD at zero and non-zero chemical potentials}.
\newblock {\em Phys. Lett. B}, 795:15--21, 2019.
\newblock \href {http://arxiv.org/abs/1812.08235} {\path{arXiv:1812.08235}},
  \href {https://doi.org/10.1016/j.physletb.2019.05.013}
  {\path{doi:10.1016/j.physletb.2019.05.013}}.

\bibitem{HotQCD:2019xnw}
H.~T. Ding et~al.
\newblock {Chiral Phase Transition Temperature in ( 2+1 )-Flavor QCD}.
\newblock {\em Phys. Rev. Lett.}, 123(6):062002, 2019.
\newblock \href {http://arxiv.org/abs/1903.04801} {\path{arXiv:1903.04801}},
  \href {https://doi.org/10.1103/PhysRevLett.123.062002}
  {\path{doi:10.1103/PhysRevLett.123.062002}}.

\bibitem{Borsanyi:2010cj}
Szabolcs Borsanyi, Gergely Endrodi, Zoltan Fodor, Antal Jakovac, Sandor~D.
  Katz, Stefan Krieg, Claudia Ratti, and Kalman~K. Szabo.
\newblock {The QCD equation of state with dynamical quarks}.
\newblock {\em JHEP}, 11:077, 2010.
\newblock \href {http://arxiv.org/abs/1007.2580} {\path{arXiv:1007.2580}},
  \href {https://doi.org/10.1007/JHEP11(2010)077}
  {\path{doi:10.1007/JHEP11(2010)077}}.

\bibitem{Borsanyi:2013bia}
Szabocls Borsanyi, Zoltan Fodor, Christian Hoelbling, Sandor~D. Katz, Stefan
  Krieg, and Kalman~K. Szabo.
\newblock {Full result for the QCD equation of state with 2+1 flavors}.
\newblock {\em Phys. Lett. B}, 730:99--104, 2014.
\newblock \href {http://arxiv.org/abs/1309.5258} {\path{arXiv:1309.5258}},
  \href {https://doi.org/10.1016/j.physletb.2014.01.007}
  {\path{doi:10.1016/j.physletb.2014.01.007}}.

\bibitem{Bazavov:2014pvz}
A.~Bazavov et~al.
\newblock {Equation of state in ( 2+1 )-flavor QCD}.
\newblock {\em Phys. Rev. D}, 90:094503, 2014.
\newblock \href {http://arxiv.org/abs/1407.6387} {\path{arXiv:1407.6387}},
  \href {https://doi.org/10.1103/PhysRevD.90.094503}
  {\path{doi:10.1103/PhysRevD.90.094503}}.

\bibitem{Bazavov:2017dus}
A.~Bazavov et~al.
\newblock {The QCD Equation of State to $\mathcal{O}(\mu_B^6)$ from Lattice
  QCD}.
\newblock {\em Phys. Rev. D}, 95(5):054504, 2017.
\newblock \href {http://arxiv.org/abs/1701.04325} {\path{arXiv:1701.04325}},
  \href {https://doi.org/10.1103/PhysRevD.95.054504}
  {\path{doi:10.1103/PhysRevD.95.054504}}.

\bibitem{Guenther:2017hnx}
J.~N. Guenther, R.~Bellwied, S.~Borsanyi, Z.~Fodor, S.~D. Katz, A.~Pasztor,
  C.~Ratti, and K.~K. Szab\'o.
\newblock {The QCD equation of state at finite density from analytical
  continuation}.
\newblock {\em Nucl. Phys. A}, 967:720--723, 2017.
\newblock \href {http://arxiv.org/abs/1607.02493} {\path{arXiv:1607.02493}},
  \href {https://doi.org/10.1016/j.nuclphysa.2017.05.044}
  {\path{doi:10.1016/j.nuclphysa.2017.05.044}}.

\bibitem{Bollweg:2022fqq}
D.~Bollweg, D.~A. Clarke, J.~Goswami, O.~Kaczmarek, F.~Karsch, Swagato
  Mukherjee, P.~Petreczky, C.~Schmidt, and Sipaz Sharma.
\newblock {Equation of state and speed of sound of (2+1)-flavor QCD in
  strangeness-neutral matter at non-vanishing net baryon-number density}.
\newblock 12 2022.
\newblock \href {http://arxiv.org/abs/2212.09043} {\path{arXiv:2212.09043}}.

\bibitem{Borsanyi:2021sxv}
S.~Bors\'anyi, Z.~Fodor, J.~N. Guenther, R.~Kara, S.~D. Katz, P.~Parotto,
  A.~P\'asztor, C.~Ratti, and K.~K. Szab\'o.
\newblock {Lattice QCD equation of state at finite chemical potential from an
  alternative expansion scheme}.
\newblock {\em Phys. Rev. Lett.}, 126(23):232001, 2021.
\newblock \href {http://arxiv.org/abs/2102.06660} {\path{arXiv:2102.06660}},
  \href {https://doi.org/10.1103/PhysRevLett.126.232001}
  {\path{doi:10.1103/PhysRevLett.126.232001}}.

\bibitem{Vovchenko:2017gkg}
Volodymyr Vovchenko, Jan Steinheimer, Owe Philipsen, and Horst Stoecker.
\newblock {Cluster Expansion Model for QCD Baryon Number Fluctuations: No Phase
  Transition at $\mu_B / T < \pi$}.
\newblock {\em Phys. Rev. D}, 97(11):114030, 2018.
\newblock \href {http://arxiv.org/abs/1711.01261} {\path{arXiv:1711.01261}},
  \href {https://doi.org/10.1103/PhysRevD.97.114030}
  {\path{doi:10.1103/PhysRevD.97.114030}}.

\bibitem{Borsanyi:2022qlh}
Szabolcs Borsanyi, Jana~N. Guenther, Ruben Kara, Zoltan Fodor, Paolo Parotto,
  Attila Pasztor, Claudia Ratti, and Kalman~K. Szabo.
\newblock {Resummed lattice QCD equation of state at finite baryon density:
  Strangeness neutrality and beyond}.
\newblock {\em Phys. Rev. D}, 105(11):114504, 2022.
\newblock \href {http://arxiv.org/abs/2202.05574} {\path{arXiv:2202.05574}},
  \href {https://doi.org/10.1103/PhysRevD.105.114504}
  {\path{doi:10.1103/PhysRevD.105.114504}}.

\bibitem{Karthein:2021nxe}
J.~M. Karthein, D.~Mroczek, A.~R. Nava~Acuna, J.~Noronha-Hostler, P.~Parotto,
  D.~R.~P. Price, and C.~Ratti.
\newblock {Strangeness-neutral equation of state for QCD with a critical
  point}.
\newblock {\em Eur. Phys. J. Plus}, 136(6):621, 2021.
\newblock \href {http://arxiv.org/abs/2103.08146} {\path{arXiv:2103.08146}},
  \href {https://doi.org/10.1140/epjp/s13360-021-01615-5}
  {\path{doi:10.1140/epjp/s13360-021-01615-5}}.

\bibitem{Ding:2012sp}
H.~T. Ding, A.~Francis, O.~Kaczmarek, F.~Karsch, H.~Satz, and W.~Soeldner.
\newblock {Charmonium properties in hot quenched lattice QCD}.
\newblock {\em Phys. Rev. D}, 86:014509, 2012.
\newblock \href {http://arxiv.org/abs/1204.4945} {\path{arXiv:1204.4945}},
  \href {https://doi.org/10.1103/PhysRevD.86.014509}
  {\path{doi:10.1103/PhysRevD.86.014509}}.

\bibitem{Francis:2015daa}
A.~Francis, O.~Kaczmarek, M.~Laine, T.~Neuhaus, and H.~Ohno.
\newblock {Nonperturbative estimate of the heavy quark momentum diffusion
  coefficient}.
\newblock {\em Phys. Rev. D}, 92(11):116003, 2015.
\newblock \href {http://arxiv.org/abs/1508.04543} {\path{arXiv:1508.04543}},
  \href {https://doi.org/10.1103/PhysRevD.92.116003}
  {\path{doi:10.1103/PhysRevD.92.116003}}.

\bibitem{Banerjee:2022uge}
D.~Banerjee, S.~Datta, and M.~Laine.
\newblock {Lattice study of a magnetic contribution to heavy quark momentum
  diffusion}.
\newblock {\em JHEP}, 08:128, 2022.
\newblock \href {http://arxiv.org/abs/2204.14075} {\path{arXiv:2204.14075}},
  \href {https://doi.org/10.1007/JHEP08(2022)128}
  {\path{doi:10.1007/JHEP08(2022)128}}.

\bibitem{Brambilla:2020siz}
Nora Brambilla, Viljami Leino, Peter Petreczky, and Antonio Vairo.
\newblock {Lattice QCD constraints on the heavy quark diffusion coefficient}.
\newblock {\em Phys. Rev. D}, 102(7):074503, 2020.
\newblock \href {http://arxiv.org/abs/2007.10078} {\path{arXiv:2007.10078}},
  \href {https://doi.org/10.1103/PhysRevD.102.074503}
  {\path{doi:10.1103/PhysRevD.102.074503}}.

\bibitem{ALCC}
https://science.osti.gov/ascr/Facilities/Accessing-ASCR-Facilities/ALCC.

\bibitem{INCITE}
https://www.alcf.anl.gov/science/incite-allocation-program.

\bibitem{SciDAC}
https://www.scidac.gov.

\bibitem{Brambilla:2016wgg}
Nora Brambilla, Miguel~A. Escobedo, Joan Soto, and Antonio Vairo.
\newblock {Quarkonium suppression in heavy-ion collisions: an open quantum
  system approach}.
\newblock {\em Phys. Rev. D}, 96(3):034021, 2017.
\newblock \href {http://arxiv.org/abs/1612.07248} {\path{arXiv:1612.07248}},
  \href {https://doi.org/10.1103/PhysRevD.96.034021}
  {\path{doi:10.1103/PhysRevD.96.034021}}.

\bibitem{Yao:2020eqy}
Xiaojun Yao and Thomas Mehen.
\newblock {Quarkonium Semiclassical Transport in Quark-Gluon Plasma:
  Factorization and Quantum Correction}.
\newblock {\em JHEP}, 02:062, 2021.
\newblock \href {http://arxiv.org/abs/2009.02408} {\path{arXiv:2009.02408}},
  \href {https://doi.org/10.1007/JHEP02(2021)062}
  {\path{doi:10.1007/JHEP02(2021)062}}.

\bibitem{Eller:2019spw}
Alexander~M. Eller, Jacopo Ghiglieri, and Guy~D. Moore.
\newblock {Thermal Heavy Quark Self-Energy from Euclidean Correlators}.
\newblock {\em Phys. Rev. D}, 99(9):094042, 2019.
\newblock [Erratum: Phys.Rev.D 102, 039901 (2020)].
\newblock \href {http://arxiv.org/abs/1903.08064} {\path{arXiv:1903.08064}},
  \href {https://doi.org/10.1103/PhysRevD.99.094042}
  {\path{doi:10.1103/PhysRevD.99.094042}}.

\bibitem{Scheihing-Hitschfeld:2022xqx}
Bruno Scheihing-Hitschfeld and Xiaojun Yao.
\newblock {Gauge Invariance of Non-Abelian Field Strength Correlators: The
  Axial Gauge Puzzle}.
\newblock {\em Phys. Rev. Lett.}, 130(5):052302, 2023.
\newblock \href {http://arxiv.org/abs/2205.04477} {\path{arXiv:2205.04477}},
  \href {https://doi.org/10.1103/PhysRevLett.130.052302}
  {\path{doi:10.1103/PhysRevLett.130.052302}}.

\bibitem{Moch:2004pa}
S.~Moch, J.~A.~M. Vermaseren, and A.~Vogt.
\newblock {The Three loop splitting functions in QCD: The Nonsinglet case}.
\newblock {\em Nucl. Phys. B}, 688:101--134, 2004.
\newblock \href {http://arxiv.org/abs/hep-ph/0403192}
  {\path{arXiv:hep-ph/0403192}}, \href
  {https://doi.org/10.1016/j.nuclphysb.2004.03.030}
  {\path{doi:10.1016/j.nuclphysb.2004.03.030}}.

\bibitem{Vogt:2004mw}
A.~Vogt, S.~Moch, and J.~A.~M. Vermaseren.
\newblock {The Three-loop splitting functions in QCD: The Singlet case}.
\newblock {\em Nucl. Phys. B}, 691:129--181, 2004.
\newblock \href {http://arxiv.org/abs/hep-ph/0404111}
  {\path{arXiv:hep-ph/0404111}}, \href
  {https://doi.org/10.1016/j.nuclphysb.2004.04.024}
  {\path{doi:10.1016/j.nuclphysb.2004.04.024}}.

\bibitem{Moch:2014sna}
S.~Moch, J.~A.~M. Vermaseren, and A.~Vogt.
\newblock {The Three-Loop Splitting Functions in QCD: The Helicity-Dependent
  Case}.
\newblock {\em Nucl. Phys. B}, 889:351--400, 2014.
\newblock \href {http://arxiv.org/abs/1409.5131} {\path{arXiv:1409.5131}},
  \href {https://doi.org/10.1016/j.nuclphysb.2014.10.016}
  {\path{doi:10.1016/j.nuclphysb.2014.10.016}}.

\bibitem{Zijlstra:1992qd}
E.~B. Zijlstra and W.~L. van Neerven.
\newblock {Order alpha-s**2 QCD corrections to the deep inelastic proton
  structure functions F2 and F(L)}.
\newblock {\em Nucl. Phys. B}, 383:525--574, 1992.
\newblock \href {https://doi.org/10.1016/0550-3213(92)90087-R}
  {\path{doi:10.1016/0550-3213(92)90087-R}}.

\bibitem{Zijlstra:1993sh}
E.~B. Zijlstra and W.~L. van Neerven.
\newblock {Order-$\alpha_s^2$ corrections to the polarized structure function
  $g_1 (x,Q^2)$}.
\newblock {\em Nucl. Phys. B}, 417:61--100, 1994.
\newblock [Erratum: Nucl.Phys.B 426, 245 (1994), Erratum: Nucl.Phys.B 773,
  105--106 (2007), Erratum: Nucl.Phys.B 501, 599--599 (1997)].
\newblock \href {https://doi.org/10.1016/0550-3213(94)90538-X}
  {\path{doi:10.1016/0550-3213(94)90538-X}}.

\bibitem{Borsa:2022irn}
Ignacio Borsa, Daniel de~Florian, and Iv\'an Pedron.
\newblock {The full set of polarized deep inelastic scattering structure
  functions at NNLO accuracy}.
\newblock {\em Eur. Phys. J. C}, 82(12):1167, 2022.
\newblock \href {http://arxiv.org/abs/2210.12014} {\path{arXiv:2210.12014}},
  \href {https://doi.org/10.1140/epjc/s10052-022-11140-z}
  {\path{doi:10.1140/epjc/s10052-022-11140-z}}.

\bibitem{Currie:2017tpe}
James Currie, Thomas Gehrmann, Alexander Huss, and Jan Niehues.
\newblock {NNLO QCD corrections to jet production in deep inelastic
  scattering}.
\newblock {\em JHEP}, 07:018, 2017.
\newblock [Erratum: JHEP 12, 042 (2020)].
\newblock \href {http://arxiv.org/abs/1703.05977} {\path{arXiv:1703.05977}},
  \href {https://doi.org/10.1007/JHEP07(2017)018}
  {\path{doi:10.1007/JHEP07(2017)018}}.

\bibitem{Currie:2018fgr}
J.~Currie, T.~Gehrmann, E.~W.~N. Glover, A.~Huss, J.~Niehues, and A.~Vogt.
\newblock {N$^{3}$LO corrections to jet production in deep inelastic scattering
  using the Projection-to-Born method}.
\newblock {\em JHEP}, 05:209, 2018.
\newblock \href {http://arxiv.org/abs/1803.09973} {\path{arXiv:1803.09973}},
  \href {https://doi.org/10.1007/JHEP05(2018)209}
  {\path{doi:10.1007/JHEP05(2018)209}}.

\bibitem{Boughezal:2018azh}
Radja Boughezal, Frank Petriello, and Hongxi Xing.
\newblock {Inclusive jet production as a probe of polarized parton distribution
  functions at a future EIC}.
\newblock {\em Phys. Rev. D}, 98(5):054031, 2018.
\newblock \href {http://arxiv.org/abs/1806.07311} {\path{arXiv:1806.07311}},
  \href {https://doi.org/10.1103/PhysRevD.98.054031}
  {\path{doi:10.1103/PhysRevD.98.054031}}.

\bibitem{Borsa:2020ulb}
Ignacio Borsa, Daniel de~Florian, and Iv\'an Pedron.
\newblock {Jet Production in Polarized Deep Inelastic Scattering at
  Next-to-Next-to-Leading Order}.
\newblock {\em Phys. Rev. Lett.}, 125(8):082001, 2020.
\newblock \href {http://arxiv.org/abs/2005.10705} {\path{arXiv:2005.10705}},
  \href {https://doi.org/10.1103/PhysRevLett.125.082001}
  {\path{doi:10.1103/PhysRevLett.125.082001}}.

\bibitem{Bodwin:1994jh}
Geoffrey~T. Bodwin, Eric Braaten, and G.~Peter Lepage.
\newblock {Rigorous QCD analysis of inclusive annihilation and production of
  heavy quarkonium}.
\newblock {\em Phys. Rev. D}, 51:1125--1171, 1995.
\newblock [Erratum: Phys.Rev.D 55, 5853 (1997)].
\newblock \href {http://arxiv.org/abs/hep-ph/9407339}
  {\path{arXiv:hep-ph/9407339}}, \href
  {https://doi.org/10.1103/PhysRevD.55.5853}
  {\path{doi:10.1103/PhysRevD.55.5853}}.

\bibitem{Brambilla:2010cs}
N.~Brambilla et~al.
\newblock {Heavy Quarkonium: Progress, Puzzles, and Opportunities}.
\newblock {\em Eur. Phys. J. C}, 71:1534, 2011.
\newblock \href {http://arxiv.org/abs/1010.5827} {\path{arXiv:1010.5827}},
  \href {https://doi.org/10.1140/epjc/s10052-010-1534-9}
  {\path{doi:10.1140/epjc/s10052-010-1534-9}}.

\bibitem{Ma:2016exq}
Yan-Qing Ma and Ramona Vogt.
\newblock {Quarkonium Production in an Improved Color Evaporation Model}.
\newblock {\em Phys. Rev. D}, 94(11):114029, 2016.
\newblock \href {http://arxiv.org/abs/1609.06042} {\path{arXiv:1609.06042}},
  \href {https://doi.org/10.1103/PhysRevD.94.114029}
  {\path{doi:10.1103/PhysRevD.94.114029}}.

\bibitem{Cheung:2017loo}
Vincent Cheung and Ramona Vogt.
\newblock {Polarized Heavy Quarkonium Production in the Color Evaporation
  Model}.
\newblock {\em Phys. Rev. D}, 95(7):074021, 2017.
\newblock \href {http://arxiv.org/abs/1702.07809} {\path{arXiv:1702.07809}},
  \href {https://doi.org/10.1103/PhysRevD.95.074021}
  {\path{doi:10.1103/PhysRevD.95.074021}}.

\bibitem{Cheung:2017osx}
Vincent Cheung and Ramona Vogt.
\newblock {Polarization of prompt $J/\psi$ and $\Upsilon$(1S) production in the
  color evaporation model}.
\newblock {\em Phys. Rev. D}, 96(5):054014, 2017.
\newblock \href {http://arxiv.org/abs/1706.07686} {\path{arXiv:1706.07686}},
  \href {https://doi.org/10.1103/PhysRevD.96.054014}
  {\path{doi:10.1103/PhysRevD.96.054014}}.

\bibitem{Cheung:2018tvq}
Vincent Cheung and Ramona Vogt.
\newblock {Production and polarization of prompt $J/\psi$ in the improved color
  evaporation model using the $k_T$-factorization approach}.
\newblock {\em Phys. Rev. D}, 98(11):114029, 2018.
\newblock \href {http://arxiv.org/abs/1808.02909} {\path{arXiv:1808.02909}},
  \href {https://doi.org/10.1103/PhysRevD.98.114029}
  {\path{doi:10.1103/PhysRevD.98.114029}}.

\bibitem{Cheung:2018upe}
Vincent Cheung and Ramona Vogt.
\newblock {Production and polarization of prompt $\varUpsilon$($n$S) in the
  improved color evaporation model using the $k_T$-factorization approach}.
\newblock {\em Phys. Rev. D}, 99(3):034007, 2019.
\newblock \href {http://arxiv.org/abs/1811.11570} {\path{arXiv:1811.11570}},
  \href {https://doi.org/10.1103/PhysRevD.99.034007}
  {\path{doi:10.1103/PhysRevD.99.034007}}.

\bibitem{Cheung:2021epq}
Vincent Cheung and Ramona Vogt.
\newblock {Production and polarization of direct J/\ensuremath{\psi} to
  O(\ensuremath{\alpha}s3) in the improved color evaporation model in collinear
  factorization}.
\newblock {\em Phys. Rev. D}, 104(9):094026, 2021.
\newblock \href {http://arxiv.org/abs/2102.09118} {\path{arXiv:2102.09118}},
  \href {https://doi.org/10.1103/PhysRevD.104.094026}
  {\path{doi:10.1103/PhysRevD.104.094026}}.

\bibitem{Vogt:2018oje}
R.~Vogt.
\newblock {Heavy Flavor Azimuthal Correlations in Cold Nuclear Matter}.
\newblock {\em Phys. Rev. C}, 98(3):034907, 2018.
\newblock \href {http://arxiv.org/abs/1806.01904} {\path{arXiv:1806.01904}},
  \href {https://doi.org/10.1103/PhysRevC.98.034907}
  {\path{doi:10.1103/PhysRevC.98.034907}}.

\bibitem{Vogt:2019xmm}
R.~Vogt.
\newblock {$b \overline b$ kinematic correlations in cold nuclear matter}.
\newblock {\em Phys. Rev. C}, 101(2):024910, 2020.
\newblock \href {http://arxiv.org/abs/1908.05320} {\path{arXiv:1908.05320}},
  \href {https://doi.org/10.1103/PhysRevC.101.024910}
  {\path{doi:10.1103/PhysRevC.101.024910}}.

\bibitem{Wu:2019mky}
Xing-Gang Wu, Jian-Ming Shen, Bo-Lun Du, Xu-Dong Huang, Sheng-Quan Wang, and
  Stanley~J. Brodsky.
\newblock {The QCD renormalization group equation and the elimination of
  fixed-order scheme-and-scale ambiguities using the principle of maximum
  conformality}.
\newblock {\em Prog. Part. Nucl. Phys.}, 108:103706, 2019.
\newblock \href {http://arxiv.org/abs/1903.12177} {\path{arXiv:1903.12177}},
  \href {https://doi.org/10.1016/j.ppnp.2019.05.003}
  {\path{doi:10.1016/j.ppnp.2019.05.003}}.

\bibitem{Braun:2017cih}
V.~M. Braun, A.~N. Manashov, S.~Moch, and M.~Strohmaier.
\newblock {Three-loop evolution equation for flavor-nonsinglet operators in
  off-forward kinematics}.
\newblock {\em JHEP}, 06:037, 2017.
\newblock \href {http://arxiv.org/abs/1703.09532} {\path{arXiv:1703.09532}},
  \href {https://doi.org/10.1007/JHEP06(2017)037}
  {\path{doi:10.1007/JHEP06(2017)037}}.

\bibitem{Braun:2020yib}
V.~M. Braun, A.~N. Manashov, S.~Moch, and J.~Schoenleber.
\newblock {Two-loop coefficient function for DVCS: vector contributions}.
\newblock {\em JHEP}, 09:117, 2020.
\newblock [Erratum: JHEP 02, 115 (2022)].
\newblock \href {http://arxiv.org/abs/2007.06348} {\path{arXiv:2007.06348}},
  \href {https://doi.org/10.1007/JHEP09(2020)117}
  {\path{doi:10.1007/JHEP09(2020)117}}.

\bibitem{Braun:2022bpn}
V.~M. Braun, Yao Ji, and Jakob Schoenleber.
\newblock {Deeply Virtual Compton Scattering at Next-to-Next-to-Leading Order}.
\newblock {\em Phys. Rev. Lett.}, 129(17):172001, 2022.
\newblock \href {http://arxiv.org/abs/2207.06818} {\path{arXiv:2207.06818}},
  \href {https://doi.org/10.1103/PhysRevLett.129.172001}
  {\path{doi:10.1103/PhysRevLett.129.172001}}.

\bibitem{Belitsky:2001ns}
Andrei~V. Belitsky, Dieter Mueller, and A.~Kirchner.
\newblock {Theory of deeply virtual Compton scattering on the nucleon}.
\newblock {\em Nucl. Phys. B}, 629:323--392, 2002.
\newblock \href {http://arxiv.org/abs/hep-ph/0112108}
  {\path{arXiv:hep-ph/0112108}}, \href
  {https://doi.org/10.1016/S0550-3213(02)00144-X}
  {\path{doi:10.1016/S0550-3213(02)00144-X}}.

\bibitem{Berthou:2015oaw}
B.~Berthou et~al.
\newblock {PARTONS: PARtonic Tomography Of Nucleon Software}: {A computing
  framework for the phenomenology of Generalized Parton Distributions}.
\newblock {\em Eur. Phys. J. C}, 78(6):478, 2018.
\newblock \href {http://arxiv.org/abs/1512.06174} {\path{arXiv:1512.06174}},
  \href {https://doi.org/10.1140/epjc/s10052-018-5948-0}
  {\path{doi:10.1140/epjc/s10052-018-5948-0}}.

\bibitem{Kumericki:2016ehc}
Kresimir Kumericki, Simonetta Liuti, and Herve Moutarde.
\newblock {GPD phenomenology and DVCS fitting}: {Entering the high-precision
  era}.
\newblock {\em Eur. Phys. J. A}, 52(6):157, 2016.
\newblock \href {http://arxiv.org/abs/1602.02763} {\path{arXiv:1602.02763}},
  \href {https://doi.org/10.1140/epja/i2016-16157-3}
  {\path{doi:10.1140/epja/i2016-16157-3}}.

\bibitem{Kriesten:2019jep}
Brandon Kriesten, Simonetta Liuti, Liliet Calero-Diaz, Dustin Keller, Andrew
  Meyer, Gary~R. Goldstein, and J.~Osvaldo Gonzalez-Hernandez.
\newblock {Extraction of generalized parton distribution observables from
  deeply virtual electron proton scattering experiments}.
\newblock {\em Phys. Rev. D}, 101(5):054021, 2020.
\newblock \href {http://arxiv.org/abs/1903.05742} {\path{arXiv:1903.05742}},
  \href {https://doi.org/10.1103/PhysRevD.101.054021}
  {\path{doi:10.1103/PhysRevD.101.054021}}.

\bibitem{Kriesten:2020wcx}
Brandon Kriesten and Simonetta Liuti.
\newblock {Theory of deeply virtual Compton scattering off the unpolarized
  proton}.
\newblock {\em Phys. Rev. D}, 105(1):016015, 2022.
\newblock \href {http://arxiv.org/abs/2004.08890} {\path{arXiv:2004.08890}},
  \href {https://doi.org/10.1103/PhysRevD.105.016015}
  {\path{doi:10.1103/PhysRevD.105.016015}}.

\bibitem{Grigsby:2020auv}
Jake Grigsby, Brandon Kriesten, Joshua Hoskins, Simonetta Liuti, Peter Alonzi,
  and Matthias Burkardt.
\newblock {Deep learning analysis of deeply virtual exclusive photoproduction}.
\newblock {\em Phys. Rev. D}, 104(1):016001, 2021.
\newblock \href {http://arxiv.org/abs/2012.04801} {\path{arXiv:2012.04801}},
  \href {https://doi.org/10.1103/PhysRevD.104.016001}
  {\path{doi:10.1103/PhysRevD.104.016001}}.

\bibitem{Kriesten:2021sqc}
Brandon Kriesten, Philip Velie, Emma Yeats, Fernanda~Yepez Lopez, and Simonetta
  Liuti.
\newblock {Parametrization of quark and gluon generalized parton distributions
  in a dynamical framework}.
\newblock {\em Phys. Rev. D}, 105(5):056022, 2022.
\newblock \href {http://arxiv.org/abs/2101.01826} {\path{arXiv:2101.01826}},
  \href {https://doi.org/10.1103/PhysRevD.105.056022}
  {\path{doi:10.1103/PhysRevD.105.056022}}.

\bibitem{Guo:2021gru}
Yuxun Guo, Xiangdong Ji, and Kyle Shiells.
\newblock {Higher-order kinematical effects in deeply virtual Compton
  scattering}.
\newblock {\em JHEP}, 12:103, 2021.
\newblock \href {http://arxiv.org/abs/2109.10373} {\path{arXiv:2109.10373}},
  \href {https://doi.org/10.1007/JHEP12(2021)103}
  {\path{doi:10.1007/JHEP12(2021)103}}.

\bibitem{Shiells:2021xqo}
Kyle Shiells, Yuxun Guo, and Xiangdong Ji.
\newblock {On extraction of twist-two Compton form factors from DVCS
  observables through harmonic analysis}.
\newblock {\em JHEP}, 08:048, 2022.
\newblock \href {http://arxiv.org/abs/2112.15144} {\path{arXiv:2112.15144}},
  \href {https://doi.org/10.1007/JHEP08(2022)048}
  {\path{doi:10.1007/JHEP08(2022)048}}.

\bibitem{Guo:2022cgq}
Yuxun Guo, Xiangdong Ji, Brandon Kriesten, and Kyle Shiells.
\newblock {Twist-three cross-sections in deeply virtual Compton scattering}.
\newblock {\em JHEP}, 06:096, 2022.
\newblock \href {http://arxiv.org/abs/2202.11114} {\path{arXiv:2202.11114}},
  \href {https://doi.org/10.1007/JHEP06(2022)096}
  {\path{doi:10.1007/JHEP06(2022)096}}.

\bibitem{Bacchetta:2017gcc}
Alessandro Bacchetta, Filippo Delcarro, Cristian Pisano, Marco Radici, and
  Andrea Signori.
\newblock {Extraction of partonic transverse momentum distributions from
  semi-inclusive deep-inelastic scattering, Drell-Yan and Z-boson production}.
\newblock {\em JHEP}, 06:081, 2017.
\newblock [Erratum: JHEP 06, 051 (2019)].
\newblock \href {http://arxiv.org/abs/1703.10157} {\path{arXiv:1703.10157}},
  \href {https://doi.org/10.1007/JHEP06(2017)081}
  {\path{doi:10.1007/JHEP06(2017)081}}.

\bibitem{Scimemi:2017etj}
Ignazio Scimemi and Alexey Vladimirov.
\newblock {Analysis of vector boson production within TMD factorization}.
\newblock {\em Eur. Phys. J. C}, 78(2):89, 2018.
\newblock \href {http://arxiv.org/abs/1706.01473} {\path{arXiv:1706.01473}},
  \href {https://doi.org/10.1140/epjc/s10052-018-5557-y}
  {\path{doi:10.1140/epjc/s10052-018-5557-y}}.

\bibitem{Scimemi:2019cmh}
Ignazio Scimemi and Alexey Vladimirov.
\newblock {Non-perturbative structure of semi-inclusive deep-inelastic and
  Drell-Yan scattering at small transverse momentum}.
\newblock {\em JHEP}, 06:137, 2020.
\newblock \href {http://arxiv.org/abs/1912.06532} {\path{arXiv:1912.06532}},
  \href {https://doi.org/10.1007/JHEP06(2020)137}
  {\path{doi:10.1007/JHEP06(2020)137}}.

\bibitem{Bacchetta:2022awv}
Alessandro Bacchetta, Valerio Bertone, Chiara Bissolotti, Giuseppe Bozzi,
  Matteo Cerutti, Fulvio Piacenza, Marco Radici, and Andrea Signori.
\newblock {Unpolarized transverse momentum distributions from a global fit of
  Drell-Yan and semi-inclusive deep-inelastic scattering data}.
\newblock {\em JHEP}, 10:127, 2022.
\newblock \href {http://arxiv.org/abs/2206.07598} {\path{arXiv:2206.07598}},
  \href {https://doi.org/10.1007/JHEP10(2022)127}
  {\path{doi:10.1007/JHEP10(2022)127}}.

\bibitem{Bailey:2020ooq}
S.~Bailey, T.~Cridge, L.~A. Harland-Lang, A.~D. Martin, and R.~S. Thorne.
\newblock {Parton distributions from LHC, HERA, Tevatron and fixed target data:
  MSHT20 PDFs}.
\newblock {\em Eur. Phys. J. C}, 81(4):341, 2021.
\newblock \href {http://arxiv.org/abs/2012.04684} {\path{arXiv:2012.04684}},
  \href {https://doi.org/10.1140/epjc/s10052-021-09057-0}
  {\path{doi:10.1140/epjc/s10052-021-09057-0}}.

\bibitem{NNPDF:2021njg}
Richard~D. Ball et~al.
\newblock {The path to proton structure at 1\% accuracy}.
\newblock {\em Eur. Phys. J. C}, 82(5):428, 2022.
\newblock \href {http://arxiv.org/abs/2109.02653} {\path{arXiv:2109.02653}},
  \href {https://doi.org/10.1140/epjc/s10052-022-10328-7}
  {\path{doi:10.1140/epjc/s10052-022-10328-7}}.

\bibitem{Barry:2021osv}
P.~C. Barry, Chueng-Ryong Ji, N.~Sato, and W.~Melnitchouk.
\newblock {Global QCD Analysis of Pion Parton Distributions with Threshold
  Resummation}.
\newblock {\em Phys. Rev. Lett.}, 127(23):232001, 2021.
\newblock \href {http://arxiv.org/abs/2108.05822} {\path{arXiv:2108.05822}},
  \href {https://doi.org/10.1103/PhysRevLett.127.232001}
  {\path{doi:10.1103/PhysRevLett.127.232001}}.

\bibitem{Barry:2018ort}
P.~C. Barry, N.~Sato, W.~Melnitchouk, and Chueng-Ryong Ji.
\newblock {First Monte Carlo Global QCD Analysis of Pion Parton Distributions}.
\newblock {\em Phys. Rev. Lett.}, 121(15):152001, 2018.
\newblock \href {http://arxiv.org/abs/1804.01965} {\path{arXiv:1804.01965}},
  \href {https://doi.org/10.1103/PhysRevLett.121.152001}
  {\path{doi:10.1103/PhysRevLett.121.152001}}.

\bibitem{Bauer:2000yr}
Christian~W. Bauer, Sean Fleming, Dan Pirjol, and Iain~W. Stewart.
\newblock {An Effective field theory for collinear and soft gluons: Heavy to
  light decays}.
\newblock {\em Phys. Rev. D}, 63:114020, 2001.
\newblock \href {http://arxiv.org/abs/hep-ph/0011336}
  {\path{arXiv:hep-ph/0011336}}, \href
  {https://doi.org/10.1103/PhysRevD.63.114020}
  {\path{doi:10.1103/PhysRevD.63.114020}}.

\bibitem{Bauer:2001yt}
Christian~W. Bauer, Dan Pirjol, and Iain~W. Stewart.
\newblock {Soft collinear factorization in effective field theory}.
\newblock {\em Phys. Rev. D}, 65:054022, 2002.
\newblock \href {http://arxiv.org/abs/hep-ph/0109045}
  {\path{arXiv:hep-ph/0109045}}, \href
  {https://doi.org/10.1103/PhysRevD.65.054022}
  {\path{doi:10.1103/PhysRevD.65.054022}}.

\bibitem{Bauer:2002nz}
Christian~W. Bauer, Sean Fleming, Dan Pirjol, Ira~Z. Rothstein, and Iain~W.
  Stewart.
\newblock {Hard scattering factorization from effective field theory}.
\newblock {\em Phys. Rev. D}, 66:014017, 2002.
\newblock \href {http://arxiv.org/abs/hep-ph/0202088}
  {\path{arXiv:hep-ph/0202088}}, \href
  {https://doi.org/10.1103/PhysRevD.66.014017}
  {\path{doi:10.1103/PhysRevD.66.014017}}.

\bibitem{Ebert:2021jhy}
Markus~A. Ebert, Anjie Gao, and Iain~W. Stewart.
\newblock {Factorization for azimuthal asymmetries in SIDIS at next-to-leading
  power}.
\newblock {\em JHEP}, 06:007, 2022.
\newblock \href {http://arxiv.org/abs/2112.07680} {\path{arXiv:2112.07680}},
  \href {https://doi.org/10.1007/JHEP06(2022)007}
  {\path{doi:10.1007/JHEP06(2022)007}}.

\bibitem{Gribov:1984tu}
L.~V. Gribov, E.~M. Levin, and M.~G. Ryskin.
\newblock {Semihard Processes in QCD}.
\newblock {\em Phys. Rept.}, 100:1--150, 1983.
\newblock \href {https://doi.org/10.1016/0370-1573(83)90022-4}
  {\path{doi:10.1016/0370-1573(83)90022-4}}.

\bibitem{Mueller:1985wy}
Alfred~H. Mueller and Jian-wei Qiu.
\newblock {Gluon Recombination and Shadowing at Small Values of x}.
\newblock {\em Nucl. Phys. B}, 268:427--452, 1986.
\newblock \href {https://doi.org/10.1016/0550-3213(86)90164-1}
  {\path{doi:10.1016/0550-3213(86)90164-1}}.

\bibitem{McLerran:1993ni}
Larry~D. McLerran and Raju Venugopalan.
\newblock {Computing quark and gluon distribution functions for very large
  nuclei}.
\newblock {\em Phys. Rev. D}, 49:2233--2241, 1994.
\newblock \href {http://arxiv.org/abs/hep-ph/9309289}
  {\path{arXiv:hep-ph/9309289}}, \href
  {https://doi.org/10.1103/PhysRevD.49.2233}
  {\path{doi:10.1103/PhysRevD.49.2233}}.

\bibitem{McLerran:1993ka}
Larry~D. McLerran and Raju Venugopalan.
\newblock {Gluon distribution functions for very large nuclei at small
  transverse momentum}.
\newblock {\em Phys. Rev. D}, 49:3352--3355, 1994.
\newblock \href {http://arxiv.org/abs/hep-ph/9311205}
  {\path{arXiv:hep-ph/9311205}}, \href
  {https://doi.org/10.1103/PhysRevD.49.3352}
  {\path{doi:10.1103/PhysRevD.49.3352}}.

\bibitem{Iancu:2003xm}
Edmond Iancu and Raju Venugopalan.
\newblock {\em {The Color glass condensate and high-energy scattering in QCD}},
  pages 249--3363.
\newblock 3 2003.
\newblock \href {http://arxiv.org/abs/hep-ph/0303204}
  {\path{arXiv:hep-ph/0303204}}, \href
  {https://doi.org/10.1142/9789812795533_0005}
  {\path{doi:10.1142/9789812795533_0005}}.

\bibitem{Kovchegov:2012mbw}
Yuri~V. Kovchegov and Eugene Levin.
\newblock {\em {Quantum Chromodynamics at High Energy}}, volume~33 of {\em
  Cambridge Monographs on Particle Physics, Nuclear Physics and Cosmology
  (33)}.
\newblock Cambridge University Press, 11 2022.
\newblock \href {https://doi.org/10.1017/9781009291446}
  {\path{doi:10.1017/9781009291446}}.

\bibitem{Balitsky:1995ub}
I.~Balitsky.
\newblock {Operator expansion for high-energy scattering}.
\newblock {\em Nucl. Phys. B}, 463:99--160, 1996.
\newblock \href {http://arxiv.org/abs/hep-ph/9509348}
  {\path{arXiv:hep-ph/9509348}}, \href
  {https://doi.org/10.1016/0550-3213(95)00638-9}
  {\path{doi:10.1016/0550-3213(95)00638-9}}.

\bibitem{Kovchegov:1998bi}
Yuri~V. Kovchegov and Alfred~H. Mueller.
\newblock {Gluon production in current nucleus and nucleon - nucleus collisions
  in a quasiclassical approximation}.
\newblock {\em Nucl. Phys. B}, 529:451--479, 1998.
\newblock \href {http://arxiv.org/abs/hep-ph/9802440}
  {\path{arXiv:hep-ph/9802440}}, \href
  {https://doi.org/10.1016/S0550-3213(98)00384-8}
  {\path{doi:10.1016/S0550-3213(98)00384-8}}.

\bibitem{JalilianMarian:1997gr}
Jamal Jalilian-Marian, Alex Kovner, Andrei Leonidov, and Heribert Weigert.
\newblock {The Wilson renormalization group for low x physics: Towards the high
  density regime}.
\newblock {\em Phys. Rev. D}, 59:014014, 1998.
\newblock \href {http://arxiv.org/abs/hep-ph/9706377}
  {\path{arXiv:hep-ph/9706377}}, \href
  {https://doi.org/10.1103/PhysRevD.59.014014}
  {\path{doi:10.1103/PhysRevD.59.014014}}.

\bibitem{JalilianMarian:1997dw}
Jamal Jalilian-Marian, Alex Kovner, and Heribert Weigert.
\newblock {The Wilson renormalization group for low x physics: Gluon evolution
  at finite parton density}.
\newblock {\em Phys. Rev. D}, 59:014015, 1998.
\newblock \href {http://arxiv.org/abs/hep-ph/9709432}
  {\path{arXiv:hep-ph/9709432}}, \href
  {https://doi.org/10.1103/PhysRevD.59.014015}
  {\path{doi:10.1103/PhysRevD.59.014015}}.

\bibitem{Iancu:2000hn}
Edmond Iancu, Andrei Leonidov, and Larry~D. McLerran.
\newblock {Nonlinear gluon evolution in the color glass condensate. 1.}
\newblock {\em Nucl. Phys. A}, 692:583--645, 2001.
\newblock \href {http://arxiv.org/abs/hep-ph/0011241}
  {\path{arXiv:hep-ph/0011241}}, \href
  {https://doi.org/10.1016/S0375-9474(01)00642-X}
  {\path{doi:10.1016/S0375-9474(01)00642-X}}.

\bibitem{Ferreiro:2001qy}
Elena Ferreiro, Edmond Iancu, Andrei Leonidov, and Larry McLerran.
\newblock {Nonlinear gluon evolution in the color glass condensate. 2.}
\newblock {\em Nucl. Phys. A}, 703:489--538, 2002.
\newblock \href {http://arxiv.org/abs/hep-ph/0109115}
  {\path{arXiv:hep-ph/0109115}}, \href
  {https://doi.org/10.1016/S0375-9474(01)01329-X}
  {\path{doi:10.1016/S0375-9474(01)01329-X}}.

\bibitem{Morreale:2021pnn}
Astrid Morreale and Farid Salazar.
\newblock {Mining for Gluon Saturation at Colliders}.
\newblock {\em Universe}, 7(8):312, 2021.
\newblock \href {http://arxiv.org/abs/2108.08254} {\path{arXiv:2108.08254}},
  \href {https://doi.org/10.3390/universe7080312}
  {\path{doi:10.3390/universe7080312}}.

\bibitem{Hatta:2016dxp}
Yoshitaka Hatta, Bo-Wen Xiao, and Feng Yuan.
\newblock {Probing the Small- x Gluon Tomography in Correlated Hard Diffractive
  Dijet Production in Deep Inelastic Scattering}.
\newblock {\em Phys. Rev. Lett.}, 116(20):202301, 2016.
\newblock \href {http://arxiv.org/abs/1601.01585} {\path{arXiv:1601.01585}},
  \href {https://doi.org/10.1103/PhysRevLett.116.202301}
  {\path{doi:10.1103/PhysRevLett.116.202301}}.

\bibitem{Kharzeev:2017qzs}
Dmitri~E. Kharzeev and Eugene~M. Levin.
\newblock {Deep inelastic scattering as a probe of entanglement}.
\newblock {\em Phys. Rev. D}, 95(11):114008, 2017.
\newblock \href {http://arxiv.org/abs/1702.03489} {\path{arXiv:1702.03489}},
  \href {https://doi.org/10.1103/PhysRevD.95.114008}
  {\path{doi:10.1103/PhysRevD.95.114008}}.

\bibitem{Hagiwara:2017uaz}
Yoshikazu Hagiwara, Yoshitaka Hatta, Bo-Wen Xiao, and Feng Yuan.
\newblock {Classical and quantum entropy of parton distributions}.
\newblock {\em Phys. Rev. D}, 97(9):094029, 2018.
\newblock \href {http://arxiv.org/abs/1801.00087} {\path{arXiv:1801.00087}},
  \href {https://doi.org/10.1103/PhysRevD.97.094029}
  {\path{doi:10.1103/PhysRevD.97.094029}}.

\bibitem{Hentschinski:2022rsa}
Martin Hentschinski, Krzysztof Kutak, and Robert Straka.
\newblock {Maximally entangled proton and charged hadron multiplicity in Deep
  Inelastic Scattering}.
\newblock {\em Eur. Phys. J. C}, 82(12):1147, 2022.
\newblock \href {http://arxiv.org/abs/2207.09430} {\path{arXiv:2207.09430}},
  \href {https://doi.org/10.1140/epjc/s10052-022-11122-1}
  {\path{doi:10.1140/epjc/s10052-022-11122-1}}.

\bibitem{Dvali:2021ooc}
Gia Dvali and Raju Venugopalan.
\newblock {Classicalization and unitarization of wee partons in QCD and
  gravity: The CGC-black hole correspondence}.
\newblock {\em Phys. Rev. D}, 105(5):056026, 2022.
\newblock \href {http://arxiv.org/abs/2106.11989} {\path{arXiv:2106.11989}},
  \href {https://doi.org/10.1103/PhysRevD.105.056026}
  {\path{doi:10.1103/PhysRevD.105.056026}}.

\bibitem{Dumitru:2022tud}
Adrian Dumitru and Eric Kolbusz.
\newblock {Quark and gluon entanglement in the proton on the light cone at
  intermediate $x$}.
\newblock {\em Phys. Rev. D}, 105:074030, 2022.
\newblock \href {http://arxiv.org/abs/2202.01803} {\path{arXiv:2202.01803}},
  \href {https://doi.org/10.1103/PhysRevD.105.074030}
  {\path{doi:10.1103/PhysRevD.105.074030}}.

\bibitem{Duan:2021clk}
Haowu Duan, Alex Kovner, and Vladimir~V. Skokov.
\newblock {Gluon quasiparticles and the CGC density matrix}.
\newblock {\em Phys. Rev. D}, 105(5):056009, 2022.
\newblock \href {http://arxiv.org/abs/2111.06475} {\path{arXiv:2111.06475}},
  \href {https://doi.org/10.1103/PhysRevD.105.056009}
  {\path{doi:10.1103/PhysRevD.105.056009}}.

\bibitem{Berges:2014bba}
J.~Berges, K.~Boguslavski, S.~Schlichting, and R.~Venugopalan.
\newblock {Universality far from equilibrium: From superfluid Bose gases to
  heavy-ion collisions}.
\newblock {\em Phys. Rev. Lett.}, 114(6):061601, 2015.
\newblock \href {http://arxiv.org/abs/1408.1670} {\path{arXiv:1408.1670}},
  \href {https://doi.org/10.1103/PhysRevLett.114.061601}
  {\path{doi:10.1103/PhysRevLett.114.061601}}.

\bibitem{Prufer:2018hto}
Maximilian Pr\"ufer, Philipp Kunkel, Helmut Strobel, Stefan Lannig, Daniel
  Linnemann, Christian-Marcel Schmied, J\"urgen Berges, Thomas Gasenzer, and
  Markus~K. Oberthaler.
\newblock {Observation of universal dynamics in a spinor Bose gas far from
  equilibrium}.
\newblock {\em Nature}, 563(7730):217--220, 2018.
\newblock \href {http://arxiv.org/abs/1805.11881} {\path{arXiv:1805.11881}},
  \href {https://doi.org/10.1038/s41586-018-0659-0}
  {\path{doi:10.1038/s41586-018-0659-0}}.

\bibitem{Bhaduri:1988gc}
R.~K. Bhaduri.
\newblock {\em {MODELS OF THE NUCLEON: FROM QUARKS TO SOLITON}}.
\newblock 1988.

\bibitem{Thomas:2001kw}
Anthony~William Thomas and Wolfram Weise.
\newblock {\em {The Structure of the Nucleon}}.
\newblock Wiley, Germany, 2001.
\newblock \href {https://doi.org/10.1002/352760314X}
  {\path{doi:10.1002/352760314X}}.

\bibitem{Maris:2003vk}
Pieter Maris and Craig~D. Roberts.
\newblock {Dyson-Schwinger equations: A Tool for hadron physics}.
\newblock {\em Int. J. Mod. Phys. E}, 12:297--365, 2003.
\newblock \href {http://arxiv.org/abs/nucl-th/0301049}
  {\path{arXiv:nucl-th/0301049}}, \href
  {https://doi.org/10.1142/S0218301303001326}
  {\path{doi:10.1142/S0218301303001326}}.

\bibitem{Cloet:2013jya}
Ian~C. Cloet and Craig~D. Roberts.
\newblock {Explanation and Prediction of Observables using Continuum Strong
  QCD}.
\newblock {\em Prog. Part. Nucl. Phys.}, 77:1--69, 2014.
\newblock \href {http://arxiv.org/abs/1310.2651} {\path{arXiv:1310.2651}},
  \href {https://doi.org/10.1016/j.ppnp.2014.02.001}
  {\path{doi:10.1016/j.ppnp.2014.02.001}}.

\bibitem{Schafer:1996wv}
Thomas Sch\"afer and Edward~V. Shuryak.
\newblock {Instantons in QCD}.
\newblock {\em Rev. Mod. Phys.}, 70:323--426, 1998.
\newblock \href {http://arxiv.org/abs/hep-ph/9610451}
  {\path{arXiv:hep-ph/9610451}}, \href
  {https://doi.org/10.1103/RevModPhys.70.323}
  {\path{doi:10.1103/RevModPhys.70.323}}.

\bibitem{Brodsky:2014yha}
Stanley~J. Brodsky, Guy~F. de~Teramond, Hans~Gunter Dosch, and Joshua Erlich.
\newblock {Light-Front Holographic QCD and Emerging Confinement}.
\newblock {\em Phys. Rept.}, 584:1--105, 2015.
\newblock \href {http://arxiv.org/abs/1407.8131} {\path{arXiv:1407.8131}},
  \href {https://doi.org/10.1016/j.physrep.2015.05.001}
  {\path{doi:10.1016/j.physrep.2015.05.001}}.

\bibitem{Lan:2019vui}
Jiangshan Lan, Chandan Mondal, Shaoyang Jia, Xingbo Zhao, and James~P. Vary.
\newblock {Parton Distribution Functions from a Light Front Hamiltonian and QCD
  Evolution for Light Mesons}.
\newblock {\em Phys. Rev. Lett.}, 122(17):172001, 2019.
\newblock \href {http://arxiv.org/abs/1901.11430} {\path{arXiv:1901.11430}},
  \href {https://doi.org/10.1103/PhysRevLett.122.172001}
  {\path{doi:10.1103/PhysRevLett.122.172001}}.

\bibitem{Xu:2021wwj}
Siqi Xu, Chandan Mondal, Jiangshan Lan, Xingbo Zhao, Yang Li, and James~P.
  Vary.
\newblock {Nucleon structure from basis light-front quantization}.
\newblock {\em Phys. Rev. D}, 104(9):094036, 2021.
\newblock \href {http://arxiv.org/abs/2108.03909} {\path{arXiv:2108.03909}},
  \href {https://doi.org/10.1103/PhysRevD.104.094036}
  {\path{doi:10.1103/PhysRevD.104.094036}}.

\bibitem{Arifi:2022pal}
Ahmad~Jafar Arifi, Ho-Meoyng Choi, Chueng-Ryong ji, and Yongseok Oh.
\newblock {Mixing effects on 1S and 2S state heavy mesons in the light-front
  quark model}.
\newblock {\em Phys. Rev. D}, 106(1):014009, 2022.
\newblock \href {http://arxiv.org/abs/2205.04075} {\path{arXiv:2205.04075}},
  \href {https://doi.org/10.1103/PhysRevD.106.014009}
  {\path{doi:10.1103/PhysRevD.106.014009}}.

\bibitem{Choi:2015ywa}
Ho-Meoyng Choi, Chueng-Ryong Ji, Ziyue Li, and Hui-Young Ryu.
\newblock {Variational analysis of mass spectra and decay constants for ground
  state pseudoscalar and vector mesons in the light-front quark model}.
\newblock {\em Phys. Rev. C}, 92(5):055203, 2015.
\newblock \href {http://arxiv.org/abs/1502.03078} {\path{arXiv:1502.03078}},
  \href {https://doi.org/10.1103/PhysRevC.92.055203}
  {\path{doi:10.1103/PhysRevC.92.055203}}.

\bibitem{Ji:2000fy}
Chueng-Ryong Ji and Ho-Meoyng Choi.
\newblock {New effective treatment of the light front nonvalence contribution
  in timelike exclusive processes}.
\newblock {\em Phys. Lett. B}, 513:330, 2001.
\newblock \href {http://arxiv.org/abs/hep-ph/0009281}
  {\path{arXiv:hep-ph/0009281}}, \href
  {https://doi.org/10.1016/S0370-2693(01)00481-6}
  {\path{doi:10.1016/S0370-2693(01)00481-6}}.

\bibitem{Pilloni:2016obd}
A.~Pilloni, C.~Fernandez-Ramirez, A.~Jackura, V.~Mathieu, M.~Mikhasenko,
  J.~Nys, and A.~P. Szczepaniak.
\newblock {Amplitude analysis and the nature of the Z$_c$(3900)}.
\newblock {\em Phys. Lett. B}, 772:200--209, 2017.
\newblock \href {http://arxiv.org/abs/1612.06490} {\path{arXiv:1612.06490}},
  \href {https://doi.org/10.1016/j.physletb.2017.06.030}
  {\path{doi:10.1016/j.physletb.2017.06.030}}.

\bibitem{COMPASS:2020yhb}
G.~D. Alexeev et~al.
\newblock {Triangle Singularity as the Origin of the $a_1(1420)$}.
\newblock {\em Phys. Rev. Lett.}, 127(8):082501, 2021.
\newblock \href {http://arxiv.org/abs/2006.05342} {\path{arXiv:2006.05342}},
  \href {https://doi.org/10.1103/PhysRevLett.127.082501}
  {\path{doi:10.1103/PhysRevLett.127.082501}}.

\bibitem{Albaladejo:2020tzt}
M.~Albaladejo, A.~N. Hiller~Blin, A.~Pilloni, D.~Winney,
  C.~Fern\'andez-Ram\'\i{}rez, V.~Mathieu, and A.~Szczepaniak.
\newblock {XYZ spectroscopy at electron-hadron facilities: Exclusive
  processes}.
\newblock {\em Phys. Rev. D}, 102:114010, 2020.
\newblock \href {http://arxiv.org/abs/2008.01001} {\path{arXiv:2008.01001}},
  \href {https://doi.org/10.1103/PhysRevD.102.114010}
  {\path{doi:10.1103/PhysRevD.102.114010}}.

\bibitem{Winney:2022tky}
D.~Winney, A.~Pilloni, V.~Mathieu, A.~N. Hiller~Blin, M.~Albaladejo, W.~A.
  Smith, and A.~Szczepaniak.
\newblock {XYZ spectroscopy at electron-hadron facilities. II. Semi-inclusive
  processes with pion exchange}.
\newblock {\em Phys. Rev. D}, 106(9):094009, 2022.
\newblock \href {http://arxiv.org/abs/2209.05882} {\path{arXiv:2209.05882}},
  \href {https://doi.org/10.1103/PhysRevD.106.094009}
  {\path{doi:10.1103/PhysRevD.106.094009}}.

\bibitem{STAR:2002ggv}
C.~Adler et~al.
\newblock {Centrality dependence of high $p_{T}$ hadron suppression in Au+Au
  collisions at $\sqrt{s}_{NN}$ = 130-GeV}.
\newblock {\em Phys. Rev. Lett.}, 89:202301, 2002.
\newblock \href {http://arxiv.org/abs/nucl-ex/0206011}
  {\path{arXiv:nucl-ex/0206011}}, \href
  {https://doi.org/10.1103/PhysRevLett.89.202301}
  {\path{doi:10.1103/PhysRevLett.89.202301}}.

\bibitem{Jeon:2003gi}
Sangyong Jeon and Guy~D. Moore.
\newblock {Energy loss of leading partons in a thermal QCD medium}.
\newblock {\em Phys. Rev. C}, 71:034901, 2005.
\newblock \href {http://arxiv.org/abs/hep-ph/0309332}
  {\path{arXiv:hep-ph/0309332}}, \href
  {https://doi.org/10.1103/PhysRevC.71.034901}
  {\path{doi:10.1103/PhysRevC.71.034901}}.

\bibitem{Blaizot:2013vha}
Jean-Paul Blaizot, Fabio Dominguez, Edmond Iancu, and Yacine Mehtar-Tani.
\newblock {Probabilistic picture for medium-induced jet evolution}.
\newblock {\em JHEP}, 06:075, 2014.
\newblock \href {http://arxiv.org/abs/1311.5823} {\path{arXiv:1311.5823}},
  \href {https://doi.org/10.1007/JHEP06(2014)075}
  {\path{doi:10.1007/JHEP06(2014)075}}.

\bibitem{Blaizot:2013hx}
Jean-Paul Blaizot, Edmond Iancu, and Yacine Mehtar-Tani.
\newblock {Medium-induced QCD cascade: democratic branching and wave
  turbulence}.
\newblock {\em Phys. Rev. Lett.}, 111:052001, 2013.
\newblock \href {http://arxiv.org/abs/1301.6102} {\path{arXiv:1301.6102}},
  \href {https://doi.org/10.1103/PhysRevLett.111.052001}
  {\path{doi:10.1103/PhysRevLett.111.052001}}.

\bibitem{Mehtar-Tani:2018zba}
Yacine Mehtar-Tani and Soeren Schlichting.
\newblock {Universal quark to gluon ratio in medium-induced parton cascade}.
\newblock {\em JHEP}, 09:144, 2018.
\newblock \href {http://arxiv.org/abs/1807.06181} {\path{arXiv:1807.06181}},
  \href {https://doi.org/10.1007/JHEP09(2018)144}
  {\path{doi:10.1007/JHEP09(2018)144}}.

\bibitem{Schlichting:2020lef}
Soeren Schlichting and Ismail Soudi.
\newblock {Medium-induced fragmentation and equilibration of highly energetic
  partons}.
\newblock {\em JHEP}, 07:077, 2021.
\newblock \href {http://arxiv.org/abs/2008.04928} {\path{arXiv:2008.04928}},
  \href {https://doi.org/10.1007/JHEP07(2021)077}
  {\path{doi:10.1007/JHEP07(2021)077}}.

\bibitem{Sievert:2019cwq}
Matthew~D. Sievert, Ivan Vitev, and Boram Yoon.
\newblock {A complete set of in-medium splitting functions to any order in
  opacity}.
\newblock {\em Phys. Lett. B}, 795:502--510, 2019.
\newblock \href {http://arxiv.org/abs/1903.06170} {\path{arXiv:1903.06170}},
  \href {https://doi.org/10.1016/j.physletb.2019.06.019}
  {\path{doi:10.1016/j.physletb.2019.06.019}}.

\bibitem{Arnold:2021pin}
Peter Arnold, Tyler Gorda, and Shahin Iqbal.
\newblock {The LPM effect in sequential bremsstrahlung: analytic results for
  sub-leading (single) logarithms}.
\newblock {\em JHEP}, 04:085, 2022.
\newblock \href {http://arxiv.org/abs/2112.05161} {\path{arXiv:2112.05161}},
  \href {https://doi.org/10.1007/JHEP04(2022)085}
  {\path{doi:10.1007/JHEP04(2022)085}}.

\bibitem{Arnold:2020uzm}
Peter Arnold, Tyler Gorda, and Shahin Iqbal.
\newblock {The LPM effect in sequential bremsstrahlung: nearly complete results
  for QCD}.
\newblock {\em JHEP}, 11:053, 2020.
\newblock [Erratum: JHEP 05, 114 (2022)].
\newblock \href {http://arxiv.org/abs/2007.15018} {\path{arXiv:2007.15018}},
  \href {https://doi.org/10.1007/JHEP11(2020)053}
  {\path{doi:10.1007/JHEP11(2020)053}}.

\bibitem{Mehtar-Tani:2019tvy}
Yacine Mehtar-Tani.
\newblock {Gluon bremsstrahlung in finite media beyond multiple soft scattering
  approximation}.
\newblock {\em JHEP}, 07:057, 2019.
\newblock \href {http://arxiv.org/abs/1903.00506} {\path{arXiv:1903.00506}},
  \href {https://doi.org/10.1007/JHEP07(2019)057}
  {\path{doi:10.1007/JHEP07(2019)057}}.

\bibitem{Barata:2021wuf}
Jo\~ao Barata, Yacine Mehtar-Tani, Alba Soto-Ontoso, and Konrad Tywoniuk.
\newblock {Medium-induced radiative kernel with the Improved Opacity
  Expansion}.
\newblock {\em JHEP}, 09:153, 2021.
\newblock \href {http://arxiv.org/abs/2106.07402} {\path{arXiv:2106.07402}},
  \href {https://doi.org/10.1007/JHEP09(2021)153}
  {\path{doi:10.1007/JHEP09(2021)153}}.

\bibitem{Tachibana:2015qxa}
Yasuki Tachibana and Tetsufumi Hirano.
\newblock {Interplay between Mach cone and radial expansion and its signal in
  \ensuremath{\gamma}-jet events}.
\newblock {\em Phys. Rev. C}, 93(5):054907, 2016.
\newblock \href {http://arxiv.org/abs/1510.06966} {\path{arXiv:1510.06966}},
  \href {https://doi.org/10.1103/PhysRevC.93.054907}
  {\path{doi:10.1103/PhysRevC.93.054907}}.

\bibitem{Casalderrey-Solana:2016jvj}
Jorge Casalderrey-Solana, Doga Gulhan, Guilherme Milhano, Daniel Pablos, and
  Krishna Rajagopal.
\newblock {Angular Structure of Jet Quenching Within a Hybrid Strong/Weak
  Coupling Model}.
\newblock {\em JHEP}, 03:135, 2017.
\newblock \href {http://arxiv.org/abs/1609.05842} {\path{arXiv:1609.05842}},
  \href {https://doi.org/10.1007/JHEP03(2017)135}
  {\path{doi:10.1007/JHEP03(2017)135}}.

\bibitem{Tachibana:2017syd}
Yasuki Tachibana, Ning-Bo Chang, and Guang-You Qin.
\newblock {Full jet in quark-gluon plasma with hydrodynamic medium response}.
\newblock {\em Phys. Rev. C}, 95(4):044909, 2017.
\newblock \href {http://arxiv.org/abs/1701.07951} {\path{arXiv:1701.07951}},
  \href {https://doi.org/10.1103/PhysRevC.95.044909}
  {\path{doi:10.1103/PhysRevC.95.044909}}.

\bibitem{KunnawalkamElayavalli:2017hxo}
Raghav Kunnawalkam~Elayavalli and Korinna~Christine Zapp.
\newblock {Medium response in JEWEL and its impact on jet shape observables in
  heavy ion collisions}.
\newblock {\em JHEP}, 07:141, 2017.
\newblock \href {http://arxiv.org/abs/1707.01539} {\path{arXiv:1707.01539}},
  \href {https://doi.org/10.1007/JHEP07(2017)141}
  {\path{doi:10.1007/JHEP07(2017)141}}.

\bibitem{Casalderrey-Solana:2020rsj}
Jorge Casalderrey-Solana, Jos\'e~Guilherme Milhano, Daniel Pablos, Krishna
  Rajagopal, and Xiaojun Yao.
\newblock {Jet Wake from Linearized Hydrodynamics}.
\newblock {\em JHEP}, 05:230, 2021.
\newblock \href {http://arxiv.org/abs/2010.01140} {\path{arXiv:2010.01140}},
  \href {https://doi.org/10.1007/JHEP05(2021)230}
  {\path{doi:10.1007/JHEP05(2021)230}}.

\bibitem{Tachibana:2020mtb}
Yasuki Tachibana, Chun Shen, and Abhijit Majumder.
\newblock {Bulk medium evolution has considerable effects on jet observables}.
\newblock {\em Phys. Rev. C}, 106(2):L021902, 2022.
\newblock \href {http://arxiv.org/abs/2001.08321} {\path{arXiv:2001.08321}},
  \href {https://doi.org/10.1103/PhysRevC.106.L021902}
  {\path{doi:10.1103/PhysRevC.106.L021902}}.

\bibitem{Yang:2021qtl}
Zhong Yang, Wei Chen, Yayun He, Weiyao Ke, Longgang Pang, and Xin-Nian Wang.
\newblock {Search for the Elusive Jet-Induced Diffusion Wake in $Z/\gamma$-Jets
  with 2D Jet Tomography in High-Energy Heavy-Ion Collisions}.
\newblock {\em Phys. Rev. Lett.}, 127(8):082301, 2021.
\newblock \href {http://arxiv.org/abs/2101.05422} {\path{arXiv:2101.05422}},
  \href {https://doi.org/10.1103/PhysRevLett.127.082301}
  {\path{doi:10.1103/PhysRevLett.127.082301}}.

\bibitem{Yang:2022nei}
Zhong Yang, Tan Luo, Wei Chen, Long-Gang Pang, and Xin-Nian Wang.
\newblock {3D Structure of Jet-Induced Diffusion Wake in an Expanding
  Quark-Gluon Plasma}.
\newblock {\em Phys. Rev. Lett.}, 130(5):052301, 2023.
\newblock \href {http://arxiv.org/abs/2203.03683} {\path{arXiv:2203.03683}},
  \href {https://doi.org/10.1103/PhysRevLett.130.052301}
  {\path{doi:10.1103/PhysRevLett.130.052301}}.

\bibitem{Mehtar-Tani:2010ebp}
Yacine Mehtar-Tani, Carlos~A. Salgado, and Konrad Tywoniuk.
\newblock {Anti-angular ordering of gluon radiation in QCD media}.
\newblock {\em Phys. Rev. Lett.}, 106:122002, 2011.
\newblock \href {http://arxiv.org/abs/1009.2965} {\path{arXiv:1009.2965}},
  \href {https://doi.org/10.1103/PhysRevLett.106.122002}
  {\path{doi:10.1103/PhysRevLett.106.122002}}.

\bibitem{Mehtar-Tani:2011hma}
Y.~Mehtar-Tani, C.~A. Salgado, and K.~Tywoniuk.
\newblock {Jets in QCD Media: From Color Coherence to Decoherence}.
\newblock {\em Phys. Lett. B}, 707:156--159, 2012.
\newblock \href {http://arxiv.org/abs/1102.4317} {\path{arXiv:1102.4317}},
  \href {https://doi.org/10.1016/j.physletb.2011.12.042}
  {\path{doi:10.1016/j.physletb.2011.12.042}}.

\bibitem{Casalderrey-Solana:2012evi}
Jorge Casalderrey-Solana, Yacine Mehtar-Tani, Carlos~A. Salgado, and Konrad
  Tywoniuk.
\newblock {New picture of jet quenching dictated by color coherence}.
\newblock {\em Phys. Lett. B}, 725:357--360, 2013.
\newblock \href {http://arxiv.org/abs/1210.7765} {\path{arXiv:1210.7765}},
  \href {https://doi.org/10.1016/j.physletb.2013.07.046}
  {\path{doi:10.1016/j.physletb.2013.07.046}}.

\bibitem{Mehtar-Tani:2017ypq}
Yacine Mehtar-Tani and Konrad Tywoniuk.
\newblock {Radiative energy loss of neighboring subjets}.
\newblock {\em Nucl. Phys. A}, 979:165--203, 2018.
\newblock \href {http://arxiv.org/abs/1706.06047} {\path{arXiv:1706.06047}},
  \href {https://doi.org/10.1016/j.nuclphysa.2018.09.041}
  {\path{doi:10.1016/j.nuclphysa.2018.09.041}}.

\bibitem{Mehtar-Tani:2021fud}
Yacine Mehtar-Tani, Daniel Pablos, and Konrad Tywoniuk.
\newblock {Cone-Size Dependence of Jet Suppression in Heavy-Ion Collisions}.
\newblock {\em Phys. Rev. Lett.}, 127(25):252301, 2021.
\newblock \href {http://arxiv.org/abs/2101.01742} {\path{arXiv:2101.01742}},
  \href {https://doi.org/10.1103/PhysRevLett.127.252301}
  {\path{doi:10.1103/PhysRevLett.127.252301}}.

\bibitem{Caucal:2020zcz}
Paul Caucal.
\newblock {\em {Jet evolution in a dense QCD medium}}.
\newblock PhD thesis, Saclay, 9 2020.
\newblock \href {http://arxiv.org/abs/2010.02874} {\path{arXiv:2010.02874}}.

\bibitem{Mehtar-Tani:2014yea}
Yacine Mehtar-Tani and Konrad Tywoniuk.
\newblock {Jet (de)coherence in Pb\textendash{}Pb collisions at the LHC}.
\newblock {\em Phys. Lett. B}, 744:284--287, 2015.
\newblock \href {http://arxiv.org/abs/1401.8293} {\path{arXiv:1401.8293}},
  \href {https://doi.org/10.1016/j.physletb.2015.03.041}
  {\path{doi:10.1016/j.physletb.2015.03.041}}.

\bibitem{Caucal:2021cfb}
Paul Caucal, Alba Soto-Ontoso, and Adam Takacs.
\newblock {Dynamically groomed jet radius in heavy-ion collisions}.
\newblock {\em Phys. Rev. D}, 105(11):114046, 2022.
\newblock \href {http://arxiv.org/abs/2111.14768} {\path{arXiv:2111.14768}},
  \href {https://doi.org/10.1103/PhysRevD.105.114046}
  {\path{doi:10.1103/PhysRevD.105.114046}}.

\bibitem{Casalderrey-Solana:2020jbx}
Jorge Casalderrey-Solana, Guilherme Milhano, Daniel Pablos, and Krishna
  Rajagopal.
\newblock {Jet substructure modification probes the QGP resolution length}.
\newblock {\em Nucl. Phys. A}, 1005:121904, 2021.
\newblock \href {http://arxiv.org/abs/2002.09193} {\path{arXiv:2002.09193}},
  \href {https://doi.org/10.1016/j.nuclphysa.2020.121904}
  {\path{doi:10.1016/j.nuclphysa.2020.121904}}.

\bibitem{Ringer:2019rfk}
Felix Ringer, Bo-Wen Xiao, and Feng Yuan.
\newblock {Can we observe jet $P_T$-broadening in heavy-ion collisions at the
  LHC?}
\newblock {\em Phys. Lett. B}, 808:135634, 2020.
\newblock \href {http://arxiv.org/abs/1907.12541} {\path{arXiv:1907.12541}},
  \href {https://doi.org/10.1016/j.physletb.2020.135634}
  {\path{doi:10.1016/j.physletb.2020.135634}}.

\bibitem{Chien:2018dfn}
Yang-Ting Chien and Raghav Kunnawalkam~Elayavalli.
\newblock {Probing heavy ion collisions using quark and gluon jet
  substructure}.
\newblock 3 2018.
\newblock \href {http://arxiv.org/abs/1803.03589} {\path{arXiv:1803.03589}}.

\bibitem{Mehtar-Tani:2016aco}
Yacine Mehtar-Tani and Konrad Tywoniuk.
\newblock {Groomed jets in heavy-ion collisions: sensitivity to medium-induced
  bremsstrahlung}.
\newblock {\em JHEP}, 04:125, 2017.
\newblock \href {http://arxiv.org/abs/1610.08930} {\path{arXiv:1610.08930}},
  \href {https://doi.org/10.1007/JHEP04(2017)125}
  {\path{doi:10.1007/JHEP04(2017)125}}.

\bibitem{Pablos:2022mrx}
Daniel Pablos and Alba Soto-Ontoso.
\newblock {Pushing forward jet substructure measurements in heavy-ion
  collisions}.
\newblock 10 2022.
\newblock \href {http://arxiv.org/abs/2210.07901} {\path{arXiv:2210.07901}}.

\bibitem{Liou:2013qya}
Tseh Liou, A.~H. Mueller, and Bin Wu.
\newblock {Radiative $p_\bot$-broadening of high-energy quarks and gluons in
  QCD matter}.
\newblock {\em Nucl. Phys. A}, 916:102--125, 2013.
\newblock \href {http://arxiv.org/abs/1304.7677} {\path{arXiv:1304.7677}},
  \href {https://doi.org/10.1016/j.nuclphysa.2013.08.005}
  {\path{doi:10.1016/j.nuclphysa.2013.08.005}}.

\bibitem{Blaizot:2014bha}
Jean-Paul Blaizot and Yacine Mehtar-Tani.
\newblock {Renormalization of the jet-quenching parameter}.
\newblock {\em Nucl. Phys. A}, 929:202--229, 2014.
\newblock \href {http://arxiv.org/abs/1403.2323} {\path{arXiv:1403.2323}},
  \href {https://doi.org/10.1016/j.nuclphysa.2014.05.018}
  {\path{doi:10.1016/j.nuclphysa.2014.05.018}}.

\bibitem{Iancu:2014kga}
Edmond Iancu.
\newblock {The non-linear evolution of jet quenching}.
\newblock {\em JHEP}, 10:095, 2014.
\newblock \href {http://arxiv.org/abs/1403.1996} {\path{arXiv:1403.1996}},
  \href {https://doi.org/10.1007/JHEP10(2014)095}
  {\path{doi:10.1007/JHEP10(2014)095}}.

\bibitem{Ghiglieri:2022gyv}
Jacopo Ghiglieri and Eamonn Weitz.
\newblock {Classical vs quantum corrections to jet broadening in a
  weakly-coupled Quark-Gluon Plasma}.
\newblock {\em JHEP}, 11:068, 2022.
\newblock \href {http://arxiv.org/abs/2207.08842} {\path{arXiv:2207.08842}},
  \href {https://doi.org/10.1007/JHEP11(2022)068}
  {\path{doi:10.1007/JHEP11(2022)068}}.

\bibitem{Arnold:2021mow}
Peter Arnold.
\newblock {Universality (beyond leading log) of soft radiative corrections to $
  \hat{q} $ in p$_\perp$ broadening and energy loss}.
\newblock {\em JHEP}, 03:134, 2022.
\newblock \href {http://arxiv.org/abs/2111.05348} {\path{arXiv:2111.05348}},
  \href {https://doi.org/10.1007/JHEP03(2022)134}
  {\path{doi:10.1007/JHEP03(2022)134}}.

\bibitem{Caucal:2022fhc}
Paul Caucal and Yacine Mehtar-Tani.
\newblock {Universality aspects of quantum corrections to transverse momentum
  broadening in QCD media}.
\newblock {\em JHEP}, 09:023, 2022.
\newblock \href {http://arxiv.org/abs/2203.09407} {\path{arXiv:2203.09407}},
  \href {https://doi.org/10.1007/JHEP09(2022)023}
  {\path{doi:10.1007/JHEP09(2022)023}}.

\bibitem{Caucal:2021lgf}
Paul Caucal and Yacine Mehtar-Tani.
\newblock {Anomalous diffusion in QCD matter}.
\newblock {\em Phys. Rev. D}, 106(5):L051501, 2022.
\newblock \href {http://arxiv.org/abs/2109.12041} {\path{arXiv:2109.12041}},
  \href {https://doi.org/10.1103/PhysRevD.106.L051501}
  {\path{doi:10.1103/PhysRevD.106.L051501}}.

\bibitem{Arnold:2016kek}
Peter Arnold, Han-Chih Chang, and Shahin Iqbal.
\newblock {The LPM effect in sequential bremsstrahlung 2: factorization}.
\newblock {\em JHEP}, 09:078, 2016.
\newblock \href {http://arxiv.org/abs/1605.07624} {\path{arXiv:1605.07624}},
  \href {https://doi.org/10.1007/JHEP09(2016)078}
  {\path{doi:10.1007/JHEP09(2016)078}}.

\bibitem{Arnold:2016jnq}
Peter Arnold, Han-Chih Chang, and Shahin Iqbal.
\newblock {The LPM effect in sequential bremsstrahlung: 4-gluon vertices}.
\newblock {\em JHEP}, 10:124, 2016.
\newblock \href {http://arxiv.org/abs/1608.05718} {\path{arXiv:1608.05718}},
  \href {https://doi.org/10.1007/JHEP10(2016)124}
  {\path{doi:10.1007/JHEP10(2016)124}}.

\bibitem{Wu:2014nca}
Bin Wu.
\newblock {Radiative energy loss and radiative $p_{\bot}$-broadening of
  high-energy partons in QCD matter}.
\newblock {\em JHEP}, 12:081, 2014.
\newblock \href {http://arxiv.org/abs/1408.5459} {\path{arXiv:1408.5459}},
  \href {https://doi.org/10.1007/JHEP12(2014)081}
  {\path{doi:10.1007/JHEP12(2014)081}}.

\bibitem{Haque:2013qta}
Najmul Haque, Munshi~G. Mustafa, and Michael Strickland.
\newblock {Quark Number Susceptibilities from Two-Loop Hard Thermal Loop
  Perturbation Theory}.
\newblock {\em JHEP}, 07:184, 2013.
\newblock \href {http://arxiv.org/abs/1302.3228} {\path{arXiv:1302.3228}},
  \href {https://doi.org/10.1007/JHEP07(2013)184}
  {\path{doi:10.1007/JHEP07(2013)184}}.

\bibitem{Haque:2013sja}
Najmul Haque, Jens~O. Andersen, Munshi~G. Mustafa, Michael Strickland, and Nan
  Su.
\newblock {Three-loop pressure and susceptibility at finite temperature and
  density from hard-thermal-loop perturbation theory}.
\newblock {\em Phys. Rev. D}, 89(6):061701, 2014.
\newblock \href {http://arxiv.org/abs/1309.3968} {\path{arXiv:1309.3968}},
  \href {https://doi.org/10.1103/PhysRevD.89.061701}
  {\path{doi:10.1103/PhysRevD.89.061701}}.

\bibitem{Andersen:2015eoa}
Jens~O. Andersen, Najmul Haque, Munshi~G. Mustafa, and Michael Strickland.
\newblock {Three-loop hard-thermal-loop perturbation theory thermodynamics at
  finite temperature and finite baryonic and isospin chemical potential}.
\newblock {\em Phys. Rev. D}, 93(5):054045, 2016.
\newblock \href {http://arxiv.org/abs/1511.04660} {\path{arXiv:1511.04660}},
  \href {https://doi.org/10.1103/PhysRevD.93.054045}
  {\path{doi:10.1103/PhysRevD.93.054045}}.

\bibitem{Haque:2020eyj}
Najmul Haque and Michael Strickland.
\newblock {Next-to-next-to leading-order hard-thermal-loop perturbation-theory
  predictions for the curvature of the QCD phase transition line}.
\newblock {\em Phys. Rev. C}, 103(3):031901, 2021.
\newblock \href {http://arxiv.org/abs/2011.06938} {\path{arXiv:2011.06938}},
  \href {https://doi.org/10.1103/PhysRevC.103.L031901}
  {\path{doi:10.1103/PhysRevC.103.L031901}}.

\bibitem{Fu:2022gou}
Wei-jie Fu.
\newblock {QCD at finite temperature and density within the fRG approach: an
  overview}.
\newblock {\em Commun. Theor. Phys.}, 74(9):097304, 2022.
\newblock \href {http://arxiv.org/abs/2205.00468} {\path{arXiv:2205.00468}},
  \href {https://doi.org/10.1088/1572-9494/ac86be}
  {\path{doi:10.1088/1572-9494/ac86be}}.

\bibitem{Horak:2021pfr}
Jan Horak, Joannis Papavassiliou, Jan~M. Pawlowski, and Nicolas Wink.
\newblock {Ghost spectral function from the spectral Dyson-Schwinger equation}.
\newblock {\em Phys. Rev. D}, 104, 2021.
\newblock \href {http://arxiv.org/abs/2103.16175} {\path{arXiv:2103.16175}},
  \href {https://doi.org/10.1103/PhysRevD.104.074017}
  {\path{doi:10.1103/PhysRevD.104.074017}}.

\bibitem{Horak:2021syv}
Jan Horak, Jan~M. Pawlowski, Jos\'e Rodr\'\i{}guez-Quintero, Jonas Turnwald,
  Julian~M. Urban, Nicolas Wink, and Savvas Zafeiropoulos.
\newblock {Reconstructing QCD spectral functions with Gaussian processes}.
\newblock {\em Phys. Rev. D}, 105(3):036014, 2022.
\newblock \href {http://arxiv.org/abs/2107.13464} {\path{arXiv:2107.13464}},
  \href {https://doi.org/10.1103/PhysRevD.105.036014}
  {\path{doi:10.1103/PhysRevD.105.036014}}.

\bibitem{Horak:2022myj}
Jan Horak, Jan~M. Pawlowski, and Nicolas Wink.
\newblock {On the complex structure of Yang-Mills theory}.
\newblock 2 2022.
\newblock \href {http://arxiv.org/abs/2202.09333} {\path{arXiv:2202.09333}}.

\bibitem{Lowdon:2021ehf}
Peter Lowdon, Ralf-Arno Tripolt, Jan~M. Pawlowski, and Dirk~H. Rischke.
\newblock {Spectral representation of the shear viscosity for local scalar QFTs
  at finite temperature}.
\newblock {\em Phys. Rev. D}, 104(6):065010, 2021.
\newblock \href {http://arxiv.org/abs/2104.13413} {\path{arXiv:2104.13413}},
  \href {https://doi.org/10.1103/PhysRevD.104.065010}
  {\path{doi:10.1103/PhysRevD.104.065010}}.

\bibitem{McLaughlin:2021dph}
Emma McLaughlin, Jacob Rose, Travis Dore, Paolo Parotto, Claudia Ratti, and
  Jacquelyn Noronha-Hostler.
\newblock {Building a testable shear viscosity across the QCD phase diagram}.
\newblock {\em Phys. Rev. C}, 105(2):024903, 2022.
\newblock \href {http://arxiv.org/abs/2103.02090} {\path{arXiv:2103.02090}},
  \href {https://doi.org/10.1103/PhysRevC.105.024903}
  {\path{doi:10.1103/PhysRevC.105.024903}}.

\bibitem{Grefa:2022sav}
Joaquin Grefa, Mauricio Hippert, Jorge Noronha, Jacquelyn Noronha-Hostler,
  Israel Portillo, Claudia Ratti, and Romulo Rougemont.
\newblock {Transport coefficients of the quark-gluon plasma at the critical
  point and across the first-order line}.
\newblock {\em Phys. Rev. D}, 106(3):034024, 2022.
\newblock \href {http://arxiv.org/abs/2203.00139} {\path{arXiv:2203.00139}},
  \href {https://doi.org/10.1103/PhysRevD.106.034024}
  {\path{doi:10.1103/PhysRevD.106.034024}}.

\bibitem{Fischer:2018sdj}
Christian~S. Fischer.
\newblock {QCD at finite temperature and chemical potential from
  Dyson\textendash{}Schwinger equations}.
\newblock {\em Prog. Part. Nucl. Phys.}, 105:1--60, 2019.
\newblock \href {http://arxiv.org/abs/1810.12938} {\path{arXiv:1810.12938}},
  \href {https://doi.org/10.1016/j.ppnp.2019.01.002}
  {\path{doi:10.1016/j.ppnp.2019.01.002}}.

\bibitem{Weil:2016zrk}
J.~Weil et~al.
\newblock {Particle production and equilibrium properties within a new hadron
  transport approach for heavy-ion collisions}.
\newblock {\em Phys. Rev. C}, 94(5):054905, 2016.
\newblock \href {http://arxiv.org/abs/1606.06642} {\path{arXiv:1606.06642}},
  \href {https://doi.org/10.1103/PhysRevC.94.054905}
  {\path{doi:10.1103/PhysRevC.94.054905}}.

\bibitem{Bleicher:2022kcu}
Marcus Bleicher and Elena Bratkovskaya.
\newblock {Modelling relativistic heavy-ion collisions with dynamical transport
  approaches}.
\newblock {\em Prog. Part. Nucl. Phys.}, 122:103920, 2022.
\newblock \href {https://doi.org/10.1016/j.ppnp.2021.103920}
  {\path{doi:10.1016/j.ppnp.2021.103920}}.

\bibitem{Aichelin:2019tnk}
J.~Aichelin, E.~Bratkovskaya, A.~Le~F\`evre, V.~Kireyeu, V.~Kolesnikov,
  Y.~Leifels, V.~Voronyuk, and G.~Coci.
\newblock {Parton-hadron-quantum-molecular dynamics: A novel microscopic $n$
  -body transport approach for heavy-ion collisions, dynamical cluster
  formation, and hypernuclei production}.
\newblock {\em Phys. Rev. C}, 101(4):044905, 2020.
\newblock \href {http://arxiv.org/abs/1907.03860} {\path{arXiv:1907.03860}},
  \href {https://doi.org/10.1103/PhysRevC.101.044905}
  {\path{doi:10.1103/PhysRevC.101.044905}}.

\bibitem{Moreau:2019vhw}
Pierre Moreau, Olga Soloveva, Lucia Oliva, Taesoo Song, Wolfgang Cassing, and
  Elena Bratkovskaya.
\newblock {Exploring the partonic phase at finite chemical potential within an
  extended off-shell transport approach}.
\newblock {\em Phys. Rev. C}, 100(1):014911, 2019.
\newblock \href {http://arxiv.org/abs/1903.10257} {\path{arXiv:1903.10257}},
  \href {https://doi.org/10.1103/PhysRevC.100.014911}
  {\path{doi:10.1103/PhysRevC.100.014911}}.

\bibitem{Soloveva:2019xph}
Olga Soloveva, Pierre Moreau, and Elena Bratkovskaya.
\newblock {Transport coefficients for the hot quark-gluon plasma at finite
  chemical potential $\mu_B$}.
\newblock {\em Phys. Rev. C}, 101(4):045203, 2020.
\newblock \href {http://arxiv.org/abs/1911.08547} {\path{arXiv:1911.08547}},
  \href {https://doi.org/10.1103/PhysRevC.101.045203}
  {\path{doi:10.1103/PhysRevC.101.045203}}.

\bibitem{Soloveva:2020hpr}
Olga Soloveva, David Fuseau, J\"org Aichelin, and Elena Bratkovskaya.
\newblock {Shear viscosity and electric conductivity of a hot and dense QGP
  with a chiral phase transition}.
\newblock {\em Phys. Rev. C}, 103(5):054901, 2021.
\newblock \href {http://arxiv.org/abs/2011.03505} {\path{arXiv:2011.03505}},
  \href {https://doi.org/10.1103/PhysRevC.103.054901}
  {\path{doi:10.1103/PhysRevC.103.054901}}.

\bibitem{Rose:2020sjv}
Jean-Bernard Rose, Moritz Greif, Jan Hammelmann, Jan~A. Fotakis, Gabriel~S.
  Denicol, Hannah Elfner, and Carsten Greiner.
\newblock {Cross-conductivity: novel transport coefficients to constrain the
  hadronic degrees of freedom of nuclear matter}.
\newblock {\em Phys. Rev. D}, 101(11):114028, 2020.
\newblock \href {http://arxiv.org/abs/2001.10606} {\path{arXiv:2001.10606}},
  \href {https://doi.org/10.1103/PhysRevD.101.114028}
  {\path{doi:10.1103/PhysRevD.101.114028}}.

\bibitem{Cassing:2021fkc}
Wolfgang Cassing.
\newblock {Transport Theories for Strongly-Interacting Systems: Applications to
  Heavy-Ion Collisions}.
\newblock {\em Lect. Notes Phys.}, 989:pp., 2021.
\newblock \href {https://doi.org/10.1007/978-3-030-80295-0}
  {\path{doi:10.1007/978-3-030-80295-0}}.

\bibitem{Fotakis:2021diq}
Jan~A. Fotakis, Olga Soloveva, Carsten Greiner, Olaf Kaczmarek, and Elena
  Bratkovskaya.
\newblock {Diffusion coefficient matrix of the strongly interacting quark-gluon
  plasma}.
\newblock {\em Phys. Rev. D}, 104(3):034014, 2021.
\newblock \href {http://arxiv.org/abs/2102.08140} {\path{arXiv:2102.08140}},
  \href {https://doi.org/10.1103/PhysRevD.104.034014}
  {\path{doi:10.1103/PhysRevD.104.034014}}.

\bibitem{Soloveva:2021quj}
Olga Soloveva, J\"org Aichelin, and Elena Bratkovskaya.
\newblock {Transport properties and equation-of-state of hot and dense QGP
  matter near the critical endpoint in the phenomenological dynamical
  quasiparticle model}.
\newblock {\em Phys. Rev. D}, 105(5):054011, 2022.
\newblock \href {http://arxiv.org/abs/2108.08561} {\path{arXiv:2108.08561}},
  \href {https://doi.org/10.1103/PhysRevD.105.054011}
  {\path{doi:10.1103/PhysRevD.105.054011}}.

\bibitem{Li:2022ozl}
Fu-Peng Li, Hong-Liang L\"u, Long-Gang Pang, and Guang-You Qin.
\newblock {Deep-learning quasi-particle masses from QCD equation of state}.
\newblock 11 2022.
\newblock \href {http://arxiv.org/abs/2211.07994} {\path{arXiv:2211.07994}}.

\bibitem{TMEP:2022xjg}
Hermann Wolter et~al.
\newblock {Transport model comparison studies of intermediate-energy heavy-ion
  collisions}.
\newblock {\em Prog. Part. Nucl. Phys.}, 125:103962, 2022.
\newblock \href {http://arxiv.org/abs/2202.06672} {\path{arXiv:2202.06672}},
  \href {https://doi.org/10.1016/j.ppnp.2022.103962}
  {\path{doi:10.1016/j.ppnp.2022.103962}}.

\bibitem{Grishmanovskii:2022tpb}
Ilia Grishmanovskii, Taesoo Song, Olga Soloveva, Carsten Greiner, and Elena
  Bratkovskaya.
\newblock {Exploring jet transport coefficients by elastic scattering in the
  strongly interacting quark-gluon plasma}.
\newblock {\em Phys. Rev. C}, 106(1):014903, 2022.
\newblock \href {http://arxiv.org/abs/2204.01561} {\path{arXiv:2204.01561}},
  \href {https://doi.org/10.1103/PhysRevC.106.014903}
  {\path{doi:10.1103/PhysRevC.106.014903}}.

\bibitem{DeWolfe:2010he}
Oliver DeWolfe, Steven~S. Gubser, and Christopher Rosen.
\newblock {A holographic critical point}.
\newblock {\em Phys. Rev. D}, 83:086005, 2011.
\newblock \href {http://arxiv.org/abs/1012.1864} {\path{arXiv:1012.1864}},
  \href {https://doi.org/10.1103/PhysRevD.83.086005}
  {\path{doi:10.1103/PhysRevD.83.086005}}.

\bibitem{Critelli:2017oub}
Renato Critelli, Jorge Noronha, Jacquelyn Noronha-Hostler, Israel Portillo,
  Claudia Ratti, and Romulo Rougemont.
\newblock {Critical point in the phase diagram of primordial quark-gluon matter
  from black hole physics}.
\newblock {\em Phys. Rev. D}, 96(9):096026, 2017.
\newblock \href {http://arxiv.org/abs/1706.00455} {\path{arXiv:1706.00455}},
  \href {https://doi.org/10.1103/PhysRevD.96.096026}
  {\path{doi:10.1103/PhysRevD.96.096026}}.

\bibitem{Ishii:2019gta}
Takaaki Ishii, Matti J\"arvinen, and Govert Nijs.
\newblock {Cool baryon and quark matter in holographic QCD}.
\newblock {\em JHEP}, 07:003, 2019.
\newblock \href {http://arxiv.org/abs/1903.06169} {\path{arXiv:1903.06169}},
  \href {https://doi.org/10.1007/JHEP07(2019)003}
  {\path{doi:10.1007/JHEP07(2019)003}}.

\bibitem{Jokela:2020piw}
Niko Jokela, Matti J\"arvinen, Govert Nijs, and Jere Remes.
\newblock {Unified weak and strong coupling framework for nuclear matter and
  neutron stars}.
\newblock {\em Phys. Rev. D}, 103(8):086004, 2021.
\newblock \href {http://arxiv.org/abs/2006.01141} {\path{arXiv:2006.01141}},
  \href {https://doi.org/10.1103/PhysRevD.103.086004}
  {\path{doi:10.1103/PhysRevD.103.086004}}.

\bibitem{Jokela:2021vwy}
Niko Jokela, Matti J\"arvinen, and Jere Remes.
\newblock {Holographic QCD in the NICER era}.
\newblock {\em Phys. Rev. D}, 105(8):086005, 2022.
\newblock \href {http://arxiv.org/abs/2111.12101} {\path{arXiv:2111.12101}},
  \href {https://doi.org/10.1103/PhysRevD.105.086005}
  {\path{doi:10.1103/PhysRevD.105.086005}}.

\bibitem{Grefa:2021qvt}
Joaquin Grefa, Jorge Noronha, Jacquelyn Noronha-Hostler, Israel Portillo,
  Claudia Ratti, and Romulo Rougemont.
\newblock {Hot and dense quark-gluon plasma thermodynamics from holographic
  black holes}.
\newblock {\em Phys. Rev. D}, 104(3):034002, 2021.
\newblock \href {http://arxiv.org/abs/2102.12042} {\path{arXiv:2102.12042}},
  \href {https://doi.org/10.1103/PhysRevD.104.034002}
  {\path{doi:10.1103/PhysRevD.104.034002}}.

\bibitem{Demircik:2021zll}
Tuna Demircik, Christian Ecker, and Matti J\"arvinen.
\newblock {Dense and Hot QCD at Strong Coupling}.
\newblock {\em Phys. Rev. X}, 12(4):041012, 2022.
\newblock \href {http://arxiv.org/abs/2112.12157} {\path{arXiv:2112.12157}},
  \href {https://doi.org/10.1103/PhysRevX.12.041012}
  {\path{doi:10.1103/PhysRevX.12.041012}}.

\bibitem{Finazzo:2014cna}
Stefano~I. Finazzo, Romulo Rougemont, Hugo Marrochio, and Jorge Noronha.
\newblock {Hydrodynamic transport coefficients for the non-conformal
  quark-gluon plasma from holography}.
\newblock {\em JHEP}, 02:051, 2015.
\newblock \href {http://arxiv.org/abs/1412.2968} {\path{arXiv:1412.2968}},
  \href {https://doi.org/10.1007/JHEP02(2015)051}
  {\path{doi:10.1007/JHEP02(2015)051}}.

\bibitem{Hoyos:2020hmq}
Carlos Hoyos, Niko Jokela, Matti Jarvinen, Javier~G. Subils, Javier Tarrio, and
  Aleksi Vuorinen.
\newblock {Transport in strongly coupled quark matter}.
\newblock {\em Phys. Rev. Lett.}, 125:241601, 2020.
\newblock \href {http://arxiv.org/abs/2005.14205} {\path{arXiv:2005.14205}},
  \href {https://doi.org/10.1103/PhysRevLett.125.241601}
  {\path{doi:10.1103/PhysRevLett.125.241601}}.

\bibitem{Hoyos:2021njg}
Carlos Hoyos, Niko Jokela, Matti J\"arvinen, Javier~G. Subils, Javier Tarrio,
  and Aleksi Vuorinen.
\newblock {Holographic approach to transport in dense QCD matter}.
\newblock {\em Phys. Rev. D}, 105(6):066014, 2022.
\newblock \href {http://arxiv.org/abs/2109.12122} {\path{arXiv:2109.12122}},
  \href {https://doi.org/10.1103/PhysRevD.105.066014}
  {\path{doi:10.1103/PhysRevD.105.066014}}.

\bibitem{Hidaka:2009hs}
Yoshimasa Hidaka and Robert~D. Pisarski.
\newblock {Hard thermal loops, to quadratic order, in the background of a
  spatial 't Hooft loop}.
\newblock {\em Phys. Rev. D}, 80(3):036004, 2009.
\newblock [Erratum: Phys.Rev.D 102, 059902 (2020)].
\newblock \href {http://arxiv.org/abs/0906.1751} {\path{arXiv:0906.1751}},
  \href {https://doi.org/10.1103/PhysRevD.80.036004}
  {\path{doi:10.1103/PhysRevD.80.036004}}.

\bibitem{Dumitru:2010mj}
Adrian Dumitru, Yun Guo, Yoshimasa Hidaka, Christiaan P.~Korthals Altes, and
  Robert~D. Pisarski.
\newblock {How Wide is the Transition to Deconfinement?}
\newblock {\em Phys. Rev. D}, 83:034022, 2011.
\newblock \href {http://arxiv.org/abs/1011.3820} {\path{arXiv:1011.3820}},
  \href {https://doi.org/10.1103/PhysRevD.83.034022}
  {\path{doi:10.1103/PhysRevD.83.034022}}.

\bibitem{Dumitru:2012fw}
Adrian Dumitru, Yun Guo, Yoshimasa Hidaka, Christiaan P.~Korthals Altes, and
  Robert~D. Pisarski.
\newblock {Effective Matrix Model for Deconfinement in Pure Gauge Theories}.
\newblock {\em Phys. Rev. D}, 86:105017, 2012.
\newblock \href {http://arxiv.org/abs/1205.0137} {\path{arXiv:1205.0137}},
  \href {https://doi.org/10.1103/PhysRevD.86.105017}
  {\path{doi:10.1103/PhysRevD.86.105017}}.

\bibitem{Pisarski:2016ixt}
Robert~D. Pisarski and Vladimir~V. Skokov.
\newblock {Chiral matrix model of the semi-QGP in QCD}.
\newblock {\em Phys. Rev. D}, 94(3):034015, 2016.
\newblock \href {http://arxiv.org/abs/1604.00022} {\path{arXiv:1604.00022}},
  \href {https://doi.org/10.1103/PhysRevD.94.034015}
  {\path{doi:10.1103/PhysRevD.94.034015}}.

\bibitem{Hidaka:2020vna}
Yoshimasa Hidaka and Robert~D. Pisarski.
\newblock {Effective models of a semi-quark-gluon plasma}.
\newblock {\em Phys. Rev. D}, 104(7):074036, 2021.
\newblock \href {http://arxiv.org/abs/2009.03903} {\path{arXiv:2009.03903}},
  \href {https://doi.org/10.1103/PhysRevD.104.074036}
  {\path{doi:10.1103/PhysRevD.104.074036}}.

\bibitem{Hidaka:2008dr}
Yoshimasa Hidaka and Robert~D. Pisarski.
\newblock {Suppression of the Shear Viscosity in a ''semi'' Quark Gluon
  Plasma}.
\newblock {\em Phys. Rev. D}, 78:071501, 2008.
\newblock \href {http://arxiv.org/abs/0803.0453} {\path{arXiv:0803.0453}},
  \href {https://doi.org/10.1103/PhysRevD.78.071501}
  {\path{doi:10.1103/PhysRevD.78.071501}}.

\bibitem{Hidaka:2009ma}
Yoshimasa Hidaka and Robert~D. Pisarski.
\newblock {Small shear viscosity in the semi quark gluon plasma}.
\newblock {\em Phys. Rev. D}, 81:076002, 2010.
\newblock \href {http://arxiv.org/abs/0912.0940} {\path{arXiv:0912.0940}},
  \href {https://doi.org/10.1103/PhysRevD.81.076002}
  {\path{doi:10.1103/PhysRevD.81.076002}}.

\bibitem{Chesler:2010bi}
Paul~M. Chesler and Laurence~G. Yaffe.
\newblock {Holography and colliding gravitational shock waves in asymptotically
  AdS$_{5}$ spacetime}.
\newblock {\em Phys. Rev. Lett.}, 106:021601, 2011.
\newblock \href {http://arxiv.org/abs/1011.3562} {\path{arXiv:1011.3562}},
  \href {https://doi.org/10.1103/PhysRevLett.106.021601}
  {\path{doi:10.1103/PhysRevLett.106.021601}}.

\bibitem{Heller:2011ju}
Michal~P. Heller, Romuald~A. Janik, and Przemyslaw Witaszczyk.
\newblock {The characteristics of thermalization of boost-invariant plasma from
  holography}.
\newblock {\em Phys. Rev. Lett.}, 108:201602, 2012.
\newblock \href {http://arxiv.org/abs/1103.3452} {\path{arXiv:1103.3452}},
  \href {https://doi.org/10.1103/PhysRevLett.108.201602}
  {\path{doi:10.1103/PhysRevLett.108.201602}}.

\bibitem{Heller:2012km}
Michal~P. Heller, David Mateos, Wilke van~der Schee, and Diego Trancanelli.
\newblock {Strong Coupling Isotropization of Non-Abelian Plasmas Simplified}.
\newblock {\em Phys. Rev. Lett.}, 108:191601, 2012.
\newblock \href {http://arxiv.org/abs/1202.0981} {\path{arXiv:1202.0981}},
  \href {https://doi.org/10.1103/PhysRevLett.108.191601}
  {\path{doi:10.1103/PhysRevLett.108.191601}}.

\bibitem{Heller:2013oxa}
Michal~P. Heller, David Mateos, Wilke van~der Schee, and Miquel Triana.
\newblock {Holographic isotropization linearized}.
\newblock {\em JHEP}, 09:026, 2013.
\newblock \href {http://arxiv.org/abs/1304.5172} {\path{arXiv:1304.5172}},
  \href {https://doi.org/10.1007/JHEP09(2013)026}
  {\path{doi:10.1007/JHEP09(2013)026}}.

\bibitem{Chesler:2013lia}
Paul~M. Chesler and Laurence~G. Yaffe.
\newblock {Numerical solution of gravitational dynamics in asymptotically
  anti-de Sitter spacetimes}.
\newblock {\em JHEP}, 07:086, 2014.
\newblock \href {http://arxiv.org/abs/1309.1439} {\path{arXiv:1309.1439}},
  \href {https://doi.org/10.1007/JHEP07(2014)086}
  {\path{doi:10.1007/JHEP07(2014)086}}.

\bibitem{Chesler:2015wra}
Paul~M. Chesler and Laurence~G. Yaffe.
\newblock {Holography and off-center collisions of localized shock waves}.
\newblock {\em JHEP}, 10:070, 2015.
\newblock \href {http://arxiv.org/abs/1501.04644} {\path{arXiv:1501.04644}},
  \href {https://doi.org/10.1007/JHEP10(2015)070}
  {\path{doi:10.1007/JHEP10(2015)070}}.

\bibitem{Keegan:2015avk}
Liam Keegan, Aleksi Kurkela, Paul Romatschke, Wilke van~der Schee, and Yan Zhu.
\newblock {Weak and strong coupling equilibration in nonabelian gauge
  theories}.
\newblock {\em JHEP}, 04:031, 2016.
\newblock \href {http://arxiv.org/abs/1512.05347} {\path{arXiv:1512.05347}},
  \href {https://doi.org/10.1007/JHEP04(2016)031}
  {\path{doi:10.1007/JHEP04(2016)031}}.

\bibitem{Spalinski:2017mel}
Micha\l{} Spali\'nski.
\newblock {On the hydrodynamic attractor of Yang\textendash{}Mills plasma}.
\newblock {\em Phys. Lett. B}, 776:468--472, 2018.
\newblock \href {http://arxiv.org/abs/1708.01921} {\path{arXiv:1708.01921}},
  \href {https://doi.org/10.1016/j.physletb.2017.11.059}
  {\path{doi:10.1016/j.physletb.2017.11.059}}.

\bibitem{Denicol:2014xca}
Gabriel~S. Denicol, Ulrich~W. Heinz, Mauricio Martinez, Jorge Noronha, and
  Michael Strickland.
\newblock {New Exact Solution of the Relativistic Boltzmann Equation and its
  Hydrodynamic Limit}.
\newblock {\em Phys. Rev. Lett.}, 113(20):202301, 2014.
\newblock \href {http://arxiv.org/abs/1408.5646} {\path{arXiv:1408.5646}},
  \href {https://doi.org/10.1103/PhysRevLett.113.202301}
  {\path{doi:10.1103/PhysRevLett.113.202301}}.

\bibitem{Denicol:2014tha}
Gabriel~S. Denicol, Ulrich~W. Heinz, Mauricio Martinez, Jorge Noronha, and
  Michael Strickland.
\newblock {Studying the validity of relativistic hydrodynamics with a new exact
  solution of the Boltzmann equation}.
\newblock {\em Phys. Rev. D}, 90(12):125026, 2014.
\newblock \href {http://arxiv.org/abs/1408.7048} {\path{arXiv:1408.7048}},
  \href {https://doi.org/10.1103/PhysRevD.90.125026}
  {\path{doi:10.1103/PhysRevD.90.125026}}.

\bibitem{Kurkela:2015qoa}
Aleksi Kurkela and Yan Zhu.
\newblock {Isotropization and hydrodynamization in weakly coupled heavy-ion
  collisions}.
\newblock {\em Phys. Rev. Lett.}, 115(18):182301, 2015.
\newblock \href {http://arxiv.org/abs/1506.06647} {\path{arXiv:1506.06647}},
  \href {https://doi.org/10.1103/PhysRevLett.115.182301}
  {\path{doi:10.1103/PhysRevLett.115.182301}}.

\bibitem{Bazow:2015dha}
D.~Bazow, G.~S. Denicol, U.~Heinz, M.~Martinez, and J.~Noronha.
\newblock {Analytic solution of the Boltzmann equation in an expanding system}.
\newblock {\em Phys. Rev. Lett.}, 116(2):022301, 2016.
\newblock \href {http://arxiv.org/abs/1507.07834} {\path{arXiv:1507.07834}},
  \href {https://doi.org/10.1103/PhysRevLett.116.022301}
  {\path{doi:10.1103/PhysRevLett.116.022301}}.

\bibitem{Romatschke:2017vte}
Paul Romatschke.
\newblock {Relativistic Fluid Dynamics Far From Local Equilibrium}.
\newblock {\em Phys. Rev. Lett.}, 120(1):012301, 2018.
\newblock \href {http://arxiv.org/abs/1704.08699} {\path{arXiv:1704.08699}},
  \href {https://doi.org/10.1103/PhysRevLett.120.012301}
  {\path{doi:10.1103/PhysRevLett.120.012301}}.

\bibitem{Kurkela:2018vqr}
Aleksi Kurkela, Aleksas Mazeliauskas, Jean-Fran\c{c}ois Paquet, S\"oren
  Schlichting, and Derek Teaney.
\newblock {Effective kinetic description of event-by-event pre-equilibrium
  dynamics in high-energy heavy-ion collisions}.
\newblock {\em Phys. Rev. C}, 99(3):034910, 2019.
\newblock \href {http://arxiv.org/abs/1805.00961} {\path{arXiv:1805.00961}},
  \href {https://doi.org/10.1103/PhysRevC.99.034910}
  {\path{doi:10.1103/PhysRevC.99.034910}}.

\bibitem{Kurkela:2019set}
Aleksi Kurkela, Wilke van~der Schee, Urs~Achim Wiedemann, and Bin Wu.
\newblock {Early- and Late-Time Behavior of Attractors in Heavy-Ion
  Collisions}.
\newblock {\em Phys. Rev. Lett.}, 124(10):102301, 2020.
\newblock \href {http://arxiv.org/abs/1907.08101} {\path{arXiv:1907.08101}},
  \href {https://doi.org/10.1103/PhysRevLett.124.102301}
  {\path{doi:10.1103/PhysRevLett.124.102301}}.

\bibitem{Denicol:2019lio}
Gabriel~S. Denicol and Jorge Noronha.
\newblock {Exact hydrodynamic attractor of an ultrarelativistic gas of hard
  spheres}.
\newblock {\em Phys. Rev. Lett.}, 124(15):152301, 2020.
\newblock \href {http://arxiv.org/abs/1908.09957} {\path{arXiv:1908.09957}},
  \href {https://doi.org/10.1103/PhysRevLett.124.152301}
  {\path{doi:10.1103/PhysRevLett.124.152301}}.

\bibitem{Brewer:2019oha}
Jasmine Brewer, Li~Yan, and Yi~Yin.
\newblock {Adiabatic hydrodynamization in rapidly-expanding
  quark\textendash{}gluon plasma}.
\newblock {\em Phys. Lett. B}, 816:136189, 2021.
\newblock \href {http://arxiv.org/abs/1910.00021} {\path{arXiv:1910.00021}},
  \href {https://doi.org/10.1016/j.physletb.2021.136189}
  {\path{doi:10.1016/j.physletb.2021.136189}}.

\bibitem{Almaalol:2020rnu}
Dekrayat Almaalol, Aleksi Kurkela, and Michael Strickland.
\newblock {Nonequilibrium Attractor in High-Temperature QCD Plasmas}.
\newblock {\em Phys. Rev. Lett.}, 125(12):122302, 2020.
\newblock \href {http://arxiv.org/abs/2004.05195} {\path{arXiv:2004.05195}},
  \href {https://doi.org/10.1103/PhysRevLett.125.122302}
  {\path{doi:10.1103/PhysRevLett.125.122302}}.

\bibitem{Mullins:2022fbx}
Nicki Mullins, Gabriel~S. Denicol, and Jorge Noronha.
\newblock {Far-from-equilibrium kinetic dynamics of
  \ensuremath{\lambda}\ensuremath{\phi}4 theory in an expanding universe}.
\newblock {\em Phys. Rev. D}, 106(5):056024, 2022.
\newblock \href {http://arxiv.org/abs/2207.07786} {\path{arXiv:2207.07786}},
  \href {https://doi.org/10.1103/PhysRevD.106.056024}
  {\path{doi:10.1103/PhysRevD.106.056024}}.

\bibitem{Brewer:2022vkq}
Jasmine Brewer, Bruno Scheihing-Hitschfeld, and Yi~Yin.
\newblock {Scaling and adiabaticity in a rapidly expanding gluon plasma}.
\newblock {\em JHEP}, 05:145, 2022.
\newblock \href {http://arxiv.org/abs/2203.02427} {\path{arXiv:2203.02427}},
  \href {https://doi.org/10.1007/JHEP05(2022)145}
  {\path{doi:10.1007/JHEP05(2022)145}}.

\bibitem{Heller:2013fn}
Michal~P. Heller, Romuald~A. Janik, and Przemyslaw Witaszczyk.
\newblock {Hydrodynamic Gradient Expansion in Gauge Theory Plasmas}.
\newblock {\em Phys. Rev. Lett.}, 110(21):211602, 2013.
\newblock \href {http://arxiv.org/abs/1302.0697} {\path{arXiv:1302.0697}},
  \href {https://doi.org/10.1103/PhysRevLett.110.211602}
  {\path{doi:10.1103/PhysRevLett.110.211602}}.

\bibitem{Buchel:2016cbj}
Alex Buchel, Michal~P. Heller, and Jorge Noronha.
\newblock {Entropy Production, Hydrodynamics, and Resurgence in the Primordial
  Quark-Gluon Plasma from Holography}.
\newblock {\em Phys. Rev. D}, 94(10):106011, 2016.
\newblock \href {http://arxiv.org/abs/1603.05344} {\path{arXiv:1603.05344}},
  \href {https://doi.org/10.1103/PhysRevD.94.106011}
  {\path{doi:10.1103/PhysRevD.94.106011}}.

\bibitem{Denicol:2016bjh}
Gabriel~S. Denicol and Jorge Noronha.
\newblock {Divergence of the Chapman-Enskog expansion in relativistic kinetic
  theory}.
\newblock 8 2016.
\newblock \href {http://arxiv.org/abs/1608.07869} {\path{arXiv:1608.07869}}.

\bibitem{Heller:2016rtz}
Michal~P. Heller, Aleksi Kurkela, Michal Spali\'nski, and Viktor Svensson.
\newblock {Hydrodynamization in kinetic theory: Transient modes and the
  gradient expansion}.
\newblock {\em Phys. Rev. D}, 97(9):091503, 2018.
\newblock \href {http://arxiv.org/abs/1609.04803} {\path{arXiv:1609.04803}},
  \href {https://doi.org/10.1103/PhysRevD.97.091503}
  {\path{doi:10.1103/PhysRevD.97.091503}}.

\bibitem{Heller:2015dha}
Michal~P. Heller and Michal Spalinski.
\newblock {Hydrodynamics Beyond the Gradient Expansion: Resurgence and
  Resummation}.
\newblock {\em Phys. Rev. Lett.}, 115(7):072501, 2015.
\newblock \href {http://arxiv.org/abs/1503.07514} {\path{arXiv:1503.07514}},
  \href {https://doi.org/10.1103/PhysRevLett.115.072501}
  {\path{doi:10.1103/PhysRevLett.115.072501}}.

\bibitem{Florkowski:2017olj}
Wojciech Florkowski, Michal~P. Heller, and Michal Spalinski.
\newblock {New theories of relativistic hydrodynamics in the LHC era}.
\newblock {\em Rept. Prog. Phys.}, 81(4):046001, 2018.
\newblock \href {http://arxiv.org/abs/1707.02282} {\path{arXiv:1707.02282}},
  \href {https://doi.org/10.1088/1361-6633/aaa091}
  {\path{doi:10.1088/1361-6633/aaa091}}.

\bibitem{Romatschke:2017ejr}
Paul Romatschke and Ulrike Romatschke.
\newblock {\em {Relativistic Fluid Dynamics In and Out of Equilibrium}}.
\newblock Cambridge Monographs on Mathematical Physics. Cambridge University
  Press, 5 2019.
\newblock \href {http://arxiv.org/abs/1712.05815} {\path{arXiv:1712.05815}},
  \href {https://doi.org/10.1017/9781108651998}
  {\path{doi:10.1017/9781108651998}}.

\bibitem{Berges:2020fwq}
J\"urgen Berges, Michal~P. Heller, Aleksas Mazeliauskas, and Raju Venugopalan.
\newblock {QCD thermalization: Ab initio approaches and interdisciplinary
  connections}.
\newblock {\em Rev. Mod. Phys.}, 93(3):035003, 2021.
\newblock \href {http://arxiv.org/abs/2005.12299} {\path{arXiv:2005.12299}},
  \href {https://doi.org/10.1103/RevModPhys.93.035003}
  {\path{doi:10.1103/RevModPhys.93.035003}}.

\bibitem{Chattopadhyay:2021ive}
Chandrodoy Chattopadhyay, Sunil Jaiswal, Lipei Du, Ulrich Heinz, and Subrata
  Pal.
\newblock {Non-conformal attractor in boost-invariant plasmas}.
\newblock {\em Phys. Lett. B}, 824:136820, 2022.
\newblock \href {http://arxiv.org/abs/2107.05500} {\path{arXiv:2107.05500}},
  \href {https://doi.org/10.1016/j.physletb.2021.136820}
  {\path{doi:10.1016/j.physletb.2021.136820}}.

\bibitem{Jaiswal:2021uvv}
Sunil Jaiswal, Chandrodoy Chattopadhyay, Lipei Du, Ulrich Heinz, and Subrata
  Pal.
\newblock {Nonconformal kinetic theory and hydrodynamics for Bjorken flow}.
\newblock {\em Phys. Rev. C}, 105(2):024911, 2022.
\newblock \href {http://arxiv.org/abs/2107.10248} {\path{arXiv:2107.10248}},
  \href {https://doi.org/10.1103/PhysRevC.105.024911}
  {\path{doi:10.1103/PhysRevC.105.024911}}.

\bibitem{Martinez:2010sc}
Mauricio Martinez and Michael Strickland.
\newblock {Dissipative Dynamics of Highly Anisotropic Systems}.
\newblock {\em Nucl. Phys. A}, 848:183--197, 2010.
\newblock \href {http://arxiv.org/abs/1007.0889} {\path{arXiv:1007.0889}},
  \href {https://doi.org/10.1016/j.nuclphysa.2010.08.011}
  {\path{doi:10.1016/j.nuclphysa.2010.08.011}}.

\bibitem{Florkowski:2010cf}
Wojciech Florkowski and Radoslaw Ryblewski.
\newblock {Highly-anisotropic and strongly-dissipative hydrodynamics for early
  stages of relativistic heavy-ion collisions}.
\newblock {\em Phys. Rev. C}, 83:034907, 2011.
\newblock \href {http://arxiv.org/abs/1007.0130} {\path{arXiv:1007.0130}},
  \href {https://doi.org/10.1103/PhysRevC.83.034907}
  {\path{doi:10.1103/PhysRevC.83.034907}}.

\bibitem{McNelis:2021zji}
Mike McNelis, Dennis Bazow, and Ulrich Heinz.
\newblock {Anisotropic fluid dynamical simulations of heavy-ion collisions}.
\newblock {\em Comput. Phys. Commun.}, 267:108077, 2021.
\newblock \href {http://arxiv.org/abs/2101.02827} {\path{arXiv:2101.02827}},
  \href {https://doi.org/10.1016/j.cpc.2021.108077}
  {\path{doi:10.1016/j.cpc.2021.108077}}.

\bibitem{Bemfica:2017wps}
F\'abio~S. Bemfica, Marcelo~M. Disconzi, and Jorge Noronha.
\newblock {Causality and existence of solutions of relativistic viscous fluid
  dynamics with gravity}.
\newblock {\em Phys. Rev. D}, 98(10):104064, 2018.
\newblock \href {http://arxiv.org/abs/1708.06255} {\path{arXiv:1708.06255}},
  \href {https://doi.org/10.1103/PhysRevD.98.104064}
  {\path{doi:10.1103/PhysRevD.98.104064}}.

\bibitem{Kovtun:2019hdm}
Pavel Kovtun.
\newblock {First-order relativistic hydrodynamics is stable}.
\newblock {\em JHEP}, 10:034, 2019.
\newblock \href {http://arxiv.org/abs/1907.08191} {\path{arXiv:1907.08191}},
  \href {https://doi.org/10.1007/JHEP10(2019)034}
  {\path{doi:10.1007/JHEP10(2019)034}}.

\bibitem{Bemfica:2019knx}
F\'abio~S. Bemfica, F\'abio~S. Bemfica, Marcelo~M. Disconzi, Marcelo~M.
  Disconzi, Jorge Noronha, and Jorge Noronha.
\newblock {Nonlinear Causality of General First-Order Relativistic Viscous
  Hydrodynamics}.
\newblock {\em Phys. Rev. D}, 100(10):104020, 2019.
\newblock [Erratum: Phys.Rev.D 105, 069902 (2022)].
\newblock \href {http://arxiv.org/abs/1907.12695} {\path{arXiv:1907.12695}},
  \href {https://doi.org/10.1103/PhysRevD.100.104020}
  {\path{doi:10.1103/PhysRevD.100.104020}}.

\bibitem{Hoult:2020eho}
Raphael~E. Hoult and Pavel Kovtun.
\newblock {Stable and causal relativistic Navier-Stokes equations}.
\newblock {\em JHEP}, 06:067, 2020.
\newblock \href {http://arxiv.org/abs/2004.04102} {\path{arXiv:2004.04102}},
  \href {https://doi.org/10.1007/JHEP06(2020)067}
  {\path{doi:10.1007/JHEP06(2020)067}}.

\bibitem{Bemfica:2020zjp}
Fabio~S. Bemfica, Marcelo~M. Disconzi, and Jorge Noronha.
\newblock {First-Order General-Relativistic Viscous Fluid Dynamics}.
\newblock {\em Phys. Rev. X}, 12(2):021044, 2022.
\newblock \href {http://arxiv.org/abs/2009.11388} {\path{arXiv:2009.11388}},
  \href {https://doi.org/10.1103/PhysRevX.12.021044}
  {\path{doi:10.1103/PhysRevX.12.021044}}.

\bibitem{Noronha:2021syv}
Jorge Noronha, Micha\l{} Spali\'nski, and Enrico Speranza.
\newblock {Transient Relativistic Fluid Dynamics in a General Hydrodynamic
  Frame}.
\newblock {\em Phys. Rev. Lett.}, 128(25):252302, 2022.
\newblock \href {http://arxiv.org/abs/2105.01034} {\path{arXiv:2105.01034}},
  \href {https://doi.org/10.1103/PhysRevLett.128.252302}
  {\path{doi:10.1103/PhysRevLett.128.252302}}.

\bibitem{Israel:1979wp}
W.~Israel and J.~M. Stewart.
\newblock {Transient relativistic thermodynamics and kinetic theory}.
\newblock {\em Annals Phys.}, 118:341--372, 1979.
\newblock \href {https://doi.org/10.1016/0003-4916(79)90130-1}
  {\path{doi:10.1016/0003-4916(79)90130-1}}.

\bibitem{Bemfica:2020xym}
F\'abio~S. Bemfica, Marcelo~M. Disconzi, Vu~Hoang, Jorge Noronha, and Maria
  Radosz.
\newblock {Nonlinear Constraints on Relativistic Fluids Far from Equilibrium}.
\newblock {\em Phys. Rev. Lett.}, 126(22):222301, 2021.
\newblock \href {http://arxiv.org/abs/2005.11632} {\path{arXiv:2005.11632}},
  \href {https://doi.org/10.1103/PhysRevLett.126.222301}
  {\path{doi:10.1103/PhysRevLett.126.222301}}.

\bibitem{Greif:2017byw}
Moritz Greif, Jan~A. Fotakis, Gabriel~S. Denicol, and Carsten Greiner.
\newblock {Diffusion of conserved charges in relativistic heavy ion
  collisions}.
\newblock {\em Phys. Rev. Lett.}, 120(24):242301, 2018.
\newblock \href {http://arxiv.org/abs/1711.08680} {\path{arXiv:1711.08680}},
  \href {https://doi.org/10.1103/PhysRevLett.120.242301}
  {\path{doi:10.1103/PhysRevLett.120.242301}}.

\bibitem{Almaalol:2022pjc}
Dekrayat Almaalol, Travis Dore, and Jacquelyn Noronha-Hostler.
\newblock {Stability of multi-component relativistic viscous hydrodynamics from
  Israel-Stewart and reproducing DNMR from maximizing the entropy}.
\newblock 9 2022.
\newblock \href {http://arxiv.org/abs/2209.11210} {\path{arXiv:2209.11210}}.

\bibitem{Martinez:2019jbu}
Mauricio Martinez, Matthew~D. Sievert, Douglas~E. Wertepny, and Jacquelyn
  Noronha-Hostler.
\newblock {Initial state fluctuations of QCD conserved charges in heavy-ion
  collisions}.
\newblock 11 2019.
\newblock \href {http://arxiv.org/abs/1911.10272} {\path{arXiv:1911.10272}}.

\bibitem{Chiu:2021muk}
Cheng Chiu and Chun Shen.
\newblock {Exploring theoretical uncertainties in the hydrodynamic description
  of relativistic heavy-ion collisions}.
\newblock {\em Phys. Rev. C}, 103(6):064901, 2021.
\newblock \href {http://arxiv.org/abs/2103.09848} {\path{arXiv:2103.09848}},
  \href {https://doi.org/10.1103/PhysRevC.103.064901}
  {\path{doi:10.1103/PhysRevC.103.064901}}.

\bibitem{Plumberg:2021bme}
Christopher Plumberg, Dekrayat Almaalol, Travis Dore, Jorge Noronha, and
  Jacquelyn Noronha-Hostler.
\newblock {Causality violations in realistic simulations of heavy-ion
  collisions}.
\newblock {\em Phys. Rev. C}, 105(6):L061901, 2022.
\newblock \href {http://arxiv.org/abs/2103.15889} {\path{arXiv:2103.15889}},
  \href {https://doi.org/10.1103/PhysRevC.105.L061901}
  {\path{doi:10.1103/PhysRevC.105.L061901}}.

\bibitem{Giacalone:2019ldn}
Giuliano Giacalone, Aleksas Mazeliauskas, and S\"oren Schlichting.
\newblock {Hydrodynamic attractors, initial state energy and particle
  production in relativistic nuclear collisions}.
\newblock {\em Phys. Rev. Lett.}, 123(26):262301, 2019.
\newblock \href {http://arxiv.org/abs/1908.02866} {\path{arXiv:1908.02866}},
  \href {https://doi.org/10.1103/PhysRevLett.123.262301}
  {\path{doi:10.1103/PhysRevLett.123.262301}}.

\bibitem{Kurkela:2018xxd}
Aleksi Kurkela and Aleksas Mazeliauskas.
\newblock {Chemical Equilibration in Hadronic Collisions}.
\newblock {\em Phys. Rev. Lett.}, 122:142301, 2019.
\newblock \href {http://arxiv.org/abs/1811.03040} {\path{arXiv:1811.03040}},
  \href {https://doi.org/10.1103/PhysRevLett.122.142301}
  {\path{doi:10.1103/PhysRevLett.122.142301}}.

\bibitem{Kurkela:2018oqw}
Aleksi Kurkela and Aleksas Mazeliauskas.
\newblock {Chemical equilibration in weakly coupled QCD}.
\newblock {\em Phys. Rev. D}, 99(5):054018, 2019.
\newblock \href {http://arxiv.org/abs/1811.03068} {\path{arXiv:1811.03068}},
  \href {https://doi.org/10.1103/PhysRevD.99.054018}
  {\path{doi:10.1103/PhysRevD.99.054018}}.

\bibitem{NunesdaSilva:2020bfs}
Tiago Nunes~da Silva, David Chinellato, Mauricio Hippert, Willian Serenone, Jun
  Takahashi, Gabriel~S. Denicol, Matthew Luzum, and Jorge Noronha.
\newblock {Pre-hydrodynamic evolution and its signatures in final-state
  heavy-ion observables}.
\newblock {\em Phys. Rev. C}, 103:054906, 2021.
\newblock \href {http://arxiv.org/abs/2006.02324} {\path{arXiv:2006.02324}},
  \href {https://doi.org/10.1103/PhysRevC.103.054906}
  {\path{doi:10.1103/PhysRevC.103.054906}}.

\bibitem{daSilva:2022xwu}
Tiago~Nunes da~Silva, David~D. Chinellato, Andr\'e~V. Giannini,
  Maur\'\i{}cio~N. Ferreira, Gabriel~S. Denicol, Maur\'\i{}cio Hippert, Matthew
  Luzum, Jorge Noronha, and Jun Takahashi.
\newblock {Pre-hydrodynamic evolution in large and small systems}.
\newblock 11 2022.
\newblock \href {http://arxiv.org/abs/2211.10561} {\path{arXiv:2211.10561}}.

\bibitem{Akamatsu:2017zzl}
Yukinao Akamatsu, Aleksas Mazeliauskas, and Derek Teaney.
\newblock {A kinetic regime of hydrodynamic fluctuations and long time tails
  for a Bjorken expansion}.
\newblock {\em Nucl. Phys. A}, 967:872--875, 2017.
\newblock \href {http://arxiv.org/abs/1705.08199} {\path{arXiv:1705.08199}},
  \href {https://doi.org/10.1016/j.nuclphysa.2017.04.029}
  {\path{doi:10.1016/j.nuclphysa.2017.04.029}}.

\bibitem{Akamatsu:2017rdu}
Yukinao Akamatsu, Aleksas Mazeliauskas, and Derek Teaney.
\newblock {Bulk viscosity from hydrodynamic fluctuations with relativistic
  hydrokinetic theory}.
\newblock {\em Phys. Rev. C}, 97(2):024902, 2018.
\newblock \href {http://arxiv.org/abs/1708.05657} {\path{arXiv:1708.05657}},
  \href {https://doi.org/10.1103/PhysRevC.97.024902}
  {\path{doi:10.1103/PhysRevC.97.024902}}.

\bibitem{Martinez:2019bsn}
M.~Martinez, T.~Sch\"afer, and V.~Skokov.
\newblock {Critical behavior of the bulk viscosity in QCD}.
\newblock {\em Phys. Rev. D}, 100(7):074017, 2019.
\newblock \href {http://arxiv.org/abs/1906.11306} {\path{arXiv:1906.11306}},
  \href {https://doi.org/10.1103/PhysRevD.100.074017}
  {\path{doi:10.1103/PhysRevD.100.074017}}.

\bibitem{Kharzeev:2015znc}
D.~E. Kharzeev, J.~Liao, S.~A. Voloshin, and G.~Wang.
\newblock {Chiral magnetic and vortical effects in high-energy nuclear
  collisions\textemdash{}A status report}.
\newblock {\em Prog. Part. Nucl. Phys.}, 88:1--28, 2016.
\newblock \href {http://arxiv.org/abs/1511.04050} {\path{arXiv:1511.04050}},
  \href {https://doi.org/10.1016/j.ppnp.2016.01.001}
  {\path{doi:10.1016/j.ppnp.2016.01.001}}.

\bibitem{Becattini:2022zvf}
Francesco Becattini.
\newblock {Spin and polarization: a new direction in relativistic heavy ion
  physics}.
\newblock {\em Rept. Prog. Phys.}, 85(12):122301, 2022.
\newblock \href {http://arxiv.org/abs/2204.01144} {\path{arXiv:2204.01144}},
  \href {https://doi.org/10.1088/1361-6633/ac97a9}
  {\path{doi:10.1088/1361-6633/ac97a9}}.

\bibitem{Son:2009tf}
Dam~T. Son and Piotr Surowka.
\newblock {Hydrodynamics with Triangle Anomalies}.
\newblock {\em Phys. Rev. Lett.}, 103:191601, 2009.
\newblock \href {http://arxiv.org/abs/0906.5044} {\path{arXiv:0906.5044}},
  \href {https://doi.org/10.1103/PhysRevLett.103.191601}
  {\path{doi:10.1103/PhysRevLett.103.191601}}.

\bibitem{Huang:2015oca}
Xu-Guang Huang.
\newblock {Electromagnetic fields and anomalous transports in heavy-ion
  collisions --- A pedagogical review}.
\newblock {\em Rept. Prog. Phys.}, 79(7):076302, 2016.
\newblock \href {http://arxiv.org/abs/1509.04073} {\path{arXiv:1509.04073}},
  \href {https://doi.org/10.1088/0034-4885/79/7/076302}
  {\path{doi:10.1088/0034-4885/79/7/076302}}.

\bibitem{Hosur:2013kxa}
Pavan Hosur and Xiaoliang Qi.
\newblock {Recent developments in transport phenomena in Weyl semimetals}.
\newblock {\em Comptes Rendus Physique}, 14:857--870, 2013.
\newblock \href {http://arxiv.org/abs/1309.4464} {\path{arXiv:1309.4464}},
  \href {https://doi.org/10.1016/j.crhy.2013.10.010}
  {\path{doi:10.1016/j.crhy.2013.10.010}}.

\bibitem{Speranza:2021bxf}
Enrico Speranza, F\'abio~S. Bemfica, Marcelo~M. Disconzi, and Jorge Noronha.
\newblock {Challenges in Solving Chiral Hydrodynamics}.
\newblock 4 2021.
\newblock \href {http://arxiv.org/abs/2104.02110} {\path{arXiv:2104.02110}}.

\bibitem{Hattori:2019lfp}
Koichi Hattori, Masaru Hongo, Xu-Guang Huang, Mamoru Matsuo, and Hidetoshi
  Taya.
\newblock {Fate of spin polarization in a relativistic fluid: An
  entropy-current analysis}.
\newblock {\em Phys. Lett. B}, 795:100--106, 2019.
\newblock \href {http://arxiv.org/abs/1901.06615} {\path{arXiv:1901.06615}},
  \href {https://doi.org/10.1016/j.physletb.2019.05.040}
  {\path{doi:10.1016/j.physletb.2019.05.040}}.

\bibitem{Montenegro:2020paq}
David Montenegro and Giorgio Torrieri.
\newblock {Linear response theory and effective action of relativistic
  hydrodynamics with spin}.
\newblock {\em Phys. Rev. D}, 102(3):036007, 2020.
\newblock \href {http://arxiv.org/abs/2004.10195} {\path{arXiv:2004.10195}},
  \href {https://doi.org/10.1103/PhysRevD.102.036007}
  {\path{doi:10.1103/PhysRevD.102.036007}}.

\bibitem{Gallegos:2020otk}
A.~D. Gallegos and U.~G\"ursoy.
\newblock {Holographic spin liquids and Lovelock Chern-Simons gravity}.
\newblock {\em JHEP}, 11:151, 2020.
\newblock \href {http://arxiv.org/abs/2004.05148} {\path{arXiv:2004.05148}},
  \href {https://doi.org/10.1007/JHEP11(2020)151}
  {\path{doi:10.1007/JHEP11(2020)151}}.

\bibitem{Hongo:2021ona}
Masaru Hongo, Xu-Guang Huang, Matthias Kaminski, Mikhail Stephanov, and Ho-Ung
  Yee.
\newblock {Relativistic spin hydrodynamics with torsion and linear response
  theory for spin relaxation}.
\newblock {\em JHEP}, 11:150, 2021.
\newblock \href {http://arxiv.org/abs/2107.14231} {\path{arXiv:2107.14231}},
  \href {https://doi.org/10.1007/JHEP11(2021)150}
  {\path{doi:10.1007/JHEP11(2021)150}}.

\bibitem{Gallegos:2022jow}
A.~D. Gallegos, U.~Gursoy, and A.~Yarom.
\newblock {Hydrodynamics, spin currents and torsion}.
\newblock 3 2022.
\newblock \href {http://arxiv.org/abs/2203.05044} {\path{arXiv:2203.05044}}.

\bibitem{Weickgenannt:2020aaf}
Nora Weickgenannt, Enrico Speranza, Xin-li Sheng, Qun Wang, and Dirk~H.
  Rischke.
\newblock {Generating Spin Polarization from Vorticity through Nonlocal
  Collisions}.
\newblock {\em Phys. Rev. Lett.}, 127(5):052301, 2021.
\newblock \href {http://arxiv.org/abs/2005.01506} {\path{arXiv:2005.01506}},
  \href {https://doi.org/10.1103/PhysRevLett.127.052301}
  {\path{doi:10.1103/PhysRevLett.127.052301}}.

\bibitem{Weickgenannt:2022zxs}
Nora Weickgenannt, David Wagner, Enrico Speranza, and Dirk~H. Rischke.
\newblock {Relativistic second-order dissipative spin hydrodynamics from the
  method of moments}.
\newblock {\em Phys. Rev. D}, 106(9):096014, 2022.
\newblock \href {http://arxiv.org/abs/2203.04766} {\path{arXiv:2203.04766}},
  \href {https://doi.org/10.1103/PhysRevD.106.096014}
  {\path{doi:10.1103/PhysRevD.106.096014}}.

\bibitem{Weickgenannt:2022qvh}
Nora Weickgenannt, David Wagner, Enrico Speranza, and Dirk~H. Rischke.
\newblock {Relativistic dissipative spin hydrodynamics from kinetic theory with
  a nonlocal collision term}.
\newblock {\em Phys. Rev. D}, 106(9):L091901, 2022.
\newblock \href {http://arxiv.org/abs/2208.01955} {\path{arXiv:2208.01955}},
  \href {https://doi.org/10.1103/PhysRevD.106.L091901}
  {\path{doi:10.1103/PhysRevD.106.L091901}}.

\bibitem{Wagner:2022amr}
David Wagner, Nora Weickgenannt, and Dirk~H. Rischke.
\newblock {Lorentz-covariant nonlocal collision term for spin-1/2 particles}.
\newblock {\em Phys. Rev. D}, 106(11):116021, 2022.
\newblock \href {http://arxiv.org/abs/2210.06187} {\path{arXiv:2210.06187}},
  \href {https://doi.org/10.1103/PhysRevD.106.116021}
  {\path{doi:10.1103/PhysRevD.106.116021}}.

\bibitem{Hartnack:1994ce}
C.~Hartnack, J.~Aichelin, Horst Stoecker, and W.~Greiner.
\newblock {Out of plane squeeze of clusters in relativistic heavy ion
  collisions}.
\newblock {\em Phys. Lett. B}, 336:131--135, 1994.
\newblock \href {https://doi.org/10.1016/0370-2693(94)90237-2}
  {\path{doi:10.1016/0370-2693(94)90237-2}}.

\bibitem{Li:1998ze}
Bao-An Li and C.~M. Ko.
\newblock {Probing the softest region of nuclear equation of state}.
\newblock {\em Phys. Rev. C}, 58:R1382--R1384, 1998.
\newblock \href {http://arxiv.org/abs/nucl-th/9807088}
  {\path{arXiv:nucl-th/9807088}}, \href
  {https://doi.org/10.1103/PhysRevC.58.R1382}
  {\path{doi:10.1103/PhysRevC.58.R1382}}.

\bibitem{Wang:2018hsw}
Yongjia Wang, Chenchen Guo, Qingfeng Li, Arnaud Le~F\`evre, Yvonne Leifels, and
  Wolfgang Trautmann.
\newblock {Determination of the nuclear incompressibility from the
  rapidity-dependent elliptic flow in heavy-ion collisions at beam energies 0.4
  A \textendash{}1.0 A GeV}.
\newblock {\em Phys. Lett. B}, 778:207--212, 2018.
\newblock \href {http://arxiv.org/abs/1804.04293} {\path{arXiv:1804.04293}},
  \href {https://doi.org/10.1016/j.physletb.2018.01.035}
  {\path{doi:10.1016/j.physletb.2018.01.035}}.

\bibitem{Nara:2021fuu}
Yasushi Nara and Akira Ohnishi.
\newblock {Mean-field update in the JAM microscopic transport model: Mean-field
  effects on collective flow in high-energy heavy-ion collisions at
  sNN=2\textendash{}20 GeV energies}.
\newblock {\em Phys. Rev. C}, 105(1):014911, 2022.
\newblock \href {http://arxiv.org/abs/2109.07594} {\path{arXiv:2109.07594}},
  \href {https://doi.org/10.1103/PhysRevC.105.014911}
  {\path{doi:10.1103/PhysRevC.105.014911}}.

\bibitem{Sorensen:2023zkk}
Agnieszka Sorensen et~al.
\newblock {Dense Nuclear Matter Equation of State from Heavy-Ion Collisions}.
\newblock 1 2023.
\newblock \href {http://arxiv.org/abs/2301.13253} {\path{arXiv:2301.13253}}.

\bibitem{NSAC-QIS-2019-QuantumInformationScience}
Nuclear {Physics} and {Quantum} {Information} {Science}: {Report} by the {NSAC}
  {QIS} {Subcommittee}.
\newblock Technical report, NSF \& DOE Office of Science, October 2019.
\newblock URL:
  \url{https://science.osti.gov/-/media/np/pdf/Reports/NSAC_QIS_Report.pdf}.

\bibitem{Detmold:2019ghl}
William Detmold, Robert~G. Edwards, Jozef~J. Dudek, Michael Engelhardt,
  Huey-Wen Lin, Stefan Meinel, Kostas Orginos, and Phiala Shanahan.
\newblock {Hadrons and Nuclei}.
\newblock {\em Eur. Phys. J. A}, 55(11):193, 2019.
\newblock \href {http://arxiv.org/abs/1904.09512} {\path{arXiv:1904.09512}},
  \href {https://doi.org/10.1140/epja/i2019-12902-4}
  {\path{doi:10.1140/epja/i2019-12902-4}}.

\bibitem{Cirigliano:2019jig}
Vincenzo Cirigliano, Zohreh Davoudi, Tanmoy Bhattacharya, Taku Izubuchi,
  Phiala~E. Shanahan, Sergey Syritsyn, and Michael~L. Wagman.
\newblock {The Role of Lattice QCD in Searches for Violations of Fundamental
  Symmetries and Signals for New Physics}.
\newblock {\em Eur. Phys. J. A}, 55(11):197, 2019.
\newblock \href {http://arxiv.org/abs/1904.09704} {\path{arXiv:1904.09704}},
  \href {https://doi.org/10.1140/epja/i2019-12889-8}
  {\path{doi:10.1140/epja/i2019-12889-8}}.

\bibitem{Kronfeld:2019nfb}
Andreas~S. Kronfeld, David~G. Richards, William Detmold, Rajan Gupta, Huey-Wen
  Lin, Keh-Fei Liu, Aaron~S. Meyer, Raza Sufian, and Sergey Syritsyn.
\newblock {Lattice QCD and Neutrino-Nucleus Scattering}.
\newblock {\em Eur. Phys. J. A}, 55(11):196, 2019.
\newblock \href {http://arxiv.org/abs/1904.09931} {\path{arXiv:1904.09931}},
  \href {https://doi.org/10.1140/epja/i2019-12916-x}
  {\path{doi:10.1140/epja/i2019-12916-x}}.

\bibitem{Bazavov:2019lgz}
Alexei Bazavov, Frithjof Karsch, Swagato Mukherjee, and Peter Petreczky.
\newblock {Hot-dense Lattice QCD: USQCD whitepaper 2018}.
\newblock {\em Eur. Phys. J. A}, 55(11):194, 2019.
\newblock \href {http://arxiv.org/abs/1904.09951} {\path{arXiv:1904.09951}},
  \href {https://doi.org/10.1140/epja/i2019-12922-0}
  {\path{doi:10.1140/epja/i2019-12922-0}}.

\bibitem{Joo:2019byq}
B\'alint Jo\'o, Chulwoo Jung, Norman~H. Christ, William Detmold, Robert
  Edwards, Martin Savage, and Phiala Shanahan.
\newblock {Status and Future Perspectives for Lattice Gauge Theory Calculations
  to the Exascale and Beyond}.
\newblock {\em Eur. Phys. J. A}, 55(11):199, 2019.
\newblock \href {http://arxiv.org/abs/1904.09725} {\path{arXiv:1904.09725}},
  \href {https://doi.org/10.1140/epja/i2019-12919-7}
  {\path{doi:10.1140/epja/i2019-12919-7}}.

\bibitem{Ciavarella:2021nmj}
Anthony Ciavarella, Natalie Klco, and Martin~J. Savage.
\newblock {Trailhead for quantum simulation of SU(3) Yang-Mills lattice gauge
  theory in the local multiplet basis}.
\newblock {\em Phys. Rev. D}, 103(9):094501, 2021.
\newblock \href {http://arxiv.org/abs/2101.10227} {\path{arXiv:2101.10227}},
  \href {https://doi.org/10.1103/PhysRevD.103.094501}
  {\path{doi:10.1103/PhysRevD.103.094501}}.

\bibitem{Ciavarella:2022zhe}
Anthony Ciavarella, Natalie Klco, and Martin~J. Savage.
\newblock {Some Conceptual Aspects of Operator Design for Quantum Simulations
  of Non-Abelian Lattice Gauge Theories}.
\newblock 3 2022.
\newblock \href {http://arxiv.org/abs/2203.11988} {\path{arXiv:2203.11988}}.

\bibitem{buluta2009quantum}
Iulia Buluta and Franco Nori.
\newblock Quantum simulators.
\newblock {\em Science}, 326(5949):108--111, 2009.

\bibitem{brown2010using}
Katherine~L Brown, William~J Munro, and Vivien~M Kendon.
\newblock Using quantum computers for quantum simulation.
\newblock {\em Entropy}, 12(11):2268--2307, 2010.

\bibitem{Georgescu:2013oza}
I.~M. Georgescu, S.~Ashhab, and Franco Nori.
\newblock {Quantum Simulation}.
\newblock {\em Rev. Mod. Phys.}, 86:153, 2014.
\newblock \href {http://arxiv.org/abs/1308.6253} {\path{arXiv:1308.6253}},
  \href {https://doi.org/10.1103/RevModPhys.86.153}
  {\path{doi:10.1103/RevModPhys.86.153}}.

\bibitem{altman2021quantum}
Ehud Altman, Kenneth~R Brown, Giuseppe Carleo, Lincoln~D Carr, Eugene Demler,
  Cheng Chin, Brian DeMarco, Sophia~E Economou, Mark~A Eriksson, Kai-Mei~C Fu,
  et~al.
\newblock Quantum simulators: Architectures and opportunities.
\newblock {\em PRX Quantum}, 2(1):017003, 2021.

\bibitem{alexeev2021quantum}
Yuri Alexeev, Dave Bacon, Kenneth~R Brown, Robert Calderbank, Lincoln~D Carr,
  Frederic~T Chong, Brian DeMarco, Dirk Englund, Edward Farhi, Bill Fefferman,
  et~al.
\newblock Quantum computer systems for scientific discovery.
\newblock {\em PRX Quantum}, 2(1):017001, 2021.

\bibitem{Elben:2022jvo}
Andreas Elben, Steven~T. Flammia, Hsin-Yuan Huang, Richard Kueng, John
  Preskill, Beno\^\i{}t Vermersch, and Peter Zoller.
\newblock {The randomized measurement toolbox}.
\newblock {\em Nature Rev. Phys.}, 5(1):9--24, 2023.
\newblock \href {http://arxiv.org/abs/2203.11374} {\path{arXiv:2203.11374}},
  \href {https://doi.org/10.1038/s42254-022-00535-2}
  {\path{doi:10.1038/s42254-022-00535-2}}.

\bibitem{martinez2016real}
Esteban~A Martinez, Christine~A Muschik, Philipp Schindler, Daniel Nigg,
  Alexander Erhard, Markus Heyl, Philipp Hauke, Marcello Dalmonte, Thomas Monz,
  Peter Zoller, et~al.
\newblock Real-time dynamics of lattice gauge theories with a few-qubit quantum
  computer.
\newblock {\em Nature}, 534(7608):516--519, 2016.

\bibitem{klco2018quantum}
N~Klco, EF~Dumitrescu, AJ~McCaskey, TD~Morris, RC~Pooser, M~Sanz, E~Solano,
  P~Lougovski, and MJ~Savage.
\newblock Quantum-classical dynamical calculations of the schwinger model using
  quantum computers.
\newblock {\em arXiv preprint arXiv:1803.03326}, 2018.

\bibitem{nguyen2022digital}
Nhung~H Nguyen, Minh~C Tran, Yingyue Zhu, Alaina~M Green, C~Huerta Alderete,
  Zohreh Davoudi, and Norbert~M Linke.
\newblock Digital quantum simulation of the schwinger model and symmetry
  protection with trapped ions.
\newblock {\em PRX Quantum}, 3(2):020324, 2022.

\bibitem{Mueller:2022xbg}
Niklas Mueller, Joseph~A. Carolan, Andrew Connelly, Zohreh Davoudi, Eugene~F.
  Dumitrescu, and K\"ubra Yeter-Aydeniz.
\newblock {Quantum computation of dynamical quantum phase transitions and
  entanglement tomography in a lattice gauge theory}.
\newblock 10 2022.
\newblock \href {http://arxiv.org/abs/2210.03089} {\path{arXiv:2210.03089}}.

\bibitem{de2021quantum}
Wibe~A de~Jong, Kyle Lee, James Mulligan, Mateusz P{\l}osko{\'n}, Felix Ringer,
  and Xiaojun Yao.
\newblock Quantum simulation of non-equilibrium dynamics and thermalization in
  the schwinger model.
\newblock {\em arXiv preprint arXiv:2106.08394}, 2021.

\bibitem{mil2020scalable}
Alexander Mil, Torsten~V Zache, Apoorva Hegde, Andy Xia, Rohit~P Bhatt,
  Markus~K Oberthaler, Philipp Hauke, J{\"u}rgen Berges, and Fred
  Jendrzejewski.
\newblock {A scalable realization of local U(1) gauge invariance in cold atomic
  mixtures}.
\newblock {\em Science}, 367(6482):1128--1130, 2020.

\bibitem{zhou2021thermalization}
Zhao-Yu Zhou, Guo-Xian Su, Jad~C Halimeh, Robert Ott, Hui Sun, Philipp Hauke,
  Bing Yang, Zhen-Sheng Yuan, J{\"u}rgen Berges, and Jian-Wei Pan.
\newblock Thermalization dynamics of a gauge theory on a quantum simulator.
\newblock {\em arXiv preprint arXiv:2107.13563}, 2021.

\bibitem{Byrnes:2005qx}
Tim Byrnes and Yoshihisa Yamamoto.
\newblock {Simulating lattice gauge theories on a quantum computer}.
\newblock {\em Phys. Rev. A}, 73:022328, 2006.
\newblock \href {http://arxiv.org/abs/quant-ph/0510027}
  {\path{arXiv:quant-ph/0510027}}, \href
  {https://doi.org/10.1103/PhysRevA.73.022328}
  {\path{doi:10.1103/PhysRevA.73.022328}}.

\bibitem{kan2021lattice}
Angus Kan and Yunseong Nam.
\newblock Lattice quantum chromodynamics and electrodynamics on a universal
  quantum computer.
\newblock {\em arXiv:2107.12769}, 2021.
\newblock URL: \url{https://arxiv.org/abs/2107.12769}.

\bibitem{Bauer:2022hpo}
Christian~W. Bauer et~al.
\newblock {Quantum Simulation for High Energy Physics}.
\newblock 4 2022.
\newblock \href {http://arxiv.org/abs/2204.03381} {\path{arXiv:2204.03381}}.

\bibitem{shaw2020quantum}
Alexander~F Shaw, Pavel Lougovski, Jesse~R Stryker, and Nathan Wiebe.
\newblock Quantum algorithms for simulating the lattice schwinger model.
\newblock {\em Quantum}, 4:306, 2020.

\bibitem{Lamm:2019bik}
Henry Lamm, Scott Lawrence, and Yukari Yamauchi.
\newblock {General Methods for Digital Quantum Simulation of Gauge Theories}.
\newblock {\em Phys. Rev. D}, 100(3):034518, 2019.
\newblock \href {http://arxiv.org/abs/1903.08807} {\path{arXiv:1903.08807}},
  \href {https://doi.org/10.1103/PhysRevD.100.034518}
  {\path{doi:10.1103/PhysRevD.100.034518}}.

\bibitem{Klco:2019evd}
Natalie Klco, Jesse~R. Stryker, and Martin~J. Savage.
\newblock {SU(2) non-Abelian gauge field theory in one dimension on digital
  quantum computers}.
\newblock {\em Phys. Rev. D}, 101(7):074512, 2020.
\newblock \href {http://arxiv.org/abs/1908.06935} {\path{arXiv:1908.06935}},
  \href {https://doi.org/10.1103/PhysRevD.101.074512}
  {\path{doi:10.1103/PhysRevD.101.074512}}.

\bibitem{Mueller:2019qqj}
Niklas Mueller, Andrey Tarasov, and Raju Venugopalan.
\newblock {Deeply inelastic scattering structure functions on a hybrid quantum
  computer}.
\newblock {\em Phys. Rev. D}, 102(1):016007, 2020.
\newblock \href {http://arxiv.org/abs/1908.07051} {\path{arXiv:1908.07051}},
  \href {https://doi.org/10.1103/PhysRevD.102.016007}
  {\path{doi:10.1103/PhysRevD.102.016007}}.

\bibitem{Barata:2020jtq}
Jo\~ao Barata, Niklas Mueller, Andrey Tarasov, and Raju Venugopalan.
\newblock {Single-particle digitization strategy for quantum computation of a
  $\phi^4$ scalar field theory}.
\newblock {\em Phys. Rev. A}, 103(4):042410, 2021.
\newblock \href {http://arxiv.org/abs/2012.00020} {\path{arXiv:2012.00020}},
  \href {https://doi.org/10.1103/PhysRevA.103.042410}
  {\path{doi:10.1103/PhysRevA.103.042410}}.

\bibitem{Atas:2021ext}
Yasar~Y. Atas, Jinglei Zhang, Randy Lewis, Amin Jahanpour, Jan~F. Haase, and
  Christine~A. Muschik.
\newblock {SU(2) hadrons on a quantum computer via a variational approach}.
\newblock {\em Nature Commun.}, 12(1):6499, 2021.
\newblock \href {http://arxiv.org/abs/2102.08920} {\path{arXiv:2102.08920}},
  \href {https://doi.org/10.1038/s41467-021-26825-4}
  {\path{doi:10.1038/s41467-021-26825-4}}.

\bibitem{rahman20212}
Sarmed~A Rahman, Randy Lewis, Emanuele Mendicelli, and Sarah Powell.
\newblock Su (2) lattice gauge theory on a quantum annealer.
\newblock {\em Physical Review D}, 104(3):034501, 2021.

\bibitem{Atas:2022dqm}
Yasar~Y. Atas, Jan~F. Haase, Jinglei Zhang, Victor Wei, Sieglinde M.~L.
  Pfaendler, Randy Lewis, and Christine~A. Muschik.
\newblock {Real-time evolution of SU(3) hadrons on a quantum computer}.
\newblock 7 2022.
\newblock \href {http://arxiv.org/abs/2207.03473} {\path{arXiv:2207.03473}}.

\bibitem{Illa:2022jqb}
Marc Illa and Martin~J. Savage.
\newblock {Basic elements for simulations of standard-model physics with
  quantum annealers: Multigrid and clock states}.
\newblock {\em Phys. Rev. A}, 106(5):052605, 2022.
\newblock \href {http://arxiv.org/abs/2202.12340} {\path{arXiv:2202.12340}},
  \href {https://doi.org/10.1103/PhysRevA.106.052605}
  {\path{doi:10.1103/PhysRevA.106.052605}}.

\bibitem{Farrell:2022vyh}
Roland~C. Farrell, Ivan~A. Chernyshev, Sarah J.~M. Powell, Nikita~A.
  Zemlevskiy, Marc Illa, and Martin~J. Savage.
\newblock {Preparations for Quantum Simulations of Quantum Chromodynamics in
  1+1 Dimensions: (II) Single-Baryon $\beta$-Decay in Real Time}.
\newblock 9 2022.
\newblock \href {http://arxiv.org/abs/2209.10781} {\path{arXiv:2209.10781}}.

\bibitem{Bedaque:2022ftd}
Paulo~F. Bedaque, Ratna Khadka, Gautam Rupak, and Muhammad Yusf.
\newblock {Radiative processes on a quantum computer}.
\newblock 9 2022.
\newblock \href {http://arxiv.org/abs/2209.09962} {\path{arXiv:2209.09962}}.

\bibitem{DeJong:2020riy}
Wibe~A. De~Jong, Mekena Metcalf, James Mulligan, Mateusz P\l{}osko\'n, Felix
  Ringer, and Xiaojun Yao.
\newblock {Quantum simulation of open quantum systems in heavy-ion collisions}.
\newblock {\em Phys. Rev. D}, 104(5):051501, 2021.
\newblock \href {http://arxiv.org/abs/2010.03571} {\path{arXiv:2010.03571}},
  \href {https://doi.org/10.1103/PhysRevD.104.L051501}
  {\path{doi:10.1103/PhysRevD.104.L051501}}.

\bibitem{Czajka:2021yll}
Alexander~M. Czajka, Zhong-Bo Kang, Henry Ma, and Fanyi Zhao.
\newblock {Quantum simulation of chiral phase transitions}.
\newblock {\em JHEP}, 08:209, 2022.
\newblock \href {http://arxiv.org/abs/2112.03944} {\path{arXiv:2112.03944}},
  \href {https://doi.org/10.1007/JHEP08(2022)209}
  {\path{doi:10.1007/JHEP08(2022)209}}.

\bibitem{Davoudi:2022uzo}
Zohreh Davoudi, Niklas Mueller, and Connor Powers.
\newblock {Toward Quantum Computing Phase Diagrams of Gauge Theories with
  Thermal Pure Quantum States}.
\newblock 8 2022.
\newblock \href {http://arxiv.org/abs/2208.13112} {\path{arXiv:2208.13112}}.

\bibitem{Cohen:2021imf}
Thomas~D. Cohen, Henry Lamm, Scott Lawrence, and Yukari Yamauchi.
\newblock {Quantum algorithms for transport coefficients in gauge theories}.
\newblock {\em Phys. Rev. D}, 104(9):094514, 2021.
\newblock \href {http://arxiv.org/abs/2104.02024} {\path{arXiv:2104.02024}},
  \href {https://doi.org/10.1103/PhysRevD.104.094514}
  {\path{doi:10.1103/PhysRevD.104.094514}}.

\bibitem{Lamm:2019uyc}
Henry Lamm, Scott Lawrence, and Yukari Yamauchi.
\newblock {Parton physics on a quantum computer}.
\newblock {\em Phys. Rev. Res.}, 2(1):013272, 2020.
\newblock \href {http://arxiv.org/abs/1908.10439} {\path{arXiv:1908.10439}},
  \href {https://doi.org/10.1103/PhysRevResearch.2.013272}
  {\path{doi:10.1103/PhysRevResearch.2.013272}}.

\bibitem{Gong:2021bcp}
Wenjie Gong, Ganesh Parida, Zhoudunming Tu, and Raju Venugopalan.
\newblock {Measurement of Bell-type inequalities and quantum entanglement from
  \ensuremath{\Lambda}-hyperon spin correlations at high energy colliders}.
\newblock {\em Phys. Rev. D}, 106(3):L031501, 2022.
\newblock \href {http://arxiv.org/abs/2107.13007} {\path{arXiv:2107.13007}},
  \href {https://doi.org/10.1103/PhysRevD.106.L031501}
  {\path{doi:10.1103/PhysRevD.106.L031501}}.

\bibitem{Yao:2022eqm}
Xiaojun Yao.
\newblock {Quantum Simulation of Light-Front QCD for Jet Quenching in Nuclear
  Environments}.
\newblock 5 2022.
\newblock \href {http://arxiv.org/abs/2205.07902} {\path{arXiv:2205.07902}}.

\bibitem{humble2022snowmass}
Travis~S Humble, Andrea Delgado, Raphael Pooser, Christopher Seck, Ryan
  Bennink, Vicente Leyton-Ortega, C-C~Joseph Wang, Eugene Dumitrescu, Titus
  Morris, Kathleen Hamilton, et~al.
\newblock Snowmass white paper: Quantum computing systems and software for
  high-energy physics research.
\newblock {\em arXiv preprint arXiv:2203.07091}, 2022.

\bibitem{Heffernan:2022swr}
Matthew Heffernan, Charles Gale, Sangyong Jeon, and Jean-Francois Paquet.
\newblock {Bayesian quantification of the Quark-Gluon Plasma: Improved design
  and closure demonstration}.
\newblock In {\em {29th International Conference on Ultra-relativistic
  Nucleus-Nucleus Collisions}}, 7 2022.
\newblock \href {http://arxiv.org/abs/2207.14751} {\path{arXiv:2207.14751}}.

\bibitem{Liyanage:2022byj}
Dananjaya Liyanage, Yi~Ji, Derek Everett, Matthew Heffernan, Ulrich Heinz,
  Simon Mak, and Jean-Francois Paquet.
\newblock {Efficient emulation of relativistic heavy ion collisions with
  transfer learning}.
\newblock {\em Phys. Rev. C}, 105(3):034910, 2022.
\newblock \href {http://arxiv.org/abs/2201.07302} {\path{arXiv:2201.07302}},
  \href {https://doi.org/10.1103/PhysRevC.105.034910}
  {\path{doi:10.1103/PhysRevC.105.034910}}.

\bibitem{Ji:2022xzo}
Yi~Ji, Henry~Shaowu Yuchi, Derek Soeder, J.~F. Paquet, Steffen~A. Bass,
  V.~Roshan Joseph, C.~F.~Jeff Wu, and Simon Mak.
\newblock {Multi-Stage Multi-Fidelity Gaussian Process Modeling, with
  Application to Heavy-Ion Collisions}.
\newblock 9 2022.
\newblock \href {http://arxiv.org/abs/2209.13748} {\path{arXiv:2209.13748}}.

\bibitem{Phillips:2020dmw}
D.~R. Phillips et~al.
\newblock {Get on the BAND Wagon: A Bayesian Framework for Quantifying Model
  Uncertainties in Nuclear Dynamics}.
\newblock {\em J. Phys. G}, 48(7):072001, 2021.
\newblock \href {http://arxiv.org/abs/2012.07704} {\path{arXiv:2012.07704}},
  \href {https://doi.org/10.1088/1361-6471/abf1df}
  {\path{doi:10.1088/1361-6471/abf1df}}.

\bibitem{Parkkila:2021yha}
J.~E. Parkkila, A.~Onnerstad, S.~F. Taghavi, C.~Mordasini, A.~Bilandzic,
  M.~Virta, and D.~J. Kim.
\newblock {New constraints for QCD matter from improved Bayesian parameter
  estimation in heavy-ion collisions at LHC}.
\newblock {\em Phys. Lett. B}, 835:137485, 2022.
\newblock \href {http://arxiv.org/abs/2111.08145} {\path{arXiv:2111.08145}},
  \href {https://doi.org/10.1016/j.physletb.2022.137485}
  {\path{doi:10.1016/j.physletb.2022.137485}}.

\bibitem{ALICE-PUBLIC-2019-001}
{ALICE upgrade physics performance studies for 2018 Report on HL/HE-LHC
  physics}.
\newblock 2019.

\bibitem{ALICE:2023udb}
{ALICE upgrades during the LHC Long Shutdown 2}.
\newblock 2 2023.
\newblock \href {http://arxiv.org/abs/2302.01238} {\path{arXiv:2302.01238}}.

\bibitem{CMS-DP-2021-037}
{New opportunities of heavy ion physics with CMS-MTD at the HL-LHC}.
\newblock 2021.
\newblock URL: \url{http://cds.cern.ch/record/2800541}.

\bibitem{CMS-PAS-FTR-22-001}
{Snowmass White Paper Contribution: Physics with the Phase-2 ATLAS and CMS
  Detectors}.
\newblock Technical report, CERN, Geneva, 2022.
\newblock URL: \url{https://cds.cern.ch/record/2806962}.

\bibitem{Krintiras:2022ohr}
Georgios~K. Krintiras and Andre G.~Stahl Leiton.
\newblock {The CMS Heavy Ion Group contribution to 2022 NSAC Long-Range Plan
  Town Hall Meeting (Hot and Cold QCD) \textendash{} Letter of Interest}.
\newblock 9 2022.
\newblock \href {http://arxiv.org/abs/2209.11564} {\path{arXiv:2209.11564}}.

\bibitem{Citron:2018lsq}
Z.~Citron et~al.
\newblock {Report from Working Group 5}: {Future physics opportunities for
  high-density QCD at the LHC with heavy-ion and proton beams}.
\newblock {\em CERN Yellow Rep. Monogr.}, 7:1159--1410, 2019.
\newblock \href {http://arxiv.org/abs/1812.06772} {\path{arXiv:1812.06772}},
  \href {https://doi.org/10.23731/CYRM-2019-007.1159}
  {\path{doi:10.23731/CYRM-2019-007.1159}}.

\bibitem{Gardim:2019brr}
Fernando~G. Gardim, Giuliano Giacalone, and Jean-Yves Ollitrault.
\newblock {The mean transverse momentum of ultracentral heavy-ion collisions: A
  new probe of hydrodynamics}.
\newblock {\em Phys. Lett. B}, 809:135749, 2020.
\newblock \href {http://arxiv.org/abs/1909.11609} {\path{arXiv:1909.11609}},
  \href {https://doi.org/10.1016/j.physletb.2020.135749}
  {\path{doi:10.1016/j.physletb.2020.135749}}.

\bibitem{Stoecker:1980vf}
Horst Stoecker, J.~A. Maruhn, and W.~Greiner.
\newblock {Collective sideward flow of nuclear matter in violent high-energy
  heavy ion collisions}.
\newblock {\em Phys. Rev. Lett.}, 44:725, 1980.
\newblock \href {https://doi.org/10.1103/PhysRevLett.44.725}
  {\path{doi:10.1103/PhysRevLett.44.725}}.

\bibitem{Ollitrault:1992bk}
Jean-Yves Ollitrault.
\newblock {Anisotropy as a signature of transverse collective flow}.
\newblock {\em Phys. Rev. D}, 46:229--245, 1992.
\newblock \href {https://doi.org/10.1103/PhysRevD.46.229}
  {\path{doi:10.1103/PhysRevD.46.229}}.

\bibitem{Rischke:1995pe}
Dirk~H. Rischke, Yaris P\"urs\"un, Joachim~A. Maruhn, Horst Stoecker, and
  Walter Greiner.
\newblock {The Phase transition to the quark - gluon plasma and its effects on
  hydrodynamic flow}.
\newblock {\em Acta Phys. Hung. A}, 1:309--322, 1995.
\newblock \href {http://arxiv.org/abs/nucl-th/9505014}
  {\path{arXiv:nucl-th/9505014}}, \href {https://doi.org/10.1007/BF03053749}
  {\path{doi:10.1007/BF03053749}}.

\bibitem{Stoecker:2004qu}
Horst Stoecker.
\newblock {Collective flow signals the quark gluon plasma}.
\newblock {\em Nucl. Phys. A}, 750:121--147, 2005.
\newblock \href {http://arxiv.org/abs/nucl-th/0406018}
  {\path{arXiv:nucl-th/0406018}}, \href
  {https://doi.org/10.1016/j.nuclphysa.2004.12.074}
  {\path{doi:10.1016/j.nuclphysa.2004.12.074}}.

\bibitem{Brachmann:1999xt}
J.~Brachmann, S.~Soff, A.~Dumitru, Horst Stoecker, J.~A. Maruhn, W.~Greiner,
  L.~V. Bravina, and D.~H. Rischke.
\newblock {Antiflow of nucleons at the softest point of the EoS}.
\newblock {\em Phys. Rev. C}, 61:024909, 2000.
\newblock \href {http://arxiv.org/abs/nucl-th/9908010}
  {\path{arXiv:nucl-th/9908010}}, \href
  {https://doi.org/10.1103/PhysRevC.61.024909}
  {\path{doi:10.1103/PhysRevC.61.024909}}.

\bibitem{Csernai:1999nf}
L.~P. Csernai and D.~Rohrich.
\newblock {Third flow component as QGP signal}.
\newblock {\em Phys. Lett. B}, 458:454, 1999.
\newblock \href {http://arxiv.org/abs/nucl-th/9908034}
  {\path{arXiv:nucl-th/9908034}}, \href
  {https://doi.org/10.1016/S0370-2693(99)00615-2}
  {\path{doi:10.1016/S0370-2693(99)00615-2}}.

\bibitem{Ivanov:2014ioa}
Yu.~B. Ivanov and A.~A. Soldatov.
\newblock {Directed flow indicates a cross-over deconfinement transition in
  relativistic nuclear collisions}.
\newblock {\em Phys. Rev. C}, 91(2):024915, 2015.
\newblock \href {http://arxiv.org/abs/1412.1669} {\path{arXiv:1412.1669}},
  \href {https://doi.org/10.1103/PhysRevC.91.024915}
  {\path{doi:10.1103/PhysRevC.91.024915}}.

\bibitem{Sorensen:2021zme}
Agnieszka Sorensen, Dmytro Oliinychenko, Volker Koch, and Larry McLerran.
\newblock {Speed of Sound and Baryon Cumulants in Heavy-Ion Collisions}.
\newblock {\em Phys. Rev. Lett.}, 127(4):042303, 2021.
\newblock \href {http://arxiv.org/abs/2103.07365} {\path{arXiv:2103.07365}},
  \href {https://doi.org/10.1103/PhysRevLett.127.042303}
  {\path{doi:10.1103/PhysRevLett.127.042303}}.

\bibitem{Vovchenko:2020tsr}
Volodymyr Vovchenko, Oleh Savchuk, Roman~V. Poberezhnyuk, Mark~I. Gorenstein,
  and Volker Koch.
\newblock {Connecting fluctuation measurements in heavy-ion collisions with the
  grand-canonical susceptibilities}.
\newblock {\em Phys. Lett. B}, 811:135868, 2020.
\newblock \href {http://arxiv.org/abs/2003.13905} {\path{arXiv:2003.13905}},
  \href {https://doi.org/10.1016/j.physletb.2020.135868}
  {\path{doi:10.1016/j.physletb.2020.135868}}.

\bibitem{Vovchenko:2020gne}
Volodymyr Vovchenko, Roman~V. Poberezhnyuk, and Volker Koch.
\newblock {Cumulants of multiple conserved charges and global conservation
  laws}.
\newblock {\em JHEP}, 10:089, 2020.
\newblock \href {http://arxiv.org/abs/2007.03850} {\path{arXiv:2007.03850}},
  \href {https://doi.org/10.1007/JHEP10(2020)089}
  {\path{doi:10.1007/JHEP10(2020)089}}.

\bibitem{Almaalol:2022xwv}
D.~Almaalol et~al.
\newblock {QCD Phase Structure and Interactions at High Baryon Density:
  Continuation of BES Physics Program with CBM at FAIR}.
\newblock 9 2022.
\newblock \href {http://arxiv.org/abs/2209.05009} {\path{arXiv:2209.05009}}.

\bibitem{Lovato:2022vgq}
Alessandro Lovato et~al.
\newblock {Long Range Plan: Dense matter theory for heavy-ion collisions and
  neutron stars}.
\newblock 11 2022.
\newblock \href {http://arxiv.org/abs/2211.02224} {\path{arXiv:2211.02224}}.

\bibitem{Nara:2020ztb}
Yasushi Nara, Tomoyuki Maruyama, and Horst Stoecker.
\newblock {Momentum-dependent potential and collective flows within the
  relativistic quantum molecular dynamics approach based on relativistic
  mean-field theory}.
\newblock {\em Phys. Rev. C}, 102(2):024913, 2020.
\newblock \href {http://arxiv.org/abs/2004.05550} {\path{arXiv:2004.05550}},
  \href {https://doi.org/10.1103/PhysRevC.102.024913}
  {\path{doi:10.1103/PhysRevC.102.024913}}.

\bibitem{Monnai:2015bca}
Akihiko Monnai.
\newblock {Chemically non-equilibrated QGP and thermal photon elliptic flow}.
\newblock In {\em {7th International Conference on Hard and Electromagnetic
  Probes of High-Energy Nuclear Collisions}}, 10 2015.
\newblock \href {http://arxiv.org/abs/1510.00539} {\path{arXiv:1510.00539}},
  \href {https://doi.org/10.1016/j.nuclphysbps.2016.05.052}
  {\path{doi:10.1016/j.nuclphysbps.2016.05.052}}.

\bibitem{Linnyk:2015rco}
O.~Linnyk, E.~L. Bratkovskaya, and W.~Cassing.
\newblock {Effective QCD and transport description of dilepton and photon
  production in heavy-ion collisions and elementary processes}.
\newblock {\em Prog. Part. Nucl. Phys.}, 87:50--115, 2016.
\newblock \href {http://arxiv.org/abs/1512.08126} {\path{arXiv:1512.08126}},
  \href {https://doi.org/10.1016/j.ppnp.2015.12.003}
  {\path{doi:10.1016/j.ppnp.2015.12.003}}.

\bibitem{Greif:2016jeb}
Moritz Greif, Florian Senzel, Heiner Kremer, Kai Zhou, Carsten Greiner, and Zhe
  Xu.
\newblock {Nonequilibrium photon production in partonic transport simulations}.
\newblock {\em Phys. Rev. C}, 95(5):054903, 2017.
\newblock \href {http://arxiv.org/abs/1612.05811} {\path{arXiv:1612.05811}},
  \href {https://doi.org/10.1103/PhysRevC.95.054903}
  {\path{doi:10.1103/PhysRevC.95.054903}}.

\bibitem{Vovchenko:2016ijt}
V.~Vovchenko, Iu.~A. Karpenko, M.~I. Gorenstein, L.~M. Satarov, I.~N.
  Mishustin, B.~K\"ampfer, and H.~Stoecker.
\newblock {Electromagnetic probes of a pure-glue initial state in
  nucleus-nucleus collisions at energies available at the CERN Large Hadron
  Collider}.
\newblock {\em Phys. Rev. C}, 94(2):024906, 2016.
\newblock \href {http://arxiv.org/abs/1604.06346} {\path{arXiv:1604.06346}},
  \href {https://doi.org/10.1103/PhysRevC.94.024906}
  {\path{doi:10.1103/PhysRevC.94.024906}}.

\bibitem{Srivastava:2016hwr}
Dinesh~K. Srivastava, Rupa Chatterjee, and Munshi~G. Mustafa.
\newblock {Initial temperature and extent of chemical equilibration of partons
  in relativistic collisions of heavy nuclei}.
\newblock {\em J. Phys. G}, 45(1):015103, 2018.
\newblock \href {http://arxiv.org/abs/1609.06496} {\path{arXiv:1609.06496}},
  \href {https://doi.org/10.1088/1361-6471/aa9421}
  {\path{doi:10.1088/1361-6471/aa9421}}.

\bibitem{Oliva:2017pri}
L.~Oliva, M.~Ruggieri, S.~Plumari, F.~Scardina, G.~X. Peng, and V.~Greco.
\newblock {Photons from the Early Stages of Relativistic Heavy Ion Collisions}.
\newblock {\em Phys. Rev. C}, 96(1):014914, 2017.
\newblock \href {http://arxiv.org/abs/1703.00116} {\path{arXiv:1703.00116}},
  \href {https://doi.org/10.1103/PhysRevC.96.014914}
  {\path{doi:10.1103/PhysRevC.96.014914}}.

\bibitem{Berges:2017eom}
Jurgen Berges, Klaus Reygers, Naoto Tanji, and Raju Venugopalan.
\newblock {Parametric estimate of the relative photon yields from the glasma
  and the quark-gluon plasma in heavy-ion collisions}.
\newblock {\em Phys. Rev. C}, 95(5):054904, 2017.
\newblock \href {http://arxiv.org/abs/1701.05064} {\path{arXiv:1701.05064}},
  \href {https://doi.org/10.1103/PhysRevC.95.054904}
  {\path{doi:10.1103/PhysRevC.95.054904}}.

\bibitem{Monnai:2019vup}
Akihiko Monnai.
\newblock {Prompt, pre-equilibrium, and thermal photons in relativistic nuclear
  collisions}.
\newblock {\em J. Phys. G}, 47(7):075105, 2020.
\newblock \href {http://arxiv.org/abs/1907.09266} {\path{arXiv:1907.09266}},
  \href {https://doi.org/10.1088/1361-6471/ab8d8c}
  {\path{doi:10.1088/1361-6471/ab8d8c}}.

\bibitem{Churchill:2020uvk}
Jessica Churchill, Li~Yan, Sangyong Jeon, and Charles Gale.
\newblock {Emission of electromagnetic radiation from the early stages of
  relativistic heavy-ion collisions}.
\newblock {\em Phys. Rev. C}, 103(2):024904, 2021.
\newblock \href {http://arxiv.org/abs/2008.02902} {\path{arXiv:2008.02902}},
  \href {https://doi.org/10.1103/PhysRevC.103.024904}
  {\path{doi:10.1103/PhysRevC.103.024904}}.

\bibitem{Garcia-Montero:2019vju}
Oscar Garcia-Montero.
\newblock {Non-equilibrium photons from the bottom-up thermalization scenario}.
\newblock {\em Annals Phys.}, 443:168984, 2022.
\newblock \href {http://arxiv.org/abs/1909.12294} {\path{arXiv:1909.12294}},
  \href {https://doi.org/10.1016/j.aop.2022.168984}
  {\path{doi:10.1016/j.aop.2022.168984}}.

\bibitem{Khachatryan:2018ori}
Vladimir Khachatryan, Bjoern Schenke, Mickey Chiu, Axel Drees, Thomas~K.
  Hemmick, and Norbert Novitzky.
\newblock {Photons from thermalizing matter in heavy ion collisions}.
\newblock {\em Nucl. Phys. A}, 978:123--159, 2018.
\newblock \href {http://arxiv.org/abs/1804.09257} {\path{arXiv:1804.09257}},
  \href {https://doi.org/10.1016/j.nuclphysa.2018.07.013}
  {\path{doi:10.1016/j.nuclphysa.2018.07.013}}.

\bibitem{Coquet:2021gms}
Maurice Coquet, Xiaojian Du, Jean-Yves Ollitrault, Soeren Schlichting, and
  Michael Winn.
\newblock {Transverse mass scaling of dilepton radiation off a quark-gluon
  plasma}.
\newblock {\em Nucl. Phys. A}, 1030:122579, 2023.
\newblock \href {http://arxiv.org/abs/2112.13876} {\path{arXiv:2112.13876}},
  \href {https://doi.org/10.1016/j.nuclphysa.2022.122579}
  {\path{doi:10.1016/j.nuclphysa.2022.122579}}.

\bibitem{Shen:2015qba}
C.~Shen, J.~F. Paquet, G.~S. Denicol, S.~Jeon, and C.~Gale.
\newblock {Thermal photon radiation in high multiplicity p+Pb collisions at the
  Large Hadron Collider}.
\newblock {\em Phys. Rev. Lett.}, 116(7):072301, 2016.
\newblock \href {http://arxiv.org/abs/1504.07989} {\path{arXiv:1504.07989}},
  \href {https://doi.org/10.1103/PhysRevLett.116.072301}
  {\path{doi:10.1103/PhysRevLett.116.072301}}.

\bibitem{Endres:2015fna}
Stephan Endres, Hendrik van Hees, Janus Weil, and Marcus Bleicher.
\newblock {Dilepton production and reaction dynamics in heavy-ion collisions at
  SIS energies from coarse-grained transport simulations}.
\newblock {\em Phys. Rev. C}, 92(1):014911, 2015.
\newblock \href {http://arxiv.org/abs/1505.06131} {\path{arXiv:1505.06131}},
  \href {https://doi.org/10.1103/PhysRevC.92.014911}
  {\path{doi:10.1103/PhysRevC.92.014911}}.

\bibitem{Galatyuk:2015pkq}
Tetyana Galatyuk, Paul~M. Hohler, Ralf Rapp, Florian Seck, and Joachim Stroth.
\newblock {Thermal Dileptons from Coarse-Grained Transport as Fireball Probes
  at SIS Energies}.
\newblock {\em Eur. Phys. J. A}, 52(5):131, 2016.
\newblock \href {http://arxiv.org/abs/1512.08688} {\path{arXiv:1512.08688}},
  \href {https://doi.org/10.1140/epja/i2016-16131-1}
  {\path{doi:10.1140/epja/i2016-16131-1}}.

\bibitem{Staudenmaier:2017vtq}
Jan Staudenmaier, Janus Weil, Vinzent Steinberg, Stephan Endres, and Hannah
  Petersen.
\newblock {Dilepton production and resonance properties within a new hadronic
  transport approach in the context of the GSI-HADES experimental data}.
\newblock {\em Phys. Rev. C}, 98(5):054908, 2018.
\newblock \href {http://arxiv.org/abs/1711.10297} {\path{arXiv:1711.10297}},
  \href {https://doi.org/10.1103/PhysRevC.98.054908}
  {\path{doi:10.1103/PhysRevC.98.054908}}.

\bibitem{HADES:2019auv}
J.~Adamczewski-Musch et~al.
\newblock {Probing dense baryon-rich matter with virtual photons}.
\newblock {\em Nature Phys.}, 15(10):1040--1045, 2019.
\newblock \href {https://doi.org/10.1038/s41567-019-0583-8}
  {\path{doi:10.1038/s41567-019-0583-8}}.

\bibitem{NA60:2008dcb}
R~Arnaldi et~al.
\newblock {Evidence for the production of thermal-like muon pairs with masses
  above 1-GeV/c**2 in 158-A-GeV Indium-Indium Collisions}.
\newblock {\em Eur. Phys. J. C}, 59:607--623, 2009.
\newblock \href {http://arxiv.org/abs/0810.3204} {\path{arXiv:0810.3204}},
  \href {https://doi.org/10.1140/epjc/s10052-008-0857-2}
  {\path{doi:10.1140/epjc/s10052-008-0857-2}}.

\bibitem{Seck:2020qbx}
Florian Seck, Tetyana Galatyuk, Ayon Mukherjee, Ralf Rapp, Jan Steinheimer,
  Joachim Stroth, and Maximilian Wiest.
\newblock {Dilepton signature of a first-order phase transition}.
\newblock {\em Phys. Rev. C}, 106(1):014904, 2022.
\newblock \href {http://arxiv.org/abs/2010.04614} {\path{arXiv:2010.04614}},
  \href {https://doi.org/10.1103/PhysRevC.106.014904}
  {\path{doi:10.1103/PhysRevC.106.014904}}.

\bibitem{Ahdida:2022avx}
C.~Ahdida et~al.
\newblock {Letter of Intent: the NA60+ experiment}.
\newblock 12 2022.
\newblock \href {http://arxiv.org/abs/2212.14452} {\path{arXiv:2212.14452}}.

\bibitem{CBM:2016kpk}
T.~Ablyazimov et~al.
\newblock {Challenges in QCD matter physics --The scientific programme of the
  Compressed Baryonic Matter experiment at FAIR}.
\newblock {\em Eur. Phys. J. A}, 53(3):60, 2017.
\newblock \href {http://arxiv.org/abs/1607.01487} {\path{arXiv:1607.01487}},
  \href {https://doi.org/10.1140/epja/i2017-12248-y}
  {\path{doi:10.1140/epja/i2017-12248-y}}.

\bibitem{ALICE:2022wwr}
{Letter of intent for ALICE 3: A next-generation heavy-ion experiment at the
  LHC}.
\newblock 11 2022.
\newblock \href {http://arxiv.org/abs/2211.02491} {\path{arXiv:2211.02491}}.

\bibitem{Low:1958sn}
F.~E. Low.
\newblock {Bremsstrahlung of very low-energy quanta in elementary particle
  collisions}.
\newblock {\em Phys. Rev.}, 110:974--977, 1958.
\newblock \href {https://doi.org/10.1103/PhysRev.110.974}
  {\path{doi:10.1103/PhysRev.110.974}}.

\bibitem{Lai:2021ckt}
Yue~Shi Lai, James Mulligan, Mateusz P\l{}osko\'n, and Felix Ringer.
\newblock {The information content of jet quenching and machine learning
  assisted observable design}.
\newblock {\em JHEP}, 10:011, 2022.
\newblock \href {http://arxiv.org/abs/2111.14589} {\path{arXiv:2111.14589}},
  \href {https://doi.org/10.1007/JHEP10(2022)011}
  {\path{doi:10.1007/JHEP10(2022)011}}.

\bibitem{Brewer:2021kiv}
Jasmine Brewer, Aleksas Mazeliauskas, and Wilke van~der Schee.
\newblock {Opportunities of OO and $p$O collisions at the LHC}.
\newblock In {\em {Opportunities of OO and pO collisions at the LHC}}, 3 2021.
\newblock \href {http://arxiv.org/abs/2103.01939} {\path{arXiv:2103.01939}}.

\bibitem{Serenone:2021zef}
Willian~Matioli Serenone, Jo\~ao Guilherme~Prado Barbon, David~Dobrigkeit
  Chinellato, Michael~Annan Lisa, Chun Shen, Jun Takahashi, and Giorgio
  Torrieri.
\newblock {\ensuremath{\Lambda} polarization from thermalized jet energy}.
\newblock {\em Phys. Lett. B}, 820:136500, 2021.
\newblock \href {http://arxiv.org/abs/2102.11919} {\path{arXiv:2102.11919}},
  \href {https://doi.org/10.1016/j.physletb.2021.136500}
  {\path{doi:10.1016/j.physletb.2021.136500}}.

\bibitem{Luo:2021voy}
Ao~Luo, Ya-Xian Mao, Guang-You Qin, En-Ke Wang, and Han-Zhong Zhang.
\newblock {Enhancement of baryon-to-meson ratios around jets as a signature of
  medium response}.
\newblock {\em Phys. Lett. B}, 837:137638, 2023.
\newblock \href {http://arxiv.org/abs/2109.14314} {\path{arXiv:2109.14314}},
  \href {https://doi.org/10.1016/j.physletb.2022.137638}
  {\path{doi:10.1016/j.physletb.2022.137638}}.

\bibitem{sPHENIXBUP2022}
{sPHENIX Collaboration}.
\newblock {sPHENIX Beam Use Proposal}, 2022.
\newblock URL:
  \url{{https://indico.bnl.gov/event/15148/attachments/40846/68568/sPHENIX_Beam_Use_Proposal_2022.pdf}}.

\bibitem{Wang:2019xph}
Feng-Tao Wang and Jun Xu.
\newblock {Hadronization using the Wigner function approach for a multiphase
  transport model}.
\newblock {\em Phys. Rev. C}, 100(6):064909, 2019.
\newblock \href {http://arxiv.org/abs/1908.04956} {\path{arXiv:1908.04956}},
  \href {https://doi.org/10.1103/PhysRevC.100.064909}
  {\path{doi:10.1103/PhysRevC.100.064909}}.

\bibitem{Chien:2015ctp}
Yang-Ting Chien, Zhong-Bo Kang, Felix Ringer, Ivan Vitev, and Hongxi Xing.
\newblock {Jet fragmentation functions in proton-proton collisions using
  soft-collinear effective theory}.
\newblock {\em JHEP}, 05:125, 2016.
\newblock \href {http://arxiv.org/abs/1512.06851} {\path{arXiv:1512.06851}},
  \href {https://doi.org/10.1007/JHEP05(2016)125}
  {\path{doi:10.1007/JHEP05(2016)125}}.

\bibitem{Anderle:2017cgl}
Daniele~P. Anderle, Tom Kaufmann, Marco Stratmann, Felix Ringer, and Ivan
  Vitev.
\newblock {Using hadron-in-jet data in a global analysis of $D^{*}$
  fragmentation functions}.
\newblock {\em Phys. Rev. D}, 96(3):034028, 2017.
\newblock \href {http://arxiv.org/abs/1706.09857} {\path{arXiv:1706.09857}},
  \href {https://doi.org/10.1103/PhysRevD.96.034028}
  {\path{doi:10.1103/PhysRevD.96.034028}}.

\bibitem{LHCb:2018qbx}
{LHCb projections for proton-lead collisions during LHC Runs 3 and 4}.
\newblock 11 2018.

\bibitem{CMS:2022cju}
{Snowmass White Paper Contribution: Physics with the Phase-2 ATLAS and CMS
  Detectors}.
\newblock 2022.

\bibitem{LHCb:2018roe}
Roel Aaij et~al.
\newblock {Physics case for an LHCb Upgrade II - Opportunities in flavour
  physics, and beyond, in the HL-LHC era}.
\newblock 8 2018.
\newblock \href {http://arxiv.org/abs/1808.08865} {\path{arXiv:1808.08865}}.

\bibitem{Butler:2019rpu}
Joel~N. Butler and Tommaso Tabarelli~de Fatis.
\newblock {A MIP Timing Detector for the CMS Phase-2 Upgrade}.
\newblock 2019.

\bibitem{Li:2020zbk}
Hai~Tao Li, Ze~Long Liu, and Ivan Vitev.
\newblock {Heavy meson tomography of cold nuclear matter at the electron-ion
  collider}.
\newblock {\em Phys. Lett. B}, 816:136261, 2021.
\newblock \href {http://arxiv.org/abs/2007.10994} {\path{arXiv:2007.10994}},
  \href {https://doi.org/10.1016/j.physletb.2021.136261}
  {\path{doi:10.1016/j.physletb.2021.136261}}.

\bibitem{Li:2021gjw}
Hai~Tao Li, Ze~Long Liu, and Ivan Vitev.
\newblock {Heavy flavor jet production and substructure in electron-nucleus
  collisions}.
\newblock {\em Phys. Lett. B}, 827:137007, 2022.
\newblock \href {http://arxiv.org/abs/2108.07809} {\path{arXiv:2108.07809}},
  \href {https://doi.org/10.1016/j.physletb.2022.137007}
  {\path{doi:10.1016/j.physletb.2022.137007}}.

\bibitem{CMS-PAS-HIN-21-007}
{Observation of the $\Upsilon\textrm{(3S)}$ meson and sequential suppression of
  $\Upsilon$ states in PbPb collisions at
  $\sqrt{\mathrm{s_{NN}}}=5.02~\mathrm{TeV}$}.
\newblock Technical report, CERN, Geneva, 2022.
\newblock URL: \url{https://cds.cern.ch/record/2805926}.

\bibitem{Binder:2021otw}
Tobias Binder, Kyohei Mukaida, Bruno Scheihing-Hitschfeld, and Xiaojun Yao.
\newblock {Non-Abelian electric field correlator at NLO for dark matter relic
  abundance and quarkonium transport}.
\newblock {\em JHEP}, 01:137, 2022.
\newblock \href {http://arxiv.org/abs/2107.03945} {\path{arXiv:2107.03945}},
  \href {https://doi.org/10.1007/JHEP01(2022)137}
  {\path{doi:10.1007/JHEP01(2022)137}}.

\bibitem{SMOG2}
{LHCb Collaboration}.
\newblock {LHCb SMOG Upgrade}.
\newblock {CERN-LHCC-2019-005}.
\newblock URL: \url{https://cds.cern.ch/record/2673690?ln=en}.

\bibitem{CMS_upgrade}
{CMS Collaboration}.
\newblock {Technical Proposal for the Phase-II Upgrade of the CMS Detector}.
\newblock {CERN-LHCC-2015-010}, {2015}.

\bibitem{Brodsky:1980pb}
S.~J. Brodsky, P.~Hoyer, C.~Peterson, and N.~Sakai.
\newblock {The Intrinsic Charm of the Proton}.
\newblock {\em Phys. Lett. B}, 93:451--455, 1980.
\newblock \href {https://doi.org/10.1016/0370-2693(80)90364-0}
  {\path{doi:10.1016/0370-2693(80)90364-0}}.

\bibitem{Brodsky:1981se}
Stanley~J. Brodsky, C.~Peterson, and N.~Sakai.
\newblock {Intrinsic Heavy Quark States}.
\newblock {\em Phys. Rev. D}, 23:2745, 1981.
\newblock \href {https://doi.org/10.1103/PhysRevD.23.2745}
  {\path{doi:10.1103/PhysRevD.23.2745}}.

\bibitem{Hobbs:2013bia}
T.~J. Hobbs, J.~T. Londergan, and W.~Melnitchouk.
\newblock {Phenomenology of nonperturbative charm in the nucleon}.
\newblock {\em Phys. Rev. D}, 89(7):074008, 2014.
\newblock \href {http://arxiv.org/abs/1311.1578} {\path{arXiv:1311.1578}},
  \href {https://doi.org/10.1103/PhysRevD.89.074008}
  {\path{doi:10.1103/PhysRevD.89.074008}}.

\bibitem{Brodsky:1991dj}
Stanley~J. Brodsky, Paul Hoyer, Alfred~H. Mueller, and Wai-Keung Tang.
\newblock {New QCD production mechanisms for hard processes at large x}.
\newblock {\em Nucl. Phys. B}, 369:519--542, 1992.
\newblock \href {https://doi.org/10.1016/0550-3213(92)90278-J}
  {\path{doi:10.1016/0550-3213(92)90278-J}}.

\bibitem{NA3:1983ltt}
J.~Badier et~al.
\newblock {Experimental J/psi Hadronic Production from 150-GeV/c to 280-GeV/c}.
\newblock {\em Z. Phys. C}, 20:101, 1983.
\newblock \href {https://doi.org/10.1007/BF01573213}
  {\path{doi:10.1007/BF01573213}}.

\bibitem{EuropeanMuon:1981obg}
J.~J. Aubert et~al.
\newblock {An Experimental Limit on the Intrinsic Charm Component of the
  Nucleon}.
\newblock {\em Phys. Lett. B}, 110:73--76, 1982.
\newblock \href {https://doi.org/10.1016/0370-2693(82)90955-8}
  {\path{doi:10.1016/0370-2693(82)90955-8}}.

\bibitem{R608:1987dyw}
P.~Chauvat et~al.
\newblock {Production of $\Lambda(c$) With Large X(f) at the {ISR}}.
\newblock {\em Phys. Lett. B}, 199:304, 1987.
\newblock \href {https://doi.org/10.1016/0370-2693(87)91379-7}
  {\path{doi:10.1016/0370-2693(87)91379-7}}.

\bibitem{LHCb:2021stx}
Roel Aaij et~al.
\newblock {Study of Z Bosons Produced in Association with Charm in the Forward
  Region}.
\newblock {\em Phys. Rev. Lett.}, 128(8):082001, 2022.
\newblock \href {http://arxiv.org/abs/2109.08084} {\path{arXiv:2109.08084}},
  \href {https://doi.org/10.1103/PhysRevLett.128.082001}
  {\path{doi:10.1103/PhysRevLett.128.082001}}.

\bibitem{Vogt:1991qd}
R.~Vogt, Stanley~J. Brodsky, and Paul Hoyer.
\newblock {Systematics of J / psi production in nuclear collisions}.
\newblock {\em Nucl. Phys. B}, 360:67--96, 1991.
\newblock \href {https://doi.org/10.1016/0550-3213(91)90435-Z}
  {\path{doi:10.1016/0550-3213(91)90435-Z}}.

\bibitem{Vogt:2021vsc}
R.~Vogt.
\newblock {Limits on Intrinsic Charm Production from the SeaQuest Experiment}.
\newblock {\em Phys. Rev. C}, 103(3):035204, 2021.
\newblock \href {http://arxiv.org/abs/2101.02858} {\path{arXiv:2101.02858}},
  \href {https://doi.org/10.1103/PhysRevC.103.035204}
  {\path{doi:10.1103/PhysRevC.103.035204}}.

\bibitem{Vogt:2022glr}
R.~Vogt.
\newblock {Energy dependence of intrinsic charm production: Determining the
  best energy for observation}.
\newblock {\em Phys. Rev. C}, 106(2):025201, 2022.
\newblock \href {http://arxiv.org/abs/2207.04347} {\path{arXiv:2207.04347}},
  \href {https://doi.org/10.1103/PhysRevC.106.025201}
  {\path{doi:10.1103/PhysRevC.106.025201}}.

\bibitem{Ball:2022qks}
Richard~D. Ball, Alessandro Candido, Juan Cruz-Martinez, Stefano Forte, Tommaso
  Giani, Felix Hekhorn, Kirill Kudashkin, Giacomo Magni, and Juan Rojo.
\newblock {Evidence for intrinsic charm quarks in the proton}.
\newblock {\em Nature}, 608(7923):483--487, 2022.
\newblock \href {http://arxiv.org/abs/2208.08372} {\path{arXiv:2208.08372}},
  \href {https://doi.org/10.1038/s41586-022-04998-2}
  {\path{doi:10.1038/s41586-022-04998-2}}.

\bibitem{Pumplin:2007wg}
J.~Pumplin, H.~L. Lai, and W.~K. Tung.
\newblock {The Charm Parton Content of the Nucleon}.
\newblock {\em Phys. Rev. D}, 75:054029, 2007.
\newblock \href {http://arxiv.org/abs/hep-ph/0701220}
  {\path{arXiv:hep-ph/0701220}}, \href
  {https://doi.org/10.1103/PhysRevD.75.054029}
  {\path{doi:10.1103/PhysRevD.75.054029}}.

\bibitem{Jimenez-Delgado:2014zga}
P.~Jimenez-Delgado, T.~J. Hobbs, J.~T. Londergan, and W.~Melnitchouk.
\newblock {New limits on intrinsic charm in the nucleon from global analysis of
  parton distributions}.
\newblock {\em Phys. Rev. Lett.}, 114(8):082002, 2015.
\newblock \href {http://arxiv.org/abs/1408.1708} {\path{arXiv:1408.1708}},
  \href {https://doi.org/10.1103/PhysRevLett.114.082002}
  {\path{doi:10.1103/PhysRevLett.114.082002}}.

\bibitem{Hou:2017khm}
Tie-Jiun Hou, Sayipjamal Dulat, Jun Gao, Marco Guzzi, Joey Huston, Pavel
  Nadolsky, Carl Schmidt, Jan Winter, Keping Xie, and C.~P. Yuan.
\newblock {CT14 Intrinsic Charm Parton Distribution Functions from CTEQ-TEA
  Global Analysis}.
\newblock {\em JHEP}, 02:059, 2018.
\newblock \href {http://arxiv.org/abs/1707.00657} {\path{arXiv:1707.00657}},
  \href {https://doi.org/10.1007/JHEP02(2018)059}
  {\path{doi:10.1007/JHEP02(2018)059}}.

\bibitem{Guzzi:2022rca}
Marco Guzzi, T.~J. Hobbs, Keping Xie, Joey Huston, Pavel Nadolsky, and C.~P.
  Yuan.
\newblock {The persistent nonperturbative charm enigma}.
\newblock 11 2022.
\newblock \href {http://arxiv.org/abs/2211.01387} {\path{arXiv:2211.01387}}.

\bibitem{SMOG}
Roel Aaij et~al.
\newblock {First Measurement of Charm Production in its Fixed-Target
  Configuration at the LHC}.
\newblock {\em Phys. Rev. Lett.}, 122(13):132002, 2019.
\newblock \href {http://arxiv.org/abs/1810.07907} {\path{arXiv:1810.07907}},
  \href {https://doi.org/10.1103/PhysRevLett.122.132002}
  {\path{doi:10.1103/PhysRevLett.122.132002}}.

\bibitem{NA60p}
Gianluca Usai et~al.
\newblock {Study of hard and electromagnetic processes at CERN-SPS energies: an
  investigation of the high-$\mu_{\mathbf{B}}$ region of the QCD phase diagram
  with NA60+}.
\newblock {\em JPS Conf. Proc.}, 33:011113, 2021.
\newblock \href {http://arxiv.org/abs/1812.07948} {\path{arXiv:1812.07948}},
  \href {https://doi.org/10.7566/JPSCP.33.011113}
  {\path{doi:10.7566/JPSCP.33.011113}}.

\bibitem{Klein:2023zlf}
Spencer~R. Klein.
\newblock {Challenges to the Good-Walker paradigm in coherent and incoherent
  photoproduction}.
\newblock 1 2023.
\newblock \href {http://arxiv.org/abs/2301.01408} {\path{arXiv:2301.01408}}.

\bibitem{Apolinario:2022vzg}
Liliana Apolin\'ario, Yen-Jie Lee, and Michael Winn.
\newblock {Heavy quarks and jets as probes of the QGP}.
\newblock {\em Prog. Part. Nucl. Phys.}, 127:103990, 2022.
\newblock \href {http://arxiv.org/abs/2203.16352} {\path{arXiv:2203.16352}},
  \href {https://doi.org/10.1016/j.ppnp.2022.103990}
  {\path{doi:10.1016/j.ppnp.2022.103990}}.

\bibitem{JETSCAPE:2022ixz}
Raymond Ehlers.
\newblock {Bayesian analysis of QGP jet transport using multi-scale modeling
  applied to inclusive hadron and reconstructed jet data}.
\newblock In {\em {29th International Conference on Ultra-relativistic
  Nucleus-Nucleus Collisions}}, 8 2022.
\newblock \href {http://arxiv.org/abs/2208.07950} {\path{arXiv:2208.07950}}.

\bibitem{Paukkunen:2015bwa}
Hannu Paukkunen.
\newblock {Neutron skin and centrality classification in high-energy heavy-ion
  collisions at the LHC}.
\newblock {\em Phys. Lett. B}, 745:73--78, 2015.
\newblock \href {http://arxiv.org/abs/1503.02448} {\path{arXiv:1503.02448}},
  \href {https://doi.org/10.1016/j.physletb.2015.04.037}
  {\path{doi:10.1016/j.physletb.2015.04.037}}.

\bibitem{Helenius:2016dsk}
Ilkka Helenius, Hannu Paukkunen, and Kari~J. Eskola.
\newblock {Neutron-skin effect in direct-photon and charged hadron-production
  in Pb+Pb collisions at the LHC}.
\newblock {\em Eur. Phys. J. C}, 77(3):148, 2017.
\newblock \href {http://arxiv.org/abs/1606.06910} {\path{arXiv:1606.06910}},
  \href {https://doi.org/10.1140/epjc/s10052-017-4709-9}
  {\path{doi:10.1140/epjc/s10052-017-4709-9}}.

\bibitem{Jonas:2021xju}
Florian Jonas and Constantin Loizides.
\newblock {Centrality dependence of electroweak boson production in PbPb
  collisions at the CERN Large Hadron Collider}.
\newblock {\em Phys. Rev. C}, 104(4):044905, 2021.
\newblock \href {http://arxiv.org/abs/2104.14903} {\path{arXiv:2104.14903}},
  \href {https://doi.org/10.1103/PhysRevC.104.044905}
  {\path{doi:10.1103/PhysRevC.104.044905}}.

\bibitem{Bylinkin:2022temp}
ALICE Collaboration.
\newblock {ALICE Forward Calorimeter (FoCal) Science Proposal}.
\newblock {to be published}, 2023.
\newblock URL: \url{https://sites.google.com/lbl.gov/alice-usa/projects}.

\bibitem{Wang:2021kxm}
Ren-jie Wang, Shi Pu, and Qun Wang.
\newblock {Lepton pair production in ultraperipheral collisions}.
\newblock {\em Phys. Rev. D}, 104(5):056011, 2021.
\newblock \href {http://arxiv.org/abs/2106.05462} {\path{arXiv:2106.05462}},
  \href {https://doi.org/10.1103/PhysRevD.104.056011}
  {\path{doi:10.1103/PhysRevD.104.056011}}.

\bibitem{Klein:2020jom}
Spencer Klein, A.~H. Mueller, Bo-Wen Xiao, and Feng Yuan.
\newblock {Lepton Pair Production Through Two Photon Process in Heavy Ion
  Collisions}.
\newblock {\em Phys. Rev. D}, 102(9):094013, 2020.
\newblock \href {http://arxiv.org/abs/2003.02947} {\path{arXiv:2003.02947}},
  \href {https://doi.org/10.1103/PhysRevD.102.094013}
  {\path{doi:10.1103/PhysRevD.102.094013}}.

\bibitem{CMS:2020skx}
Albert~M Sirunyan et~al.
\newblock {Observation of Forward Neutron Multiplicity Dependence of Dimuon
  Acoplanarity in Ultraperipheral Pb-Pb Collisions at
  $\sqrt{s_{NN}}$=5.02\,\,TeV}.
\newblock {\em Phys. Rev. Lett.}, 127(12):122001, 2021.
\newblock \href {http://arxiv.org/abs/2011.05239} {\path{arXiv:2011.05239}},
  \href {https://doi.org/10.1103/PhysRevLett.127.122001}
  {\path{doi:10.1103/PhysRevLett.127.122001}}.

\bibitem{Sun:2020ygb}
Ze-hao Sun, Du-xin Zheng, Jian Zhou, and Ya-jin Zhou.
\newblock {Studying Coulomb correction at EIC and EicC}.
\newblock {\em Phys. Lett. B}, 808:135679, 2020.
\newblock \href {http://arxiv.org/abs/2002.07373} {\path{arXiv:2002.07373}},
  \href {https://doi.org/10.1016/j.physletb.2020.135679}
  {\path{doi:10.1016/j.physletb.2020.135679}}.

\bibitem{Zha:2021jhf}
Wangmei Zha and Zebo Tang.
\newblock {Discovery of higher-order quantum electrodynamics effect for the
  vacuum pair production}.
\newblock {\em JHEP}, 08:083, 2021.
\newblock \href {http://arxiv.org/abs/2103.04605} {\path{arXiv:2103.04605}},
  \href {https://doi.org/10.1007/JHEP08(2021)083}
  {\path{doi:10.1007/JHEP08(2021)083}}.

\bibitem{ATLAS:2018pfw}
Morad Aaboud et~al.
\newblock {Observation of centrality-dependent acoplanarity for muon pairs
  produced via two-photon scattering in Pb+Pb collisions at
  $\sqrt{s_{\mathrm{NN}}}=5.02$ TeV with the ATLAS detector}.
\newblock {\em Phys. Rev. Lett.}, 121(21):212301, 2018.
\newblock \href {http://arxiv.org/abs/1806.08708} {\path{arXiv:1806.08708}},
  \href {https://doi.org/10.1103/PhysRevLett.121.212301}
  {\path{doi:10.1103/PhysRevLett.121.212301}}.

\bibitem{Klein:2018fmp}
Spencer Klein, A.~H. Mueller, Bo-Wen Xiao, and Feng Yuan.
\newblock {Acoplanarity of a Lepton Pair to Probe the Electromagnetic Property
  of Quark Matter}.
\newblock {\em Phys. Rev. Lett.}, 122(13):132301, 2019.
\newblock \href {http://arxiv.org/abs/1811.05519} {\path{arXiv:1811.05519}},
  \href {https://doi.org/10.1103/PhysRevLett.122.132301}
  {\path{doi:10.1103/PhysRevLett.122.132301}}.

\bibitem{Wang:2021oqq}
Zeyan Wang, Jiaxing Zhao, Carsten Greiner, Zhe Xu, and Pengfei Zhuang.
\newblock {Incomplete electromagnetic response of hot QCD matter}.
\newblock {\em Phys. Rev. C}, 105(4):L041901, 2022.
\newblock \href {http://arxiv.org/abs/2110.14302} {\path{arXiv:2110.14302}},
  \href {https://doi.org/10.1103/PhysRevC.105.L041901}
  {\path{doi:10.1103/PhysRevC.105.L041901}}.

\bibitem{Klusek-Gawenda:2018zfz}
Mariola K\l{}usek-Gawenda, Ralf Rapp, Wolfgang Sch\"afer, and Antoni Szczurek.
\newblock {Dilepton Radiation in Heavy-Ion Collisions at Small Transverse
  Momentum}.
\newblock {\em Phys. Lett. B}, 790:339--344, 2019.
\newblock \href {http://arxiv.org/abs/1809.07049} {\path{arXiv:1809.07049}},
  \href {https://doi.org/10.1016/j.physletb.2019.01.035}
  {\path{doi:10.1016/j.physletb.2019.01.035}}.

\bibitem{Wang:2022ihj}
Xiaofeng Wang, James~Daniel Brandenburg, Lijuan Ruan, Fenglan Shao, Zhangbu Xu,
  Chi Yang, and Wangmei Zha.
\newblock {Energy Dependence of the Breit-Wheeler process in Heavy-Ion
  Collisions and its Application to Nuclear Charge Radius Measurements}.
\newblock 7 2022.
\newblock \href {http://arxiv.org/abs/2207.05595} {\path{arXiv:2207.05595}}.

\bibitem{Budker:2021fts}
Dmitry Budker et~al.
\newblock {Expanding Nuclear Physics Horizons with the Gamma Factory}.
\newblock 6 2021.
\newblock \href {http://arxiv.org/abs/2106.06584} {\path{arXiv:2106.06584}},
  \href {https://doi.org/10.1002/andp.202100284}
  {\path{doi:10.1002/andp.202100284}}.

\bibitem{Xing:2020hwh}
Hongxi Xing, Cheng Zhang, Jian Zhou, and Ya-Jin Zhou.
\newblock {The cos 2$\phi$ azimuthal asymmetry in $\rho^{0}$ meson production
  in ultraperipheral heavy ion collisions}.
\newblock {\em JHEP}, 10:064, 2020.
\newblock \href {http://arxiv.org/abs/2006.06206} {\path{arXiv:2006.06206}},
  \href {https://doi.org/10.1007/JHEP10(2020)064}
  {\path{doi:10.1007/JHEP10(2020)064}}.

\bibitem{Bor:2022fga}
Jelle Bor and Dani\"el Boer.
\newblock {TMD evolution study of the cos2\ensuremath{\phi} azimuthal asymmetry
  in unpolarized J/\ensuremath{\psi} production at EIC}.
\newblock {\em Phys. Rev. D}, 106(1):014030, 2022.
\newblock \href {http://arxiv.org/abs/2204.01527} {\path{arXiv:2204.01527}},
  \href {https://doi.org/10.1103/PhysRevD.106.014030}
  {\path{doi:10.1103/PhysRevD.106.014030}}.

\bibitem{Zha:2018jin}
Wangmei Zha, Lijuan Ruan, Zebo Tang, Zhangbu Xu, and Shuai Yang.
\newblock {Double-slit experiment at fermi scale: coherent photoproduction in
  heavy-ion collisions}.
\newblock {\em Phys. Rev. C}, 99(6):061901, 2019.
\newblock \href {http://arxiv.org/abs/1810.10694} {\path{arXiv:1810.10694}},
  \href {https://doi.org/10.1103/PhysRevC.99.061901}
  {\path{doi:10.1103/PhysRevC.99.061901}}.

\bibitem{Zha:2020cst}
Wangmei Zha, James~Daniel Brandenburg, Lijuan Ruan, Zebo Tang, and Zhangbu Xu.
\newblock {Exploring the double-slit interference with linearly polarized
  photons}.
\newblock {\em Phys. Rev. D}, 103(3):033007, 2021.
\newblock \href {http://arxiv.org/abs/2006.12099} {\path{arXiv:2006.12099}},
  \href {https://doi.org/10.1103/PhysRevD.103.033007}
  {\path{doi:10.1103/PhysRevD.103.033007}}.

\bibitem{Dyndal:2020yen}
Mateusz Dyndal, Mariola Klusek-Gawenda, Matthias Schott, and Antoni Szczurek.
\newblock {Anomalous electromagnetic moments of $\tau$ lepton in $\gamma \gamma
  \to \tau^+ \tau^-$ reaction in Pb+Pb collisions at the LHC}.
\newblock {\em Phys. Lett. B}, 809:135682, 2020.
\newblock \href {http://arxiv.org/abs/2002.05503} {\path{arXiv:2002.05503}},
  \href {https://doi.org/10.1016/j.physletb.2020.135682}
  {\path{doi:10.1016/j.physletb.2020.135682}}.

\bibitem{Xu:2022qme}
Isabel Xu, Nicole Lewis, Xiaofeng Wang, James~Daniel Brandenburg, and Lijuan
  Ruan.
\newblock {Search for Dark Photons in $\gamma\gamma\rightarrow e^+e^-$ at
  RHIC}.
\newblock 11 2022.
\newblock \href {http://arxiv.org/abs/2211.02132} {\path{arXiv:2211.02132}}.

\bibitem{Brandenburg:2022hrp}
James~Daniel Brandenburg, Nicole Lewis, Prithwish Tribedy, and Zhangbu Xu.
\newblock {Search for baryon junctions in photonuclear processes and isobar
  collisions at RHIC}.
\newblock 5 2022.
\newblock \href {http://arxiv.org/abs/2205.05685} {\path{arXiv:2205.05685}}.

\bibitem{Gayoso:2021rzj}
C.~Ayerbe Gayoso et~al.
\newblock {Progress and opportunities in backward angle (u-channel) physics}.
\newblock {\em Eur. Phys. J. A}, 57(12):342, 2021.
\newblock \href {http://arxiv.org/abs/2107.06748} {\path{arXiv:2107.06748}},
  \href {https://doi.org/10.1140/epja/s10050-021-00625-2}
  {\path{doi:10.1140/epja/s10050-021-00625-2}}.

\bibitem{Cebra:2022avc}
Daniel Cebra, Zachary Sweger, Xin Dong, Yuanjing Ji, and Spencer~R. Klein.
\newblock {Backward-angle (u-channel) production at an electron-ion collider}.
\newblock {\em Phys. Rev. C}, 106(1):015204, 2022.
\newblock \href {http://arxiv.org/abs/2204.07915} {\path{arXiv:2204.07915}},
  \href {https://doi.org/10.1103/PhysRevC.106.015204}
  {\path{doi:10.1103/PhysRevC.106.015204}}.

\bibitem{Huang:2019tgz}
Shengli Huang, Zhenyu Chen, Jiangyong Jia, and Wei Li.
\newblock {Disentangling contributions to small-system collectivity via scans
  of light nucleus-nucleus collisions}.
\newblock {\em Phys. Rev. C}, 101(2):021901, 2020.
\newblock \href {http://arxiv.org/abs/1904.10415} {\path{arXiv:1904.10415}},
  \href {https://doi.org/10.1103/PhysRevC.101.021901}
  {\path{doi:10.1103/PhysRevC.101.021901}}.

\bibitem{Zhao:2022ugy}
Wenbin Zhao, Sangwook Ryu, Chun Shen, and Bj\"orn Schenke.
\newblock {3D structure of anisotropic flow in small collision systems at
  energies available at the BNL Relativistic Heavy Ion Collider}.
\newblock {\em Phys. Rev. C}, 107(1):014904, 2023.
\newblock \href {http://arxiv.org/abs/2211.16376} {\path{arXiv:2211.16376}},
  \href {https://doi.org/10.1103/PhysRevC.107.014904}
  {\path{doi:10.1103/PhysRevC.107.014904}}.

\bibitem{Liu:2021izt}
Yu-Fei Liu, Wen-Jing Xing, Xiang-Yu Wu, Guang-You Qin, Shanshan Cao, and Hongxi
  Xing.
\newblock {Heavy and light flavor jet quenching in different collision systems
  at energies available at the CERN Large Hadron Collider}.
\newblock {\em Phys. Rev. C}, 105(4):044904, 2022.
\newblock \href {http://arxiv.org/abs/2107.01522} {\path{arXiv:2107.01522}},
  \href {https://doi.org/10.1103/PhysRevC.105.044904}
  {\path{doi:10.1103/PhysRevC.105.044904}}.

\bibitem{Katz:2019qwv}
Roland Katz, Caio A.~G. Prado, Jacquelyn Noronha-Hostler, and Alexandre A.~P.
  Suaide.
\newblock {System-size scan of $D$ meson $R_{AA}$ and $v_n$ using PbPb, XeXe,
  ArAr, and OO collisions at energies available at the CERN Large Hadron
  Collider}.
\newblock {\em Phys. Rev. C}, 102(4):041901, 2020.
\newblock \href {http://arxiv.org/abs/1907.03308} {\path{arXiv:1907.03308}},
  \href {https://doi.org/10.1103/PhysRevC.102.041901}
  {\path{doi:10.1103/PhysRevC.102.041901}}.

\bibitem{Ke:2022gkq}
Weiyao Ke and Ivan Vitev.
\newblock {Searching for QGP droplets with high-$p_T$ hadrons and heavy
  flavor}.
\newblock 4 2022.
\newblock \href {http://arxiv.org/abs/2204.00634} {\path{arXiv:2204.00634}}.

\bibitem{Lonardoni:2014bwa}
Diego Lonardoni, Alessandro Lovato, Stefano Gandolfi, and Francesco Pederiva.
\newblock {Hyperon Puzzle: Hints from Quantum Monte Carlo Calculations}.
\newblock {\em Phys. Rev. Lett.}, 114(9):092301, 2015.
\newblock \href {http://arxiv.org/abs/1407.4448} {\path{arXiv:1407.4448}},
  \href {https://doi.org/10.1103/PhysRevLett.114.092301}
  {\path{doi:10.1103/PhysRevLett.114.092301}}.

\bibitem{ALICE:2020mfd}
Alice Collaboration et~al.
\newblock {Unveiling the strong interaction among hadrons at the LHC}.
\newblock {\em Nature}, 588:232--238, 2020.
\newblock [Erratum: Nature 590, E13 (2021)].
\newblock \href {http://arxiv.org/abs/2005.11495} {\path{arXiv:2005.11495}},
  \href {https://doi.org/10.1038/s41586-020-3001-6}
  {\path{doi:10.1038/s41586-020-3001-6}}.

\bibitem{NA60:2022sze}
C.~Ahdida et~al.
\newblock {Letter of Intent: the NA60+ experiment}.
\newblock 12 2022.
\newblock \href {http://arxiv.org/abs/2212.14452} {\path{arXiv:2212.14452}}.

\bibitem{NuPECC}
http://www.nupecc.org.
\newblock 2017.

\bibitem{NuPECCLRP2017}
http://nupecc.org/pub/lrp17/lrp2017.pdf.
\newblock 2017.

\bibitem{Galatyuk:2019lcf}
Tetyana Galatyuk.
\newblock {Future facilities for high $\mu_B$ physics}.
\newblock {\em Nucl. Phys. A}, 982:163--169, 2019.
\newblock \href {https://doi.org/10.1016/j.nuclphysa.2018.11.025}
  {\path{doi:10.1016/j.nuclphysa.2018.11.025}}.

\bibitem{Fukushima:2020yzx}
Kenji Fukushima, Bedangadas Mohanty, and Nu~Xu.
\newblock {Little-Bang and Femto-Nova in Nucleus-Nucleus Collisions}.
\newblock {\em AAPPS Bull.}, 31:1, 2021.
\newblock \href {http://arxiv.org/abs/2009.03006} {\path{arXiv:2009.03006}},
  \href {https://doi.org/10.1007/s43673-021-00002-7}
  {\path{doi:10.1007/s43673-021-00002-7}}.

\bibitem{Kharzeev:2022hqz}
Dmitri~E. Kharzeev, Jinfeng Liao, and Shuzhe Shi.
\newblock {Implications of the isobar-run results for the chiral magnetic
  effect in heavy-ion collisions}.
\newblock {\em Phys. Rev. C}, 106(5):L051903, 2022.
\newblock \href {http://arxiv.org/abs/2205.00120} {\path{arXiv:2205.00120}},
  \href {https://doi.org/10.1103/PhysRevC.106.L051903}
  {\path{doi:10.1103/PhysRevC.106.L051903}}.

\bibitem{starbur25}
STAR Collaboration.
\newblock {STAR Beam Use Request 23-25}.
\newblock {BNL PAC}, 2022.
\newblock URL:
  \url{hhttps://indico.bnl.gov/event/15148/attachments/40846/68609/STAR_BUR_Runs23_25___2022\%20\%281\%29.pdf}.

\bibitem{Ikeda:2020agk}
Kazuki Ikeda, Dmitri~E. Kharzeev, and Yuta Kikuchi.
\newblock {Real-time dynamics of Chern-Simons fluctuations near a critical
  point}.
\newblock {\em Phys. Rev. D}, 103(7):L071502, 2021.
\newblock \href {http://arxiv.org/abs/2012.02926} {\path{arXiv:2012.02926}},
  \href {https://doi.org/10.1103/PhysRevD.103.L071502}
  {\path{doi:10.1103/PhysRevD.103.L071502}}.

\bibitem{Cartwright:2021maz}
Casey Cartwright, Matthias Kaminski, and Bjoern Schenke.
\newblock {Energy dependence of the chiral magnetic effect in expanding
  holographic plasma}.
\newblock {\em Phys. Rev. C}, 105(3):034903, 2022.
\newblock \href {http://arxiv.org/abs/2112.13857} {\path{arXiv:2112.13857}},
  \href {https://doi.org/10.1103/PhysRevC.105.034903}
  {\path{doi:10.1103/PhysRevC.105.034903}}.

\bibitem{STAR:2022ahj}
{Search for the Chiral Magnetic Effect in Au+Au collisions at
  $\sqrt{s_{_{{NN}}}}=27$ GeV with the STAR forward Event Plane Detectors}.
\newblock 9 2022.
\newblock \href {http://arxiv.org/abs/2209.03467} {\path{arXiv:2209.03467}}.

\bibitem{Milton:2021wku}
Ryan Milton, Gang Wang, Maria Sergeeva, Shuzhe Shi, Jinfeng Liao, and
  Huan~Zhong Huang.
\newblock {Utilization of event shape in search of the chiral magnetic effect
  in heavy-ion collisions}.
\newblock {\em Phys. Rev. C}, 104(6):064906, 2021.
\newblock \href {http://arxiv.org/abs/2110.01435} {\path{arXiv:2110.01435}},
  \href {https://doi.org/10.1103/PhysRevC.104.064906}
  {\path{doi:10.1103/PhysRevC.104.064906}}.

\bibitem{Feng:2021oub}
Yicheng Feng, Yufu Lin, Jie Zhao, and Fuqiang Wang.
\newblock {Revisit the chiral magnetic effect expectation in isobaric
  collisions at the relativistic heavy ion collider}.
\newblock {\em Phys. Lett. B}, 820:136549, 2021.
\newblock \href {http://arxiv.org/abs/2103.10378} {\path{arXiv:2103.10378}},
  \href {https://doi.org/10.1016/j.physletb.2021.136549}
  {\path{doi:10.1016/j.physletb.2021.136549}}.

\bibitem{STAR:2022hfy}
{Search for the chiral magnetic wave using anisotropic flow of identified
  particles at RHIC}.
\newblock 10 2022.
\newblock \href {http://arxiv.org/abs/2210.14027} {\path{arXiv:2210.14027}}.

\bibitem{Finch:2017xiz}
L.~E. Finch and S.~J. Murray.
\newblock {Investigating local parity violation in heavy-ion collisions using
  \ensuremath{\Lambda} helicity}.
\newblock {\em Phys. Rev. C}, 96(4):044911, 2017.
\newblock \href {http://arxiv.org/abs/1801.06476} {\path{arXiv:1801.06476}},
  \href {https://doi.org/10.1103/PhysRevC.96.044911}
  {\path{doi:10.1103/PhysRevC.96.044911}}.

\bibitem{Jiang:2016woz}
Yin Jiang, Zi-Wei Lin, and Jinfeng Liao.
\newblock {Rotating quark-gluon plasma in relativistic heavy ion collisions}.
\newblock {\em Phys. Rev. C}, 94(4):044910, 2016.
\newblock [Erratum: Phys.Rev.C 95, 049904 (2017)].
\newblock \href {http://arxiv.org/abs/1602.06580} {\path{arXiv:1602.06580}},
  \href {https://doi.org/10.1103/PhysRevC.94.044910}
  {\path{doi:10.1103/PhysRevC.94.044910}}.

\bibitem{Liang:2019pst}
Zuo-Tang Liang, Jun Song, Isaac Upsal, Qun Wang, and Zhang-Bu Xu.
\newblock {Rapidity dependence of global polarization in heavy ion collisions}.
\newblock {\em Chin. Phys. C}, 45(1):014102, 2021.
\newblock \href {http://arxiv.org/abs/1912.10223} {\path{arXiv:1912.10223}},
  \href {https://doi.org/10.1088/1674-1137/abc065}
  {\path{doi:10.1088/1674-1137/abc065}}.

\bibitem{Ivanov:2018eej}
Yu~B. Ivanov and A.~A. Soldatov.
\newblock {Vortex rings in fragmentation regions in heavy-ion collisions at
  $\sqrt{s_{NN}}=$ 39 GeV}.
\newblock {\em Phys. Rev. C}, 97(4):044915, 2018.
\newblock \href {http://arxiv.org/abs/1803.01525} {\path{arXiv:1803.01525}},
  \href {https://doi.org/10.1103/PhysRevC.97.044915}
  {\path{doi:10.1103/PhysRevC.97.044915}}.

\bibitem{Wu:2019eyi}
Hong-Zhong Wu, Long-Gang Pang, Xu-Guang Huang, and Qun Wang.
\newblock {Local spin polarization in high energy heavy ion collisions}.
\newblock {\em Phys. Rev. Research.}, 1:033058, 2019.
\newblock \href {http://arxiv.org/abs/1906.09385} {\path{arXiv:1906.09385}},
  \href {https://doi.org/10.1103/PhysRevResearch.1.033058}
  {\path{doi:10.1103/PhysRevResearch.1.033058}}.

\bibitem{Ivanov:2019ern}
Yu~B. Ivanov, V.~D. Toneev, and A.~A. Soldatov.
\newblock {Estimates of hyperon polarization in heavy-ion collisions at
  collision energies $\sqrt{s_{NN}}=$ 4--40 GeV}.
\newblock {\em Phys. Rev. C}, 100(1):014908, 2019.
\newblock \href {http://arxiv.org/abs/1903.05455} {\path{arXiv:1903.05455}},
  \href {https://doi.org/10.1103/PhysRevC.100.014908}
  {\path{doi:10.1103/PhysRevC.100.014908}}.

\bibitem{Ivanov:2020wak}
Yu.~B. Ivanov and A.~A. Soldatov.
\newblock {Correlation between global polarization, angular momentum, and flow
  in heavy-ion collisions}.
\newblock {\em Phys. Rev. C}, 102(2):024916, 2020.
\newblock \href {http://arxiv.org/abs/2004.05166} {\path{arXiv:2004.05166}},
  \href {https://doi.org/10.1103/PhysRevC.102.024916}
  {\path{doi:10.1103/PhysRevC.102.024916}}.

\bibitem{Deng:2016gyh}
Wei-Tian Deng and Xu-Guang Huang.
\newblock {Vorticity in Heavy-Ion Collisions}.
\newblock {\em Phys. Rev. C}, 93(6):064907, 2016.
\newblock \href {http://arxiv.org/abs/1603.06117} {\path{arXiv:1603.06117}},
  \href {https://doi.org/10.1103/PhysRevC.93.064907}
  {\path{doi:10.1103/PhysRevC.93.064907}}.

\bibitem{Wei:2018zfb}
De-Xian Wei, Wei-Tian Deng, and Xu-Guang Huang.
\newblock {Thermal vorticity and spin polarization in heavy-ion collisions}.
\newblock {\em Phys. Rev. C}, 99(1):014905, 2019.
\newblock \href {http://arxiv.org/abs/1810.00151} {\path{arXiv:1810.00151}},
  \href {https://doi.org/10.1103/PhysRevC.99.014905}
  {\path{doi:10.1103/PhysRevC.99.014905}}.

\bibitem{Xie:2019jun}
Yilong Xie, Dujuan Wang, and Laszlo~Pal Csernai.
\newblock {Fluid dynamics study of the $\varLambda $ polarization for Au + Au
  collisions at $\sqrt{s_{NN}}=200$ GeV}.
\newblock {\em Eur. Phys. J. C}, 80(1):39, 2020.
\newblock \href {http://arxiv.org/abs/1907.00773} {\path{arXiv:1907.00773}},
  \href {https://doi.org/10.1140/epjc/s10052-019-7576-8}
  {\path{doi:10.1140/epjc/s10052-019-7576-8}}.

\bibitem{Baznat:2015eca}
Mircea~I. Baznat, Konstantin~K. Gudima, Alexander~S. Sorin, and O.~V. Teryaev.
\newblock {Femto-vortex sheets and hyperon polarization in heavy-ion
  collisions}.
\newblock {\em Phys. Rev. C}, 93(3):031902, 2016.
\newblock \href {http://arxiv.org/abs/1507.04652} {\path{arXiv:1507.04652}},
  \href {https://doi.org/10.1103/PhysRevC.93.031902}
  {\path{doi:10.1103/PhysRevC.93.031902}}.

\bibitem{Teryaev:2015gxa}
Oleg Teryaev and Rahim Usubov.
\newblock {Vorticity and hydrodynamic helicity in heavy-ion collisions in the
  hadron-string dynamics model}.
\newblock {\em Phys. Rev. C}, 92(1):014906, 2015.
\newblock \href {https://doi.org/10.1103/PhysRevC.92.014906}
  {\path{doi:10.1103/PhysRevC.92.014906}}.

\bibitem{Xia:2018tes}
Xiao-Liang Xia, Hui Li, Ze-Bo Tang, and Qun Wang.
\newblock {Probing vorticity structure in heavy-ion collisions by local
  $\Lambda$ polarization}.
\newblock {\em Phys. Rev. C}, 98:024905, 2018.
\newblock \href {http://arxiv.org/abs/1803.00867} {\path{arXiv:1803.00867}},
  \href {https://doi.org/10.1103/PhysRevC.98.024905}
  {\path{doi:10.1103/PhysRevC.98.024905}}.

\bibitem{Betz:2007kg}
Barbara Betz, Miklos Gyulassy, and Giorgio Torrieri.
\newblock {Polarization probes of vorticity in heavy ion collisions}.
\newblock {\em Phys. Rev. C}, 76:044901, 2007.
\newblock \href {http://arxiv.org/abs/0708.0035} {\path{arXiv:0708.0035}},
  \href {https://doi.org/10.1103/PhysRevC.76.044901}
  {\path{doi:10.1103/PhysRevC.76.044901}}.

\bibitem{Tachibana:2012sa}
Yasuki Tachibana and Tetsufumi Hirano.
\newblock {Emission of Low Momentum Particles at Large Angles from Jet}.
\newblock {\em Nucl. Phys. A}, 904-905:1023c--1026c, 2013.
\newblock \href {http://arxiv.org/abs/1210.5567} {\path{arXiv:1210.5567}},
  \href {https://doi.org/10.1016/j.nuclphysa.2013.02.189}
  {\path{doi:10.1016/j.nuclphysa.2013.02.189}}.

\bibitem{ALICE:2019aid}
Shreyasi Acharya et~al.
\newblock {Evidence of Spin-Orbital Angular Momentum Interactions in
  Relativistic Heavy-Ion Collisions}.
\newblock {\em Phys. Rev. Lett.}, 125(1):012301, 2020.
\newblock \href {http://arxiv.org/abs/1910.14408} {\path{arXiv:1910.14408}},
  \href {https://doi.org/10.1103/PhysRevLett.125.012301}
  {\path{doi:10.1103/PhysRevLett.125.012301}}.

\bibitem{STAR:2022fan}
M.~S. Abdallah et~al.
\newblock {Pattern of global spin alignment of \ensuremath{\phi} and K$^{*0}$
  mesons in heavy-ion collisions}.
\newblock {\em Nature}, 614(7947):244--248, 2023.
\newblock \href {http://arxiv.org/abs/2204.02302} {\path{arXiv:2204.02302}},
  \href {https://doi.org/10.1038/s41586-022-05557-5}
  {\path{doi:10.1038/s41586-022-05557-5}}.

\bibitem{Lisa:2021zkj}
Michael~Annan Lisa, Jo\~ao Guilherme~Prado Barbon, David~Dobrigkeit Chinellato,
  Willian~Matioli Serenone, Chun Shen, Jun Takahashi, and Giorgio Torrieri.
\newblock {Vortex rings from high energy central p+A collisions}.
\newblock {\em Phys. Rev. C}, 104(1):011901, 2021.
\newblock \href {http://arxiv.org/abs/2101.10872} {\path{arXiv:2101.10872}},
  \href {https://doi.org/10.1103/PhysRevC.104.L011901}
  {\path{doi:10.1103/PhysRevC.104.L011901}}.

\bibitem{StarBur2022}
{STAR Collaboration}.
\newblock The {STAR} beam use request for run-23-25 -
  https://drupal.star.bnl.gov/star/starnotes/public/sn0793, 2022.
\newblock URL: \url{https://drupal.star.bnl.gov/STAR/starnotes/public/SN0793}.

\bibitem{BNLPAC22}
{BNL Program Advisory Committee Report}, 2022.
\newblock URL:
  \url{https://www.bnl.gov/npp/docs/2022-npp-pac-recommendations-final.pdf}.

\bibitem{ALICE:2012dtf}
B~Abelev et~al.
\newblock {Upgrade of the ALICE Experiment: Letter Of Intent}.
\newblock {\em J. Phys. G}, 41:087001, 2014.
\newblock \href {https://doi.org/10.1088/0954-3899/41/8/087001}
  {\path{doi:10.1088/0954-3899/41/8/087001}}.

\bibitem{LHCb:2012doh}
I.~Bediaga et~al.
\newblock {Framework TDR for the LHCb Upgrade}: {Technical Design Report}.
\newblock 4 2012.

\bibitem{LHCbCollaboration:2014vzo}
{LHCb Trigger and Online Upgrade Technical Design Report}.
\newblock 5 2014.

\bibitem{ALICE:2020mso}
{Letter of Intent: A Forward Calorimeter (FoCal) in the ALICE experiment}.
\newblock 6 2020.

\bibitem{Alizadehvandchali:2022ier}
N.~Alizadehvandchali et~al.
\newblock {Hot and Cold QCD White Paper from ALICE-USA: Input for 2023 U.S.
  Long Range Plan for Nuclear Science}.
\newblock 12 2022.
\newblock \href {http://arxiv.org/abs/2212.00512} {\path{arXiv:2212.00512}}.

\bibitem{LHCb:2022ine}
{Future physics potential of LHCb}.
\newblock 2022.

\bibitem{Arrington:2021alx}
J.~Arrington et~al.
\newblock {Physics with CEBAF at 12 GeV and future opportunities}.
\newblock {\em Prog. Part. Nucl. Phys.}, 127:103985, 2022.
\newblock \href {http://arxiv.org/abs/2112.00060} {\path{arXiv:2112.00060}},
  \href {https://doi.org/10.1016/j.ppnp.2022.103985}
  {\path{doi:10.1016/j.ppnp.2022.103985}}.

\bibitem{JeffersonLabSoLID:2022iod}
J.~Arrington et~al.
\newblock {The Solenoidal Large Intensity Device (SoLID) for JLab 12 GeV}.
\newblock 9 2022.
\newblock \href {http://arxiv.org/abs/2209.13357} {\path{arXiv:2209.13357}}.

\bibitem{PRad:2020oor}
A.~Gasparian et~al.
\newblock {PRad-II: A New Upgraded High Precision Measurement of the Proton
  Charge Radius}.
\newblock 9 2020.
\newblock \href {http://arxiv.org/abs/2009.10510} {\path{arXiv:2009.10510}}.

\bibitem{Melendez:2020ikd}
J.~A. Melendez, R.~J. Furnstahl, H.~W. Griesshammer, J.~A. McGovern, D.~R.
  Phillips, and M.~T. Pratola.
\newblock {Designing Optimal Experiments: An Application to Proton Compton
  Scattering}.
\newblock {\em Eur. Phys. J. A}, 57(3):81, 2021.
\newblock \href {http://arxiv.org/abs/2004.11307} {\path{arXiv:2004.11307}},
  \href {https://doi.org/10.1140/epja/s10050-021-00382-2}
  {\path{doi:10.1140/epja/s10050-021-00382-2}}.

\bibitem{Pasquini:2021qdi}
Barbara Pasquini and Marc Vanderhaeghen.
\newblock {Virtual Compton scattering at low energies with a positron beam}.
\newblock {\em Eur. Phys. J. A}, 57(11):316, 2021.
\newblock \href {http://arxiv.org/abs/2106.05683} {\path{arXiv:2106.05683}},
  \href {https://doi.org/10.1140/epja/s10050-021-00630-5}
  {\path{doi:10.1140/epja/s10050-021-00630-5}}.

\bibitem{Esser:2020vjb}
A.~Esser et~al.
\newblock {Beam-normal single spin asymmetry in elastic electron scattering off
  $^{28}$Si and $^{90}$Zr}.
\newblock {\em Phys. Lett. B}, 808:135664, 2020.
\newblock \href {http://arxiv.org/abs/2004.14682} {\path{arXiv:2004.14682}},
  \href {https://doi.org/10.1016/j.physletb.2020.135664}
  {\path{doi:10.1016/j.physletb.2020.135664}}.

\bibitem{Rios:2017vsw}
D.~Balaguer R\'\i{}os et~al.
\newblock {New Measurements of the Beam Normal Spin Asymmetries at Large
  Backward Angles with Hydrogen and Deuterium Targets}.
\newblock {\em Phys. Rev. Lett.}, 119(1):012501, 2017.
\newblock \href {https://doi.org/10.1103/PhysRevLett.119.012501}
  {\path{doi:10.1103/PhysRevLett.119.012501}}.

\bibitem{Gou:2020viq}
B.~Gou et~al.
\newblock {Study of Two-Photon Exchange via the Beam Transverse Single Spin
  Asymmetry in Electron-Proton Elastic Scattering at Forward Angles over a Wide
  Energy Range}.
\newblock {\em Phys. Rev. Lett.}, 124(12):122003, 2020.
\newblock \href {http://arxiv.org/abs/2002.06252} {\path{arXiv:2002.06252}},
  \href {https://doi.org/10.1103/PhysRevLett.124.122003}
  {\path{doi:10.1103/PhysRevLett.124.122003}}.

\bibitem{QWeak:2020fih}
D.~Androi\'c et~al.
\newblock {Precision Measurement of the Beam-Normal Single-Spin Asymmetry in
  Forward-Angle Elastic Electron-Proton Scattering}.
\newblock {\em Phys. Rev. Lett.}, 125(11):112502, 2020.
\newblock \href {http://arxiv.org/abs/2006.12435} {\path{arXiv:2006.12435}},
  \href {https://doi.org/10.1103/PhysRevLett.125.112502}
  {\path{doi:10.1103/PhysRevLett.125.112502}}.

\bibitem{QWeak:2021jew}
D.~Androi\'c et~al.
\newblock {Measurement of the Beam-Normal Single-Spin Asymmetry for Elastic
  Electron Scattering from $^{12}$C and $^{27}$Al}.
\newblock {\em Phys. Rev. C}, 104:014606, 2021.
\newblock \href {http://arxiv.org/abs/2103.09758} {\path{arXiv:2103.09758}},
  \href {https://doi.org/10.1103/PhysRevC.104.014606}
  {\path{doi:10.1103/PhysRevC.104.014606}}.

\bibitem{PREX:2021uwt}
D.~Adhikari et~al.
\newblock {New Measurements of the Beam-Normal Single Spin Asymmetry in Elastic
  Electron Scattering over a Range of Spin-0 Nuclei}.
\newblock {\em Phys. Rev. Lett.}, 128(14):142501, 2022.
\newblock \href {http://arxiv.org/abs/2111.04250} {\path{arXiv:2111.04250}},
  \href {https://doi.org/10.1103/PhysRevLett.128.142501}
  {\path{doi:10.1103/PhysRevLett.128.142501}}.

\bibitem{HERMES:2009hsi}
A.~Airapetian et~al.
\newblock {Search for a Two-Photon Exchange Contribution to Inclusive
  Deep-Inelastic Scattering}.
\newblock {\em Phys. Lett. B}, 682:351--354, 2010.
\newblock \href {http://arxiv.org/abs/0907.5369} {\path{arXiv:0907.5369}},
  \href {https://doi.org/10.1016/j.physletb.2009.11.041}
  {\path{doi:10.1016/j.physletb.2009.11.041}}.

\bibitem{Katich:2013atq}
J.~Katich et~al.
\newblock {Measurement of the Target-Normal Single-Spin Asymmetry in
  Deep-Inelastic Scattering from the Reaction
  $^{3}\mathrm{He}^{\uparrow}(e,e')X$}.
\newblock {\em Phys. Rev. Lett.}, 113(2):022502, 2014.
\newblock \href {http://arxiv.org/abs/1311.0197} {\path{arXiv:1311.0197}},
  \href {https://doi.org/10.1103/PhysRevLett.113.022502}
  {\path{doi:10.1103/PhysRevLett.113.022502}}.

\bibitem{Zhang:2015kna}
Y.~W. Zhang et~al.
\newblock {Measurement of the Target-Normal Single-Spin Asymmetry in
  Quasielastic Scattering from the Reaction $^3$He$^\uparrow(e,e^\prime)$}.
\newblock {\em Phys. Rev. Lett.}, 115(17):172502, 2015.
\newblock \href {http://arxiv.org/abs/1502.02636} {\path{arXiv:1502.02636}},
  \href {https://doi.org/10.1103/PhysRevLett.115.172502}
  {\path{doi:10.1103/PhysRevLett.115.172502}}.

\bibitem{CDF:2022hxs}
T.~Aaltonen et~al.
\newblock {High-precision measurement of the $W$ boson mass with the CDF II
  detector}.
\newblock {\em Science}, 376(6589):170--176, 2022.
\newblock \href {https://doi.org/10.1126/science.abk1781}
  {\path{doi:10.1126/science.abk1781}}.

\bibitem{Belitsky:2003nz}
Andrei~V. Belitsky, Xiang-dong Ji, and Feng Yuan.
\newblock {Quark imaging in the proton via quantum phase space distributions}.
\newblock {\em Phys. Rev. D}, 69:074014, 2004.
\newblock \href {http://arxiv.org/abs/hep-ph/0307383}
  {\path{arXiv:hep-ph/0307383}}, \href
  {https://doi.org/10.1103/PhysRevD.69.074014}
  {\path{doi:10.1103/PhysRevD.69.074014}}.

\bibitem{Moutarde:2019tqa}
H.~Moutarde, P.~Sznajder, and J.~Wagner.
\newblock {Unbiased determination of DVCS Compton Form Factors}.
\newblock {\em Eur. Phys. J. C}, 79(7):614, 2019.
\newblock \href {http://arxiv.org/abs/1905.02089} {\path{arXiv:1905.02089}},
  \href {https://doi.org/10.1140/epjc/s10052-019-7117-5}
  {\path{doi:10.1140/epjc/s10052-019-7117-5}}.

\bibitem{JeffersonLabHallA:2015dwe}
M.~Defurne et~al.
\newblock {E00-110 experiment at Jefferson Lab Hall A: Deeply virtual Compton
  scattering off the proton at 6 GeV}.
\newblock {\em Phys. Rev. C}, 92(5):055202, 2015.
\newblock \href {http://arxiv.org/abs/1504.05453} {\path{arXiv:1504.05453}},
  \href {https://doi.org/10.1103/PhysRevC.92.055202}
  {\path{doi:10.1103/PhysRevC.92.055202}}.

\bibitem{HERMES:2004mhh}
A.~Airapetian et~al.
\newblock {Single-spin asymmetries in semi-inclusive deep-inelastic scattering
  on a transversely polarized hydrogen target}.
\newblock {\em Phys. Rev. Lett.}, 94:012002, 2005.
\newblock \href {http://arxiv.org/abs/hep-ex/0408013}
  {\path{arXiv:hep-ex/0408013}}, \href
  {https://doi.org/10.1103/PhysRevLett.94.012002}
  {\path{doi:10.1103/PhysRevLett.94.012002}}.

\bibitem{COMPASS:2014kcy}
C.~Adolph et~al.
\newblock {Measurement of azimuthal hadron asymmetries in semi-inclusive deep
  inelastic scattering off unpolarised nucleons}.
\newblock {\em Nucl. Phys. B}, 886:1046--1077, 2014.
\newblock \href {http://arxiv.org/abs/1401.6284} {\path{arXiv:1401.6284}},
  \href {https://doi.org/10.1016/j.nuclphysb.2014.07.019}
  {\path{doi:10.1016/j.nuclphysb.2014.07.019}}.

\bibitem{JeffersonLabHallA:2011ayy}
X.~Qian et~al.
\newblock {Single Spin Asymmetries in Charged Pion Production from
  Semi-Inclusive Deep Inelastic Scattering on a Transversely Polarized $^3$He
  Target}.
\newblock {\em Phys. Rev. Lett.}, 107:072003, 2011.
\newblock \href {http://arxiv.org/abs/1106.0363} {\path{arXiv:1106.0363}},
  \href {https://doi.org/10.1103/PhysRevLett.107.072003}
  {\path{doi:10.1103/PhysRevLett.107.072003}}.

\bibitem{Ye:2016prn}
Zhihong Ye, Nobuo Sato, Kalyan Allada, Tianbo Liu, Jian-Ping Chen, Haiyan Gao,
  Zhong-Bo Kang, Alexei Prokudin, Peng Sun, and Feng Yuan.
\newblock {Unveiling the nucleon tensor charge at Jefferson Lab: A study of the
  SoLID case}.
\newblock {\em Phys. Lett. B}, 767:91--98, 2017.
\newblock \href {http://arxiv.org/abs/1609.02449} {\path{arXiv:1609.02449}},
  \href {https://doi.org/10.1016/j.physletb.2017.01.046}
  {\path{doi:10.1016/j.physletb.2017.01.046}}.

\bibitem{Aguilar:2019teb}
Arlene~C. Aguilar et~al.
\newblock {Pion and Kaon Structure at the Electron-Ion Collider}.
\newblock {\em Eur. Phys. J. A}, 55(10):190, 2019.
\newblock \href {http://arxiv.org/abs/1907.08218} {\path{arXiv:1907.08218}},
  \href {https://doi.org/10.1140/epja/i2019-12885-0}
  {\path{doi:10.1140/epja/i2019-12885-0}}.

\bibitem{Arrington:2021biu}
J.~Arrington et~al.
\newblock {Revealing the structure of light pseudoscalar mesons at the
  electron\textendash{}ion collider}.
\newblock {\em J. Phys. G}, 48(7):075106, 2021.
\newblock \href {http://arxiv.org/abs/2102.11788} {\path{arXiv:2102.11788}},
  \href {https://doi.org/10.1088/1361-6471/abf5c3}
  {\path{doi:10.1088/1361-6471/abf5c3}}.

\bibitem{Carmignotto:2018uqj}
M.~Carmignotto et~al.
\newblock {Separated Kaon Electroproduction Cross Section and the Kaon Form
  Factor from 6 GeV JLab Data}.
\newblock {\em Phys. Rev. C}, 97(2):025204, 2018.
\newblock \href {http://arxiv.org/abs/1801.01536} {\path{arXiv:1801.01536}},
  \href {https://doi.org/10.1103/PhysRevC.97.025204}
  {\path{doi:10.1103/PhysRevC.97.025204}}.

\bibitem{JeffersonLabFpi-2:2006ysh}
T.~Horn et~al.
\newblock {Determination of the Charged Pion Form Factor at Q**2 = 1.60 and
  2.45-(GeV/c)**2}.
\newblock {\em Phys. Rev. Lett.}, 97:192001, 2006.
\newblock \href {http://arxiv.org/abs/nucl-ex/0607005}
  {\path{arXiv:nucl-ex/0607005}}, \href
  {https://doi.org/10.1103/PhysRevLett.97.192001}
  {\path{doi:10.1103/PhysRevLett.97.192001}}.

\bibitem{Horn:2007ug}
T.~Horn et~al.
\newblock {Scaling study of the pion electroproduction cross sections and the
  pion form factor}.
\newblock {\em Phys. Rev. C}, 78:058201, 2008.
\newblock \href {http://arxiv.org/abs/0707.1794} {\path{arXiv:0707.1794}},
  \href {https://doi.org/10.1103/PhysRevC.78.058201}
  {\path{doi:10.1103/PhysRevC.78.058201}}.

\bibitem{JeffersonLab:2008jve}
G.~M. Huber et~al.
\newblock {Charged pion form-factor between Q**2 = 0.60-GeV**2 and 2.45-GeV**2.
  II. Determination of, and results for, the pion form-factor}.
\newblock {\em Phys. Rev. C}, 78:045203, 2008.
\newblock \href {http://arxiv.org/abs/0809.3052} {\path{arXiv:0809.3052}},
  \href {https://doi.org/10.1103/PhysRevC.78.045203}
  {\path{doi:10.1103/PhysRevC.78.045203}}.

\bibitem{Horn:2012zza}
T.~Horn.
\newblock {Global analysis of exclusive kaon and pion electroproduction}.
\newblock {\em Phys. Rev. C}, 85:018202, 2012.
\newblock \href {https://doi.org/10.1103/PhysRevC.85.018202}
  {\path{doi:10.1103/PhysRevC.85.018202}}.

\bibitem{JeffersonLabFpi:2014ykr}
G.~M. Huber et~al.
\newblock {Separated Response Function Ratios in Exclusive, Forward $\pi^{\pm}$
  Electroproduction}.
\newblock {\em Phys. Rev. Lett.}, 112(18):182501, 2014.
\newblock \href {http://arxiv.org/abs/1404.3985} {\path{arXiv:1404.3985}},
  \href {https://doi.org/10.1103/PhysRevLett.112.182501}
  {\path{doi:10.1103/PhysRevLett.112.182501}}.

\bibitem{JeffersonLabFpi:2014tbe}
G.~M. Huber et~al.
\newblock {Separated Response Functions in Exclusive, Forward $\pi^{\pm}$
  Electroproduction on Deuterium}.
\newblock {\em Phys. Rev. C}, 91(1):015202, 2015.
\newblock \href {http://arxiv.org/abs/1412.5140} {\path{arXiv:1412.5140}},
  \href {https://doi.org/10.1103/PhysRevC.91.015202}
  {\path{doi:10.1103/PhysRevC.91.015202}}.

\bibitem{Benesch:2022xmb}
J.~Benesch et~al.
\newblock {Jefferson Lab Hall C: Precision Physics at the Luminosity Frontier}.
\newblock 9 2022.
\newblock \href {http://arxiv.org/abs/2209.11838} {\path{arXiv:2209.11838}}.

\bibitem{Godfrey:1985xj}
S.~Godfrey and Nathan Isgur.
\newblock {Mesons in a Relativized Quark Model with Chromodynamics}.
\newblock {\em Phys. Rev. D}, 32:189--231, 1985.
\newblock \href {https://doi.org/10.1103/PhysRevD.32.189}
  {\path{doi:10.1103/PhysRevD.32.189}}.

\bibitem{Capstick:1986ter}
Simon Capstick and Nathan Isgur.
\newblock {Baryons in a relativized quark model with chromodynamics}.
\newblock {\em Phys. Rev. D}, 34(9):2809--2835, 1986.
\newblock \href {https://doi.org/10.1103/physrevd.34.2809}
  {\path{doi:10.1103/physrevd.34.2809}}.

\bibitem{Proceedings:2020fyd}
S.~J. Brodsky et~al.
\newblock {Strong QCD from Hadron Structure Experiments}: {Newport News, VA,
  USA, November 4-8, 2019}.
\newblock {\em Int. J. Mod. Phys. E}, 29(08):2030006, 2020.
\newblock \href {http://arxiv.org/abs/2006.06802} {\path{arXiv:2006.06802}},
  \href {https://doi.org/10.1142/S0218301320300064}
  {\path{doi:10.1142/S0218301320300064}}.

\bibitem{Aznauryan:2011qj}
I.~G. Aznauryan and V.~D. Burkert.
\newblock {Electroexcitation of nucleon resonances}.
\newblock {\em Prog. Part. Nucl. Phys.}, 67:1--54, 2012.
\newblock \href {http://arxiv.org/abs/1109.1720} {\path{arXiv:1109.1720}},
  \href {https://doi.org/10.1016/j.ppnp.2011.08.001}
  {\path{doi:10.1016/j.ppnp.2011.08.001}}.

\bibitem{Carman:2020qmb}
D.S. Carman, K.~Joo, and V.I. Mokeev.
\newblock {Strong QCD Insights from Excited Nucleon Structure Studies with CLAS
  and CLAS12}.
\newblock {\em Few Body Syst.}, 61(3):29, 2020.

\bibitem{Mokeev:2022xfo}
V.I. Mokeev and D.S. Carman.
\newblock {Photo- and Electrocouplings of Nucleon Resonances}.
\newblock {\em Few Body Syst.}, 63(3):59, 2022.

\bibitem{EuropeanMuon:1983wih}
J.~J. Aubert et~al.
\newblock {The ratio of the nucleon structure functions $F2_n$ for iron and
  deuterium}.
\newblock {\em Phys. Lett. B}, 123:275--278, 1983.
\newblock \href {https://doi.org/10.1016/0370-2693(83)90437-9}
  {\path{doi:10.1016/0370-2693(83)90437-9}}.

\bibitem{sPHENIXBUP}
sPHENIX Collaboration.
\newblock {sPHENIX Beam Use Proposal}.
\newblock {BNL PAC}, 2022.
\newblock URL:
  \url{https://indico.bnl.gov/event/15148/attachments/40846/68568/sPHENIX_Beam_Use_Proposal_2022.pdf}.

\bibitem{LHCb:2014vhh}
Roel Aaij et~al.
\newblock {Precision luminosity measurements at LHCb}.
\newblock {\em JINST}, 9(12):P12005, 2014.
\newblock \href {http://arxiv.org/abs/1410.0149} {\path{arXiv:1410.0149}},
  \href {https://doi.org/10.1088/1748-0221/9/12/P12005}
  {\path{doi:10.1088/1748-0221/9/12/P12005}}.

\bibitem{LHCb:2018jry}
Roel Aaij et~al.
\newblock {First Measurement of Charm Production in its Fixed-Target
  Configuration at the LHC}.
\newblock {\em Phys. Rev. Lett.}, 122(13):132002, 2019.
\newblock \href {http://arxiv.org/abs/1810.07907} {\path{arXiv:1810.07907}},
  \href {https://doi.org/10.1103/PhysRevLett.122.132002}
  {\path{doi:10.1103/PhysRevLett.122.132002}}.

\bibitem{LHCb:2018ygc}
Roel Aaij et~al.
\newblock {Measurement of Antiproton Production in ${\rm p He}$ Collisions at
  $\sqrt{s_{NN}}=110$ GeV}.
\newblock {\em Phys. Rev. Lett.}, 121(22):222001, 2018.
\newblock \href {http://arxiv.org/abs/1808.06127} {\path{arXiv:1808.06127}},
  \href {https://doi.org/10.1103/PhysRevLett.121.222001}
  {\path{doi:10.1103/PhysRevLett.121.222001}}.

\bibitem{LHCb:2022lnf}
{Open charm production and asymmetry in $p$Ne collisions at
  $\sqrt{s_{\scriptscriptstyle\rm NN}} =$ 68.5 GeV}.
\newblock 11 2022.
\newblock \href {http://arxiv.org/abs/2211.11633} {\path{arXiv:2211.11633}}.

\bibitem{LHCb:2022tum}
{Charmonium production in $p$Ne collisions at $\sqrt{s_{\rm NN}}=68.5$ GeV}.
\newblock 11 2022.
\newblock \href {http://arxiv.org/abs/2211.11645} {\path{arXiv:2211.11645}}.

\bibitem{Loizides:2020tey}
Constantin Loizides.
\newblock {\textquotedblleft{}QM19 summary talk\textquotedblright{}: Outlook
  and future of heavy-ion collisions}.
\newblock {\em Nucl. Phys. A}, 1005:121964, 2021.
\newblock \href {http://arxiv.org/abs/2007.00710} {\path{arXiv:2007.00710}},
  \href {https://doi.org/10.1016/j.nuclphysa.2020.121964}
  {\path{doi:10.1016/j.nuclphysa.2020.121964}}.

\bibitem{Aidala:2019pit}
C.~A. Aidala et~al.
\newblock {The LHCSpin Project}.
\newblock 1 2019.
\newblock \href {http://arxiv.org/abs/1901.08002} {\path{arXiv:1901.08002}}.

\bibitem{Santimaria:2022tss}
M.~Santimaria, V.~Carassiti, G.~Ciullo, P.~Di~Nezza, P.~Lenisa, S.~Mariani,
  L.~L. Pappalardo, and E.~Steffens.
\newblock {The LHCspin project}.
\newblock In {\em {20th International Conference on Strangeness in Quark Matter
  2022}}, 10 2022.
\newblock \href {http://arxiv.org/abs/2210.13997} {\path{arXiv:2210.13997}}.

\bibitem{Puckett:2017flj}
A.~J.~R. Puckett et~al.
\newblock {Polarization Transfer Observables in Elastic Electron Proton
  Scattering at $Q^2 = $2.5, 5.2, 6.8, and 8.5 GeV$^2$}.
\newblock {\em Phys. Rev. C}, 96(5):055203, 2017.
\newblock [Erratum: Phys.Rev.C 98, 019907 (2018)].
\newblock \href {http://arxiv.org/abs/1707.08587} {\path{arXiv:1707.08587}},
  \href {https://doi.org/10.1103/PhysRevC.96.055203}
  {\path{doi:10.1103/PhysRevC.96.055203}}.

\bibitem{PEPPo:2016saj}
D.~Abbott et~al.
\newblock {Production of Highly Polarized Positrons Using Polarized Electrons
  at MeV Energies}.
\newblock {\em Phys. Rev. Lett.}, 116(21):214801, 2016.
\newblock \href {http://arxiv.org/abs/1606.08877} {\path{arXiv:1606.08877}},
  \href {https://doi.org/10.1103/PhysRevLett.116.214801}
  {\path{doi:10.1103/PhysRevLett.116.214801}}.

\bibitem{Habet:2022fch}
Sami Habet et~al.
\newblock {Concept of a Polarized Positron Source for CEBAF}.
\newblock {\em JACoW}, IPAC2022:457--460, 2022.
\newblock \href {https://doi.org/10.18429/JACoW-IPAC2022-MOPOTK012}
  {\path{doi:10.18429/JACoW-IPAC2022-MOPOTK012}}.

\bibitem{Talman:2018voj}
Richard Talman, B.~L. Roberts, J.~Grames, A.~Hofler, R.~Kazimi, M.~Poelker, and
  R.~Suleiman.
\newblock {Resonant (Longitudinal and Transverse) Electron Polarimetry}.
\newblock {\em PoS}, PSTP2017:028, 2018.

\bibitem{Bartnik:2020pos}
A.~Bartnik et~al.
\newblock {CBETA: First Multipass Superconducting Linear Accelerator with
  Energy Recovery}.
\newblock {\em Phys. Rev. Lett.}, 125(4):044803, 2020.
\newblock \href {https://doi.org/10.1103/PhysRevLett.125.044803}
  {\path{doi:10.1103/PhysRevLett.125.044803}}.

\bibitem{Gasser:2020hzn}
J.~Gasser, H.~Leutwyler, and A.~Rusetsky.
\newblock {Sum rule for the Compton amplitude and implications for the
  proton\textendash{}neutron mass difference}.
\newblock {\em Eur. Phys. J. C}, 80(12):1121, 2020.
\newblock \href {http://arxiv.org/abs/2008.05806} {\path{arXiv:2008.05806}},
  \href {https://doi.org/10.1140/epjc/s10052-020-08615-2}
  {\path{doi:10.1140/epjc/s10052-020-08615-2}}.

\bibitem{MUSE:2017dod}
R.~Gilman et~al.
\newblock {Technical Design Report for the Paul Scherrer Institute Experiment
  R-12-01.1: Studying the Proton ''Radius'' Puzzle with $\mu p$ Elastic
  Scattering}.
\newblock 9 2017.
\newblock \href {http://arxiv.org/abs/1709.09753} {\path{arXiv:1709.09753}}.

\bibitem{Cline:2021vlw}
E.~Cline et~al.
\newblock {Characterization of muon and electron beams in the Paul Scherrer
  Institute PiM1 channel for the MUSE experiment}.
\newblock {\em Phys. Rev. C}, 105(5):055201, 2022.
\newblock \href {http://arxiv.org/abs/2109.09508} {\path{arXiv:2109.09508}},
  \href {https://doi.org/10.1103/PhysRevC.105.055201}
  {\path{doi:10.1103/PhysRevC.105.055201}}.

\bibitem{Belle-II:2018jsg}
W.~Altmannshofer et~al.
\newblock {The Belle II Physics Book}.
\newblock {\em PTEP}, 2019(12):123C01, 2019.
\newblock [Erratum: PTEP 2020, 029201 (2020)].
\newblock \href {http://arxiv.org/abs/1808.10567} {\path{arXiv:1808.10567}},
  \href {https://doi.org/10.1093/ptep/ptz106} {\path{doi:10.1093/ptep/ptz106}}.

\bibitem{Accardi:2022oog}
A.~Accardi et~al.
\newblock {Opportunities for precision QCD physics in hadronization at Belle II
  -- a snowmass whitepaper}.
\newblock In {\em {2022 Snowmass Summer Study}}, 4 2022.
\newblock \href {http://arxiv.org/abs/2204.02280} {\path{arXiv:2204.02280}}.

\bibitem{Adams:2018pwt}
B.~Adams et~al.
\newblock {Letter of Intent: A New QCD facility at the M2 beam line of the CERN
  SPS (COMPASS++/AMBER)}.
\newblock 8 2018.
\newblock \href {http://arxiv.org/abs/1808.00848} {\path{arXiv:1808.00848}}.

\bibitem{Symons:2002}
R.~Alarcon et~al.
\newblock {Opportunities in Nuclear Science: A Long-Range Plan for the Next
  Decade.}
\newblock {DOE/NSF Nuclear Science Advisory Panel Report}, 2002.
\newblock URL:
  \url{https://science.osti.gov/np/nsac/Reports/Reports-Archive#2002}.

\bibitem{Tribble:2007}
D.~Bryman et~al.
\newblock {The Frontiers of Nuclear Science, A Long Range Plan}.
\newblock {DOE/NSF Nuclear Science Advisory Panel Report}, 2007.
\newblock URL:
  \url{https://science.osti.gov/np/nsac/Reports/Reports-Archive#2007}.

\bibitem{EIC-WP-2007LRP}
{A High Luminosity, High Energy Electron-Ion-Collider : A new experimental
  quest to study the glue that binds us all }.
\newblock \url{http://web.mit.edu/eicc/DOCUMENTS/EIC_LRP-20070424.pdf}, 2007.

\bibitem{Boer:2011fh}
Daniel Boer et~al.
\newblock {Gluons and the quark sea at high energies: Distributions,
  polarization, tomography}.
\newblock 8 2011.
\newblock \href {http://arxiv.org/abs/1108.1713} {\path{arXiv:1108.1713}}.

\bibitem{Accardi:2012qut}
A.~Accardi et~al.
\newblock {Electron Ion Collider: The Next QCD Frontier}: {Understanding the
  glue that binds us all}.
\newblock {\em Eur. Phys. J. A}, 52(9):268, 2016.
\newblock \href {http://arxiv.org/abs/1212.1701} {\path{arXiv:1212.1701}},
  \href {https://doi.org/10.1140/epja/i2016-16268-9}
  {\path{doi:10.1140/epja/i2016-16268-9}}.

\bibitem{NAP25171}
{National Academies of Sciences, Engineering, and Medicine}.
\newblock {\em An Assessment of U.S.-Based Electron-Ion Collider Science}.
\newblock The National Academies Press, Washington, DC, 2018.
\newblock \href {https://doi.org/10.17226/25171} {\path{doi:10.17226/25171}}.

\bibitem{EIC-WP-2023LRP}
EIC~User Group.
\newblock {White Paper on the Electron-Ion Collider in Preparation for the NSAC
  Long Range Plan}.
\newblock \url{https://zenodo.org/record/7500024#.Y_41WBzMJkg}, 2023.

\bibitem{Borsa:2020lsz}
Ignacio Borsa, Gonzalo Lucero, Rodolfo Sassot, Elke~C. Aschenauer, and Ana~S.
  Nunes.
\newblock {Revisiting helicity parton distributions at a future electron-ion
  collider}.
\newblock {\em Phys. Rev. D}, 102(9):094018, 2020.
\newblock \href {http://arxiv.org/abs/2007.08300} {\path{arXiv:2007.08300}},
  \href {https://doi.org/10.1103/PhysRevD.102.094018}
  {\path{doi:10.1103/PhysRevD.102.094018}}.

\bibitem{deFlorian:2019zkl}
Daniel De~Florian, Gonzalo~Agust\'\i{}n Lucero, Rodolfo Sassot, Marco
  Stratmann, and Werner Vogelsang.
\newblock {Monte Carlo sampling variant of the DSSV14 set of helicity parton
  densities}.
\newblock {\em Phys. Rev. D}, 100(11):114027, 2019.
\newblock \href {http://arxiv.org/abs/1902.10548} {\path{arXiv:1902.10548}},
  \href {https://doi.org/10.1103/PhysRevD.100.114027}
  {\path{doi:10.1103/PhysRevD.100.114027}}.

\bibitem{Adamiak:2021ppq}
Daniel Adamiak, Yuri~V. Kovchegov, W.~Melnitchouk, Daniel Pitonyak, Nobuo Sato,
  and Matthew~D. Sievert.
\newblock {First analysis of world polarized DIS data with small-x helicity
  evolution}.
\newblock {\em Phys. Rev. D}, 104(3):L031501, 2021.
\newblock \href {http://arxiv.org/abs/2102.06159} {\path{arXiv:2102.06159}},
  \href {https://doi.org/10.1103/PhysRevD.104.L031501}
  {\path{doi:10.1103/PhysRevD.104.L031501}}.

\bibitem{Bartels:1996wc}
Jochen Bartels, B.~I. Ermolaev, and M.~G. Ryskin.
\newblock {Flavor singlet contribution to the structure function G(1) at small
  x}.
\newblock {\em Z. Phys. C}, 72:627--635, 1996.
\newblock \href {http://arxiv.org/abs/hep-ph/9603204}
  {\path{arXiv:hep-ph/9603204}}, \href {https://doi.org/10.1007/BF02909194}
  {\path{doi:10.1007/BF02909194}}.

\bibitem{Kovchegov:2015pbl}
Yuri~V. Kovchegov, Daniel Pitonyak, and Matthew~D. Sievert.
\newblock {Helicity Evolution at Small-x}.
\newblock {\em JHEP}, 01:072, 2016.
\newblock [Erratum: JHEP 10, 148 (2016)].
\newblock \href {http://arxiv.org/abs/1511.06737} {\path{arXiv:1511.06737}},
  \href {https://doi.org/10.1007/JHEP01(2016)072}
  {\path{doi:10.1007/JHEP01(2016)072}}.

\bibitem{Hatta:2016aoc}
Yoshitaka Hatta, Yuya Nakagawa, Feng Yuan, Yong Zhao, and Bowen Xiao.
\newblock {Gluon orbital angular momentum at small-$x$}.
\newblock {\em Phys. Rev. D}, 95(11):114032, 2017.
\newblock \href {http://arxiv.org/abs/1612.02445} {\path{arXiv:1612.02445}},
  \href {https://doi.org/10.1103/PhysRevD.95.114032}
  {\path{doi:10.1103/PhysRevD.95.114032}}.

\bibitem{Boussarie:2019icw}
Renaud Boussarie, Yoshitaka Hatta, and Feng Yuan.
\newblock {Proton Spin Structure at Small-$x$}.
\newblock {\em Phys. Lett. B}, 797:134817, 2019.
\newblock \href {http://arxiv.org/abs/1904.02693} {\path{arXiv:1904.02693}},
  \href {https://doi.org/10.1016/j.physletb.2019.134817}
  {\path{doi:10.1016/j.physletb.2019.134817}}.

\bibitem{Chirilli:2021lif}
Giovanni~Antonio Chirilli.
\newblock {High-energy operator product expansion at sub-eikonal level}.
\newblock {\em JHEP}, 06:096, 2021.
\newblock \href {http://arxiv.org/abs/2101.12744} {\path{arXiv:2101.12744}},
  \href {https://doi.org/10.1007/JHEP06(2021)096}
  {\path{doi:10.1007/JHEP06(2021)096}}.

\bibitem{Cougoulic:2022gbk}
Florian Cougoulic, Yuri~V. Kovchegov, Andrey Tarasov, and Yossathorn Tawabutr.
\newblock {Quark and gluon helicity evolution at small x: revised and updated}.
\newblock {\em JHEP}, 07:095, 2022.
\newblock \href {http://arxiv.org/abs/2204.11898} {\path{arXiv:2204.11898}},
  \href {https://doi.org/10.1007/JHEP07(2022)095}
  {\path{doi:10.1007/JHEP07(2022)095}}.

\bibitem{Ji:2016jgn}
Xiangdong Ji, Feng Yuan, and Yong Zhao.
\newblock {Hunting the Gluon Orbital Angular Momentum at the Electron-Ion
  Collider}.
\newblock {\em Phys. Rev. Lett.}, 118(19):192004, 2017.
\newblock \href {http://arxiv.org/abs/1612.02438} {\path{arXiv:1612.02438}},
  \href {https://doi.org/10.1103/PhysRevLett.118.192004}
  {\path{doi:10.1103/PhysRevLett.118.192004}}.

\bibitem{Bhattacharya:2022vvo}
Shohini Bhattacharya, Renaud Boussarie, and Yoshitaka Hatta.
\newblock {Signature of the Gluon Orbital Angular Momentum}.
\newblock {\em Phys. Rev. Lett.}, 128(18):182002, 2022.
\newblock \href {http://arxiv.org/abs/2201.08709} {\path{arXiv:2201.08709}},
  \href {https://doi.org/10.1103/PhysRevLett.128.182002}
  {\path{doi:10.1103/PhysRevLett.128.182002}}.

\bibitem{Gribov:1972ri}
V.~N. Gribov and L.~N. Lipatov.
\newblock {Deep inelastic e p scattering in perturbation theory}.
\newblock {\em Sov. J. Nucl. Phys.}, 15:438--450, 1972.

\bibitem{Altarelli:1977zs}
Guido Altarelli and G.~Parisi.
\newblock {Asymptotic Freedom in Parton Language}.
\newblock {\em Nucl. Phys. B}, 126:298--318, 1977.
\newblock \href {https://doi.org/10.1016/0550-3213(77)90384-4}
  {\path{doi:10.1016/0550-3213(77)90384-4}}.

\bibitem{Dokshitzer:1977sg}
Yuri~L. Dokshitzer.
\newblock {Calculation of the Structure Functions for Deep Inelastic Scattering
  and e+ e- Annihilation by Perturbation Theory in Quantum Chromodynamics.}
\newblock {\em Sov. Phys. JETP}, 46:641--653, 1977.

\bibitem{Kuraev:1977fs}
E.~A. Kuraev, L.~N. Lipatov, and Victor~S. Fadin.
\newblock {The Pomeranchuk Singularity in Nonabelian Gauge Theories}.
\newblock {\em Sov. Phys. JETP}, 45:199--204, 1977.

\bibitem{Balitsky:1978ic}
I.~I. Balitsky and L.~N. Lipatov.
\newblock {The Pomeranchuk Singularity in Quantum Chromodynamics}.
\newblock {\em Sov. J. Nucl. Phys.}, 28:822--829, 1978.

\bibitem{Weigert:2005us}
Heribert Weigert.
\newblock {Evolution at small x(bj): The Color glass condensate}.
\newblock {\em Prog. Part. Nucl. Phys.}, 55:461--565, 2005.
\newblock \href {http://arxiv.org/abs/hep-ph/0501087}
  {\path{arXiv:hep-ph/0501087}}, \href
  {https://doi.org/10.1016/j.ppnp.2005.01.029}
  {\path{doi:10.1016/j.ppnp.2005.01.029}}.

\bibitem{Jalilian-Marian:2005ccm}
Jamal Jalilian-Marian and Yuri~V. Kovchegov.
\newblock {Saturation physics and deuteron-Gold collisions at RHIC}.
\newblock {\em Prog. Part. Nucl. Phys.}, 56:104--231, 2006.
\newblock \href {http://arxiv.org/abs/hep-ph/0505052}
  {\path{arXiv:hep-ph/0505052}}, \href
  {https://doi.org/10.1016/j.ppnp.2005.07.002}
  {\path{doi:10.1016/j.ppnp.2005.07.002}}.

\bibitem{Gelis:2010nm}
Francois Gelis, Edmond Iancu, Jamal Jalilian-Marian, and Raju Venugopalan.
\newblock {The Color Glass Condensate}.
\newblock {\em Ann. Rev. Nucl. Part. Sci.}, 60:463--489, 2010.
\newblock \href {http://arxiv.org/abs/1002.0333} {\path{arXiv:1002.0333}},
  \href {https://doi.org/10.1146/annurev.nucl.010909.083629}
  {\path{doi:10.1146/annurev.nucl.010909.083629}}.

\bibitem{Albacete:2014fwa}
Javier~L. Albacete and Cyrille Marquet.
\newblock {Gluon saturation and initial conditions for relativistic heavy ion
  collisions}.
\newblock {\em Prog. Part. Nucl. Phys.}, 76:1--42, 2014.
\newblock \href {http://arxiv.org/abs/1401.4866} {\path{arXiv:1401.4866}},
  \href {https://doi.org/10.1016/j.ppnp.2014.01.004}
  {\path{doi:10.1016/j.ppnp.2014.01.004}}.

\bibitem{Balitsky:1998ya}
Ian Balitsky.
\newblock {Factorization and high-energy effective action}.
\newblock {\em Phys. Rev. D}, 60:014020, 1999.
\newblock \href {http://arxiv.org/abs/hep-ph/9812311}
  {\path{arXiv:hep-ph/9812311}}, \href
  {https://doi.org/10.1103/PhysRevD.60.014020}
  {\path{doi:10.1103/PhysRevD.60.014020}}.

\bibitem{Kovchegov:1999yj}
Yuri~V. Kovchegov.
\newblock {Small x F(2) structure function of a nucleus including multiple
  pomeron exchanges}.
\newblock {\em Phys. Rev. D}, 60:034008, 1999.
\newblock \href {http://arxiv.org/abs/hep-ph/9901281}
  {\path{arXiv:hep-ph/9901281}}, \href
  {https://doi.org/10.1103/PhysRevD.60.034008}
  {\path{doi:10.1103/PhysRevD.60.034008}}.

\bibitem{Kovchegov:1999ua}
Yuri~V. Kovchegov.
\newblock {Unitarization of the BFKL pomeron on a nucleus}.
\newblock {\em Phys. Rev. D}, 61:074018, 2000.
\newblock \href {http://arxiv.org/abs/hep-ph/9905214}
  {\path{arXiv:hep-ph/9905214}}, \href
  {https://doi.org/10.1103/PhysRevD.61.074018}
  {\path{doi:10.1103/PhysRevD.61.074018}}.

\bibitem{Jalilian-Marian:1997jhx}
Jamal Jalilian-Marian, Alex Kovner, Andrei Leonidov, and Heribert Weigert.
\newblock {The Wilson renormalization group for low x physics: Towards the high
  density regime}.
\newblock {\em Phys. Rev. D}, 59:014014, 1998.
\newblock \href {http://arxiv.org/abs/hep-ph/9706377}
  {\path{arXiv:hep-ph/9706377}}, \href
  {https://doi.org/10.1103/PhysRevD.59.014014}
  {\path{doi:10.1103/PhysRevD.59.014014}}.

\bibitem{Jalilian-Marian:1997qno}
Jamal Jalilian-Marian, Alex Kovner, Andrei Leonidov, and Heribert Weigert.
\newblock {The BFKL equation from the Wilson renormalization group}.
\newblock {\em Nucl. Phys. B}, 504:415--431, 1997.
\newblock \href {http://arxiv.org/abs/hep-ph/9701284}
  {\path{arXiv:hep-ph/9701284}}, \href
  {https://doi.org/10.1016/S0550-3213(97)00440-9}
  {\path{doi:10.1016/S0550-3213(97)00440-9}}.

\bibitem{Weigert:2000gi}
Heribert Weigert.
\newblock {Unitarity at small Bjorken x}.
\newblock {\em Nucl. Phys. A}, 703:823--860, 2002.
\newblock \href {http://arxiv.org/abs/hep-ph/0004044}
  {\path{arXiv:hep-ph/0004044}}, \href
  {https://doi.org/10.1016/S0375-9474(01)01668-2}
  {\path{doi:10.1016/S0375-9474(01)01668-2}}.

\bibitem{Iancu:2001ad}
Edmond Iancu, Andrei Leonidov, and Larry~D. McLerran.
\newblock {The Renormalization group equation for the color glass condensate}.
\newblock {\em Phys. Lett. B}, 510:133--144, 2001.
\newblock \href {http://arxiv.org/abs/hep-ph/0102009}
  {\path{arXiv:hep-ph/0102009}}, \href
  {https://doi.org/10.1016/S0370-2693(01)00524-X}
  {\path{doi:10.1016/S0370-2693(01)00524-X}}.

\bibitem{Mueller:1989st}
Alfred~H. Mueller.
\newblock {Small x Behavior and Parton Saturation: A QCD Model}.
\newblock {\em Nucl. Phys. B}, 335:115--137, 1990.
\newblock \href {https://doi.org/10.1016/0550-3213(90)90173-B}
  {\path{doi:10.1016/0550-3213(90)90173-B}}.

\bibitem{McLerran:1994vd}
Larry~D. McLerran and Raju Venugopalan.
\newblock {Green's functions in the color field of a large nucleus}.
\newblock {\em Phys. Rev. D}, 50:2225--2233, 1994.
\newblock \href {http://arxiv.org/abs/hep-ph/9402335}
  {\path{arXiv:hep-ph/9402335}}, \href
  {https://doi.org/10.1103/PhysRevD.50.2225}
  {\path{doi:10.1103/PhysRevD.50.2225}}.

\bibitem{Hauenstein:2021zql}
F.~Hauenstein et~al.
\newblock {Measuring recoiling nucleons from the nucleus with the future
  Electron Ion Collider}.
\newblock {\em Phys. Rev. C}, 105(3):034001, 2022.
\newblock \href {http://arxiv.org/abs/2109.09509} {\path{arXiv:2109.09509}},
  \href {https://doi.org/10.1103/PhysRevC.105.034001}
  {\path{doi:10.1103/PhysRevC.105.034001}}.

\bibitem{Tu:2020ymk}
Zhoudunming Tu, Alexander Jentsch, Mark Baker, Liang Zheng, Jeong-Hun Lee, Raju
  Venugopalan, Or~Hen, Douglas Higinbotham, Elke-Caroline Aschenauer, and
  Thomas Ullrich.
\newblock {Probing short-range correlations in the deuteron via incoherent
  diffractive J/$\psi$ production with spectator tagging at the EIC}.
\newblock {\em Phys. Lett. B}, 811:135877, 2020.
\newblock \href {http://arxiv.org/abs/2005.14706} {\path{arXiv:2005.14706}},
  \href {https://doi.org/10.1016/j.physletb.2020.135877}
  {\path{doi:10.1016/j.physletb.2020.135877}}.

\bibitem{Jentsch:2021qdp}
Alexander Jentsch, Zhoudunming Tu, and Christian Weiss.
\newblock {Deep-inelastic electron-deuteron scattering with spectator nucleon
  tagging at the future Electron Ion Collider: Extracting free nucleon
  structure}.
\newblock {\em Phys. Rev. C}, 104(6):065205, 2021.
\newblock \href {http://arxiv.org/abs/2108.08314} {\path{arXiv:2108.08314}},
  \href {https://doi.org/10.1103/PhysRevC.104.065205}
  {\path{doi:10.1103/PhysRevC.104.065205}}.

\bibitem{Friscic:2021oti}
Ivica Friscic et~al.
\newblock {Neutron spin structure from e-3He scattering with double spectator
  tagging at the electron-ion collider}.
\newblock {\em Phys. Lett. B}, 823:136726, 2021.
\newblock \href {http://arxiv.org/abs/2106.08805} {\path{arXiv:2106.08805}},
  \href {https://doi.org/10.1016/j.physletb.2021.136726}
  {\path{doi:10.1016/j.physletb.2021.136726}}.

\bibitem{CiofidegliAtti:2010uwl}
C.~Ciofi~degli Atti and L.~P. Kaptari.
\newblock {Semi-inclusive Deep Inelastic Scattering off Few-Nucleon Systems:
  Tagging the EMC Effect and Hadronization Mechanisms with Detection of Slow
  Recoiling Nuclei}.
\newblock {\em Phys. Rev. C}, 83:044602, 2011.
\newblock \href {http://arxiv.org/abs/1011.5960} {\path{arXiv:1011.5960}},
  \href {https://doi.org/10.1103/PhysRevC.83.044602}
  {\path{doi:10.1103/PhysRevC.83.044602}}.

\bibitem{CiofidegliAtti:2011rm}
Claudio Ciofi~degli Atti, Leonid~P. Kaptari, and Chiara~Benedetta Mezzetti.
\newblock {Tagging emc effects and hadronization mechanisms by semi-inclusive
  deep inelastic scattering off nuclei}.
\newblock {\em Fizika B}, 20:161--172, 2011.
\newblock \href {http://arxiv.org/abs/1103.3674} {\path{arXiv:1103.3674}}.

\bibitem{Strikman:2017koc}
M.~Strikman and C.~Weiss.
\newblock {Electron-deuteron deep-inelastic scattering with spectator nucleon
  tagging and final-state interactions at intermediate x}.
\newblock {\em Phys. Rev. C}, 97(3):035209, 2018.
\newblock \href {http://arxiv.org/abs/1706.02244} {\path{arXiv:1706.02244}},
  \href {https://doi.org/10.1103/PhysRevC.97.035209}
  {\path{doi:10.1103/PhysRevC.97.035209}}.

\bibitem{Cosyn:2019hem}
W.~Cosyn and C.~Weiss.
\newblock {Neutron spin structure from polarized deuteron DIS with proton
  tagging}.
\newblock {\em Phys. Lett. B}, 799:135035, 2019.
\newblock \href {http://arxiv.org/abs/1906.11119} {\path{arXiv:1906.11119}},
  \href {https://doi.org/10.1016/j.physletb.2019.135035}
  {\path{doi:10.1016/j.physletb.2019.135035}}.

\bibitem{Cosyn:2020kwu}
W.~Cosyn and C.~Weiss.
\newblock {Polarized electron-deuteron deep-inelastic scattering with spectator
  nucleon tagging}.
\newblock {\em Phys. Rev. C}, 102:065204, 2020.
\newblock \href {http://arxiv.org/abs/2006.03033} {\path{arXiv:2006.03033}},
  \href {https://doi.org/10.1103/PhysRevC.102.065204}
  {\path{doi:10.1103/PhysRevC.102.065204}}.

\bibitem{Li:2021lbj}
Hai~Tao Li, Ze~Long Liu, and Ivan Vitev.
\newblock {Nuclear matter effects on jet production at electron-ion colliders}.
\newblock {\em SciPost Phys. Proc.}, 8:134, 2022.
\newblock \href {http://arxiv.org/abs/2110.04858} {\path{arXiv:2110.04858}},
  \href {https://doi.org/10.21468/SciPostPhysProc.8.134}
  {\path{doi:10.21468/SciPostPhysProc.8.134}}.

\bibitem{Davoudiasl:2021mjy}
Hooman Davoudiasl, Roman Marcarelli, and Ethan~T. Neil.
\newblock {Lepton-flavor-violating ALPs at the Electron-Ion Collider: a golden
  opportunity}.
\newblock {\em JHEP}, 02:071, 2023.
\newblock \href {http://arxiv.org/abs/2112.04513} {\path{arXiv:2112.04513}},
  \href {https://doi.org/10.1007/JHEP02(2023)071}
  {\path{doi:10.1007/JHEP02(2023)071}}.

\bibitem{Vidovic:1992ik}
M.~Vidovic, M.~Greiner, C.~Best, and G.~Soff.
\newblock {Impact parameter dependence of the electromagnetic particle
  production in ultrarelativistic heavy ion collisions}.
\newblock {\em Phys. Rev. C}, 47:2308--2319, 1993.
\newblock \href {https://doi.org/10.1103/PhysRevC.47.2308}
  {\path{doi:10.1103/PhysRevC.47.2308}}.

\bibitem{Klein:2018cjh}
Spencer~R. Klein.
\newblock {Two-photon production of dilepton pairs in peripheral heavy ion
  collisions}.
\newblock {\em Phys. Rev. C}, 97(5):054903, 2018.
\newblock \href {http://arxiv.org/abs/1801.04320} {\path{arXiv:1801.04320}},
  \href {https://doi.org/10.1103/PhysRevC.97.054903}
  {\path{doi:10.1103/PhysRevC.97.054903}}.

\bibitem{Brandenburg:2022tna}
James~Daniel Brandenburg, Janet Seger, Zhangbu Xu, and Wangmei Zha.
\newblock {Report on Progress in Physics: Observation of the Breit-Wheeler
  Process and Vacuum Birefringence in Heavy-Ion Collisions}.
\newblock 8 2022.
\newblock \href {http://arxiv.org/abs/2208.14943} {\path{arXiv:2208.14943}}.

\bibitem{Adams:2004rz}
J.~Adams et~al.
\newblock {Production of e+ e- pairs accompanied by nuclear dissociation in
  ultra-peripheral heavy ion collision}.
\newblock {\em Phys. Rev. C}, 70:031902, 2004.
\newblock \href {http://arxiv.org/abs/nucl-ex/0404012}
  {\path{arXiv:nucl-ex/0404012}}, \href
  {https://doi.org/10.1103/PhysRevC.70.031902}
  {\path{doi:10.1103/PhysRevC.70.031902}}.

\bibitem{Afanasiev:2009hy}
S.~Afanasiev et~al.
\newblock {Photoproduction of J/psi and of high mass e+e- in ultra-peripheral
  Au+Au collisions at s**(1/2) = 200-GeV}.
\newblock {\em Phys. Lett. B}, 679:321--329, 2009.
\newblock \href {http://arxiv.org/abs/0903.2041} {\path{arXiv:0903.2041}},
  \href {https://doi.org/10.1016/j.physletb.2009.07.061}
  {\path{doi:10.1016/j.physletb.2009.07.061}}.

\bibitem{Adam:2018tdm}
Jaroslav Adam et~al.
\newblock {Low-$p_T$ $e^{+}e^{-}$ pair production in Au$+$Au collisions at
  $\sqrt{s_{NN}}$ = 200 GeV and U$+$U collisions at $\sqrt{s_{NN}}$ = 193 GeV
  at STAR}.
\newblock {\em Phys. Rev. Lett.}, 121(13):132301, 2018.
\newblock \href {http://arxiv.org/abs/1806.02295} {\path{arXiv:1806.02295}},
  \href {https://doi.org/10.1103/PhysRevLett.121.132301}
  {\path{doi:10.1103/PhysRevLett.121.132301}}.

\bibitem{STAR:2019wlg}
Jaroslav Adam et~al.
\newblock {Measurement of $e^+e^-$ Momentum and Angular Distributions from
  Linearly Polarized Photon Collisions}.
\newblock {\em Phys. Rev. Lett.}, 127(5):052302, 2021.
\newblock \href {http://arxiv.org/abs/1910.12400} {\path{arXiv:1910.12400}},
  \href {https://doi.org/10.1103/PhysRevLett.127.052302}
  {\path{doi:10.1103/PhysRevLett.127.052302}}.

\bibitem{Abbas:2013oua}
E.~Abbas et~al.
\newblock {Charmonium and $e^+e^-$ pair photoproduction at mid-rapidity in
  ultra-peripheral Pb-Pb collisions at $\sqrt{s_{\rm NN}}$=2.76 TeV}.
\newblock {\em Eur. Phys. J. C}, 73(11):2617, 2013.
\newblock \href {http://arxiv.org/abs/1305.1467} {\path{arXiv:1305.1467}},
  \href {https://doi.org/10.1140/epjc/s10052-013-2617-1}
  {\path{doi:10.1140/epjc/s10052-013-2617-1}}.

\bibitem{ATLAS:2020epq}
Georges Aad et~al.
\newblock {Exclusive dimuon production in ultraperipheral Pb+Pb collisions at
  $\sqrt{s_{\mathrm{NN}}} = 5.02$ TeV with ATLAS}.
\newblock {\em Phys. Rev. C}, 104:024906, 2021.
\newblock \href {http://arxiv.org/abs/2011.12211} {\path{arXiv:2011.12211}},
  \href {https://doi.org/10.1103/PhysRevC.104.024906}
  {\path{doi:10.1103/PhysRevC.104.024906}}.

\bibitem{ATLAS:2022srr}
{Exclusive dielectron production in ultraperipheral Pb+Pb collisions at
  $\sqrt{s_{_\text{NN}}} = 5.02$ TeV with ATLAS}.
\newblock 7 2022.
\newblock \href {http://arxiv.org/abs/2207.12781} {\path{arXiv:2207.12781}}.

\bibitem{ATLAS:2022ryk}
{Observation of the $\gamma\gamma\to\tau\tau$ process in Pb+Pb collisions and
  constraints on the $\tau$-lepton anomalous magnetic moment with the ATLAS
  detector}.
\newblock 4 2022.
\newblock \href {http://arxiv.org/abs/2204.13478} {\path{arXiv:2204.13478}}.

\bibitem{ATLAS:2022vbe}
{Measurement of muon pairs produced via $\gamma\gamma$ scattering in
  non-ultraperipheral Pb+Pb collisions at $\sqrt{s_{_\mathrm{NN}}} = 5.02$ TeV
  with the ATLAS detector}.
\newblock 6 2022.
\newblock \href {http://arxiv.org/abs/2206.12594} {\path{arXiv:2206.12594}}.

\bibitem{Acharya:2018nxm}
Shreyasi Acharya et~al.
\newblock {Measurement of dielectron production in central Pb-Pb collisions at
  $\sqrt{{\textit{s}}_{\mathrm{NN}}}$ = 2.76 TeV}.
\newblock {\em Phys. Rev. C}, 99(2):024002, 2019.
\newblock \href {http://arxiv.org/abs/1807.00923} {\path{arXiv:1807.00923}},
  \href {https://doi.org/10.1103/PhysRevC.99.024002}
  {\path{doi:10.1103/PhysRevC.99.024002}}.

\bibitem{Aaboud:2018eph}
Morad Aaboud et~al.
\newblock {Observation of centrality-dependent acoplanarity for muon pairs
  produced via two-photon scattering in Pb+Pb collisions at
  $\sqrt{s_{\mathrm{NN}}}=5.02$ TeV with the ATLAS detector}.
\newblock {\em Phys. Rev. Lett.}, 121(21):212301, 2018.
\newblock \href {http://arxiv.org/abs/1806.08708} {\path{arXiv:1806.08708}},
  \href {https://doi.org/10.1103/PhysRevLett.121.212301}
  {\path{doi:10.1103/PhysRevLett.121.212301}}.

\bibitem{Klein:2016yzr}
Spencer~R. Klein, Joakim Nystrand, Janet Seger, Yuri Gorbunov, and Joey
  Butterworth.
\newblock {STARlight: A Monte Carlo simulation program for ultra-peripheral
  collisions of relativistic ions}.
\newblock {\em Comput. Phys. Commun.}, 212:258--268, 2017.
\newblock \href {http://arxiv.org/abs/1607.03838} {\path{arXiv:1607.03838}},
  \href {https://doi.org/10.1016/j.cpc.2016.10.016}
  {\path{doi:10.1016/j.cpc.2016.10.016}}.

\bibitem{Suranyi:2021ssd}
O.~Sur\'anyi et~al.
\newblock {Performance of the CMS Zero Degree Calorimeters in pPb collisions at
  the LHC}.
\newblock {\em JINST}, 16(05):P05008, 2021.
\newblock \href {http://arxiv.org/abs/2102.06640} {\path{arXiv:2102.06640}},
  \href {https://doi.org/10.1088/1748-0221/16/05/P05008}
  {\path{doi:10.1088/1748-0221/16/05/P05008}}.

\bibitem{SantosDiaz:2022hxl}
P.~Santos~D\'\i{}az et~al.
\newblock {Mechanical and thermal design of the target neutral beam absorber
  for the High-Luminosity LHC upgrade}.
\newblock {\em Phys. Rev. Accel. Beams}, 25(5):053001, 2022.
\newblock \href {https://doi.org/10.1103/PhysRevAccelBeams.25.053001}
  {\path{doi:10.1103/PhysRevAccelBeams.25.053001}}.

\bibitem{Li:2019yzy}
Cong Li, Jian Zhou, and Ya-Jin Zhou.
\newblock {Probing the linear polarization of photons in ultraperipheral heavy
  ion collisions}.
\newblock {\em Phys. Lett. B}, 795:576--580, 2019.
\newblock \href {http://arxiv.org/abs/1903.10084} {\path{arXiv:1903.10084}},
  \href {https://doi.org/10.1016/j.physletb.2019.07.005}
  {\path{doi:10.1016/j.physletb.2019.07.005}}.

\bibitem{Li:2019sin}
Cong Li, Jian Zhou, and Ya-Jin Zhou.
\newblock {Impact parameter dependence of the azimuthal asymmetry in lepton
  pair production in heavy ion collisions}.
\newblock {\em Phys. Rev. D}, 101(3):034015, 2020.
\newblock \href {http://arxiv.org/abs/1911.00237} {\path{arXiv:1911.00237}},
  \href {https://doi.org/10.1103/PhysRevD.101.034015}
  {\path{doi:10.1103/PhysRevD.101.034015}}.

\bibitem{Brandenburg:2021lnj}
James~Daniel Brandenburg, Wangmei Zha, and Zhangbu Xu.
\newblock {Mapping the electromagnetic fields of heavy-ion collisions with the
  Breit-Wheeler process}.
\newblock {\em Eur. Phys. J. A}, 57(10):299, 2021.
\newblock \href {http://arxiv.org/abs/2103.16623} {\path{arXiv:2103.16623}},
  \href {https://doi.org/10.1140/epja/s10050-021-00595-5}
  {\path{doi:10.1140/epja/s10050-021-00595-5}}.

\bibitem{ATLAS-CONF-2022-021}
{Photo-nuclear jet production in ultra-peripheral Pb+Pb collisions at
  $\sqrt{s}_\text{NN} = 5.02$ TeV with the ATLAS detector}.
\newblock 2022.

\bibitem{TheLIGOScientific:2017qsa}
B.~P. Abbott et~al.
\newblock {GW170817: Observation of Gravitational Waves from a Binary Neutron
  Star Inspiral}.
\newblock {\em Phys. Rev. Lett.}, 119(16):161101, 2017.
\newblock \href {http://arxiv.org/abs/1710.05832} {\path{arXiv:1710.05832}},
  \href {https://doi.org/10.1103/PhysRevLett.119.161101}
  {\path{doi:10.1103/PhysRevLett.119.161101}}.

\bibitem{Riley:2021pdl}
Thomas~E. Riley et~al.
\newblock {A NICER View of the Massive Pulsar PSR J0740+6620 Informed by Radio
  Timing and XMM-Newton Spectroscopy}.
\newblock {\em Astrophys. J. Lett.}, 918(2):L27, 2021.
\newblock \href {http://arxiv.org/abs/2105.06980} {\path{arXiv:2105.06980}},
  \href {https://doi.org/10.3847/2041-8213/ac0a81}
  {\path{doi:10.3847/2041-8213/ac0a81}}.

\bibitem{Miller:2021qha}
M.~C. Miller et~al.
\newblock {The Radius of PSR J0740+6620 from NICER and XMM-Newton Data}.
\newblock {\em Astrophys. J. Lett.}, 918(2):L28, 2021.
\newblock \href {http://arxiv.org/abs/2105.06979} {\path{arXiv:2105.06979}},
  \href {https://doi.org/10.3847/2041-8213/ac089b}
  {\path{doi:10.3847/2041-8213/ac089b}}.

\bibitem{Fonseca:2021wxt}
E.~Fonseca et~al.
\newblock {Refined Mass and Geometric Measurements of the High-mass PSR
  J0740+6620}.
\newblock {\em Astrophys. J. Lett.}, 915(1):L12, 2021.
\newblock \href {http://arxiv.org/abs/2104.00880} {\path{arXiv:2104.00880}},
  \href {https://doi.org/10.3847/2041-8213/ac03b8}
  {\path{doi:10.3847/2041-8213/ac03b8}}.

\bibitem{Pineda:2021shy}
Skyy~V. Pineda et~al.
\newblock {Charge Radius of Neutron-Deficient Ni54 and Symmetry Energy
  Constraints Using the Difference in Mirror Pair Charge Radii}.
\newblock {\em Phys. Rev. Lett.}, 127(18):182503, 2021.
\newblock \href {http://arxiv.org/abs/2106.10378} {\path{arXiv:2106.10378}},
  \href {https://doi.org/10.1103/PhysRevLett.127.182503}
  {\path{doi:10.1103/PhysRevLett.127.182503}}.

\bibitem{Alford:2017rxf}
Mark~G. Alford, Luke Bovard, Matthias Hanauske, Luciano Rezzolla, and Kai
  Schwenzer.
\newblock {Viscous Dissipation and Heat Conduction in Binary Neutron-Star
  Mergers}.
\newblock {\em Phys. Rev. Lett.}, 120(4):041101, 2018.
\newblock \href {http://arxiv.org/abs/1707.09475} {\path{arXiv:1707.09475}},
  \href {https://doi.org/10.1103/PhysRevLett.120.041101}
  {\path{doi:10.1103/PhysRevLett.120.041101}}.

\bibitem{Perego:2019adq}
Albino Perego, Sebastiano Bernuzzi, and David Radice.
\newblock {Thermodynamics conditions of matter in neutron star mergers}.
\newblock {\em Eur. Phys. J. A}, 55(8):124, 2019.
\newblock \href {http://arxiv.org/abs/1903.07898} {\path{arXiv:1903.07898}},
  \href {https://doi.org/10.1140/epja/i2019-12810-7}
  {\path{doi:10.1140/epja/i2019-12810-7}}.

\bibitem{Arras:2018fxj}
Phil Arras and Nevin~N. Weinberg.
\newblock {Urca reactions during neutron star inspiral}.
\newblock {\em Mon. Not. Roy. Astron. Soc.}, 486(1):1424--1436, 2019.
\newblock \href {http://arxiv.org/abs/1806.04163} {\path{arXiv:1806.04163}},
  \href {https://doi.org/10.1093/mnras/stz880}
  {\path{doi:10.1093/mnras/stz880}}.

\bibitem{Most:2021zvc}
Elias~R. Most, Steven~P. Harris, Christopher Plumberg, Mark~G. Alford, Jorge
  Noronha, Jacquelyn Noronha-Hostler, Frans Pretorius, Helvi Witek, and
  Nicol\'as Yunes.
\newblock {Projecting the likely importance of weak-interaction-driven bulk
  viscosity in neutron star mergers}.
\newblock {\em Mon. Not. Roy. Astron. Soc.}, 509(1):1096--1108, 2021.
\newblock \href {http://arxiv.org/abs/2107.05094} {\path{arXiv:2107.05094}},
  \href {https://doi.org/10.1093/mnras/stab2793}
  {\path{doi:10.1093/mnras/stab2793}}.

\bibitem{Hammond:2022uua}
Peter Hammond, Ian Hawke, and Nils Andersson.
\newblock {Detecting the impact of nuclear reactions on neutron star mergers
  through gravitational waves}.
\newblock 5 2022.
\newblock \href {http://arxiv.org/abs/2205.11377} {\path{arXiv:2205.11377}}.

\bibitem{Most:2022yhe}
Elias~R. Most, Alexander Haber, Steven~P. Harris, Ziyuan Zhang, Mark~G. Alford,
  and Jorge Noronha.
\newblock {Emergence of microphysical viscosity in binary neutron star
  post-merger dynamics}.
\newblock 7 2022.
\newblock \href {http://arxiv.org/abs/2207.00442} {\path{arXiv:2207.00442}}.

\bibitem{Gavassino:2020kwo}
L.~Gavassino, M.~Antonelli, and B.~Haskell.
\newblock {Bulk viscosity in relativistic fluids: from thermodynamics to
  hydrodynamics}.
\newblock {\em Class. Quant. Grav.}, 38(7):075001, 2021.
\newblock \href {http://arxiv.org/abs/2003.04609} {\path{arXiv:2003.04609}},
  \href {https://doi.org/10.1088/1361-6382/abe588}
  {\path{doi:10.1088/1361-6382/abe588}}.

\bibitem{Camelio:2022ljs}
Giovanni Camelio, Lorenzo Gavassino, Marco Antonelli, Sebastiano Bernuzzi, and
  Brynmor Haskell.
\newblock {Simulating bulk viscosity in neutron stars I: formalism}.
\newblock 4 2022.
\newblock \href {http://arxiv.org/abs/2204.11809} {\path{arXiv:2204.11809}}.

\bibitem{Denicol:2012cn}
G.~S. Denicol, H.~Niemi, E.~Molnar, and D.~H. Rischke.
\newblock {Derivation of transient relativistic fluid dynamics from the
  Boltzmann equation}.
\newblock {\em Phys. Rev. D}, 85:114047, 2012.
\newblock [Erratum: Phys.Rev.D 91, 039902 (2015)].
\newblock \href {http://arxiv.org/abs/1202.4551} {\path{arXiv:1202.4551}},
  \href {https://doi.org/10.1103/PhysRevD.85.114047}
  {\path{doi:10.1103/PhysRevD.85.114047}}.

\bibitem{PREX:2021umo}
D.~Adhikari et~al.
\newblock {Accurate Determination of the Neutron Skin Thickness of $^{208}$Pb
  through Parity-Violation in Electron Scattering}.
\newblock {\em Phys. Rev. Lett.}, 126(17):172502, 2021.
\newblock \href {http://arxiv.org/abs/2102.10767} {\path{arXiv:2102.10767}},
  \href {https://doi.org/10.1103/PhysRevLett.126.172502}
  {\path{doi:10.1103/PhysRevLett.126.172502}}.

\bibitem{CREX:2022kgg}
D.~Adhikari et~al.
\newblock {Precision Determination of the Neutral Weak Form Factor of Ca48}.
\newblock {\em Phys. Rev. Lett.}, 129(4):042501, 2022.
\newblock \href {http://arxiv.org/abs/2205.11593} {\path{arXiv:2205.11593}},
  \href {https://doi.org/10.1103/PhysRevLett.129.042501}
  {\path{doi:10.1103/PhysRevLett.129.042501}}.

\bibitem{IceCube:2021ixw}
Stef Verpoest et~al.
\newblock {Testing Hadronic Interaction Models with Cosmic Ray Measurements at
  the IceCube Neutrino Observatory}.
\newblock {\em PoS}, ICRC2021:357, 2021.
\newblock \href {http://arxiv.org/abs/2107.09387} {\path{arXiv:2107.09387}},
  \href {https://doi.org/10.22323/1.395.0357} {\path{doi:10.22323/1.395.0357}}.

\bibitem{PierreAuger:2022atd}
Adila Abdul~Halim et~al.
\newblock {Constraining the sources of ultra-high-energy cosmic rays across and
  above the ankle with the spectrum and composition data measured at the Pierre
  Auger Observatory}.
\newblock 11 2022.
\newblock \href {http://arxiv.org/abs/2211.02857} {\path{arXiv:2211.02857}}.

\bibitem{Hanlon:2018hlv}
William Hanlon.
\newblock {Measurements of UHECR Mass Composition by Telescope Array}.
\newblock {\em EPJ Web Conf.}, 210:01008, 2019.
\newblock \href {http://arxiv.org/abs/1812.05688} {\path{arXiv:1812.05688}},
  \href {https://doi.org/10.1051/epjconf/201921001008}
  {\path{doi:10.1051/epjconf/201921001008}}.

\bibitem{Albrecht:2021cxw}
Johannes Albrecht et~al.
\newblock {The Muon Puzzle in cosmic-ray induced air showers and its connection
  to the Large Hadron Collider}.
\newblock {\em Astrophys. Space Sci.}, 367(3):27, 2022.
\newblock \href {http://arxiv.org/abs/2105.06148} {\path{arXiv:2105.06148}},
  \href {https://doi.org/10.1007/s10509-022-04054-5}
  {\path{doi:10.1007/s10509-022-04054-5}}.

\bibitem{Pajares:2000sn}
C.~Pajares, D.~Sousa, and R.~A. Vazquez.
\newblock {Consequences of parton's saturation and string's percolation on the
  developments of cosmic ray showers}.
\newblock {\em Phys. Rev. Lett.}, 86:1674--1677, 2001.
\newblock \href {http://arxiv.org/abs/astro-ph/0005588}
  {\path{arXiv:astro-ph/0005588}}, \href
  {https://doi.org/10.1103/PhysRevLett.86.1674}
  {\path{doi:10.1103/PhysRevLett.86.1674}}.

\bibitem{Alvarez-Muniz:2006wca}
J.~Alvarez-Muniz, P.~Brogueira, R.~Conceicao, J.~Dias~de Deus, M.~C.
  Espirito~Santo, and M.~Pimenta.
\newblock {Percolation and high energy cosmic rays above 10**17 eV}.
\newblock {\em Astropart. Phys.}, 27:271--277, 2007.
\newblock \href {http://arxiv.org/abs/hep-ph/0608050}
  {\path{arXiv:hep-ph/0608050}}, \href
  {https://doi.org/10.1016/j.astropartphys.2006.11.006}
  {\path{doi:10.1016/j.astropartphys.2006.11.006}}.

\bibitem{Baur:2019cpv}
Sebastian Baur, Hans Dembinski, Matias Perlin, Tanguy Pierog, Ralf Ulrich, and
  Klaus Werner.
\newblock {Core-corona effect in hadron collisions and muon production in air
  showers}.
\newblock 2 2019.
\newblock \href {http://arxiv.org/abs/1902.09265} {\path{arXiv:1902.09265}}.

\bibitem{Hentschinski:2022xnd}
Martin Hentschinski et~al.
\newblock {White Paper on Forward Physics, BFKL, Saturation Physics and
  Diffraction}.
\newblock 3 2022.
\newblock \href {http://arxiv.org/abs/2203.08129} {\path{arXiv:2203.08129}}.

\bibitem{CMS:2020ldm}
Vardan Khachatryan et~al.
\newblock {The very forward CASTOR calorimeter of the CMS experiment}.
\newblock {\em JINST}, 16(02):P02010, 2021.
\newblock \href {http://arxiv.org/abs/2011.01185} {\path{arXiv:2011.01185}},
  \href {https://doi.org/10.1088/1748-0221/16/02/P02010}
  {\path{doi:10.1088/1748-0221/16/02/P02010}}.

\bibitem{Feng:2022inv}
Jonathan~L. Feng et~al.
\newblock {The Forward Physics Facility at the High-Luminosity LHC}.
\newblock {\em J. Phys. G}, 50(3):030501, 2023.
\newblock \href {http://arxiv.org/abs/2203.05090} {\path{arXiv:2203.05090}},
  \href {https://doi.org/10.1088/1361-6471/ac865e}
  {\path{doi:10.1088/1361-6471/ac865e}}.

\bibitem{Bhattacharya:2020hfs}
Atri Bhattacharya, Rikard Enberg, Mary~Hall Reno, Ina Sarcevic, and Anna
  Stasto.
\newblock {Perturbative Charm Production and the Prompt Atmospheric Neutrino
  Flux in light of RHIC and LHC}.
\newblock {\em PoS}, ICHEP2020:116, 2021.
\newblock \href {http://arxiv.org/abs/2012.15190} {\path{arXiv:2012.15190}},
  \href {https://doi.org/10.22323/1.390.0116} {\path{doi:10.22323/1.390.0116}}.

\bibitem{IceCube:2017roe}
M.~G. Aartsen et~al.
\newblock {Measurement of the multi-TeV neutrino cross section with IceCube
  using Earth absorption}.
\newblock {\em Nature}, 551:596--600, 2017.
\newblock \href {http://arxiv.org/abs/1711.08119} {\path{arXiv:1711.08119}},
  \href {https://doi.org/10.1038/nature24459} {\path{doi:10.1038/nature24459}}.

\bibitem{IceCube:2018pgc}
M.~G. Aartsen et~al.
\newblock {Measurements using the inelasticity distribution of multi-TeV
  neutrino interactions in IceCube}.
\newblock {\em Phys. Rev. D}, 99(3):032004, 2019.
\newblock \href {http://arxiv.org/abs/1808.07629} {\path{arXiv:1808.07629}},
  \href {https://doi.org/10.1103/PhysRevD.99.032004}
  {\path{doi:10.1103/PhysRevD.99.032004}}.

\bibitem{Klein:2019nbu}
Spencer~R. Klein.
\newblock {\em {Probing high-energy interactions of atmospheric and
  astrophysical neutrinos}}.
\newblock 2020.
\newblock \href {http://arxiv.org/abs/1906.02221} {\path{arXiv:1906.02221}},
  \href {https://doi.org/10.1142/9789813275027_0004}
  {\path{doi:10.1142/9789813275027_0004}}.

\bibitem{NuSTEC:2017hzk}
L.~Alvarez-Ruso et~al.
\newblock {NuSTEC White Paper: Status and challenges of
  neutrino\textendash{}nucleus scattering}.
\newblock {\em Prog. Part. Nucl. Phys.}, 100:1--68, 2018.
\newblock \href {http://arxiv.org/abs/1706.03621} {\path{arXiv:1706.03621}},
  \href {https://doi.org/10.1016/j.ppnp.2018.01.006}
  {\path{doi:10.1016/j.ppnp.2018.01.006}}.

\bibitem{T2K:2018rhz}
K.~Abe et~al.
\newblock {Search for CP Violation in Neutrino and Antineutrino Oscillations by
  the T2K Experiment with $2.2\times10^{21}$ Protons on Target}.
\newblock {\em Phys. Rev. Lett.}, 121(17):171802, 2018.
\newblock \href {http://arxiv.org/abs/1807.07891} {\path{arXiv:1807.07891}},
  \href {https://doi.org/10.1103/PhysRevLett.121.171802}
  {\path{doi:10.1103/PhysRevLett.121.171802}}.

\bibitem{T2K:2019bcf}
K.~Abe et~al.
\newblock {Constraint on the matter\textendash{}antimatter symmetry-violating
  phase in neutrino oscillations}.
\newblock {\em Nature}, 580(7803):339--344, 2020.
\newblock [Erratum: Nature 583, E16 (2020)].
\newblock \href {http://arxiv.org/abs/1910.03887} {\path{arXiv:1910.03887}},
  \href {https://doi.org/10.1038/s41586-020-2177-0}
  {\path{doi:10.1038/s41586-020-2177-0}}.

\bibitem{NOvA:2018gge}
M.~A. Acero et~al.
\newblock {New constraints on oscillation parameters from $\nu_e$ appearance
  and $\nu_\mu$ disappearance in the NOvA experiment}.
\newblock {\em Phys. Rev. D}, 98:032012, 2018.
\newblock \href {http://arxiv.org/abs/1806.00096} {\path{arXiv:1806.00096}},
  \href {https://doi.org/10.1103/PhysRevD.98.032012}
  {\path{doi:10.1103/PhysRevD.98.032012}}.

\bibitem{Ankowski:2015kya}
Artur~M. Ankowski, Pilar Coloma, Patrick Huber, Camillo Mariani, and Erica
  Vagnoni.
\newblock {Missing energy and the measurement of the CP-violating phase in
  neutrino oscillations}.
\newblock {\em Phys. Rev. D}, 92(9):091301, 2015.
\newblock \href {http://arxiv.org/abs/1507.08561} {\path{arXiv:1507.08561}},
  \href {https://doi.org/10.1103/PhysRevD.92.091301}
  {\path{doi:10.1103/PhysRevD.92.091301}}.

\bibitem{CLAS:2021neh}
M.~Khachatryan et~al.
\newblock {Electron-beam energy reconstruction for neutrino oscillation
  measurements}.
\newblock {\em Nature}, 599(7886):565--570, 2021.
\newblock \href {https://doi.org/10.1038/s41586-021-04046-5}
  {\path{doi:10.1038/s41586-021-04046-5}}.

\bibitem{Dai:2018xhi}
H.~Dai et~al.
\newblock {First Measurement of the Ti$(e,e^\prime){\rm X}$ Cross Section at
  Jefferson Lab}.
\newblock {\em Phys. Rev. C}, 98(1):014617, 2018.
\newblock \href {http://arxiv.org/abs/1803.01910} {\path{arXiv:1803.01910}},
  \href {https://doi.org/10.1103/PhysRevC.98.014617}
  {\path{doi:10.1103/PhysRevC.98.014617}}.

\bibitem{Dai:2018gch}
H.~Dai et~al.
\newblock {First measurement of the Ar$(e,e')X$ cross section at Jefferson
  Laboratory}.
\newblock {\em Phys. Rev. C}, 99(5):054608, 2019.
\newblock \href {http://arxiv.org/abs/1810.10575} {\path{arXiv:1810.10575}},
  \href {https://doi.org/10.1103/PhysRevC.99.054608}
  {\path{doi:10.1103/PhysRevC.99.054608}}.

\bibitem{Murphy:2019wed}
M.~Murphy et~al.
\newblock {Measurement of the cross sections for inclusive electron scattering
  in the E12-14-012 experiment at Jefferson Lab}.
\newblock {\em Phys. Rev. C}, 100(5):054606, 2019.
\newblock \href {http://arxiv.org/abs/1908.01802} {\path{arXiv:1908.01802}},
  \href {https://doi.org/10.1103/PhysRevC.100.054606}
  {\path{doi:10.1103/PhysRevC.100.054606}}.

\bibitem{Gu:2020rcp}
L.~Gu et~al.
\newblock {Measurement of the Ar(e,e$^\prime$ p) and Ti(e,e$^\prime$ p) cross
  sections in Jefferson Lab Hall A}.
\newblock {\em Phys. Rev. C}, 103(3):034604, 2021.
\newblock \href {http://arxiv.org/abs/2012.11466} {\path{arXiv:2012.11466}},
  \href {https://doi.org/10.1103/PhysRevC.103.034604}
  {\path{doi:10.1103/PhysRevC.103.034604}}.

\bibitem{JeffersonLabHallA:2022cit}
L.~Jiang et~al.
\newblock {Determination of the argon spectral function from (e,e'p) data}.
\newblock {\em Phys. Rev. D}, 105(11):112002, 2022.
\newblock \href {http://arxiv.org/abs/2203.01748} {\path{arXiv:2203.01748}},
  \href {https://doi.org/10.1103/PhysRevD.105.112002}
  {\path{doi:10.1103/PhysRevD.105.112002}}.

\bibitem{JeffersonLabHallA:2022msz}
L.~Jiang et~al.
\newblock {Determination of the titanium spectral function from (e,e'p) data}.
\newblock 9 2022.
\newblock \href {http://arxiv.org/abs/2209.14108} {\path{arXiv:2209.14108}}.

\bibitem{Benhar:2010nx}
Omar Benhar, Pietro Coletti, and Davide Meloni.
\newblock {Electroweak nuclear response in quasi-elastic regime}.
\newblock {\em Phys. Rev. Lett.}, 105:132301, 2010.
\newblock \href {http://arxiv.org/abs/1006.4783} {\path{arXiv:1006.4783}},
  \href {https://doi.org/10.1103/PhysRevLett.105.132301}
  {\path{doi:10.1103/PhysRevLett.105.132301}}.

\bibitem{Benhar:2015wva}
Omar Benhar, Patrick Huber, Camillo Mariani, and Davide Meloni.
\newblock {Neutrino\textendash{}nucleus interactions and the determination of
  oscillation parameters}.
\newblock {\em Phys. Rept.}, 700:1--47, 2017.
\newblock \href {http://arxiv.org/abs/1501.06448} {\path{arXiv:1501.06448}},
  \href {https://doi.org/10.1016/j.physrep.2017.07.004}
  {\path{doi:10.1016/j.physrep.2017.07.004}}.

\bibitem{Ankowski:2016jdd}
Artur~M Ankowski and Camillo Mariani.
\newblock {Systematic uncertainties in long-baseline neutrino-oscillation
  experiments}.
\newblock {\em J. Phys. G}, 44(5):054001, 2017.
\newblock \href {http://arxiv.org/abs/1609.00258} {\path{arXiv:1609.00258}},
  \href {https://doi.org/10.1088/1361-6471/aa61b2}
  {\path{doi:10.1088/1361-6471/aa61b2}}.

\bibitem{Rocco:2018mwt}
Noemi Rocco, Carlo Barbieri, Omar Benhar, Arturo De~Pace, and Alessandro
  Lovato.
\newblock {Neutrino-Nucleus Cross Section within the Extended Factorization
  Scheme}.
\newblock {\em Phys. Rev. C}, 99(2):025502, 2019.
\newblock \href {http://arxiv.org/abs/1810.07647} {\path{arXiv:1810.07647}},
  \href {https://doi.org/10.1103/PhysRevC.99.025502}
  {\path{doi:10.1103/PhysRevC.99.025502}}.

\bibitem{electronsforneutrinos:2020tbf}
A.~Papadopoulou et~al.
\newblock {Inclusive Electron Scattering And The GENIE Neutrino Event
  Generator}.
\newblock {\em Phys. Rev. D}, 103:113003, 2021.
\newblock \href {http://arxiv.org/abs/2009.07228} {\path{arXiv:2009.07228}},
  \href {https://doi.org/10.1103/PhysRevD.103.113003}
  {\path{doi:10.1103/PhysRevD.103.113003}}.

\bibitem{Bruce:2018yzs}
Roderik Bruce et~al.
\newblock {New physics searches with heavy-ion collisions at the CERN Large
  Hadron Collider}.
\newblock {\em J. Phys. G}, 47(6):060501, 2020.
\newblock \href {http://arxiv.org/abs/1812.07688} {\path{arXiv:1812.07688}},
  \href {https://doi.org/10.1088/1361-6471/ab7ff7}
  {\path{doi:10.1088/1361-6471/ab7ff7}}.

\bibitem{dEnterria:2022sut}
David d'Enterria et~al.
\newblock {Opportunities for new physics searches with heavy ions at
  colliders}.
\newblock In {\em {2022 Snowmass Summer Study}}, 3 2022.
\newblock \href {http://arxiv.org/abs/2203.05939} {\path{arXiv:2203.05939}}.

\bibitem{dEnterria:2013zqi}
David d'Enterria and Gustavo~G. da~Silveira.
\newblock {Observing light-by-light scattering at the Large Hadron Collider}.
\newblock {\em Phys. Rev. Lett.}, 111:080405, 2013.
\newblock [Erratum: Phys.Rev.Lett. 116, 129901 (2016)].
\newblock \href {http://arxiv.org/abs/1305.7142} {\path{arXiv:1305.7142}},
  \href {https://doi.org/10.1103/PhysRevLett.111.080405}
  {\path{doi:10.1103/PhysRevLett.111.080405}}.

\bibitem{ATLAS:2017fur}
Morad Aaboud et~al.
\newblock {Evidence for light-by-light scattering in heavy-ion collisions with
  the ATLAS detector at the LHC}.
\newblock {\em Nature Phys.}, 13(9):852--858, 2017.
\newblock \href {http://arxiv.org/abs/1702.01625} {\path{arXiv:1702.01625}},
  \href {https://doi.org/10.1038/nphys4208} {\path{doi:10.1038/nphys4208}}.

\bibitem{CMS:2018erd}
Albert~M Sirunyan et~al.
\newblock {Evidence for light-by-light scattering and searches for axion-like
  particles in ultraperipheral PbPb collisions at $\sqrt{s_\mathrm{NN}} =$ 5.02
  TeV}.
\newblock {\em Phys. Lett. B}, 797:134826, 2019.
\newblock \href {http://arxiv.org/abs/1810.04602} {\path{arXiv:1810.04602}},
  \href {https://doi.org/10.1016/j.physletb.2019.134826}
  {\path{doi:10.1016/j.physletb.2019.134826}}.

\bibitem{ATLAS:2019azn}
Georges Aad et~al.
\newblock {Observation of light-by-light scattering in ultraperipheral Pb+Pb
  collisions with the ATLAS detector}.
\newblock {\em Phys. Rev. Lett.}, 123(5):052001, 2019.
\newblock \href {http://arxiv.org/abs/1904.03536} {\path{arXiv:1904.03536}},
  \href {https://doi.org/10.1103/PhysRevLett.123.052001}
  {\path{doi:10.1103/PhysRevLett.123.052001}}.

\bibitem{delAguila:1991rm}
F.~del Aguila, F.~Cornet, and Jose~I. Illana.
\newblock {The possibility of using a large heavy-ion collider for measuring
  the electromagnetic properties of the tau lepton}.
\newblock {\em Phys. Lett. B}, 271:256--260, 1991.
\newblock \href {https://doi.org/10.1016/0370-2693(91)91309-J}
  {\path{doi:10.1016/0370-2693(91)91309-J}}.

\bibitem{Muong-2:2021ojo}
B.~Abi et~al.
\newblock {Measurement of the Positive Muon Anomalous Magnetic Moment to 0.46
  ppm}.
\newblock {\em Phys. Rev. Lett.}, 126(14):141801, 2021.
\newblock \href {http://arxiv.org/abs/2104.03281} {\path{arXiv:2104.03281}},
  \href {https://doi.org/10.1103/PhysRevLett.126.141801}
  {\path{doi:10.1103/PhysRevLett.126.141801}}.

\bibitem{CMS:2022arf}
{Observation of $\tau$ lepton pair production in ultraperipheral lead-lead
  collisions at $\sqrt{s_\mathrm{NN}}$ = 5.02 TeV}.
\newblock 6 2022.
\newblock \href {http://arxiv.org/abs/2206.05192} {\path{arXiv:2206.05192}}.

\bibitem{Burmasov:2022gnl}
Nazar Burmasov, Evgeny Kryshen, Paul Buehler, and Roman Lavicka.
\newblock {Feasibility of tau g-2 measurements in ultra-peripheral collisions
  of heavy ions}.
\newblock In {\em {16th International Workshop on Tau Lepton Physics~}}, 3
  2022.
\newblock \href {http://arxiv.org/abs/2203.00990} {\path{arXiv:2203.00990}}.

\bibitem{DELPHI:2003nah}
J.~Abdallah et~al.
\newblock {Study of tau-pair production in photon-photon collisions at LEP and
  limits on the anomalous electromagnetic moments of the tau lepton}.
\newblock {\em Eur. Phys. J. C}, 35:159--170, 2004.
\newblock \href {http://arxiv.org/abs/hep-ex/0406010}
  {\path{arXiv:hep-ex/0406010}}, \href
  {https://doi.org/10.1140/epjc/s2004-01852-y}
  {\path{doi:10.1140/epjc/s2004-01852-y}}.

\bibitem{Androic:2013rhu}
D.~Androic et~al.
\newblock {First Determination of the Weak Charge of the Proton}.
\newblock {\em Phys. Rev. Lett.}, 111(14):141803, 2013.
\newblock \href {http://arxiv.org/abs/1307.5275} {\path{arXiv:1307.5275}},
  \href {https://doi.org/10.1103/PhysRevLett.111.141803}
  {\path{doi:10.1103/PhysRevLett.111.141803}}.

\bibitem{Androic:2018kni}
D.~Androi\'c et~al.
\newblock {Precision measurement of the weak charge of the proton}.
\newblock {\em Nature}, 557(7704):207--211, 2018.
\newblock \href {http://arxiv.org/abs/1905.08283} {\path{arXiv:1905.08283}},
  \href {https://doi.org/10.1038/s41586-018-0096-0}
  {\path{doi:10.1038/s41586-018-0096-0}}.

\bibitem{Wood:1997zq}
C.~S. Wood, S.~C. Bennett, D.~Cho, B.~P. Masterson, J.~L. Roberts, C.~E.
  Tanner, and Carl~E. Wieman.
\newblock {Measurement of parity nonconservation and an anapole moment in
  cesium}.
\newblock {\em Science}, 275:1759--1763, 1997.
\newblock \href {https://doi.org/10.1126/science.275.5307.1759}
  {\path{doi:10.1126/science.275.5307.1759}}.

\bibitem{Guena:2005uj}
Jocelyne Guena, Michel Lintz, and Marie-Anne Bouchiat.
\newblock {Atomic parity violation: Principles, recent results, present
  motivations}.
\newblock {\em Mod. Phys. Lett. A}, 20:375--390, 2005.
\newblock \href {http://arxiv.org/abs/physics/0503143}
  {\path{arXiv:physics/0503143}}, \href
  {https://doi.org/10.1142/S0217732305016853}
  {\path{doi:10.1142/S0217732305016853}}.

\bibitem{Toh:2019iro}
George Toh, Amy Damitz, Carol~E. Tanner, W.~R. Johnson, and D.~S. Elliott.
\newblock {Determination of the scalar and vector polarizabilities of the
  cesium $6s \ ^2S_{1/2} \rightarrow 7s \ ^2S_{1/2}$ transition and
  implications for atomic parity non-conservation}.
\newblock {\em Phys. Rev. Lett.}, 123(7):073002, 2019.
\newblock \href {http://arxiv.org/abs/1905.02768} {\path{arXiv:1905.02768}},
  \href {https://doi.org/10.1103/PhysRevLett.123.073002}
  {\path{doi:10.1103/PhysRevLett.123.073002}}.

\bibitem{Becker:2018ggl}
Dominik Becker et~al.
\newblock {The P2 experiment}.
\newblock {\em Eur. Phys. J. A}, 54(11):208, 2018.
\newblock \href {http://arxiv.org/abs/1802.04759} {\path{arXiv:1802.04759}},
  \href {https://doi.org/10.1140/epja/i2018-12611-6}
  {\path{doi:10.1140/epja/i2018-12611-6}}.

\bibitem{Wang:2014bba}
D.~Wang et~al.
\newblock {Measurement of parity violation in electron\textendash{}quark
  scattering}.
\newblock {\em Nature}, 506(7486):67--70, 2014.
\newblock \href {https://doi.org/10.1038/nature12964}
  {\path{doi:10.1038/nature12964}}.

\bibitem{Wang:2014guo}
D.~Wang et~al.
\newblock {Measurement of Parity-Violating Asymmetry in Electron-Deuteron
  Inelastic Scattering}.
\newblock {\em Phys. Rev. C}, 91(4):045506, 2015.
\newblock \href {http://arxiv.org/abs/1411.3200} {\path{arXiv:1411.3200}},
  \href {https://doi.org/10.1103/PhysRevC.91.045506}
  {\path{doi:10.1103/PhysRevC.91.045506}}.

\bibitem{Barzi:2022edg}
Emanuela Barzi, Simonetta Liuti, Christine Nattrass, Roxanne Springer, and
  Charles~H. Bennett.
\newblock {How Community Agreements Can Improve Workplace Culture in Physics}.
\newblock 9 2022.
\newblock \href {http://arxiv.org/abs/2209.06755} {\path{arXiv:2209.06755}}.

\bibitem{Diefenthaler2022}
Markus Diefenthaler, Abdullah Farhat, Andrii Verbytskyi, and Yuesheng Xu.
\newblock {Deeply learning deep inelastic scattering kinematics}.
\newblock {\em Eur. Phys. J. C}, 82(11):1064, 2022.
\newblock \href {http://arxiv.org/abs/2108.11638} {\path{arXiv:2108.11638}},
  \href {https://doi.org/10.1140/epjc/s10052-022-10964-z}
  {\path{doi:10.1140/epjc/s10052-022-10964-z}}.

\bibitem{ATLAS_Collaboration_2022}
ATLAS Collaboration.
\newblock Atlas software and computing hl-lhc roadmap.
\newblock Technical report, CERN, Geneva, 2022.
\newblock URL: \url{http://cds.cern.ch/record/2802918}.

\bibitem{boccali2019computing}
T~Boccali.
\newblock Computing models in high energy physics.
\newblock {\em Rev Phys}, 4:100034, 2019.
\newblock URL: \url{https://doi.org/10.1016/j.revip.2019.100034}.

\bibitem{future_trends_in_nuclear_physics_computing}
Future trends in nuclear physics computing.
\newblock URL: \url{https://www.jlab.org/FTNPC}.

\bibitem{Boehnlein2020}
Amber Boehnlein et~al.
\newblock {Colloquium: Machine learning in nuclear physics}.
\newblock {\em Rev. Mod. Phys.}, 94(3):031003, 2022.
\newblock \href {http://arxiv.org/abs/2112.02309} {\path{arXiv:2112.02309}},
  \href {https://doi.org/10.1103/RevModPhys.94.031003}
  {\path{doi:10.1103/RevModPhys.94.031003}}.

\bibitem{AI4EIC}
{A.I. For the Electron Ion Colider}.
\newblock URL: \url{https://eic.ai}.

\bibitem{CSSI}
{NSF Cyberinfrastruction for Sustained Scientific Innovation}.
\newblock URL:
  \url{https://beta.nsf.gov/funding/opportunities/cyberinfrastructure-sustained-scientific}.

\bibitem{AI_institutes}
{NSF National Artificial Intelligence Research Institutes}.
\newblock URL:
  \url{https://beta.nsf.gov/funding/opportunities/national-artificial-intelligence-research}.

\bibitem{Gavalian2020}
Polykarpos Thomadakis, Angelos Angelopoulos, Gagik Gavalian, and Nikos
  Chrisochoides.
\newblock {Using Machine Learning for Particle Track Identification in the
  CLAS12 Detector}.
\newblock 8 2020.
\newblock \href {http://arxiv.org/abs/2008.12860} {\path{arXiv:2008.12860}}.

\bibitem{Barsotti2020}
Rebecca Barsotti and Matthew~R. Shepherd.
\newblock Using machine learning to separate hadronic and electromagnetic
  interactions in the gluex forward calorimeter.
\newblock {\em Journal of Instrumentation}, 15(05):P05021, 2020.

\bibitem{Bielčíková2021}
J.~Bielčíková et~al.
\newblock {\em Journal of Instrumentation}, 16(03017), 2021.

\bibitem{Lee2022}
Kyle Lee, James Mulligan, Mateusz P\l{}osko\'n, Felix Ringer, and Feng Yuan.
\newblock {Machine learning-based jet and event classification at the
  Electron-Ion Collider with applications to hadron structure and spin
  physics}.
\newblock 10 2022.
\newblock \href {http://arxiv.org/abs/2210.06450} {\path{arXiv:2210.06450}}.

\bibitem{Fanelli2022a}
Cristiano Fanelli, J.~Giroux, and Z.~Papandreou.
\newblock {‘Flux+ Mutability’: a conditional generative approach to
  one-class classification and anomaly detection}.
\newblock {\em Machine Learning: Science and Technology}, 3(4):045012, 2022.

\bibitem{Andreassen2020}
Anders Andreassen, Patrick~T. Komiske, Eric~M. Metodiev, Benjamin Nachman, and
  Jesse Thaler.
\newblock {OmniFold: A Method to Simultaneously Unfold All Observables}.
\newblock {\em Phys. Rev. Lett.}, 124(18):182001, 2020.
\newblock \href {http://arxiv.org/abs/1911.09107} {\path{arXiv:1911.09107}},
  \href {https://doi.org/10.1103/PhysRevLett.124.182001}
  {\path{doi:10.1103/PhysRevLett.124.182001}}.

\bibitem{Andreev2020}
V.~Andreev et~al.
\newblock {Measurement of Lepton-Jet Correlation in Deep-Inelastic Scattering
  with the H1 Detector Using Machine Learning for Unfolding}.
\newblock {\em Phys. Rev. Lett.}, 128(13):132002, 2022.
\newblock \href {http://arxiv.org/abs/2108.12376} {\path{arXiv:2108.12376}},
  \href {https://doi.org/10.1103/PhysRevLett.128.132002}
  {\path{doi:10.1103/PhysRevLett.128.132002}}.

\bibitem{Fanelli2020}
Cristiano Fanelli and Jary Pomponi.
\newblock {DeepRICH: learning deeply Cherenkov detectors}.
\newblock {\em Machine Learning: Science and Technology}, 1(1):015010, 2020.

\bibitem{Aad2022}
G.~Aad et~al.
\newblock {AtlFast3: the next generation of fast simulation in ATLAS}.
\newblock {\em Computing and Software for Big Science}, 6(1):1--54, 2022.

\bibitem{Nachman2020}
Benjamin Nachman.
\newblock {A guide for deploying Deep Learning in LHC searches: How to achieve
  optimality and account for uncertainty}.
\newblock {\em SciPost Physics}, 8(6):090, 2020.

\bibitem{Schram2022}
Malachi Schram, Kishansingh Rajput, Karthik~Somayaji NS, Peng Li, Jason
  St.~John, and Himanshu Sharma.
\newblock {Uncertainty Aware ML-based surrogate models for particle
  accelerators: A Study at the Fermilab Booster Accelerator Complex}.
\newblock 9 2022.
\newblock \href {http://arxiv.org/abs/2209.07458} {\path{arXiv:2209.07458}}.

\bibitem{Kitouni2021}
Ouail Kitouni, Niklas Nolte, and Mike Williams.
\newblock {Robust and Provably Monotonic Networks}.
\newblock 11 2021.
\newblock \href {http://arxiv.org/abs/2112.00038} {\path{arXiv:2112.00038}}.

\bibitem{Cisbani2020}
Evaristo Cisbani et~al.
\newblock {AI-optimized detector design for the future Electron-Ion Collider:
  the dual-radiator RICH case}.
\newblock {\em Journal of Instrumentation}, 15(05):P05009, 2020.

\bibitem{Fanelli2022b}
Cristiano Fanelli.
\newblock Design of detectors at the electron ion collider with artificial
  intelligence.
\newblock {\em Journal of Instrumentation}, 17(04):C04038, 2022.

\bibitem{Liu2022}
Ming~Xiong Liu.
\newblock {Intelligent Experiment Through Real-Time AI: Fast Data Processing
  and Autonomous Detector Control for sPHENIX and Future EIC Detectors}.
\newblock Technical Report LA-UR-22-28083, Los Alamos National Lab., Los
  Alamos, NM, 2022.

\bibitem{Ameli2022}
F.~Ameli, M.~Battaglieri, V.V. Berdnikov, et~al.
\newblock {Streaming readout for next generation electron scattering
  experiments}.
\newblock {\em European Physical Journal Plus}, 137:958, 2022.
\newblock \href {https://doi.org/10.1140/epjp/s13360-022-03146-z}
  {\path{doi:10.1140/epjp/s13360-022-03146-z}}.

\bibitem{Jeske2022}
T.~Jeske et~al.
\newblock {AI for Experimental Controls at Jefferson Lab}.
\newblock {\em Journal of Instrumentation}, 17:C03043, 2022.

\bibitem{PhysRevAccelBeams.25.104604}
David Meier, Luis~Vera Ramirez, Jens V\"olker, Jens Viefhaus, Bernhard Sick,
  and Gregor Hartmann.
\newblock Optimizing a superconducting radio-frequency gun using deep
  reinforcement learning.
\newblock {\em Phys. Rev. Accel. Beams}, 25:104604, Oct 2022.
\newblock URL:
  \url{https://link.aps.org/doi/10.1103/PhysRevAccelBeams.25.104604}, \href
  {https://doi.org/10.1103/PhysRevAccelBeams.25.104604}
  {\path{doi:10.1103/PhysRevAccelBeams.25.104604}}.

\bibitem{PhysRevAccelBeams.25.104601}
Christoph Obermair, Thomas Cartier-Michaud, Andrea Apollonio, William Millar,
  Lukas Felsberger, Lorenz Fischl, Holger~Severin Bovbjerg, Daniel Wollmann,
  Walter Wuensch, Nuria Catalan-Lasheras, Mar\ifmmode
  \mbox{\c{c}}\else~\c{c}\fi{}\`a Boronat, Franz Pernkopf, and Graeme Burt.
\newblock Explainable machine learning for breakdown prediction in high
  gradient rf cavities.
\newblock {\em Phys. Rev. Accel. Beams}, 25:104601, Oct 2022.
\newblock URL:
  \url{https://link.aps.org/doi/10.1103/PhysRevAccelBeams.25.104601}, \href
  {https://doi.org/10.1103/PhysRevAccelBeams.25.104601}
  {\path{doi:10.1103/PhysRevAccelBeams.25.104601}}.

\bibitem{PhysRevAccelBeams.25.064402}
A.~H. Lumpkin, R.~Thurman-Keup, D.~Edstrom, P.~Prieto, J.~Ruan, B.~Jacobson,
  J.~Sikora, J.~Diaz-Cruz, A.~Edelen, and F.~Zhou.
\newblock Submacropulse electron-beam dynamics correlated with higher-order
  modes in a tesla-type cryomodule.
\newblock {\em Phys. Rev. Accel. Beams}, 25:064402, Jun 2022.
\newblock URL:
  \url{https://link.aps.org/doi/10.1103/PhysRevAccelBeams.25.064402}, \href
  {https://doi.org/10.1103/PhysRevAccelBeams.25.064402}
  {\path{doi:10.1103/PhysRevAccelBeams.25.064402}}.

\bibitem{PhysRevAccelBeams.25.014601}
Y.~Gao, W.~Lin, K.~A. Brown, X.~Gu, G.~H. Hoffstaetter, J.~Morris, and
  S.~Seletskiy.
\newblock Bayesian optimization experiment for trajectory alignment at the low
  energy rhic electron cooling system.
\newblock {\em Phys. Rev. Accel. Beams}, 25:014601, Jan 2022.
\newblock URL:
  \url{https://link.aps.org/doi/10.1103/PhysRevAccelBeams.25.014601}, \href
  {https://doi.org/10.1103/PhysRevAccelBeams.25.014601}
  {\path{doi:10.1103/PhysRevAccelBeams.25.014601}}.

\bibitem{PhysRevAccelBeams.24.114601}
Aashwin~Ananda Mishra, Auralee Edelen, Adi Hanuka, and Christopher Mayes.
\newblock Uncertainty quantification for deep learning in particle accelerator
  applications.
\newblock {\em Phys. Rev. Accel. Beams}, 24:114601, Nov 2021.
\newblock URL:
  \url{https://link.aps.org/doi/10.1103/PhysRevAccelBeams.24.114601}, \href
  {https://doi.org/10.1103/PhysRevAccelBeams.24.114601}
  {\path{doi:10.1103/PhysRevAccelBeams.24.114601}}.

\bibitem{PhysRevAccelBeams.24.104601}
Jason St.~John, Christian Herwig, Diana Kafkes, Jovan Mitrevski, William~A.
  Pellico, Gabriel~N. Perdue, Andres Quintero-Parra, Brian~A. Schupbach, Kiyomi
  Seiya, Nhan Tran, Malachi Schram, Javier~M. Duarte, Yunzhi Huang, and Rachael
  Keller.
\newblock Real-time artificial intelligence for accelerator control: A study at
  the fermilab booster.
\newblock {\em Phys. Rev. Accel. Beams}, 24:104601, Oct 2021.
\newblock URL:
  \url{https://link.aps.org/doi/10.1103/PhysRevAccelBeams.24.104601}, \href
  {https://doi.org/10.1103/PhysRevAccelBeams.24.104601}
  {\path{doi:10.1103/PhysRevAccelBeams.24.104601}}.

\bibitem{PhysRevAccelBeams.24.082802}
Louis Emery, Hairong Shang, Yipeng Sun, and Xiaobiao Huang.
\newblock Application of a machine learning based algorithm to online
  optimization of the nonlinear beam dynamics of the argonne advanced photon
  source.
\newblock {\em Phys. Rev. Accel. Beams}, 24:082802, Aug 2021.
\newblock URL:
  \url{https://link.aps.org/doi/10.1103/PhysRevAccelBeams.24.082802}, \href
  {https://doi.org/10.1103/PhysRevAccelBeams.24.082802}
  {\path{doi:10.1103/PhysRevAccelBeams.24.082802}}.

\bibitem{PhysRevAccelBeams.24.072802}
Adi Hanuka, X.~Huang, J.~Shtalenkova, D.~Kennedy, A.~Edelen, Z.~Zhang, V.~R.
  Lalchand, D.~Ratner, and J.~Duris.
\newblock Physics model-informed gaussian process for online optimization of
  particle accelerators.
\newblock {\em Phys. Rev. Accel. Beams}, 24:072802, Jul 2021.
\newblock URL:
  \url{https://link.aps.org/doi/10.1103/PhysRevAccelBeams.24.072802}, \href
  {https://doi.org/10.1103/PhysRevAccelBeams.24.072802}
  {\path{doi:10.1103/PhysRevAccelBeams.24.072802}}.

\bibitem{PhysRevAccelBeams.23.114601}
Chris Tennant, Adam Carpenter, Tom Powers, Anna Shabalina~Solopova, Lasitha
  Vidyaratne, and Khan Iftekharuddin.
\newblock Superconducting radio-frequency cavity fault classification using
  machine learning at jefferson laboratory.
\newblock {\em Phys. Rev. Accel. Beams}, 23:114601, Nov 2020.
\newblock URL:
  \url{https://link.aps.org/doi/10.1103/PhysRevAccelBeams.23.114601}, \href
  {https://doi.org/10.1103/PhysRevAccelBeams.23.114601}
  {\path{doi:10.1103/PhysRevAccelBeams.23.114601}}.

\bibitem{PhysRevAccelBeams.23.044601}
Auralee Edelen, Nicole Neveu, Matthias Frey, Yannick Huber, Christopher Mayes,
  and Andreas Adelmann.
\newblock Machine learning for orders of magnitude speedup in multiobjective
  optimization of particle accelerator systems.
\newblock {\em Phys. Rev. Accel. Beams}, 23:044601, Apr 2020.
\newblock URL:
  \url{https://link.aps.org/doi/10.1103/PhysRevAccelBeams.23.044601}, \href
  {https://doi.org/10.1103/PhysRevAccelBeams.23.044601}
  {\path{doi:10.1103/PhysRevAccelBeams.23.044601}}.

\bibitem{PhysRevAccelBeams.23.032805}
Xingyi Xu, Yimei Zhou, and Yongbin Leng.
\newblock Machine learning based image processing technology application in
  bunch longitudinal phase information extraction.
\newblock {\em Phys. Rev. Accel. Beams}, 23:032805, Mar 2020.
\newblock URL:
  \url{https://link.aps.org/doi/10.1103/PhysRevAccelBeams.23.032805}, \href
  {https://doi.org/10.1103/PhysRevAccelBeams.23.032805}
  {\path{doi:10.1103/PhysRevAccelBeams.23.032805}}.

\bibitem{PhysRevAccelBeams.22.093001}
Gabriella Azzopardi, Belen Salvachua, Gianluca Valentino, Stefano Redaelli, and
  Adrian Muscat.
\newblock Operational results on the fully automatic lhc collimator alignment.
\newblock {\em Phys. Rev. Accel. Beams}, 22:093001, Sep 2019.
\newblock URL:
  \url{https://link.aps.org/doi/10.1103/PhysRevAccelBeams.22.093001}, \href
  {https://doi.org/10.1103/PhysRevAccelBeams.22.093001}
  {\path{doi:10.1103/PhysRevAccelBeams.22.093001}}.

\bibitem{PhysRevAccelBeams.22.052801}
D.~Vilsmeier, M.~Sapinski, and R.~Singh.
\newblock Space-charge distortion of transverse profiles measured by
  electron-based ionization profile monitors and correction methods.
\newblock {\em Phys. Rev. Accel. Beams}, 22:052801, May 2019.
\newblock URL:
  \url{https://link.aps.org/doi/10.1103/PhysRevAccelBeams.22.052801}, \href
  {https://doi.org/10.1103/PhysRevAccelBeams.22.052801}
  {\path{doi:10.1103/PhysRevAccelBeams.22.052801}}.

\bibitem{PhysRevAccelBeams.21.112802}
C.~Emma, A.~Edelen, M.~J. Hogan, B.~O'Shea, G.~White, and V.~Yakimenko.
\newblock Machine learning-based longitudinal phase space prediction of
  particle accelerators.
\newblock {\em Phys. Rev. Accel. Beams}, 21:112802, Nov 2018.
\newblock URL:
  \url{https://link.aps.org/doi/10.1103/PhysRevAccelBeams.21.112802}, \href
  {https://doi.org/10.1103/PhysRevAccelBeams.21.112802}
  {\path{doi:10.1103/PhysRevAccelBeams.21.112802}}.

\bibitem{PhysRevAccelBeams.21.054601}
Yongjun Li, Weixing Cheng, Li~Hua Yu, and Robert Rainer.
\newblock Genetic algorithm enhanced by machine learning in dynamic aperture
  optimization.
\newblock {\em Phys. Rev. Accel. Beams}, 21:054601, May 2018.
\newblock URL:
  \url{https://link.aps.org/doi/10.1103/PhysRevAccelBeams.21.054601}, \href
  {https://doi.org/10.1103/PhysRevAccelBeams.21.054601}
  {\path{doi:10.1103/PhysRevAccelBeams.21.054601}}.

\bibitem{SBIRFY2022PhaseIRelease2}
{SBIR FY 2022 Phase I Release 2, version 5}.
\newblock \\
  \url{https://science.osti.gov/-/media/sbir/pdf/TechnicalTopics/FY22-Phase-I-Release-2-Combined-TopicsV512012021.pdf},
  2021.
\newblock Accessed: 2022-12-10.

\bibitem{SEABAIMLWorkingGroup2020}
{Preliminary Findings: SEAB AIML Working Group}.
\newblock \\
  \url{https://www.energy.gov/sites/prod/files/2020/04/f73/SEAB%20AI%20WG%20PRELIMI
  NARY%20FINDINGS_0.pdf}, 2020.
\newblock Accessed: 2022-12-10.

\bibitem{AIforNuclearPhysicsWorkshop2020}
{AI for Nuclear Physics Workshop}.
\newblock In {\em AI2020}, 2020.
\newblock URL: \url{https://www.jlab.org/conference/AI2020}.

\bibitem{AI4EICWorkshop}
{AI4EIC Workshop}.
\newblock In {\em AI4EIC}, 2022.
\newblock URL: \url{https://eic.ai/workshops}.

\bibitem{AIforNuclearPhysicsWinterSchool}
{AI for Nuclear Physics WinterSchool}.
\newblock URL: \url{https://www.jlab.org/remote-ai4np-winter-school}.

\bibitem{AI4EICHackathon}
{AI4EIC Hackathon}.
\newblock URL: \url{https://eic.ai/hackathons}.

\bibitem{AIHackathonJLab}
{AI Hackathon}.
\newblock URL: \url{https://www.jlab.org/ai-town-hall-0}.

\bibitem{GEANT4:2002zbu}
S.~Agostinelli et~al.
\newblock {GEANT4--a simulation toolkit}.
\newblock {\em Nucl. Instrum. Meth. A}, 506:250--303, 2003.
\newblock \href {https://doi.org/10.1016/S0168-9002(03)01368-8}
  {\path{doi:10.1016/S0168-9002(03)01368-8}}.

\bibitem{Ahdida:2022gjl}
C.~Ahdida et~al.
\newblock {New Capabilities of the FLUKA Multi-Purpose Code}.
\newblock {\em Front. in Phys.}, 9:788253, 2022.
\newblock \href {https://doi.org/10.3389/fphy.2021.788253}
  {\path{doi:10.3389/fphy.2021.788253}}.

\bibitem{Tepel:1984}
J.W. Tepel.
\newblock Ensdf - the evaluated nuclear structure data file.
\newblock {\em Computer Physics Communications}, 33(1):129--146, 1984.
\newblock URL:
  \url{https://www.sciencedirect.com/science/article/pii/0010465584901152},
  \href {https://doi.org/https://doi.org/10.1016/0010-4655(84)90115-2}
  {\path{doi:https://doi.org/10.1016/0010-4655(84)90115-2}}.

\bibitem{NNDC}
National nuclear data center.
\newblock URL: \url{https://www.nndc.bnl.gov/}.

\bibitem{Plompen:2020due}
A.~J.~M. Plompen et~al.
\newblock {The joint evaluated fission and fusion nuclear data library,
  JEFF-3.3}.
\newblock {\em Eur. Phys. J. A}, 56(7):181, 2020.
\newblock \href {https://doi.org/10.1140/epja/s10050-020-00141-9}
  {\path{doi:10.1140/epja/s10050-020-00141-9}}.

\bibitem{Chadwick:2011xwu}
M.~B. Chadwick et~al.
\newblock {ENDF/B-VII.1 Nuclear Data for Science and Technology: Cross
  Sections, Covariances, Fission Product Yields and Decay Data}.
\newblock {\em Nucl. Data Sheets}, 112(12):2887--2996, 2011.
\newblock \href {https://doi.org/10.1016/j.nds.2011.11.002}
  {\path{doi:10.1016/j.nds.2011.11.002}}.

\bibitem{Koning:2019qbo}
A.~J. Koning, D.~Rochman, J.~Ch. Sublet, N.~Dzysiuk, M.~Fleming, and S.~van~der
  Marck.
\newblock {TENDL: Complete Nuclear Data Library for Innovative Nuclear Science
  and Technology}.
\newblock {\em Nucl. Data Sheets}, 155:1--55, 2019.
\newblock \href {https://doi.org/10.1016/j.nds.2019.01.002}
  {\path{doi:10.1016/j.nds.2019.01.002}}.

\bibitem{Arndt:2007qn}
R.~A. Arndt, W.~J. Briscoe, I.~I. Strakovsky, and R.~L. Workman.
\newblock {Updated analysis of NN elastic scattering to 3-GeV}.
\newblock {\em Phys. Rev. C}, 76:025209, 2007.
\newblock \href {http://arxiv.org/abs/0706.2195} {\path{arXiv:0706.2195}},
  \href {https://doi.org/10.1103/PhysRevC.76.025209}
  {\path{doi:10.1103/PhysRevC.76.025209}}.

\bibitem{ALICE:2022zuz}
Shreyasi Acharya et~al.
\newblock {First measurement of the absorption of $^{3}\overline{\rm He}$
  nuclei in matter and impact on their propagation in the galaxy}.
\newblock 2 2022.
\newblock \href {http://arxiv.org/abs/2202.01549} {\path{arXiv:2202.01549}}.

\bibitem{Nepali:2006ep}
C.~Nepali, G.~Fai, and D.~Keane.
\newblock {Advantage of U+U over Au+Au collisions at constant beam energy}.
\newblock {\em Phys. Rev. C}, 73:034911, 2006.
\newblock \href {https://doi.org/10.1103/PhysRevC.73.034911}
  {\path{doi:10.1103/PhysRevC.73.034911}}.

\bibitem{Badhwar:1992}
G.~D. Badhwar and P.~M. O'Neill.
\newblock {An improved model of galactic cosmic radiation for space exploration
  missions}.
\newblock {\em Int. J. Rad. Applications Instrumentation. Part D. Nuclear
  Tracks and Radiation Measurements}, 20:403--410, 1992.

\bibitem{Grupen:2003ih}
C.~Grupen et~al.
\newblock {Measurements of the muon component of extensive air showers at 320
  m.w.e. underground}.
\newblock {\em Nucl. Instrum. Meth. A}, 510:190--193, 2003.
\newblock \href {https://doi.org/10.1016/S0168-9002(03)01697-8}
  {\path{doi:10.1016/S0168-9002(03)01697-8}}.

\bibitem{Ridky:2005mx}
J.~Ridky and P.~Travnicek.
\newblock {Detection of muon bundles from cosmic ray showers by the DELPHI
  experiment}.
\newblock {\em Nucl. Phys. B Proc. Suppl.}, 138:295--298, 2005.
\newblock \href {https://doi.org/10.1016/j.nuclphysbps.2004.11.066}
  {\path{doi:10.1016/j.nuclphysbps.2004.11.066}}.

\bibitem{L3:2004sed}
P.~Achard et~al.
\newblock {Measurement of the atmospheric muon spectrum from 20-GeV to
  3000-GeV}.
\newblock {\em Phys. Lett. B}, 598:15--32, 2004.
\newblock \href {http://arxiv.org/abs/hep-ex/0408114}
  {\path{arXiv:hep-ex/0408114}}, \href
  {https://doi.org/10.1016/j.physletb.2004.08.003}
  {\path{doi:10.1016/j.physletb.2004.08.003}}.

\bibitem{ALICE:2015wfa}
Jaroslav Adam et~al.
\newblock {Study of cosmic ray events with high muon multiplicity using the
  ALICE detector at the CERN Large Hadron Collider}.
\newblock {\em JCAP}, 01:032, 2016.
\newblock \href {http://arxiv.org/abs/1507.07577} {\path{arXiv:1507.07577}},
  \href {https://doi.org/10.1088/1475-7516/2016/01/032}
  {\path{doi:10.1088/1475-7516/2016/01/032}}.

\bibitem{FernandezTellez:2007yxa}
Arturo Fernandez~Tellez.
\newblock {ACORDE, The ALICE Cosmic Ray Detector}.
\newblock In {\em {30th International Cosmic Ray Conference}}, volume~5, pages
  1201--1204, 7 2007.

\bibitem{Finckenor:2018}
Miria~M. Finckenor.
\newblock {\em Materials for Spacecraft}, chapter~6, pages 403--434.
\newblock American Inst. Aeronautics Astronautics, 2018.
\newblock URL:
  \url{https://arc.aiaa.org/doi/abs/10.2514/5.9781624104893.0403.0434}, \href
  {http://arxiv.org/abs/https://arc.aiaa.org/doi/pdf/10.2514/5.9781624104893.0403.0434}
  {\path{arXiv:https://arc.aiaa.org/doi/pdf/10.2514/5.9781624104893.0403.0434}},
  \href {https://doi.org/10.2514/5.9781624104893.0403.0434}
  {\path{doi:10.2514/5.9781624104893.0403.0434}}.

\bibitem{Hoeffgen:2020}
Stefan~K. Höeffgen, Stefan Metzger, and Michael Steffens.
\newblock Investigating the effects of cosmic rays on space electronics.
\newblock {\em Frontiers in Physics}, 8, 2020.
\newblock URL:
  \url{https://www.frontiersin.org/articles/10.3389/fphy.2020.00318}, \href
  {https://doi.org/10.3389/fphy.2020.00318}
  {\path{doi:10.3389/fphy.2020.00318}}.

\bibitem{Durante:2011}
Marco Durante and Francis~A. Cucinotta.
\newblock Physical basis of radiation protection in space travel.
\newblock {\em Rev. Mod. Phys.}, 83:1245--1281, Nov 2011.
\newblock URL: \url{https://link.aps.org/doi/10.1103/RevModPhys.83.1245}, \href
  {https://doi.org/10.1103/RevModPhys.83.1245}
  {\path{doi:10.1103/RevModPhys.83.1245}}.

\bibitem{45}
Workshop for applied nuclear data activities (wanda 2022), 2022.
\newblock URL:
  \url{https://conferences.lbl.gov/event/880/contributions/5586/attachments/3823/
  3098/ND_at_RHIC_WANDA2022_Cebra.pdf}.

\bibitem{Norbury:2021coa}
John~W. Norbury.
\newblock {Double-Differential FRaGmentation (DDFRG) models for proton and
  light ion production in high energy nuclear collisions}.
\newblock {\em Nucl. Instrum. Meth. A}, 986:164681, 2021.
\newblock \href {https://doi.org/10.1016/j.nima.2020.164681}
  {\path{doi:10.1016/j.nima.2020.164681}}.

\bibitem{Hufner:1975zz}
J.~Hufner, K.~Schafer, and B.~Schurmann.
\newblock {Abrasion-ablation in reactions between relativistic heavy ions}.
\newblock {\em Phys. Rev. C}, 12:1888--1898, 1975.
\newblock \href {https://doi.org/10.1103/PhysRevC.12.1888}
  {\path{doi:10.1103/PhysRevC.12.1888}}.

\bibitem{Werneth:2021}
C.M. Werneth, W.C. {de Wet}, L.W. Townsend, K.M. Maung, J.W. Norbury, T.C.
  Slaba, R.B. Norman, S.R. Blattnig, and W.P. Ford.
\newblock Relativistic abrasion–ablation de-excitation fragmentation
  (raadfrg) model.
\newblock {\em Nuclear Instruments and Methods in Physics Research Section B:
  Beam Interactions with Materials and Atoms}, 502:118--135, 2021.
\newblock URL:
  \url{https://www.sciencedirect.com/science/article/pii/S0168583X21002275},
  \href {https://doi.org/https://doi.org/10.1016/j.nimb.2021.06.016}
  {\path{doi:https://doi.org/10.1016/j.nimb.2021.06.016}}.

\bibitem{Luoni:2021bne}
F.~Luoni et~al.
\newblock {Total nuclear reaction cross-section database for radiation
  protection in space and heavy-ion therapy applications}.
\newblock {\em New J. Phys.}, 23(10):101201, 2021.
\newblock \href {http://arxiv.org/abs/2105.11981} {\path{arXiv:2105.11981}},
  \href {https://doi.org/10.1088/1367-2630/ac27e1}
  {\path{doi:10.1088/1367-2630/ac27e1}}.

\bibitem{Sombun:2018yqh}
Sukanya Sombun, Kristiya Tomuang, Ayut Limphirat, Paula Hillmann, Christoph
  Herold, Jan Steinheimer, Yupeng Yan, and Marcus Bleicher.
\newblock {Deuteron production from phase-space coalescence in the UrQMD
  approach}.
\newblock {\em Phys. Rev. C}, 99(1):014901, 2019.
\newblock \href {http://arxiv.org/abs/1805.11509} {\path{arXiv:1805.11509}},
  \href {https://doi.org/10.1103/PhysRevC.99.014901}
  {\path{doi:10.1103/PhysRevC.99.014901}}.

\bibitem{E-802:1991unu}
T.~Abbott et~al.
\newblock {Measurement of particle production in proton induced reactions at
  14.6-GeV/c}.
\newblock {\em Phys. Rev. D}, 45:3906--3920, 1992.
\newblock \href {https://doi.org/10.1103/PhysRevD.45.3906}
  {\path{doi:10.1103/PhysRevD.45.3906}}.

\bibitem{Kolos:2022stv}
Karolina Kolos et~al.
\newblock {Current nuclear data needs for applications}.
\newblock {\em Phys. Rev. Res.}, 4(2):021001, 2022.
\newblock \href {https://doi.org/10.1103/PhysRevResearch.4.021001}
  {\path{doi:10.1103/PhysRevResearch.4.021001}}.

\bibitem{Smith:2022xyz}
M.~S. Smith, R.~Vogt, and K.~LaBel.
\newblock Nuclear data for high energy ion interactions and secondary particle
  production, 2022.
\newblock \href {https://doi.org/10.2172/1883853} {\path{doi:10.2172/1883853}}.

\end{thebibliography}

\end{document}